\documentclass[12pt]{report}
\usepackage[a4paper,width=150mm,top=25mm,bottom=35mm, footskip=15mm]{geometry}
\usepackage{slashed}
\usepackage{bm}
\usepackage{dcolumn}
\usepackage{multirow}
\usepackage{ifpdf}
\usepackage{url}
\usepackage{util}
\usepackage{hic}
\usepackage{here}
\usepackage{graphicx}
\usepackage{indentfirst}
\usepackage{listings}

\usepackage{sectsty}
\usepackage{ifthen}
\usepackage{kpfonts}
\usepackage{tgtermes}
\usepackage{amsmath,amssymb}
\usepackage{fancyhdr}
\usepackage{titletoc}
\usepackage{xcolor}
\usepackage{color}

\usepackage{mathtools}
\definecolor{myblue}{rgb}{0.27, 0.37, 0.39}
\definecolor{myheadcol}{rgb}{0.87, 0.87, 0.87}

\definecolor{codegreen}{rgb}{0,0.6,0}
\definecolor{codepastelred}{rgb}{1.0, 0.41, 0.38}
\definecolor{codegray}{rgb}{0.5,0.5,0.5}
\definecolor{codepurple}{rgb}{0.58,0,0.82}
\definecolor{codeolivine}{rgb}{0.6, 0.73, 0.45}
\definecolor{background}{rgb}{1.0, 0.94, 0.84}
\lstdefinestyle{mystyle}{
language=C++,
backgroundcolor=\color{background},
commentstyle=\color{codepastelred},
keywordstyle=\color{codepastelred},
numberstyle=\tiny\color{codegray},
stringstyle=\color{codeolivine},
basicstyle=\fontfamily{pxtt}\selectfont\footnotesize,
breakatwhitespace=false,
breaklines=true,
captionpos=b,
keepspaces=true,
numbers=left,
numbersep=2pt,
showspaces=false,
showstringspaces=false,
showtabs=false,
tabsize=3,
}

\titlecontents{section}[3.8em]{\bfseries\color{myblue}}{\contentslabel{2.3em}}{\hspace*{-2.3em}}{\hfill\contentspage}
\dottedcontents{subsection}[3.8em]{}{3.2em}{1pc}
\subsectionfont{\color{myblue}}  % sets colour of sections

\usepackage{caption}
\renewcommand{\headrulewidth}{1.5pt}

\renewcommand{\headrule}{\hbox to\headwidth{%
  \color{myheadcol}\leaders\hrule height \headrulewidth\hfill}}

\title{
\vspace{30pt}

\LARGE{Unified description of high-energy nuclear collisions based on dynamical core--corona picture}\\
\LARGE{}

\vspace{20pt}

\Large{Yuuka Kanakubo}\\

\vspace{20pt}

{\large Ph.D. in Physics}\\
{\large Sophia University}

}

\date{\today}

\begin{document}

\maketitle

\chapter*{Abstract}
One of the primary goals of nuclear/elementary particle physicists is to fully understand the properties of a matter governed by Quantum Chromodynamics (QCD).
Towards the primary goal, the properties of a state called Quark-Gluon Plasma (QGP), the many-body system of quarks and gluons under thermal and chemical equilibrium at high- temperature and baryon density regime of the QCD phase diagram, have been explored through high-energy heavy-ion collision experiments.
The Large Hadron Collider (LHC) at CERN in-between France and Switzerland and the Relativistic Heavy Ion Collider (RHIC) at Brookhaven National Laboratory (BNL) in New York, USA, are the two main facilities carrying out the high-energy heavy-ion collision experiment. 
The collision energy is reached up to $\sim 10^{-1}-10^{1}$ TeV in the center-of-mass of nucleons at RHIC and LHC. 
Such highly accelerated nuclei induce the many-body scatterings and productions of quarks and gluons in a collision. Once they become a locally equilibrated state, the state can be called the QGP.

In order to understand properties of the QGP from heavy-ion collision experimental data, multi-stage dynamical frameworks based on relativistic hydrodynamics have been established and utilized as a powerful tool.
This is because a high-energy heavy-ion collision is a dynamical reaction, and the state of the QGP is transiently achieved in the middle of the reaction.
In order to build a theoretical framework that is capable of being compared with experimental data, one needs to describe all the stages of the reaction, such as a collision of nuclei, evolution of the QGP, and evolution of hadron gas.
The multi-stage dynamical frameworks are nowadays regarded as a standard model of relativistic heavy-ion collisions
and widely used to extract information on the QGP from comparisons with experimental data.

Despite this conventional picture that the QGP is formed in relativistic heavy-ion collisions ($A$+$A$),
recent experimental data suggest that there is a possibility of the QGP formation in proton+proton ($p$+$p$) or proton-nucleus collisions ($p$+$A$).
Motivated by this, it becomes necessary to extend the applicability of the multi-stage dynamical framework to collisions of smaller colliding system sizes compared to heavy ions.
The key experimental data for the extended framework building is the particle yield ratio as a function of multiplicity.
The ALICE Collaboration, an experimental group at the LHC, reported that strange hadron yields relative to those of pions show enhancement with increasing multiplicity even in $p$+$p$ collisions, which smoothly connected to the results obtained in $Pb$+$Pb$ collisions where the QGP is regarded to be formed.
From this surprising result, it can be interpreted as follows: in averaged-multiplicity events of $p$+$p$ collisions, particle production would be dominated by the one of non-equilibrated matter in a vacuum.
However, if one sees particle productions indifferent multiplicity,
more particles are produced from equilibrated matter in higher-multiplicity events.

Based on this picture, I extend the hydro-based framework {\it{partially}} incorporating non-equilibrated components. 
It has been widely accepted that relativistic hydrodynamics well describes the dynamics of the QGP at low transverse momentum ($p_T$) regimes in heavy-ion collisions. In contrast, particle productions in small colliding systems have been studied through QCD-motivated
phenomenological models such as perturbative QCD (semi-)hard processes followed
by string fragmentation. The smooth connection of relativistic hydrodynamics and QCD-motivated phenomenological framework by keeping these pictures in each regime is indispensable to exploring the dynamics of the wide range of colliding systems.

In this extension of the hydro-based framework, 
I also aim to reconcile one of the major issues that inhere in conventional hydro-based dynamical models, 
which is the breakdown of the energy and momentum conservation of the entire colliding system including low to high momentum particle productions.
In conventional hydro-based dynamical models, usually initial conditions of hydrodynamic equations are parametrized so that multiplicity is described by the models.
However, as Monte-Carlo (MC) event generators respect the conservation of total energy and momentum of the system,
initial conditions of hydrodynamics should be put with respecting that.

In this thesis, I establish the updated version of the dynamical core--corona initialization framework (DCCI2) as an MC event generator based on relativistic hydrodynamics to simulate high-energy nuclear collisions from $p$+$p$ to $A$+$A$ collisions. 
In DCCI, QGP fluids are generated from initial partons considering the total energy and momentum of incoming nuclei. We phenomenologically and dynamically describe the formation of QGP from the initial partons based on the so-called core--corona picture.
Partons with sufficient secondary scatterings tend to generate QGP fluids (core) as equilibrated matter. On the other hand, partons with insufficient secondary scatterings tend to
survive as non-equilibrated matter (corona). By treating both locally equilibrated QGP fluids and non-equilibrated matter, the DCCI, as a hydro-based Monte Carlo event generator, is capable of describing from low to high transverse momentum, $p_T$, from backward to forward rapidity, and from small to large colliding systems.
The update from DCCI to DCCI2 includes sophistication of four-momentum deposition of initial partons in dynamical core--corona initialization, samplings of hadrons from hypersurface of fluids with \ISthreeD,
hadronic afterburner for final hadrons from core and corona with a hadronic transport model \jam,
and modification on color string structures due to the co-existence with fluids in coordinate space.

With DCCI2 established in this Ph.D. thesis,
I perform analysis roughly dividing into three sections: particle yields (momentum integrated particle distributions), momentum distributions, and anisotropic flows.
In the followings, I highlight the main results shown in this thesis.

{\it{Particle yields --}} Major parameters in DCCI2 are determined via multi-strange particle yields compared to charged pions as a function of multiplicity.
The fractions of core and corona components in the final hadronic productions are extracted as functions of multiplicity from $p$+$p$ at \snn[proton] = 7 and 13 TeV and $Pb$+$Pb$ collisions at \snn = 2.76 TeV.
I found that the core components become dominant at \dndeta $\approx20$, which roughly corresponds to the highest multiplicity classes in $p$+$p$ collisions,
and $\sim80\%$ centrality class in $Pb$+$Pb$ collisions.

{\it{Momentum distributions --}} The interplay between core and corona components is investigated by showing each contribution as a function of $p_T$.
Non-trivial interplay is seen: there is a tendency that a fraction of corona components enhances at very low $p_T$ such as $p_T<1$ GeV and
core contribution dominates the hadronic productions in intermediate $p_T$, which is around $1<p_T<3$ GeV.
The former corona enhancement originates from the feed-down of partons from intermediate to high $p_T$ to the very low $p_T$ due to the energy and momentum deposition in dynamical initialization.
On the other hand, the latter originates from the radial flow in the core components.
Centrality classification and particle identification reveal that the above tendency becomes 
clearer in more central events and/or in heavier particles likewise protons.
The comparisons between DCCI2 results and experimental data are made to show the possibility of existence of the corona components at very low $p_T$.
The results show that the slopes of $p_T$ spectra at the very low $p_T$ observed in experiments cannot be reproduced only with core components
with hadronic rescatterings, and the full simulation results of DCCI2 show better description by including the corona component.

{\it{Anisotropic flows --}}
The effects of the co-existence of core and corona components are investigated via anisotropic flows. Especially, the correction due to the existence of the corona components at the very low $p_T$ regime is explored in $Pb$+$Pb$ collision results.
The second order of anisotropic flow coefficients, $\vtwtw$, as a function of multiplicity shows $\approx15$-$30\%$ of dilution due to the existence of corona components below $\approx400$.
I also show the results of the second-order four-particle cumulant, $\ctwofour$, 
as a function of multiplicity in $Pb$+$Pb$ collisions.
The results show that $\ctwofour$ obtained only from core components is diluted by the existence of the corona components while the corona components show zero-consistent $\ctwofour$.
To investigate the origin of anisotropic flows in small systems, ridge structures are explored in $p$+$p$ collisions with the two different events classification and kinematics used in the ALICE and CMS experimental data where the ridge structure is observed.
The ridge-like structure is seen in high-multiplicity events originating from the core components in the method used in the ALICE experiment.
However, the ridge structure seen in the CMS experiment is not described with DCCI2 results.

All the results obtained in this thesis indicate that
both equilibrated matter (core) and non-equilibrated matter (corona) 
exist in final hadronic productions in both $p$+$p$ and $A$+$A$ collisions.
Hence, one should interpret experimental data of $p$+$p$ and $A$+$A$ collisions
by considering both core and corona components
in order to understand the QCD dynamics behind them and extract the properties of the QGP.
Although there is still room for more improvement in DCCI2, such as the introduction of viscosity and sophisticated jet quenching mechanism for a better description of experimental data,
dynamical modelings containing the core--corona picture such as DCCI2 could become the next-generation model inevitably
needed for the precision study of the QGP properties.

\chapter*{Dedication}
Dedicated to the memory of my grandpa, Iwazo Kanakubo.
\begin{figure}
    \centering
    \includegraphics[bb=0 0 595 842, width=0.7\textwidth, angle=90]{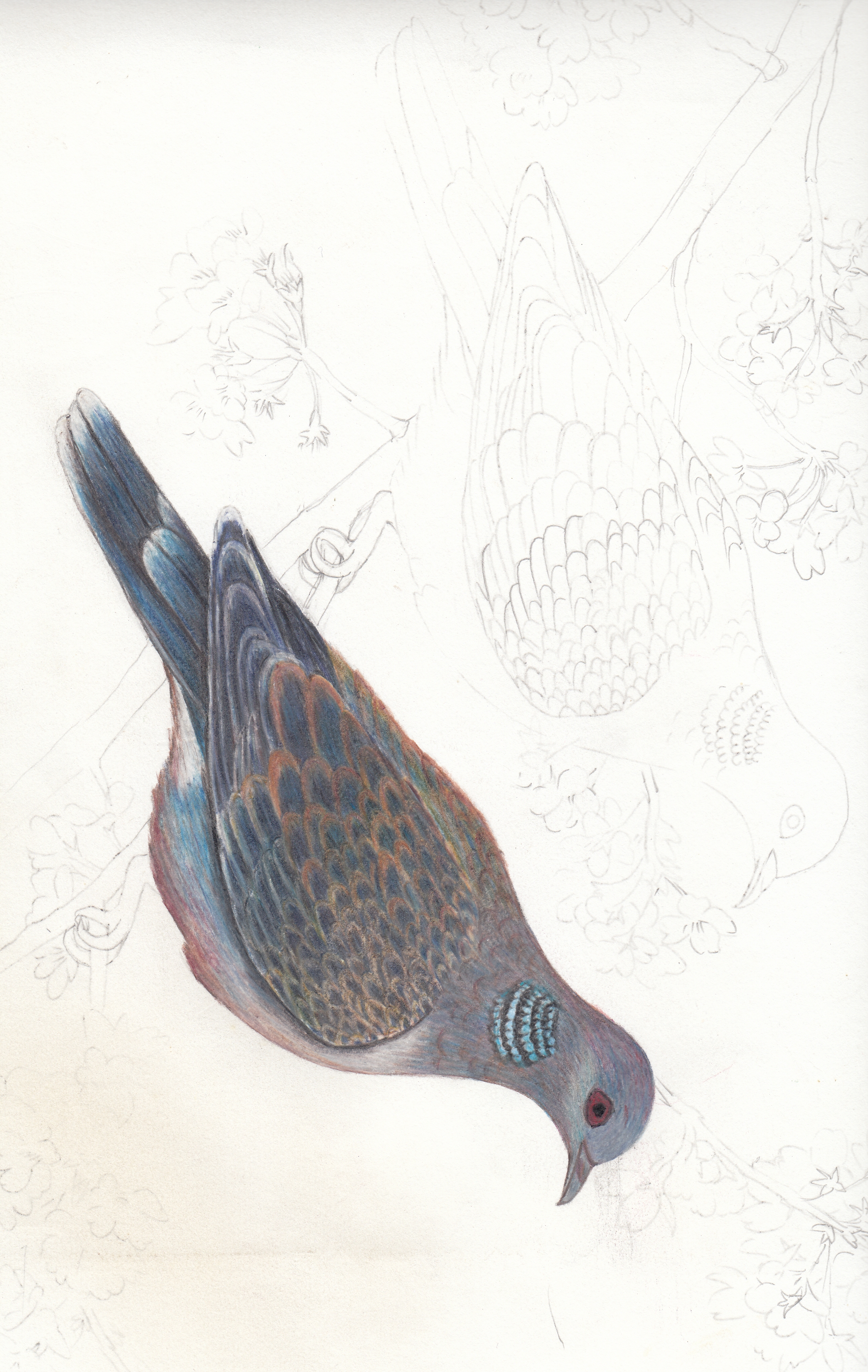}
    \caption*{A turtle dove with sakura drawn by my grandpa.}
    \label{fig:Abird}
\end{figure}

%\chapter*{Declaration}
%I declare that..

\chapter*{Acknowledgements}

First of all, my biggest thank-you is to my supervisor, Tetsufumi Hirano,
who makes me dive into such an exciting research field.
His awesome insight which extracts an essential physics from experimental data or theory
inspires me a lot!!
It was a lot of fun for me to have discussions/small chats that I've had with Tetsu on new experimental data or theoretical calculations popping up on conferences or arXiv.
%Without the millions of discussions/small chats with him,
%I couldn't ``steal'' his such an insight (still trying to do so though).
Also, he has strongly believed me 
that I am able to succeed as a researcher at some day.
Because of that, I am here finishing my Ph.D. and am about to start a new career. Thank you.

I would also like to thank Yasushi Nara, a professor at Akita International University for serving as my Ph.D. dissertation committee member and for discussions on developing the DCCI2.
As a long-term collaborator of Tetsu,
thankfully, I recently have opportunities to have discussions
and to work together with Yasushi.
His powerful and moving-forward research style is exactly what I aim to have!!

I could not finish my Ph.D. without enormous help of professors at Sophia physics theory group, Kazuo Takayanagi and Tomi Ohtsuki, especially for being my Ph.D. dissertation committee member.
While the areas of expertise are different,
both two have refereed my work from the bachelor, master, and to doctoral dissertations.
There is a big thank-you to two of you for being a part of my long journey and for taking care of me as a science and engineering (under)graduate student at Sophia University.

It's super lucky for me to have Yasuki Tachibana, who is an associate professor at Akita International University, as my collaborator. Thank you for guiding me to
my results with tons of insightful and fruitful comments as a collaborator, and not to mention, for being an excellent senpai!!
Yasuki is also one of them who makes me want to survive in academia as a researcher like him, so thank you so much.

Needless to say, I would like to thank all of my colleagues at the Hirano lab. I got inspired by great seminars or presentations by several students.
Moreover, there are a lot of awesome professors, postdocs, or students all over the world who are committing to heavy-ion research and whom I met at workshops or conferences.
I want to say here a lot of thanks to them
for keeping me inspired and encouraged.

Finally, also cannot forget to say, my work has been partly supported by JSPS KAKENHI Grant No.~20J20401.

\tableofcontents

\chapter{Introduction}
\thispagestyle{fancy}

\section{Quark-gluon plasma (QGP)}
{\it{Quark-gluon plasma (QGP)}} \cite{Shuryak:1977ut} is a state in the {\it{QCD (quantum chromodynamics) phase diagram}} \cite{Cabibbo:1975ig} which is characterized by macroscopic variables such as temperature and baryon chemical potential.
Thus, the QGP is an equilibrium state of a many-body system of quarks and gluons.
It is proposed that the QGP phase is achieved in the high-temperature \cite{Cabibbo:1975ig} and/or high-density \cite{Collins:1974ky} regimes in the diagram.
Elucidating the nature of QGP is the ultimate goal in particle and nuclear physics, and has been studied through relativistic heavy-ion collision experiments as well as theoretical interpretations and predictions.
While thousands of studies have been conducted towards revealing the properties of matter governed by QCD, only a tip of the iceberg has been understood. I summarize the past achievement and the current status of studies on relativistic heavy-ion collisions in Sec.~\ref{sec:INTRO_TopicalReviewOnQGPStudies} of this chapter.

In the sections below, I briefly explain some basic characteristics of the QGP as a state of QCD phase diagram.

\subsection{Symmetry}
States of matter are classified according to the symmetry that they have in their structures. In other words, a phase diagram is a map of phases classified by symmetries.
To understand the QGP from the perspective of the phase diagram, I briefly explain symmetries lying on QCD.
There are two types of symmetries in QCD: local and global symmetries. The local symmetry is the gauge symmetry $\rm{SU_c(3)}$.
On the other hand, the global symmetries are the chiral symmetry $\rm{SU_R(2)\times SU_L(2)}$,
and the baryon number symmetry $\rm{U}_B(1)$.
In the following sections, I briefly mention the {\it{gauge symmetry}} and {\it{chiral symmetry}}, which can be 
especially important to capture the characteristics of the QGP.
Note that there are more symmetries in QCD, such as Lorentz symmetry, spin symmetry, isospin ($u$ and $d$) symmetry, or flavor symmetry.
Some of them are strict and others are approximate: for example, Lorentz, spin, and gauge symmetries are exact, while isospin, flavor, and chiral symmetries are approximate ones
because of the finite quark mass.

\subsubsection{Gauge symmetry}
A theory of strong interaction which has the symmetry of $\mathrm{SU_c(3)}$ is called QCD, where $3$ is the number of the fundamental representation of the color of quarks such as $(R,G,B)$
\footnote{The color symmetry can be generally expressed as $\mathrm{SU_c}(N_c)$ where $N_c$ is the degree of freedom of the color, which can be larger than 3.}.
This internal degree of freedom was introduced so that baryons can be formed with three quarks in an $s$-wave state.
The Lagrangian density of the classical QCD is given by
\begin{align}
\label{eq:LagrangianQCD}
    \mathcal{L} &= \bar{q}^{\alpha i} \left( i \slashed{\mathcal{D}}_{\alpha \beta ij} - \bm{m} \right) q^{\beta j} -\frac{1}{4}F^{ai}_{\mu\nu} F_{ai}^{\mu\nu},
\end{align}
where $\mathcal{D}_\mu$ and $F^a_{\mu\nu}$ are covariant derivative and field strength given as
\begin{align}
\label{eq:CovariantDerivativeQCD}
    \mathcal{D}_\mu &\equiv \partial_\mu + igt^aA^a_\mu, \\
\label{eq:F_QCD}
    F^a_{\mu\nu} &=\partial_\mu A^a_\nu -\partial_\nu A^a_\mu + gf_{abc} A^b_\mu A^c_\nu. 
\end{align}
In Eq.~\refbra{eq:LagrangianQCD}, the first term sandwiched with quark fields, $q^{\alpha i}$, is the fermion itself and the interaction term of Fermion and gauge fields, and the second term is the kinetic term of gauge fields. Here, $g$ is the dimensionless coupling constant in QCD.
The indices $i, j \cdots$ of the quark and gluon fields denotes flavors.
With today's standard model, where six flavors appear as the internal degree of freedom of a quark, the quark field contains six components as $q^i=^t(u,d,s,c,b,t)$. At the same time, the corresponding mass matrix can be expressed as $\bm{m}=\diag(m_u, m_d, m_s, m_c, m_b, m_t)$.
The indices $\alpha, \beta \cdots$ of the quark field, $q^{\alpha}$, and $a, b, \cdots$ of the gluon field, $A^a_\mu$, correspond to the degree of freedom of the color. The colors of quarks and gluons are expressed as the fundamental and adjoint representation of $\mathrm{SU_c}(3)$, respectively.
Hence, each index of $\alpha, \beta, \cdots$ runs from 1 to 3, and $a, b, \cdots$ runs from 1 to 8 which originates from the degree of freedom of the adjoint representation of $\mathrm{SU_c}(3)$, {\it{i.e.,}} $3^2-1$.
Then, the full expressions of the quark field can be explicitly represented as,
\begin{align}
\label{eq:quark_f}
q^{\alpha i}&=^t(u^\alpha,d^\alpha,s^\alpha,c^\alpha,b^\alpha,t^\alpha), \\
\label{eq:quark_c}
u^\alpha = ^t(u_R, u_G, u_B), \hspace{10pt}& d^\alpha = ^t(d_R, d_G, d_B), \hspace{10pt} s^\alpha = ^t(s_R, s_G, s_B), \cdots,
\end{align}
where each component of Eq.~\refbra{eq:quark_c} is a Dirac spinor.
The corresponding mass matrix is,
\begin{align}
 \bm{m}=
\begin{pmatrix}
 \bm{m_u} & 0 & 0 & \cdots \\
 0 & \bm{m_d} & 0 & \cdots \\
 0 & 0   & \bm{m_s}  & \cdots \\
 \vdots & \vdots & \vdots & \ddots
\end{pmatrix}
,
\end{align}
where each diagonal component is
\begin{align}
    \bm{m_u} = 
    \begin{pmatrix}
   m_{u_R} & 0 & 0 \\
   0 & m_{u_G} & 0 \\
   0 & 0 & m_{u_B}
    \end{pmatrix},
%    \vspace{10pt}
 \enskip
    \bm{m_d} = 
    \begin{pmatrix}
   m_{d_R} & 0 & 0 \\
   0 & m_{d_G} & 0 \\
   0 & 0 & m_{d_B}
    \end{pmatrix},
%    \vspace{10pt}
\enskip
    \bm{m_s} = 
    \begin{pmatrix}
   m_{s_R} & 0 & 0 \\
   0 & m_{s_G} & 0 \\
   0 & 0 & m_{s_B}
    \end{pmatrix} \cdots
\end{align}
Note that the indices of flavors are eliminated in Eq.~\refbra{eq:CovariantDerivativeQCD} and \refbra{eq:F_QCD} to avoid busy notations.

The subscript $\mu$ is the index of the Lorentz matrix which runs from 0 to 3.
The $t^a$ and $f_{abc}$ denote the generators and structure constants of $\mathrm{SU_c}(3)$, which as a group belonging to the Lie algebra satisfy the following relations,
\begin{align}
    [t^a, t^b] = if_{abc}t^c, \hspace{10pt} \mathrm{tr}(t^a t^b)=\frac{1}{2}\delta^{ab}.
\end{align}
As a $\mathrm{SU}(3)$ matrix, the generator $t^a$ is given with the Gell-Mann matrix $\lambda^a$ as $t^a=\lambda^a/2$.
The Lagrangian density is made to be invariant under the following local continuum rotation (gauge transformations):
\begin{align}
    \psi(x) \rightarrow e^{-i\theta^a(x)t^a} \psi(x).
\end{align}
As a requirement of the gauge invariance, gluon mass terms such as $A_\mu A^\mu$ do not appear in the Lagrangian density. This leads to the fact that gluons should be massless in QCD.
%On the other hand, quarks have finite masses while their respective masses are not constrained only by this gauge symmetry.
On the other hand, quark mass is not constrained only by this gauge symmetry, likewise gluon mass.
The quark mass is, in fact, finite.

\subsubsection{Chiral symmetry}
Another symmetry of the QGP is the so-called chiral symmetry, where the Lagrangian remains invariant even when the right- and left-handed quarks are globally rotated in flavor space, respectively.
It should be noted here again that the chiral symmetry only holds for the massless QCD Lagrangian ($m\rightarrow0$). This is because in the Lagrangian denoted in Eq.~\refbra{eq:LagrangianQCD}, there is a term with quark mass that clearly breaks the chiral symmetry. Such an approximation of the QCD Lagrangian is called the {\it{chiral limit}}.

The fact that the Lagrangian is invariant under the global transformation of the fields is equivalent to the existence of a corresponding conserved charge.
Once one assumes that the number of degenerate flavors of quarks is 2 corresponding to the light quark ($ud$) sector
\footnote{Although here I only considers $u$ and $d$, in general, one can express the symmetry group as $\mathrm{SU_L}(N_f)$ or $\mathrm{SU_L}(N_f)$ depending on what kind of the rotating-space one considers.}
,
the chiral symmetry is expressed as $\mathrm{SU_L}(2)\times \mathrm{SU_R}(2)$.
The chiral symmetry is formulated from the fact that generators of $\mathrm{SU_L}(2)$ and $\mathrm{SU_R}(2)$ can be constructed as a linear combination of generators of $\mathrm{SU_V}(2)$ and $\mathrm{SU_A}(2)$ that give vector and axial-vector currents, respectively, according to Noether's theorem.
If one prepare the state of right-handed and left-handed quarks by projecting the Fermion field as
\begin{align}
    \psi_R(x) \equiv \frac{1}{2} (1+\gamma^5) \psi(x), \hspace{10pt} \psi_L(x) \equiv \frac{1}{2} (1-\gamma^5) \psi(x),
\end{align}
the right- and left-handed Fermion fields, $\psi_R$ and $\psi_L$, can be rotated independently as follows
\footnote{With the relation, $\psi=\psi_R+\psi_L$, each transformation of Eq.~\refbra{eq:chiralTrans} acts on each Fermion filed like $e^{i(\bm{\epsilon} + \bm{\epsilon}_5)\cdot \frac{\bm{\tau}}{2}} e^{i(\bm{\epsilon} - \bm{\epsilon}_5)\cdot \frac{\bm{\tau}}{2}} \psi = e^{i(\bm{\epsilon} - \bm{\epsilon}_5)\cdot \frac{\bm{\tau}}{2}} \psi_L + e^{i(\bm{\epsilon} + \bm{\epsilon}_5)\cdot \frac{\bm{\tau}}{2}} \psi_R$.
This is the meaning of the independent rotation of right- and left-handed fields.
}
:
\begin{align}
\label{eq:chiralTrans}
    \psi_R(x)\rightarrow e^{i(\bm{\epsilon} + \bm{\epsilon}_5)\cdot \frac{\bm{\tau}}{2}} \psi_R(x), \hspace{10pt}
    \psi_L(x)\rightarrow e^{i(\bm{\epsilon} - \bm{\epsilon}_5)\cdot \frac{\bm{\tau}}{2}} \psi_L(x).
\end{align}
Here $\tau$ is the Pauli matrix that corresponds to the generator of $\mathrm{SU}(2)$.

To illustrate the spontaneous chiral symmetry breaking, let us consider the linear $\sigma$ model, which is a very simple effective theory.
The Lagrangian of the $\sigma$ model starts with a state under the chiral symmetry.
%The Lagrangian hold a kinetic term of mass-less Dirac Fermion,
%coupling term of scalar and pseudoscalar fields, kinetic terms of scalar and pseudoscalar fields, and a potential term. 
However, the chiral symmetry is spontaneously broken with an emergence of the Dirac Fermion mass term which is not invariant under the chiral transformation given in Eq.~\refbra{eq:chiralTrans}.
Thus, the spontaneous symmetry breaking allows massless nucleons/quarks to have mass. 
%This is regarded as one of the possible origins of the mass of hadrons.
The massless boson appearing here is the so-called {\it{Nambu--Goldstone boson}}, which corresponds to pions with the spontaneous chiral symmetry breaking under $\mathrm{SU}(2)$ with $ud$ sector.
Note that after the symmetry breaking, only the isospin symmetry remains. 
Therefore, the following change of the symmetry occurs:
$\mathrm{SU_L}(2)\times \mathrm{SU_R}(2)\rightarrow \mathrm{SU_V}(2)$.

In the {\it{Nambu--Jona-Lasinio model}} which is the effective theory to describe the spontaneous chiral symmetry breaking in QCD, 
the vacuum expectation value of a quark and an anti-quark condensate, $\langle \bar{q}q \rangle$, gives the quark-mass
\cite{Nambu:1961fr, Nambu:1961tp,Goldstone:1961eq, Goldstone:1962es, Kunihiro:1983ej,Hatsuda:1994pi}.
From the above discussions, one is able to tell that $\langle \bar{q}q \rangle$ serves as an order parameter to describe the chiral symmetry breaking and is called chiral condensate. 
%Thus, chiral phase transition means that the chiral condensate melts and becomes zero at high- temperature and baryon density regimes.

For more details about the phase structure, see {\it{i.e.,}}, for example, Refs.~\cite{Baym:2017whm,Pisarski:2006ie,Fukushima:2013rx,Sogabe:2020xsp,Sorensen:2021zxd}.

\subsubsection{Hadron-QGP phase transition}
As discussed above, the chiral phase transition lies between the hadron and the QGP state.
Another phase transition has actually been suggested in QCD, which is the {\it{(de)confinement transition}}. Let me briefly introduce it in the followings.
If one considers a world without dynamical quarks ($N_f=0$), which means quarks are extremely heavy ($m\rightarrow\infty$) so that they are static, another important symmetry appears: the $\mathrm{Z}(3)$ center symmetry \cite{McLerran:1981pb}.
The $\mathrm{Z}(3)$ is a descretized symmetry.
Here the quantity called Polyakov loop, $\langle L \rangle$, \cite{McLerran:1981pb} becomes an exact order parameter with the given situation of $m\rightarrow\infty$. Polyakov loop $\langle L \rangle$ can be interpreted as a partition function when an extremely heavy quark is placed in the system \cite{book:YagiHatdudaMiake}. Thus, when one requires infinitely large free energy to place a free single quark in the system, $\langle L \rangle$ becomes zero. This means that when $\langle L \rangle=0$, no single quark is allowed in the system, which exactly means the confinement of quarks.

\vspace{35pt}

Now, there are two possible phase transitions: the chiral and (de)confinement. 
\begin{figure}[tbp]
\begin{center}
\includegraphics[bb=0 0 960 540, width=0.7\textwidth]{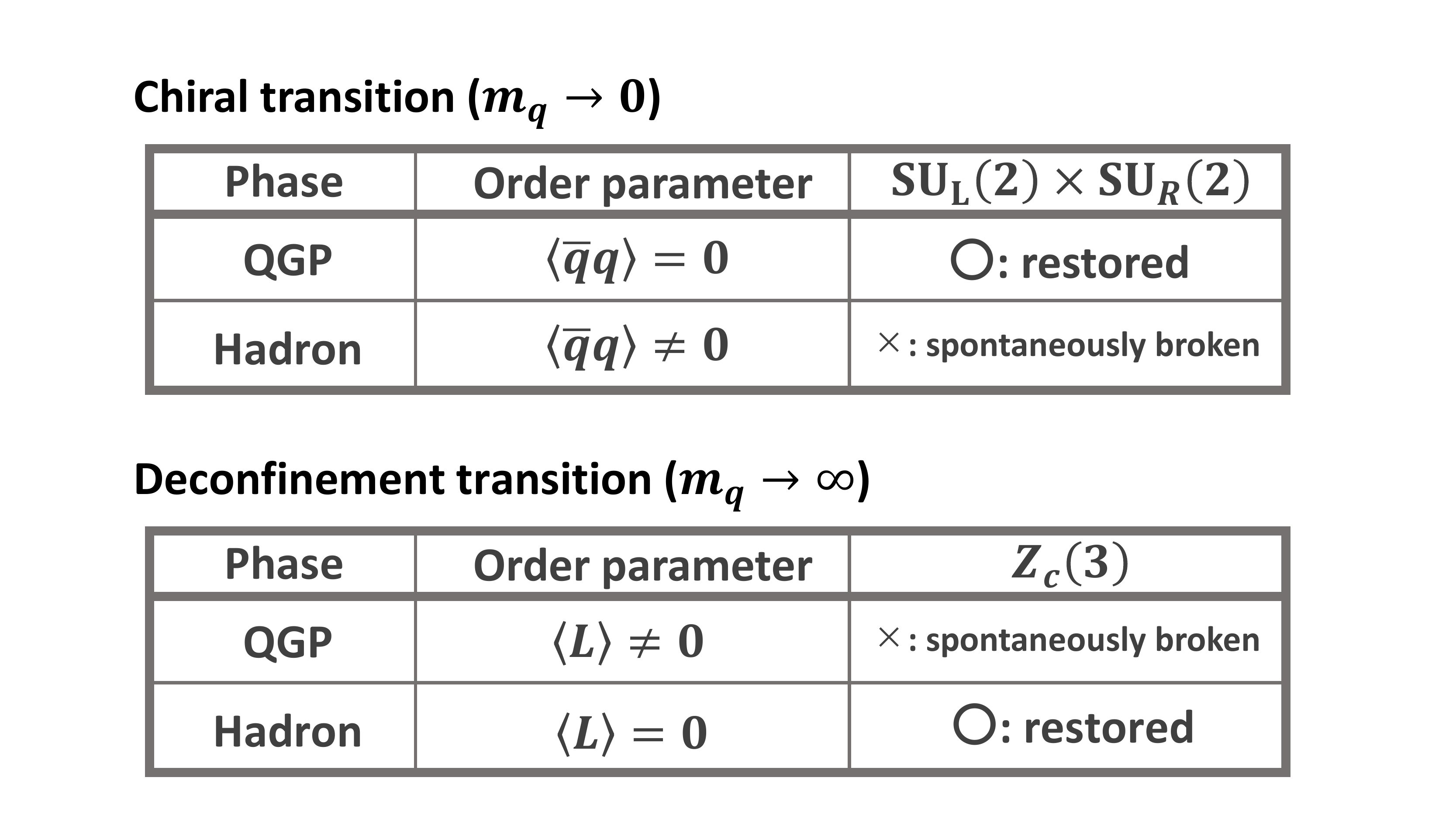}
\caption{Symmetry braking/restoration in chiral and confinement--deconfinement phase transition in QCD. }
\label{fig:Symmetry}
\end{center}
\end{figure}
The pattern of symmetry breaking in phase transitions is summarized in Fig.~\ref{fig:Symmetry}.
In the real world, quark masses are neither zero nor infinite. Therefore, both $\langle \bar{q}q\rangle$ and $\langle L \rangle$ order parameters are only approximate ones.
Therefore, how the real world should be described and understood is a very difficult problem in QCD.

On the other hand, it has been shown that both chiral condensation and Polyakov loops exhibit smooth crossovers over a narrow enough temperature range to see phase changes \cite{Kogut:1982rt, Fukugita:1986rr}.
Therefore, the QGP can be explained as follows: {\it{it is a deconfined and chiral symmetric phase at high temperatures}} \cite{Pisarski:2006ie}.
It is also worth mentioning that there is an effective theoretical model that can handle both
chiral and (de)confinement phase transition, the so-called {\it{the Polyakov-loop Nambu–-Jona-Lasinio model (PNJL)}} \cite{Fukushima:2003fw,Fukushima:2008wg}
\footnote{
Note that the above statements in this subsection is written referring Ref.~\cite{Fukushima:2009xxx}, so see the reference for more detailed discussions.
}
.

In the next subsection, I will deepen the definition of the QGP from a different perspective than symmetry.

\subsection{Plasma-ness of the QGP}
Speaking of the definition of QGP, it is necessary to focus on the basic concept of ``{\it{plasma}}''.
The etymology of the word plasma comes from the ancient Greek word for ``moldable matter'' \cite{wiki:Plasma_(physics)}.
The word has been used in the fields of biology and medicine, but it came into use in the field of discharge physics after Langmuir, who won the Nobel Prize in Chemistry in 1932, defined plasma as a uniformly discharging substance in equilibrium consisting of many-body charged particles \cite{web:Plasma}.

Looking around us, we can find plasma inside a fluorescent light. A filament implanted at the end of the tube of a fluorescent light emits electrons. Atoms of noble gas are hit by those electrons and electrically dissociate. This state where positive and negative charges are dissociated is called plasma.
Another place where one can find plasma is the aurora (the northern lights). The origin of the aurora is the hot and ionized solar winds from the Sun. Once the solar winds sneak into the crack of the shield of the magnetic field of the Earth, the plasma is accelerated. The collision with oxygen or nitrogen atom in the air produces beautiful lights in the sky.

Now, considering the name of QGP, it should be a state of dissociated quarks and gluons from the analogy of a word, plasma.
Now the question is, how can we find its ``plasma-ness''?

\subsubsection{Basics of plasma}
\begin{figure}[tbp]
\begin{center}
\includegraphics[bb=0 0 960 540, width=0.7\textwidth]{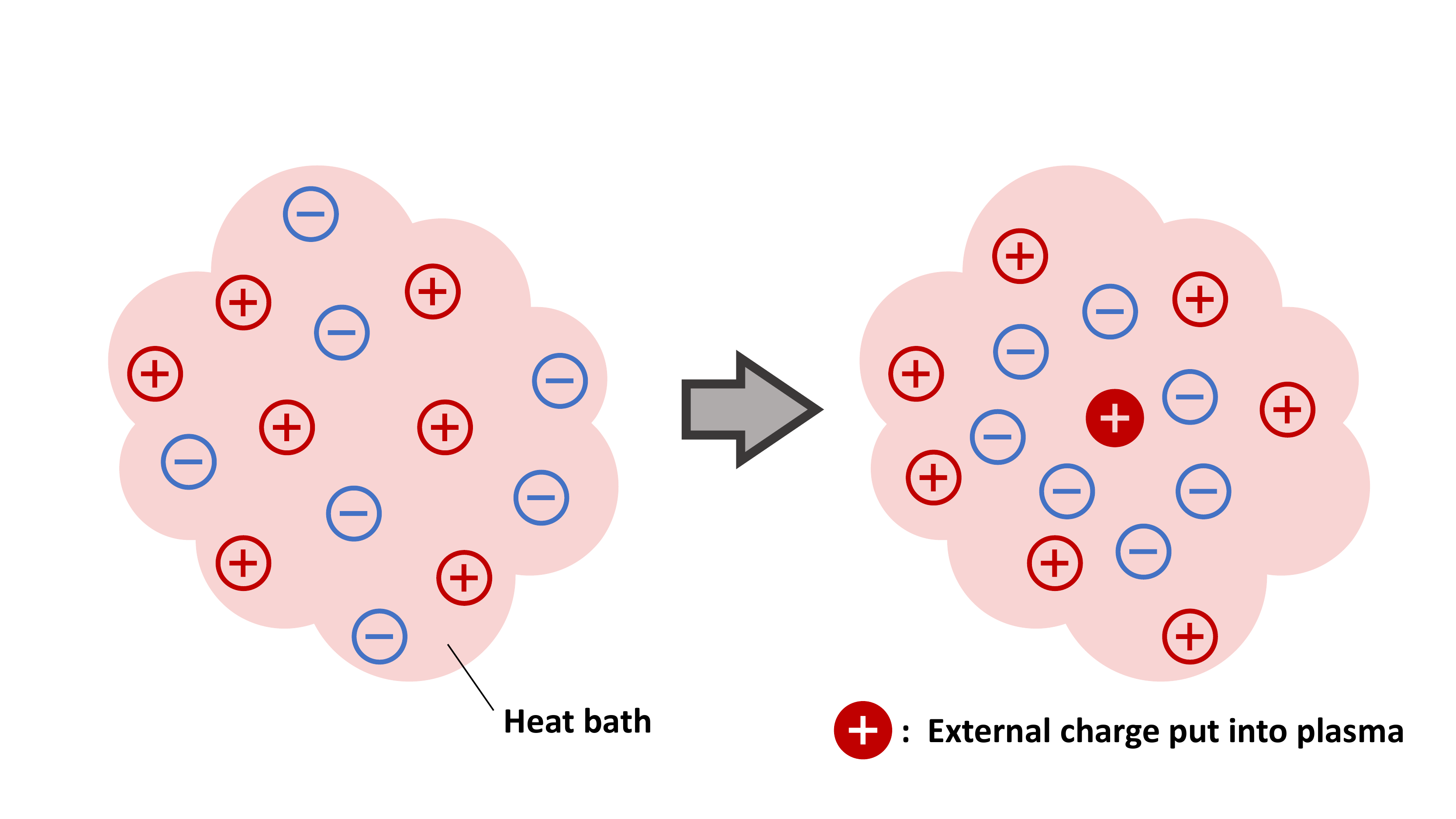}
\caption{How an external charge is Debye-screened in plasma. (left) Uniformly dissociated charged particles. (right) After an external positive charge is added at the center of plasma, the potential of external charge is Debye screened due to surrounding negative charges.}
\label{fig:DebyeScreening}
\end{center}
\end{figure}

The main features of plasma are as follows: 1) charge-neutral as a whole system, 2) under equilibrium condition (finite temperature) where some constituent charges are moving, 3) shorter Debye screening length than the system size.
One can discuss the Debye screening length assuming a system satisfying conditions 1) and 2) with semi-classical {\it{Quantum Electrodynamics}} (QED).
Suppose that one puts an external charged particle by hand into a uniformly ionized plasma consisting of electrons and positrons in a heat bath.
Here, induced and external electric density will be a source of electric field for Gauss's law.
Integrating Fermi-Dirac distribution to get the induced charge density, one can get potential with the following form,
\begin{align}
    \label{eq:DebyeLengthinV}
    V(r)&\propto-\frac{e^{-r/\lambda_D}}{r} \\
    \label{eq:DebyeMassinV}
    &= \frac{{e}^{-m_D r}}{r},
\end{align}
where $\lambda_D$ is the Debye screening length, and $m_D$ is the Debye screening mass which is an inverse of $\lambda_D$. Because $V(r)$ decreases dramatically when $r$ is smaller than $\lambda_D$, it can be interpreted as the length that potential is able to reach.
Thus, the Debye screening length is one of the main scales which characterize plasma.
Figure~\ref{fig:DebyeScreening} shows an image of how the Debye screening happens in the given situation.
In addition, it should be mentioned that
Eq.~\refbra{eq:DebyeMassinV} shows that
the Debye mass can be understood as an effective mass of the charge caused by the shielding of the bare charge due to the (vacuum) polarization
\footnote{
In quantum field theory (QFT), one can obtain the Debye mass as an effective mass in the denominator of the propagator for vacuum polarization.
}. 
From this fact, it is often said that the external charge is dressed up and acquires effective mass due to the interaction with the polarization.

Now, let us come back to the discussion of the scale which characterizes the plasma-ness.
Considering the condition 2) and 3), plasma should satisfy the following two conditions:
\begin{align}
\label{eq:PlasmaParameter}
 \Lambda \gg 1 & \hspace{10pt} \Lambda=n\lambda_D^3,\\
 \label{eq:PlasmaLength}
 L & \gg \lambda_D.
\end{align}
Here $\Lambda$ is often called the plasma parameter, and $n$ is the number density of charged particles per volume.
Thus, $\Lambda$ represents the number of charged particles in the screened area with a length of $\lambda_D$, which should be much larger than $1$ to satisfy the condition 2); it should be a many-body system under equilibrium.
Relation of Eq.~\refbra{eq:PlasmaLength} should be satisfied so that the potential of the external charge is screened.
If the system size $L$ is shorter than the $\lambda_D$, one would see the bare potential rather than the screened one, which does not correspond to the plasma-ness.

In the following subsections, I would like to pick up two important papers that calculated Debye screening/mass/length from vacuum polarization of quarks or gluons in QCD.

\subsubsection{Superdense Matter: Neutrons or Asymptotically Free Quarks? [Collins \& Perry (1975)]}
This paper \cite{Collins:1974ky} discusses QCD matter in the high baryon density region.
They proposed that the plasma-like QCD matter would appear in the neutron star's core, expanding black holes and the early universe. The big motivation for this work is the existence of Bjorken scaling. The Bjorken scaling suggests that the quarks inside of hadrons would behave in-coherently.
Based on this discovery, one can expect that weakly interacting quarks would be achieved at a large energy scale. Collins and Perry called this state ``quark soup''.
One of the main results presented in this paper is the Debye screening length $\lambda_D$ from the renormalization-group equation considering a gluon propagator polarized in a vacuum.
To the best of my knowledge, this is the first calculation deriving the plasma-ness of QCD matter at a large energy scale.

\subsubsection{Theory of hadron plasma [Shuryak (1978)]}
The name ``quark gluon plasma'' appeared for the first time on this paper \cite{Shuryak:1977ut}.
Likewise the work by Collins and Perry, the Debye screening length is obtained.
The results are obtained considering situations of both cold and hot plasma.
The former is applied to the neutron star merger and the core of neutron stars,
while the latter is applied to the early universe and high-energy collisions.

\subsection{QGPs in nature}
Let me first define the system of units that I use throughout this thesis which is needed to discuss the scales of physical variables.
I use the natural unit, $\hbar = c = k_{B} =1$, and the Minkowski metric,
$g_{\mu \nu} = \mathrm{diag}(1, -1, -1, -1)$.

\begin{figure}
    \centering
    \includegraphics[bb=0 0 960 540, width=1.0\textwidth]{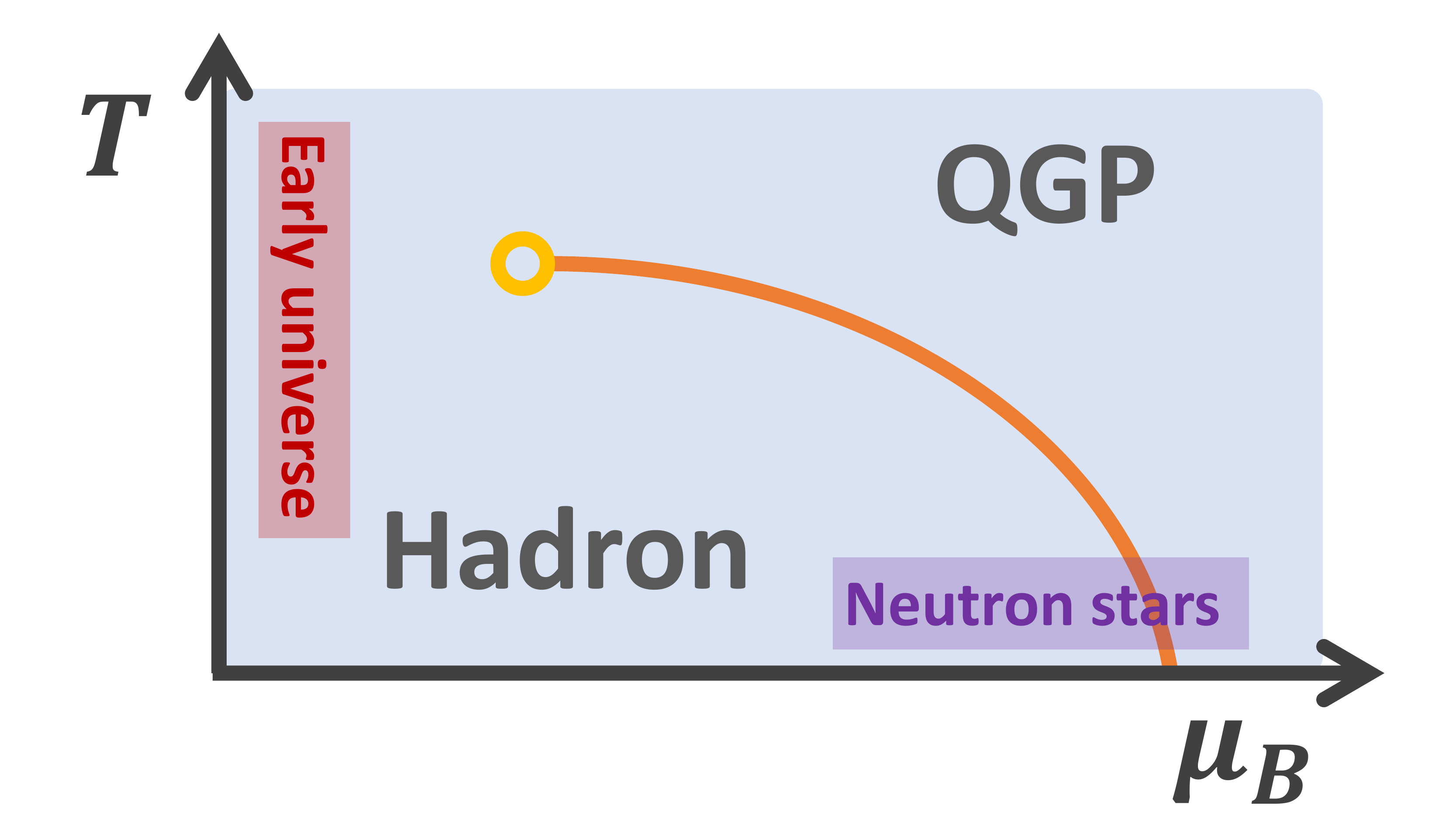}
    \caption{Schematic QCD phase diagram. Critical point (yellow open circle) at the end of the first order phase transition (orange line) is shown.}
    \label{fig:QCDphasediagram}
\end{figure}

To show the situations where we can find a QGP state in nature, 
I would like to start by showing the QCD phase diagram in Fig.~\ref{fig:QCDphasediagram}.
The QCD diagram is characterized by two variables, temperature $T$ and baryon chemical potential $\mu_B$ \cite{Cabibbo:1975ig}.
At high temperature and/or large baryon chemical potential, the QGP phase is realized where the chiral symmetry is restored,
and quarks and gluons are deconfined, as shown in Fig.~\ref{fig:Symmetry}.
On the other hand, at low temperature and/or small baryon chemical potential, the degree of freedom of a system is governed by hadrons.
Several effective theories suggest the existence of the QCD critical point
\cite{Asakawa:1989bq,Halasz:1998qr,Stephanov:2004wx} that is the endpoint of the first-order phase transition line in the QCD phase diagram.
The critical point is shown with the yellow open circle, and the first-order phase transition is shown with the orange line in Fig.~\ref{fig:QCDphasediagram}.
However, 
it should be noted that, actually, the full structure of the QCD phase diagram including the locations of the critical point and the first phase transition line has not been still unrevealed.

From the first principal calculation based on the {\it{lattice QCD}}, which is the method to perform QCD calculation by latticizing QCD in Euclidean space \cite{Wilson:1974sk}, only the region with the limit of $\mu_B \rightarrow 0$ is revealed
\footnote{There have been several attempts to adopt lattice QCD calculations to a finite baryon density regime \cite{Steinbrecher:2018phh,Borsanyi:2020fev}.}
. 
The lattice QCD calculation shows that the transition between hadron and the QGP is a smooth crossover around $T\approx 150$-$170$ MeV rather than a phase transition \cite{Borsanyi:2010cj,HotQCD:2014kol,Bellwied:2015rza}.

For the large baryon chemical potential regimes,
there is a difficulty in the numerical simulation of lattice QCD, which is known as notorious {\it{sign problem}}. Thus, one needs to guess what is happening with phenomenological discussion.
Several descriptions of the phase in this region have been proposed. For example, the {\it{color super conductor}} \cite{Bailin:1983bm, Iwasaki:1994ij,Berges:1998rc} has been studies as a state where color charged quarks form Cooper pairs \cite{Cooper:1956zz, Bardeen:1957kj} which leads to {\it{diquark condensation}}.
While no one knows when the phase change from hadron to the QGP at low temperature and large baryon chemical potential area happens, one can estimate its value from a simple math.
The typical number density of nucleons inside of nucleus is $n_0 \approx 0.16 \ \mathrm{fm}^{-3}$, and the typical radius of a nucleon is $r_B\approx 0.5$-$0.7$ fm.
The more one squeezes a nucleus, the larger the density of nucleon becomes.
Eventually, nucleons start overlapping each other and become a matter whose degree of freedom is governed by quarks.
Given the above typical values of $n_0$ and $r_B$, the nucleon density when nucleons inside of a nucleus start overlapping can be estimated as $n_B\approx4n_0$-$12n_0$
\footnote{The first phase transition line becomes a band because of the existence of the mixed state if one maps the QCD phase diagram onto $T-n_B$ instead of $T-\mu_B$.}
.

Let me come back to the discussion of where we can find the QGP state in nature.
Here, those two regimes: 1) $T\approx 150$-$170 \ \mathrm{MeV}$ with $n_B \rightarrow 0$ and 2) $n_B \approx 4n_0$-$12n_0, $ with $T\rightarrow 0$, are where the evolution of the early universe went through and the area that is possibly achieved inside of the inner structures of neutron stars, respectively, as they are depicted in Fig.~\ref{fig:QCDphasediagram}.

Just after the Big Bang, it is considered that the temperature of the universe was large enough to archive the grand unification of theories.
Because the temperature of today's universe is low enough for us to stay in the hadron phase, there should be a QGP state at some point in the evolution of the universe.
Thus, as the universe cools down in the expansion, 
QCD phase transition from QGP to hadron takes place \cite{Olive:1980dy, Kolb:1981yj, Suhonen:1982ee, Crawford:1982yz,Schramm:1983vz}.
The transition from the QGP to hadrons is considered to take place $t\approx10^{-5}$-$10^{-4}$ s after the Big Bang \cite{book:YagiHatdudaMiake}.

Neutron stars are extremely high-density matters.
According to the recent studies in astrophysics, 
several neutron stars with about two times larger mass than the solar mass have been found \cite{Demorest:2010bx,Antoniadis:2013pzd,Fonseca:2016tux}.
Thinking that mass of one nucleon is $\approx 1$ GeV, one can simply estimate the baryon density of the neutron star by dividing mass by volume, which is $n_B \approx n_0$.
This simple estimation assumes the case that a neutron star is a uniform sphere without any inner structures. 
However, in fact, the more interior of the neutron star you look, the more pressure you would expect.
Therefore, it is considered to exist a change of a phase from hadronic to quark matter in the middle of the inner structure of a neutron star.
For instance, the study on the enormously massive neutron star found in 2010 indicates a central baryon density of the star is around $2$-$5n_0$, and a radius stays between $11$-$15$ km \cite{Demorest:2010bx}.

\section{Relativistic heavy-ion collisions}
%Multiplicity \dndeta.
%Collision energy is \snn $= 2.76$.
%Proton-proton \snn[proton] $=7$ TeV.
Motivated by some theoretical proposals, the relativistic heavy-ion collision experiment started in the late 70s to investigate the properties of QCD matter by realizing an extremely hot state. In this chapter, I would like to summarize the historical path of relativistic heavy-ion collision experiments.

\subsection{History of relativistic heavy-ion collision experiments}
\label{sec:HistoryOfRelativisticHICexperiment}
%\begin{figure}
%    \centering
%    \includegraphics[bb=0 0 936 504, width=1.0\textwidth]{chapters/figure/intro/20220228_HICHistory.pdf}
%    \caption{Transition of collision energy per nucleon pair at center-of-mass frame in heavy-ion collisions. Only symmetric heavy-ion collisions are shown. Colors are assigned to classify experiments performed with different colliders.}
%    \label{fig:HICHistory}
%\end{figure}

The starting point backs to the discovery of the asymptotic freedom of QCD in 1973 by David Gross and Frank Wilczek \cite{Gross:1973id}, and Hugh David Politzer \cite{Politzer:1973fx}.
In 1974, a workshop called ``Report of the workshop on BeV/nucleon collisions of heavy ions -- how and why'' organized by T.~D.~Lee {\it{et al.}} was held in Bear Mountain in New York, where the future prospect of physics on the hadronic matter of extremely high- temperature/baryon density and future plans for developments of the new accelerator was discussed.
Around this time (after 1975), the probability was proposed that there could exist an equilibrium state in which quarks and gluons were deconfined from hadrons.

In 1974, the first heavy-ion collision experiment started at a collider called BEVALAC in Lawrence Berkeley Laboratory (LBL) \cite{osti_937059} in California
while the collision energy was not enough to generate the QGP.

In the middle of the 80s, instead of Berkeley Lab., the center of relativistic heavy-ion study moved to the Brookhaven National Laboratory (BNL) in New York, U.S., and the CERN (the European Organization for Nuclear Research) located around the boundary between France and Switzerland.
The collision energy at the Alternating Gradient Synchrotron (AGS) at the BNL was \snn = 4-5 GeV, while it was \snn = 17-20 GeV at the Super Proton Synchrotron (SPS) at the CERN.
In these experiments, the novel results were discovered -- 
especially, the significant suppression of $J/\Psi$ yields, which Matsui and Satz predicted in 1986 as a signal of the QGP formation in heavy-ion collisions \cite{Matsui:1986dk}.
It should also be noted that in 1982, Bjorken established relativistic hydrodynamics.
In the paper, the picture of ``central plateau'' of a structure of produced particles as a function rapidity at relativistic collision energy \cite{Bjorken:1982qr}, which currently is the so-called {\it{boost invariant}} picture.
This replaced the particle production picture first time in almost three decades since Fermi \cite{Fermi:1950jd} and Landau \cite{Landau:1953gs} proposed the multi-particle production picture called ``{\it{fireball}}'' in the 50s.

In 2000, the Relativistic Heavy Ion Collider (RHIC) at the BNL, one of the accelerators still being used today, started running experiments. At RHIC, four experiments: STAR, PHENIX, PHOBOS, and BRAHMS, were launched with different purposes. 
The collision energies reached up to \snn = 200 GeV, which led to the declaration of discovery of the perfect fluidity of the QGP in relativistic heavy-ion collisions \cite{Heinz:2001xi,Gyulassy:2004zy,Shuryak:2004cy,Hirano:2005wx}, which was reported as a press release by BNL on 9 AM in EST, April 18, 2005 at American Physical Society meeting held in Tampa, Florida.

In 2010, the first heavy-ion experiment was realized at the Large Hadron Collider (LHC) at CERN with $Pb$+$Pb$ collisions at \snn = 2.76 TeV.
Still today, the LHC has been running heavy-ion collision experiments at the highest order of collision energies.
For the LHC, while most experiments with proton--proton collisions are intended for the discovery of new particles such as Higgs and SUSY, the heavy-ion collision experiment was planned to be about 1/10 of the running time of the LHC. 
The ALICE experiment was specifically launched for heavy-ion collisions.
The temperature achieved at the ALICE experiments is estimated as $\sim 5.5$ trillion K, which is recorded as the highest artificial temperature by the Guinness record.

Since around 2010,
a novel project, the Beam Energy Scan (BES), to search a critical point of the QCD phase diagram has been carried out at the RHIC in BNL.
The collision energy will cover \snn=3.0-62.4 GeV, and which would expect to scan the regime on diagrams where a critical point and a phase transition line are thought to be located.
The BES program is still ongoing, and hopes for revealing the QCD phase diagram are increasing.

For more details on the history of the relativistic heavy-ion experiments, for instance, see Ref.~\cite{book:Genshikakukenkyu2,book:YagiHatdudaMiake}

\subsection{Space-time picture of relativistic heavy-ion collisions}
\label{subsec:SpaceTimePicureOfHIC}

\begin{figure}
    \centering
    \includegraphics[bb=0 0 720 540, width=0.9\textwidth]{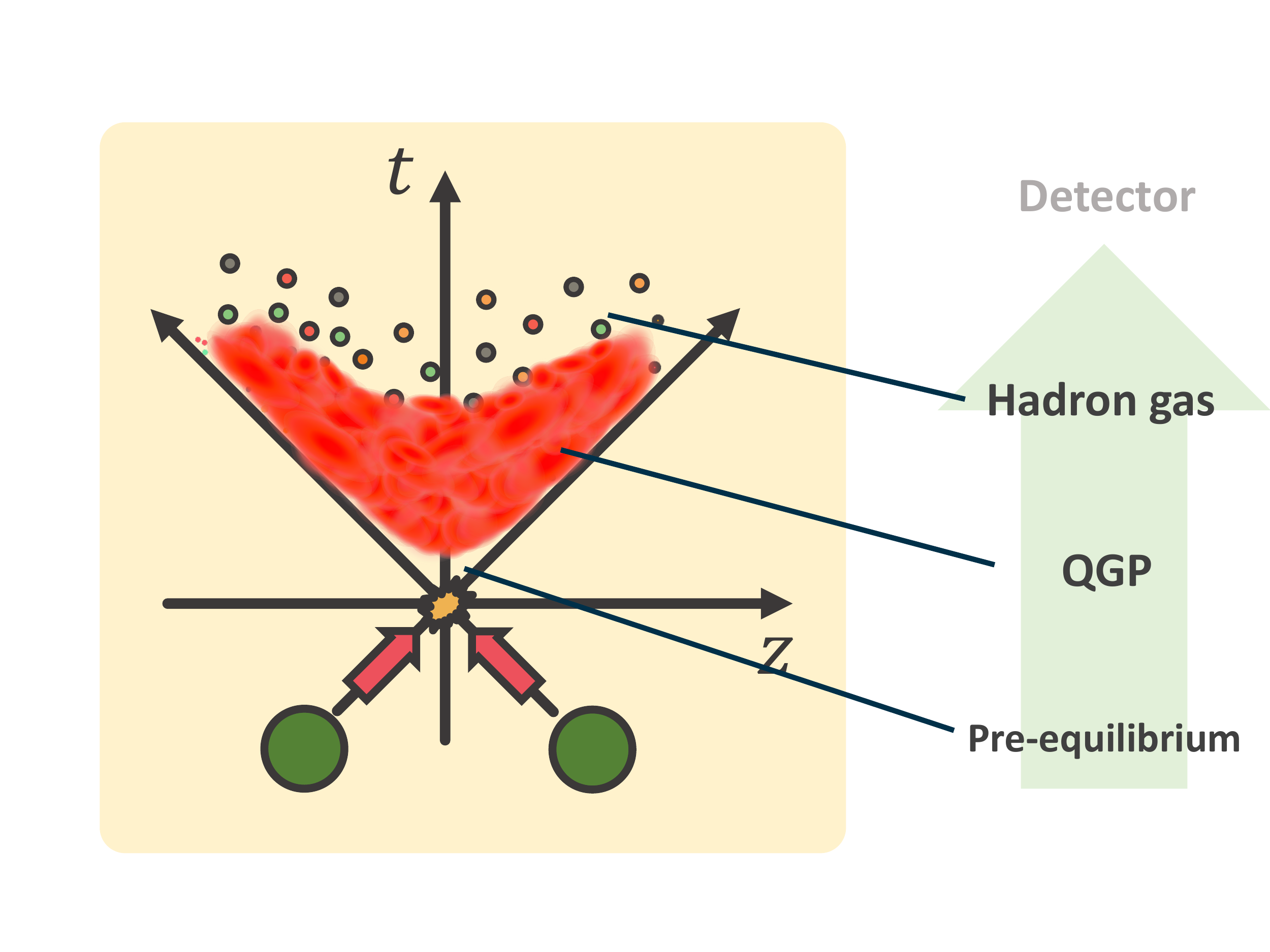}
    \caption{Schematic picture of space-time evolution of a relativistic heavy-ion collision depicted in $t-z$ plane where $z$ axis is directed to a beam direction.}
    \label{fig:SpaceTimeHIC}
\end{figure}

As we have seen so far, several experimental projects have been carried out to realize the QGP state in relativistic heavy-ion collisions.
Then, how is it realized and how does it evolve in a reaction? Before seeing what kinds of observables are analyzed in experiments, I describe the general picture of the space-time evolution of relativistic heavy-ion collisions.

The dynamical reaction of the relativistic heavy-ion collisions is described with several consequent stages: pre-equilibrium stage, hydrodynamic evolution of the QGP, and hadron gas with interactions. 
Figure \ref{fig:SpaceTimeHIC} shows the schematic picture of the space-time evolution of a relativistic heavy-ion collision.
I will follow up on each stage below.

\subsubsection{Pre-equilibrium stage}
\begin{figure}
    \centering
    \includegraphics[bb = 0 0 765 612, width=0.85\textwidth]{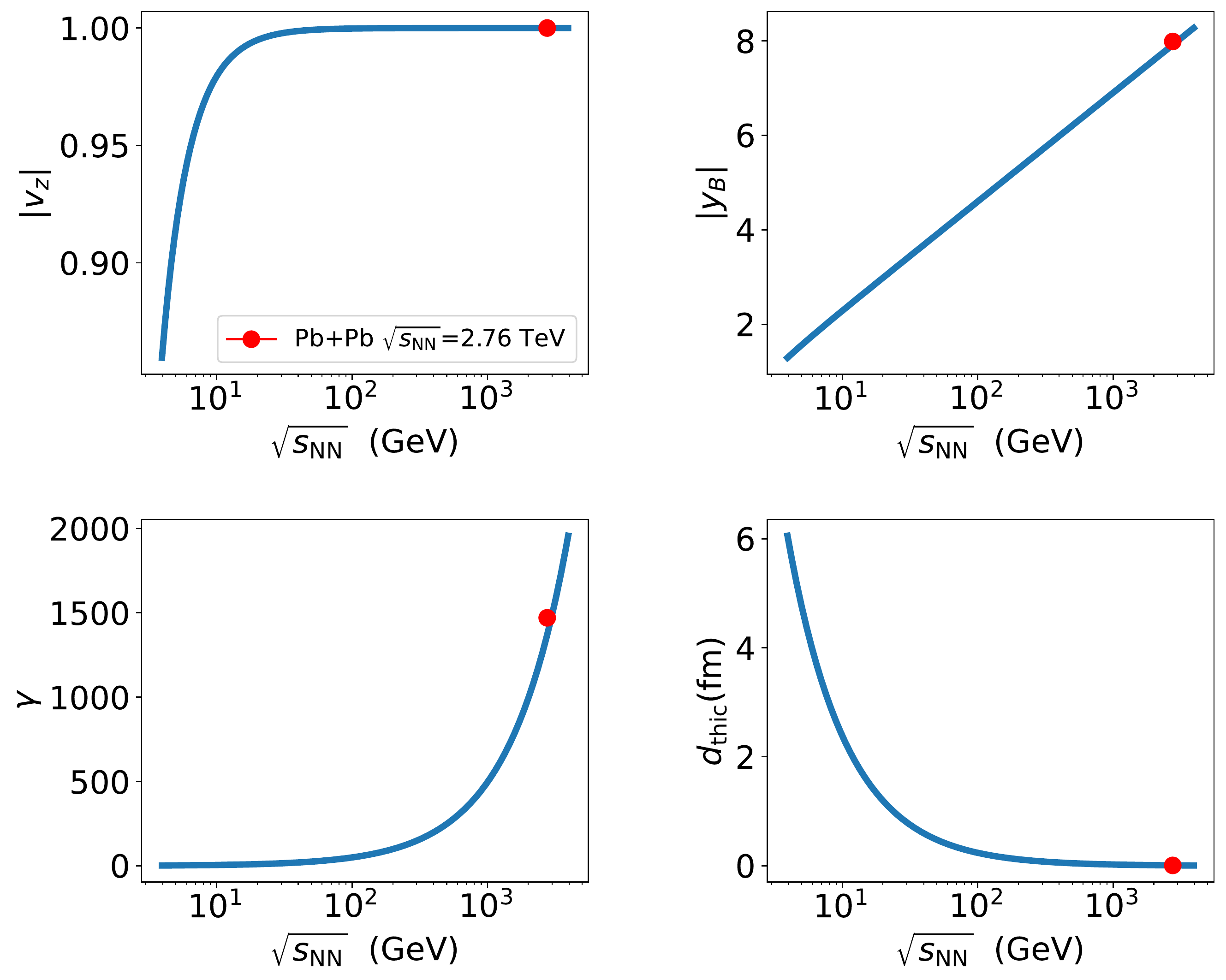}
    \caption{Kinematic variables of beam particles seen at laboratory frame as functions of center-of-mass collision energy per nucleon pair in $Pb$+$Pb$ collisions. Corresponding values in $Pb$+$Pb$ collisions at \snn = 2.76 TeV are shown as red circles.}
    \label{fig:HICEnergyBasics}
\end{figure}

Nuclei are accelerated nearly to the speed of light
\footnote{
For example, protons and lead ions are accelerated to $\approx0.9999999$\% and $\approx0.999999$\% of the speed of light at \snn[proton] = 7 TeV and \snn = 2.76 TeV, respectively at the LHC.
}
in accelerators of relativistic heavy-ion collision experiments.
Figure \ref{fig:HICEnergyBasics} shows various kinematic variables of beam particles seen at the laboratory frame as functions of center-of-mass collisions energy per nucleon pair in $Pb$+$Pb$ collisions. From left to right, the absolute value of velocity in beam direction, the absolute value of beam rapidity, Lorentz gamma factor, and thickness of Lorentz contracted nuclei are shown.
Corresponding values at \snn = 2.76 TeV are also shown for references.

As one sees, at LHC energies where the collision energy is around \snn $\approx 10^3$ GeV,
colliding nuclei are highly Lorentz contracted ($\gamma \approx \mathcal{O}(10^3)$).
With a simple estimation of overlap time of a collision with $(2A^{-3}/\gamma)/v_z$ where $A$ is the mass number, 
once can see the overlap time becomes an order of $\tau \approx \mathcal{O}(10^{-3}) $ fm which can be expressed as a point on $t-z$ plane likewise shown as an origin of the $t-z$ plane in Fig.~\ref{fig:SpaceTimeHIC}. After the collision of nuclei, baryons initially inside nuclei go through each other. 
In between the outgoing nuclei, generation of matter takes place.

Here, it should be noted that this is not the case in low-energy heavy-ion collisions such as performed at BES program at RHIC where collision energies span around \snn $\approx 10^1$ GeV \cite{Shen:2017bsr}.
A nucleon--nucleon collision happens in a region with finite width in $z$ and $t$ direction. 
Under such low collision energies, baryons are no longer be able to completely go through, which is so-called {\it{baryon stopping}} \cite{Busza:1983rj,Videbaek:1995mf},
and matter comes to be generated with finite baryon density.

The matter generated in a collision is a many-body system of quarks and gluons
.
The initially generated matter is far from local equilibrium at first.
Specifically, due to strong longitudinal expansion, the phase space distribution of matter is highly anisotropic.
Because of following soft radiation of gluons and
thermalization of such radiated soft gluons
\footnote{
Here, I explain the thermalization process based on so-called, {\it{bottom-up thermalization}} \cite{Baier:2000sb}.
The reason that I only mention gluons not quarks is that, in this theory,
there is only a degree of freedom of gluons.
}
,
more energies of semi-hard gluons are deposited into
that of soft gluons, leading to isotropization and thermalization
of the matter.
For reviews, see, for example, 
Refs.~\cite{Schlichting:2019abc,Berges:2020fwq}.

\subsubsection{Hydrodynamic evolution of the QGP}
Once matter reaches local equilibrium
\footnote{
In QGP, there should be chemical equilibrium of quarks and gluons
besides thermalization that I do not mention here.
},
the matter starts hydrodynamic evolution.
Relativistic nuclear collisions take place in a vacuum,
and which makes a large pressure gradient.
The rapid evolution of the matter induced by the large pressure gradient 
cause cool-down of the temperature.
As a matter obeys hydrodynamics, flows are generated
as a response to the pressure gradient in the initial geometry of the matter.

\subsubsection{Hadron gas with interactions}
Once the temperature of the matter cools below $T\approx150$-$170$ MeV,
the transition from the QGP to hadrons takes place.
If those hadrons interact highly and local thermal equilibrium is kept, the matter can still evolve hydrodynamically.
It is not the case when
collision rate among hadrons is less than that of the expansion rate.
The matter starts to behave as a gas consisting of ``free'' hadrons rather than a fluid.

However, it is not that those hadrons
instantly become completely free.
There are still, for instance, $2$-body scatterings or resonance decays in the hadron gas.
Once the collision rate among hadrons becomes much smaller than the expansion rate of the system, and they finally become actually free,
the chemical composite of hadrons and momentum distribution is fixed
\footnote{
These are conventionally called {\it{chemical freeze-out}} and {\it{kinematic freeze-out}}.
Note that kinematic freeze-out occurs after chemical freeze-out \cite{Hirano:2002ds}.
}
.

\section{Experimental observables}

\subsection{Geometrical setup}
\begin{figure}
    \centering
    \includegraphics[bb = 0 0 790 427, width = 1.0\textwidth]{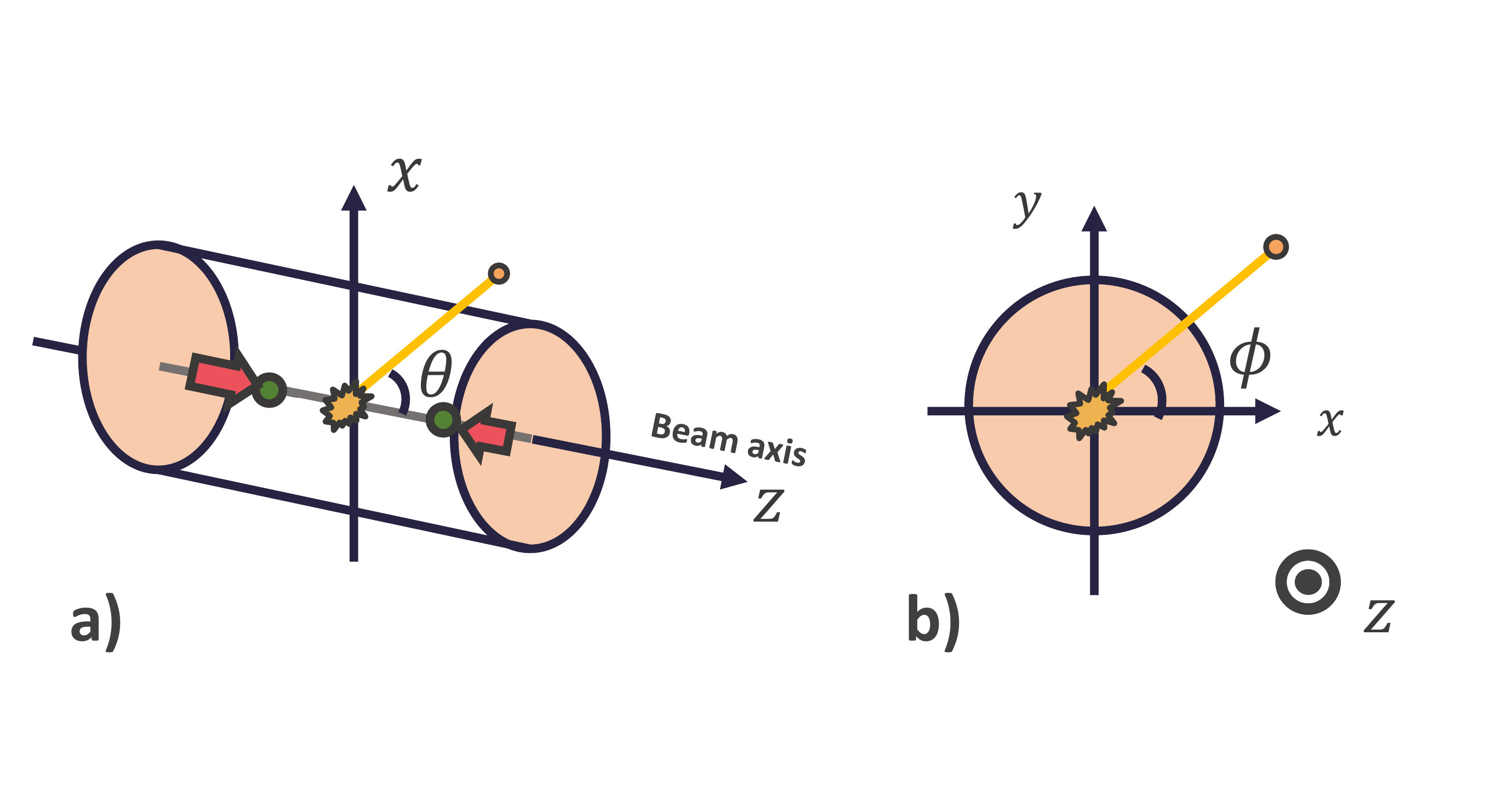}
    \caption{Geometrical setup commonly used in high energy nuclear collisions. a) Beam axis is shown horizontally. The theta, $\theta$, is the polar angle of the direction of produced particle momentum measured from the beam axis. b) Transverse plane of b) at a certain $z$ is shown. The azimuthal angle, $\phi$, is the angle around the beam axis.}
    \label{fig:GeometryHIC}
\end{figure}
Before taking a look at the experimental settings and actual observables measured in relativistic heavy-ion collision experiments, 
I would like to introduce kinetic variables commonly used in relativistic nuclear collisions for later explanations.
Figure \ref{fig:GeometryHIC} shows 
geometrical setup commonly used in high energy nuclear collisions. Beam axis is shown horizontally in Fig.~\ref{fig:GeometryHIC} a). 
The theta, $\theta$, is the polar angle of the direction of produced particle momentum measured from the beam axis, which is obtained with $\theta = \cos^{-1}(p_z/|\bm{p}|)$ where $p_z$ and $|\bm{p}|=\sqrt{p_x^2+p_y^2+p_z^2}$ are momentum of a produced particle in a direction of beam axis and magnitude of momentum in 3-dimension.
On the other hand,
Fig.~\ref{fig:GeometryHIC} b) shows transverse plane at a certain $z$ is shown. The azimuthal angle, $\phi$, is the angle around the beam axis,
and which is calculated as $\phi = \tan^{-1}(p_y/p_x)$.
Note that while $z$ is fixed to a direction of the beam, $x$ and $y$ axes can be taken freely.
Thus, there is no physical meaning in the absolute value of $\phi$.

Produced particle distribution is usually described as a function of {\it{rapidity}}, $y$, (or {\it{pseudorapidity}}, $\eta$,) or {\it{transverse momentum}}, $p_T=\sqrt{p_x^2+p_y^2}$.
Definitions of $y$ and $\eta$ are given as follows:
\begin{align}
\label{eq:rapidity}
    y = \tanh^{-1} \left(\frac{p_z}{E}\right),  \enskip \eta = \tanh^{-1}\left(\frac{p_z}{|\bm{p}|}\right).
\end{align}
The $p_z/E$ and $p_z/|\bm{p}|$ take values from $-1$ to $1$, and for massless particle, both identically means velocity in beam axis.
In relativistic nuclear collisions, beam particles run with a velocity nearly the speed of light.  Also, it is often the case that produced particles with a small scattering angle have a velocity nearly the speed of light too.
In these cases, one needs to deal with the values such as, for example, $v_z = 0.999971\cdots$ or $0.999952\cdots$ with taking care of the precision.
This is reconciled by parametrizing $p_z/E$ or $p_z/|\bm{p}|$ with hyperbolic tangent as a function of $y$ or $\eta$.
In addition to this, Lorentz transformation in beam direction can be expressed as a parallel shift with rapidity, which is another reason to use rapidity in high-energy nuclear collisions.
Conventionally, rapidity direction is called {\it{longitudinal}} direction.

Rapidity and pseudorapidity can be expressed in a different form.
From easy calculations, one can show Eqs.~\ref{eq:rapidity} can be turned into
\begin{align}
    y = \frac{1}{2} \log \frac{E+p_z}{E-p_z} = \frac{1}{2} \log \frac{1+v_z}{1-v_z},
\end{align}
and 
\begin{align}
    \eta = \frac{1}{2} \log \frac{|\bm{p}|+p_z}{|\bm{p}|-p_z} = - \log \left[ \tan \left( \frac{\theta}{2}\right) \right],
\end{align}
respectively.
Note that $y$ and $\eta$ are equal when the particle is massless and when $v_z=0$.

\subsection{Experimental setup}
Here, I briefly explain the experimental setup of relativistic heavy-ion collisions by using the ALICE experiment at CERN LHC as an example.
For more details, see Ref.~\cite{ALICE:2014sbx}. 
The detector of each experimental collaboration is located along a beam line of the LHC accelerator with about $27$ km circumference, and the ALICE experiment, which is dedicated to heavy-ion collisions, is one of them.

\begin{figure}
    \centering
    \includegraphics[bb=0 0 960 540, width=1.0\textwidth]{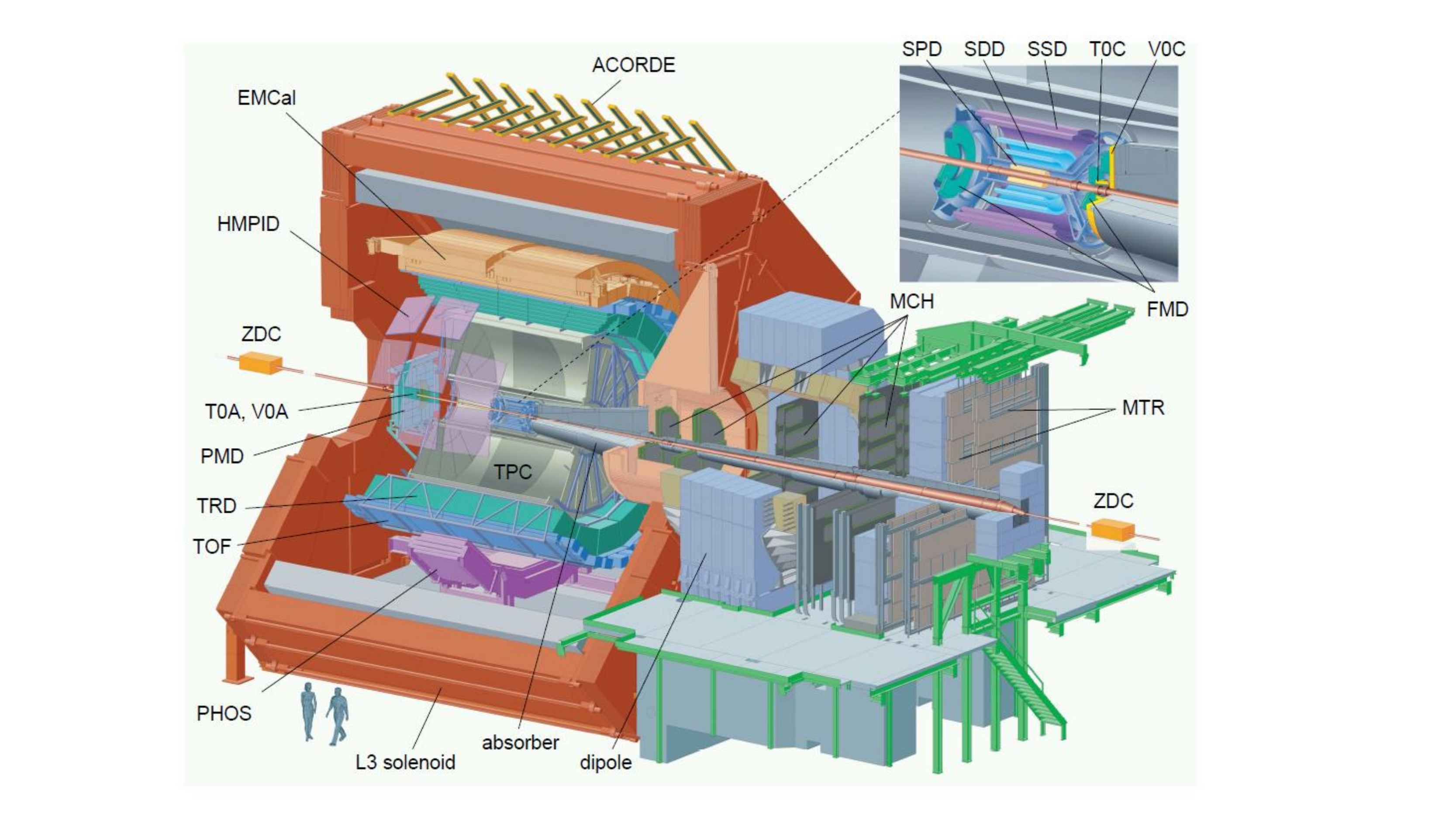}
    \caption{The ALICE experiment at the CERN LHC. The beam pipe is surrounded by the barrel-shape detector. Each name of apparatus/detector is shown. Human silhouettes are illustrated to show the size difference. The figure is taken from Ref.~\cite{ALICE:2014sbx}}
    \label{fig:ALICEsetup}
\end{figure}

Figure \ref{fig:ALICEsetup} illustrates the apparatus of the ALICE experiment. The beam pipe running horizontally is surrounded by the barrel-shaped detector. As one sees from the human silhouettes illustrated in Fig.~\ref{fig:ALICEsetup}, the apparatus is huge -- its overall dimension is $16\times16\times26 \ \mathrm{m}^3$ and weight is about $10$ kt.
The apparatus consists of various types of detectors, which can be categorized into three: 1) central-barrel detectors, 2) forward detectors, and 3) the forward muon spectrometer.
In the following, I briefly explain about 1) and 2) which are related to the analysis in this thesis:
to make a proper comparison between experimental data and results from theoretical models, one should perform the same/equivalent analysis technique.

The central-barrel detectors are, as the name tells, located central around a collision point where the QGP is expected to be generated.
Thus, they measure particles that is directly used for QGP studies, which means that they are the heart of the ALICE experiment.
Their main functions are the measure of momentum and particle identification.

The forward detectors are located along beam line but in forward and backward regions apart from the collision point.
They are used for the determination of centrality, i.e., how central/peripheral a heavy-ion collision happens, by measuring particle productions at forward and backward regions.
For more detailed discussions, see Sec.~\ref{subsec:ParticleProductions}.

\subsection{Particle productions}
\label{subsec:ParticleProductions}
In the relativistic heavy-ion experiments, the simplest observables are counts of particle productions and energy produced in a reaction.
While it is a simple and basic measurement, they provide invaluable information. Actually, the reproduction of particle production in relativistic heavy-ion collision experiments contains a significant difficulty. Most of the theoretical models, which are even built with an intention to describe experimental data, likewise Monte-Carlo event generators, do not {\it{predict}} multiplicity reported by experiments. Rather, multiplicity is an input for those models; the initial energy/entropy density profile of the QGP is often parametrized so that multiplicity is described by the models.

There are two stages of difficulty in the description of particle productions in relativistic heavy-ion collision experiments; heavy-ion and nucleon--nucleon collision levels.
First of all, a heavy-ion consists of a lot of nucleons, and a heavy-ion collision is often considered as a superposition of nucleon--nucleon collisions where interactions of nucleons inside of nuclei are neglected from the point of view of eikonal approximation \cite{Glauber:1955qq}. This picture is classically and numerically realized in, so-called, the Glauber model \cite{Miller:2007ri,Broniowski:2007nz, Loizides:2014vua}
\footnote{Note that in the original paper by R.~Glauber, the picture was not this simple and classical, which was critically pointed out by himself at Quark Matter 2005 \cite{Glauber:2006gd}
}
.
Under this model, one assumes that the cross section of several nucleon-nucleon collisions in a single nucleus-nucleus collision is given by the
cross section of {\it{inelastic}} collisions which can be experimentally obtained from $p$+$p$ collisions.
With a smooth parametrized nucleon distribution inside of the nucleus, one can obtain the transverse distribution of the number of binary (inelastic) collisions of nucleons $N_{\mathrm{coll}}$ and the number of nucleons participating in a collision $N_{\mathrm{part}}$.
The former contributes to the relatively high-$p_T$ ({\it{hard}}) productions while the latter contributes to the low-$p_T$ ({\it{soft}}) productions.
The important thing to note is that both soft and hard productions contribute to the particle production at the high-energy collisions \cite{Eskola:1988yh}. 
Assuming the energy or entropy density is proportional to $N_{\mathrm{coll}}$ or $N_{\mathrm{part}}$, one obtains the initial profile of hydrodynamics.

However, the Glauber model mainly has two issues. First, the model just tells you which nucleon collides with which projecting every nucleon onto the transverse plane at a given impact parameter. There is no information for longitudinal direction. Second, in the model, all nucleons collide at the same energy that initially the nucleon has at every collision. Thus, the energy-momentum conservation of the system is explicitly violated.
While the Glauber model mentioned above has been widely used in the heavy-ion community, it is an inevitable fact that the model has much room to be sophisticated. 
There is another picture for the initial stage of relativistic heavy-ion collisions, called the color glass condensate (CGC). In the CGC, there is no picture such as a superposition of nucleon--nucleon collisions. Instead, it is assumed that a colliding nucleus is highly occupied by gluons, so a heavy-ion collision is described as a collision of a chunk of gluons that produces fluxes of color field in-between two outgoing nuclei.
In contrast to the Glauber model, those gluons collide coherently --
those gluons strongly interact with each other.
Here, although I introduced two frequently employed initial-state pictures, it is still challenging to discriminate the pictures with current knowledge.

%Under this picture, fluctuations of cross-sections of each nucleon--nucleon collision plays an important role in description of multiplicity distribution in experiments \cite{Bierlich:2016smv}.
The second stage of the difficulty is in a level of one nucleon--nucleon collision. 
Until around the late 80s, it had been considered that the constituent quarks contribute to particle production. However, as collision energies increase in experiments, it started to be considered that sea quarks and gluons would contribute to particle production as well: more than two parton--parton scatterings tend to occur in one nucleon--nucleon collision at large collision energies.
The above picture was induced by the experimental observation of four-jet events \cite{AxialFieldSpectrometer:1986dfj}, and theoretically proposed in Ref.~\cite{Sjostrand:1987su} as {\it{multi-parton interactions (MPI)}}.
Around the late 90s, when the LHC experiment started, 
it was pointed out that the event distribution as a function of multiplicity measured in $p$+$p$ collisions at \snn[nucleon]=$540$ GeV cannot be described without multi-parton interactions in a nucleon--nucleon collision
\cite{Sjostrand:1987su}.

Therefore, the number of particle production is very simple observable, in the sense that one just needs to count the particles, but a rich physics lies behind it. 
In the following subsections, I summarize each observable related to particle production.

\begin{figure}[htpb]
    \centering
    \includegraphics[bb=0 0 960 540, width=1.0\textwidth]{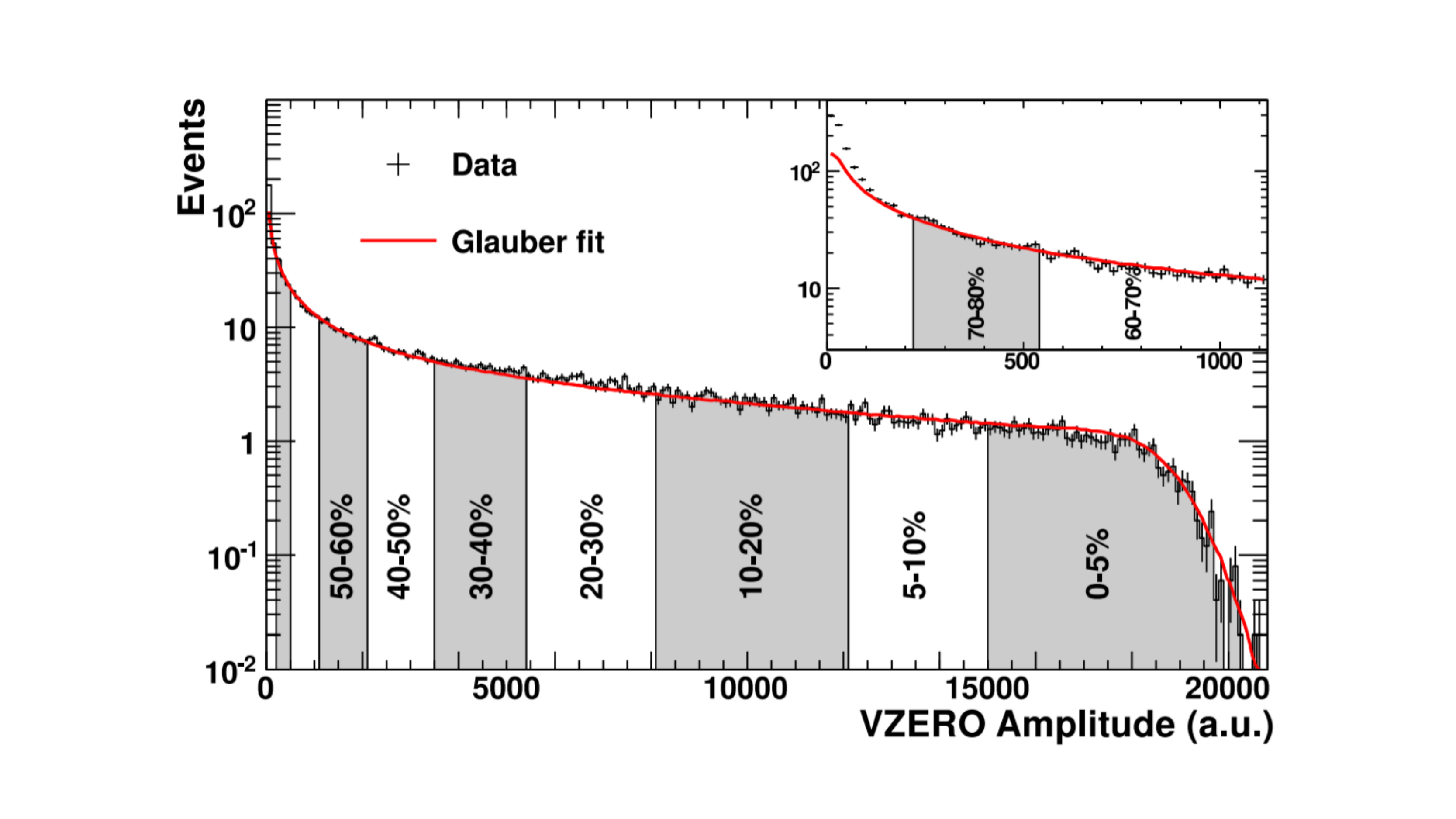}
    \caption{
    Event distribution of the total charge deposited in both of the V0 detectors (VZERO Amplitude) in $Pb$+$Pb$ collisions at \snn = 2.76 TeV reported by the ALICE experiment \cite{ALICE:2010mlf} (black histogram).
    The red curve obtained by the Glauber model is being compared.
    }
    \label{fig:VZEROAmp}
\end{figure}

\subsubsection{Multiplicity distribution}
\label{subsubsec:MultiplicityDistribution}
The word {\it{multiplicity}} means the number of final state hadrons
\footnote{
It depends on the definition of whether the hadrons should have electric charges or not.
}
produced within a certain rapidity range in a single collision.
In heavy-ion collisions, the multiplicity is controlled, roughly speaking, by the impact parameters of the colliding nuclei.
The higher the centrality of the collision, the more energy of the colliding nuclei will be dropped into the midrapidity region. In other words, more thermal matter is expected to be produced.
Therefore, in order to see physics resulting from the formation of QGP, 
observables are often calculated by dividing events into different {\it{centrality}}.

The definition of centrality differs among experiments by experiments while the concept that I mentioned above is the same.
Here, I explain the centrality classification of events that is performed by the ALICE experiments. 
In Fig.~\ref{fig:VZEROAmp}, the event distribution of the total charge deposited in both of the V0 detectors (VZERO Amplitude) in $Pb$+$Pb$ collisions at \snn = 2.76 TeV reported by the ALICE experiment \cite{ALICE:2010mlf}.
The centrality defined here is the percentage of events by sorting in descending order of the VZERO Amplitude as labeled in Fig.~\ref{fig:VZEROAmp}.
There is a reason for using VZERO Amplitude located in forward and backward rapidity regions instead of directly sorting events with multiplicity at midrapidity,
which is to avoid self-correlation.
The main interests of the QGP production are usually in physics at the mid-rapidity.
In order to see physics without having events biased, the centrality classification is performed with particle productions at forward and backward rapidity regions
\footnote{For instance, more charged particles than neutral ones would be obtained if one selects events with large number of charged particles at mid-rapidity even if physics say there is a charge symmetry.
There is a nice explanation of self-correlation in Fig.~1 of Ref.~\cite{ALICE:2018pal}.
}
.

\subsubsection{Pseudorapidity distribution}
After classifying events into different centrality bins, the next simplest observable would be pseudorapidity distribution of charged particles.
The benefit of plotting data with pseudorapidity, $\eta$, rather than rapidity, $y$, is that one does not need to identify species of particles; the former can be obtained merely from the polar angle from the beam axis while the latter requires to know the mass of the particle.

\begin{figure}
    \centering
    \includegraphics[bb=0 0 960 540, width=1.0\textwidth]{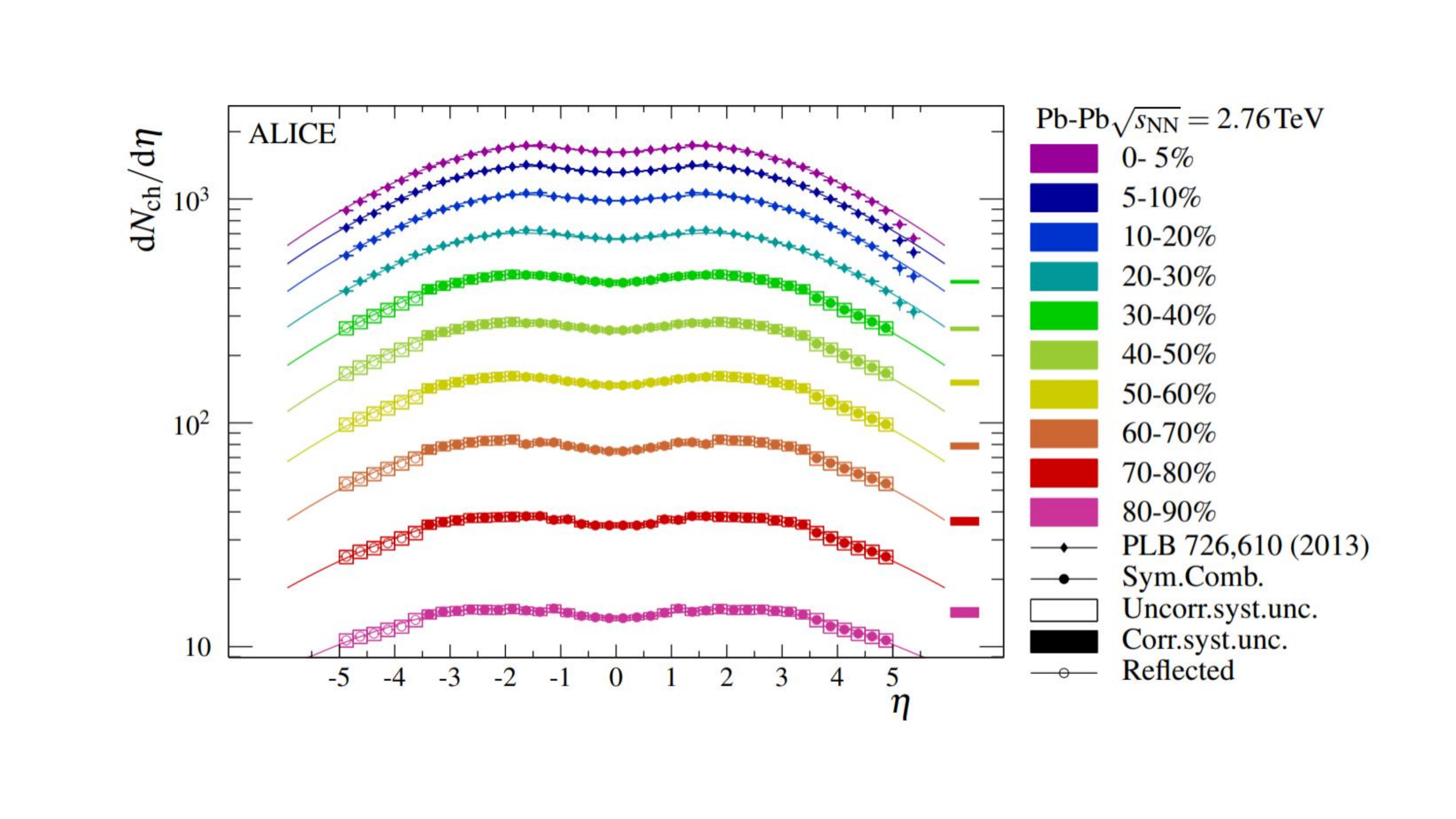}
    \vspace{-50pt}
    \caption{Pseudorapidity distribution of charged particles around mid-rapidity with different centrality obtained in $Pb$+$Pb$ collisions at \snn=2.76 TeV reported by the ALICE collaboration \cite{ALICE:2015bpk}.}
    \label{fig:Pseudorapidity}
\end{figure}

Figure \ref{fig:Pseudorapidity} shows the pseudorapidity distribution of charged particles at mid-rapidity with different centrality obtained in $Pb$+$Pb$ collisions at \snn=2.76 TeV reported by the ALICE experiments \cite{ALICE:2015bpk}.
The vertical axis, \dndeta, is the event average of the number density of charged particle produced at midrapidity, which corresponds to a mathematical notation of multiplicity.
One sees that the more central the collisions are, the more particles are produced.
At the central events, multiplicity reaches up to $\approx10^3$; most of them are charged pions which are the lightest hadrons.
It should be also noted that the beam rapidity is $|y_{\mathrm{beam}}|\approx|\eta_{\mathrm{beam}}|\approx8$ with the given collision system. The rapidity distribution spans within the beam (pseudo)rapidity.

\subsubsection{Particle ratios}
Once identification of particle species is performed,
one sees more detailed information about the generated system.
The thermal origin of particles can be assessed with the change of particle ratios with particle mass. This is because,
if the system is under equilibrium, particle production can be expressed with the Bose-Einstein or Fermi-Dirac distribution, $1/\left[\exp(\sqrt{m^2+p^2}/{T}) \pm 1\right]$.
\begin{figure}
    \centering
    \includegraphics[bb=0 0 960 540, width=1.0\textwidth]{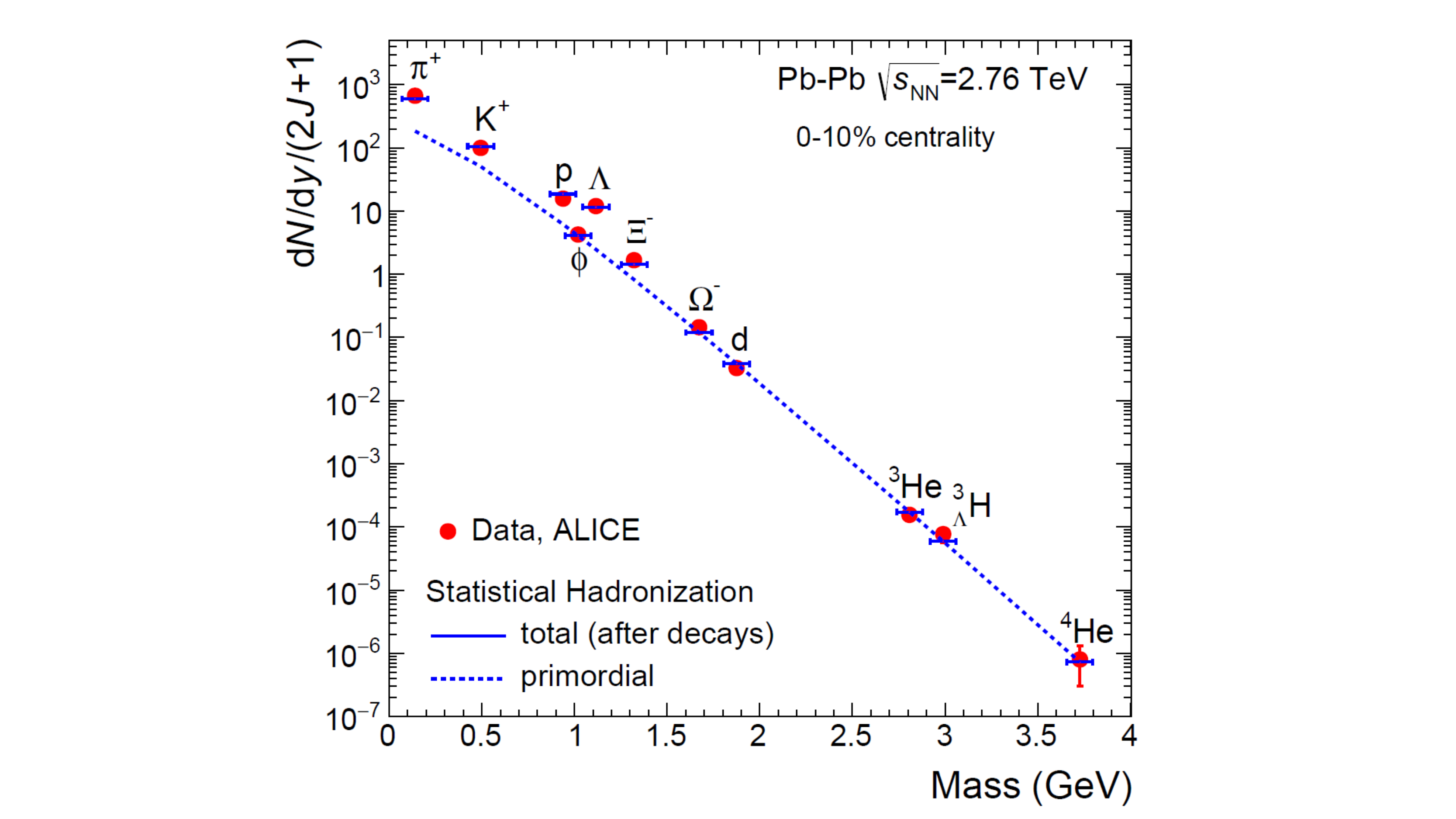}
    \vspace{-20pt}
    \caption{Multiplicity divided by the spin degeneracy, (2J+1), of various hadron species. The experimental results of centrality 0-10\% in $Pb$+$Pb$ collisions at \snn=2.76 TeV (red circles) are being compared with a statistical hadronization models (blue lines). Figure is taken from Ref.~\cite{Andronic:2017pug}.}
    \label{fig:RatioStatistic}
\end{figure}
Once the particle ratios are described by integrating the Boltzmann distribution, one can also extract the temperature of the system.
Therefore, the particle ratio is one of the most important fundamental observables to study the equilibrium matter, QGP.

Figure \ref{fig:RatioStatistic} shows 
multiplicity divided by the spin degeneracy, $(2J+1)$, as a function of mass. The experimental results of centrality 0-10\% in $Pb$+$Pb$ collisions at \snn=2.76 TeV are compared with a statistical hadronization model \cite{Andronic:2017pug}.
Under the statistical hadronization model, the picture of particle production is very simple: it is assumed that all particles are thermally produced and the particle yields are controlled by particle mass and system temperature.

Despite its simplicity, the model nicely describes the experimental data very well \cite{Andronic:2005yp, Andronic:2017pug} as one sees in Fig.~\ref{fig:RatioStatistic}.
The temperature obtained in the model here is $156.5\pm1.5$ MeV \cite{Andronic:2017pug}.
Therefore, this clearly demonstrates the possibility of that 
generated matter in relativistic heavy-ion collisions is under thermal and chemical equilibrium.

\subsubsection{Transverse momentum distribution}
\label{subsec:INTRO_TransverseMomentumDistribution}
\begin{figure}
    \centering
    \includegraphics[bb=0 0 960 540, width=1.0\textwidth]{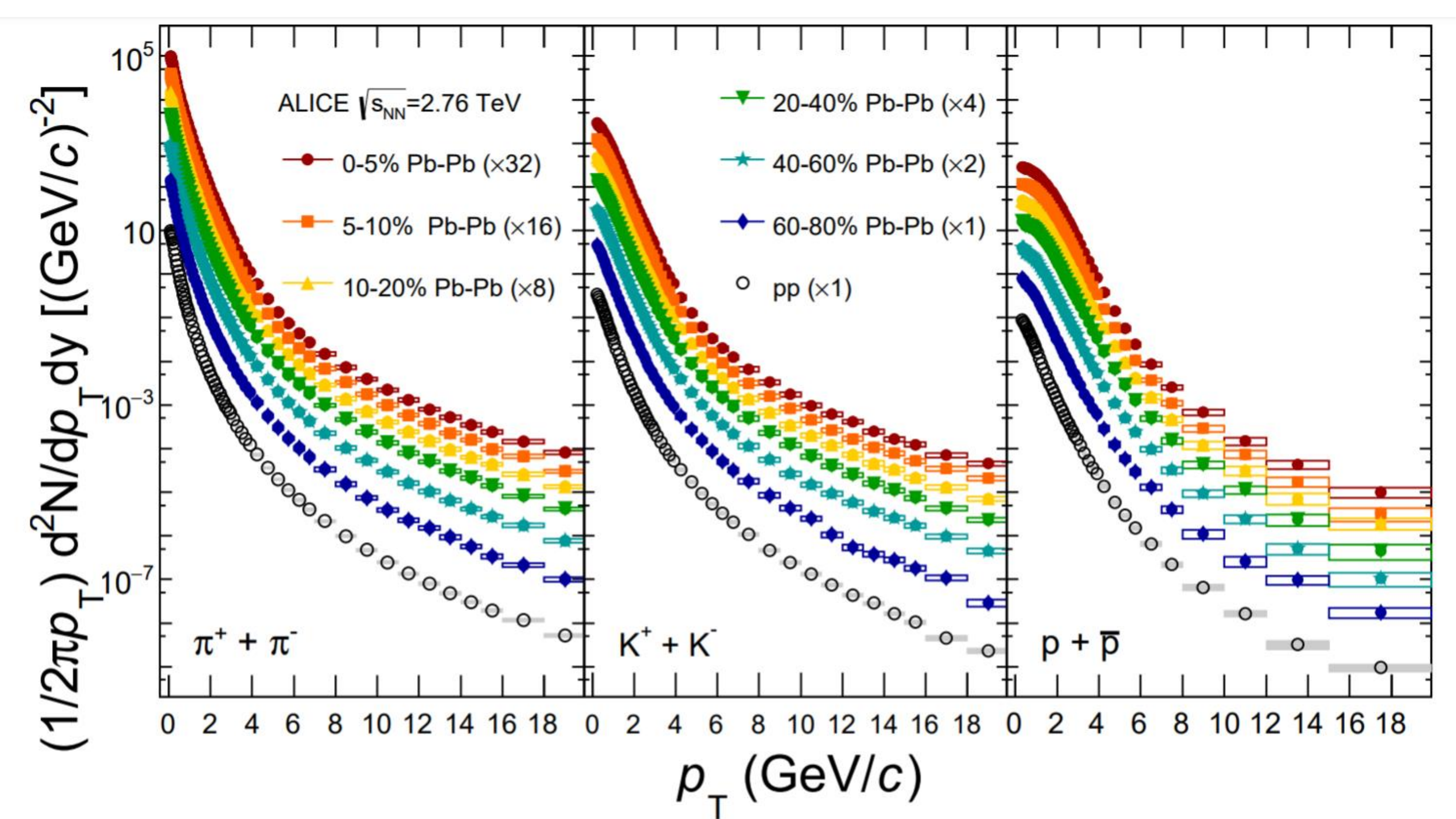}
    \vspace{-20pt}
    \caption{Transverse momentum distribution of charged pions (left), kaons (middle), and
    (anti)protons (right) measured in $Pb$+$Pb$ at \snn=2.76 TeV reported by the ALICE experiment \cite{ALICE:2015dtd}.}
    \label{fig:PTspectra}
\end{figure}

Transverse momentum, $p_T$, is the basic quantity to investigate in high-energy physics because that is almost acquired after the collision of nuclei and reflects the dynamics of the initial and final state of the reaction.
Figure \ref{fig:PTspectra} shows $p_T$ distributions of pions, kaons, and protons with different centrality in $Pb$+$Pb$ collisions at \snn=2.76 TeV reported by the ALICE experimental data \cite{ALICE:2015dtd}.
Results of $p$+$p$ collisions are also plotted as a reference.
The vertical axis is $(1/2\pi p_T)~d^2N/dp_T dy \ \mathrm{GeV^{-2}}$ which is the event average of the number of particles counted in $p_T$ bins in a certain mid-rapidity range.
The factor of $2\pi$ originates from integration for azimuthal angle and $p_T$ in the denominator is a Jacobian of the expression in the radial direction.

Looking at the $p_T$ spectra of pions in Fig.~\ref{fig:PTspectra}, the shapes of the spectra for $p_T\leq3$ look straight while the shapes seem to change for $p_T\geq3$ on the semi-log plot.
The linear shape on semi-log plots where the y-axis has a log scale indicates that the spectra are almost exponential functions.
Thus, it can be interpreted that the low-$p_T$ particles are produced from thermal matter, QGP, while the high-$p_T$ ones are produced from different production mechanisms such as hard scattering of partons at the initial state of a reaction.

It should also be noted that the shape of spectra differs for different particle species: spectra of heavier particles ($m_\pi < m_{K} < m_p$) have a thicker shoulder at low $p_T$.
This can be understood that if all particles are thermally produced from a QGP fluid with a certain velocity, heavier particles acquire more momentum. Thus, due to the existence of flow in expanding QGP, $p_T$ spectra in the low $p_T$ are pushed slightly towards high $p_T$ and result in the thick shoulder around $p_T\approx1$-$2$ GeV
.
Needless to say, the thickness of shoulder also depends on centrality because more flows are produced at more central events.
The above behaviors are realized in several hydrodynamic models
\cite{Kolb:2001qz,Hirano:2002ds,Schenke:2010nt,Pang:2012he,Shen:2014vra,Niemi:2015qia}
.

Compared to the $Pb$+$Pb$ collision results, one would notice that spectra from $p$+$p$ collisions are qualitatively different from those in $Pb$+$Pb$ collisions.
Especially, there is no shoulder seen for $p_T$ spectra of protons which indicates that there is not so much flow originating from the QGP in $p$+$p$ collisions.

%\subsubsection{Mean transverse momentum}
%\begin{figure}
%    \centering
%    \includegraphics[bb=0 0 960 540, width=1.0\textwidth]{chapters/figure/intro/MeanPT.pdf}
%    \vspace{-20pt}
%    \caption{Caption}
%    \label{fig:MeanPT}
%\end{figure}

\subsection{Collectivity}
\label{subsec:Collectivity}
Under the picture that QGP fluids are generated in high-energy heavy-ion collisions,
collectivity of produced particles arises as a result of hydrodynamic behavior of the QGP.
Because final hadronic productions from the QGP is considered to reflect that behavior,
collectivity is said to be one of the important proxies of QGP formation.

When one says ``collectivity is seen in particle productions'', it means that there is a correlation in momentum space for produced particles.
In intermediate central events (40-60 \% in centrality), generated QGP fluids are expected to have an almond shape when one sees them in  the transverse plane.
The almond-shaped QGP has an azimuthally anisotropic geometry, which leads to azimuthal anisotropy of pressure.
Since fluid velocity is generated due to the pressure gradient, the azimuthal momentum anisotropy is inherited by the emitted particles from the pushed-out fluids.

However, the origin of the collectivity is not limited to the QGP fluids. There are various factors that would produce the correlation in momentum, which needs some effort to distinguish from the one that originating from hydrodynamics.
For example, high $p_T$ productions produced from a single hard $2\rightarrow2$ scattering called jets have a back-to-back correlation, i.e., $\Delta\phi\approx\pi$.
This is because of the momentum conservation where the initial partons inside of nuclei have very small $p_T$ compared to the scattered out partons.

\subsubsection{Longitudinal correlation}
\label{subsec:LongitudinalCorrelation}
\begin{figure}
    \centering
    \includegraphics[bb=0 0 960 540, width=1.0\textwidth]{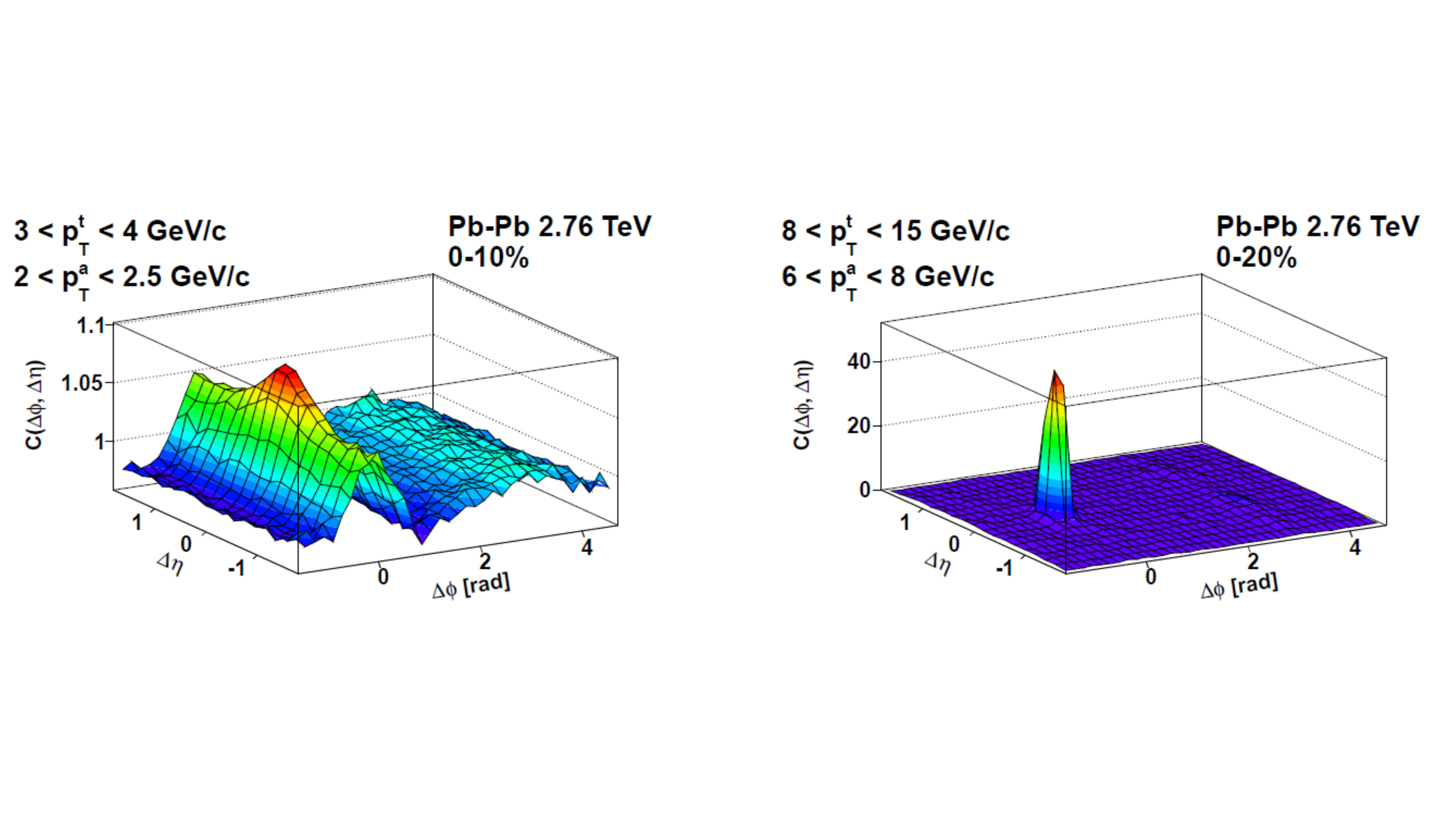}
    \vspace{-50pt}
    \caption{Two particle correlation functions for central $Pb$+$Pb$ collisions at \snn=2.76 TeV in low to intermediate (left) and in high transverse momentum regimes (right). Figure is taken from Ref.~\cite{ALICE:2011svq}.}
    \label{fig:Ridge}
\end{figure}
I would like to start by introducing the observable which has been analyzed in relativistic nuclear collisions as one of the powerful techniques to characterize the properties of particle productions, 
that is the two-particle correlation.
The two-particle correlation measures the distribution of differences in angles, $\Delta \phi$ and $\Delta \eta$, between particle pairs.
It is often the case that the particle pairs consists of a trigger and an associate where each of them is picked up from certain $p_T$ ranges.
Before seeing actual experimental data, here I explain the general view of two particle correlation. 
In $p$+$p$ collisions, two-particle correlation has a strong peak around $\Delta \phi\approx0$ and $\Delta \eta\approx0$, which is almost dominated by the ``near-side'' jet peak.
Because one hardly scattered parton fragments into hadrons inheriting their mother parton's momentum, the resultant momentum correlation becomes very strong.
There appears another peak around $\Delta \phi \approx \pi$ dominated by the ``away-side'' jet produced back-to-back against the near-side one. The latter has a weaker correlation and broadened to wide $\Delta \eta$. 
One of the main factors is the fluctuations in the longitudinal momentum distributions of partons inside of the nuclei.
If one sees two particle correlations in heavy-ion collisions, an additional structure called ``ridge'' \cite{STAR:2009ngv} (which exactly looks like so) appears around $\Delta \phi\approx0$ along $\Delta \eta$.
Since its first report of the ridge structure from an experiment, several theoretical interpretations have been made.
While there exist various interpretations,
the key factor in producing the structure is a homogeneous initial state of QGP fluids along a wide range in the longitudinal direction such as color flux tube followed by hydrodynamic expansion.

Figure \ref{fig:Ridge} shows examples of two particle correlations for central $Pb$+$Pb$ collisions at \snn=2.76 TeV \cite{ALICE:2011svq}. Right and left results correspond to two particle correlations in low to intermediate ($3<p_T^t<4$ GeV, \enskip $2<p_T^a<2.5$ GeV, where $p_T^{t}(p_T^{a})$ is $p_T$ of trigger (associate) particle) and high $p_T$ ($8<p_T^t<15$ GeV, \enskip$6<p_T^a<8$ GeV) regimes, respectively.
The former is the regime where particle production from QGP fluids is assumed to be dominated while the latter is the one where jets are dominated.
The long-ranged ridge is seen in the low $p_T$ while the structure disappears in the high $p_T$. 
It should be also noted that the scale of the vertical axis is different by one order,
which means that a much stronger correlation is caused by the near-side jets in the right case.

\subsubsection{Fourier coefficient in azimuthal angle}

\begin{figure}
    \centering
    \includegraphics[bb=0 0 960 369, width=1.0\textwidth]{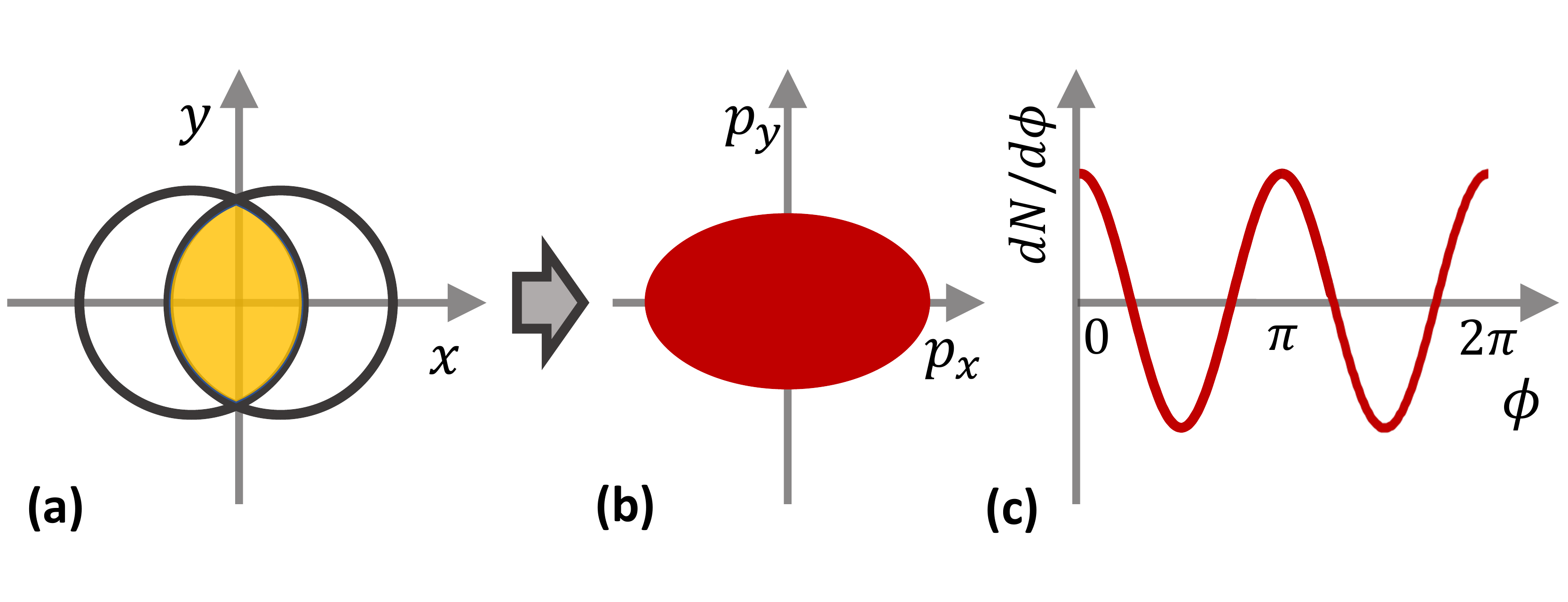}
    \vspace{-30pt}
    \caption{(a) Geometrical anisotropy in initial profile of the QGP due to a non-central collision. (b) Momentum anisotropy generated from pressure gradient in initial profile due to hydrodynamic evolution. (c) Amplitude of particle production expanded as a function of azimuthal angle.}
    \label{fig:QGPanisotropy}
\end{figure}
As we saw in the previous subsection, one can visualize momentum correlations by plotting two particle correlations.
In the following, I explain the frequently used technique to quantitatively evaluate strength of momentum correlations.
The idea of the method is as follows.
Figure \ref{fig:QGPanisotropy} shows the schematic picture of the transfer of anisotropy.
Due to the non-central collision of nuclei, generated matter in the overlap region has a geometrical anisotropy as shown in (a).
The geometrical anisotropy gives anisotropical pressure gradient as it is depicted in (b).
Pressure gradient produces velocity in hydrodynamic expansion.
Final particle productions inherit the anisotropically generated velocity of the QGP fluids.
When one expands the particle productions as a function of azimuthal angle, one would see the cosine-shape frequency if $\phi=0$ is taken to be the impact parameter axis as shown in (c). 

When one sees a frequency, quantification can be done by performing Fourier expansion:
\begin{align}
    \frac{dN}{d\phi} \propto \left[ 1+2v_1\cos{(\phi-\Psi_1)} + 2v_2 \cos{(2(\phi-\Psi_2))} + \cdots\right],
    \label{eq:Fourier}
\end{align}
where $v_1, v_2, \cdots$ are Fourier coefficients of $n$th harmonics, and $\Psi_1$, $\Psi_2$, $\cdots$ are angles of event planes which ideally correspond to the axis of impact parameter
\footnote{
In Fig.~\ref{fig:QGPanisotropy}, event plane is in the direction of beam axis, i.e., $\Psi=0$.
}
.
The sine terms do not appear because of the symmetry with respect to $x$-axis.
In the case of a non-central events shown in Fig.~\ref{fig:QGPanisotropy},
$v_2$ component, so-called {\it{elliptic flow}} \cite{Ollitrault:1992bk} becomes dominant,
and, because of its clear rotational symmetry of order 2 where the situation is too much idealized, odd orders of anisotropic coefficients cannot be finite.
Odd orders or higher orders than 2 become finite when fluctuations inhere in systems.
For instance, it has been understood that finite $v_3$ is experimentally obtained due to geometrical fluctuations in initially produced matter. 

Because the number of produced particles is finite, note that one cannot take the integral to obtain the Fourier coefficients. Instead, one needs to estimate anisotropy by taking averages against the produced particles in events.
Historically, so-called {\it{event plane method}} \cite{Poskanzer:1998yz} to obtain Fourier coefficients
has been used as a conventional method to quantify anisotropy.
Under this method, anisotropic flow coefficients are
defined as
\begin{align}
    v_n  = \frac{\langle \sum_i^{M_j} \cos{(n(\phi_{i,j} - \Psi_{\mathrm{EP}, j}))}  \rangle}{\langle M_j \rangle},
    \label{eq:EventPlaneMethod}
\end{align}
where the index of $i$ is for particles produced in a single event and
$j$ stands for events. The angle bracket, $\langle \cdot \rangle$, means event average,
$\phi_{ij}$ is an azimuthal angle of $i$-th particle generated in $j$-th event,
and $M_j$ is the number of particles produced in $j$-th event.
Here, note that $\Psi_{\mathrm{EP}, j}$ stands for the event plane
estimated event-by-event,
thus, the inside of the bracket of Eq.~\refbra{eq:EventPlaneMethod} is constructed to be invariant event-by-event.
However, there is a vulnerability in the estimation of $\Psi_{\mathrm{EP},j}$
from experimental data
because there can be momentum correlations of produced particles
which originates from something else rather than the geometry of a collision.

Another method to measure strength of anisotropy proposed after the event-plane method is called {\it{multi-particle cumulant method}} \cite{Borghini:2000sa,Borghini:2001vi}.
With this method, the event-by-event estimation of event planes are not required.
In general expression, event average of $m$-particle correlation is defined as
\begin{align}
\label{eq:multi_particle_correlation}
 \llangle m \rrangle = \llangle e^{i(n_1\phi_1 + n_2\phi_2 + n_3\phi_3 + \cdots + n_m\phi_m)} \rrangle.
\end{align}
The outer bracket of $\llangle \cdot \rrangle$ denotes average by events and particle pairs. 
A possible but time consuming way to calculate Eq.~\refbra{eq:multi_particle_correlation} is taking all possible $m$-particle combinations from one events. 
To calculate this on a code, one can naively construct $m$ loops for $N$-particle produced in a single event. However, when it comes to central heavy-ion collision events where $N\sim \mathcal{O}(10^3)$ particles are produced,
$10^3m$ loops are required.
The numerical costs of this are enormous for 
higher-order correlation ($m>2$).

To circumvent the above situation, a method called {\it{Q-cumulant}} \cite{Borghini:2000sa} is employed.
Under this method, one can calculate particle correlations via the Q-vector: 
\begin{align}
\label{eq:Qvector}
    Q_n=\sum_{j}^N e^{in\phi_j}.
\end{align}
For instance, $2$-particle correlation per particle pair from one single event, $\langle 2 \rangle $, can be expressed as,
\begin{align}
\label{eq:TwoParticleCorrelation}
    \langle 2 \rangle = \frac{1}{{}_\mathrm{N}P_2}\sum_{i\neq j}^N e^{in(\phi_i -\phi_j )} = \frac{|Q_n|^2 - N}{{}_\mathrm{N}P_2} ,
\end{align}
where ${}_NP_2$ is a permutation of 2 out of $N$ and $|Q_n|^2 = Q_n^* Q_n = \sum_{k,l}^N e^{in(\phi_k - \phi_l)}$. From the first to the second equality, the substitution of $N$ appears to subtract self-correlation such as $k=l$ in $|Q_n|^2$. 
Finally, event average of $\langle 2 \rangle$ is obtained as
\begin{align}
    \llangle 2 \rrangle = \frac{\sum_i^{N_{\mathrm{ev}}} {}_{N_i} P_2 \enskip \langle 2 \rangle_i }{\sum_i^{N_{\mathrm{ev}}} {}_{N_i} P_2},
\end{align}
where index $i$ stands for events.
Thus, one can interpret that $\llangle 2 \rrangle$ is obtained from the 2-particle correlation from one single event which is weighted by the number of particle pairs constructed in the single event.

Once one obtains $m$-particle correlation, one can relate them with {\it{cumulant}}.
In general, one can decompose $m$-particle correlation functions into
factorized terms (independent one particle function) and terms of all possible correlations of fewer ($l<m$) particles, and genuine $m$-particle in which all $m$-particles are correlated.
The last term is the so-called cumulant.
For the case of $2$- and $4$- particle correlation, cumulants are obtained as
\begin{align}
    \label{eq:Cumulant_C22}
    c_n\{2\} &= \llangle 2 \rrangle, \\
    \label{eq:Cumulant_C24}
    c_n\{4\} &= \llangle 4 \rrangle - 2 \llangle 2 \rrangle^2,
\end{align}
respectively.
In order to relate between Fourier coefficients and the cumulants,
anisotropic flow coefficients calibrated from the Q-cumulant method are defined as:
\begin{align}
\label{eq:MultiParticle_vn2}
    v_n\{2\} & \equiv \sqrt{c_n\{2\}}, \\
\label{eq:MultiParticle_vn4}
    v_n\{4\} & \equiv   \sqrt[\leftroot{-1}\uproot{2}4]{-c_n\{4\}}.
\end{align}
Note that, in reality, those anisotropic flows are not equivalent to those obtained from event plane method defined in Eq.~\refbra{eq:EventPlaneMethod}.
These definitions are coming from the following relations, for example the case of 2-particle correlation,
\begin{align}
    \llangle 2 \rrangle = \llangle  e^{in(\phi_i - \phi_j)} \rrangle = \langle \langle e^{in(\phi_i - \Psi)} \rangle  \langle e^{-in(\phi_j - \Psi )} \rangle \rangle \approx \langle v_n^2 \rangle.
\end{align}
Thus, the {\it{order}} of $v_n$ can be estimated as $\llangle 2 \rrangle \approx \langle v_n \rangle^2$
.
With the same analogy, $\llangle 4 \rrangle \approx \langle v_n \rangle^4$ with the assumptions that the event-by-event fluctuations are small enough compared to flows
\footnote{
Because the variance of $v_n$ is $\sigma = \langle v_n^2 \rangle - \langle v_n \rangle^2$, anisotropic coefficients obtained from 2-particle and 4-particle correlation contain the absolute value of $\sigma$ \cite{Bilandzic:2012wva} while it is ignored in this simple explanation.
}
.
With these relations, one can easily see the Eq.~\refbra{eq:MultiParticle_vn2} and \refbra{eq:MultiParticle_vn4} can estimate the value of original $v_n$ obtained in the event plane method.
For more details, see Refs.~\cite{Bilandzic:2012wva,Murase:2012xxx,Gajdosova:2018zrt}.

\begin{figure}
    \centering
    \includegraphics[bb=0 0 960 540, width=1.0\textwidth]{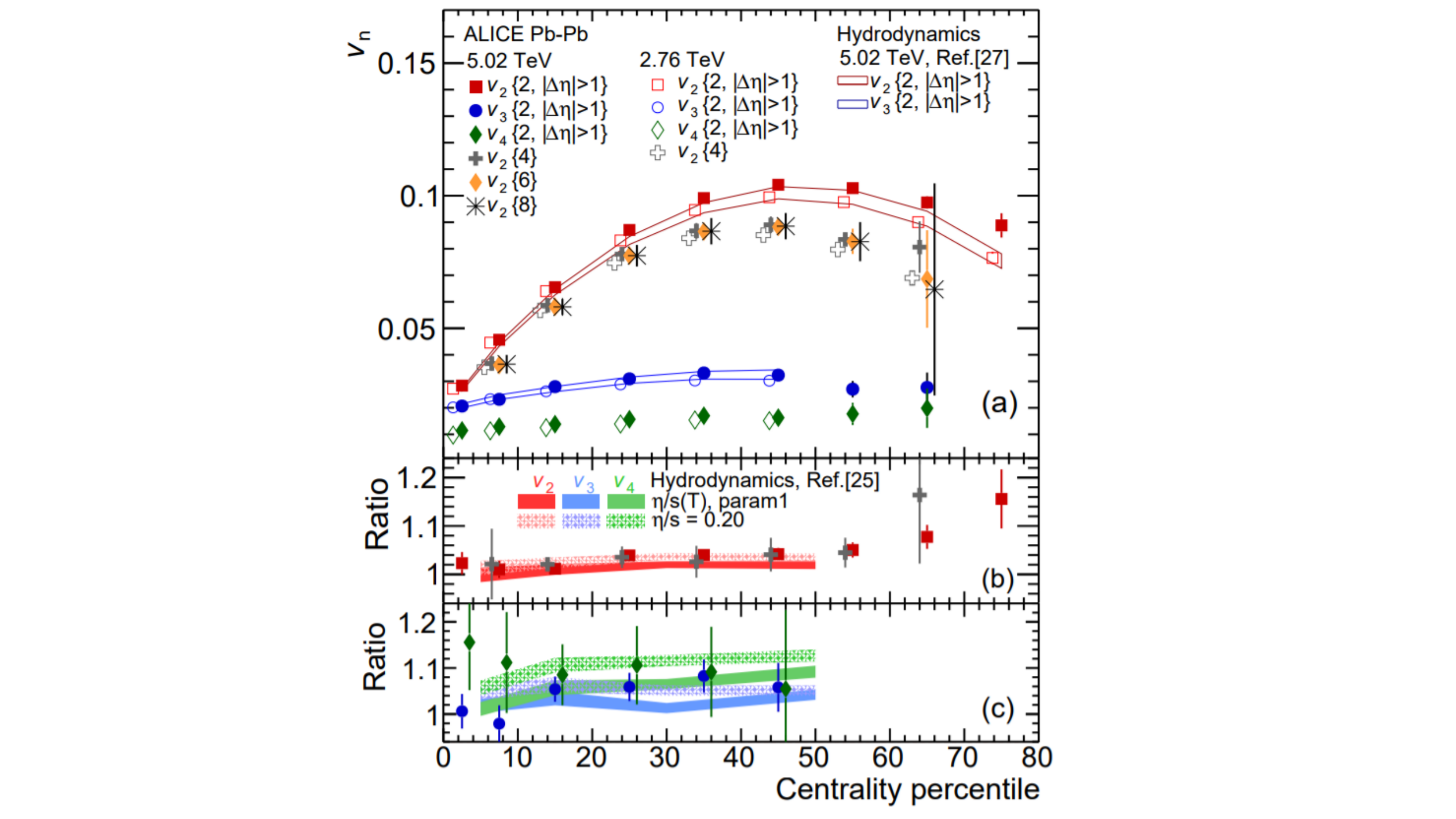}
    \vspace{-20pt}
    \caption{(a) Fourier coefficients $v_n$ from Q-cumulant method as a function of centrality in $Pb$+$Pb$ collisions at \snn = 2.76 (open) and 5.02 (filled) TeV reported from the ALICE collaboration. Two particle correlations with eta gap ($\Delta \eta>1$) and multi-particle correlations are shown.
    Fig. (b) and (c) show ratios of $v_n$ between 2.76 and 5.02 TeV.
    Ratios of $v_2\{2, \Delta \eta>1\}$ (red), $v_2\{4\}$ (gray), $v_3\{2, \Delta \eta > 1\}$ (blue), and $v_4\{2, \Delta \eta > 1\}$ (green) are plotted.
    Comparisons with theoretical calculations based on hydrodynamics (bands) are also made for references. Figure is taken from Ref.~\cite{ALICE:2016ccg}.
    }
    \label{fig:Anisotropy}
\end{figure}

Figure \ref{fig:Anisotropy} (a) shows 
Fourier coefficients $v_n$ from Q-cumulant method as a function of centrality in $Pb$+$Pb$ collisions at \snn = 2.76 and 5.02 TeV reported from the ALICE collaboration \cite{ALICE:2016ccg}. Two particle correlations with eta gap ($\Delta \eta>1$) and multi-particle correlations are shown. Comparisons with theoretical calculations based on hydrodynamics are made for references. 
I explain the meaning of imposing eta gap in Sec.~\ref{sec:RESULTS_AnisotropicFlows} and Appendix.~\ref{sec:APPENDX_SubEvent},
and I do not touch them here since it is not essential in this result.
First, one sees that $v_2\{2\}$ show maximum around centrality $40$-$60$\%,
which is consistent with the picture that the almond shape initial matter generated in peripheral collisions produce large 2nd order Fourier coefficients.
Second, one would see that there is an ordering of $v_n$ such as $v_2>v_3>v_4$ for the entire centrality bins,
and which is quite universally observed behavior for heavy-ion collisions
\footnote{
Note that this is not an obvious result.
For instance, it is known that the values of $v_2$ and $v_3$ become comparable at ultra central collisions events \cite{CMS:2013bza,Shen:2015qta,Luzum:2012wu,Carzon:2020xwp}
.
}
.

Figure \ref{fig:Anisotropy} (b) shows ratios of $v_2\{2, \Delta \eta>1\}$ and $v_2\{4\}$ between 2.76 and 5.02 TeV.
Figure \ref{fig:Anisotropy} (c) shows the corresponding ratios of $v_3\{2, \Delta \eta > 1\}$ and $v_4\{2, \Delta \eta > 1\}$.
One sees that each coefficient slightly increase with increasing collision energy as an overall tendency.
However, it should be noted that the increase is largest for $v_4$ among $v_2$, $v_3$, and $v_4$.
The same behavior is seen in the hydrodynamic calculations, and it is expected that the increase of higher order anisotropic coefficients such as $v_4$ is more sensitive to shear viscosity of QGP fluids \cite{Niemi:2015voa}.
This is one of the profits to see higher order Fourier coefficients in heavy-ion collision analysis.

\subsection{High-$p_T$ productions}
\label{subsec:HighPTProductions}
\begin{figure}
    \centering
    \includegraphics[bb=0 0 960 540, width=1.0\textwidth]{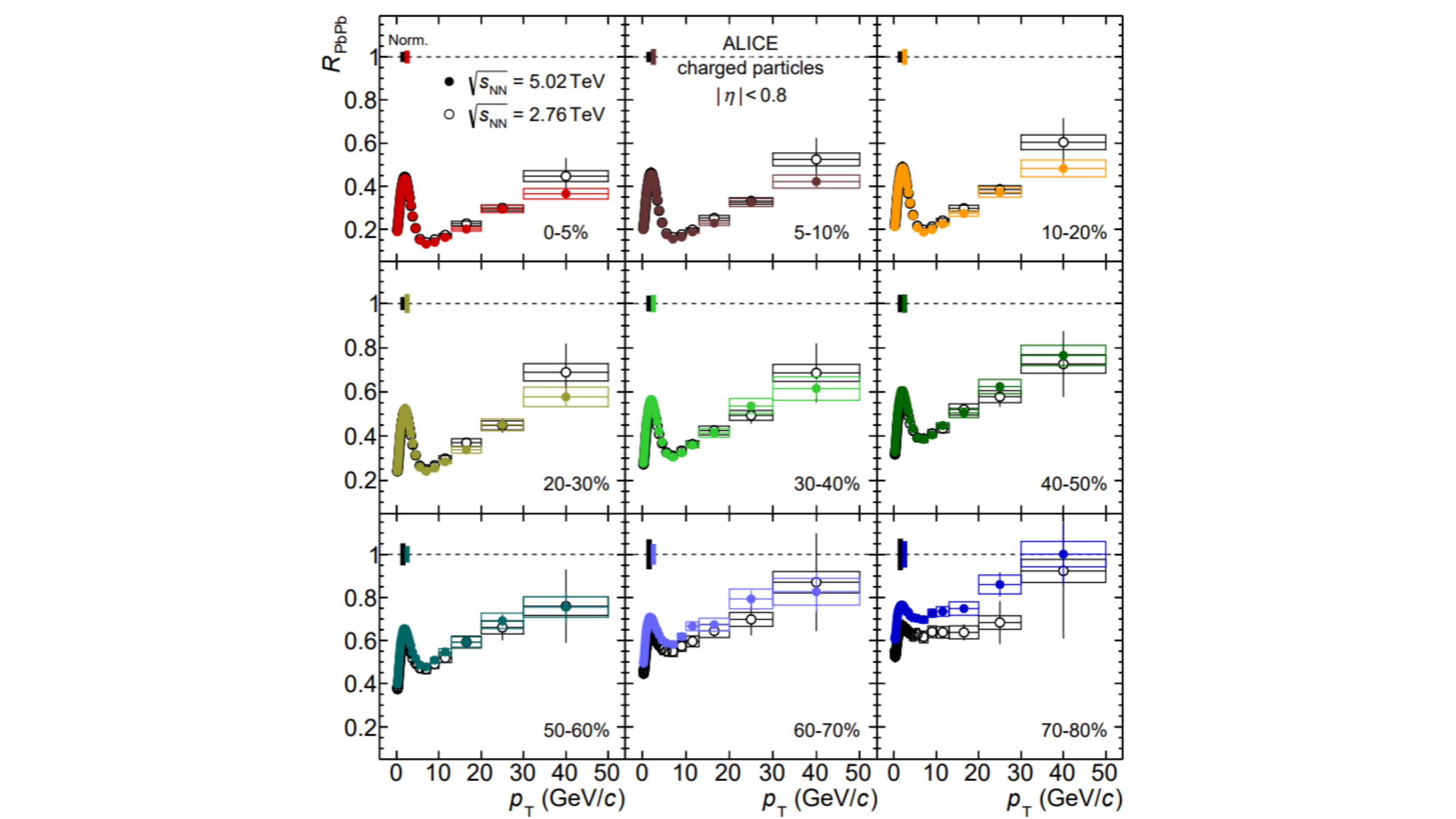} 
    \caption{Transverse momentum dependence of nuclear modification factors in various centralities reported by the ALICE Collaboration in $Pb$+$Pb$ collisions at \snn = 5.02 (filled) and 2.76 (open) TeV \cite{ALICE:2018vuu}.}
    \label{fig:NuclearModification}
\end{figure}

In relativistic heavy-ion collisions, high $p_T$ particles are produced from other sources too, which is different from particle productions from the QGP. The high $p_T$ particles originate from hard scatterings (mostly 2$\rightarrow$ 2) of partons inside of nucleons at the first contact of nuclei.
In such a scattering, large momentum transfer occurs, and quarks are scattered out into large polar angles. 
The energetic partons scattered out cascadingly emit partons traversing a vacuum or the QGP.
At some point, the collimated partons turn into hadrons. Hadron distributions produced from one energetic parton can be described with so-called fragmentation function.
The partonic or hadronic collimation is called {\it{jet}}. The particle productions or parton shower of jets are well studied in QCD because, unlike the QGP, a large energy scale of jets makes coupling constant small enough to be calculated by pQCD.
It is generally considered that there would be some contributions of jets above $p_T\sim 3$-$4$ GeV in final state hadrons.

%1. Jets are prove of QGP
%2. What is the Jet quenching -- how they energy-loss
%2. What can one get seeing jet quenching/modifications.
Jets as energetic partons produced in heavy-ion collisions play a role in probing the QGP.
Seeing the modification of jet spectra or structures due to the existence of the QGP fluids, 
one can obtain information on the transport properties of the QGP. 
This sort of study of QGP properties with jets is often called a {\it{jet tomography}} from an analogy of tomography used in medical fields -- human body is the QGP fluids, and X-rays are jets.

There are two main factors for the modification of jet spectra or structures: 1) radiative energy loss induced by the medium, and 2) collisional energy loss with particles constituting the medium.
The former is due to the Bremsstrahlung (radiation of a particle), and the latter is due to $2\rightarrow2$ scattering process.
It should be worth emphasizing that these are strong interactions described by QCD -- jets as energetic partons energy-loss seeing the degree of freedom of quarks and gluons in QGP medium.
As a consequence of these processes, jets lose their energy and momentum in the medium, and this is the very phenomenon often referred to as {\it{jet quenching}} and considered to be one of the signals of the formation of QGP
\footnote{It should be noted that there is another source of energy loss of jets due to strong interactions, which is color confinement.
When there are one quark and one anti-quark in a vacuum with a certain distance between them, attractive force works originating from color confinement. 
Thus, if either of the ones runs apart from another, they are decelerated. This kind of phenomenon should usually happen in jets in the middle of hadronization processes. However, because this should happen in every collision system including $Pb$+$Pb$, $p$+$p$ or even $e^+$+$e^-$ collisions, one can discriminate the jet quenching caused by the QGP medium by comparing results one to another system.
}
.

Thinking that the modification of jets is caused by the partial thermalization of jets due to the interaction with QGP medium, for instance, one can obtain thermalization time, diffusion coefficient, or typical energy scale of the thermalization of jets etc. which can be associated with transport coefficients of the QGP.

\subsubsection{Energy loss of jets, $R_{\mathrm{AA}}$}
One of the simplest observables of jet modification is {\it{nuclear modification factor}}, $R_{\mathrm{AA}}$,
\begin{align}
\label{eq:Definition_RAA}
    R_{\mathrm{AA}} = \frac{\langle d^2N^{AA}/dp_T d\eta \rangle}{ \langle N_{\mathrm{coll}}\rangle \langle d^2N^{pp}/dp_Td\eta \rangle},
\end{align}
where $\langle \cdot \rangle$ is an event average with in a certain centrality bin, $d^2N^{AA}/dp_T d\eta$ ($d^2N^{pp}/dp_T d\eta$) is $p_T$ distribution from heavy-ion (proton--proton) collisions, $\langle N_{\mathrm{coll}}\rangle$ is the number of binary collisions of nucleons estimated with the Glauber model.
As one can read off from the definition in Eq.~\refbra{eq:Definition_RAA}, 
 $R_{\mathrm{AA}}$ expresses how $p_T$ spectra of particles from heavy-ion collisions are deviated from the superposition of nucleon--nucleon collisions.

%HERE

Figure \ref{fig:NuclearModification} shows $R_{\mathrm{AA}}$ of charged particles reported by the ALICE Collaboration in $Pb$+$Pb$ collisions at \snn = 5.02 and 2.76 TeV from 0-5\% to 70-80\% of centrality \cite{ALICE:2018vuu}.
One would notice that the above $p_T \approx 3$-$4$ where jet productions are dominating $R_{\mathrm{AA}}$ shows a significantly smaller value than unity in $0$-$5\%$ of centrality in $Pb$+$Pb$ collisions.
On the other hand, $R_{\mathrm{AA}}$ recovers towards unity as one sees the results from central($0$-$5\%$) to peripheral ($70$-$80\%$) collisions.
This can be interpreted that the energy loss of jet partons due to interaction with QGP medium is larger in central collisions because the size and temperature of the generated QGP are expected to be larger than in peripheral collisions.

It should be also noted that the $R_{\mathrm{AA}}$
at the very low $p_T$ regions ($p_T\rightarrow 0$) reflects $\langle N_{\mathrm{part}} \rangle$ scaling of produced particles in contrast to the $\langle N_{\mathrm{coll}} \rangle$ scaling at high $p_T$.
The small peak around $p_T\approx2$-$3$ GeV appears as a consequence of the interplay between generated flow and jet quenching where the former enhances and the latter suppresses $R_{\mathrm{AA}}$.

\begin{figure}
    \centering
    \includegraphics[bb=0 0 960 540, width=1.0\textwidth]{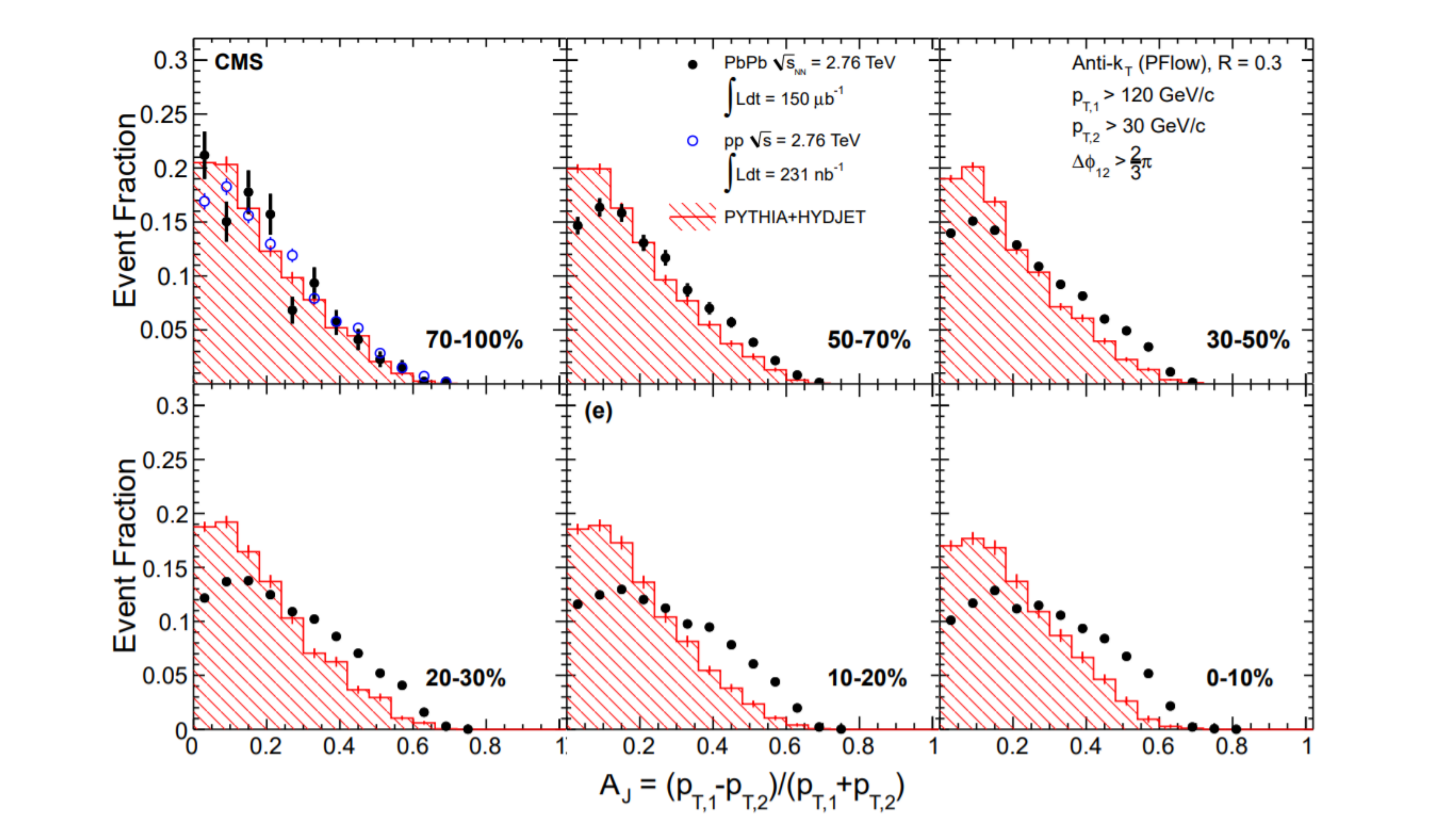}
    \caption{Distribution of asymmetry ratio (black dots) between leading and sub-leading jets in $Pb$+$Pb$ collisions at \snn = 2.76 TeV reported by the CMS collaboration \cite{CMS:2012ulu}. Results are shown for 6 different centrality. Results from $p$+$p$ collisions at \snn[proton] = 2.76 TeV (blue dots) are shown in upper right figure for a reference. Theoretical calculation of jet embedded in hydrodynamic background without jet quenching in $Pb$+$Pb$ collisions (red histograms) are shown as another reference.}
    \label{fig:Aj}
\end{figure}

\subsubsection{Momentum imbalance of jets}
To investigate the jet quenching effects on jet structure, one can see the momentum imbalance between leading and sub-leading jets.
As I mentioned at the beginning of Sec.~\ref{subsec:HighPTProductions}, jets are usually produced from hard parton--parton scattering in the first contact of nuclei and, as a result of momentum conservation, two scattered-out partons produce two back-to-back jets having a strong azimuthal correlation at $\Delta \phi \approx \pi$.
Here, the jet with larger $p_T$ is referred to as  
leading and the one with second larger $p_T$ is referred to as sub-leading jets.
When there is no QGP medium,
these two jets have almost equal $p_T$ and are said to be ``balanced''.
On the other hand, when the QGP medium exists, there appears the momentum imbalance of jets due to the jet quenching traversing the different length of the medium
\footnote{
Note that the momentum can be balanced if a hard scattering takes place at the very central part of the QGP medium because two jets are expected to deposit their energy equally.
However, because the probability that such a case happens is small, results show momentum imbalance when there is QGP medium.
}.

To characterize the momentum imbalance, an asymmetry ratio, $A_j$, is defined as
\begin{align}
    A_j = \frac{p_{T, 1} - p_{T, 2}}{p_{T, 1} + p_{T, 2}},
\end{align}
where $p_{T, 1}$ and $p_{T, 2}$ are transverse momenta of leading and sub-leading jets. Note that $A_j$ takes a value between $0$ and $1$ by definition. The more balanced the back-to-back jet is, the closer to $0$ the $A_j$ becomes, and vice versa.

Figure \ref{fig:Aj} shows momentum imbalance between leading and sub-leading jets reported by the CMS Collaboration in $Pb$+$Pb$ collisions at \snn=2.76 TeV \cite{CMS:2012ulu}. They are compared with the results from the theoretical calculation of jets embedded in the hydrodynamic background without jet quenching for $Pb$+$Pb$ collisions so that one can take into account of the existence of soft-particle background in finding jets.
The results from the most peripheral collision events ($70-100\%$) are compared with ones in $p$+$p$ collisions at the same collision energy.
One sees that the modification of the shape of $A_j$ distribution is seen only in experimental data -- $A_j$ distribution is smeared as one sees more central events losing momentum balance between leading and sub-leading jets.
On the other hand, the theoretical results only taking account of the background effect show weak broadening compared to the experimental data and the momentum balance.
It should be also noted that the results in $p$+$p$ collisions and ones in the most peripheral $Pb$+$Pb$ collisions give a similar behavior showing momentum balance.
The above can be understood as proof of the existence of the jet quenching effect.
In peripheral $Pb$+$Pb$ and $p$+$p$ collisions, momentum imbalance due to jet quenching is not observed because generated QGP medium is expected to be too small. 
In central collisions, momentum imbalance appears as a consequence of the interaction between jets and generated large QGP medium, not just an effect of the soft-particle background.

\section{Topical review of QGP studies}
\label{sec:INTRO_TopicalReviewOnQGPStudies}

\begin{figure}
    \centering
    \includegraphics[bb=0 0 1134 624, width=1.5\textwidth,angle=-90]{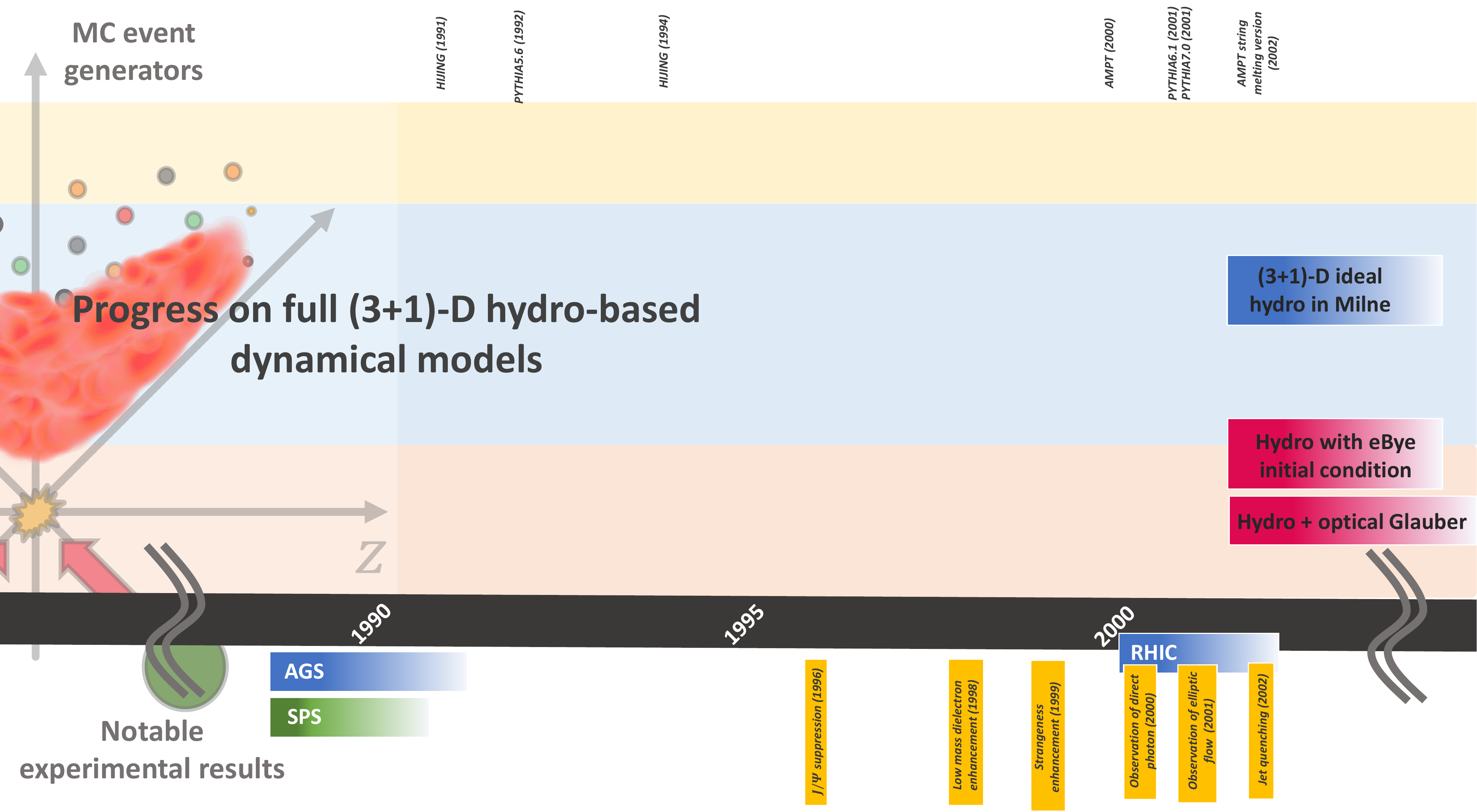}
    \caption{Notable experimental results, developments of MC event generators, and  historical progress on full (3+1)-D hydro-based dynamical models around from 1980 to 2003.}
    \label{fig:TopicalReview1}
\end{figure}
\begin{figure}
    \centering
    \includegraphics[bb=0 0 1134 624, width=1.5\textwidth,angle=-90]{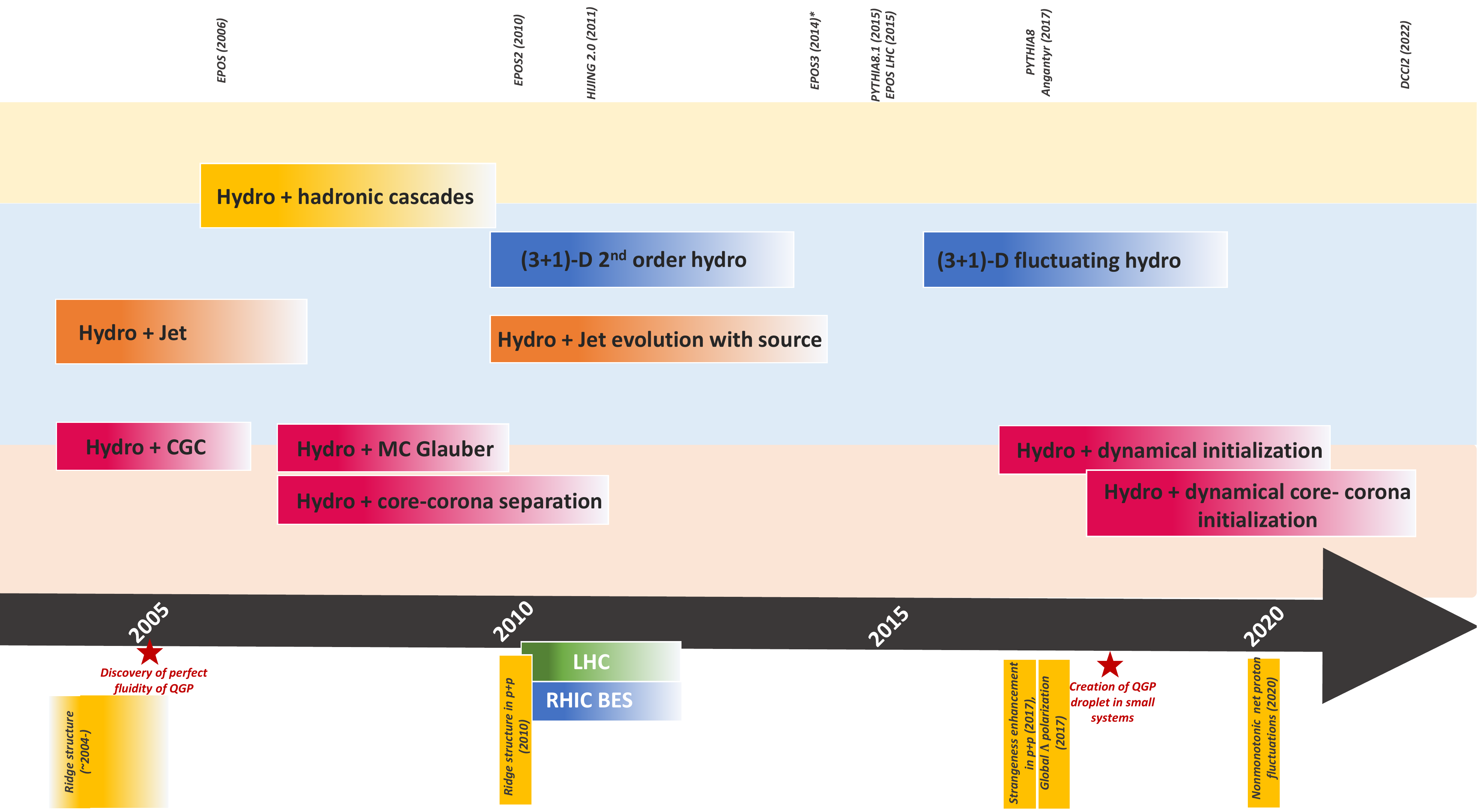}
    \caption{Notable experimental results, developments of MC event generators, and  historical progress on full (3+1)-D hydro-based dynamical models around from 2004 to present.}
    \label{fig:TopicalReview2}
\end{figure}

In this section, let me briefly summarize the historical path of QGP studies from around 1980 to present
with a particular focus on the development on full {\it{(3+1)-D}} hydro-based dynamics model with three dimensions in space and one dimension in time, to clarify the position and role of this thesis.
Figures \ref{fig:TopicalReview1} and \ref{fig:TopicalReview2} show the historical path  from around 1980 to 2003 and from 2004 to present, respectively.
In the left column, notable experimental results are shown including the starting year of the first heavy-ion beam at each experiment.
In the mid column, how full (3+1)-D hydro-based dynamical models have been developed is shown.
The color difference show description of which stage is improved.
In the right column, development of MC event generators which are explained in Sec.~\ref{sec:MCEventGenerators} is shown
\footnote{
Here, I pick up four well-known MC event generators: \pythia \ Angantyr, \textsc{Hijing}, \textsc{Ampt}, and \textsc{Epos3},
which are commonly used to explain heavy-ion collisions at LHC energies.
}
, which I leave the explanation in the next section.

\subsection{Notable experimental results}
I already explained the history of the high-energy heavy-ion collisions
in Sec.~\ref{sec:HistoryOfRelativisticHICexperiment}.
Here, I would like to focus on the physics results that have been obtained by the experiments.
The start-up of the AGS at RHIC and SPS at CERN backs into 1987 \cite{book:YagiHatdudaMiake}.
At the SPS, three significant physical results were reported,
which are anomalous $J/\Psi$ suppression in 1996 \cite{NA50:1996lag},
low mass dilepton pair production in 1998 \cite{CERESNA45:1997tgc}, and strangeness enhancement 
\footnote{
Despite that the enhancement of strange baryon productions compared to $e^-$+$e^-$ or $p$+$p$ was experimentally observed,
there was a discussion whether the strangeness enhancement observed at SPS was a consequence of QGP formation or not \cite{Koch:2017pda}.
}
in 1999 \cite{WA97:1999uwz}.
The $J/\Psi$ suppression was theoretically predicted by Matsui and Satz \cite{Matsui:1986dk} in 1986.
The $J/\Psi$ is a bound state of charm and anti-charm quark.
This kind of bound state consisting of a heavy quark and its own anti quark is called {\it{quarkonia}}.
If this bound state bounded with a linear confinement color potential is under finite temperature, the interaction of gluons is Debye screened,
and finally the dissociation of the bound state takes place at a sufficiently large temperature.
Thus, the suppression of $J/\Psi$ indicates 
the formation of finite temperature matter, 
and the magnitude of the suppression would infer the magnitude of temperature that is achieved in the generated matter.
This kind of particle production is often called {\it{thermometers}} from its such properties.
For instance, $J/\Psi$, $\psi(3686)$ \cite{Abrams:1974yy} are both quarkonia made of charm quarks, and the latter is a first excited state of the former, {\it{i.e.,}} $J/\Psi(1S)$ and $\psi(2S)$.
Thus, the binding energy is larger in $J/\Psi(1S)$ than $\psi(2S)$,
and the sequential color charge screening in the medium due to this sequential binding energy is expected.
The name, thermometer, originates from the above reason.
The low mass dilepton enhancement compared to expectation from hadronic decays is considered as the extra emission of dileptons from thermalized matter 
\cite{Feinberg:1976ua, Shuryak:1978ij, McLerran:1984ay, Gale:1987ki}.
The {\it{strangeness enhancement}} was theoretically proposed as a signal of QGP \cite{Rafelski:1982pu,Koch:1986ud}.
The idea is that once thermal matter is generated in the system and relatively light $u$ and $d$ quarks are under equilibrium,
processes of $gg\rightarrow s\bar{s}$ and $u\bar{u}$, $d\bar{d}\rightarrow s\bar{s}$ can lead to the chemical equilibrium of strangeness within a reasonable time scale of the life time of QGP.

After the RHIC launched its operation in 2000,
due to the high collision energy of the beam,
several signals of QGP formation had been observed, 
and which led to the consensus within a community to make the press release of the discovery of the perfect fluidity of QGP at RHIC in 2005.
Here, let me pick up three notable experimental results:
elliptic flow in 2001 \cite{STAR:2000ekf}, jet quenching in 2002 \cite{PHENIX:2001hpc},
and observation of direct photon in 2010 \cite{PHENIX:2008uif}.
Note that the observation of direct photon
was also reported at the SPS in 2000 \cite{WA98:2000vxl}.
Since 2005, the distribution of dihadron correlation in the azimuthal and longitudinal space has been measured.
It was observed that the dihadrons have a strong correlation at {\it{near side}} ($\Delta \phi \approx0$) at a certain pair separation in longitudinal direction
and the structure is broadening in the longitudinal direction \cite{STAR:2004wfp, STAR:2005ryu}.
In 2009, such a structure observed in heavy-ion collisions, especially consisting of low $p_T$ particles, was named as ``{\it{ridge}}'' in Ref.~\cite{STAR:2009ngv} and has been understood as a signal of flows as explained in Sec.~\ref{subsec:LongitudinalCorrelation}.

\begin{figure}
    \centering
    \includegraphics[bb= 0 0 540 850, width = 0.5\textwidth]{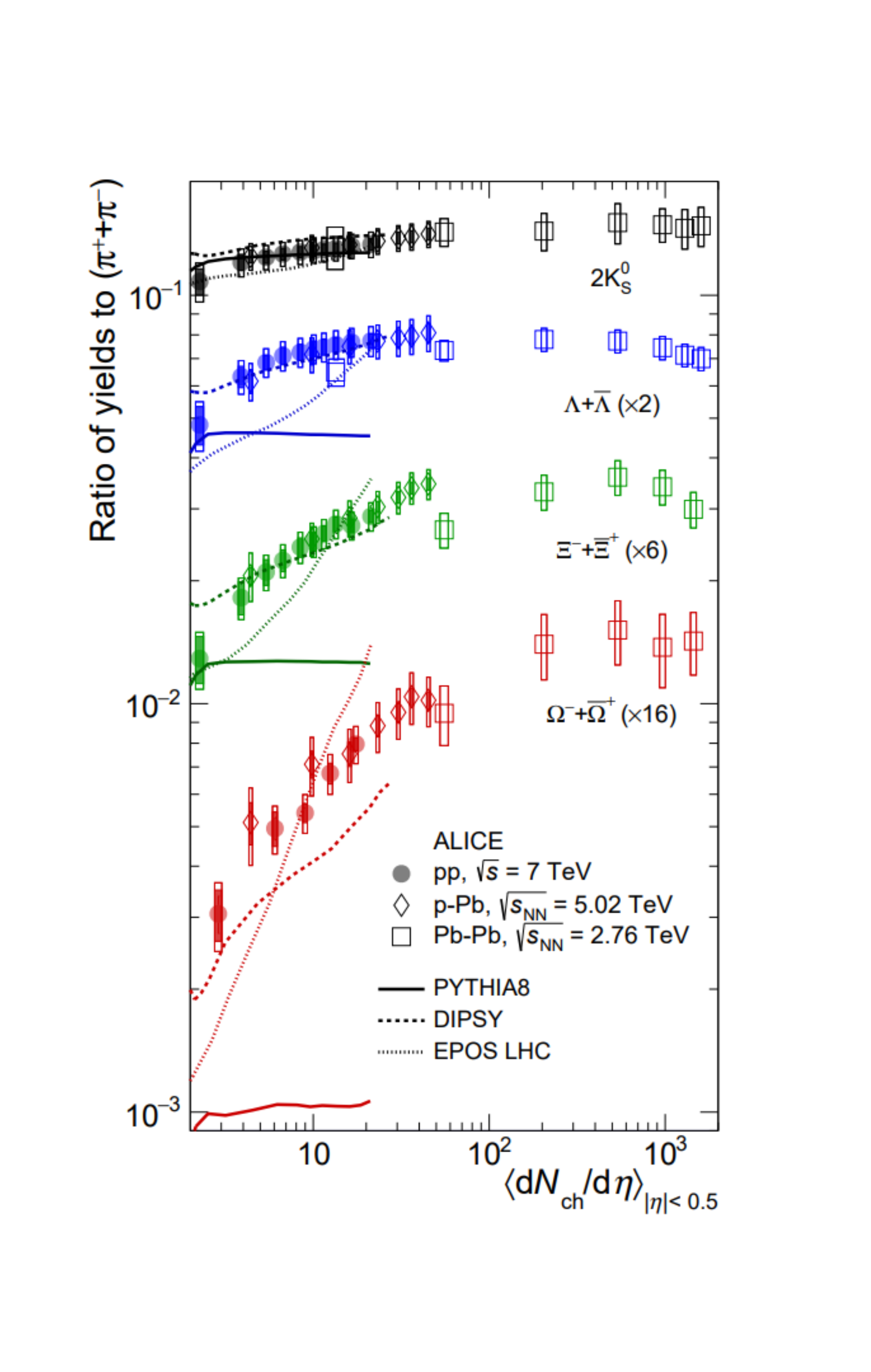}
    \caption{Strange hadron yield ratios to charged pions as a function of multiplicity in $p$+$p$ at \snn[proton]=7 TeV, $p+Pb$ at \snn = 5.02 TeV, and $Pb$+$Pb$ collisions at \snn = 2.76 TeV \cite{ALICE:2016fzo}.}
    \label{fig:ALICEStrangenessEnhancement}
\end{figure}

The first heavy-ion collision was achieved at the LHC in 2010 \cite{Muller:2012zq}.
Thanks to the increase in the collision energy,
a paradigm shift took place in the QGP study: 
although it has been understood that the QGP is formed only in heavy-ion collisions, signals of QGP formation could be observed in small colliding systems such as $p$+$p$ or proton+heavy-ion (p+A) collisions.
Starting from the ridge structure from high-multiplicity $p$+$p$ collisions reported in 2010 \cite{CMS:2010ifv}, several investigations on momentum correlation of produced particles in high-multiplicity events have been performed.
Strangeness enhancement in $p$+$p$ and $p+Pb$ collisions as a function of multiplicity shown in Fig.~\ref{fig:ALICEStrangenessEnhancement} was reported in 2017 \cite{ALICE:2016fzo}.
The smooth enhancements of strange hadron yield ratios from small systems to heavy-ion collisions indicate that the strangeness in the generated matter in small systems is partially chemically equilibrated, and the fraction of the matter increases as a function of multiplicity.
According to the scaling behavior of particle yield with multiplicity
regardless of collision system size or energy,
multiplicity is expected to be a key variable for the equilibration of the system.
In 2018, 
the experimental data suggesting the QGP formation was also observed at the RHIC.
With a result of flow coefficients observed in small systems as a signal of hydrodynamic response \cite{PHENIX:2018lia}, 
a press release of ``Creation of quark-gluon plasma droplets'' was reported by RIKEN \cite{web:QGPdroplet}.

Besides the QGP formation in small systems, as a novel feature of QGP, global $\Lambda$ polarization due to {\it{vorticity}} generated in heavy-ion collisions \cite{STAR:2017ckg} was reported in 2017.
The BES program at RHIC was launched in 2010 to explore the QCD phase diagram with low collision energy.
In 2020, the first evidence of a non-monotonic behavior of the kurtosis times variance of the net proton number as a function of collision energy was observed \cite{STAR:2020tga},
which indicates a signal of the existence of QCD critical point \cite{Gupta:2011wh, Stephanov:1999zu}.

\subsection{Progress on full (3+1)-D hydro-based dynamical models}
Hydro-based dynamical models have been adopted to
macroscopically describe the dynamics of QGP.
The first application of relativistic hydrodynamics to high-energy collisions was made by Landau \cite{Landau:1953gs}.
Around the 1950s, formulations of hydrodynamics from the point of view of QFT were on-trend \cite{10.1143/PTP.18.591,10.1143/PTP.22.403} being pushed by the establishment of linear response theory \cite{Kubo:1957mj}.
However, in order to extract the interaction of quarks and gluons under finite temperature, which gives transport coefficients, it is required to extract the systematic motion to purely discriminate the fluctuation from the kinematic motion \cite{book:Genshikakukenkyu} (need check).
After that, hydrodynamics have been developed as phenomenology, which can describe phenomena in high-energy collisions.
In the 1980s, starting with relativistic ideal hydrodynamics formulated by Bjorken \cite{Bjorken:1982qr},
several hydrodynamic models were developed \cite{book:Genshikakukenkyu} in (1+1)-D \cite{Baym:1984hpn,Chu:1986fu} and in ansatz of spherical symmetry around the beam axis \cite{Baym:1983amj,VonGersdorff:1986tqh,Kajantie:1986cu,Bialas:1983dn,Wang:1986nt,Kagiyama:1986hx,Ornik:1989jp,Akase:1990yd,Sollfrank:1996hd,Hung:1997du}.
Also, there were works so-called {\it{three fluid dynamics}} developed at that time \cite{Dumitru:1994vc}.
The first full (3+1)-D hydrodynamic model in Milne coordinate was developed in 2002 \cite{Hirano:2001eu},
and this type of hydrodynamic model is still commonly used nowadays.

Initial conditions of hydrodynamic simulation have been developed along with hydrodynamic models.
Since we do not have any access to the initial state of the QGP, it was common to put a smooth initial condition at a fixed initial time for hydrodynamic evolution.
While the Glauber model was adopted to explain high-energy collisions from around the late 1960s to the 1970s \cite{Czyz:1969jg,Bialas:1976ed}, 
the optical Glauber model was started to be applied to the initial condition of hydrodynamics in 2001 \cite{Kolb:2001qz,Heinz:2001xi}.
However, it was pointed out that event-by-event fluctuation of the initial condition is necessary in Ref.~\cite{Aguiar:2001ac} in 2002.
After that, event-by-event fluctuation of positions of nucleons inside of nuclei was incorporated as MC Glauber model   
\cite{Miller:2007ri,Broniowski:2007nz, Loizides:2014vua},
and it has been adopted for the initial condition.
While the Glauber type initial conditions are described as superposition of nucleon+nucleon collisions, the {\it{color glass condensate (CGC)}} which can describe a universal saturation behavior of gluons inside a colliding nucleons/nuclei at extremely high energy
\cite{McLerran:2001sr,Iancu:2003xm,Gelis:2010nm,Lappi:2010ek}.
The CGC type initial conditions also have been incorporated in Ref.~\cite{Hirano:2004en,Schenke:2012wb}.

After the experimental observation of jet quenching in 2002,
people started to try to describe the high $p_T$ suppression by incorporating jets in fully evolving hydrodynamics.
The first jet quenching calculation inside the hydrodynamically evolving medium was done in Ref.~\cite{Hirano:2003pw} in 2002, while no back reaction from jets to fluids is absent.
Introduction of source terms in hydrodynamics to consider the back reaction in the full (3+1)-D was done in Ref.~\cite{Betz:2010qh,Tachibana:2014lja}.

Around 2000, it started being realized that hadronic cascade, where the system has large viscosity \cite{Hirano:2005wx} and no longer be described within hydrodynamics, is important to explain high-energy heavy-ion collisions \cite{Dumitru:1999sf,Bass:1999tu,Teaney:2000cw}.
The first hydro+cascade calculation in (3+1)-D was done in 2006 \cite{Hirano:2005xf}.
Until present, it has been now standard modeling to adopt hadronic cascade after the hydrodynamics evolution of QGP fluids \cite{Petersen:2008dd,Hirano:2012kj,Shen:2014vra,Schenke:2020mbo}.

%Viscosity
%Ideal hydrodynamics had been developed in a context of numerical simulation, meanwhile, there was a development in the formalism 
The descriptions of stages outside of QGP fluids have been improved in hydro-based dynamical models, but the formulation of hydrodynamics itself have also been developed.
Hydrodynamics can describe the small deviation from local equilibrium by expanding them in the gradients of thermodynamic state.
The viscosity gives the response to the thermodynamic force which contributes as a correction to the local equilibrium,
and that is why such hydrodynamics is called {\it{viscous hydrodynamics}}.
Navier-Stokes equation is a constitutive equation which describes up to the first order of the gradient while causality is violated because the equation is parabolic.
This should be reconciled by introducing relaxation time \cite{c.cattaneo1948}.
The second order constitutive equations were obtained from the second law of thermodynamics \cite{m.kranys1967,Muller:1967zza}, the first relativistic version was derived by Israel \cite{Israel:1976tn},
and more extension were made \cite{Israel:1979wp,Hiscock:1983zz} by the 1980s.
Note that the numerical analysis of the difference between the first and second-order hydrodynamic equation is made in Ref.~\cite{Huovinen:2008te}.
Because of an ambiguity in how to formulate constitutive equations of hydrodynamics, until now, there have been several attempts to formulate second-order constitutive equations \cite{Betz:2008me,Betz:2009zz,Denicol:2012cn,Denicol:2012cn,Niemi:2011ix,Monnai:2010qp,Baier:2006um,Baier:2007ix,Tsumura:2006hnr,Tsumura:2009vm,Tsumura:2012ss,Tsumura:2015,Bhattacharyya:2007vjd,Natsuume:2007ty,Natsuume:2008ha}.
The (2+1)-D viscous hydrodynamic models have been established since the mid 2000s \cite{Heinz:2005bw,Baier:2006gy,Dusling:2007gi,Noronha-Hostler:2013gga,Niemi:2015qia,}.
The first full (3+1)-D viscous hydrodynamics was developed in 2010 \cite{Schenke:2010nt}.
Currently there are various full (3+1)-D viscous hydrodynamic codes exist
\cite{Karpenko:2013wva,Pang:2014ipa,Murase:2015oie,Schenke:2020mbo}
, and some of them are extended to non-zero baryon density \cite{Du:2019obx}.
The so-called {\it{fluctuating hydrodynamics}} also exists where 
hydrodynamic fluctuations are encoded in 2nd order constitutive equation \cite{Murase:2015oie,Murase:2019cwc}.
It should be also noted that {\it{anisotropic hydrodynamics}}
that treats anisotropy of local momentum distribution has been established in Ref.~\cite{Martinez:2012tu,Bazow:2013ifa,Florkowski:2013lya}.

Until 2007, initial conditions of hydrodynamics were given with an assumption that the entire system was under local equilibrium.
The {\it{core--corona picture}} proposed by Werner \cite{Werner:2007bf} made waves in that situation.
Under the core--corona picture, a lumpy initial condition given event-by-event
can be separated into high and low number density regions.
The former is called core which experiences hydrodynamic evolution while the latter undergoes string fragmentation without being equilibrated.
This core--corona initial condition adopted in hydrodynamics was found to be succeeded in describing ridge structure from $p$+$p$ collisions \cite{Werner:2010ss}.
In 2017, hydrodynamic calculations with {\it{dynamical initialization}} utilizing source terms of hydrodynamics appeared \cite{Okai:2017ofp}.
In the same year, its application to low energy heavy-ion collisions was done in Ref.~\cite{Shen:2017bsr}.
One year later, hydrodynamic calculations with dynamical initialization based on the core--corona picture was done at low energy
\cite{Akamatsu:2018olk},
and the first application of the dynamical initialization based on the core--corona picture at the LHC energy from small systems to heavy-ion collisions was done in \cite{Kanakubo:2018vkl,Kanakubo:2019ogh}.
With the trend to investigate high-density region of the QCD phase diagram, 
recent applications of dynamical initialization to low energy were made in Ref.~\cite{Du:2018mpf,Schafer:2021csj}.

%Initial condition

%

%In this section, I briefly summarize out standing works that has pushed us forward for deeper understanding QGP properties.

%\subsection{Relativistic Hydrodynamics}
%
%\subsection{Hadronic sampling via Cooper--Frye formula}
%
%\subsection{String fragmentation}
%
%\subsection{Hadronic afterburner}
%
%\section{Multi-stage dynamical framework}
%
%\subsection{Existing frameworks}
%
%\subsection{Existing frameworks}

%---------------------------------

\section{Monte Carlo Event Generators}
\label{sec:MCEventGenerators}

Monte Carlo (MC) event generators are indispensable for high-energy nuclear collision studies.
Especially, several MC event generators
\footnote{
Besides \pythia \ that I explain in this thesis, \textsc{Herwig++} \cite{Bellm:2015jjp} developed from its original version
\cite{Marchesini:1987ce}, and \textsc{Ariadne} 
\cite{Lonnblad:1992tz} based on the color dipole model \cite{Gustafson:1987rq} have long histories as MC event generators.
Also, \textsc{Sherpa} 
\cite{Gleisberg:2003xi,Gleisberg:2008ta} has been developed from the beginning of the 2000s.

Note that there is an MC event generator called \textsc{Vni} (Vincent Le CuCurullo Con GiGinello) \cite{Geiger:1997pf} that 
simulates from lepton-associated collision, hadronic collisions, and heavy-ion collisions.
The desciption is based on parton cascade. Implementation of hadronic cascades to \textsc{Vni} is done in Ref.~\cite{Bass:1998ca}.
}
have been established for elementary particles or hadronic collisions at high collision energies such as the LHC energies \cite{Buckley:2011ms}.
In contrast, dynamical models based on hydrodynamic evolution have been used for data to model comparisons to extract properties of the QGP in heavy-ion collisions while most of the dynamical models based on hydrodynamics are not MC event generators, which I will mention details in Sec.~\ref{sec:INTRO_ProblemsInExistingHydroModels} in this chapter.
For a proper comparison with experimental data and quantitative study of the QGP,
MC event generators that are capable of simulating heavy-ion collisions are required.

In this section, I summarize the criteria of MC event generators, existing MC event generators for heavy-ion collisions, and their characteristics.
With this review of MC event generators,
the position of this thesis should be clarified.

\subsection{Criteria}
\label{sec:MCEventGenerator_Criteria}

MC event generators simulate {\it{events}} of high-energy collisions.
The word, events, stands for an outcome of one single collision of incoming beam particles, 
such as electrons, protons, or heavy ions.
The number of out-coming particles from an event varies event-by-event, thus,
the process happening in high-energy collisions is $2\rightarrow X$ where
typical scales of $X$ is $X\approx \mathcal{O}(10)$ in $e^+$+$e^-$ or $p$+$p$ while $X\approx \mathcal{O}(10^2)-\mathcal{O}(10^4)$ in heavy-ion collisions.
An outcome of a single event in MC event generators is a set of information of out-coming particles, which are basically the numbering for the identification of a particle (ID), energy and momentum, and production point in coordinate space referred to as production vertices
\footnote{
Information of production vertices are not necessary for all event generators because
some of them describe particle production processes only in momentum space.
In high-energy experiments, production points are not observable, therefore it depends on models whether an outcome contains information of production vertices or not. 
}
.
Once one obtains sets of out-coming particles, one can design an observable in which one wishes to extract physics from events,
and statistical analysis of it enables us to investigate the characteristics of generated particles in events and, in turn, the properties of matter created in collisions.

Here, I would like to mention the reason why event generators should be Monte Carlo.
The Monte Carlo is a computational method that produces results by utilizing multiple generations of random numbers. 
The high-energy collisions should be described with this sort of method with random numbers
because the collision itself is governed by {\it{quantum}} physics.
Basically, we know the scattering and particle production probabilities, which can be obtained from perturbative calculations or modeled based on phenomenology, and these are the inputs to an MC event generator.
Random numbers determined by throwing a dice decide whether the process happens or not according to the probability.

The MC event generators, which have been explained above, have the following criteria that must be held:
energy-momentum conservation, the ability to simulate all types of collisions, and description of the whole momentum space of particle productions, 
which enables the event generator to give sets of out-coming particles that can be directly
compared to experimental data.

Energy-momentum conservation is the most indispensable criterion in event simulations.
The energy and momentum of incoming two particles should be identical to the summation of those of final out-coming $X$ particles within a single event.
It should be emphasized that in order to hold the conservation, one needs to take care whole phase space of a collision, including a ``trash'', such as remnants of beams 
which are basically not produced within the rapidity regime in which experiments mainly investigate physics.

The ability to simulate all types of collisions is required to describe event distributions of generated events.
For example, in $p$+$p$ collisions, the types of collisions can be classified into elastic or inelastic collisions 
where inelastic collisions can be further discriminated as single diffractive, double diffractive, and non-diffractive collisions.
On the other hand, for heavy-ion collisions, as we see in Sec.~\ref{subsec:ParticleProductions},
events are classified with centrality.
Without simulating all types of those collisions, one never knows which subsets of events should be compared to experimental data for a particular multiplicity/centrality class because  multiplicity/centrality class is a relative quantity compared to the parent population.
Hence, the ability to simulate all types of collisions is required to describe event distributions that correspond to a distribution of the parent population of events and to make a proper comparison between results from an MC event generator and experimental data.

Description of whole momentum space of particle productions is highly related to the above two points.
The particle production mechanism differs depending on the scale of momentum in both $p$+$p$ and heavy-ion collisions.
Simply speaking, the probability of particle productions at low momentum scales is hardly obtained in perturbative calculations because the low momentum regime is a non-perturbative one: a coupling constant of QCD becomes too large to obtain the probability of a process by perturbation.
Thus, one needs to describe this regime with a phenomenology based on perturbative QCD (pQCD) 
or effective description at a long wavelength limit, which is hydrodynamics.
Generally, no one knows the separation scale of those, 
therefore tuning a parameter that determines the scale is necessary.
It would be now clearly noticeable that the description of the whole momentum space of particle productions with consideration of 
energy-momentum conservation and generation of all types of collisions
are not obvious.
However, that is an MC event generator that realizes this within a single framework.

\subsection{Problems in existing hydro models}
\label{sec:INTRO_ProblemsInExistingHydroModels}
Here, it should be worth emphasizing that most dynamical models based on hydrodynamics do not satisfy the above criteria: they are not MC event generators.
In the following, 
I discuss problems in existing general dynamical models based on hydrodynamics
focusing on the case of simulations of heavy-ion collisions.
First, it is mostly the case that initial conditions, energy, or entropy density distributions of an event are parametrized or multiplied by a controllable parameter
so that models can reproduce particle yields:
the total energy and momentum of the incoming beam particles are not taken seriously.
This is due to the remaining conventional motivation that one just needs to describe the midrapidity where QGP is expected to be formed. Thus, the total energy of hydrodynamics is not necessary to be consistent with those of an entire collision system. 
Second, an adopted range of hydrodynamic models is limited from semi-central to central collisions because one knows that description of hydrodynamics breaks down in very peripheral collisions while no one knows to which extent hydrodynamic description works.
Third, hydrodynamic models only describe particle productions in low momentum regime, not the entire momentum space.

Therefore, it is now obvious that dynamical models based only on hydrodynamics
just describe a portion of a collision system.
In order to have quantitative discussions, 
establishment of an MC event generator based on hydrodynamics for heavy-ion collisions should be achieved
by following the examples in $e^+$+$e^-$ or $p$+$p$ collisions.

\subsection{Existing MC event generators for heavy-ion collisions at the LHC energies}
While most existing dynamical models based on hydrodynamics are not MC event generators, there are some models generally used in describing heavy-ion collisions at the LHC energies as an MC event generator.
Here, I pick up four well-known MC event generators: \pythia8 Angantyr \cite{Bierlich:2018xfw}, \textsc{Hijing} \cite{Wang:1991hta,Deng:2010mv}, \textsc{Ampt} \cite{Zhang:1999bd,Lin:2004en}, and \textsc{Epos3}\cite{Werner:2013tya}.
Because a summary of those would be instructive to spot the position of this thesis,
in this section, I would like to summarize each model briefly.

\subsubsection{Short descriptions of each MC event generator}
\pythia8 Angantyr is established in Ref.~\cite{Bierlich:2016smv,Bierlich:2018xfw}.
Since I utilize this model in the framework that I have been 
established, further details are explained in Sec.~\ref{subsection:Generating_initial_partons}.
The basis of this model is \pythia \ which simulates $e^+$+$e^-$ or $p$+$p$ collisions
at the LHC energies
where partonic production is described by pQCD and phenomenology based on pQCD. The partonic production is followed by a hadronization model called the Lund string fragmentation.
Here, it should be noted that there is no picture of the local equilibrium of the produced matter.
\pythia \ is a widely known MC event generator having a long history in high-energy reactions. For the history of the development of \pythia, see Ref.~\cite{Sjostrand:2019zhc}. 
In \pythia8 Angantyr, the above picture is extended to heavy-ion collisions with wounded nucleon model \cite{Bialas:1976ed} and Glauber models to describe the initial state of a collision in longitudinal and transverse directions of heavy-ion collisions:
a heavy-ion collision is described as multiple collisions of nucleons.
It should also be noted that there is no {\it{cascade}}, dynamics of collisions among particles,
in the default options,
while the new option for cascades in hadron level is already introduced in Ref.~\cite{Bierlich:2021poz}.
The relations with other models are mentioned in Ref.~\cite{Bierlich:2018xfw}.

{\textsc{Hijing}} (Heavy Ion Jet INteraction Generator) has been established in Refs.~\cite{Wang:1991hta,Deng:2010mv}.
{\textsc{Hijing}}2.0 was published in Ref.~\cite{Deng:2010mv}, and the latest version is {\textsc{Hijing}}++ written with C++ \cite{Barnafoldi:2017jiz,Biro:2019ijx}.
In the sense that {\textsc{Hijing}} is built based on partonic and hadronic productions implemented in \pythia, it has some similarity with \pythia \ Angnatyr.
As other similarities, the initial state of a heavy-ion collision is modeled with wounded nucleon and Glauber models and 
there are no cascades in parton and hadron levels.
One of the significant features in {\textsc{Hijing}} is 
that the impact parameter dependent nuclear shadowing is introduced as a nuclear PDF where hard particle productions are determined through this PDF.

{\textsc{Ampt}} (A Multi-Phase Transport) \cite{Zhang:1999bd,Lin:2004en} is the MC event generator including both partonic and hadronic cascades.
Initial conditions are partons obtained from {\textsc{Hijing}}.
In the latest version of this model, {\it{string melting}}
-- conversion of strings obtained from {\textsc{Hijing}} into partons via string fragmentation --
is incorporated \cite{Lin:2001zk}.
Under the string melting, strings obtained from {\textsc{Hijing}} are 
temporary fragmented into hadrons with the Lund string fragmentation, 
and those hadrons are replaced with constituent partons.
Once the string melting is done,
the Zhang's parton cascade (ZPC) \cite{Zhang:1997ej} is performed for those partons.
Finally partons are turned into hadrons via coalescence model,
and those hadrons experience hadronic interactions with A Relativistic Transport model {\textsc{(ART)}} \cite{Li:1995pra,Sa:1995fj}.

%{\textsc{UrQMD}} contains description of QGP medium with relativistic hydrodynamics to be applied for the LHC energies.
%Initial condition is generated via binary collisions of nucleons 
%distributed inside of nuclei.
%NOTICED THAT THERE IS NO JET IN URQMD

{\textsc{Epos3}} is the latest version of the series of {\textsc{Epos}} (Energy conserving quantum mechanical
multiple scattering approach based on Partons
(parton ladders), Off-shell remnants, Splitting of parton ladders) models which has been established in Ref.~\cite{Werner:2005jf,Werner:2010aa,Pierog:2013ria,Werner:2013tya}.
Especially in this latest version of the model \cite{Werner:2013tya},
the core--corona picture, which is proposed by the leading author \cite{Werner:2007bf}, is incorporated.
Being different from \pythia8 Angantyr,
the initial state of a heavy-ion collision is modeled
with collisions of partons called Gribov--Regge theory \cite{Drescher:2000ha} rather than nucleons.
After the formation of strings, strings break down into string segments.
Segments in a high number density regime are converted into hydrodynamics and evolve as QGP fluids, which is referred to as the core.
Particle production from the core is obtained from a sampling of thermal hadrons from the Cooper-Frye formula.
On the other hand, those in a low number density regime experience string fragmentation without interacting with core segments/QGP fluids.
The hadronic transport model, UrQMD \cite{Bleicher:1999xi,Petersen:2008dd}, is incorporated \cite{Werner:2010aa} to describe interactions among hadrons in the late stage.

\subsubsection{Comparisons of existing MC event generators}
\begin{figure}
    \centering
    \includegraphics[bb= 0 0 960 540, width=0.8\textwidth]{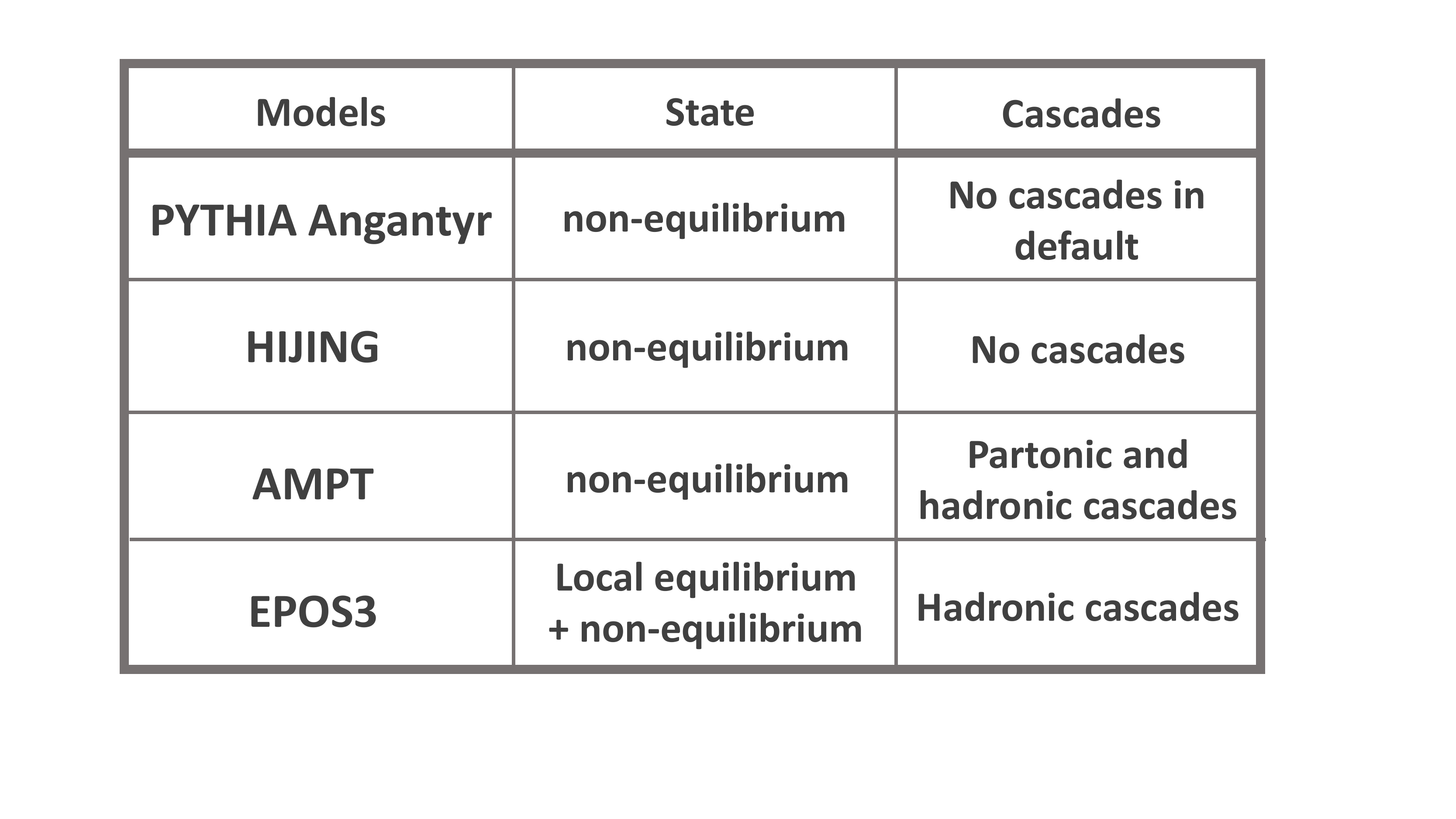}
    \caption{Comparisons of features of MC event generators commonly used for heavy-ion collisions at the LHC energies.}
    \label{fig:MCeventGen}
\end{figure}
To clarify the differences among those existing MC event generators,
I show in Fig.~\ref{fig:MCeventGen} summarizing the features of each MC event generator
introduced in the last subsection.
Here, I picked up two points for comparison: whether the matter generated in the model is assumed to be under local equilibrium or non-equilibrium,
and whether there is cascade in parton or hadron level,
which are essential features describing the dynamics of the matter.

For the former point, 
it is only {\textsc{Epos3}} that
considers the existence of locally equilibrated matter.
Thus, although the properties of the QGP have been studied in relativistic heavy-ion collisions, there are not many MC event generators available yet that can describe the dynamics of the QGP.

As for the latter point, 
as one sees in the examples of \pythia8 Angantyr or {\textsc{Epos}},
the importance of hadronic rescatterings at the late stage of a collision
has been gradually recognized in these days.
Apparently, there is a rough consensus in this field of heavy-ion collisions that
there are hadronic interactions and which is indispensable to understanding experimental data.
On the other hand, partonic interactions are only incorporated in {\textsc{Ampt}}
in which one describes the dynamics of matter with the partonic degree of freedom
with cascade instead of hydrodynamics.

\chapter{The DCCI framework}
\thispagestyle{fancy}

%--------------------------------------
\section{Motivation}
\label{sec:MODEL_Motivation}
Despite the great success of dynamical models based on relativistic hydrodynamics in describing a vast body of experimental data, it poses some open issues for a comprehensive description of the whole reaction in high-energy nuclear collisions as I mentioned in the previous section.
Here, I would like to make the motivation of my work clear by tracing the history of our group's works.
Note that the following sections ( Sec.~\ref{sec:MODEL_Motivation} in chapter 2 to chapter \ref{sec:CONCLUSION}) are written based on Ref.~\cite{Kanakubo:2021qcw}.

One of the major issues is an initial condition of relativistic hydrodynamic equations which does not respect the total energy of the colliding systems.
Initial conditions have been parametrized and put to reproduce centrality dependence of multiplicity or pseudorapidity distributions in a conventional hydrodynamic approach
\footnote{
For instance, in the early full (3+1)-D hydrodynamic models, initial energy density distribution of hydrodynamics is parametrized as follows:
flat distribution as a function of rapidity is adopted around midrapidity regions, which is smoothly connected with half of Gaussian functions at forward and backward rapidity \cite{Kagiyama:1986hx,Hirano:2001eu}.
}
.
As a result, the total energy of the initial hydrodynamic fields does not exactly match the collision energy of the system. 
Even when some outputs from event generators with a given collision system and energy are utilized for initial conditions in hydrodynamic models, an additional scale parameter is commonly introduced to adjust the model outputs of multiplicity. 
%Reference for AMPT initial conditions. How about IP-Glasma case???
One might not think it is necessary for the energy of the initial hydrodynamic fields to be the same as the total energy of the system.
This is exactly a starting point of the discussion in a series of papers \cite{Okai:2017ofp,Kanakubo:2018vkl,Kanakubo:2019ogh}: The relativistic hydrodynamics merely describes a part of system, namely, matter in local equilibrium, while other parts of the system such as propagating jets and matter out of equilibrium are to be described at the same time.

First attempts of simultaneous description of both the QGP fluids in equilibrium and the energetic partons out of equilibrium had been made in Refs.~\cite{Hirano:2002sc,Hirano:2003hq,Hirano:2003yp,Hirano:2003pw,Hirano:2004en} \footnote{Note that the very first study to utilize the hydrodynamic solutions in quantitative analysis of parton energy loss was done in Ref.~\cite{Gyulassy:2001kr}.}.
Initial conditions in those studies were still either parametrized via an optical Glauber model \cite{Glauber:2006gd} or taken from a saturation model \cite{Kharzeev:2000ph,Kharzeev:2001gp,Kharzeev:2001yq,Kharzeev:2002ei} so as to reproduce yields of low $p_T$ hadrons, while hard partons which undergo energy loss during traversing QGP fluids were supplemented to successfully reproduce the hadron spectra from low to high $p_{T}$ regions \cite{Hirano:2002sc,Hirano:2003yp,Hirano:2004en}.
It was found that an intriguing interplay between soft and hard components brought ones to interpretation of the proton yield anomaly in $p_{T}$ spectra \cite{Hirano:2003yp}.
However, the model lacked back reactions from quenching partons to the QGP fluids and a contribution from fragmentation was cut in low $p_T$ regions, both of which obviously violate the energy-momentum conservation law.

%Hydro + source terms
Medium responses to propagating energetic partons have been modeled within hydrodynamic equations with source terms by assuming the instantaneous equilibration of the deposited energy and momentum from partons \cite{Stoecker:2004qu,CasalderreySolana:2004qm,Satarov:2005mv,Renk:2005si,Chaudhuri:2005vc,Chaudhuri:2006qk,Chaudhuri:2007vc,Betz:2010qh,Tachibana:2014lja,Tachibana:2015qxa,Tachibana:2017syd,Chen:2017zte,Chang:2019sae,Tachibana:2020mtb,Zhao:2021vmu}.
Within this approach, the sum of the energy and momentum of fluids and those of traversing partons is conserved as a whole. However, it is not clear how to divide the initial system just after the collision into soft (fluids in equilibrium) and hard (partons out of equilibrium) parts.

%Dynamical initialization
To remedy this issue, a dynamical initialization model \cite{Okai:2017ofp} was proposed to describe the dynamics of gradually forming QGP fluids phenomenologically \footnote{Dynamical initialization is essential in describing the formation of fluids in lower collision energies in which secondary hadrons are gradually produced in finite time duration due to insufficient Lorentz contraction of colliding nuclei \cite{Shen:2017bsr,Akamatsu:2018olk}.
}.
In contrast to the conventional hydrodynamic models in which initial conditions are put at a fixed initial time, the QGP fluids are generated locally in time and space in the dynamical initialization framework.
Under this framework, all the input of energy-momentum of QGP fluids is the one of partons produced just after the nuclear collisions. 
Starting with vacuum, energy-momentum of the QGP fluids is dynamically generated by solving hydrodynamic equations with source terms. 
% \sout{, where the source term is defined as a phenomenological fluidization rate, \textit{i.e.}, the four-momentum deposition per unit time for a parton}
Consequently, fluids in local equilibrium are generated from the initial partons by depositing the energy and the momentum and surviving partons are considered to remain out of equilibrium.
When initial partons are taken, \textit{e.g.}, from event generators, the total energy keeps its value of the colliding two nuclei all the way through the dynamical initialization. 
Although I successfully separated matter in local equilibrium from initially produced partons in the dynamical initialization framework \cite{Okai:2017ofp}, the fluidization scheme was too simple and phenomenological to describe the transverse momentum spectra and the particle ratios.
Then, the core--corona picture is incorporated into the dynamical initialization.

The conventional core--corona picture was proposed to explain centrality dependence of strange hadron yield ratios~\cite{Werner:2007bf}. 
As multiplicity increases, the high-density region, in which the matter is mostly thermalized, is supposed to become larger. 
As a result, the final hadron yields become dominated by the hadrons from thermalized matter rather than non-thermalized matter created in low-density regions. 
The former component is referred to as core, while the latter one is referred to as corona. 

A Monte-Carlo event generator, 
\textsc{Epos} (Energy conserving quantum mechanical
multiple scattering approach based on Partons
(parton ladders), Off-shell remnants, Splitting of parton ladders) \cite{Werner:2005jf,Pierog:2013ria,Werner:2018yad} is widely accepted for its implementation of the core--corona picture. In the latest study in Refs.~\cite{Werner:2013tya,Werner:2019aqa}, string segments produced in a collision are separated sharply into the core and the corona components depending on their density and transverse momentum at a fixed time. Low-$p_T$ string segments in dense regions are fully converted into the thermalized medium fluid, while string segments with high momentum or dilute regions are directly hadronized.

On the other hand, I model the dynamical aspects of the core--corona separation introducing the particle density dependence of the dynamical initialization scheme, which is called the \textit{dynamical core--corona initialization (DCCI)}  \cite{Kanakubo:2018vkl,Kanakubo:2019ogh} \footnote{In fact, this idea was first implemented in Ref.~\cite{Akamatsu:2018olk} to describe excitation functions of particle ratios at lower collision energies. However, it was applied to the secondary produced hadrons rather than partons.}.
One of the key features of the DCCI is to deal with dynamics of the core (equilibrium) and the corona (non-equilibrium) at the same time.
With the description of the dynamics, gradual formation of core and corona in spatial and momentum space is achieved.
In the DCCI framework, the multiplicity dependence of the hadron yield ratios of multi-strange baryons to pions from small to large colliding systems in a wide range of collision energy is attributed to a continuous change of the fractions of the core and the corona components as multiplicity increases \cite{Kanakubo:2018vkl,Kanakubo:2019ogh}.
It should be noted here that the ``corona" is referred not only to an outer layer in the coordinate space but also to the one in the momentum space:
The lower $p_T$ partons are more likely to deposit their energy and momentum to form the fluids and the higher $p_T$ partons are less likely to be equilibrated during the DCCI processes.

%Summary of new features in DCCI2
%-------------------------------
In this paper, I update the DCCI framework towards a more comprehensive description of dynamics in full phase space from small to large colliding systems in a unified manner.
Hereafter I call this updated DCCI the \textit{DCCI2}.
In comparison with the previous work \cite{Kanakubo:2018vkl,Kanakubo:2019ogh}, several crucial updates have been made in this new version including a more sophisticated formula for four-momentum deposition of initial partons, particlization of the fluids on the switching hypersurface through a Monte-Carlo sampler \ISthreeD\  \cite{McNelis:2019auj}, hadronic rescatterings through a hadron cascade model \jam\  \cite{Nara:1999dz}, and modification of structure of color strings inside the fluids.
With these updates, the DCCI2 is capable of describing high-energy nuclear collisions from low to high $p_T$ region with particle identification in various colliding systems.

I generate initial partons from a general-purpose event generator \pythia8 \cite{Sjostrand:2007gs, Bierlich:2014xba}
switching off hadronization and make all of them sources of both the core and the corona parts. 
Here a special emphasis is put on to discriminate between the two terms, ``soft--hard" and ``core--corona". I call the core when it composes the matter in equilibrium. 
In DCCI2, the fluids generated through the dynamical initialization correspond to the core part and hadrons particlized on switching hypersurface are regarded as the core components.
On the other hand, I call the corona when it composes matter completely out of equilibrium. 
In DCCI2, partons in the dilute regions and/or surviving even through the dynamical initialization correspond to the corona part, and hadrons from string decays are regarded as the corona components.
Although the core (corona) component is sometimes identified with the soft (hard) component, it is not the case in the DCCI2: The hadrons from string fragmentation are distributed all the way down to very low $p_T$ region, which one cannot consider as ``hard" components.
To the best of my knowledge, it has been believed so far without any strong justification that the corona components would be negligible in low $p_T$ region in heavy-ion collisions.
In this paper, I scrutinize 
the size of the contribution from the corona components 
in soft observables (``soft-from-corona") 
and how the fraction of the corona components evolves as multiplicity increases.

One of the main interests in this field is to constrain the transport coefficients of the QGP through the hydrodynamic analysis of anisotropic flow data in low $p_T$ regions.
It is conventionally assumed that the low $p_T$ hadrons are completely dominated by the core components. 
What if the corona components, whose contribution is often considered to be very small, affect the bulk observables in heavy-ion collisions?  
The non-equilibrium contribution is at most taken through small corrections to thermodynamic quantities such as shear stress and bulk pressure. 
%Since the corona components are far-from-equilibrium distributions, these must be more important than those dissipative corrections in the hydrodynamic analysis if these are smaller than the core components, but of the same order of the magnitude.
The corona partons obey non-equilibrium distributions which are, in general, far from equilibrium distributions. 
While one deals with merely a small deviation from equilibrium distributions through the dissipative corrections from the point of view of the gradient expansion in hydrodynamic framework.
Thus the corona components are more important than the dissipative corrections in hydrodynamic analysis when the amount of corona components is non-negligible.
Thus, dynamical modeling containing the core--corona picture could become the next-generation model inevitably needed for the precision study of the QGP properties.

%------------------------------------
\section{The DCCI framework}

The DCCI2 framework as a multi-stage dynamical model describes high-energy nuclear reactions from $p$+$p$ to $A$+$A$ collisions.
Before going into the details of the modeling of each stage, I briefly summarize the entire model flow of the DCCI2 framework.

\begin{figure}[htbp]
\begin{center}
\includegraphics[bb=0 0 576 1080, width=0.8\textwidth]{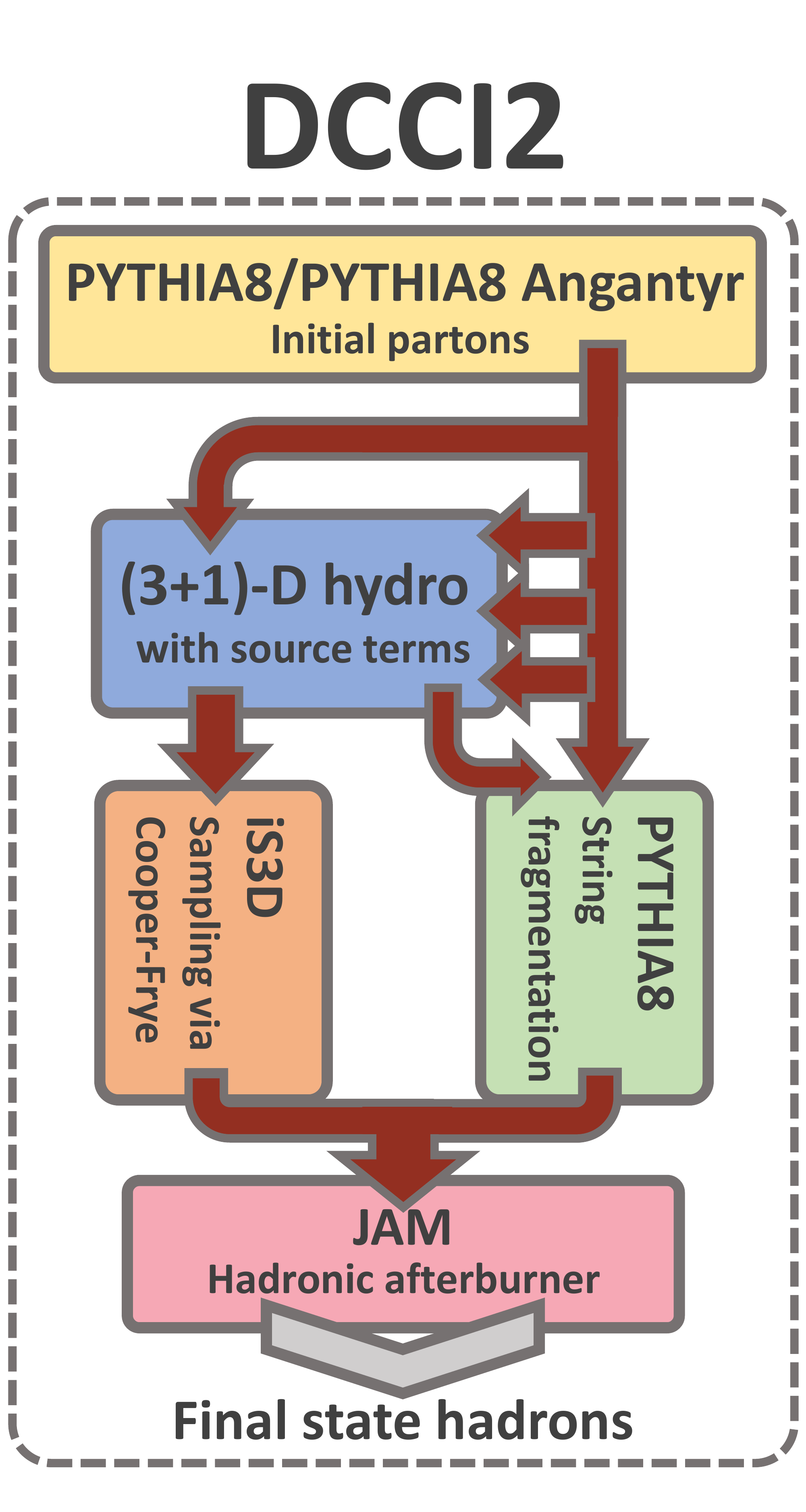}
\caption{(Color Online) Flowchart of the DCCI2 framework.
}
\label{fig:flow_chart}
\end{center}
\end{figure}

Figure~\ref{fig:flow_chart} represents the flowchart of the DCCI2 framework.
%Initial partons
%-----------------------------
First, I obtain event-by-event phase-space distributions of initial partons produced just after the first contact of incoming nuclei using \pythia8.244 or its heavy-ion mode, Angantyr model \cite{Sjostrand:2007gs, Bierlich:2014xba}.
Hereafter, I call \pythia8 and \pythia8 Angantyr, respectively.
Those initial partons are assumed to be generated at a formation time, $\tau_{0}$.
%Dynamical initialization
%-----------------------------
Under the dynamical initialization framework,
the QGP fluids are generated via energy-momentum deposition from those initial partons by solving the relativistic hydrodynamic equations with source terms from $\tau = \tau_{0}$ to the end of the hydrodynamic evolution.
%About dynamical core--corona
%-----------------------------
The energy-momentum deposition rate of partons is formulated based on the dynamical core--corona picture.
Partons which experience sufficient secondary interactions with surrounding partons are likely to deposit their energy-momentum and form QGP fluids. 
In contrast, partons that do not experience such sufficient secondary interactions give less contribution to the medium formation. 
%Hydrodynamics evolutions
%--------------------------
Hydrodynamic simulations are performed in the (3+1)-dimensional Milne coordinates originally written by Tachibana in Ref.~\cite{Tachibana:2014yai}. The hydrodynamic equations are solved with the Piecewise Parabolic Method (PPM).
For more details, see Ref.~\cite{Tachibana:2014yai}.
The $s95p$-v1.1 \cite{Huovinen:2009yb} equation of state (EoS) in incorporated in DCCI2.
In the original $s95p$-v1, an EoS of the (2+1)-flavor lattice QCD at high temperature from HotQCD Collaboration \cite{Bazavov:2009zn} is smoothly connected to that from a hadron resonance gas, whose list is taken from Particle Data group as of 2004 \cite{Eidelman:2004wy}, at low temperature.
The particular version of EoS, $s95p$-v1.1, which e employ in the present calculations, is tuned to match the EoS of the hadron resonance gas with the resonances implemented in a hadronic cascade model, JAM, below a temperature of $184$~MeV.
The fluid elements below the switching temperature $T_{\mathrm{sw}}$
can be regarded as hadron gases whose evolution is described by the hadronic cascade model to be mentioned later.
%Hadronization of core
%-----------------------
Once the temperature of the fluid element goes down to $T(x) = T_{\mathrm{sw}}$, I switch the description from hydrodynamics to hadronic transport.
For the switch of the description, I use a Monte-Carlo sampler, \ISthreeD  \ \cite{McNelis:2019auj}, to convert hydrodynamic fields at the switching hypersurface to particles, which I call direct hadrons, in the EoS by sampling based on the Cooper-Frye formula \cite{Cooper:1974mv}.
%String cutting
%-----------------------
Hadronization of non-equilibrated partons 
is performed by the string fragmentation in \pythia8. 
When a color string connecting partons from \pythia8 has a spatial overlap with the medium fluid, I assume that the string is cut and reconnected to 
partons sampled from the medium due to the screening effect of the medium.
%Hadronic afterburner
%-----------------------
The direct hadrons obtained from both \pythia8 and \ISthreeD \ are handed to the hadronic cascade model, \jam \ \cite{Nara:1999dz}, to perform hadronic rescatterings among them and resonance decays. 
In the following subsections, I explain the details of each stage.

\subsection{Generating initial partons}
\label{subsection:Generating_initial_partons}
The initially produced partons, \textit{i.e.}, all partons I use as an input of dynamical initialization, are obtained with \pythia8 or \pythia8 Angantyr.
The \pythia \ program is an open source MC event generator that simulates events in high-energy collisions
widely used among the physics communities working on the LHC experiments, {\it{i.e}.,} $e^+$+$e^-$ or $p$+$p$ collisions.
The back-bone physics which describe the particle production is perturbative QCD based phenomenology
followed by a hadronization called string fragmentation.
In this section, I briefly summarize particle production mechanism in \pythia.

\subsubsection{General picture of PYTHIA}
I would like to first illustrate the general picture of \pythia \ while excellent summary of physics incorporated in \pythia8 and its technical manual is in Ref.~\cite{Bierlich:2022pfr}. 
The flow of a high-energy collision process can be depicted as follows:
First, a {\it{hard scattering}} takes place between two partons selected from a parton distribution function (PDF) of a nucleon
of each beam hadron.
The kinematics of out-coming partons in hard scattering is calibrated from a matrix element based on perturbative QCD.
Meanwhile, those running partons can radiate particles in a vacuum,
which is a process so-called {\it{parton shower}}.
The radiation can take place before and after a hard scattering. It is referred to as {\it{intial state radiation (ISR)}} and {\it{final state radiation (FSR)}}, respectively.
In addition to the ISR and FSR, additional parton-parton scattering can take place in this stage, which is called {\it{multi parton interaction (MPI)}}.
After such partonic processes, due to the color confinement in QCD,
color singlet object, so-called {\it{strings}} in which the stored energy between those quarks linearly increases with the separation of the quark and anti-quark \cite{Neveu:1971rx,Nambu:1974zg,Chodos:1974pn},
is formed. 
Just after the partonic process, several strings coming from different MPIs are mixed up. This interaction among color flow is called {\it{color reconnection}}.
Next, a hadronization process called string fragmentation follows.
Since this is a non-perturbative process, a phenomenological model called Lund string fragmentation is incorporated. Under its description, color strings consist of partons are fragmented into color singlet hadrons.
Finally, resonance hadrons produced from those strings cascadingly decay until all particles become stable.

\subsubsection{Lund string fragmentation}
As I mentioned above, the picture of the string
is coming from the color confinement in QCD.
The confinement potential can be characterized by the linear potential
$V(r) = \kappa r $ where $\kappa$ is a string tension and its typical value is $1$ GeV/fm.
The linearity of the potential can be numerically obtained from lattice QCD calculation \cite{Bali:1994de}.
As a quark and an antiquark become far apart, energy stored in-between these partons linearly increases.
In other words, the leading partons are decelerated because energy is converted into a string in-between the leading partons.  
Thus, equation of motion is
\begin{align}
    \left|  \frac{dp_z}{dt}  \right| = \left|  \frac{dp_z}{dz}  \right| = \kappa,
    \label{eq:EOM_PARTONS_STRING}
\end{align}
which gives the picture of {\it{yo-yo}} motion of a quark and antiquark connected with a string.
Under the yo-yo motion,
if two partons having center-of-mass energy $\sqrt{s}$ form a string when both they are at $z=0$, 
those partons keep running apart until $z=\sqrt{s}/(2\kappa)$ in which
energy of both partons is completely converted into the string
\footnote{
Massless partons are assumed for a simple explanation
while partons have finite masses in actual simulations.
}.
After the stop of leading partons, they are pulled by a string likewise a mass on a spring.
This a pair of quark and antiquark constitutes a single meson
\footnote{
For baryon production in string fragmentation model, 
one needs a special treatment which I do not explain here for simplicity.
}
.
In the case that a formed string has too large energy to produce a single mass-on-shell meson,
new quark and anti-quark pairs are generated and fragment the original string into short pieces that will be hadrons by forming yo-yo as I explained above.

The tunneling probability of a quark-antiquark pair in a strong color field is given based on the WKB approximation as
\begin{align}
    \frac{1}{\kappa} \propto \exp\left( -\frac{\pi m^2_\perp}{\kappa}\right),
    \label{eq:TUNNELING}
\end{align}
where $m_\perp$ is a transverse mass defined as $m_\perp = \sqrt{m^2 + p_T^2}$.
Here, however, it should be noted that the quark flavor dependence of tunneling probability is given by a parameter rather than fixing quark masses and adopting to Eq.~\refbra{eq:TUNNELING}.
The ratio of the probability to pick up $u$, $d$, and $s$ quark is set to be
1:1:0.217 as default in \pythia8.

\subsubsection{PYTHIA Angantyr}
The Angantyr model in \pythia8 was established in Ref.~\cite{Bierlich:2016smv,Bierlich:2017sxk} which describes high-energy heavy-ion collision events based on the wounded nucleon model and string fragmentation.
This is a rich extension from elementally particle collisions or hadronic collisions.
The event generation process can be simply illustrated as follows:
nucleon distributions inside of nuclei are described with a Woods-Saxon parametrization.
With the information of nucleons inside of the nuclei, for instance, transverse positions, participant nucleons in a collision, and types of each collision are assigned: elastic, non-diffractive (absorptive), single diffractive, or double diffractive, based on the Good-Walker formalism \cite{Good:1960ba}.
Once the type of each nucleon-nucleon collision is determined, 
each nucleon-nucleon collision is performed with \pythia8.
Partonic productions produced from each collision are finally combined into one single event to describe a heavy-ion collision.

\subsubsection{Space-time picture of particle productions}

\begin{figure}
    \centering
    \includegraphics[bb=0 0 548 445, width=0.49\textwidth]{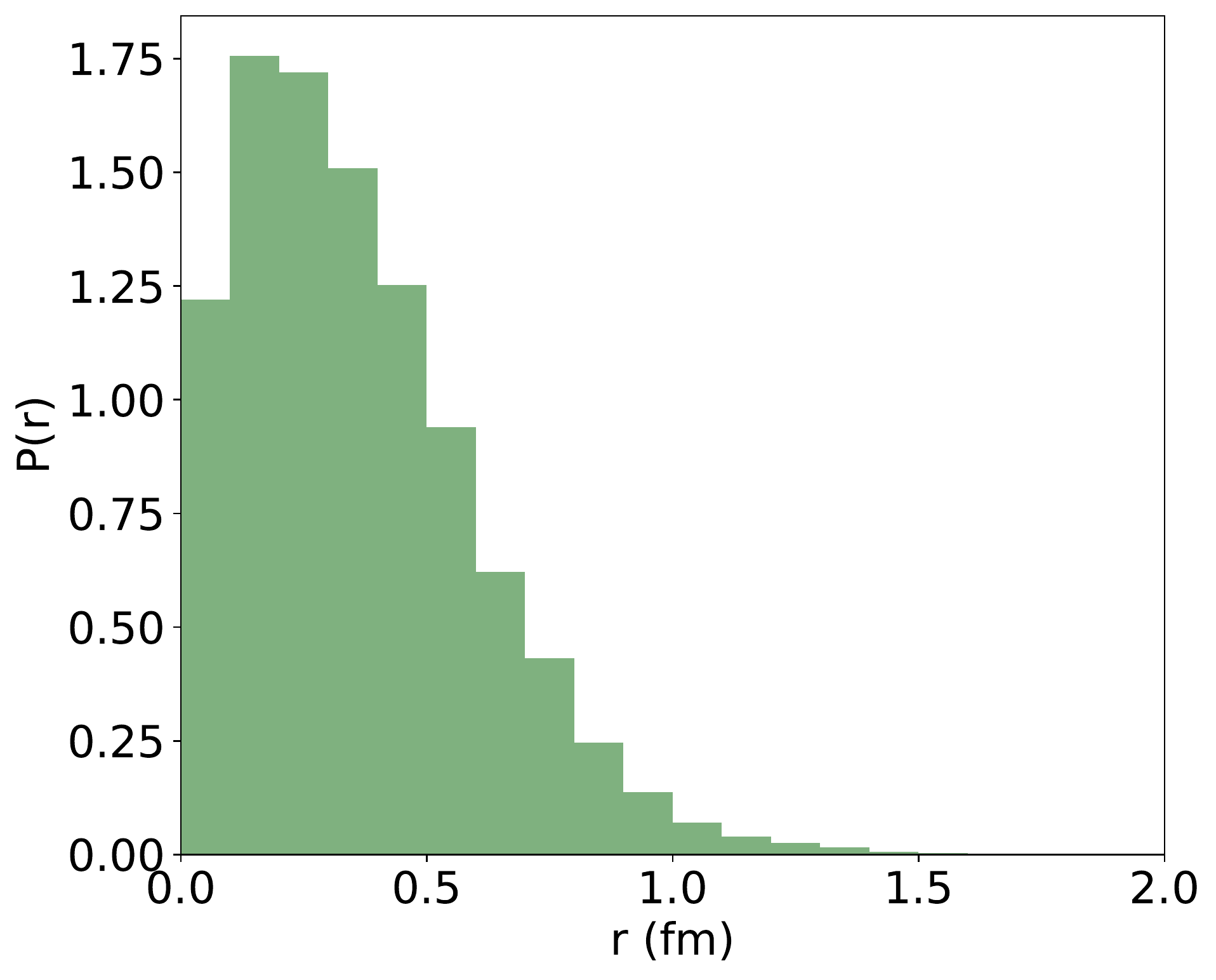}
    \includegraphics[bb=0 0 548 445, width=0.49\textwidth]{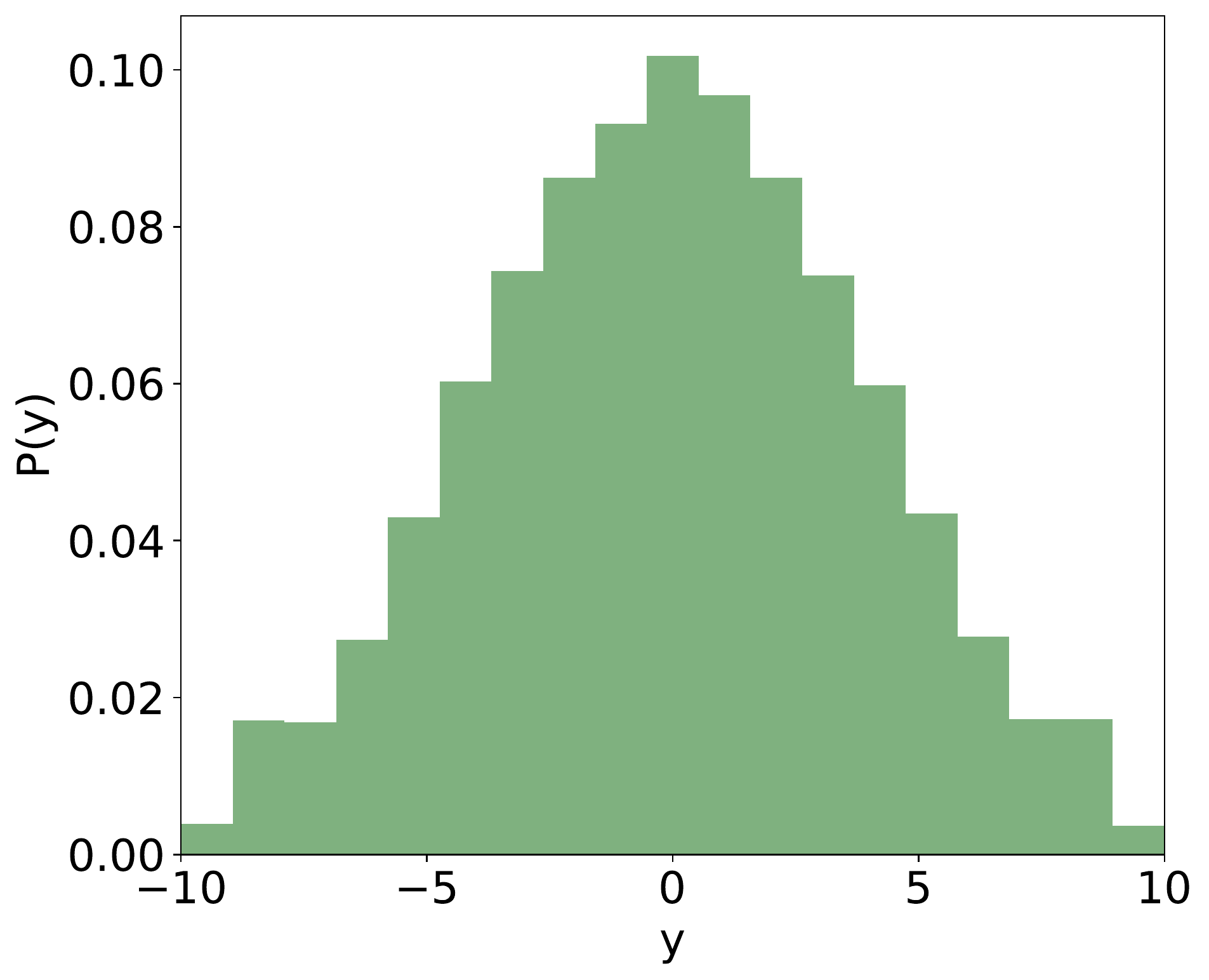}
    \caption{(Left) Normalized distribution of partons as a function of $r=\sqrt{x^2 + y^2}$ where $x$ and $y$ are parton vertices in transverse direction obtained from $p$+$p$ collisions at \snn[proton] = 7 TeV with the parameter set used in DCCI2. (Right) Corresponding normalized distributions of partons as a function of rapidity.}
    \label{fig:PP7_INITPARTON}
\end{figure}

\begin{figure}
    \centering
    \includegraphics[bb=0 0 548 445, width=0.49\textwidth]{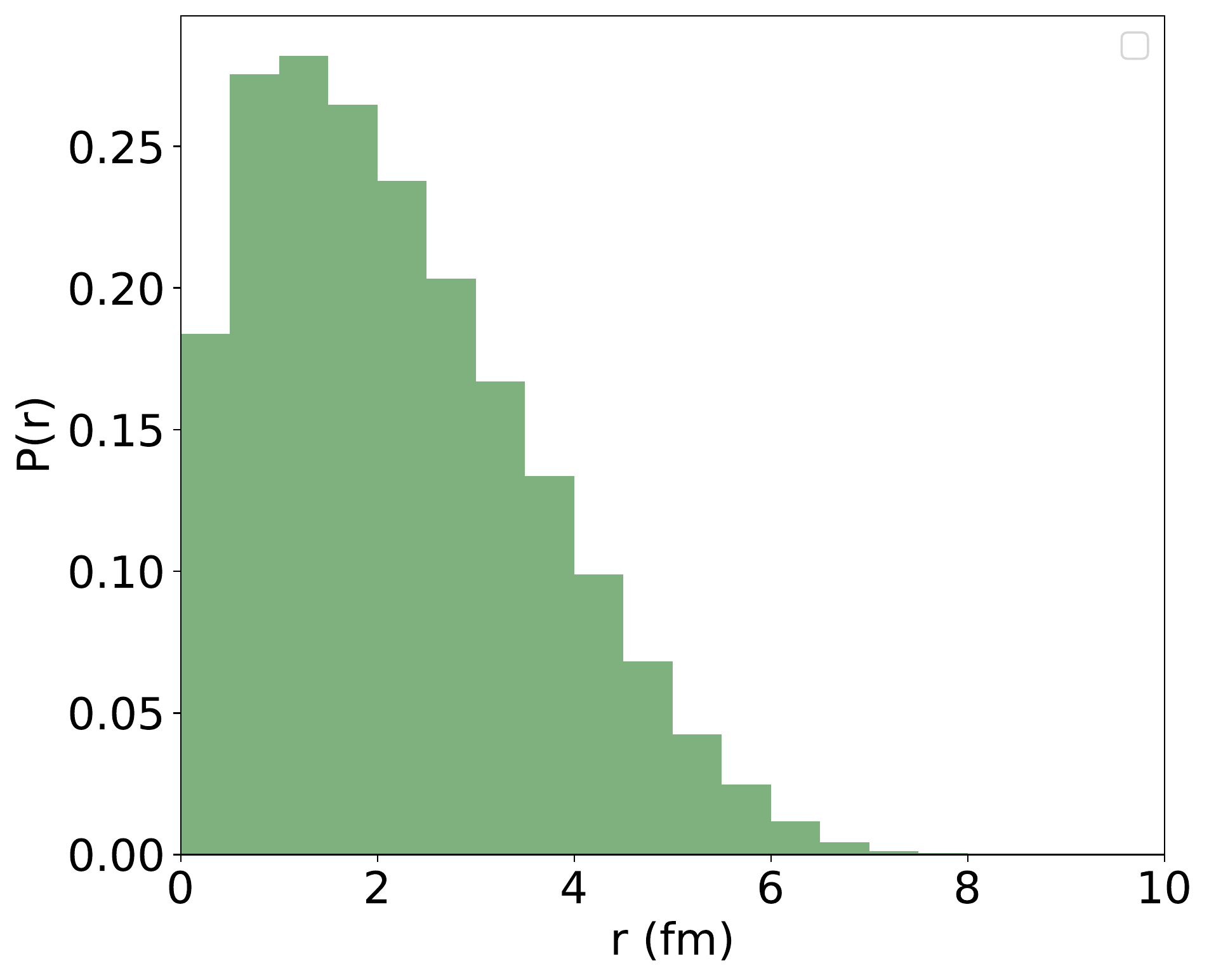}
    \includegraphics[bb=0 0 548 445, width=0.49\textwidth]{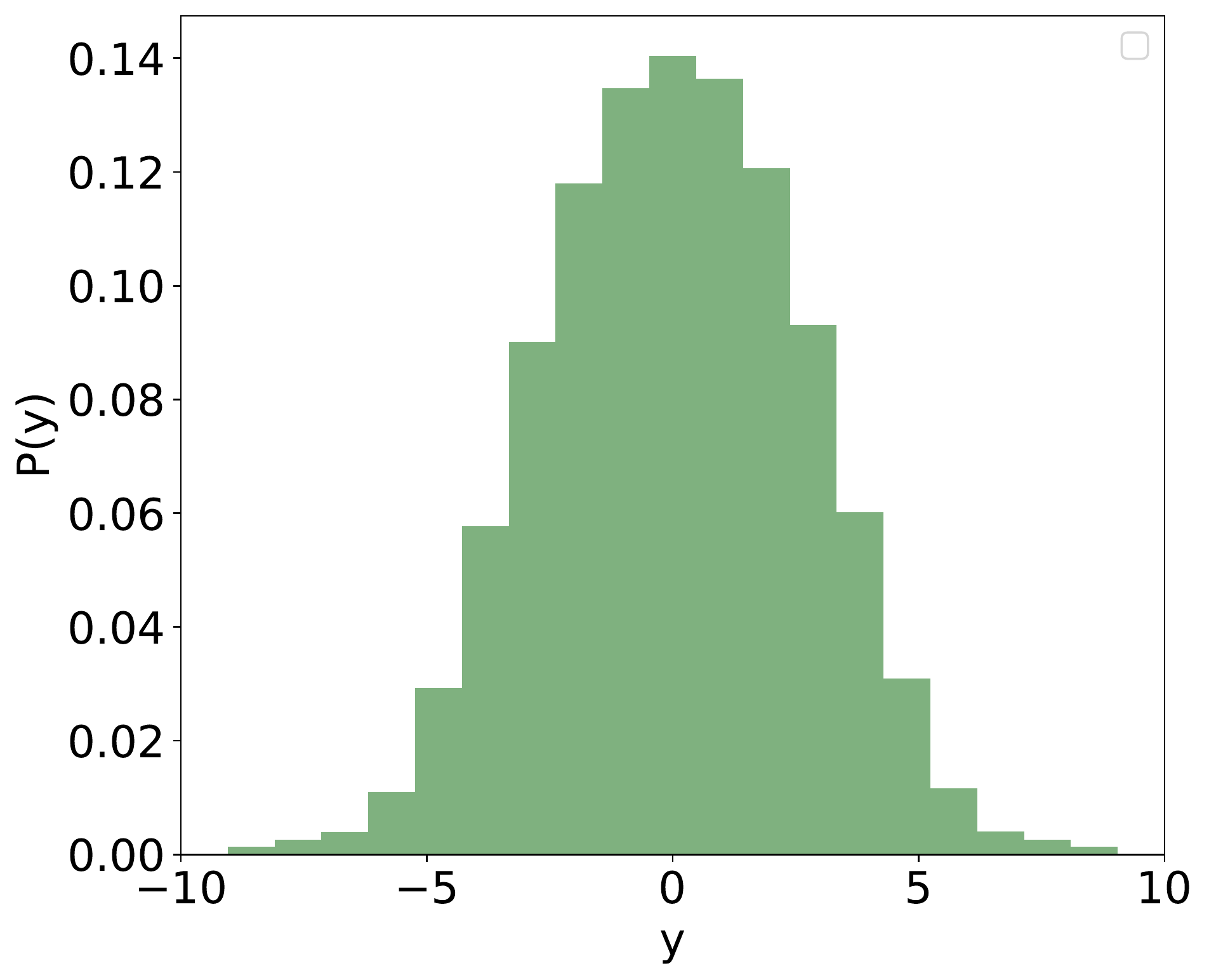}
    \caption{(Left) Normalized distribution of partons as a function of $r=\sqrt{x^2 + y^2}$ where $x$ and $y$ are parton vertices in transverse direction obtained from $Pb$+$Pb$ collisions at \snn = 2.76 TeV with the parameter set used in DCCI2. (Right) Corresponding normalized distributions of partons as a function of rapidity.}
    \label{fig:PBPB2760_INITPARTON}
\end{figure}
In \pythia, all the particle productions scheme originally is formulated in momentum space.
However, space-time pictures of particle productions have been available 
due to recent updates in parton and hadron levels.
At the parton level,
the parton vertex at a hard scattering is sampled randomly.
By default, by considering the probability of picking up a parton from a single proton as a Lorentz-contracted Gaussian function spread from the center of the proton,
the parton vertex of a hard scattering is given by a Gaussian obtained from a product of two Gaussian functions centered at two beam protons.
Once the vertices of mother partons are set, ones of daughter partons produced in parton showers are given by Gaussian distributions with a width of $\sigma_{\mathrm{v}}/\pT$.

The space-time picture of hadronic productions is developed in Ref.~\cite{Ferreres-Sole:2018vgo}
Given the equation of motion of partons described as Eq.~\refbra{eq:EOM_PARTONS_STRING},
once one knows the vertices at the endpoints, which is given by $z_0 = t_0 = \pm \sqrt{s}/(2\kappa)$,
each vertex of breakup points is recursively obtained.
Since a produced meson consisting of a quark and an antiquark originates from two adjacent breakup points,
hadron vertex is obtained from a midpoint between those.

\subsubsection{PYTHIA settings for DCCI2}
Here I summarize settings that I use to generate initial partons.
\begin{itemize}
    \item {\texttt{PartonVertex:setVertex=on}}
    \item {\texttt{HadronLevel:all=off}}
    \item {\texttt{MultipartonInteractions:pT0Ref}}
    \item {\texttt{SpaceShower:pT0Ref}}
\end{itemize}
%Special parameter pT0Ref
%---------------------------
I basically use the default settings in \pythia8 and \pythia8 Angantyr to obtain phase-space distributions of partons except the two parameters, {\texttt{MultipartonInteractions:pT0Ref}}  and {\texttt{SpaceShower:pT0Ref}}.
These parameters regularize cross sections of multiparton interactions and infrared QCD emissions \cite{Pythia82OnlineManual}.
The same value of $p_{\mathrm{T0Ref}}$ is used for both parameters just for simplicity.
Detailed discussion is given in Sec.~\ref{subsection:Evolution_of_transverse_energy} in chapter 3.

%Color information
%-------------------------
The information of color strings is given by \pythia8 and \pythia8 Angantyr besides phase-space information.
In order to respect the configuration of initially produced color strings, I keep this information for dynamical core--corona initialization.
Technically speaking, color and anti-color tags are given to each parton so that one is able to see the configuration of the color strings by tracing the tags.
Note that if there exist junctions, which are Y-shaped objects in which three-string pieces are converged, 
I keep this information as well to trace all strings generated in the event.

%What we actually get.
%-------------------------------------
Eventually, I obtained a particle list for each event including particle IDs, phase-space information, and color and anti-color tags.
Junctions are added to the particle list if they exist in the event.
For heavy-ion collisions obtained with \pythia8 Angantyr, 
weighted events are generated \cite{Bierlich:2018xfw}: The impact parameter is distributed in a way that more central collisions and fewer peripheral collisions are generated than in the minimum bias cases. The corresponding weights are stored to be used in statistical analysis.

Figures \ref{fig:PP7_INITPARTON} and \ref{fig:PBPB2760_INITPARTON} show
initial parton distributions obtained with the parameter set that I mention in Sec.~\ref{subsection:Parameter_set_in_DCCI2} from minimum-bias $p$+$p$ collisions at \snn[proton] = 7 TeV and $Pb$+$Pb$ collisions at \snn = 2.76 TeV, respectively.
As one sees, the parton distributions in radial direction
reflects the size of incoming beam:
parton vertices from $p$+$p$ collisions extends around $r \approx 1$-$2$ fm 
while ones from $Pb$+$Pb$ collisions extend around $r \approx  6$-$8$ fm.
On the other hand, normalized rapidity distributions of initial partons 
show a probability having relatively large value at mid-rapidity in $Pb$+$Pb$ collisions compared to $p$+$p$ collisions.
This is actually related with the parameter 
$p_{\mathrm{T0Ref}}$.
Detailed discussion is given in Sec.~\ref{subsection:Evolution_of_transverse_energy} in chapter 3.

\subsection{Dynamical initialization}
\label{sebsec:DYNAMICAL_INITIALIZATION}
I phenomenologically and dynamically 
describe the initial stage of high-energy nuclear collisions through \textit{dynamical initialization}. 
Just after the first contact of incoming nuclei or nucleons, quarks and gluons are produced through hard scatterings or initial or final state radiations. 
Subsequently, some of those initially produced partons experience the secondary scatterings and contribute to forming the equilibrated matter.
On the other hand, partons that do not experience the interactions and partons surviving even after the secondary interactions contribute as the non-equilibrated matter.

To describe this stage, I start from the continuity equations of the entire system generated in a single collision event
\begin{align}
\label{eq:continuum_tot}
    \partial_\mu T^{\mu\nu}_{\mathrm{tot}} (x)=0.
\end{align}
If I assume that the entire system can be decomposed into equilibrated matter (fluids) and non-equilibrated matter (partons), Eq.~\refbra{eq:continuum_tot} can be written as 
\begin{align}
\label{eq:continuum_eq}
   \partial_\mu T^{\mu\nu}_{\mathrm{tot}}(x) =    \partial_\mu \left[ T^{\mu\nu}_{\mathrm{fluids}}(x) + T^{\mu\nu}_{\mathrm{partons}} (x)\right] = 0,
\end{align}
where $T^{\mu\nu}_{\mathrm{fluids}}$ and $ T^{\mu\nu}_{\mathrm{partons}}$ are energy-momentum tensors of equilibrated matter (fluids) and non-equilibrated matter (partons), respectively.
Then the space-time evolution of fluids can be expressed as a form of hydrodynamic equations with source terms
\begin{align}
\label{eq:hydro_with_source_terms}
\partial_\mu T^{\mu\nu}_{\mathrm{fluids}}(x) &= J^{\nu} (x),
\end{align}
where the source terms are written as
\begin{align}
\label{eq:source_terms}
J^\nu  = - \partial_\mu & T_{\mathrm{partons}}^{\mu\nu}.
\end{align}
Here I assume the energy and momentum deposited from non-equilibrated partons instantly reach a state under local thermal and chemical equilibrium. 

The exact form of the source terms is obtained by defining phase-space distributions and kinematics of initial partons \cite{Kanakubo:2019ogh}.
For the phase-space distributions, I assume
\begin{align}
\label{eq:phasespace-distribution-of-partons}
    f (\bm{x}, \bm{p}; t) d^3\!xd^3\!p = \sum_{i} G(\bm{x}\!-\!\bm{x}_i(t))  \delta^{(3)}\! (\bm{p}-\bm{p}_i(t))d^3\!xd^3\!p,
\end{align}
where $G(\bm{x}\!-\!\bm{x}_i(t))$ is a three-dimensional Gaussian distribution centered at a position of the $i$th parton, $\bm{x}_i(t)$, generated in one single event. 
I assume that each parton traverses along a straight trajectory. 
Under this assumption, the position of a parton at an arbitrary time is obtained as
\begin{align}
\label{eq:trajectories_of_partons}
    \bm{x}_{i} (t) = \frac{\bm{p}_{i}(t)}{p^0_{i}(t)}(t-t_{\mathrm{form},i})+\bm{x}_{\mathrm{form},i},
\end{align}
where 
$t_{\mathrm{form},i}$ and $\bm{x}_{\mathrm{form},i}$ are a formation time and a formation position, respectively. The $i$th parton is assumed to be formed at a common proper time, $\tau = \tau_{0}$. 
With these assumptions, the explicit form of the source terms in Eq.~\refbra{eq:source_terms} is obtained as \cite{Kanakubo:2019ogh}
\begin{align}
\label{eq:source_terms_dpdt}
    J^\nu &= - \partial_\mu T_{\mathrm{partons}}^{\mu\nu} \nonumber \\
    &=-\sum_{i}\int d^3 p \frac{p^\mu p^\nu}{p^0} \partial_\mu f(\bm{x},\bm{p}; t) \nonumber \\
    &=-\sum_{i}\frac{dp_i^\nu(t)}{dt}G(\bm{x}\!-\!\bm{x}_i(t)).
\end{align}
As one reads from the last line of Eq.~\refbra{eq:source_terms_dpdt}, 
the source terms of fluids are described as a summation of deposited energy-momentum of initial partons.

Space-time evolution of fluids
is described by ideal hydrodynamics.
This does not mean I do not describe any non-equilibrium components within DCCI2: Dissipative hydrodynamics deals only with small non-equilibrium corrections to equilibrium components, while the corona components, which are far from equilibrium, are taken into account in DCCI2. 
By neglecting dissipative terms for simplicity,
energy-momentum tensor of fluids is expressed as 
\begin{align}
    \label{eq:EMtensor_idealfluids}
    T_{\mathrm{fluids}}^{\mu\nu} = (e + P) u^{\mu}u^{\nu} - Pg^{\mu\nu},
\end{align}
where $e$, $P$, and $u^{\mu}$ are energy density, hydrostatic pressure, and four-velocity of fluids, respectively.
In this study, I do not solve conserved charges such as baryon number, strangeness, and electric charges.
It would be interesting to investigate them considering the initial distribution of the charges \cite{Martinez:2019jbu,Martinez:2019rlp}
.

As I mentioned at the beginning of this section, I perform actual hydrodynamic simulations in the (3+1)-dimensional Milne coordinates, $\tau = \sqrt{t^2-z^2}$, $\bm{x}_\perp = (x, y)$, and $\eta_s = (1/2) \ln{\left[(t+z)/(t-z)\right]}$.
In this case, the Gaussian distribution in Eq.~\refbra{eq:phasespace-distribution-of-partons} is replaced with
\begin{align}
 &G(\bm{x}_\perp\!-\!\bm{x}_{\perp, i}, \eta_{s}\!-\!\eta_{s,i}) d^2\!x_\perp \tau d\eta_{s}\nonumber \\ 
    & = \frac{1}{2\pi\sigma_{\perp}^2} \exp{\left[ -\frac{(\bm{x}_\perp - \bm{x}_{\perp,i} )^2}{2\sigma_\perp^2} \right]} \nonumber \\
& \times \frac{1}{\sqrt{2\pi\sigma_{\eta_s}^2 \tau^2}} \exp{\left[ -\frac{(\eta_{s} - \eta_{s,i})^2}{2\sigma_{\eta_s}^2} \right]}d^2\!x_\perp \tau d\eta_{s},
\end{align}
where $\sigma_{\perp}$ and $\sigma_{\eta_s}$ are transverse and longitudinal widths of the Gaussian distribution, respectively.
In the longitudinal direction, a straight trajectory implies $\eta_{s,i}=y_{p,i}$, where 
$\eta_{s,i} = (1/2) \ln{\left[(t+z_i)/(t-z_i)\right]}$ and $ y_{p,i}= (1/2)\ln{\left[(E_i+p_{z,i})/(E_i-p_{z,i})\right]}$ are space-time rapidity and momentum rapidity of the $i$th parton, respectively. 
In the following, I show some formulas in the Cartesian coordinates to avoid the complex notation of them. 
In any case, all the hydrodynamic simulations are performed in the Milne coordinates.

\subsection{Dynamical core--corona initialization}
\label{subsec:DYNAMICAL_CORECORONA_SEPARATION}
I establish the dynamical aspect of core--corona picture by
modeling the four-momentum deposition of partons $dp_i^\mu (t)/dt$ in Eq.~\refbra{eq:source_terms_dpdt} as an extension of the conventional core--corona picture.
The dynamical aspect of the core--corona picture, which I are going to model, is as follows:
Partons which are to experience sufficient secondary scatterings with others are likely to deposit their energy-momentum and form equilibrated matter (QGP fluids), while partons which are to rarely interact with others are likely to be free from depositing their energy-momentum.
To model the dynamical energy-momentum deposition under the core--corona picture, I invoke the equation of motion with a drag force caused by microscopic interactions with other particles.

I define the four-momentum deposition rate of the $i$th parton generated initially at a co-moving frame along $\eta_{s,i} = y_{p, i}$, space-time rapidity of the $i$th parton, as 
\begin{align}
\label{eq:four-momentum-deposition}
    \frac{d\tilde{p}_i^\mu}{d\tau} = - \sum_j^{\mathrm{coll}} \sigma_{ij}
     \tilde{\rho}_{ij}
    |\tilde{v}_{\mathrm{rel},ij}|\tilde{p}^\mu_i,
\end{align}
where $\sigma_{ij}$ is a cross section of the collision between the $i$th and $j$th partons, $ \tilde{\rho}_{ij}$ is an effective density of the $j$th parton seen from the $i$th parton which is normalized to be unity, $|\tilde{v}_{\mathrm{rel},ij}|$ is relative velocity between the $i$th and the $j$th partons. 
Variables with tilde are defined at each co-moving frame along $\eta_{s,i}$. 
The Lorentz transformation from laboratory frame to a co-moving frame along $\eta_{s,i}$ is given as, 
\begin{eqnarray} 
\label{eq:boost_eta} 
\Lambda^\mu_{\enskip \nu} (\eta_{s,i}) &=& \left( 
\begin{array}{cccc} 
\cosh{\eta_{s,i}} & 0 & 0 & -\sinh{\eta_{s,i}}\\ 
0 & 1 & 0 & 0\\ 
0 & 0 & 1 & 0\\ 
-\sinh{\eta_{s,i}} & 0 & 0 & \cosh{\eta_{s,i}} 
\end{array} 
\right). 
\end{eqnarray}

The summation in Eq.~\refbra{eq:four-momentum-deposition} is taken for all partons with non-zero energy that the $i$th parton will collide.
The candidate partons include not only initially produced ones but also ones in the thermalized medium.
I explain details on the treatment to pick up the thermalized partons in Sec.~\ref{subsection:Sampling_of_thermalized_partons}.
I employ an algorithm \cite{Hirano:2012yy} to evaluate the number of partonic scatterings that a parton undergoes along its trajectory.
Under the geometrical interpretation of cross sections, two partons, $i$ and $j$, are supposed to collide when the closest distance of them is smaller than $\sqrt{\sigma_{ij}/\pi}$ where $\sigma_{ij}$ is the same variable used in  Eq.~\refbra{eq:four-momentum-deposition}.
The cross section $\sigma_{ij}$ is defined as 
\begin{align}
\label{eq:parton_cross_section}
    \sigma_{ij}= \min\left\{ \frac{\sigma_0}{s_{ij}/\mathrm{GeV^2}}, \pi b_{\mathrm{cut}}^2 \right\},
\end{align}
where $\sigma_0$ is a parameter with a dimension of area, $s_{ij}$ is a Mandelstam variable $s_{ij} = (\tilde{p}_i^\mu + \tilde{p}_j^\mu)^2$, and $b_{\mathrm{cut}}$ is a parameter to avoid infrared divergence of the cross section when $s_{ij}$ becomes too small.
I neglect possible color Casimir factors in the cross section, and this parametrization is applied for all quarks, anti-quarks, and gluons.
It should be emphasized that this energy dependence of the cross section captures the core--corona picture in the \textit{momentum} space: the rare and the high-energy partons are not likely to deposit the four-momentum during the dynamical initialization process.

The effective density of the $j$th parton that is seen 
from the position of the $i$th one is defined as follows. The value of Gaussian
distribution centered at $\tilde{\bm{x}}_j$ is obtained at $\tilde{\bm{x}}_i$,
\begin{align}
   \tilde{\rho}_{ij}
  & =\left. G(\tilde{\bm{x}}_{\perp}, \tilde{\eta}_{s}; \ \tilde{\bm{x}}_{\perp,j}, \tilde{\eta}_{s,j})\right|_{\tilde{\bm{x}}_{\perp}=\tilde{\bm{x}}_{\perp,i},  \tilde{\eta}_{s}= \tilde{\eta}_{s,i}}\nonumber \\
    & = \frac{1}{2\pi\tilde{w}_{\perp}^2} \exp{\left[ -\frac{(\tilde{\bm{x}}_{\perp,i} - \tilde{\bm{x}}_{\perp,j} )^2}{2\tilde{w}_\perp^2} \right]}  \times \frac{1}{\sqrt{2\pi\tilde{w}_{\eta_s}^2 \tau^2}} \exp{\left[ -\frac{(\tilde{\eta}_{s,i} - \tilde{\eta}_{s,j})^2
    }{2\tilde{w}_{\eta_s}^2} \right]}.
\end{align}
Note that $\tilde{\eta}_{s,i} - \tilde{\eta}_{s,j} = \eta_{s,i} - \eta_{s,j}$, $\tilde{\bm{x}}_{\perp,i} = \bm{x}_{\perp,i}$, $\tilde{w}_\perp = w_{\perp}$, and $\tilde{w}_{\eta_{s}} = w_{\eta_{s}}$.
The relative velocity is calculated as
\begin{align}
    |\tilde{\bm{v}}_{\mathrm{rel}, \it{ij}}| = \left| \frac{\tilde{\bm{p}}_i}{\tilde{p}_i^0} - \frac{\tilde{\bm{p}}_j}{\tilde{p}_j^0} \right|.
\end{align}

As a consequence of the modeling for the four-momentum deposition rate, 
initial partons traversing dense regions with low energy and momentum tend to deposit their energy-momentum and generate QGP fluids.
On the other hand, initial partons traversing dilute regions with high energy tend to relatively keep their initial energy and momentum.
Here the factor $\sum_j^{\mathrm{coll}} \sigma_{ij}\tilde{\rho}_{ij} |\tilde{v}_{\mathrm{rel},ij}| d\tau$ can be regarded as the number of scattering that the $i$th parton experiences during $d\tau$.

During the DCCI processes, I monitor the change of the invariant mass of a string which is composed of the color-singlet combination of initial partons provided by \pythia8.
It should be noted that once the invariant mass of a string becomes smaller than a threshold to be hadronized via string fragmentation in \pythia8, energy-momentum of all the partons that compose the string is dumped into fluids.
In this model, I use $m_{\mathrm{th}} = m_1 + m_2 +1.0$ in units of $\mathrm{GeV}$ for the threshold, where $m_1 $ and $m_2$ are masses of each parton at both ends of the string. 

I emphasize here that the formulation of the four-momentum deposition rate of a parton is largely sophisticated from the one introduced in the previous work \cite{Kanakubo:2019ogh} although the basic concept of the core--corona picture is the same in both cases.
Under the previous work, there was a problem that high $p_T$ partons suffer from unexpected large suppression even in $p$+$p$ collisions.
The reason is that, since partons in parton showers in a high $p_T$ jet are collimated and close to each other in both coordinate and momentum spaces, they had to deposit large four-momentum in our previous prescription in which only density and transverse momentum of the $i$th parton are taken into account \cite{Kanakubo:2019ogh}.
This problem is reconciled in this sophisticated modeling by considering trajectories of partons and relative velocities of parton pairs $|\tilde{v}_{\mathrm{rel},ij}|$.
Since trajectories of shower partons in a jet are supposed not to cross each other, the four-momentum deposition due to collisions among the shower partons is not likely to be counted in the summation in Eq.~\refbra{eq:four-momentum-deposition}.
Even if one consider trajectories of shower partons at the early time of dynamical core--corona initialization, they are close in space-time coordinates and would unreasonably deposit their four-momentum. 
The relative velocity avoids this issue because shower partons are supposed to have small relative velocities. 
Thus, because of the two factors, the dynamical core--corona initialization with Eq.~\refbra{eq:four-momentum-deposition} does not cause the unreasonable four-momentum deposition for partons in jets in DCCI2.

\begin{figure}
    \centering
    \includegraphics[bb = 0 0 545 448, width=0.49\textwidth]{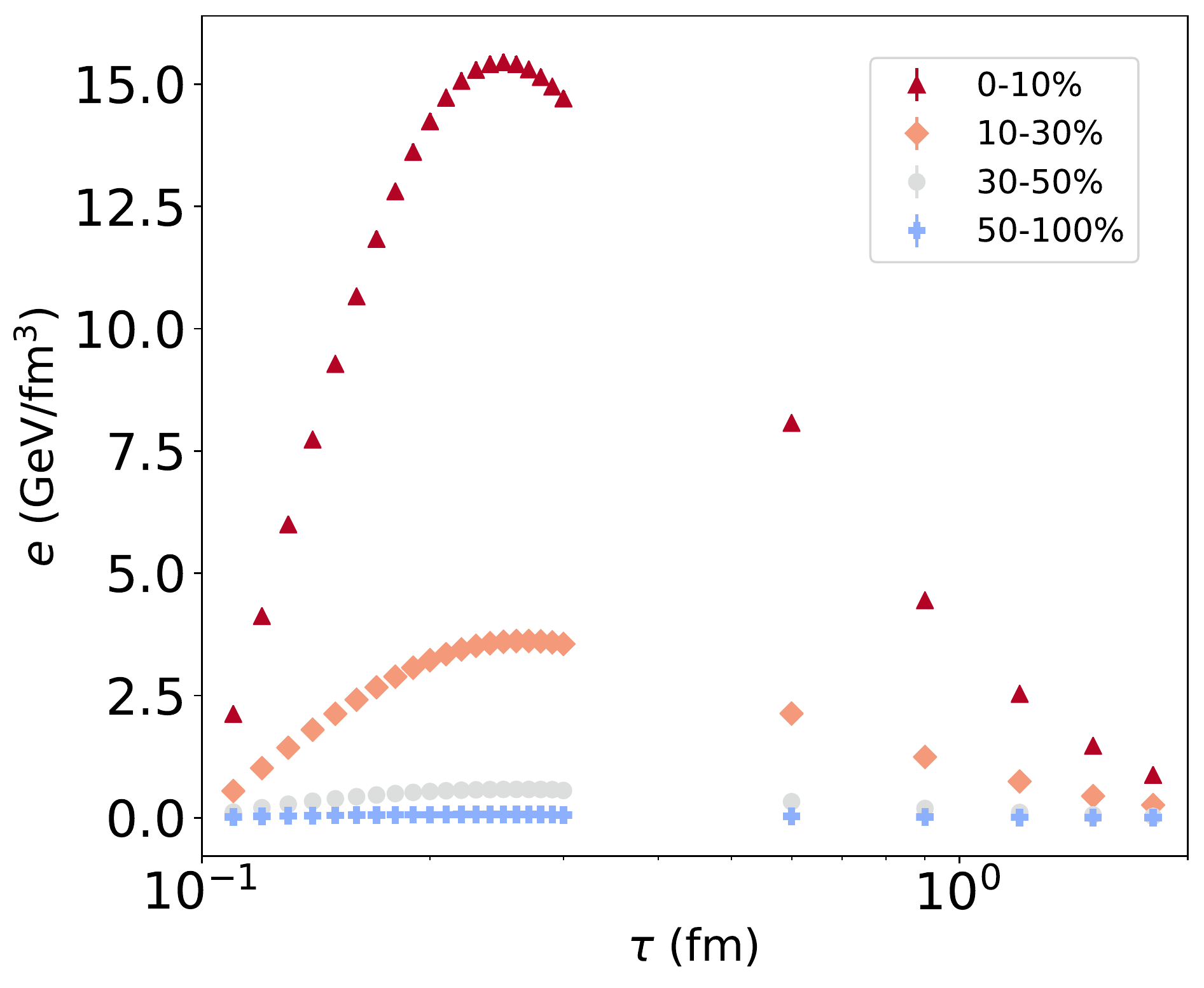}
    \includegraphics[bb = 0 0 545 448, width=0.49\textwidth]{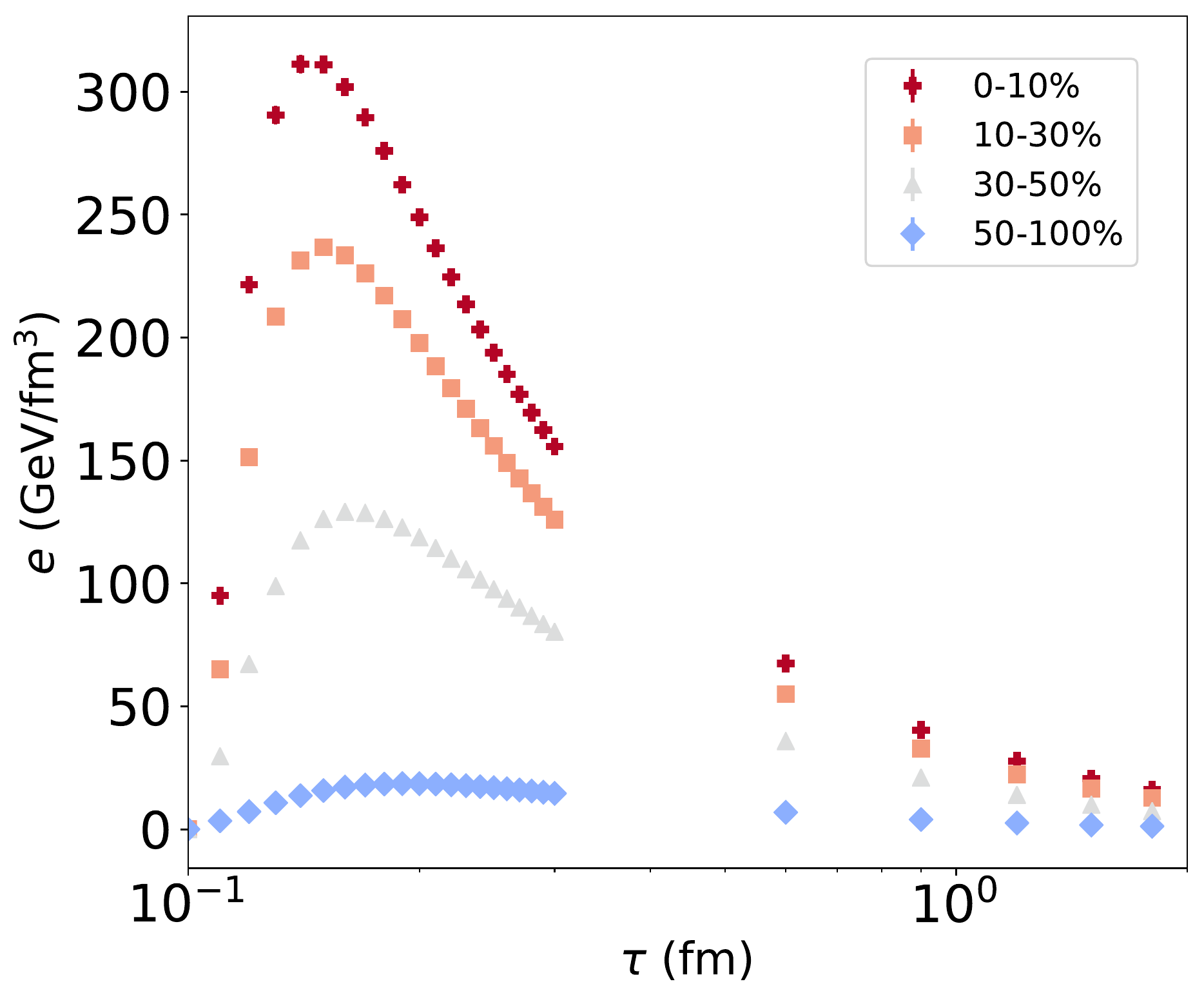}
    \caption{Averaged time evolution of energy density at $(x, y, \eta_s) = (0, 0, 0)$ in hydrodynamic simulations in DCCI2 from $p$+$p$ collisions at \snn[proton] = 13 TeV (left) and $Pb$+$Pb$ collisions at \snn= 2.76 TeV (right) in different in different multiplicity/centrality classes.}
    \label{fig:ENERGYDENS_vs_tau}
\end{figure}
Figure \ref{fig:ENERGYDENS_vs_tau} shows averaged time evolution of energy density at $(x, y, \eta_s) = (0, 0, 0)$ in hydrodynamic simulations of DCCI2 from $p$+$p$ collisions at \snn[proton] = 13 TeV and $Pb$+$Pb$ collisions at \snn= 2.76 TeV in different multiplicity/centrality classes,
which are shown in left and right figures, respectively.
Note that the width of time step of hydrodynamic simulations changes at $\tau=0.3$ fm.
Because of source terms in hydrodynamic equations given to hydrodynamics as Eq.~\refbra{eq:hydro_with_source_terms},
the enhancement of energy stored as fluids in hydrodynamics is seen.
It should be noted that there are two dynamics competing here, one is the energy deposition as sources and the other is longitudinal expansion of the system
\footnote{
The decrease of energy happens mainly because of the strong longitudinal expansion
which originate from the momentum in beam direction
while there is a certain effect of transverse expansion as well.
}
where the former increases energy density and the latter decreases.
The peak of energy density appears as a consequence of the competition.

\begin{figure}
    \centering
    \includegraphics[bb=0 0 410 324, width=0.49\textwidth]{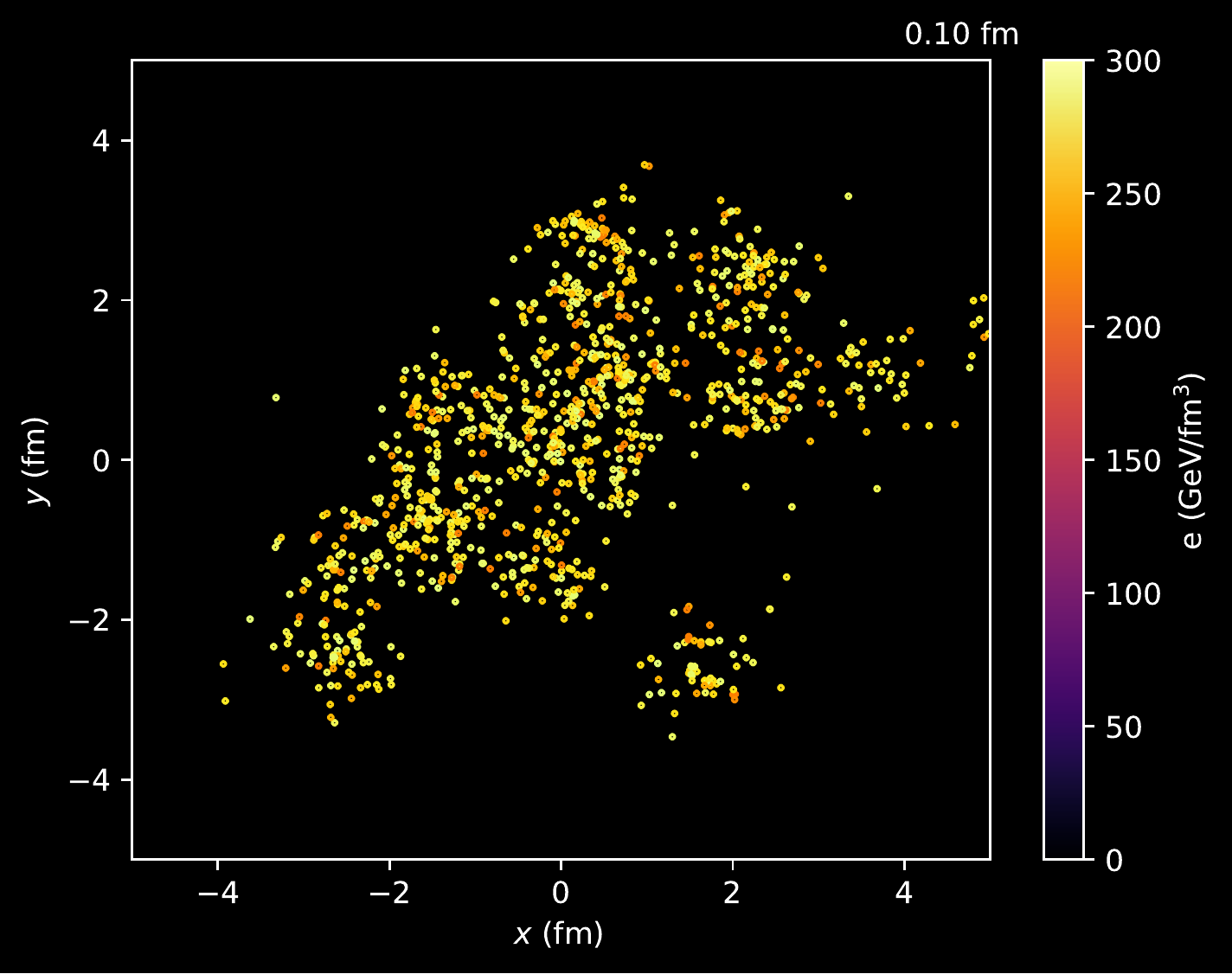}
    \includegraphics[bb=0 0 410 324, width=0.49\textwidth]{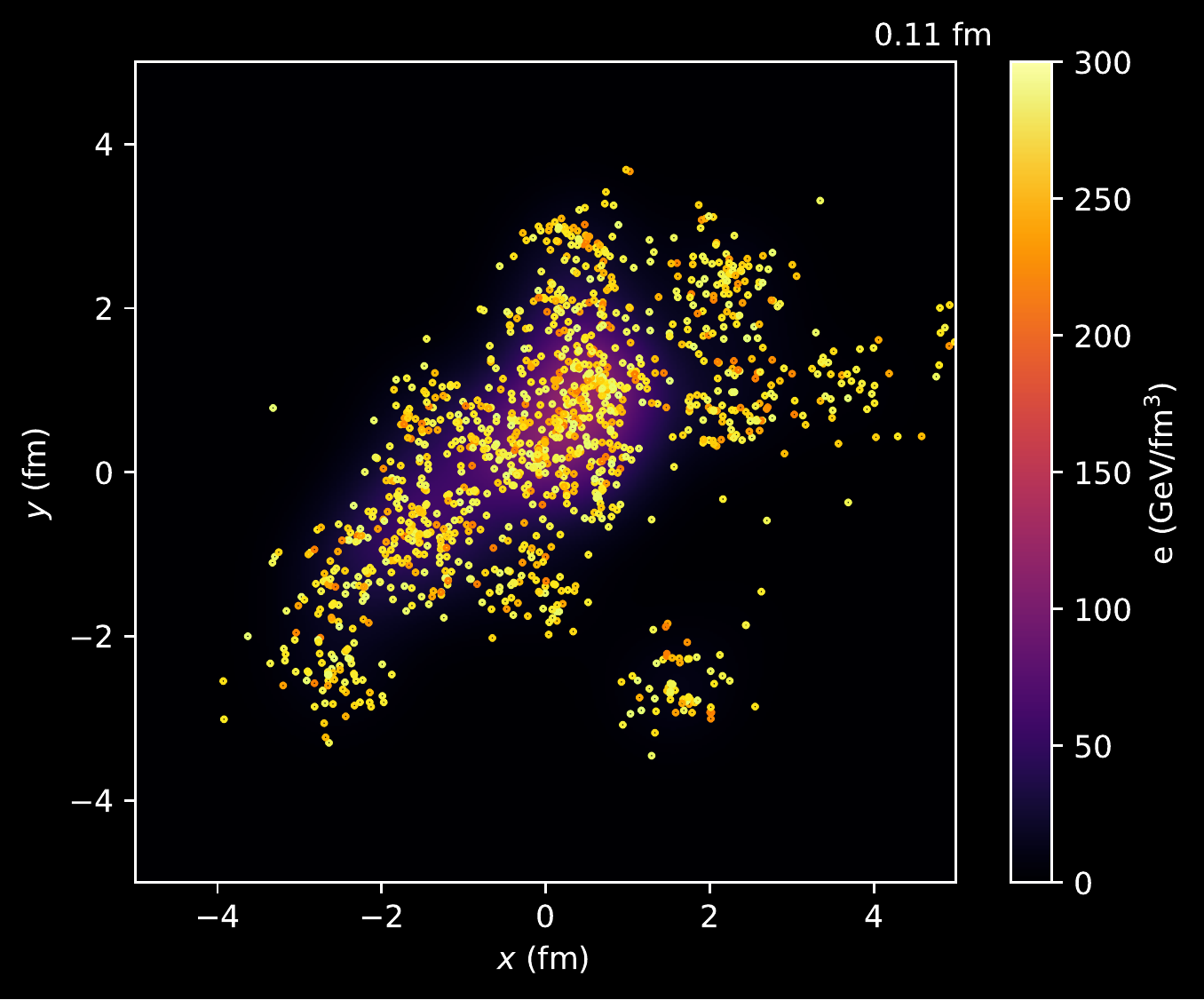}
    \includegraphics[bb=0 0 410 324, width=0.49\textwidth]{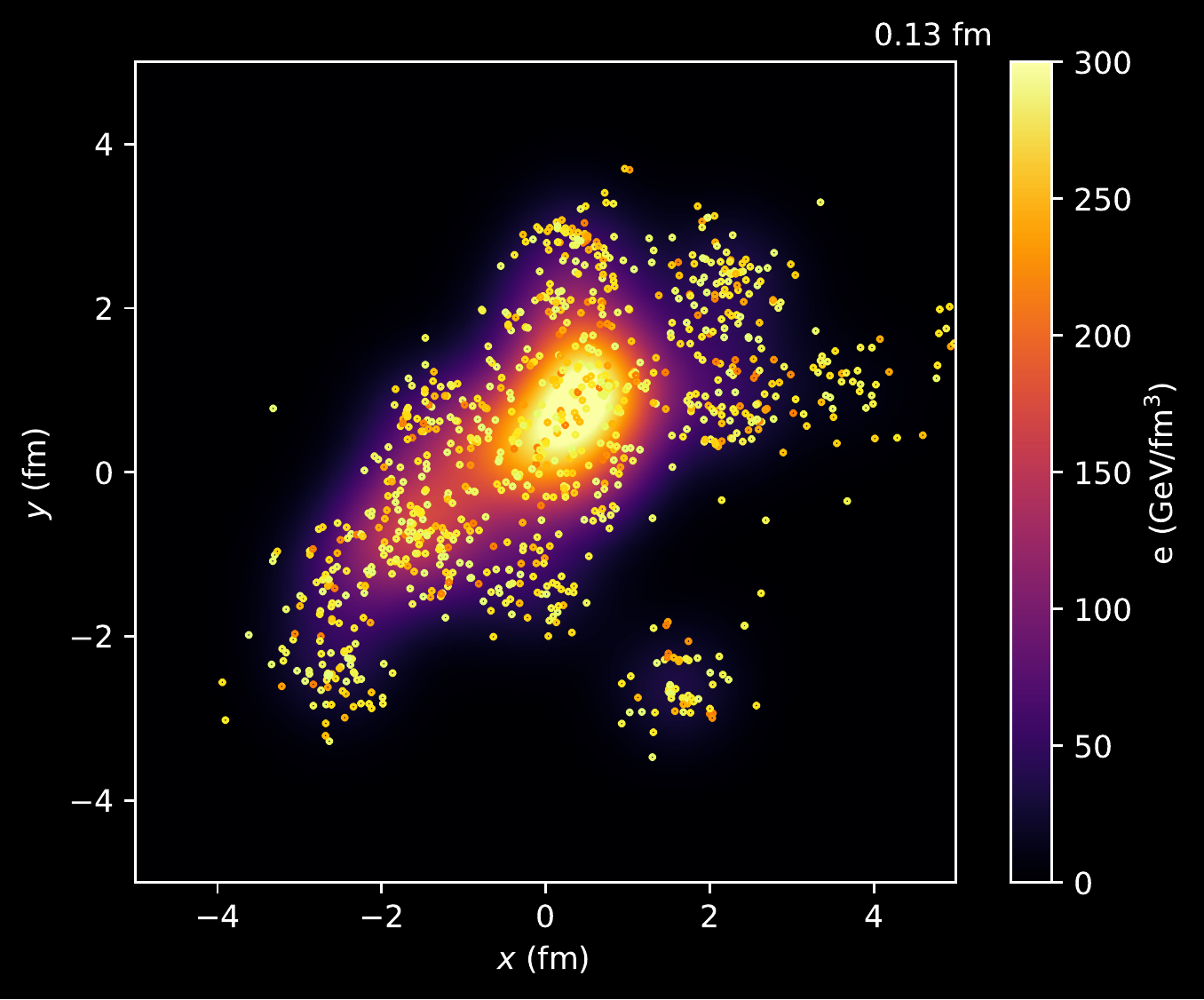}
    \includegraphics[bb=0 0 410 324, width=0.49\textwidth]{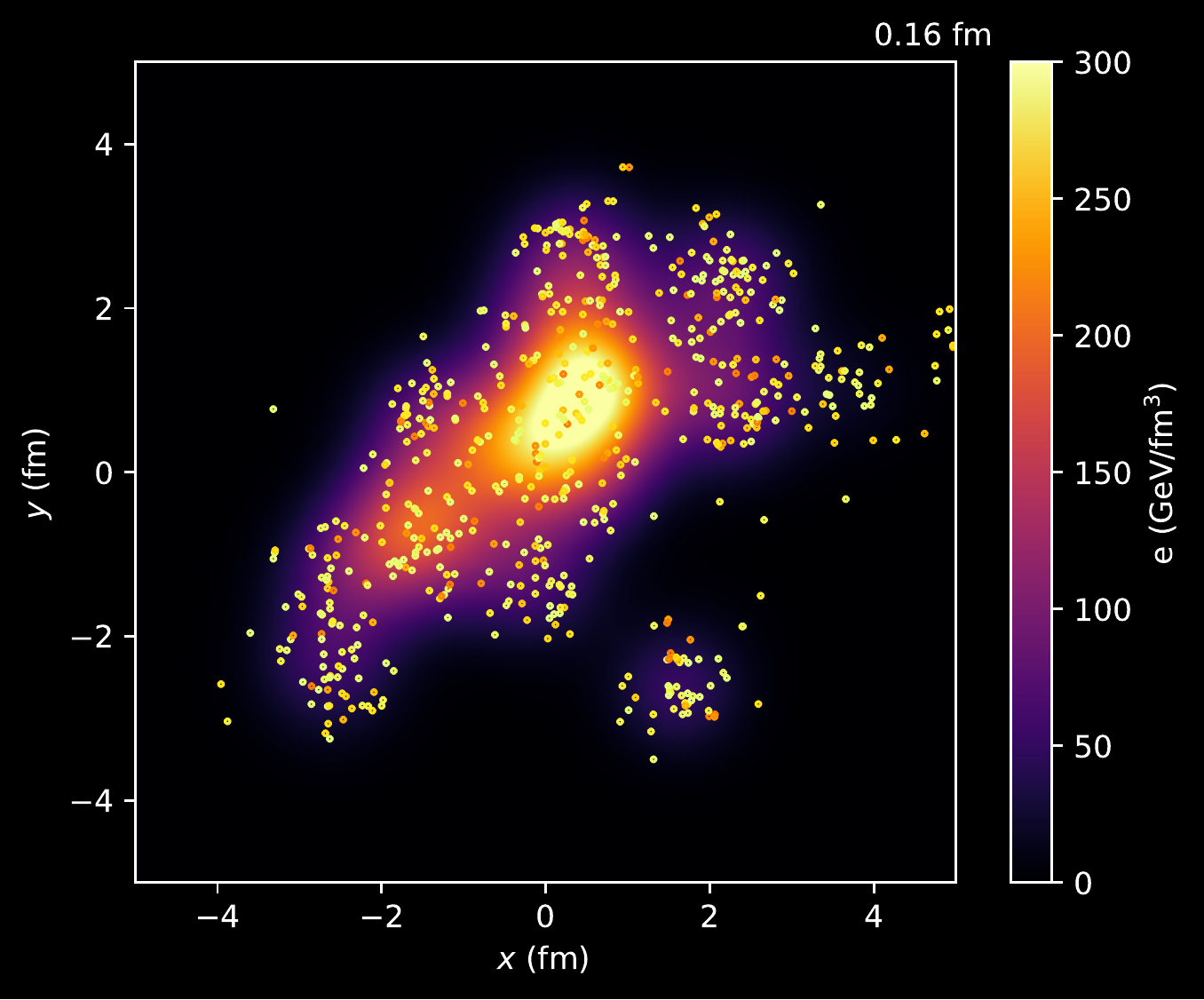}
    \includegraphics[bb=0 0 410 324, width=0.49\textwidth]{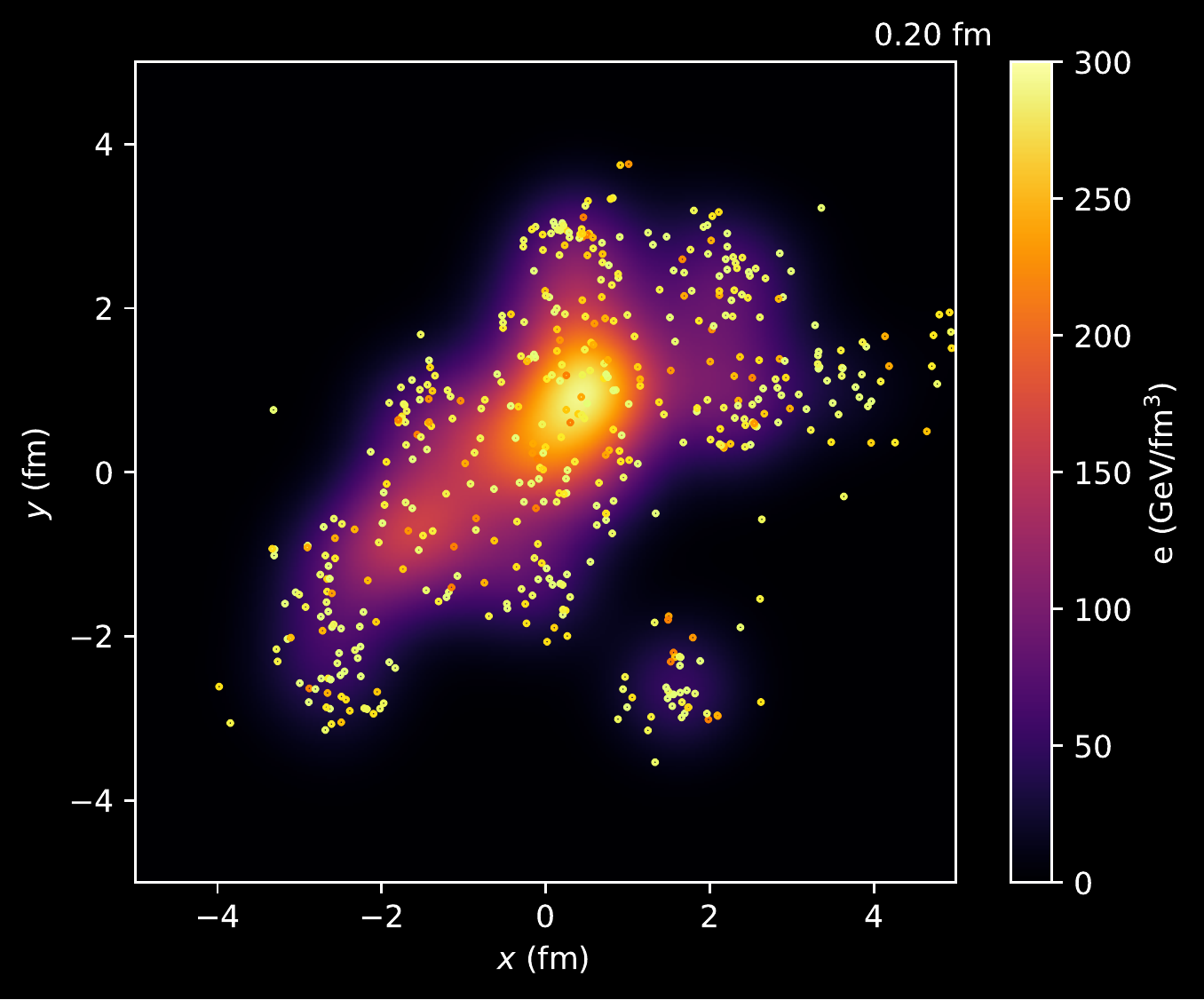}
    \includegraphics[bb=0 0 410 324, width=0.49\textwidth]{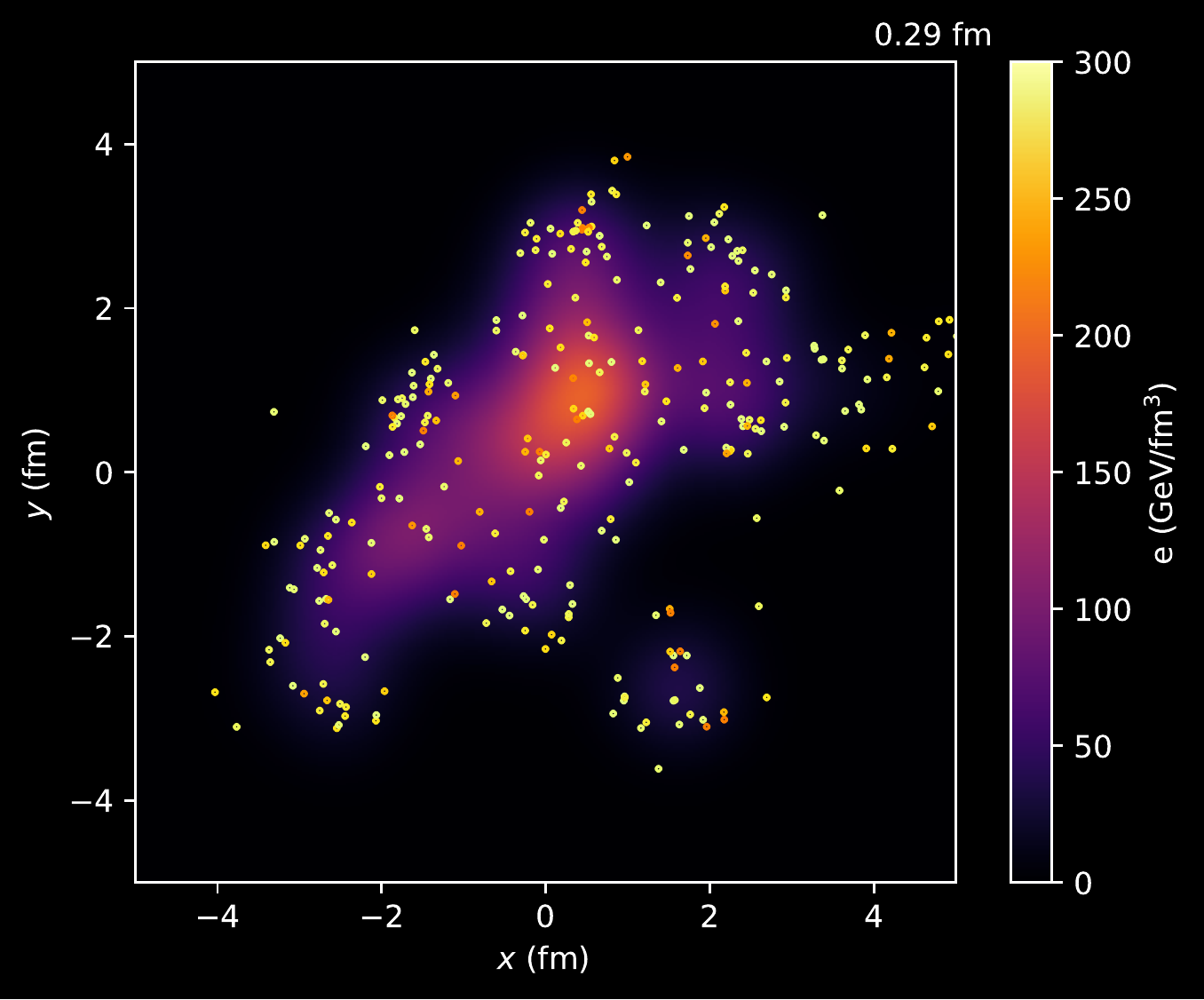}
    \caption{Snapshots of dynamical core--corona initialization from $\tau=0.10$ to $0.29$ fm in a single $Pb$+$Pb$ collision events at \snn = 2.76 TeV in transverse plane at midrapidity. Energy density of generated QGP fluids at $\eta_s=0$ is shown in color map and non-equilibrated partons within $|\eta_s|<1$ are plotted with yellow filled circles.}
    \label{fig:DCCI_DEMO_xy}
\end{figure}

\begin{figure}
    \centering
    \includegraphics[bb=0 0 409 290, width=0.49\textwidth]{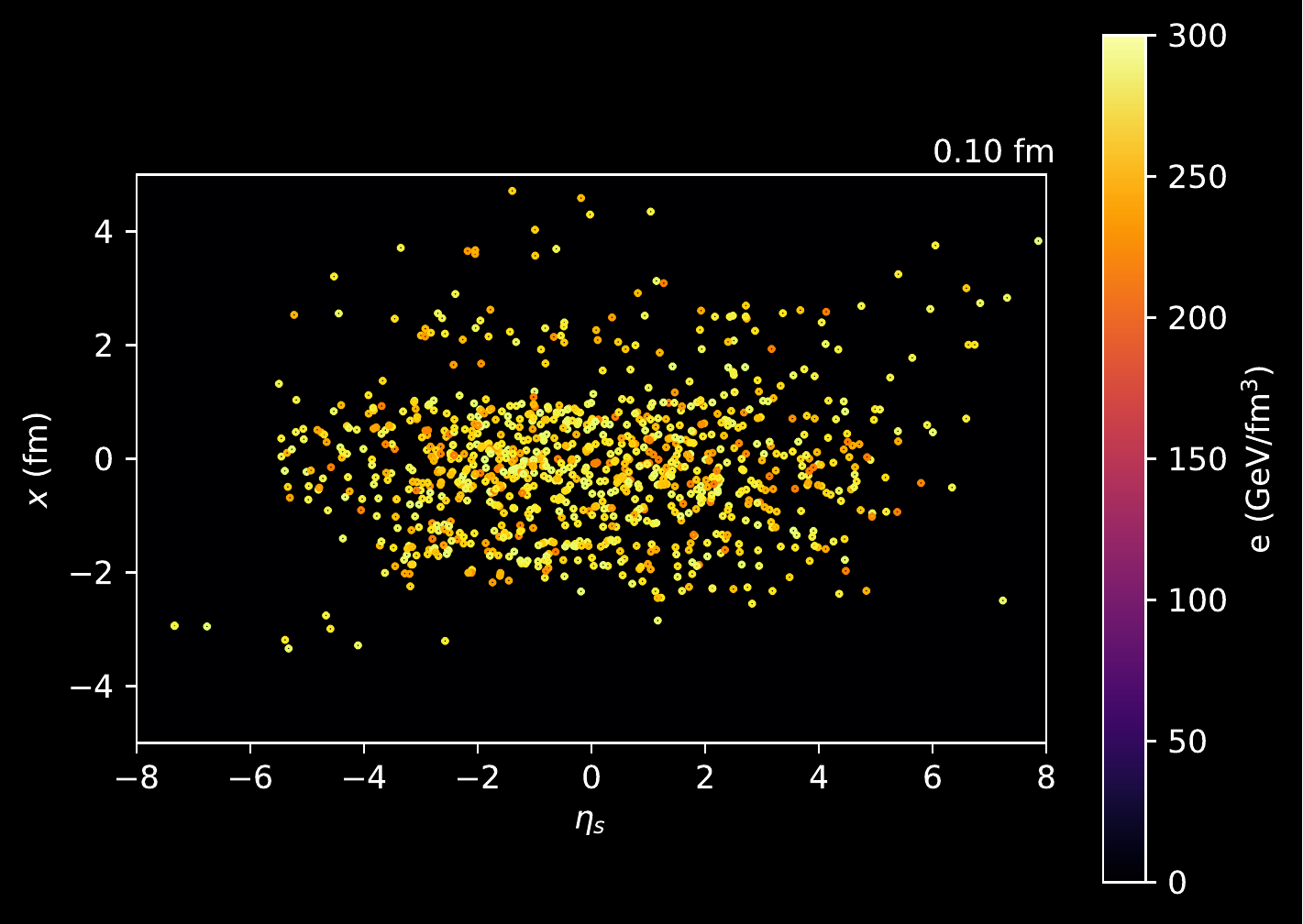}
    \includegraphics[bb=0 0 409 290, width=0.49\textwidth]{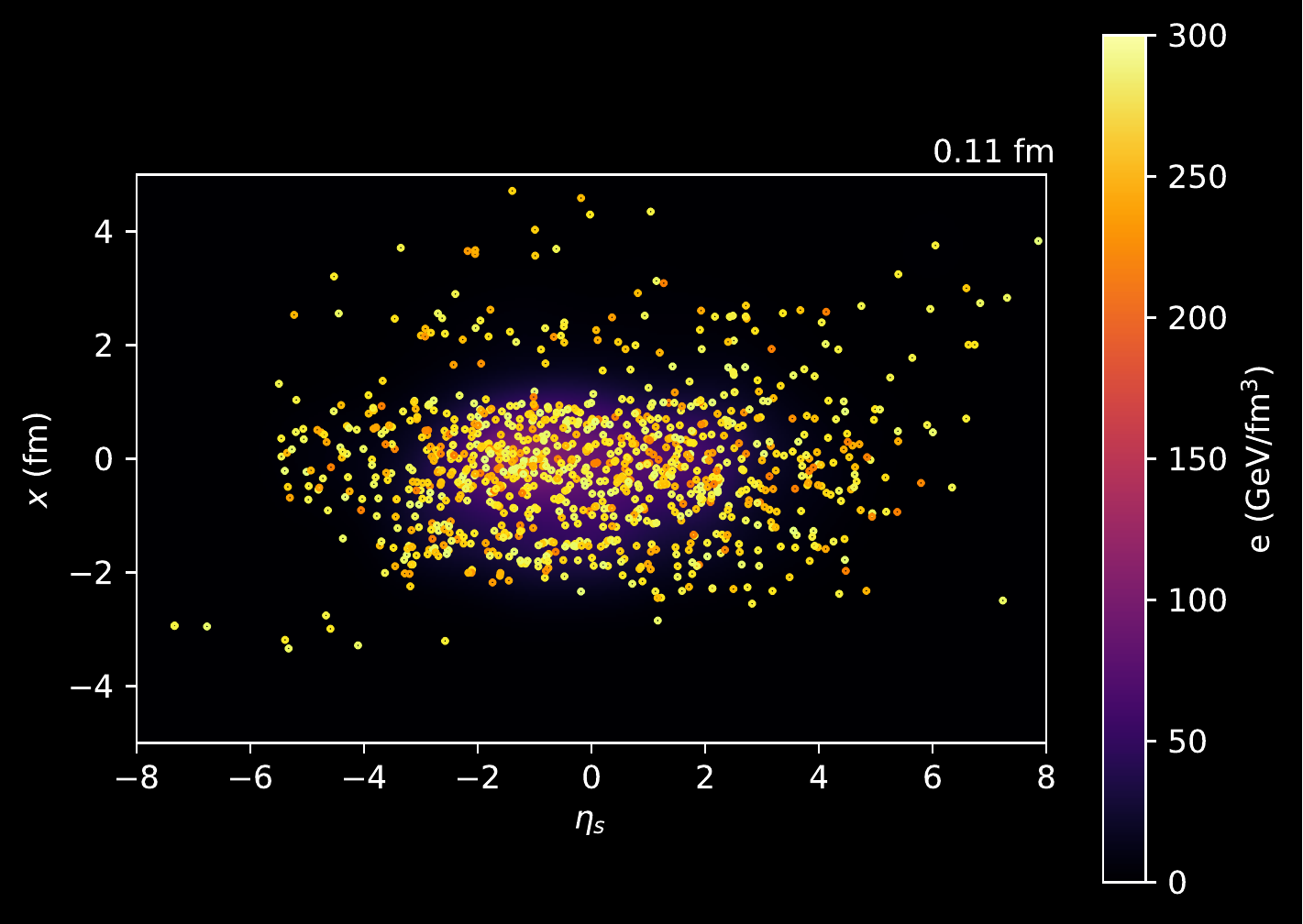}
    \includegraphics[bb=0 0 409 290, width=0.49\textwidth]{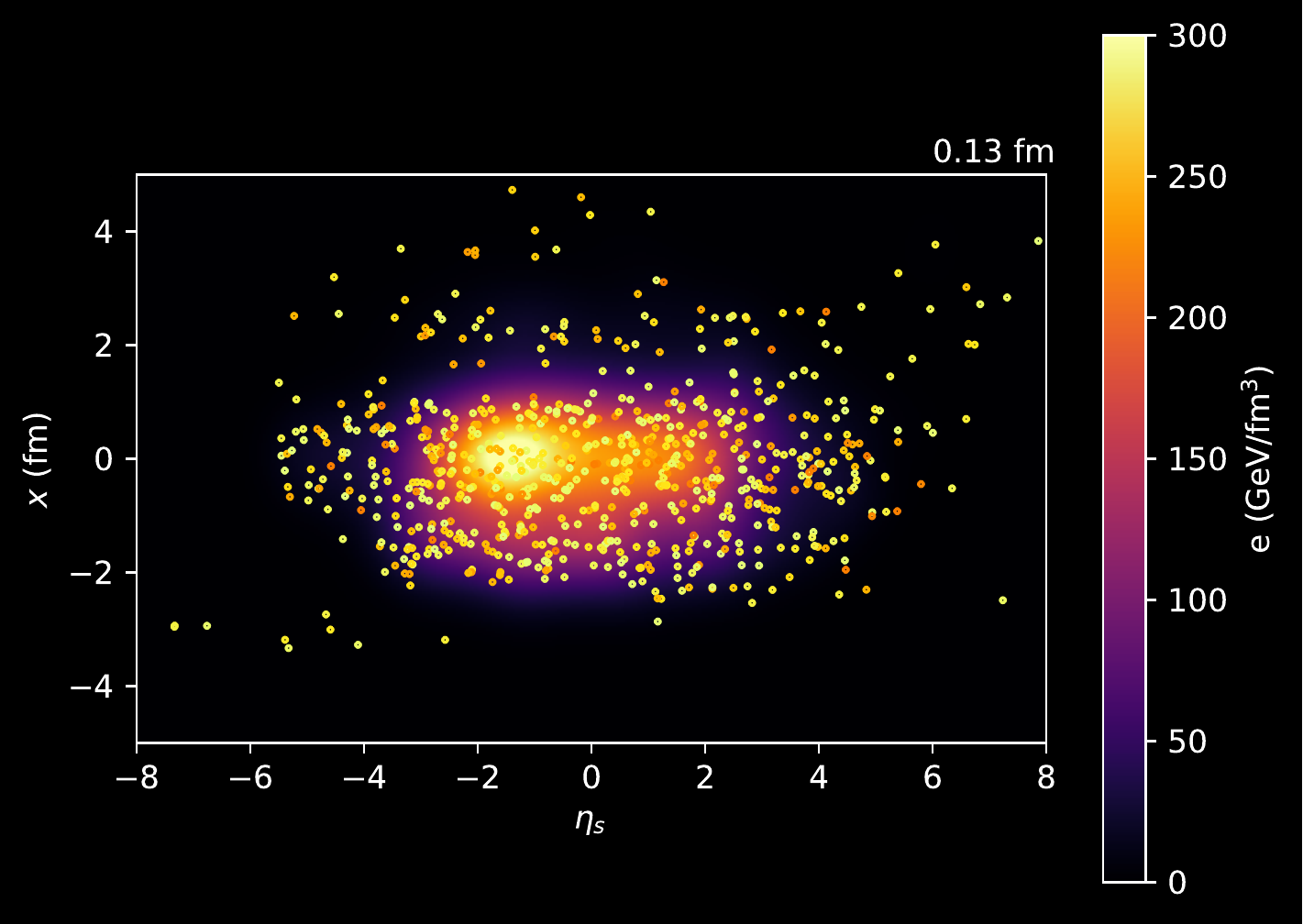}
    \includegraphics[bb=0 0 409 290, width=0.49\textwidth]{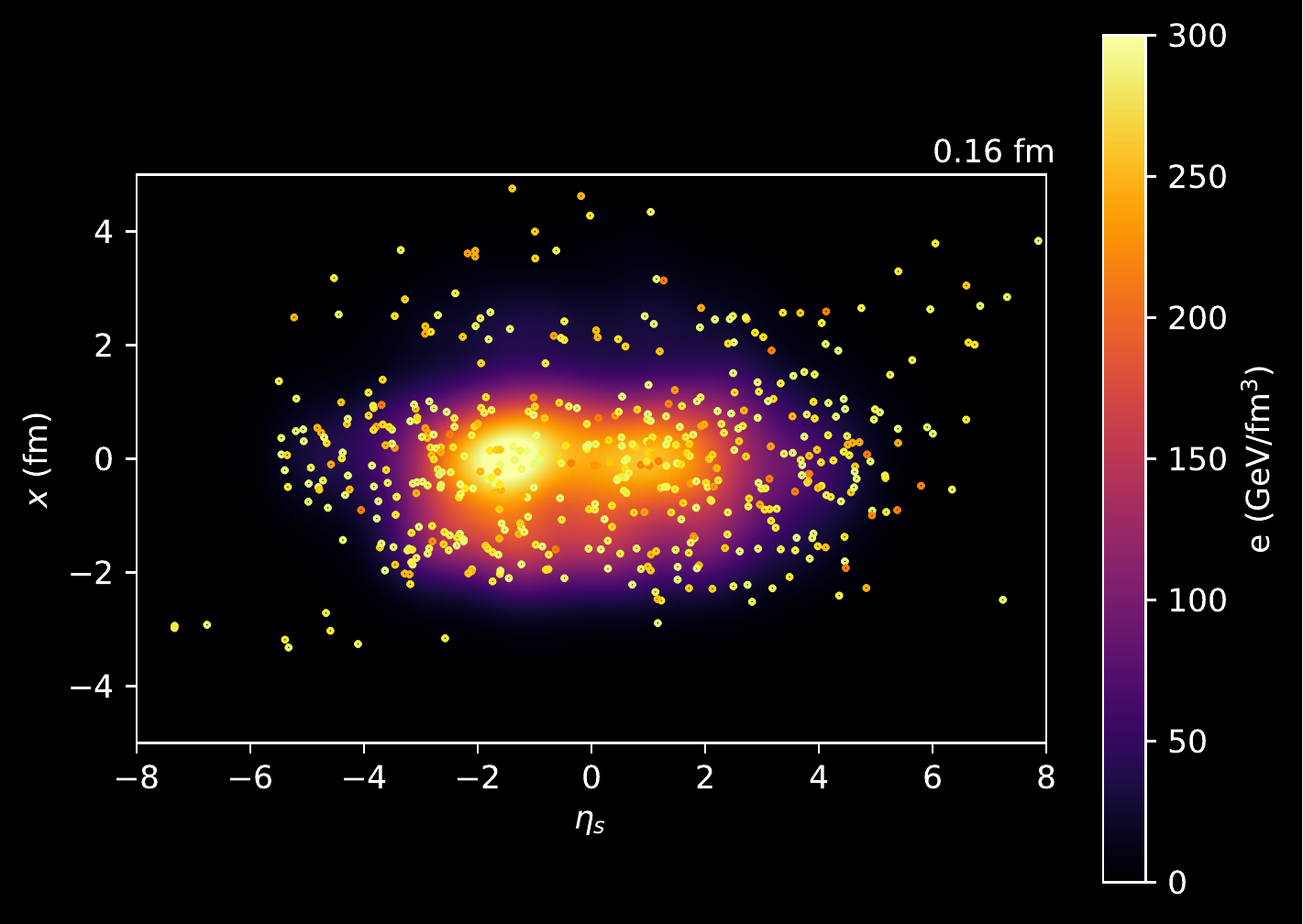}
    \includegraphics[bb=0 0 409 290, width=0.49\textwidth]{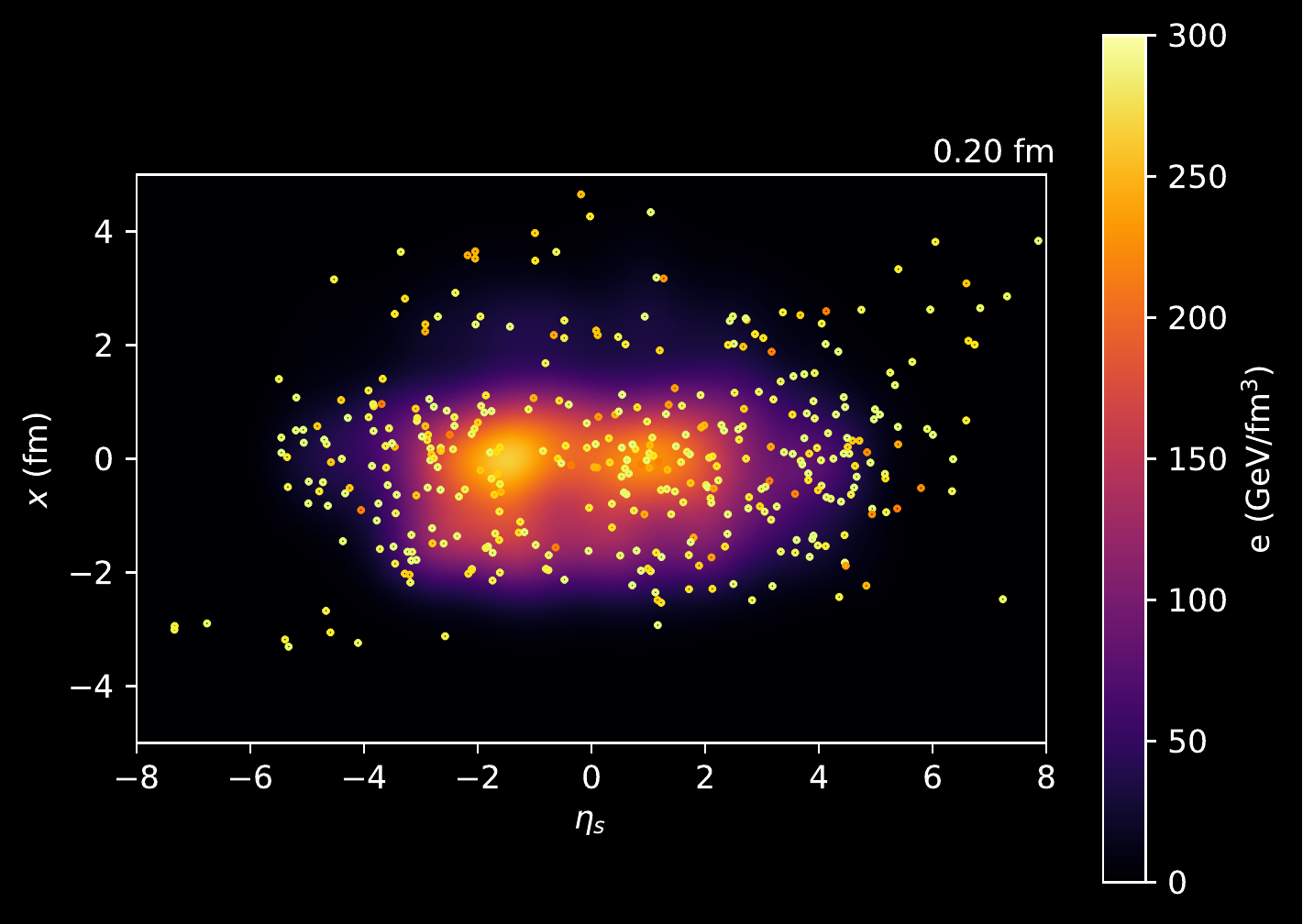}
    \includegraphics[bb=0 0 409 290, width=0.49\textwidth]{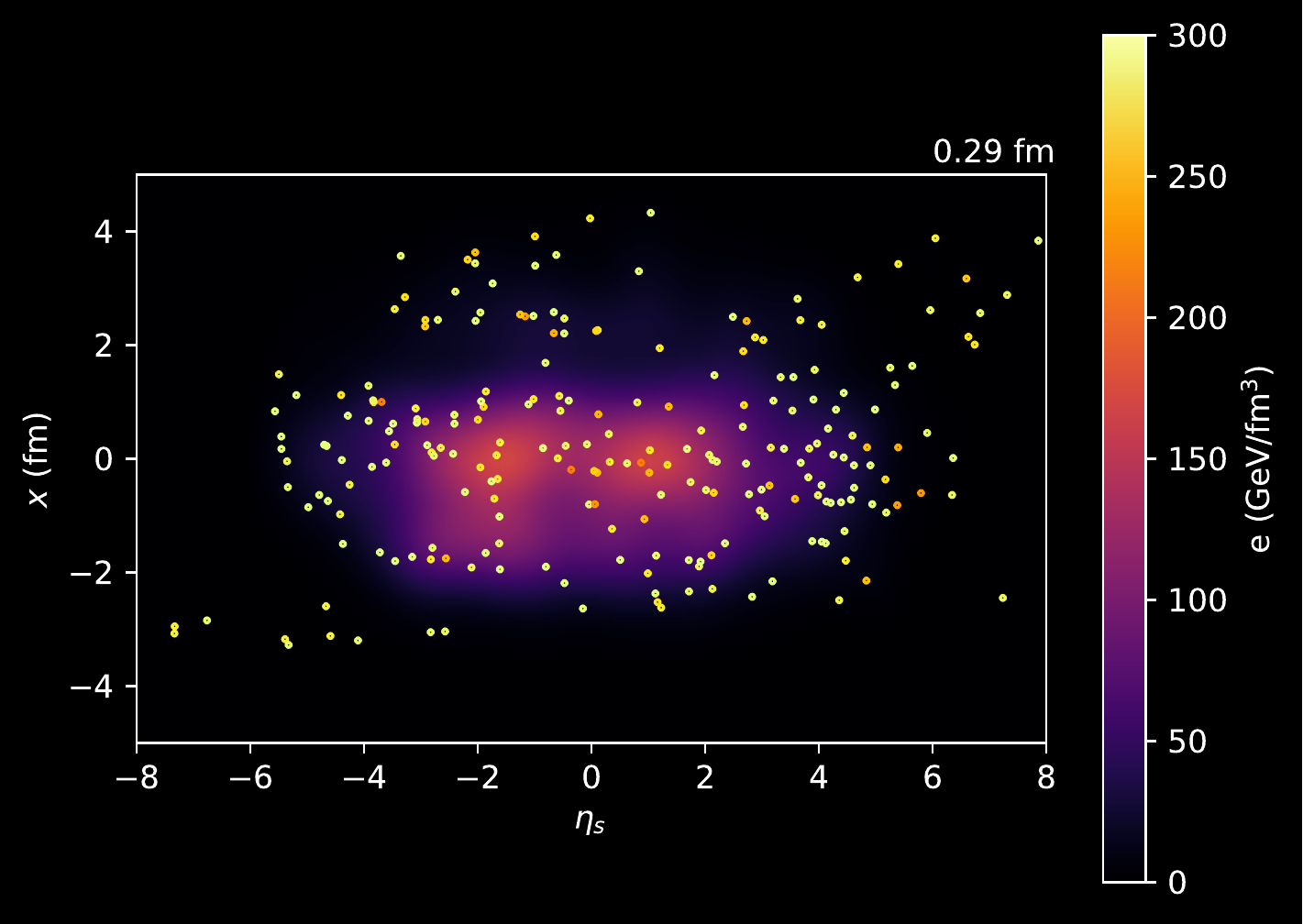}
    \caption{Snapshots of dynamical core--corona initialization from $\tau=0.10$ to $0.29$ fm in a single $Pb$+$Pb$ collision events at \snn = 2.76 TeV in x-$\eta_s$ plane at $y\approx0$ fm. Energy density of generated QGP fluids at $y=0$ is shown in color map and non-equilibrated partons within $|y|<0.5$ fm are plotted with yellow circles.}
    \label{fig:DCCI_DEMO_xetas}
\end{figure}

\begin{figure}
    \centering
    \includegraphics[bb=0 0 410 324, width=0.49\textwidth]{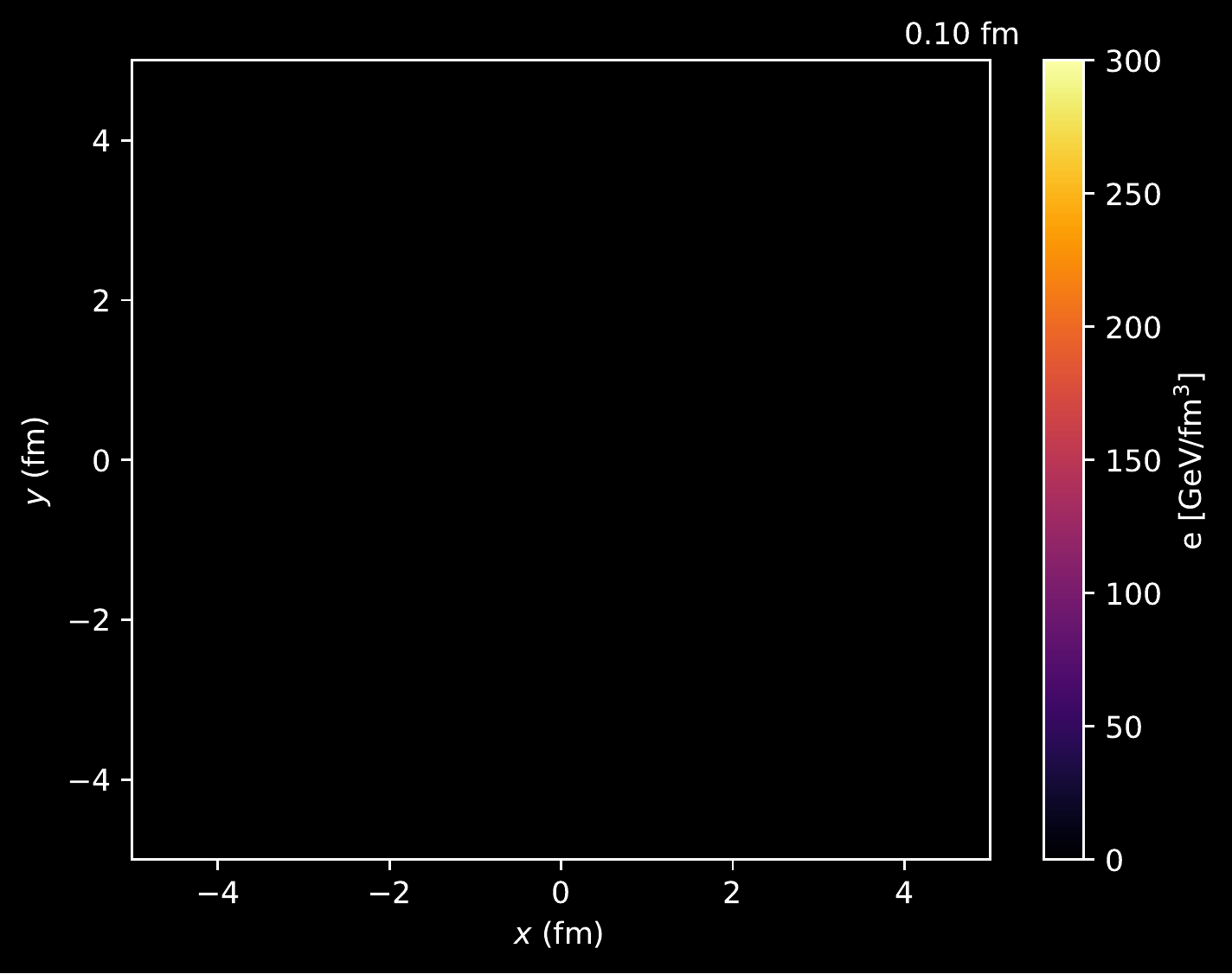}
    \includegraphics[bb=0 0 410 324, width=0.49\textwidth]{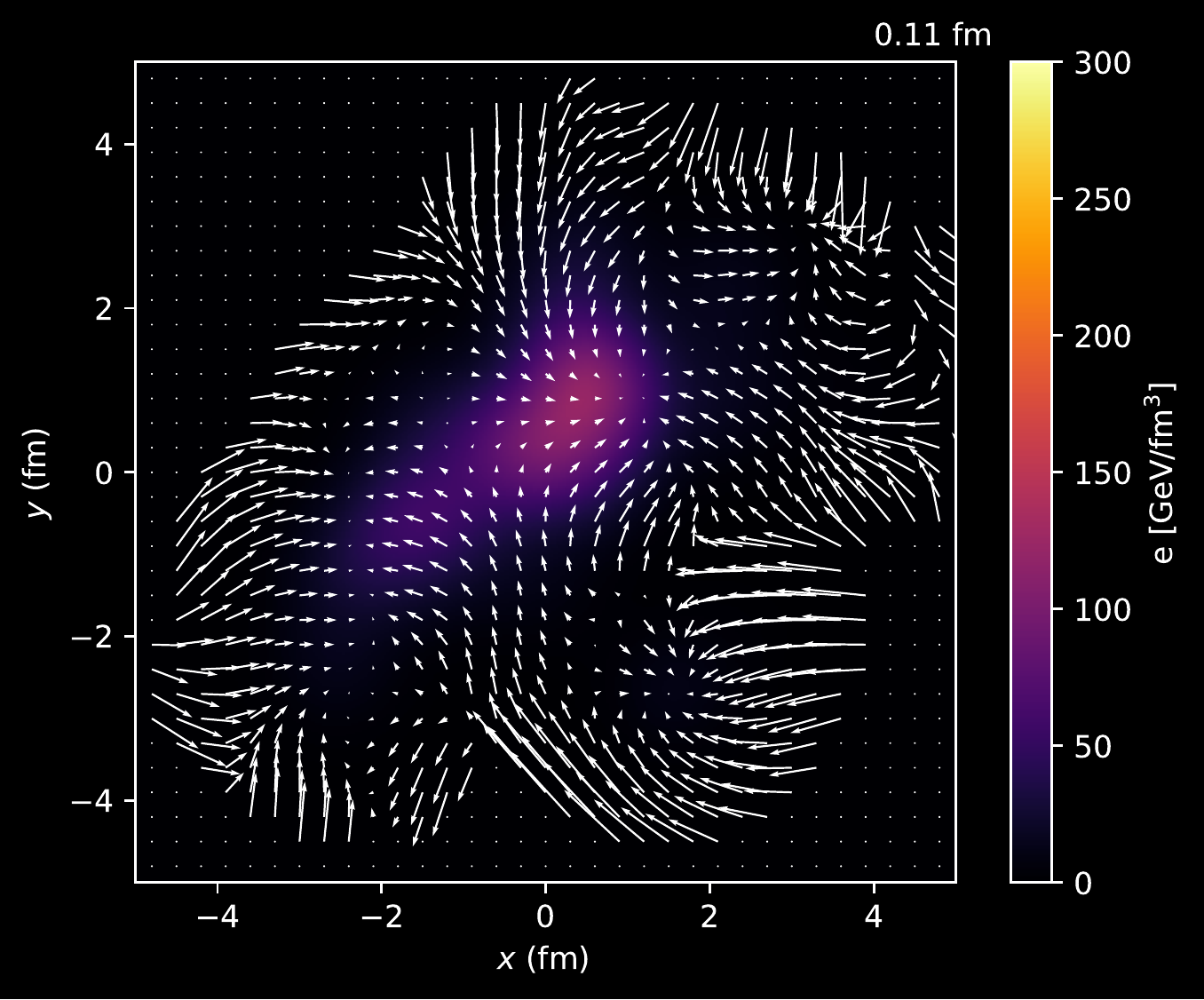}
    \includegraphics[bb=0 0 410 324, width=0.49\textwidth]{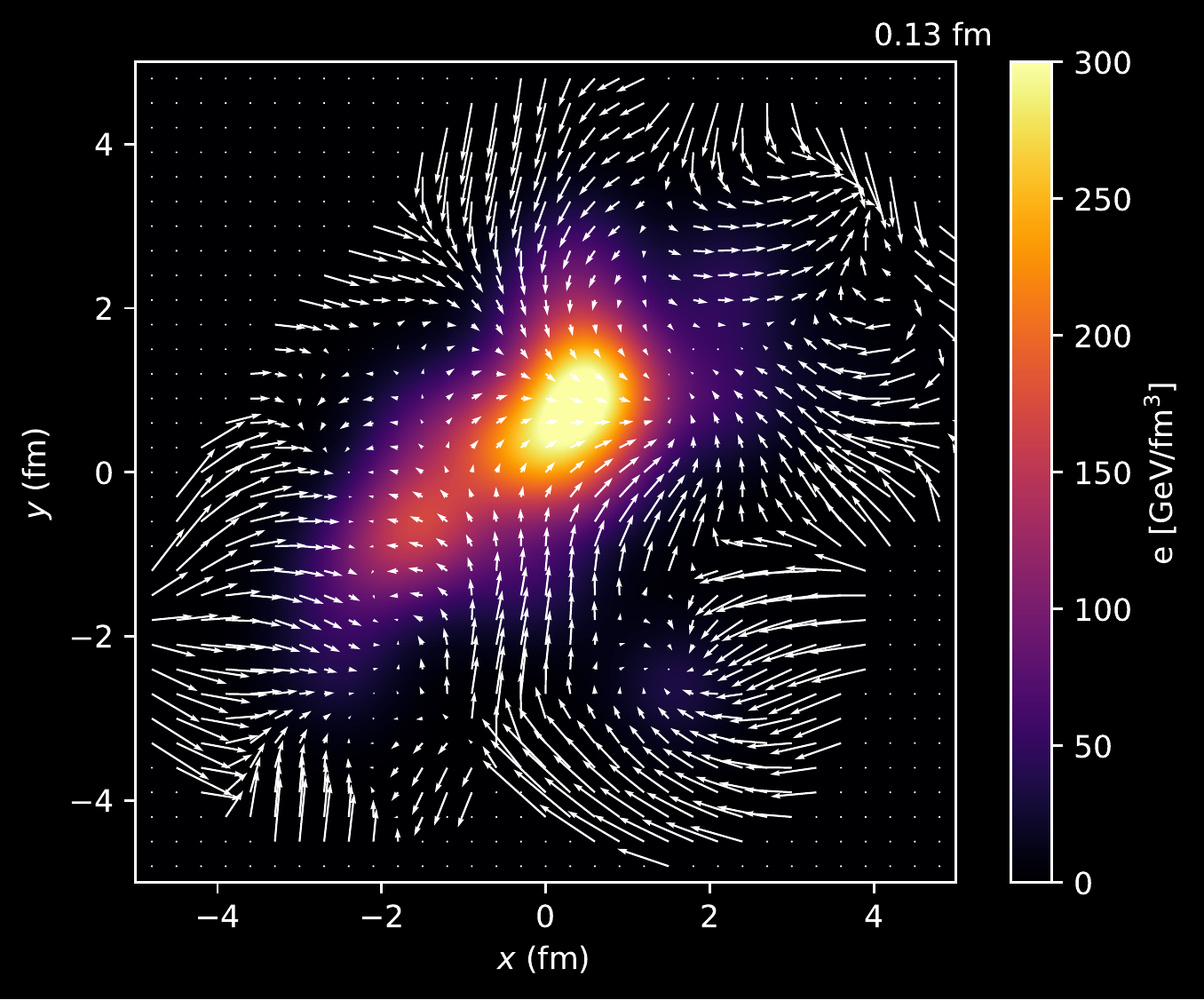}
    \includegraphics[bb=0 0 410 324, width=0.49\textwidth]{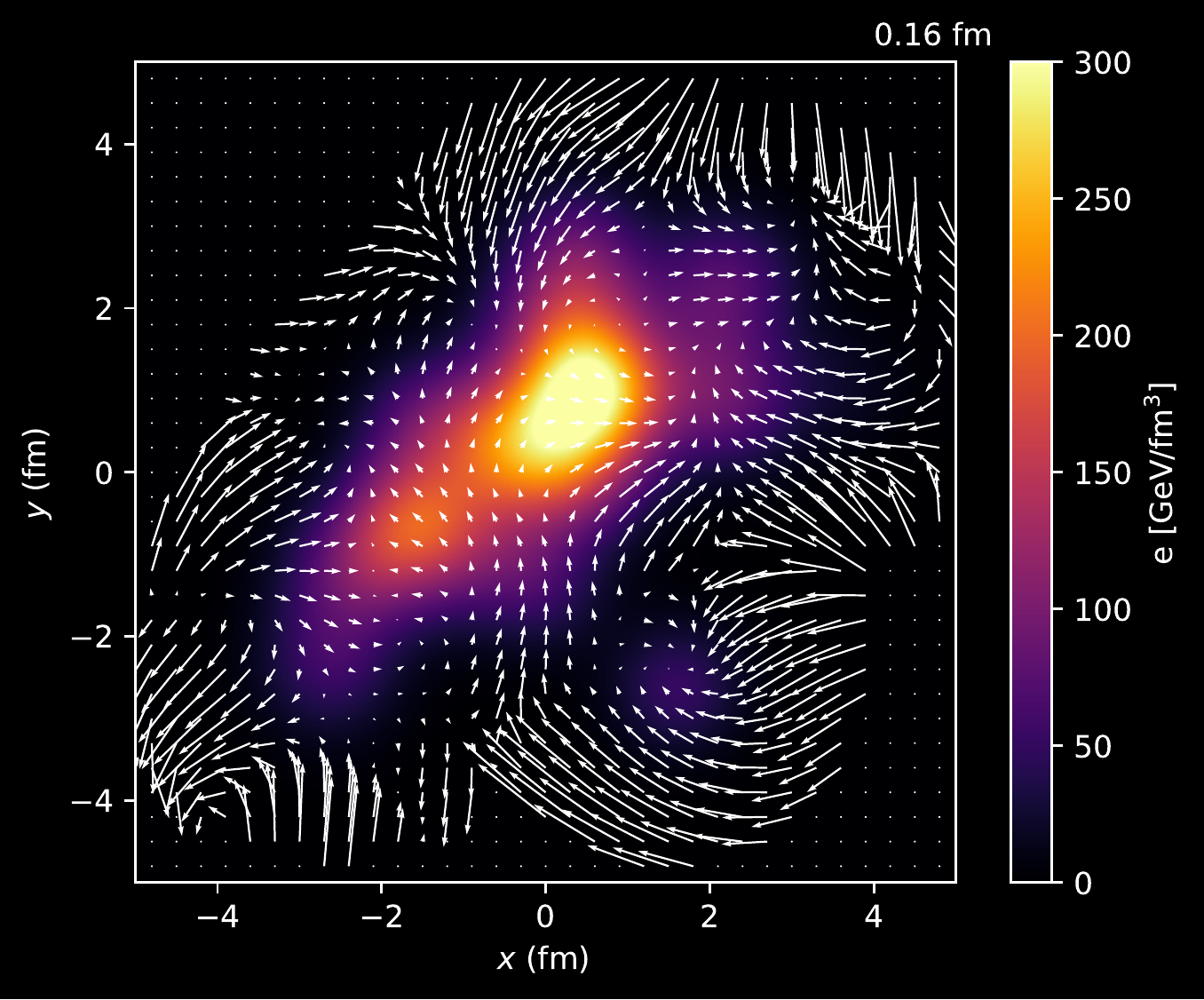}
    \includegraphics[bb=0 0 410 324, width=0.49\textwidth]{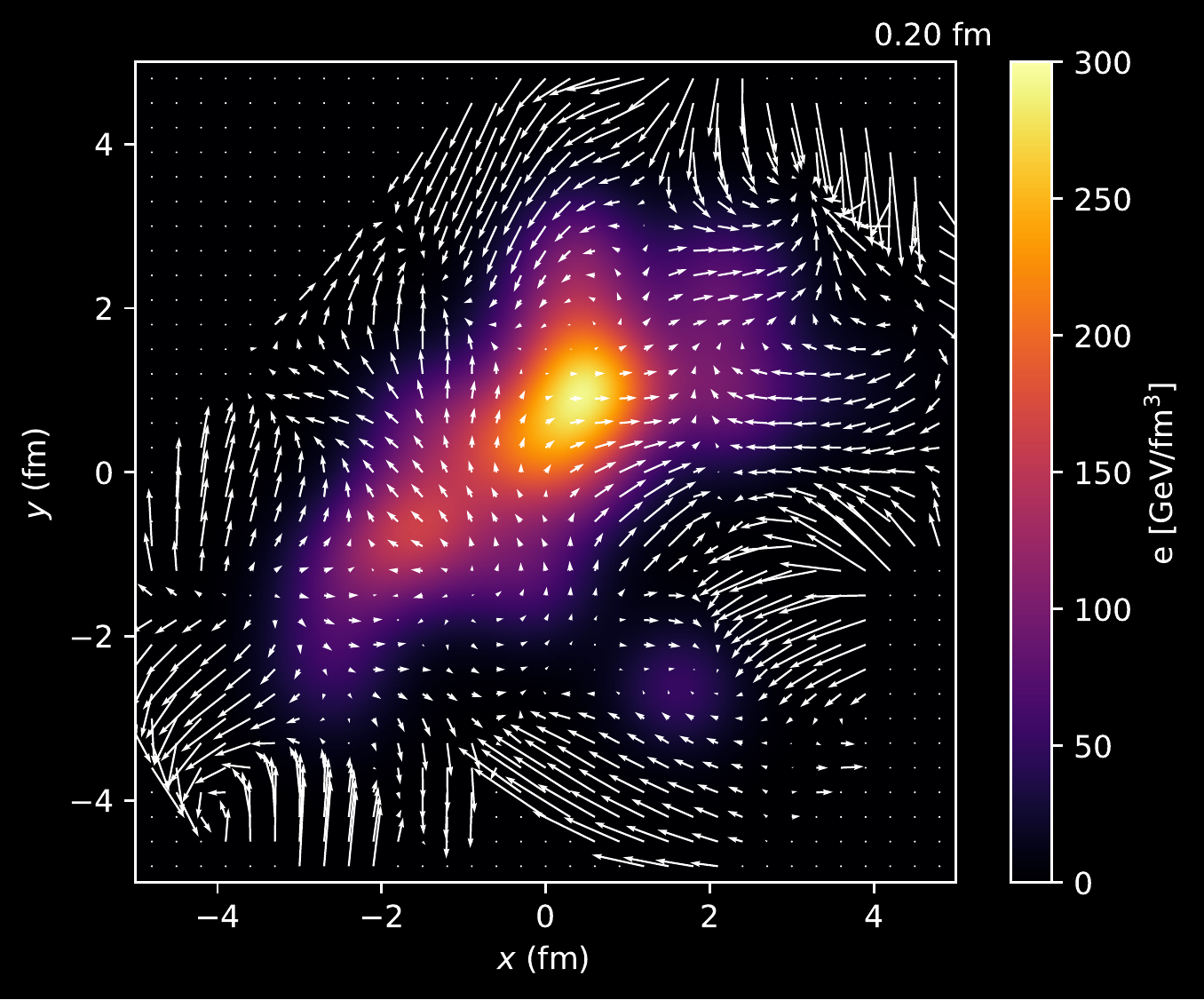}
    \includegraphics[bb=0 0 410 324, width=0.49\textwidth]{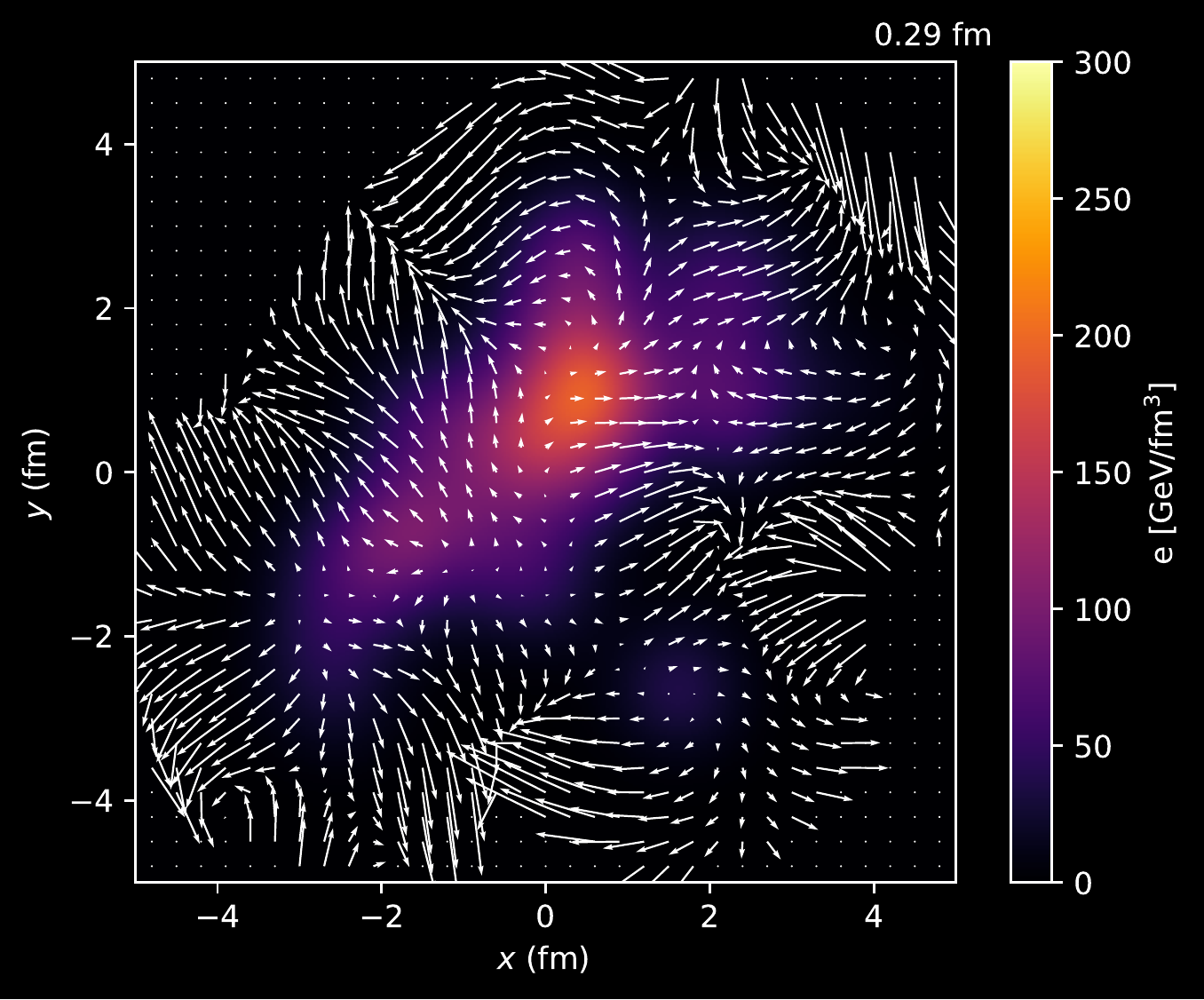}
    \caption{Snapshots of generated transverse flow velocity in dynamical core--corona initialization from $\tau=0.10$ to $0.29$ fm from a single $Pb$+$Pb$ collision events at \snn = 2.76 TeV in transverse plane at midrapidity. Energy density of generated QGP fluids at $\eta_s=0$ is shown in color map and flow velocity is depicted as white arrows.}
    \label{fig:DCCI_DEMO_xyvec}
\end{figure}

For a further clear explanation, 
I show snapshots of the dynamical core--corona initialization
from $\tau=0.10$ to $0.29$ fm in a single $Pb$+$Pb$ collision event at \snn = 2.76 TeV.
In Figs.~\ref{fig:DCCI_DEMO_xy}, \ref{fig:DCCI_DEMO_xetas}, 
distributions of energy density of generated QGP fluids and ones of non-equilibrated partons are shown
in transverse and $x$-$\eta_s$ plane, respectively.
At the initial time of the dynamical core--corona initialization, 
only the non-equilibrated partons exist in vacuum:
our hydrodynamic simulations start from vacuum.
In Fig.~\ref{fig:DCCI_DEMO_xy}, 
one sees that 
the the distribution of partons in transverse plane is not isotropic in coordinate space
which is because the event plane is tilted on the $x$-$y$ plane.
One also sees that there is an small isolated 
spots where non-equilibrated partons are highly populated.
This is coming from the Glauber-model like 
random nucleon distributions in a nuclei.
As time step proceeds, QGP fluids start to be formed from the area where the number density of non-equilibrated partons is large.
In this event, the energy density seems to reach maximum values around $\tau \approx 0.13$-$0.16$ fm, and 
the energy density decreases after that,
This is exactly what we saw in Fig.~\ref{fig:ENERGYDENS_vs_tau}:
the competition between heat-up due to dynamical initialization and cool-down due to evolution of fluids.
One would also sees that the number of non-equilibrated partons decrease because some of them deplete energy and momentum.
Figure \ref{fig:DCCI_DEMO_xetas} show the dynamical generation of initial condition of hydrodynamics equations in $x$-$\eta_s$ plane.
At the very beginning, non-equilibrated partons distribution is stretched out in longitudinal direction, which originates from the initially produced color string structure obtained from \pythia.
Due to the discrete distribution of non-equilibrated partons, which is obvious since I deal with initial partons in particle picture,
one sees the matter has inhomogeneous profile along longitudinal axis.
Figure \ref{fig:DCCI_DEMO_xyvec} shows generated transverse velocity of QGP fluids in transverse plane.
As a result of dynamical initialization,
transverse velocity is randomly embedded in the initial condition.
It should be also noted that since non-equilibrated partons dynamically deposit their energy and momentum,
the transverse velocity distribution is transient.

\subsection{Sampling of thermalized partons}
\label{subsection:Sampling_of_thermalized_partons}

%Description of this model
%-----------------------------------
As I mentioned in the previous subsection, the summation in the right hand side of  Eq.~\refbra{eq:four-momentum-deposition} is taken for all partons in the system including not only initially produced partons but also thermalized partons which are constituents of the QGP fluids.
This enables one to consider the four-momentum deposition due to scatterings with thermalized partons while traversing the medium.
%How we do in an actual simulation
%-----------------------------------
In order to consider the scatterings with thermalized partons, whose evolution is described by hydrodynamics, I sample partons from all fluid elements
and obtain phase-space distributions of them
at each time step.

Although I employ the lattice EoS, I make a massless ideal gas approximation on fluid elements for simplicity.
In this approximation, the number density of partons in a fluid element can be estimated as 
\begin{align}
    \label{eq:NumberDensity}
    n = \frac{90d^\prime \zeta(3)}{4\pi^4d}s_{\mathrm{EoS}}(T),
\end{align}
where $s_{\mathrm{EoS}}(T)$ is the entropy density obtained from the EoS via temperature $T$ at the fluid element.
The effective degeneracies of the QGP, $d$ and $d^\prime$, are defined as
\begin{align}
    \label{eq:dof_PartitionFunction}
    d &= d_{F} \times \frac{7}{8} + d_{B}, \\
    \label{eq:dof_NumberDensity}
    d^\prime & = d_{F} \times \frac{3}{4} + d_{B}.
\end{align}
The factors $\frac{7}{8}$ in Eq.~\refbra{eq:dof_PartitionFunction} and $\frac{3}{4}$ in Eq.~\refbra{eq:dof_NumberDensity} originate from differences between Fermi-Dirac and Bose-Einstein statistics in the entropy density and the number density.
The degrees of freedom of fermion $d_F$ and boson $d_B$ are obtains as
\begin{align}
    d_{F} &=d_{\rm{c}} \times d_{f} \times d_{s} \times d_{q\bar{q}} = 3\times3\times 2\times 2 = 36, \\
d_{B}&=d_{c} \times d_{s} = 8 \times 2 = 16,
\end{align}
where $d_{c}$, $d_{f}$, $d_{s}$, and $d_{q\bar{q}}$ represent the degrees of freedom of color, flavor, spin, and particle-antiparticle, respectively.

The number of partons in a fluid element is then, $\Delta N_0 = n \Delta x \Delta y  \tau \Delta \eta_s$,
where $n$ is the number density of partons obtained in Eq.~\refbra{eq:NumberDensity},
$\Delta x$, $\Delta y$, and $\Delta \eta_s$ are the widths of one fluid element in the Milne coordinates.
One can interpret $\Delta N_0$ as a mean value of Poisson distribution and sample the number of partons $N$ with
\begin{align}
 \dis P(N) = \exp{(-\Delta N_0)}\frac{\Delta N_0^{N}}{N!}.
\end{align}
For sampled $N$ partons, I stochastically assign species of them.
I pick up a quark or an anti-quark with a probability,
\begin{align}
 \label{eq:probability_of_samplingquark}
 P_{q/\bar{q}} &= \frac{(3/4) d_{F}}{(3/4)d_{F} + d_{B}},
\end{align}
which corresponds to a fraction of the degree of freedom of Fermi particles,
while a gluon is picked up with a probability, 
\begin{align}
 P_{g} &= 1- P_{q/\bar{q}}.
\end{align}
%\comm{[YT: How can Casimir factor affect the procedure here?]}
%Note here again that we neglect color Casimir factors in Eq.~\refbra{eq:parton_cross_section} and mass of partons sampled from fluid elements. 

The three-dimensional momentum $\bm{k}$ of (anti-)quarks or gluons 
in the local rest frame of the fluid element is assigned according to the normalized massless Fermi or Bose distribution, 
\begin{align}
\label{eq:mass-lessBFdistribution_for_sampling}
    \dis P(\bm{k})d^3\!k = \frac{1}{N_{\mathrm{nom}}} \frac{1}{ \exp{(k / T)} \mp_{\mathrm{B, F}} 1}d^3\!k,
\end{align}
where $N_{\mathrm{nom}}$ is a normalization factor, $T$ is temperature of the fluid element, and $\mp_{\mathrm{B, F}}$ is a sign for Bose ($-$) and Fermi ($+$) statistics. Then, the energy and momentum in the lab frame is obtained by performing Lorentz transformation on $k^\mu=(\left|\bm{k}\right|,\bm{k})$ with the velocity of the fluid element.

Space coordinates are assigned with a uniform distribution within each fluid element.
For partons sampled from a fluid element centered at $\bm{x} = \bm{x}_i = \left(x_i, y_i, \eta_{s,i} \right)$, where the index $i$ stands for the numbering of fluid elements, I assign their coordinates with
\begin{align}
\label{eq:uniform_distribution}
& P_{\mathrm{uni}}(\bm{x})\tau \Delta x \Delta y \Delta \eta_{s} \nonumber \\
& = \left\{
    \begin{aligned}
         \  &0 \quad (\bm{x}<\bm{x}_i - \Delta \bm{x}/2, \ \bm{x}_i + \Delta \bm{x}/2  < \bm{x} )  \\
            &1 \quad (\bm{x}_i - \Delta \bm{x}/2 \leq \bm{x} \leq
            \ \bm{x}_i + \Delta \bm{x}/2 ) 
    \end{aligned}
    \right. .
\end{align}
%where $\Delta \bm{x} = \left(\Delta x, \Delta y, \Delta \eta_s \right)$ is a discrete size of a fluid element in $x$, $y$, and $\eta_s$ directions, respectively.

As discussed in this subsection, the four-momentum deposition caused by collisions between a traversing parton as a corona part and thermalized partons as core parts could be regarded as a toy model of jet quenching.
I note that, although the implementation of a more sophisticated jet-quenching mechanism is the future work, the energy loss of traversing partons in the medium is phenomenologically introduced via the dynamical core--corona initialization in Eq.~\refbra{eq:four-momentum-deposition}.

\begin{figure}
    \centering
    \includegraphics[bb = 0 0 329 324, width=0.49\textwidth]{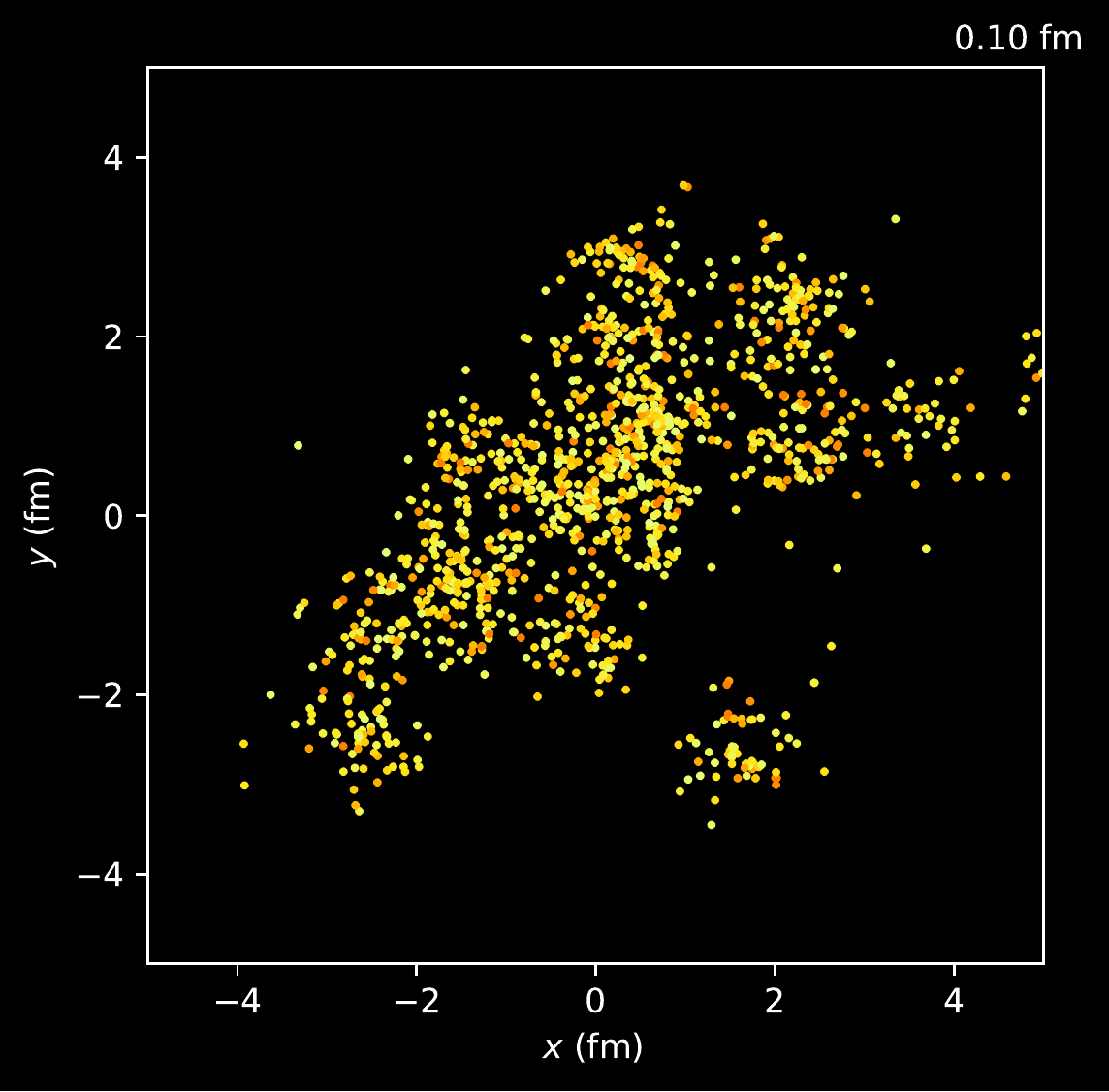}
    \includegraphics[bb = 0 0 329 324, width=0.49\textwidth]{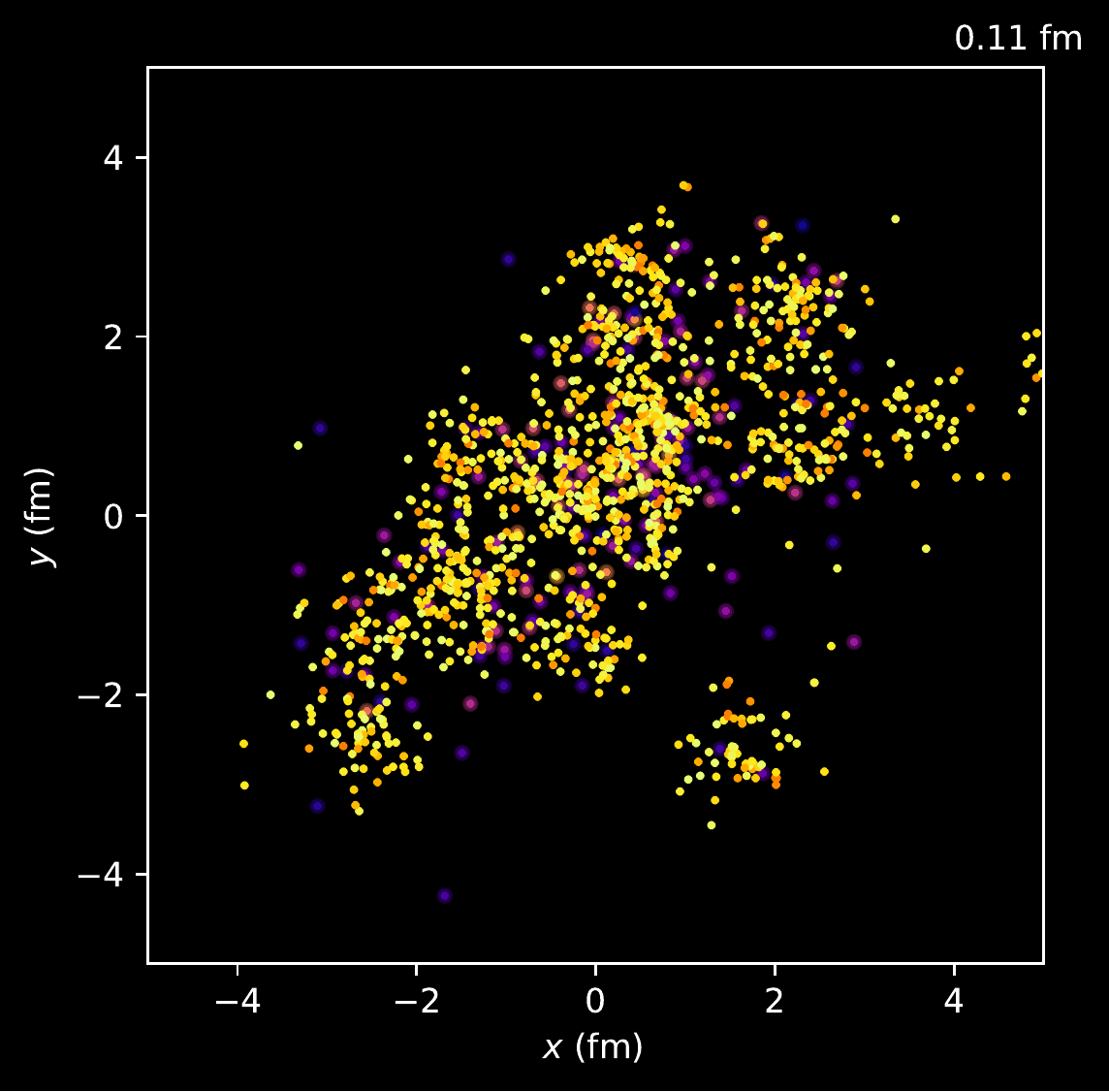}
    \includegraphics[bb = 0 0 329 324, width=0.49\textwidth]{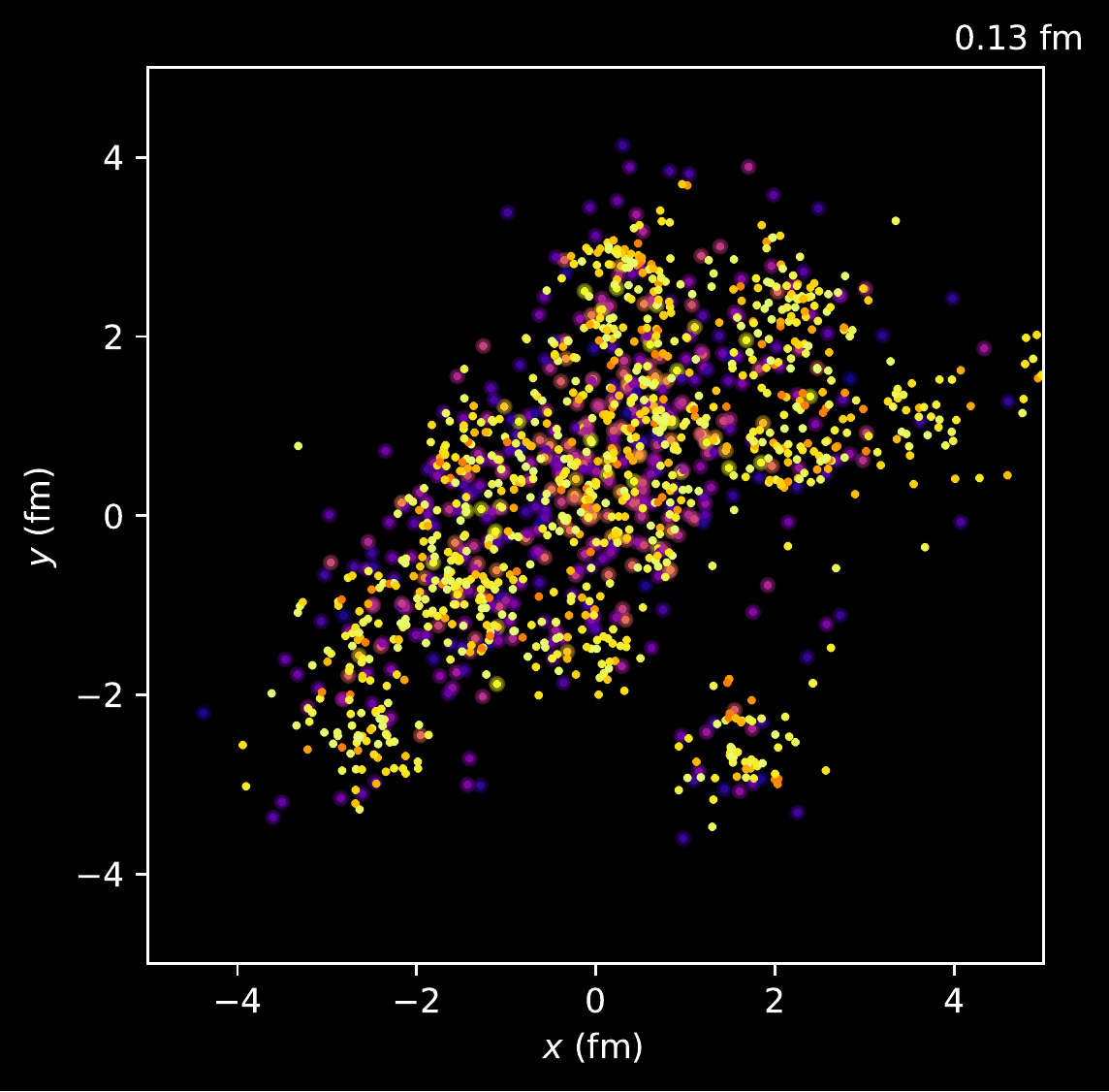}
    \includegraphics[bb = 0 0 329 324, width=0.49\textwidth]{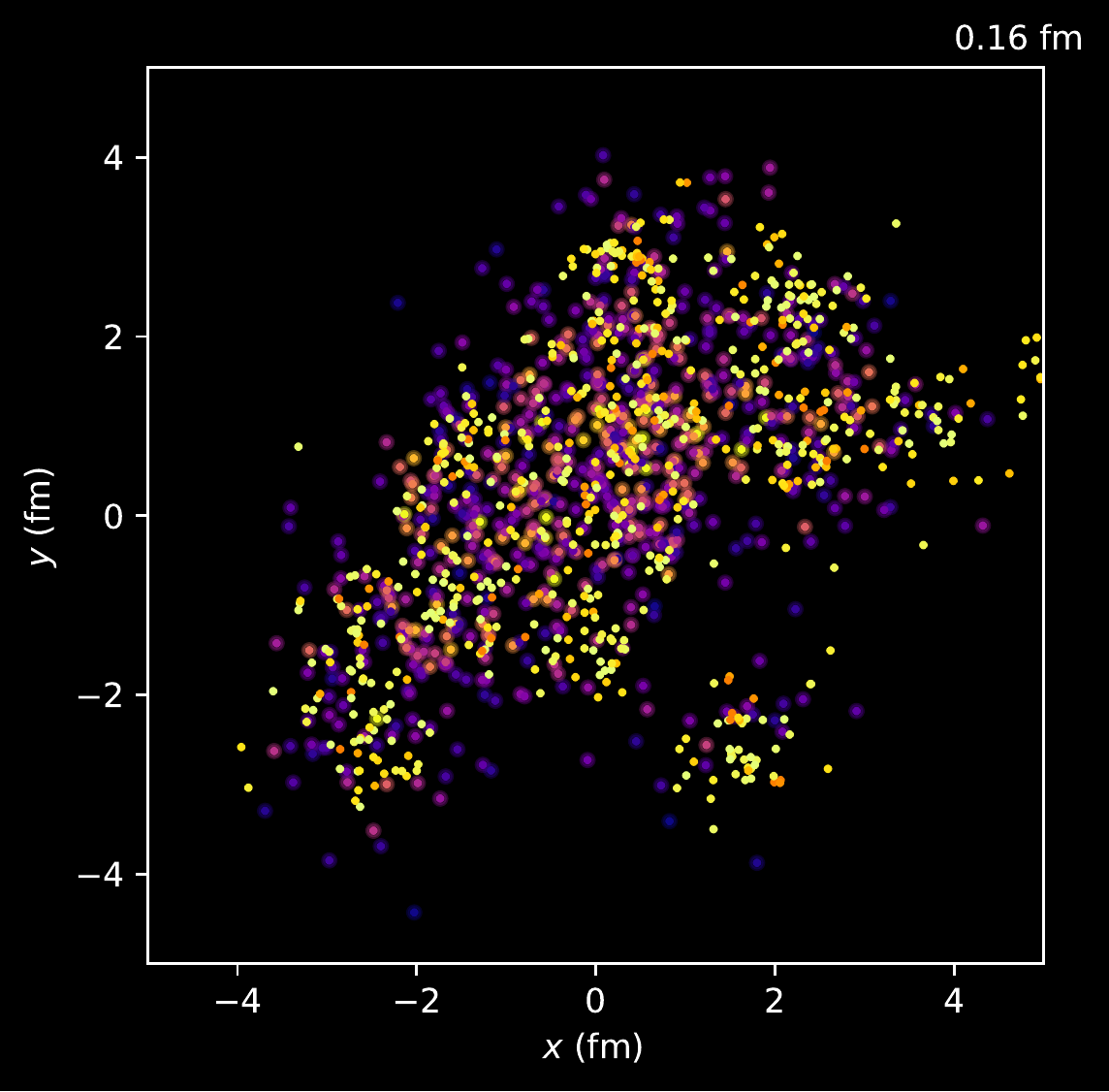}
    \includegraphics[bb = 0 0 329 324, width=0.49\textwidth]{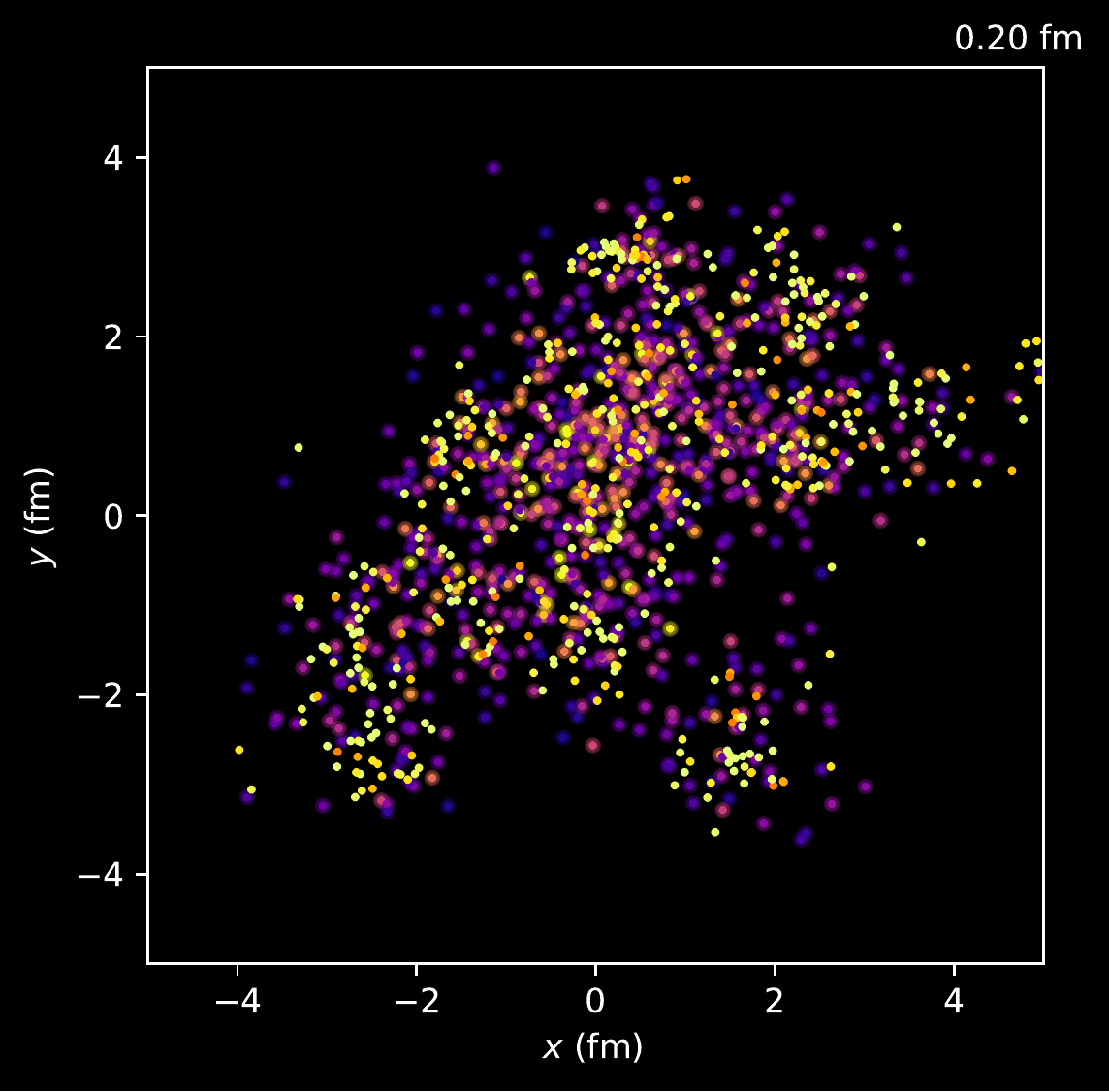}
    \includegraphics[bb = 0 0 329 324, width=0.49\textwidth]{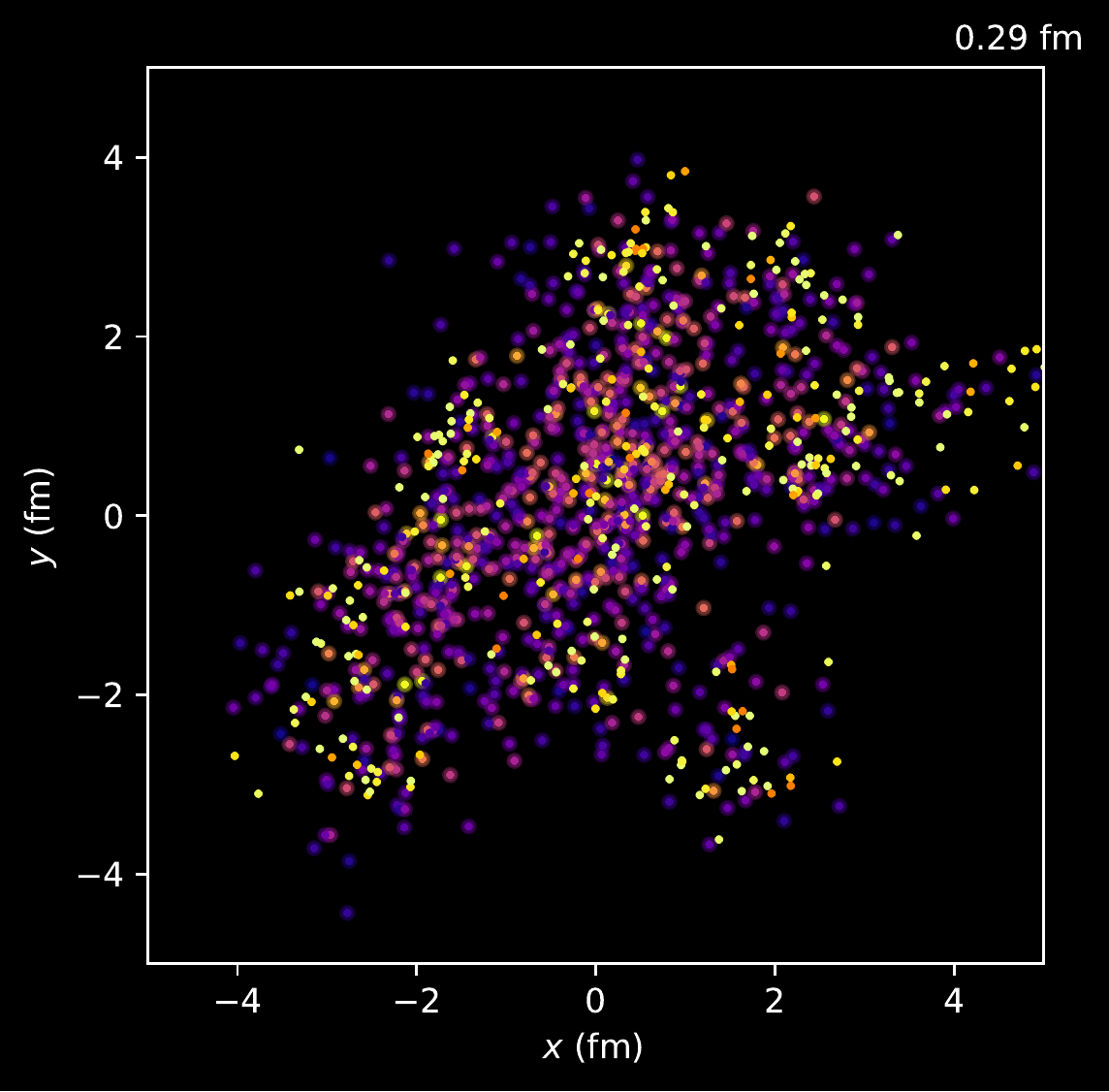}
    \caption{Snapshots of picked up thermal partons in dynamical core--corona initialization from $\tau=0.10$ to $0.29$ fm from a single $Pb$+$Pb$ collision events at \snn = 2.76 TeV in transverse plane at midrapidity. Sampled thermal partons and non-equilibrated partons picked up within $|\eta_s|<1$ are plotted with purple and yellow circles.}
    \label{fig:DCCI_DEMO_THERMALPARTONS}
\end{figure}

Figure \ref{fig:DCCI_DEMO_THERMALPARTONS} shows snapshots of distribution of sampled thermal partons and ones of non-equilibrated partons obtained from a single $Pb$+$Pb$ collision at \snn = 2.76 TeV which is identical with the event shown in Figs.~\ref{fig:DCCI_DEMO_xy} to \ref{fig:DCCI_DEMO_xyvec}.
As I explained above, four-momentum deposition of partons defined in Eq.~\refbra{eq:four-momentum-deposition} is calculated by sampled thermal partons.
Thus, these snapshots illustrate how non-equilibrated partons ``see'' a system.
The fraction of the number of sampled thermal partons against that of non-equilibrated partons increases as the time step progresses.

\subsection{Modification of color strings}

\label{subsec:STRING_CUTTING}
The Lund string model is based on a linear confinement picture of color degree of freedom \cite{Andersson:1983ia,Sjostrand:2007gs}.
Energy stored between a quark and an anti-quark linearly increases with the separation length of the quark and anti-quark in the vacuum. 
However, this picture should be modified if I put them into the QGP at finite temperature.
It is known that, at high temperature, the string tension becomes so small that color strings would disappear \cite{Kaczmarek:1999mm}.
Since the input of DCCI2 is initially produced partons connected with color strings and I generate fluids through their energy-momentum depositions, some color strings should overlap with the fluids in the coordinate space. 
I phenomenologically incorporate the modification of the color string configuration due to the finite temperature effect in DCCI2.

At $\tau = \tau_s(>\tau_{0})$, I assume that the string fragmentation happens when the entire color string is outside the fluids.
Here, ``a color string'' means chained partons as a color singlet object. 
When a color string is entirely inside the fluids at $\tau=\tau_s$, I discard the information of its color configuration and let its constituent partons evolve as individual non-equilibrated partons according to Eq.~\refbra{eq:trajectories_of_partons}. 
If a color string is partly inside and partly outside the fluids, the color string is subject to be cut off at the boundary of the fluids. The boundary here is identified with a contour of $T(x)=T_{\mathrm{sw}}$.
The color string in a vacuum cut off at the boundary is reattached to a thermal parton picked up from the hypersurface and forms a color-singlet object again to be hadronized via string fragmentation. The thermal parton is sampled in the hypersurface of the fluids. The details of the above treatment of color strings at $\tau = \tau_s$ are explained in the following subsection, {\bf{string cutting at $\tau = \tau_s$}}.
As for the rest of the color string left inside the fluids,
I discard the information of its color configuration and let its constituent partons evolve as individual non-equilibrated partons likewise the above case.

During the evolution of the fluids, the individual non-equilibrated partons come out from the hypersurface at some point. 
I assume that such a parton forms parton pairs to become a color-singlet object by picking up a thermalized parton from the hypersurface of the fluids. This prescription is based on exactly the concept of the coalescence models 
\cite{Hwa:2003ic,Hwa:2004ng,Greco:2003xt,Isse:2007pa,Fries:2008hs,Han:2016uhh,Fries:2019vws,Zhao:2019ehg,Kordell:2020wqi}. 
I will explain the parton-pairing treatment at $\tau > \tau_s$ in the following subsection, {\bf{parton-pairing for surviving partons}}.

I perform the modification only for color strings in which the transverse momentum of all partons forming that color string is less than a cut-off parameter, $p_T<p_{T, \mathrm{cut}}$, at $\tau=\tau_{0}$. Notice that this is merely a criterion of whether the modification of color string is performed and that all initially produced partons, including very high $p_T$ ones, nonetheless 
experience the dynamical core--corona initialization regardless of $p_{T,\mathrm{cut}}$.
This treatment avoids modification on $p_T$ spectra of final hadrons generated from intermediate to high $p_T$ partons which would less interact with fluids rather than low $p_T$ partons. 
Since the modification on the structure of color strings sensitively affect the final hadron distribution in momentum space, I should make a more quantitative discussion on the parameter $p_{T, \mathrm{cut}}$ as a future work.

It should be also noted that when I sample thermalized partons, energy-momentum is not subtracted from fluids just for simplicity.

\begin{figure}
    \centering
    \includegraphics[bb=0 0 410 311, width=0.49\textwidth]{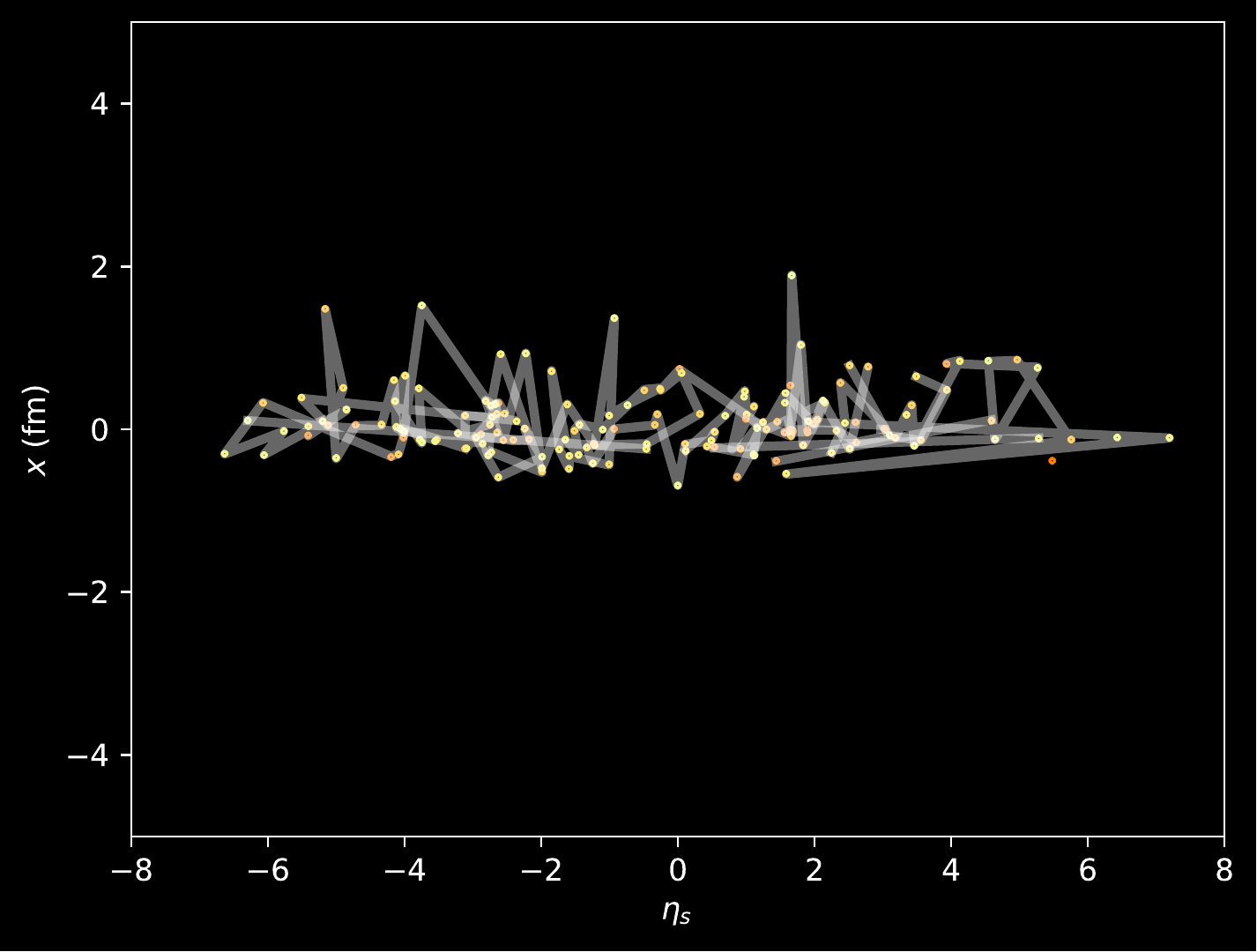}
    \includegraphics[bb=0 0 410 311, width=0.49\textwidth]{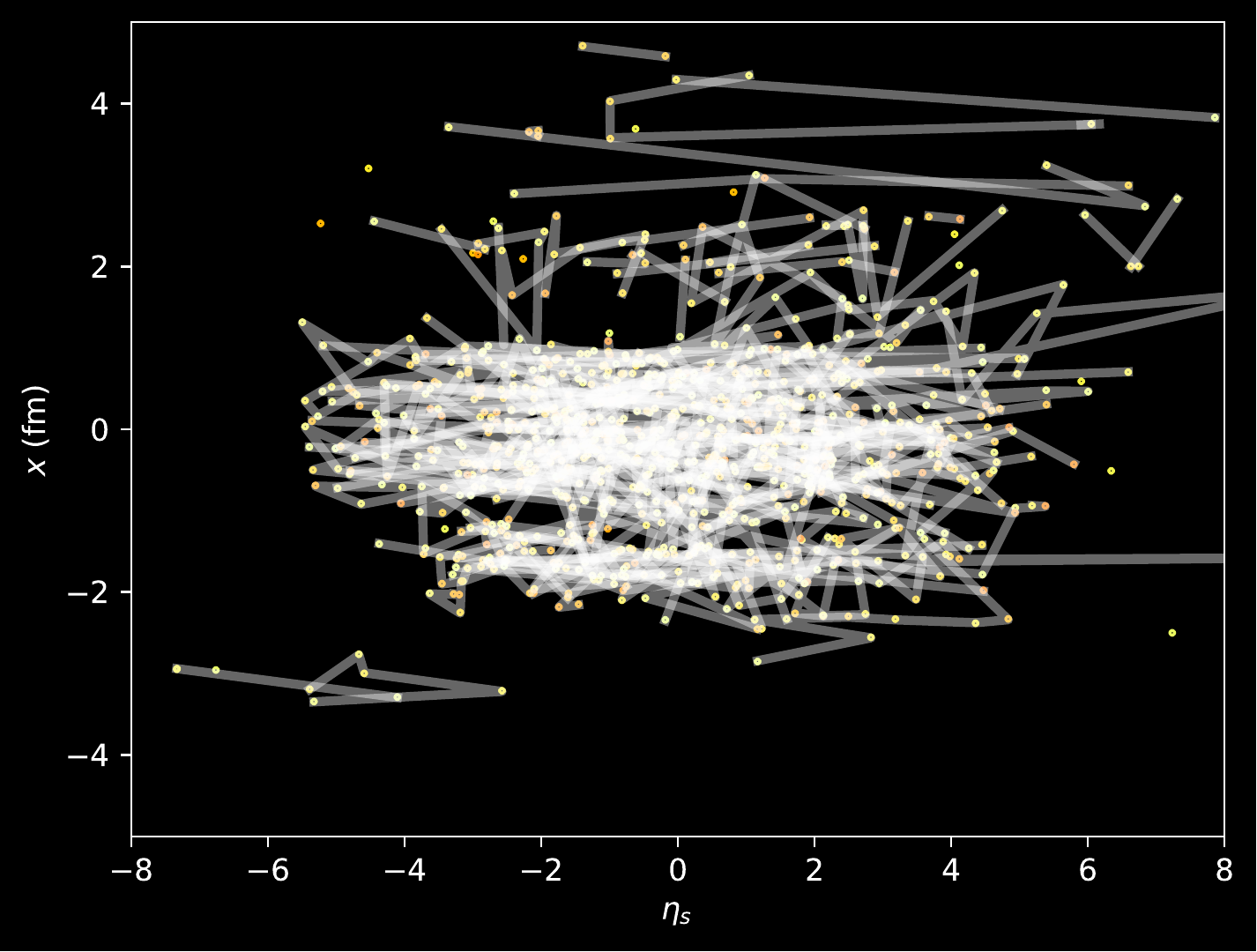}
    \caption{Distribution of color strings (white lines) generated in one single events of $p$+$p$ at \snn[proton] = 7 TeV (left) and $Pb$+$Pb$ (right) collisions at  \snn = 2.76 TeV shown in $x$-$\eta_s$ plane.}
    \label{fig:INIT_STRING}
\end{figure}

\subsubsection{String cutting at $\tau=\tau_s$} 
\label{subsec:COLORSTRING_TREATMENT_TAU0}
%Color tracing
%------------------
I find a crossing point between a color string and the hypersurface of the fluids by tracing partons chained as color strings one by one. 
Since \pythia8 and \pythia8 Angantyr give us information of structure of color strings, I respect the initially produced color flow.
Example of distributions of initially produced color flow from \pythia \ is shown in Fig.~\ref{fig:INIT_STRING} in $p$+$p$ at \snn[nucleon] = 7 TeV and $Pb$+$Pb$ collisions at \snn = 2.76 TeV.

At $\tau = \tau_s$, all initial partons are classified into four types: ``hard" partons, dead partons inside fluids,  surviving partons inside fluids, and partons outside fluids.
I regard partons which is chained with at least one high $p_T (> p_{T,\mathrm{cut}})$ parton
as ``hard" partons.
I do not modify color strings composed of these hard partons to keep the initial color flow and hadronize them in a usual way discussed in Sec.~\ref{subsection:Direct_hadrons_from_core--corona_and_hadronic_afterburner}. 
During dynamical core--corona initialization, some partons
lose their initial energy completely inside fluids.
I regard those partons as dead ones and remove them from a list of partons.
These dead partons are no longer considered to be hadronized through string fragmentation.
For the other partons, I check whether they are inside the fluids one by one and regard them as surviving partons if it is the case.
These surviving partons are to be hadronized at $\tau>\tau_s$ if they have sufficient energy to come out from the fluids.
This will be explained later in the subsection, {\bf{parton-pairing for surviving partons}}.
The rest of the partons are considered to be partons outside fluids.
Since partons outside the fluids cannot form color-singlet strings by themselves, I need to cut the original color strings at crossing points between the hypersurfaces and the color strings by sampling thermal partons.

In the following, I explain how to find crossing points and how to sample thermal partons.
I first assume that two adjacent partons in a color string, regardless of their status (dead, surviving, or outside fluids), are chained with linearly stretched color strings between the $i$th and the $(i+1)$th partons in coordinate space.
As a simple case, suppose that $[T(\bm{x}_{i})-T_{\mathrm{sw}}][T(\bm{x}_{i+1})-T_{\mathrm{sw}}] <0$,
where $\bm{x}_{i}= (x_{i}, y_{i}, \eta_{s,i})$ and $\bm{x}_{i+1}= (x_{i+1}, y_{i+1}, \eta_{s,i+1})$ are the positions of the $i$th and the $(i+1)$th partons, respectively, there exists 
a hypersurface of fluids between the $i$th and the ($i+1$)th partons.  
I scan temperature at all fluid elements along the linearly stretched color string from the $i$th to the $(i+1)$th parton to
find a crossing point.

%Picking up thermalized partons
%---------------------------------
Once the crossing point is found, a thermalized parton is picked up to form a color-singlet string.
The thermalized partons are sampled by using the information of the crossing point on the hypersurface such as velocity $\bm{v}_{\mathrm{hyp}}$, temperature $T_{\mathrm{hyp}}$, and coordinates $\bm{x}_{\mathrm{hyp}}$, which are obtained by taking an average of that of two adjacent fluid elements crossing the hypersurface. For instance, if the two adjacent fluid elements (symbolically denoted as the $j$th and the ($j+1$)th fluid elements) have temperature $T_j>T_{\mathrm{sw}}$ and $T_{j+1}<T_{\mathrm{sw}}$, respectively, the temperature at the crossing point is obtained as $T_{\mathrm{hyp}} = (T_{j+1}+T_{j})/2$.
When the hypersurface of the fluids and the configuration of the color string are highly complicated, there could exist more than one crossing point between two adjacent partons. 
In such a case, I pick up a thermal parton from the closest crossing point for each parton in the string.

%The species of picked up partons.
%-----------------------------------
The species of the picked-up parton, whether if it is a quark, an anti-quark, or a gluon, is fixed by the configuration of color strings. 
For a color string that has a quark and an anti-quark at its ends, if string cutting removes the quark(anti-quark) side of the color string, an anti-quark(a quark) is picked up from the crossing point to form a color-singlet string in the vacuum.
When there is a color string that consists of two gluons while the only one of the gluons is inside of fluids, I pick up a gluon to make a color-singlet object.
On the other hand, for color strings with more than two gluons and no quarks or anti-quarks as their components, the so-called gluon loops, I cut the loop to open and pick up two gluons from the crossing points to make this a color-singlet object again.

A momentum of a picked-up parton is sampled with a normalized Fermi or Bose distribution,
\begin{align}
\label{eq:massive_bose_fermi}
    &P(\bm{p};m)d^3\!p  \nonumber \\ &=\frac{1}{N_{\mathrm{norm}}(m)} \frac{1}{\exp{\left[ \sqrt{\bm{p}^2 +m^2 } / T_{\mathrm{hyp}} \right] } \pm_{\mathrm{B,F}} 1}d^3\!p,
\end{align}
where 
\begin{align}
\label{eq:NormFactor}
   N_{\mathrm{norm}} (m)
    ={\dis\int \frac{1}{\exp{\left[ \sqrt{\bm{p}^2 +m^2 } / T_{\mathrm{hyp}} \right] } \pm_{\mathrm{B,F}} 1}d^3\!p}.
\end{align}
The energy of the (anti-)quarks is assigned so that they are mass-on-shell, which I require to perform hadronization via string fragmentation in \pythia8.
Four-momentum of these partons is Lorentz-boosted by using fluid velocity at the crossing point, $\bm{v}_{\mathrm{hyp}}$.

I stochastically assign flavors $f=u$, $d$,  or $s$ for each quark or anti-quark with the following probability,
\begin{align}
\label{eq:probability_flavor}
P_f & = N_{\mathrm{norm}} (m_f)/N_{\mathrm{sum}},\\
N_{\mathrm{sum}} & =N_{\mathrm{norm}}(m_u) +N_{\mathrm{norm}}(m_d)+N_{\mathrm{norm}}(m_s),
\end{align}
where the mass values of these quarks are taken from general settings in \pythia8.

As I mentioned in Sec.~\ref{subsec:DYNAMICAL_CORECORONA_SEPARATION}, there is a threshold of invariant mass of a color string to be hadronized via string fragmentation in \pythia8. If the invariant mass of a modified color string is smaller than the threshold, I remove partons forming the color string from a list of partons.

\subsubsection{Parton-pairing for surviving partons}
\label{subsec:PARTON_PAIRING} 
At $\tau>\tau_s$, I hadronize ``surviving partons traversing inside of the fluids'' when each of them comes out from fluids. To make the parton color-singlet to hadronize via string fragmentation, the parton picks up a thermal parton around the hypersurface.

Whether a parton comes out from medium or not is determined by the temperature of a fluid element at which the parton is currently located.
A surviving parton traverses a fluid according to Eq.~\refbra{eq:trajectories_of_partons}.

At the $k$th proper time step $\tau=\tau_k$, suppose that
a parton is at $\bm{x} = \bm{x}(\tau_k)$, where its temperature is $T(\bm{x}(\tau_k), \tau_k) > T_{\mathrm{sw}}$, and will move to $\bm{x} = \bm{x}(\tau_{k+1})$ at the next time step.

Simply assuming that the hypersurface does not change between the $k$th and the $(k+1)$th time step and see if the temperature satisfies $T(\bm{x}(\tau_{k+1}), \tau_k) < T_{\mathrm{sw}}$ by checking hypersurface only at the $k$th time step.
If the above condition is satisfied, the parton is regarded as coming out from the fluids at $\bm{x}_{\mathrm{hyp}} = [\bm{x}(\tau_k) + \bm{x}(\tau_{k+1})]/2$ at the $k$th time step.
For a quark (an anti-quark) coming out from medium, an anti-quark (a quark) is picked up to form a color-singlet string. On the other hand, for a gluon, a gluon is picked up to do so. A momentum is again sampled by using Eq.~\refbra{eq:massive_bose_fermi}, while its flavor is sampled with Eq.~\refbra{eq:probability_flavor}.

Note that, if a surviving parton fails to escape from the fluids by losing its initial energy completely, I regard that parton as a dead one  and remove it from a list of surviving partons.
Here again, if the invariant mass of the pair of partons is smaller than the threshold to be hadronized via string fragmentation in \pythia8, I remove them from the list of partons.

\begin{figure}
    \centering
    \includegraphics[bb=0 0 517 385, width=0.49\textwidth]{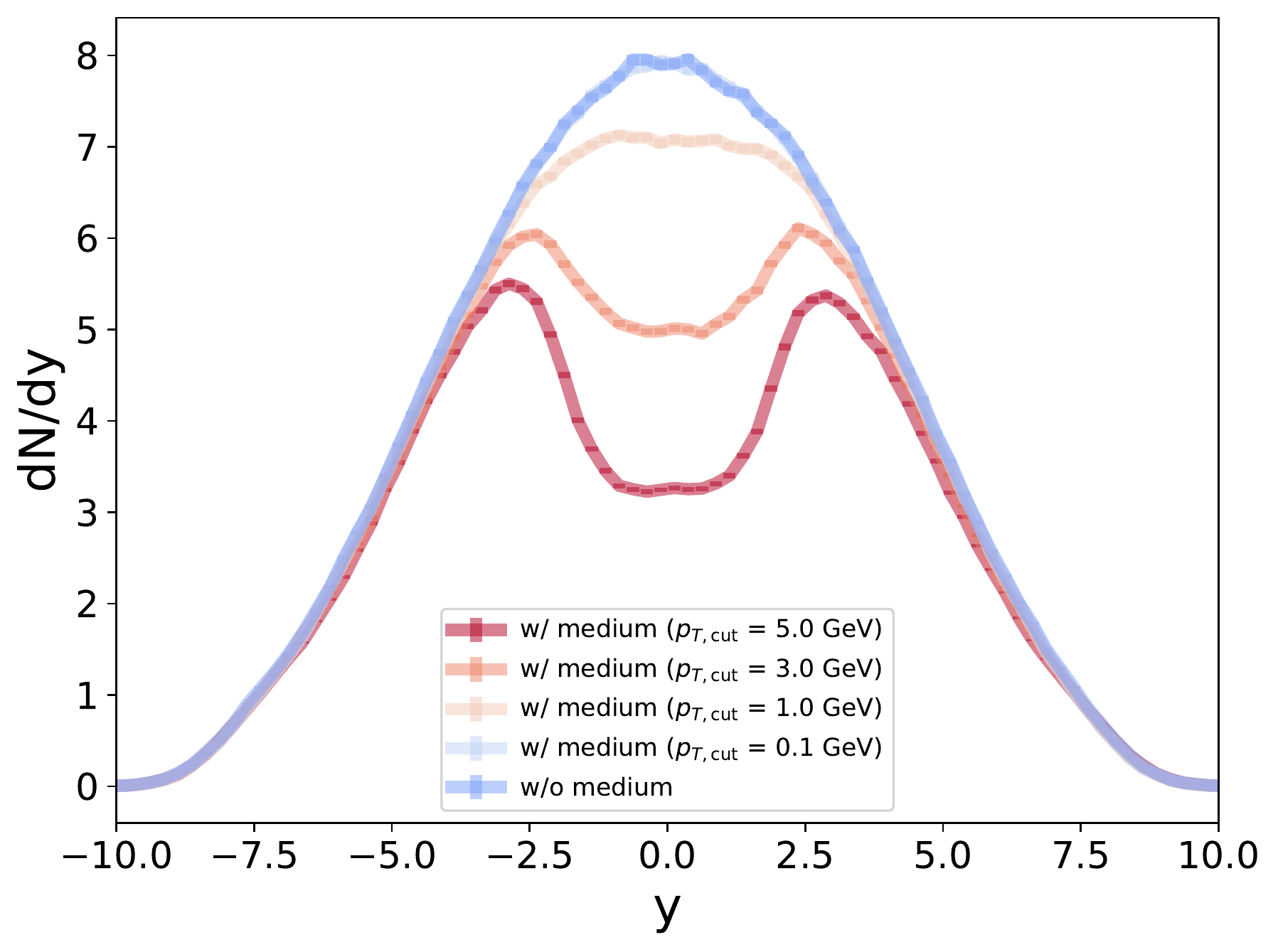}
    \includegraphics[bb=0 0 517 385, width=0.49\textwidth]{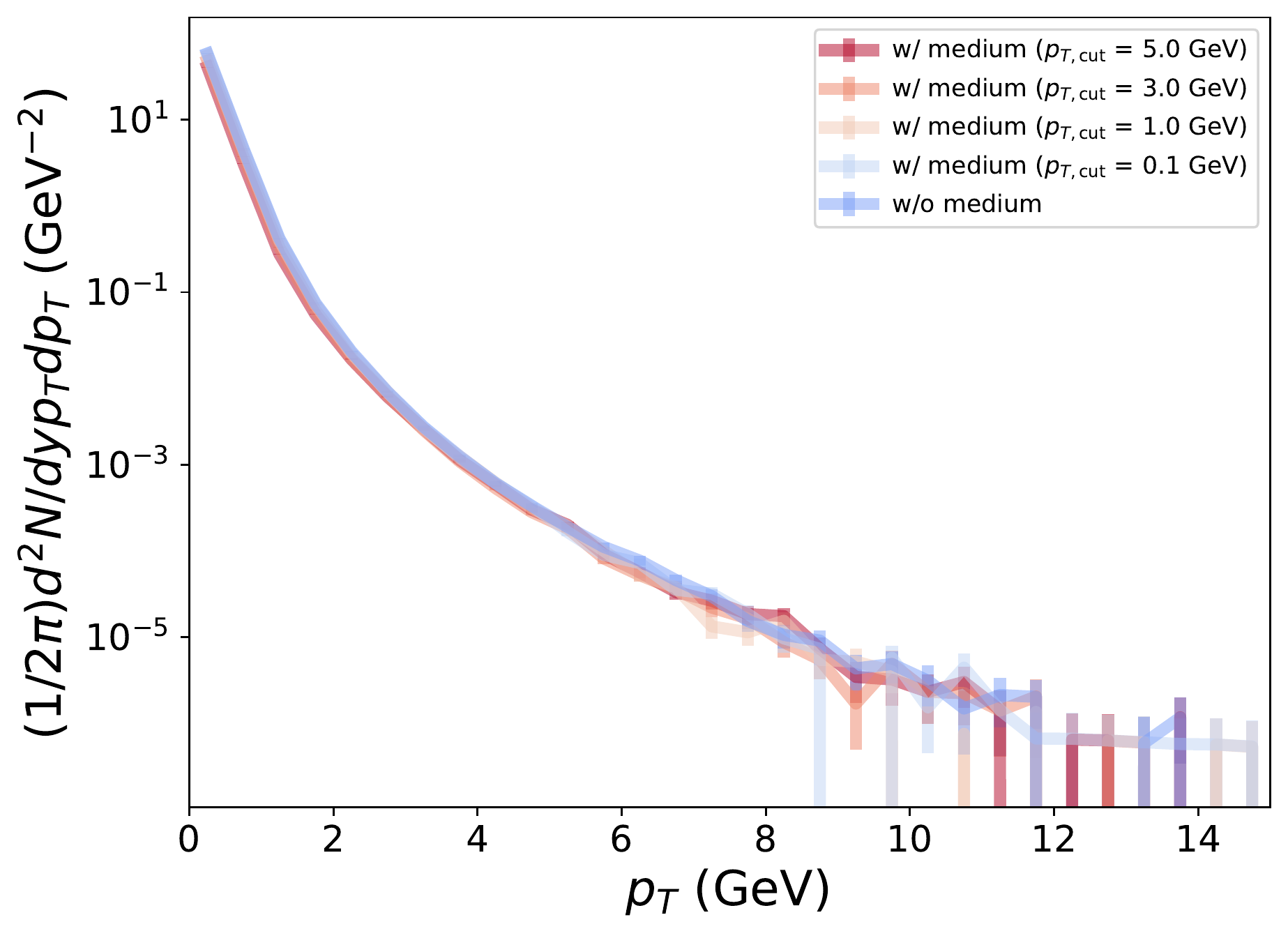}
    \caption{Effects of string modification due to overlaps between medium and color strings within toy model calculations. Left: Rapidity distribution of final charged hadrons produced from modified strings in averaged $p$+$p$ collisions at \snn[proton]=7 TeV with different $p_{T, \mathrm{cut}}$. Right: $p_T$ spectra of corresponding results.
    }
    \label{fig:STRINGCUT_EFFECT}
\end{figure}

To illustrate the effects of string modification that I have explained throughout this section,
I show results of string modification within a toy model calculation where a static temperature distribution is used instead of full simulations of DCCI2 in Fig.~\ref{fig:STRINGCUT_EFFECT}.
In this calculation, it is assumed that
partons inside of medium with $T>T_{\mathrm{sw}}$ and having low $p_T$ ($p_T<3.0$ GeV) completely deposit their energy and momentum until $\tau=\tau_0$.
Flow velocity of medium is parametrized independently of the temperature
\footnote{
Four velocity is parametrized as follows:
\begin{align}
    u^\mu = \cosh{\rho} \left( \cosh{\eta_s}, \ \tanh{\rho}\cos{\phi}, \ \tanh{\rho} \sin{\phi}, \ \sinh{\eta_s} \right), 
\end{align}
where $\phi = \arctan{y/x}$ and $\rho=\sqrt{x^2+y^2}/a$ \ $(a=3.0 \ \mathrm{fm})$.
}
.
The temperature distribution is assumed to be a Gaussian distribution with widths of $1.0$ fm transverse and $3.0$ in $\eta_s$ directions
centered at $(x, y, \eta_s)=(0,0,0)$.
The maximum temperature is set to be $T_{\mathrm{max}}=300$ GeV.
The left figure shows $p_{T, \mathrm{cut}}$ dependence of rapidity distribution of charged particles produced from modified strings in $p$+$p$ collisions at \snn[proton] = 7 TeV.
It should be noted that the size of medium is fixed.
The dip at midrapidity becomes large with larger $p_{T, \mathrm{cut}}$
because, with the larger cut off, more strings become candidates of string modifications.
The right figure shows $p_T$ distribution from the corresponding calculations to the left figure.
As I explained at the beginning of this section, I perform the modification only for color strings in which the transverse momentum of all partons forming that color string is less than a cut-off parameter $p_{T, \mathrm{cut}}$ at $\tau=\tau_0$.
This treatment avoids significant modification on $p_T$ spectra of final hadrons generated from semi hard $p_T$ partons which less interact with fluids,
and which is demonstrated in the right figure of Fig.~\ref{fig:STRINGCUT_EFFECT}.

\subsection{Direct hadrons from core--corona and hadronic afterburner}
\label{subsection:Direct_hadrons_from_core--corona_and_hadronic_afterburner}
%General
%-----------------------------
% Direct hadrons from the core parts come from the at $T(x)=T_{\mathrm{sw}}$. 
I switch the description 
of the hadrons in the core parts
from hydrodynamics to particle picture at the 
$T(x)=T_{\mathrm{sw}}$ hypersurface. 
The particlization of fluids is performed with \ISthreeD \ \cite{McNelis:2019auj}, which is an open-source code to perform conversion of hypersurface information of fluids into phase-space distributions of hadrons based on Monte-Carlo sampling of the Cooper-Frye formula \cite{Cooper:1974mv}.
Since the original \ISthreeD \ \cite{McNelis:2019auj} is not intended for event-by-event particlization,  
% \comm{[YT: What does ``event-by-event particlization'' mean?]}
% \com{[YK: The original iS3D was made for oversampling of 1 fluid event.]}
I extend the code so that this can be utilized for event-by-event analysis.
I also change the list of hadrons in \ISthreeD \ to the one from the hadronic cascade model, \jam \ \cite{Nara:1999dz},  which is employed for the hadronic afterburner in the DCCI2 framework.
%Hyper-surface findings 
%------------------------
The hypersurface information is stored from $\tau = \tau_{0}$, the beginning of dynamical core--corona initialization, to the end of hydrodynamic evolution at which temperature of all fluid elements goes below $T_{\mathrm{sw}}$.
Fluid elements with $\bm{p}\cdot  d\bm{\sigma}<0$, which is known as the negative contribution in the Cooper-Frye formula, and those with $T<0.1 \ \mathrm{GeV}$ 
are ignored in \ISthreeD.
Note that ignoring the negative contributions makes it possible to count all flux generated via source terms in dynamical initialization.
In other words, if one integrates all flux including negative ones and neglects deposition of energy inside the fluids, total flux becomes zero due to Gauss's theorem
\footnote{In the conventional hydrodynamic models, initial conditions of hydrodynamic fields are put at a fixed initial time, $\tau = \tau_{\mathrm{init}}$, which can be regarded as a negative (in-coming) energy-momentum flux from the hypersurface  $T(\bm{x},\tau=\tau_{\mathrm{init}})=T_{\mathrm{sw}}$. Thus, thanks to the Gauss's theorem, the sum of out-going energy-momentum fluxes from the hypersurface $T(\bm{x}, \tau>\tau_{\mathrm{init}})=T_{\mathrm{sw}}$ is exactly the same as that of in-coming fluxes at $\tau = \tau_{\mathrm{init}}$ when there are no source terms in hydrodynamic equations.
}.
This is because, under the dynamical initialization framework, the simulation starts from the vacuum, and the deposited energy and momentum are regarded as incoming flux into the hypersurface.
I admit that energy-momentum conservation should be improved in a better treatment while I checked the conservation is satisfied within a reasonable range. 
%Extra model added in iS3D.
%--------------------------------
The space-time coordinates of sampled hadrons $(x_i, y_i, \eta_{s,i})$ are assigned stochastically in the same way as one used for picking up thermalized partons explained in Sec.~\ref{subsection:Sampling_of_thermalized_partons}.

%General 
%--------------
Regarding the corona parts, partons out of equilibrium undergo hadronization through string fragmentation.
Until $\tau=\tau_{s}$, I assume no hadronization occurs. At $\tau=\tau_{s}$, partons outside the fluids hadronize via string fragmentation. If a part of the string is inside the fluids, I modify the color string by cutting it at the crossing point as explained in Sec.~\ref{subsec:COLORSTRING_TREATMENT_TAU0} and hadronize the modified color string.
Once surviving partons come out from the hypersurface of fluids after $\tau=\tau_{s}$, 
those partons are hadronized by picking up a thermal parton to form a string as discussed in Sec.~\ref{subsec:PARTON_PAIRING}.
The string composed of at least one high $p_{T} (> p_{T,\mathrm{cut}})$ parton is hadronized when all the partons chained with this high $p_{T}$ parton come out from the fluids. 

The string fragmentation is performed by utilizing the Lund string fragmentation which can be called in \pythia8.
The flag \texttt{ProcessLevel:all=off} is set to stop generating events and \texttt{forceHadronLevel()} is called to perform hadronization against the partons manually added as an input.
The information of input partons handed to \pythia8 is,
particle ID, four-momentum, coordinates, color and anti-color index. 
As for coordinates, only transverse coordinates, $x$ and $y$, of partons are handed while $t$ and $z$ are set to be zeros.
This is because assigning $t$ and $z$ may cause violation of causality and should be treated carefully in \pythia8.
% \sout{
% %in \pythia8 \footnote{T.~Sjostrand, Private communication}.
% Since hadron vertices are sensitive in hadronic cascades, further improvement on this treatment is a future work.} \com{[TH: What kind of imrovement? Really need to mention this?]} \com{[YK: Deleted the sentence.]}
I correct the energy of the partons to be mass-on-shell using their momenta and rest masses to perform string fragmentation in \pythia8. This is the same procedure as I did in the previous work \cite{Kanakubo:2019ogh} since quarks or anti-quarks that lose their energy in dynamical initialization are mass-off-shell due to the four-momentum deposition of Eq.~\refbra{eq:four-momentum-deposition}. 
Information of vertices of generated hadrons are obtained with an option \texttt{Fragmentation:setVertices = on} based on the model proposed in Ref.~\cite{Ferreres-Sole:2018vgo}. I use this information for initial conditions in \jam.
%While we see when each parton comes out from the medium during dynamical initialization, the coordinates of the partons at the time are not used in string fragmentation.
%This means that vertices of hadronic productions are assigned as if the partons are initially produced in \pythia8. Considering hadronic production with given vertices of partons, which is currently not available, is a future work.

In both the particlization by \ISthreeD \ 
and the string fragmentation by \pythia8, 
I turn off decays of unstable hadrons. 
Instead, JAM handles decays of the unstable hadrons together with rescatterings while describing their space-time evolution in the late stage. 
The hadrons obtained from both \ISthreeD \ and \pythia8 are put into \jam \ all together to perform the hadronic cascade since both components should interact with each other.
In \jam, an option to switch on or off hadronic rescatterings is used to see its effect on final hadrons. It should be also noted that I turn off electroweak decays except $\Sigma^0 \rightarrow \Lambda + \gamma$ to directly compare the results from DCCI2 with experimental observable.

\begin{figure}
    \centering
    \includegraphics[bb= 0 0 1094 336, width=1.0\textwidth]{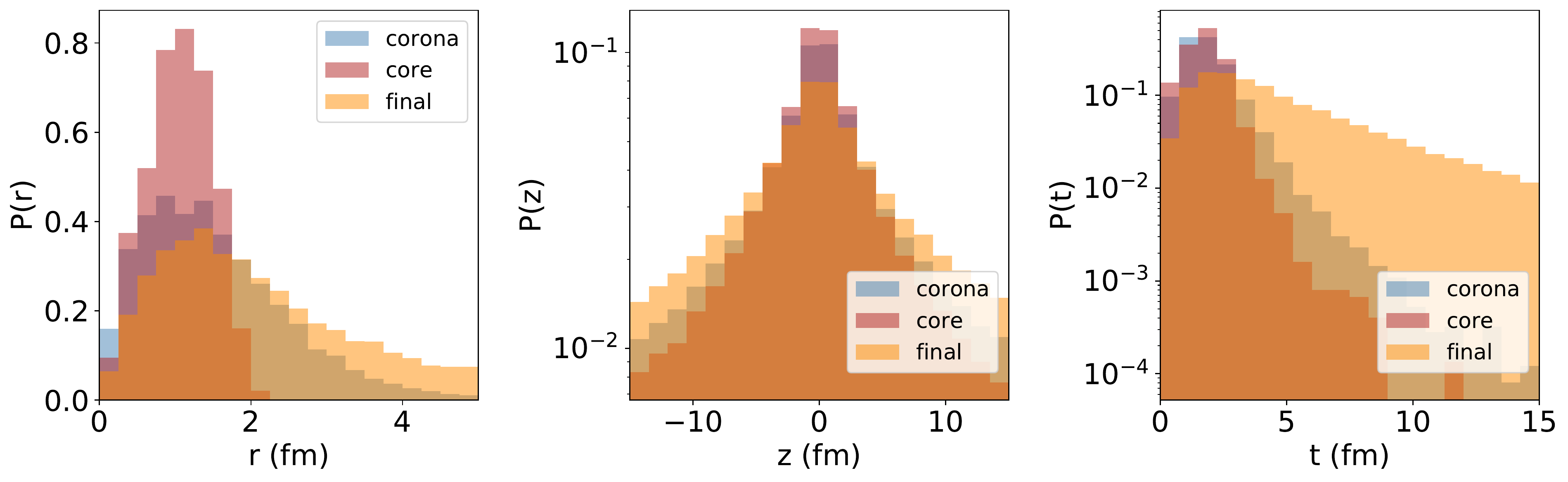}
    \includegraphics[bb= 0 0 1094 336, width=1.0\textwidth]{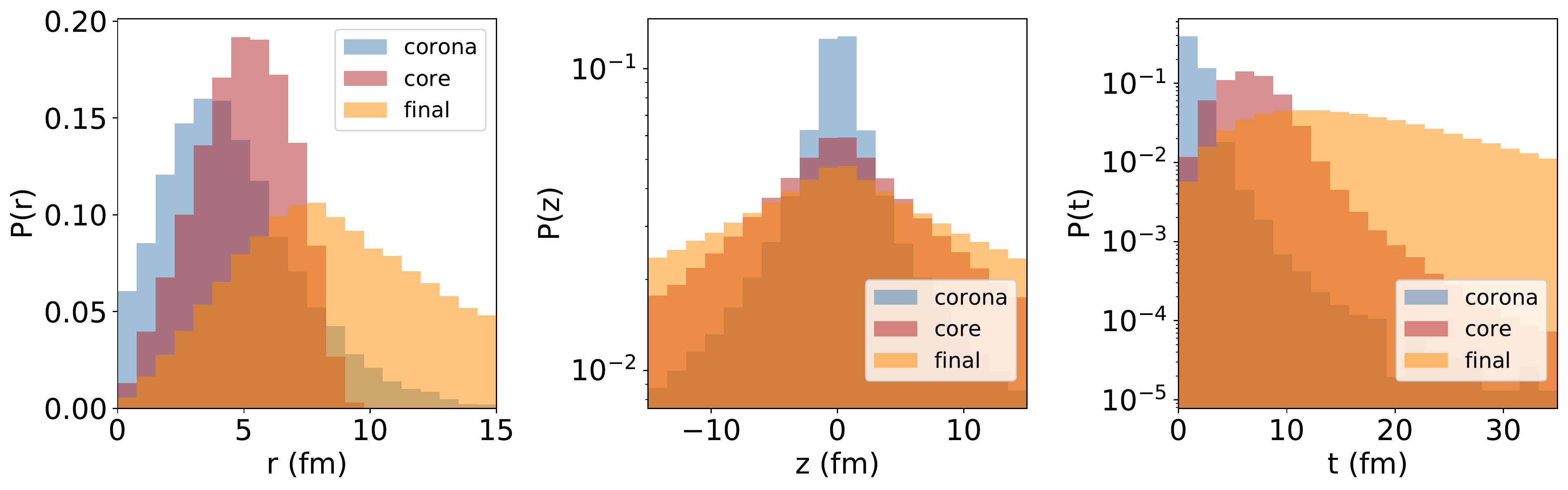}
    \caption{Normalized probability distribution of hadron vertices obtained from minimum-bias $p$+$p$ collisions at \snn[proton] = 7 TeV (upper) and $Pb$+$Pb$ collisions \snn=  2.76 TeV (lower) as functions of $r$, $z$, and $t$ shown in left, middle, and right. Comparisons of distribution of vertices between core (red) and corona (blue) components directly obtained from hydro and string fragmentation and final hadrons after hadronic rescatterings (orange).
    }
    \label{fig:DIRECT_VERTICES_SCATEFFECT}
\end{figure}

Figure \ref{fig:DIRECT_VERTICES_SCATEFFECT} shows normalized probability distribution of hadron vertices obtained from $p$+$p$ collisions at \snn[proton] = 7 TeV and $Pb$+$Pb$ collisions \snn=  2.76 TeV in upper and lower rows as functions of $r$, $z$, and $t$ shown in left, middle, and right columns. Comparisons of distribution of vertices between core and corona components directly obtained from hydro and string fragmentation and final hadrons after hadronic rescatterings are shown.
Thus, the former is an input while the latter is the output of hadronic rescatterings.
It should be emphasized that these are normalized distributions.
In minimum-bias, as I will see later, corona components are dominant for $p$+$p$ collisions while core components are dominant for $Pb$+$Pb$ collisions.

\subsection{Parameter set in DCCI2}
\label{subsection:Parameter_set_in_DCCI2}
Here I summarize all the parameters that I use throughout this thesis.
\begin{table}[htpb]
\caption{Parameter set in DCCI2 used throughout this paper.
% \com{[YK: The default value of pT0Ref used in pythia is mentioned in Sec.~\ref{subsection:Evolution_of_transverse_energy}]}
}
\begin{center}
\begin{tabular}{cc} 
\hline \hline 
 \hspace{0.5cm} Parameters \hspace{0.5cm} & \hspace{0.5cm} values  \hspace{0.5cm} \\
\hline
$p_{\mathrm{T0Ref}}$ ($p$+$p$) & 1.9 GeV \\
$p_{\mathrm{T0Ref}}$ ($Pb$+$Pb$) & 1.0 GeV \\
$\tau_{0}$ & 0.1 fm \\
$\tau_{s}$  & 0.3 fm \\
$T_{\mathrm{sw}}$  & 0.165 GeV \\
$\sigma_0$ & 0.3 $\mathrm{fm}^2$ \\
$b_{\mathrm{cut}}$ & 1.0 fm \\
$p_{T,\mathrm{cut}}$ & 3.0 GeV \\
$\sigma_\perp$  & 0.6 fm \\
$\sigma_{\eta_s}$  & 0.5 \\
$w_\perp$  & 0.5 fm \\
$w_{\eta_s}$  & 0.5 \\
$\Delta x$ & 0.3 fm \\
$\Delta y$ & 0.3 fm \\
$\Delta \eta_s $ & 0.15 \\ 
\hline \hline
\end{tabular}
\end{center}
\label{tab:PARAMETERSET}
\end{table}

Note that I use the same parameters for both $p$+$p$ and $Pb$+$Pb$ collisions except $p_{\mathrm{T0Ref}}$, which do not violate any consistencies.
The reason to use the different value of $p_{\mathrm{T0Ref}}$ for $p$+$p$ and $Pb$+$Pb$ collisions will become clear in Sec.~\ref{subsection:Evolution_of_transverse_energy} in chapter 3.
In the conventional hydrodynamic models, several parameters are used to directly parametrize the initial profiles of hydrodynamic fields. 
In contrast in DCCI2, how many initial partons are generated is determined in 
\pythia8 or \pythia8 Angantyr and how much the energy-momentum of these initial partons are converted to the hydrodynamic fields is controlled through the parameters, $\sigma_0$, $b_{\mathrm{cut}}$, $\sigma_\perp$, and $\sigma_{\eta_s}$, in Eq.~\refbra{eq:four-momentum-deposition}.
More details on how to fix these parameters are discussed in Sec.~\ref{subsec:ParameterDetermination} in chapter 3.

\chapter{Results}
\thispagestyle{fancy}

%Simulation settings
%=====================
I simulate $p$+$p$ collisions at $\sqrt{s}  = 5$, $7$, and $13$ TeV and $Pb$+$Pb$ collisions at \snn = $2.76$ TeV with DCCI2.
The following results are obtained from full simulations of about 800K, 500K, and 2M events for each $p$+$p$ collision energy, and 100K for $Pb$+$Pb$ collision.
The results shown in this chapter are basic observables that are seen in relativistic nuclear collisions and considered that should be investigated through one dynamical model. 

%Structure of this chapter
%===========================
This chapter is organized as follows. In Sec.~\ref{section:p_TintegratedObservables} and \ref{sec:MomentumDistrubutions}, I first discuss basic results -- $p_T$ integrated and differential observables,
which would clarify the tendency of particle productions within DCCI2.
In Sec.~\ref{sec:RESULTS_AnisotropicFlows}, I mainly discuss the momentum correlation in produced particles by seeing Fourier coefficients and ridge structures.
In each section, results from $p$+$p$ collisions are shown first and ones from $Pb$+$Pb$ collisions follows.

%In Sec.~\ref{subsection:PARAMETER_DETERMINATION},
%we start with fixing some major parameters in DCCI2 to reproduce the experimental data of the charged particle multiplicity as functions of multiplicity ($p$+$p$) or centrality ($Pb$+$Pb$) classes and
%the multiplicity dependence in particle yield ratios of omega baryons to charged pions.
%As a result of the parameter determination, fractions of core and corona components to final hadronic productions as a function of charged particle multiplicity at midrapidity are extracted. 
%Next, we show the transverse momentum spectrum in $p$+$p$ and $Pb$+$Pb$ collisions and its breakdown into core and corona components in Sec.~\ref{sec:TRANSVERSE_MOMENTUM_CORECORONA}.
%In order to see the interplay between core and corona components on observable obtained from final hadrons, we analyze the mean transverse momentum and the second-order anisotropic flow coefficients 
%as functions of the number of produced charged particles in certain kinematic windows in Sec.~\ref{sec:CORECORONA_CONTRIBUTION}.
%Finally, we show multiplicity dependence of radial flow effects based on violation of the mean transverse mass scaling
%and discuss if the effect can be discriminated from the one originating from pure string fragmentation with color reconnection \cite{Sjostrand:2004pf} in Sec.~\ref{sec:MTSCALING}.
%Due to the two competing particle production mechanisms,
%it is not trivial to reproduce the multiplicity within a 
%two-component model like DCCI2. We discuss details of this issue in Sec.~\ref{subsection:Evolution_of_transverse_energy}.

\begin{figure*}[htbp]
 \begin{center}
\includegraphics[bb=0 0 967 1047, width=1.0\textwidth]{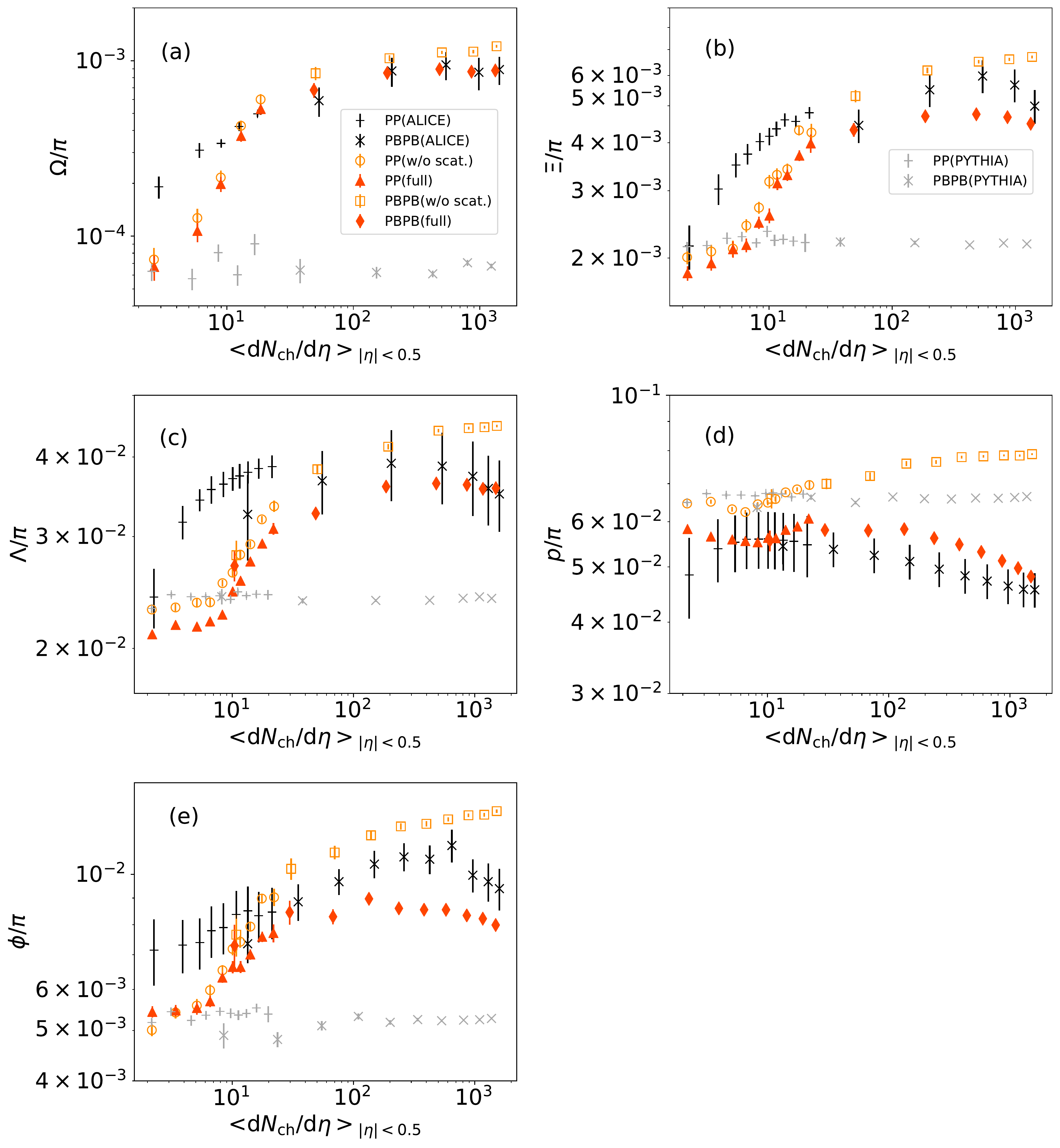}
  \caption{(Color Online)
  Particle yield ratios of (a) omegas ($\Omega^-$ and $\bar{\Omega}^{+}$), (b) cascades ($\Xi^{-}$ and $\bar{\Xi}^{+}$), (c) lambdas ($\Lambda$ and $\bar{\Lambda}$), (d) protons ($p$ and $\bar{p}$), (e) phi mesons ($\phi$) to charged pions ($\pi^+$ and $\pi^-$) as functions of charged particle multiplicity at midrapidity in $p$+$p$ and $Pb$+$Pb$ collisions.
  Results from full simulations of DCCI2 in $p$+$p$ collisions at $\sqrt{s}=7$ TeV (closed red triangles) and $Pb$+$Pb$ collisions at \snn = $2.76$ TeV (closed red diamonds) collisions are compared with the ALICE experimental data in $p$+$p$ (black pluses) and $Pb$+$Pb$ (black crosses) collisions
  \cite{ALICE:2017jyt, ABELEV:2013zaa, Abelev:2013haa, Acharya:2018orn, Abelev:2013vea, Abelev:2014uua}.  
  The $\Lambda/\pi$ ratio in $Pb$+$Pb$ collisions at \snn = $2.76 \ \mathrm{TeV}$
   reported by the ALICE Collaboration in Ref.~\cite{Abelev:2014uua}
  is plotted as a function of the number of participants $N_{\mathrm{part}}$ rather than charged particle multiplicity.
  The corresponding charged particle multiplicity at midrapidity is taken from Ref.~\cite{Abelev:2013vea}.
  Results without hadronic rescatterings are also plotted in $p$+$p$ (open orange circles) and $Pb$+$Pb$ (open orange squared)  collisions.
  Results from \pythia8 in $p$+$p$ collisions (gray pluses) and from \pythia8 Angantyr in $Pb$+$Pb$ (gray crosses) collisions are plotted as references.}
  \label{fig:PARTICLRRATIO_PP_PBPB}
 \end{center}
\end{figure*}

\section{Particle yields}
\label{section:p_TintegratedObservables}
%Dynamical models have parameters to be fixed. We do that with strangeness enhancement from pp to AA.
%==================================================================================================
Every model has parameters. They are usually chosen so that the model can describe phenomena and/or that the parameters are physically reliable. 
In DCCI2, I fix some major parameters in the former way.
The major parameter of DCCI2 is the one which controls how likely partons are fluidized, which is the constant coefficient of cross section of parton--parton scattering, $\sigma_0$, in Eq.~\refbra{eq:four-momentum-deposition} explained in Sec.~\ref{subsec:DYNAMICAL_CORECORONA_SEPARATION}, 
and I determine this so that DCCI2 gives a good description for the multi-strange hadron yield ratios as a function of multiplicity from $p$+$p$ to $Pb$+$Pb$ collisions.

\subsection{Parameter determination with multi-strange hadron yield ratios}
\label{subsec:ParameterDetermination}

%Strangeness enhancement tells us fraction of core and corona.
%================================================================
The multi-strange hadron yield ratios tell us fractions of contributions from thermalized (core) and non-thermalized (corona) matter to total final hadron yields \cite{Kanakubo:2018vkl,Kanakubo:2019ogh}. 
The idea is that yield ratios among different species of hadrons are the simplest observable that reflects particle production mechanism inherits a matter.
Particle production mechanism in each thermalized and non-thermalized QCD matter differs as follows. From the core, hadronic productions are obtained via Cooper--Frye formula, which is the Bose--Fermi distribution extended to the relativistic form.
On the other hand, particle production from corona is so called string fragmentation.
The direct hadronic productions both core and corona are followed by hadronic rescatterings and resonance decays,
so final hadronic productions should deviate from those directly obtained from statistical thermodynamics and string fragmentation. However, the modification of particle yield ratios caused by strong and weak decays are universal in both core and corona.
Thus, the difference of particle yield ratios from direct productions is not washed out.
In addition, effects of hadronic rescatterings are not significant in $p_T$ integrated productions especially for multi-strange hadrons and can be considered as a correction to the direct hadronic productions. With these reasons, we can consider that the difference of hadron yields reflects difference of particle production mechanisms.

The key differences are the scales that determines relative probability of particle productions of different hadron species, which are temperature, $T$, in the core and string tension, $\kappa$, in the corona.
Their absolute values are unique in each component in any system generated in collisions: the temperature in core productions is around $\approx 160$ MeV because QGP fluids turn into hadrons as expected in lattice QCD while string tension in string fragmentation is $\approx 1$ fm/GeV as discussed in the previous chapter.
This means that the hadron yield ratios obtained from core and corona are unique, and only the fraction of core and corona controls the hadron yield ratios.

%How we do in practical
%===========================
Here I focus on two main parameters in DCCI2, 
$\sigma_0$ to scale the magnitude of cross sections in Eq.~\refbra{eq:four-momentum-deposition}, 
and $p_{\mathrm{T0Ref}}$ used in the generation of initial partons in \pythia8 and \pythia8 Angantyr.
I determine these parameters to reasonably describe both 
the charged particle multiplicity as a function of multiplicity ($p$+$p$) or centrality ($Pb$+$Pb$) classes at midrapidity
and particle yield ratios of omega baryons to charged pions as functions of charged particle multiplicity.
The reason why I need to consider $p_{\mathrm{T0Ref}}$  not only $\sigma_0$ is because these two parameters are highly sensitive to both charged particle multiplicity and particle yield ratios and are strongly correlated. 
Detailed discussion on this issue is made in Sec.~\ref{subsection:Evolution_of_transverse_energy}. 
The resultant parameter values are summarized in Table \ref{tab:PARAMETERSET} in Sec.~\ref{subsection:Parameter_set_in_DCCI2}.

%We don't talk about string cutting
%==================================
Let me note that the results of the investigation on effects of string cutting explained in Sec.~\ref{subsec:STRING_CUTTING} are not discussed throughout this thesis.
The main reason is that it costs significant efforts and computational cost if I include the major parameter in string cutting, $p_{T, \mathrm{cut}}$, in parameter determination. Further discussion on the parameter is on a list of the future works.

%FIGURE EXP
%===================
Figure~\ref{fig:PARTICLRRATIO_PP_PBPB} shows particle yield ratios to charged pions produced in $|y|<0.5$ as functions of charged particle multiplicity $|\eta|<0.5$ in $p$+$p$ and $Pb$+$Pb$ collisions compared with the ALICE experimental data \cite{ALICE:2017jyt, ABELEV:2013zaa, Abelev:2013haa, Acharya:2018orn, Abelev:2013vea, Abelev:2014uua}.
Note that the charged particle multiplicity at midrapidity $\langle dN_{\mathrm{ch}}/d\eta \rangle_{|\eta|<0.5}$ in the horizontal axes is obtained by using V0M ($-3.7< \eta < -1.7$ and $2.8 <\eta < 5.1$) multiplicity ($p$+$p$) or centrality ($Pb$+$Pb$) class, which is the same procedure as used in the ALICE data \cite{Abelev:2014ffa}.
Determining these classes by using the multiplicity in forward and backward rapidity regions is essential even in theoretical analysis to avoid the effect of self-correlation on observables at midrapidity \cite{Acharya:2018orn}.
Throughout this thesis, the ``charged particles'' mean the sum of charged pions, charged kaons, protons, and antiprotons, which do not contain contributions from weak decays.
To obtain particle ratios of primary strange hadrons, which are stable against strong decays, I switch off their weak decays in JAM. 
Note that I take into account a particular electromagnetic decay, $\Sigma^0 \rightarrow \Lambda + \gamma$, in the presented results of $\Lambda$ yields \cite{ALICEPrimaryParticle}.
Results with switching off hadronic rescatterings are shown to reveal the effect of hadronic rescatterings \cite{Hirano:2005xf,Hirano:2007ei,Takeuchi:2015ana} on both core and corona components in the late stage.
Results from \pythia8 for $p$+$p$ collisions and \pythia8 Angantyr for $Pb$+$Pb$ are also plotted as references.

Overall, smooth changes of the particle yield ratios are observed along charged particle multiplicity, which is consistent with the previous studies \cite{Kanakubo:2018vkl,Kanakubo:2019ogh}. 
Due to the implementation of the core--corona picture in the dynamical initialization framework, particle productions from corona components with string fragmentation are dominant in final hadron yields in low-multiplicity events, while those from core components produced from equilibrated matter are dominant in high-multiplicity events.
Thus the overall tendency is that the particle yield ratio at low-multiplicity events almost reflects its value obtained from string fragmentation, while the one at high-multiplicity events reflects the value obtained only from hadronic productions from hydrodynamics.
Notice that the particle yield ratios of all hadronic species are almost independent of multiplicity from $p$+$p$ to $Pb$+$Pb$ collisions with default \pythia8 and \pythia8 Angantyr respectively, which is one of the manifestations of  ``jet universality", namely, 
the string fragmentation being independent of how the string is formed from $e^+$+$e^{-}$ to $Pb$+$Pb$ collisions \cite{Buckley:2011ms} 
\footnote{Note that \pythia 8 and \pythia 8 Angantyr with rope hadronization show enhancement of strange hadron yield ratios as a function of multiplicity
\cite{Bierlich:2014xba,Bierlich:2017sxk}.}.

I tune the parameters in the full simulations of DCCI2 to reasonably reproduce
the particle yield ratios of omega baryons to pions, $\Omega/\pi$,
reported by the ALICE Collaboration \cite{ALICE:2017jyt,ABELEV:2013zaa}.
Although I have to admit that the results from DCCI2 do not perfectly describe the experimental data as one sees in Fig.~\ref{fig:PARTICLRRATIO_PP_PBPB} (a), fine-tuning of the parameters is beyond the scope in this paper.
In Figs.~\ref{fig:PARTICLRRATIO_PP_PBPB} (b)-(e),
I also show results of cascades, lambdas, protons, and phi mesons, respectively. 
For the ratios of cascades to pions, $\Xi/\pi$, in Fig.~\ref{fig:PARTICLRRATIO_PP_PBPB} (b), results from DCCI2 underestimate the experimental data except the lowest- and the highest-multiplicity classes in $p$+$p$ collisions.
For the ratios of lambdas to pions, $\Lambda/\pi$,  in Fig.~\ref{fig:PARTICLRRATIO_PP_PBPB} (c), results from full simulations show smaller values than the experimental data in $p$+$p$ collisions for almost the entire charged particle multiplicity, while it shows good agreement with the data in $Pb$+$Pb$ collisions. For the ratios of protons to pions, $p/\pi$, in Fig.~\ref{fig:PARTICLRRATIO_PP_PBPB} (d), 
full results including hadronic rescatterings through JAM 
qualitatively describe the decreasing behavior along the charged particle multiplicity in the experimental data in $Pb$+$Pb$ collisions. 
This is consistent with a perspective of proton-antiproton annihilation \cite{Werner:2018yad,Abelev:2013vea}. 
The annihilation effect is seen even in $p$+$p$ collisions, which leads to a better agreement with the experimental data. 
For the ratios of phi mesons to pions, $\phi/\pi$,  in 
Fig.~\ref{fig:PARTICLRRATIO_PP_PBPB} (e), the tendency in the experimental data above $\langle dN_{\mathrm{ch}}/d\eta \rangle_{|\eta|<0.5} \approx 7$ is well captured by full result. 
In particular, the dissociation of phi mesons in hadronic rescatterings plays an important role to describe the suppression at high-multiplicity as observed in the experimental data.

Notably, the increasing behavior along charged particle multiplicity in $p$+$p$ collisions is achieved in results of DCCI2 with the core--corona picture.
It is also discussed that canonical suppression models, which are commonly used in the discussion on multiplicity dependence of particle yield ratios,
do not describe the increasing tendency of the phi meson yield ratios along with multiplicity due to the hidden strangeness nature of phi mesons.
One needs to incorporate additionally  incompleteness of chemical equilibrium for strangeness, which is also known as strangeness saturation factor $\gamma_{S}$, to describe the data within this framework \cite{Sollfrank:1997bq, Vovchenko:2019kes}.
Thus, both the core--corona and canonical suppression models suggest that the matter formed in $p$+$p$ collisions does not reach chemical equilibrium for strangeness.

\begin{figure*}[htbp]
  \begin{center}
 \includegraphics[bb=0 0 552 537, width=0.49\textwidth]{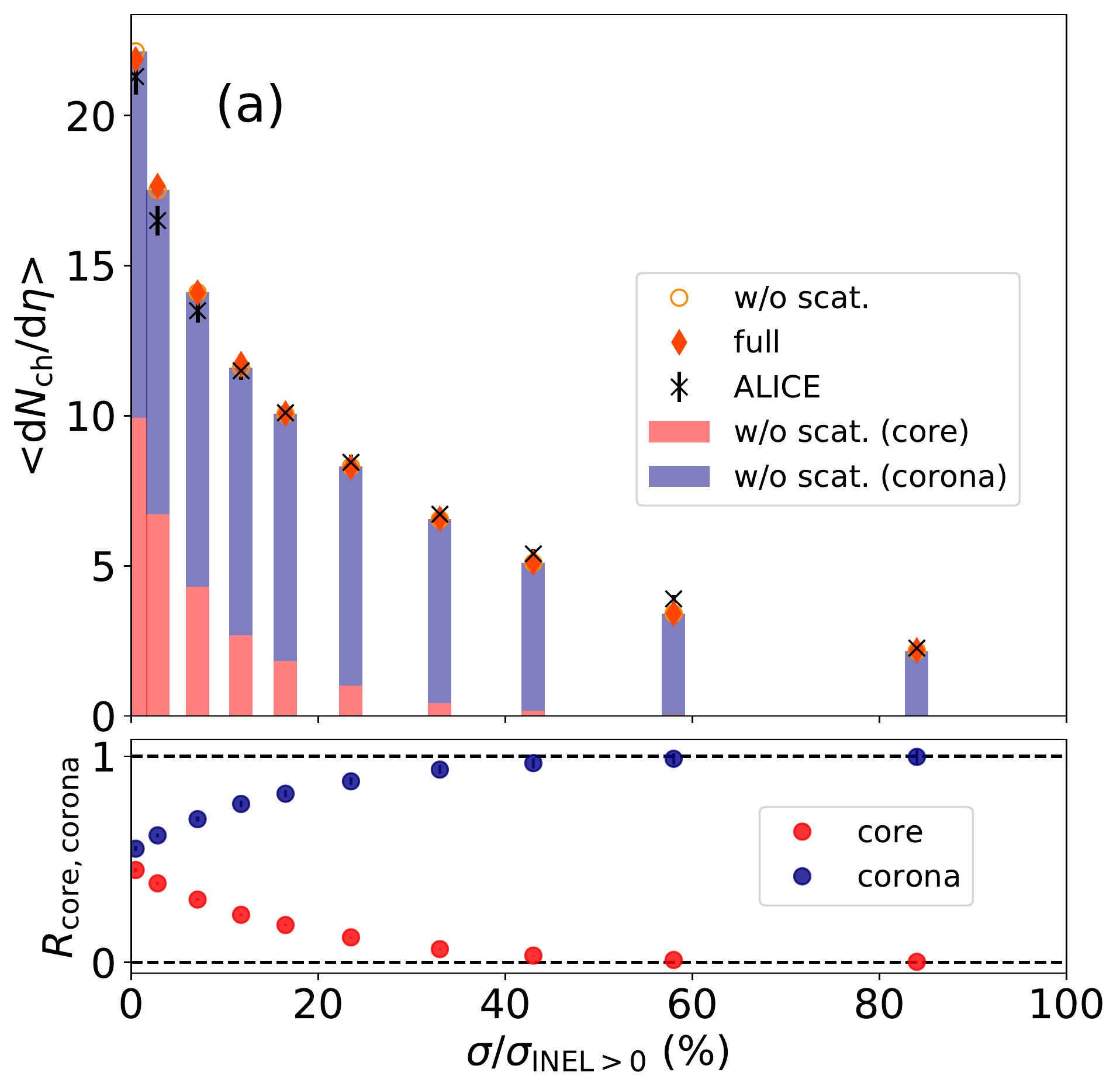}
 \includegraphics[bb=0 0 552 537, width=0.49\textwidth]{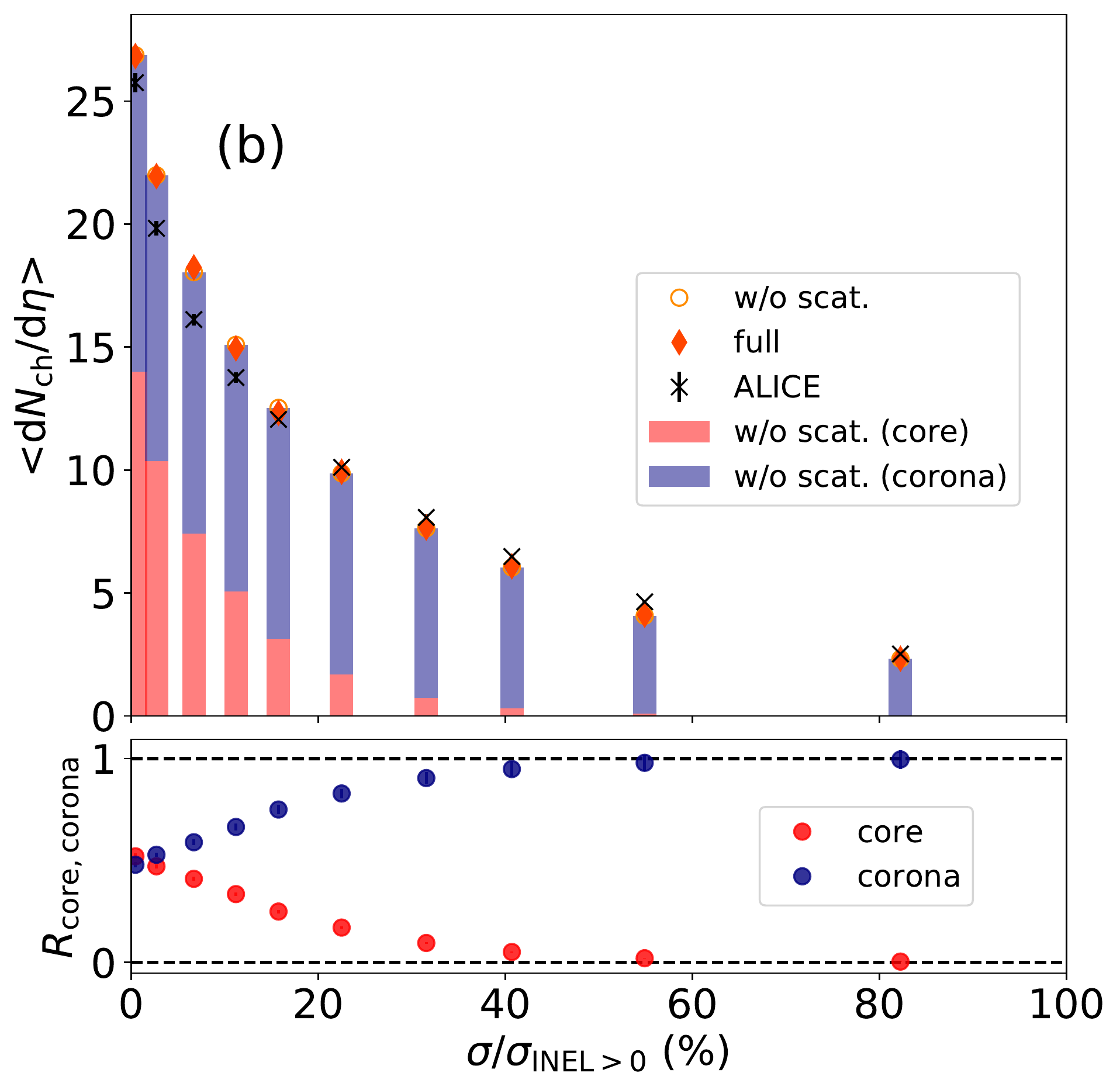}
\includegraphics[bb=0 0 552 537, width=0.49\textwidth]{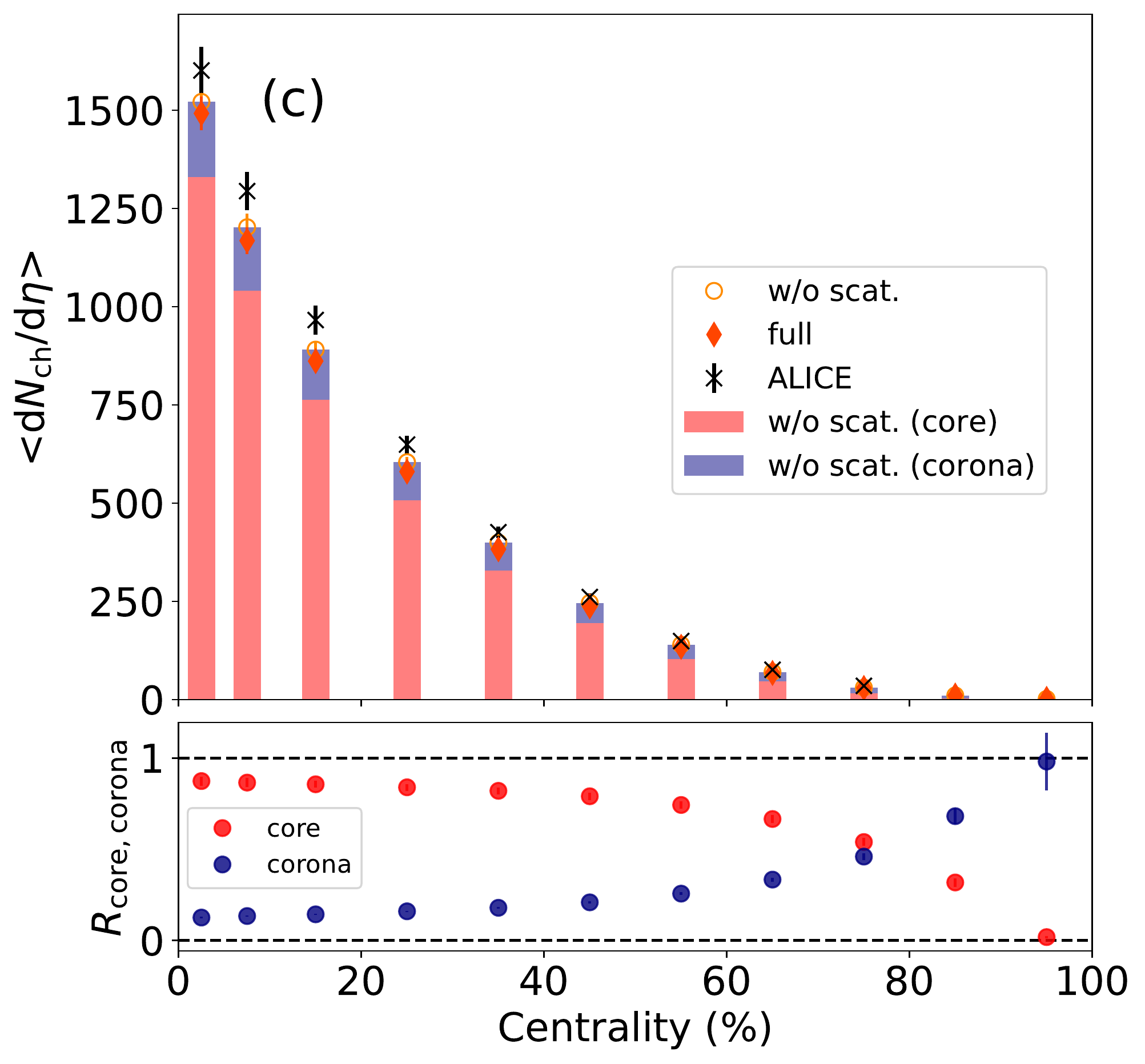}
\caption{(Color Online)
(Upper panels) Charged particle multiplicity at midrapidity as a function of the fraction of the INEL$>0$ cross sections in $p$+$p$ collisions at (a) $\sqrt{s} =7$ and (b) $13$ TeV.
(c) Charged particle multiplicity at midrapidity as a function of centrality from $Pb$+$Pb$ collisions at \snn = $2.76 \ \mathrm{TeV}$.
Results from full simulations (orange diamonds) and simulations without hadronic rescatterings (open orange circles) are compared with the ALICE experimental data (black crosses) \cite{Acharya:2018orn,Aamodt:2010cz}.
Core and corona contributions without hadronic rescatterings are shown in red and blue stacked bars, respectively.
(Lower panels) Fractions of core (red) and corona (blue) components 
without hadronic rescatterings are plotted in $p$+$p$ (a) 7 and (b) 13 TeV, and (c) $Pb$+$Pb$ collisions.
}
\label{fig:MULTIPLICITY_PP_PBPB}
 \end{center}
\end{figure*}

\subsection{Fraction of core/corona in final hadronic productions}

%FIGURE EXP
%===================
The upper panel of Fig.~\ref{fig:MULTIPLICITY_PP_PBPB} (a) and (b) show charged particle multiplicity at midrapidity $\langle  dN_{\mathrm{ch}}/d\eta \rangle_{|\eta|<0.5}$ as a function of multiplicity class $\sigma/\sigma_{\mathrm{INEL}>0}$ 
in $p$+$p$ events at \snn[proton] = 7 and 13 TeV, respectively. Here, I take into account only $\mathrm{INEL} >0$
events in which at least one charged particle is produced within a pseudorapidity range $|\eta|<1.0$ defined in the ALICE experimental analysis \cite{Acharya:2018orn}.
The upper panel of Fig.~\ref{fig:MULTIPLICITY_PP_PBPB} (c) shows the same observable but as a function of centrality class in $Pb$+$Pb$ collisions.
Here I again note that each multiplicity or centrality class is obtained with V0M multiplicity.
In these figures, results from simulations with and without hadronic rescatterings are compared with the ALICE experimental data \cite{Acharya:2018orn,Aamodt:2010cz}.
Each contribution from core and corona components is separately shown as stacked bars for the case without hadronic rescatterings.
It should be noted that the separation of core and corona components in DCCI2 is attained only by switching off hadronic rescatterings in JAM.
This is because hadronic rescatterings mix those two components up by causing parton exchange between hadrons or formation of excited states. 
Both of the results from DCCI2 show the reasonable description of the ALICE experimental data in $p$+$p$ and $Pb$+$Pb$ collisions.
From the comparison between with and without hadronic rescatterings, (quasi-)elastic scatterings would be dominant and, as a result, the effect of hadronic rescatterings on multiplicity turns out not to be significant.

The lower panels of Figs.~\ref{fig:MULTIPLICITY_PP_PBPB} (a), (b), and (b) show the yield fractions of core and corona components to the total from results without hadronic rescatterings, $R_{\mathrm{core}}$ and $R_{\mathrm{corona}}$, respectively, as functions of multiplicity ($p$+$p$) and centrality ($Pb$+$Pb$) classes.
Smooth changes along multiplicity and centrality classes are observed in all $p$+$p$ and $Pb$+$Pb$ collisions. 
In Fig.~\ref{fig:MULTIPLICITY_PP_PBPB} (a),
the fraction of core components in $p$+$p$ collisions almost vanishes for 48-68\% and 68-100\% multiplicity classes, in which $\langle dN_{\mathrm{ch}}/d\eta \rangle_{|\eta|<0.5}$ is less than $\approx5$.
Then, it increases along multiplicity and reaches $R_{\mathrm{core}} \approx 0.5$ in the highest multiplicity class 0.0-0.95\% in which $\langle dN_{\mathrm{ch}}/d\eta \rangle_{|\eta|<0.5} \approx 21$. 
It should be also noted that the fraction of core components shows $R_{\mathrm{core}}\approx0.1$ at $\langle dN_{\mathrm{ch}}/d\eta \rangle_{|\eta|<0.5} \approx 7$ which is minimum-bias multiplicity for $\mathrm{INEL} >0$ events \cite{Adam:2015gka} \footnote{
The result from EPOS 3.210 shows $\approx 30\%$ at the same multiplicity \cite{Werner:2019aqa}.}.
In the $p$+$p$ collision results at \snn[proton]=13 TeV shown in Fig.~\ref{fig:MULTIPLICITY_PP_PBPB} (b), 
while the overall tendency is the same as at 7 TeV,
one sees that the contribution of core components overtakes that of corona components in only 0.0-0.90\%  multiplicity class
within DCCI2 with the current parameter set.
This supports a perspective that recent observations of collectivity in high-multiplicity small colliding systems at the LHC energies result from the (partial) formation of the QGP fluids.
The lower panel of Fig.~\ref{fig:MULTIPLICITY_PP_PBPB} (c) shows results in $Pb$+$Pb$ collisions.
The core components highly dominate, $R_{\mathrm{core}} \gtrsim 0.90$, from 0 to 10\% centrality classes where their corresponding multiplicities are above $\langle dN_{\mathrm{ch}}/d\eta \rangle_{|\eta|<0.5} \approx 10^3$. 
The corona components become dominant around at $80$\% centrality class towards peripheral events.
It should also be mentioned that the contribution of corona components remains $R_{\mathrm{corona}}\approx 0.17$-$0.22$ at midrapidity in intermediate centrality classes ($\approx40$-$60\%$) where the whole systems is often assumed to be described by hydrodynamics.

\begin{figure}[tbp]
\begin{center}
\includegraphics[bb=0 0 645 447, width=0.8\textwidth]{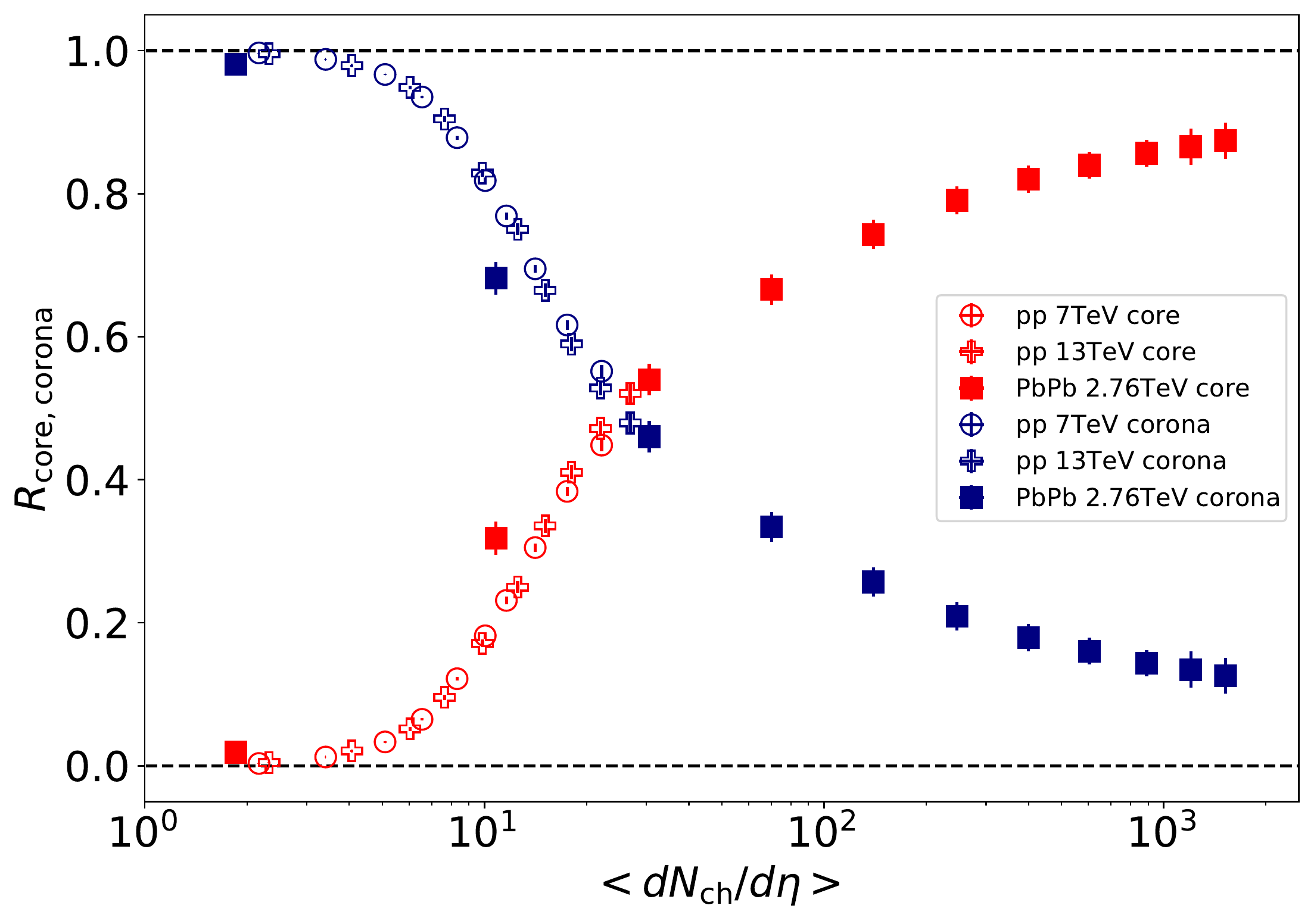}
\caption{(Color Online)
Fractions of core and corona components in the final hadron yields as functions of charged particle multiplicity at midrapidity.
Smooth behaviors of fractions of core (open red circles (crosses)) and corona (open blue circles (crosses)) contributions in $p$+$p$ collisions at $\sqrt{s} = 7$ (13) TeV are taken over by
those of core (closed red squares) and corona (closed blue squares) contributions in $Pb$+$Pb$ collisions at \snn = $2.76$ TeV, respectively.
}
\label{fig:FRACTION_CORECORONA}
\end{center}
\end{figure}

%FIGURE EXP
%==============================
Figure~\ref{fig:FRACTION_CORECORONA} shows the fractions of core and corona components in $p$+$p$ and $Pb$+$Pb$ collisions simultaneously, which are identical to the results in the lower panels in Fig.~\ref{fig:MULTIPLICITY_PP_PBPB} but as functions of charged particle multiplicity at midrapidity.
Smooth crossover from corona dominance to core dominance
appears along multiplicity from $p$+$p$ to $Pb$+$Pb$ collisions.
The dominant contribution flips at $\langle dN_{\mathrm{ch}}/d\eta \rangle_{|\eta|<0.5} \approx 20$.
It should be also noted that both results of $p$+$p$ collisions at \snn[proton] = 7 and 13 TeV are top of each other.
These results clearly demonstrate that the fractions of contribution from core and corona components are 
scaled with charged particle multiplicity in DCCI2, 
regardless of differences in the system size or collision energy
between $p$+$p$ and $Pb$+$Pb$ collisions.
Here I emphasize that, interestingly, the fraction of corona still remains $\approx 10$\% at the most central events in $Pb$+$Pb$ collisions.
This also implies that both core and corona components should be implemented even in dynamical modeling of high-energy heavy-ion collisions towards precision studies on properties of QCD matter. 

The origin of this multiplicity scaling of the fraction of core and corona within DCCI2 is coming from the similarity in phase-space distribution of initial partons from different systems at the same multiplicity.
With the fact, it would be interesting to see that the multiplicity from $p$+$p$ collisions spans around \dndeta $\approx1$-$30$ and reaches to the multiplicity of 70-80\% centrality class of $Pb$+$Pb$ collisions.
This wide range of multiplicities from $p$+$p$ collisions originates from multi-parton interactions implemented in \pythia.
Thus, the multi-parton interaction is the one that produces similarity in phase-space distribution of initial partons from different systems at the same multiplicity and that gives the seamless connection from $p$+$p$ to $Pb$+$Pb$ collisions.

\subsection{Particle productions in longitudinal direction}

So far, I have investigated particle production at midrapidity.
From now on, I investigate the dynamics of particle production in rapidity direction.

%Meaning of seeing rapidity direction
%=================================
Especially, it is worth exploring particle production in longitudinal direction within DCCI2.
Most full (3+1)-D dynamical models based on hydrodynamics often used to make data--model comparisons, {\it{e.g.,}} pseudorapidity distribution, not only at mid-rapidity but also off-midrapidity as well.
However, it can be naively considered that local equilibration of matter is more likely reached at midrapidity rather than forward and backward rapidity regimes
because the number of particles produced in collisions is largest at mid-rapidity and particles produced at mid-rapidity are the most likely to exchange momenta in secondary scatterings.
Thus, investigating results from DCCI2 which generates QGP fluids based on the core--corona picture can provide some inferences on phenomena in forward and backward rapidity regimes.

\begin{figure}
    \centering
    \includegraphics[bb=0 0 639 452, width=0.8\textwidth]{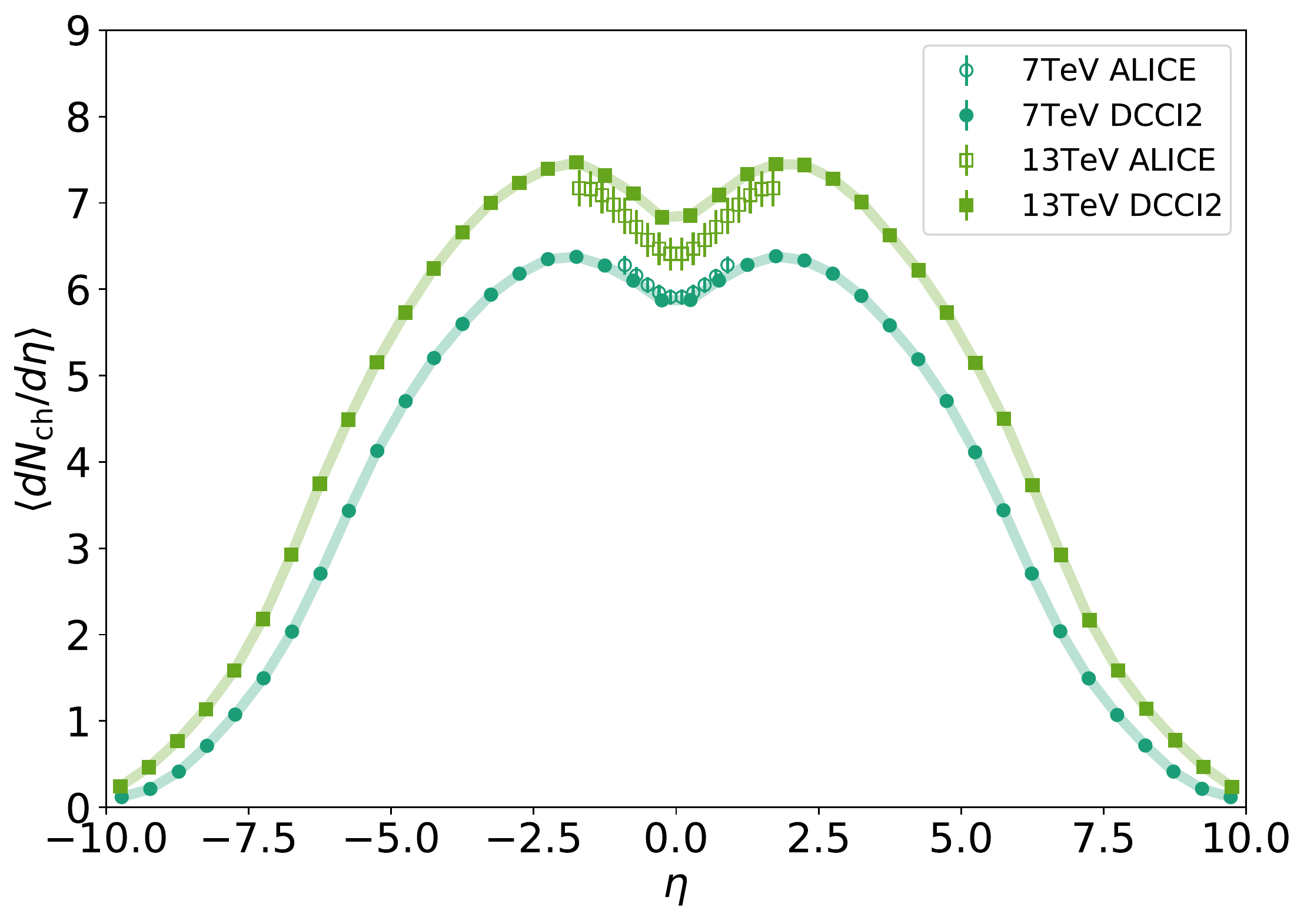}
    \caption{Charged particle distribution as a function of pseudorapidity in INEL$>$0 $p$+$p$ collisions at \snn[proton] = 7 (closed circles) and 13 (closed squares) TeV from DCCI2. The ALICE experimental data at the same collision energies are compared as open symbols.}
    \label{fig:DNDETA_DCCI2_PP}
\end{figure}

%FIGURE EXP
%================
Figure.~\ref{fig:DNDETA_DCCI2_PP} shows charged particle distribution as a function of pseudorapidity in INEL$>$0 $p$+$p$ collisions at \snn[proton]=7 and 13 TeV.
Results from DCCI2 are compared to the ALICE experimental data.
Results from DCCI2 at 7 TeV nicely describe the experimental data while ones at 13 TeV slightly overshoot at mid-pseudorapidity.
This overshooting can be understood as the fail in describing event distribution of experimental data as one can notice from Fig.~\ref{fig:MULTIPLICITY_PP_PBPB} (b).
There are obvious overestimation of multiplicity in DCCI2 around $1$-$10$\% in multiplicity classes.

%The shape of rapidity distribution
%===================================
Note that the small bump at mid-pseudorapidity is the result of kinematic effect: it is caused by Jacobian in a conversion from rapidity to pseudorapidity. 
Relation between infinitesimal values of rapidity and pseudorapidity is 
\begin{align}
     dy =\frac{|\bm{p}|}{E} d\eta = \sqrt{1-\frac{m^2}{m_T^2\cosh{(y)}}} d\eta.
\end{align}
Thus, difference on histogram between rapidity and pseudorapidity takes maximum at mid-(pseudo)rapidity,
and the effect of Jacobian becomes negligible in forward and backward.

\begin{figure}
    \centering
    \includegraphics[bb=0 0 789 452, width=0.9\textwidth]{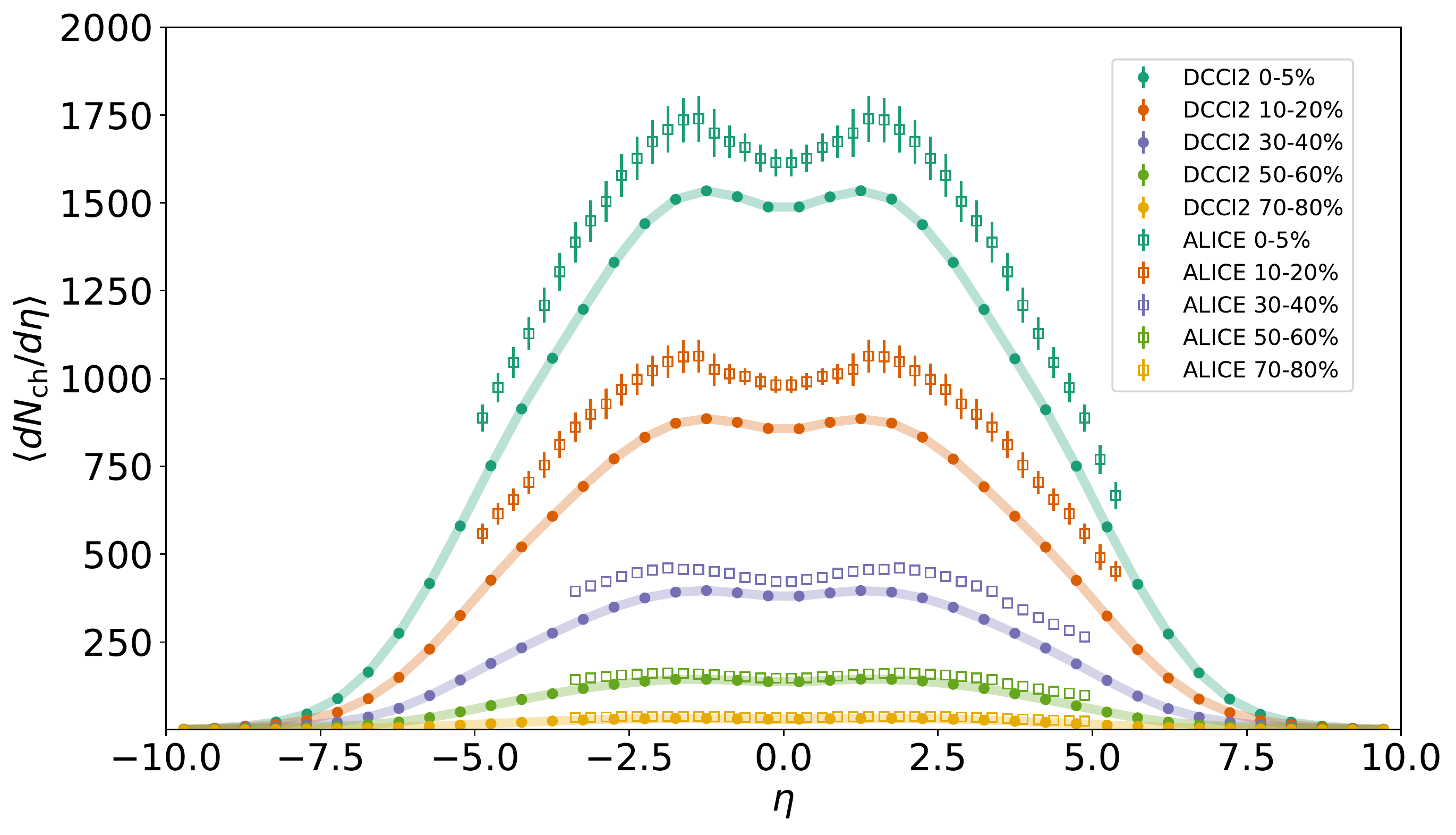}
    \caption{Centrality dependence of charged particle distribution as a function of pseudorapidity in $Pb$+$Pb$ collisions at \snn = 2.76 TeV from DCCI2 (closed circles). Comparisons to the ALICE experimental data at corresponding centrality classes (open squares) are made.}
    \label{fig:DNDETA_DCCI2_PBPB}
\end{figure}

%FIGURE EXP
%==================
Figure \ref{fig:DNDETA_DCCI2_PBPB} shows centrality dependence of charged particle distribution as a function of pseudorapidity in $Pb$+$Pb$ collisions at \snn = 2.76 TeV from DCCI2 (closed circles). Comparisons to the ALICE experimental data at corresponding centrality classes (open squares) are made.
As an overall tendency, results from DCCI2 underestimate multiplicity from experimental data.
As I discuss in Sec.~\ref{subsection:Evolution_of_transverse_energy},
it is not straightforward to reproduce multiplicity within DCCI2 as a two-component model where beam energy conservation is respected.
Further improvements should be done in future work.

\begin{figure}
    \centering
    \includegraphics[bb=0 0 551 493, width=0.49\textwidth]{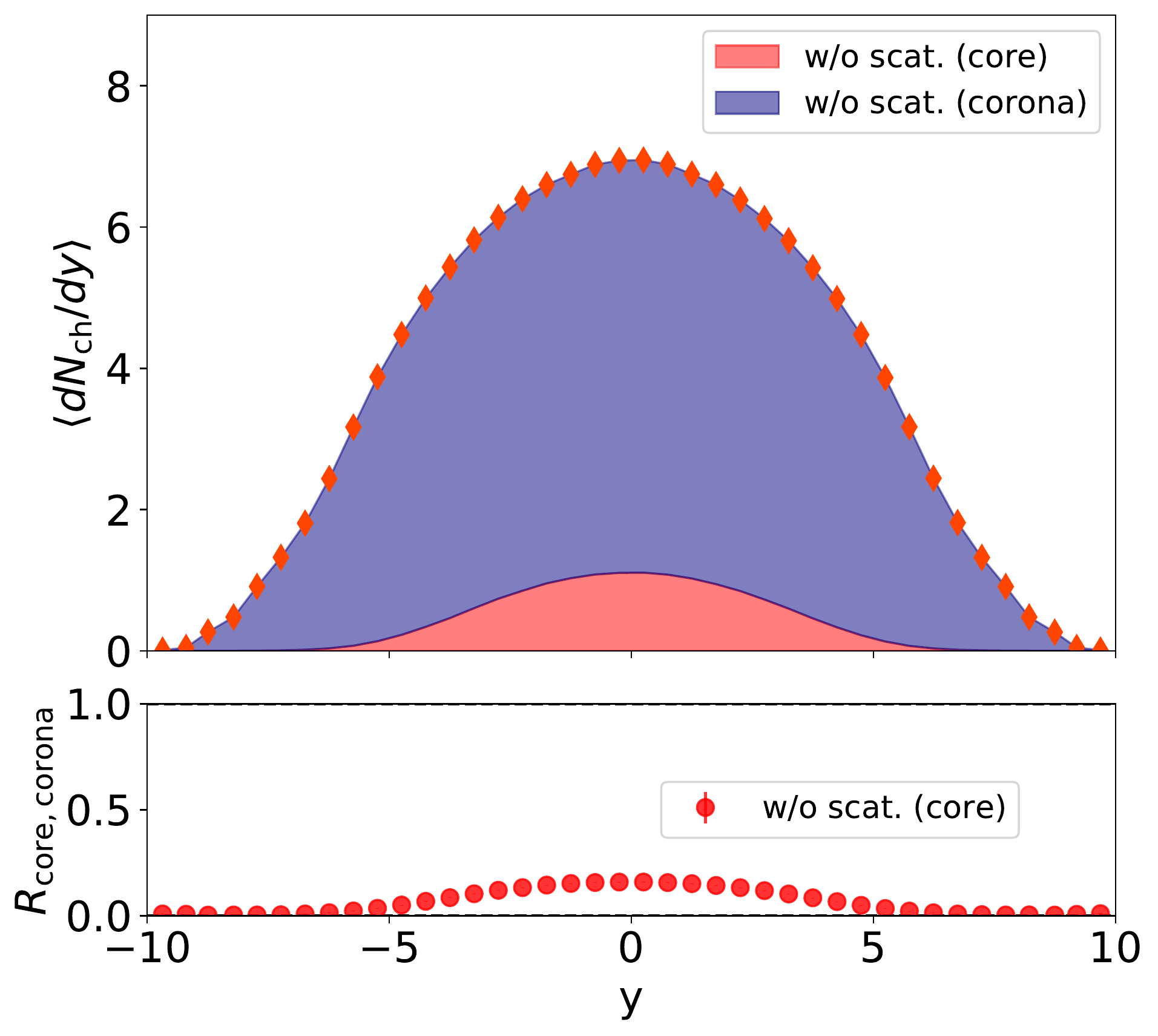}
    \includegraphics[bb=0 0 551 493, width=0.49\textwidth]{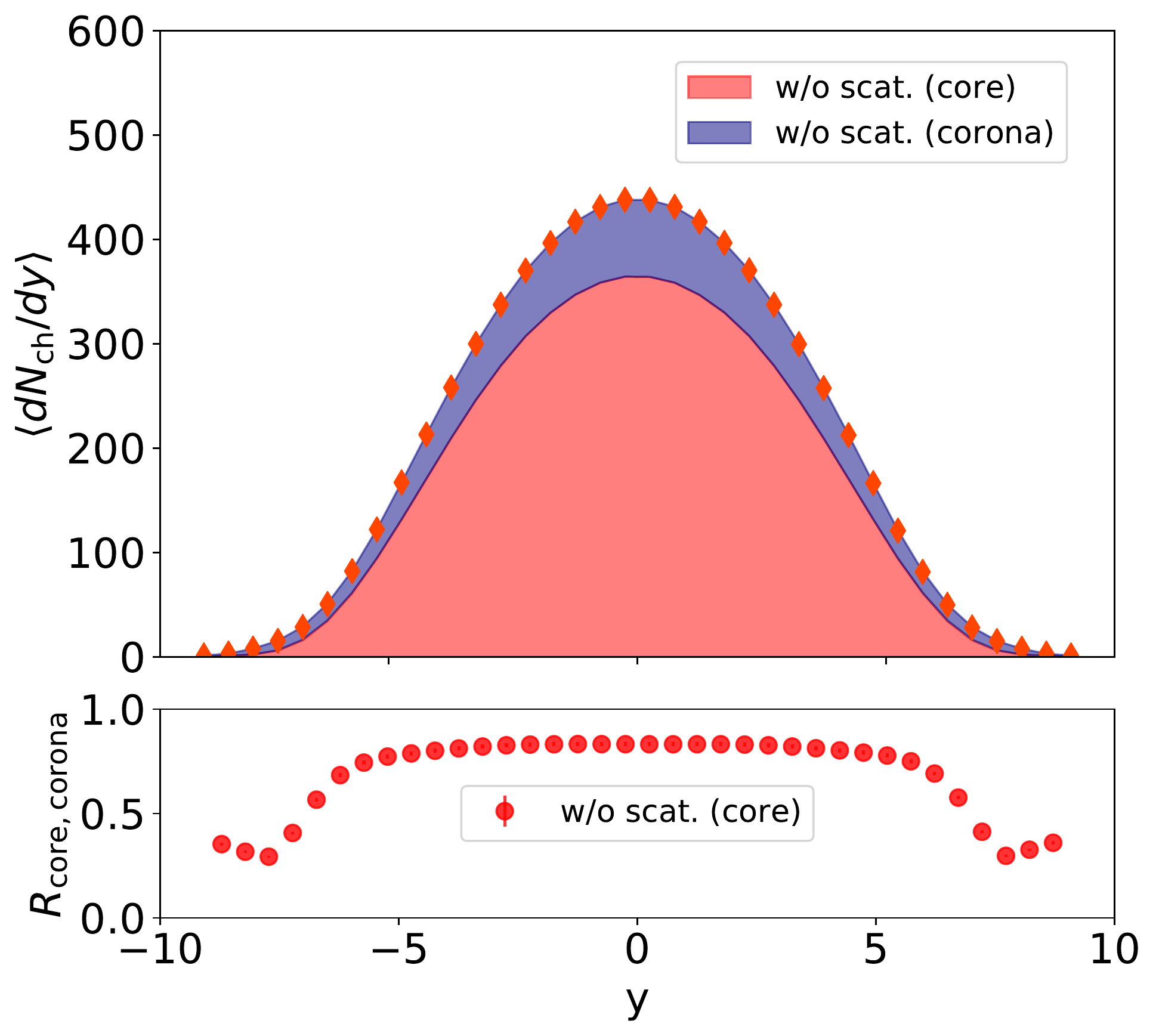}
    \caption{(Left) Charged particle distribution as a function of rapidity from INEL$>$0 $p$+$p$ collisions at \snn[proton] =7 TeV without hadronic rescatterings (orange diamonds) and its breakdown into core (red) and corona (blue) is shown in upper panel. Fraction of core component corresponding above panel is shown in lower panel. 
    (Right) Charged particle distribution as a function of rapidity from minimum-bias $Pb$+$Pb$ collisions at \snn=2.76 TeV without hadronic rescatterings is shown in upper panel. Meaning of each plot is the same as left figure. 
    }
    \label{fig:DNDY_CORECORONA}
\end{figure}

%FIGURE EXP
%=============
As I mentioned at the beginning of this subsection, it is worth investigating particle productions in longitudinal directions with DCCI2 where the core--corona picture is introduced.
Figure \ref{fig:DNDY_CORECORONA} shows 
charged particle distribution as a function of rapidity
\footnote{
It is often the case with that charged (identified) particle distributions are shown as a function of pseudorapidity (rapidity).
This is because, especially in experiments,
one needs to know particle mass to obtain rapidity, which is a function of energy and momentum, and one does not need to do so to obtain pseudorapidity, which is only a function of momentum.
}
without hadronic rescatterings and its breakdown into core and corona contributions is shown in upper panel simultaneously.
The left and right figures correspond to results from INEL$>$0 $p$+$p$ collisions at \snn[proton] =7 TeV and minimum-bias $Pb$+$Pb$ collisions at \snn = 2.76 TeV.
In lower panels, fractions of core component corresponding to the upper results are shown as a function of rapidity.

As an overall tendency, core contribution takes its maximum at mid-rapidity.
In INEL$>$0 $p$+$p$ collision results, core contribution becomes visible in rapidity distribution within around $|y|<5$, and reaches up to $\approx 10$\% at mid-rapidity.
On the other hand,
$R_{\mathrm{core}}$ in $Pb$+$Pb$ collision results are almost flat as a function of rapidity around $|y|<5$.
This means that the almost all ``soft'' partons are fluidized and reaches saturation of $R_{\mathrm{core}}$:
as one sees in Fig.~\ref{fig:FRACTION_CORECORONA},
multiplicity dependence of $R_{\mathrm{core}}$ becomes flatter in higher multiplicity events in $Pb$+$Pb$ collisions.

\begin{figure}
    \centering
    \includegraphics[bb=0 0 765 445, width=1.0\textwidth]{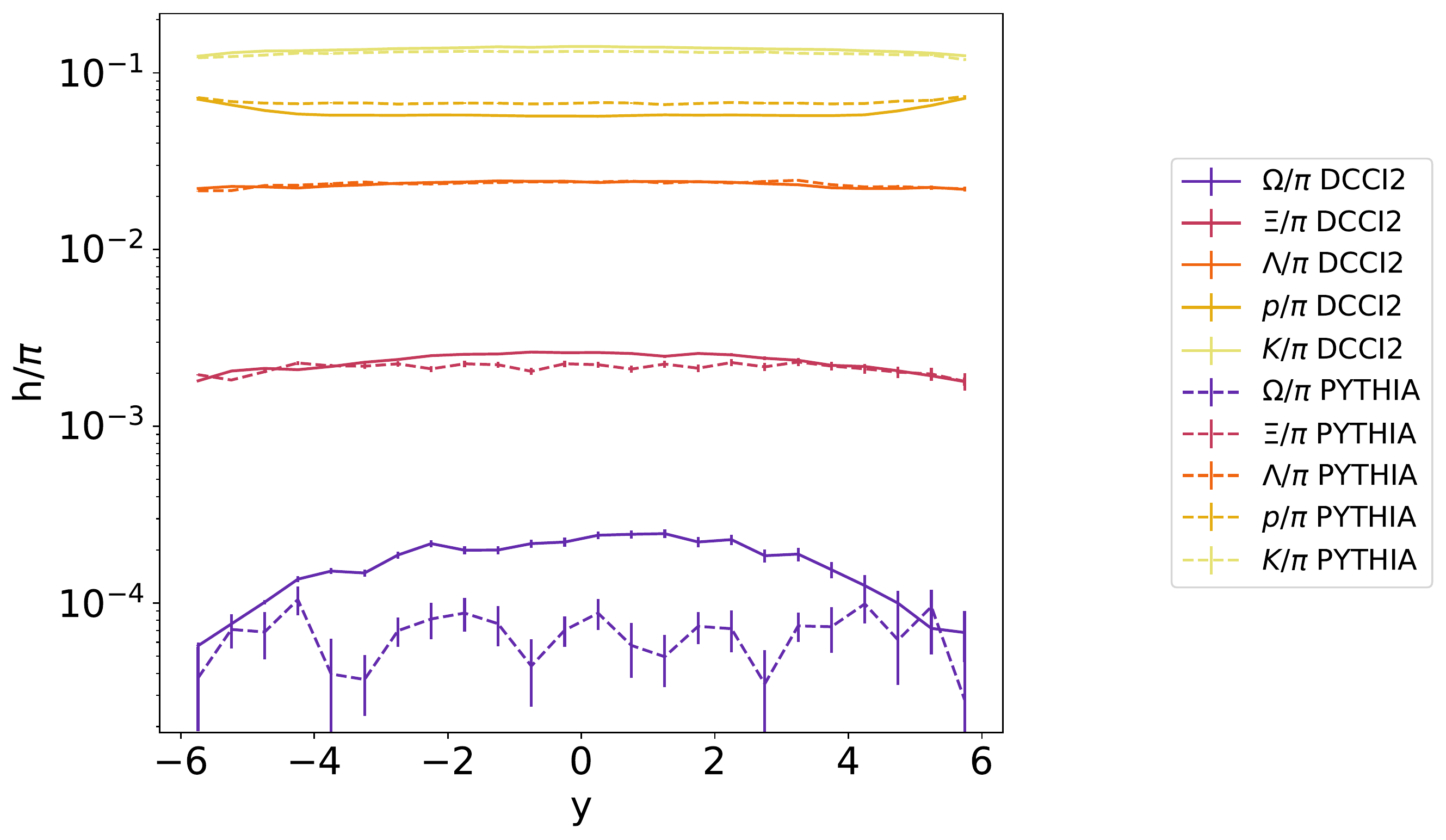}
    \caption{Particle yield ratios of omegas ($\Omega^-$ and $\bar{\Omega}^{+}$), cascades ($\Xi^{-}$ and $\bar{\Xi}^{+}$), lambdas ($\Lambda$ and $\bar{\Lambda}$), protons ($p$ and $\bar{p}$), charged kaons ($K^+$ and $K^-$) to charged pions ($\pi^+$ and $\pi^-$) from full simulations as functions of rapidity from INEL$>$0 $p$+$p$ collisions at \snn[proton]=7 TeV. Comparisons between results from DCCI2 (solid lines) and \pythia \ (dashed lines) are made.}
    \label{fig:PARTICLERATIO_LONG_PP}
\end{figure}

\begin{figure}
    \centering
    \includegraphics[bb=0 0 765 445, width=1.0\textwidth]{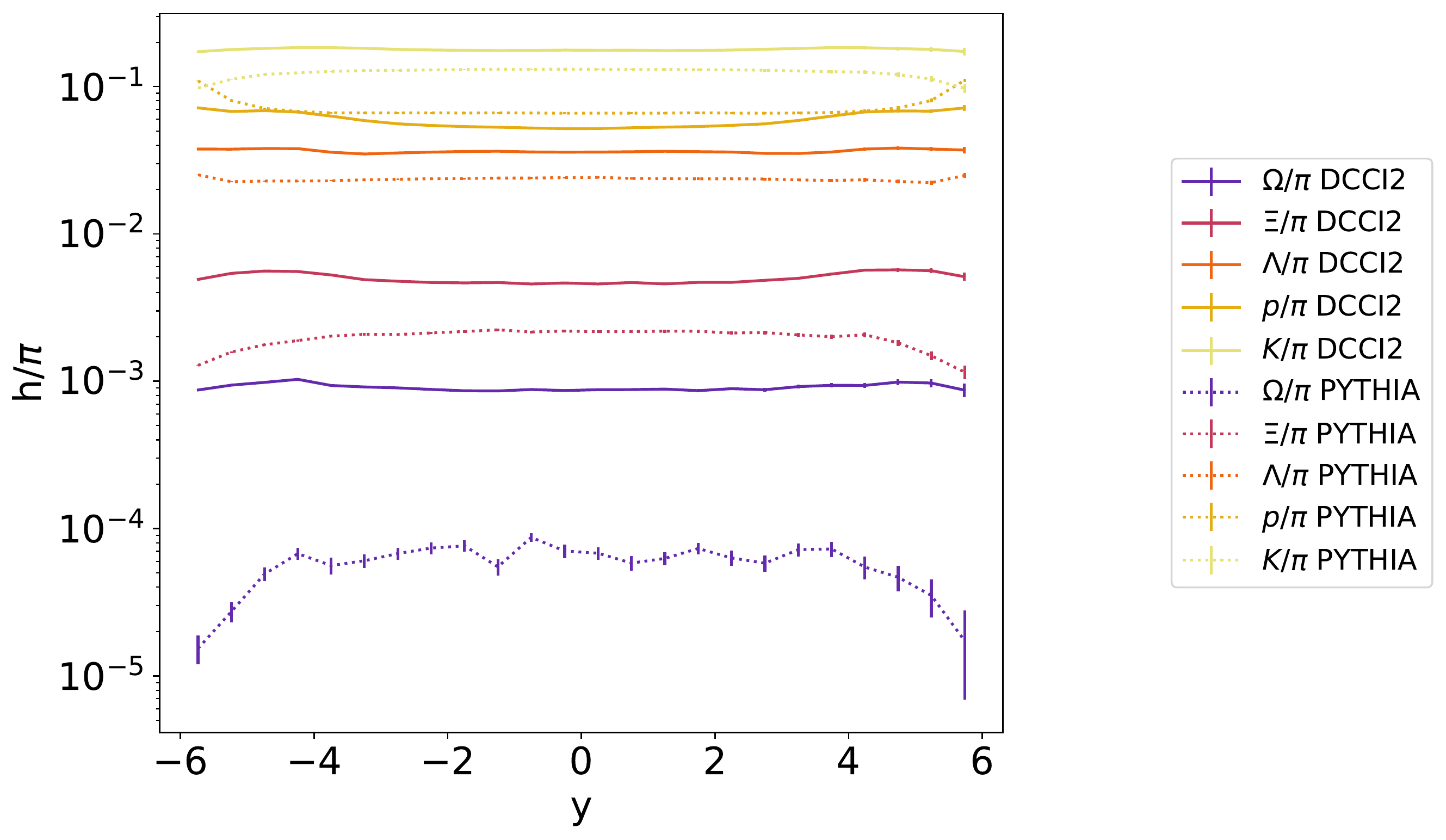}
    \caption{Particle yield ratios of omegas ($\Omega^-$ and $\bar{\Omega}^{+}$), cascades ($\Xi^{-}$ and $\bar{\Xi}^{+}$), lambdas ($\Lambda$ and $\bar{\Lambda}$), protons ($p$ and $\bar{p}$), charged kaons ($K^+$ and $K^-$) to charged pions ($\pi^+$ and $\pi^-$) from full simulations as functions of rapidity from minimum-bias $Pb$+$Pb$ collisions at \snn = 2.76 TeV. Comparisons between results from DCCI2 (solid lines) and \pythia \ Angantyr (dotted lines) are made.}
    \label{fig:PARTICLERATIO_PBPB}
\end{figure}

%FIGURE EXP
%==============
As I discussed in Sec.~\ref{subsec:ParameterDetermination} in this chapter, particle yield ratios, especially multi-strange hadrons, reflect the fraction of core and corona.
Now, it would be interesting to see particle yield ratios as functions of rapidity with keeping in mind the rapidity dependence of core contribution seen in Fig.~\ref{fig:DNDY_CORECORONA}.
Figure \ref{fig:PARTICLERATIO_LONG_PP} shows 
particle yield ratios from full simulations as functions of rapidity from INEL$>$0 $p$+$p$ collisions at \snn[proton]=7 TeV. Comparisons between results from DCCI2 and \pythia \ are made to see effects of core component.
For all hadron species, difference between DCCI2 and \pythia \ results is comparatively large at mid-rapidity, which is a consequence of rapidity dependence of the particle yield ratios with partial core productions.
Similar to the results in Fig.~\ref{fig:PARTICLRRATIO_PP_PBPB}, omega baryon yield ratios show largest enhancement in the listed results:
the ratio becomes $3$-$4$ times larger at mid-rapidity compared to string fragmentation.
This result suggests that clear enhancement of omega baryon yield ratios along rapidity can be seen even in almost minimum-bias events in $p$+$p$ collisions
if local equilibration is achieved based on the core--corona picture.
It should also be noted that the proton yield ratios show decrease compared to results from \pythia \ at mid-rapidity caused by baryon--antibaryon annihilation as seed in Fig.~\ref{fig:PARTICLRRATIO_PP_PBPB}.

Figure \ref{fig:PARTICLRRATIO_PP_PBPB} shows particle yield ratios from full simulation as a function of rapidity from minimum-bias $Pb$+$Pb$ collisions at \snn = 2.76 TeV from DCCI2. The results are compared to those from \pythia \ Angantyr for the same reason as done in Fig.~\ref{fig:PARTICLERATIO_LONG_PP}.
Difference of particle yield ratios between DCCI2 and \pythia \ is larger in $Pb$+$Pb$ collisions.
Compared to the results in $p$+$p$ collisions, one would notice that the particle yield ratios from DCCI2 show flat dependence on rapidity because of the flat $R_{\mathrm{core}}$ distribution along rapidity within $|y|<5$ seen in Fig.~\ref{fig:DNDY_CORECORONA}.
It should be noted that the all hadron yield ratios except kaons show, more or less, small decrease at mid-rapidity compared to the ratios at forward/backward in DCCI2 because of the baryon--antibaryon annihilation in hadronic rescatterings.

\begin{figure}
    \centering
    \includegraphics[bb=0 0 557 452, width=0.49\textwidth]{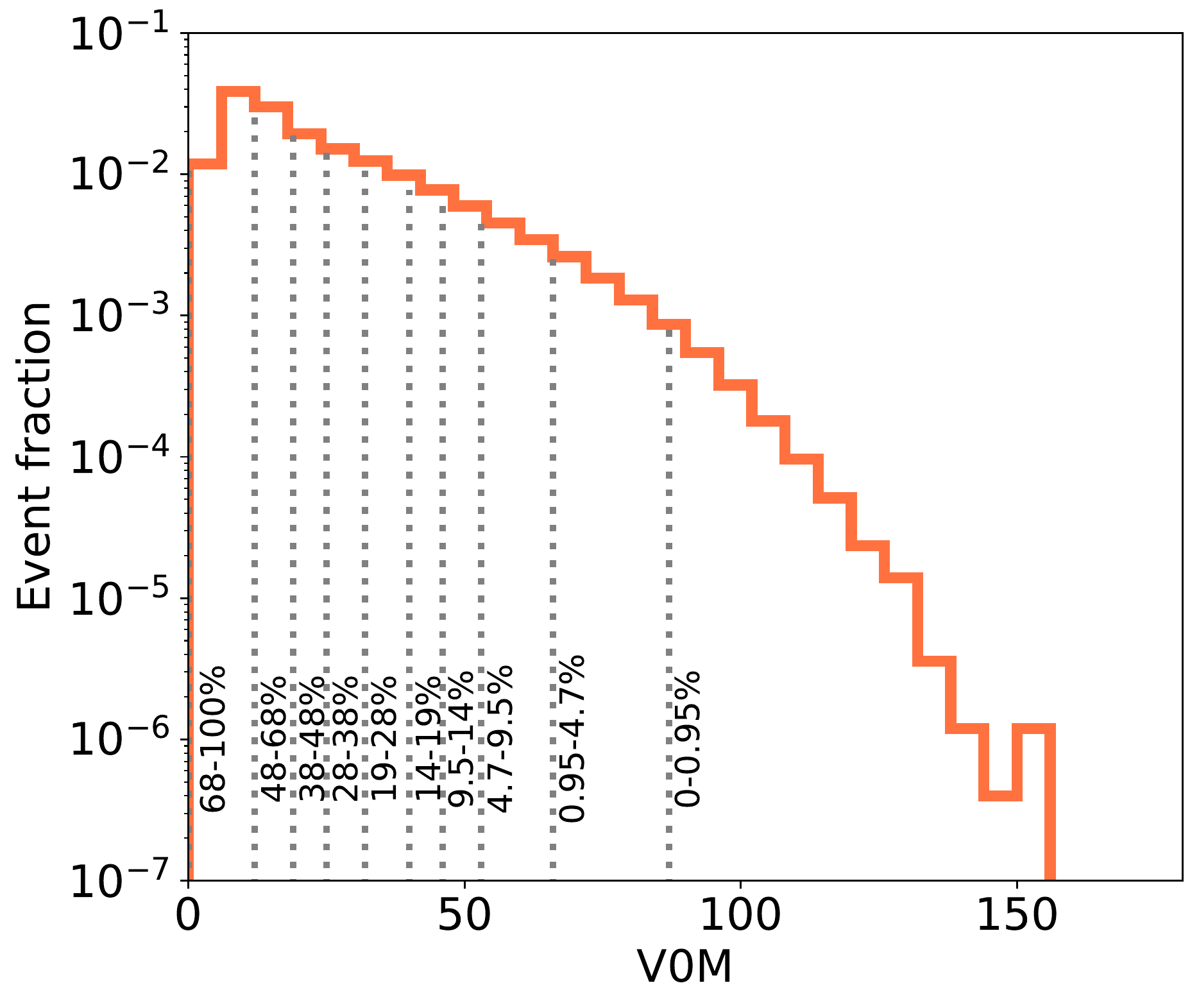}
    \includegraphics[bb=0 0 538 448, width=0.49\textwidth]{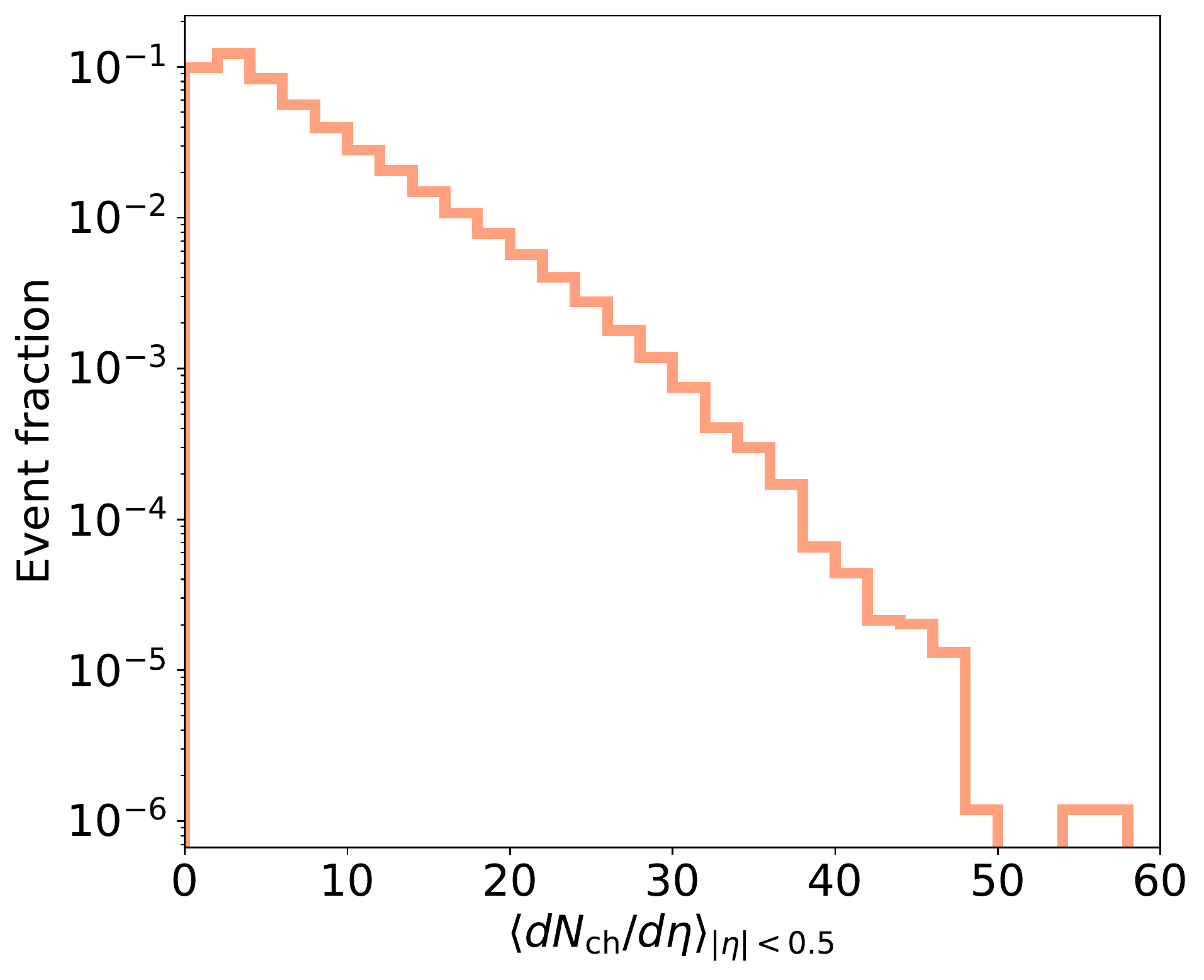}
    \caption{(Left) Normalized event distribution as a function of V0M amplitude in DCCI2. Vertical dashed lines and the percentages show slices of events into each centrality class.
    (Right) Normalized event distribution as a function of multiplicity in DCCI2. 
    Both left and right figures are obtained from $p$+$p$ collisions at \snn[proton]=7 TeV.
    }
    \label{fig:EVENT_DISTRIBUTION_PP7}
\end{figure}

\begin{figure}
    \centering
    \includegraphics[bb=0 0 538 452, width=0.49\textwidth]{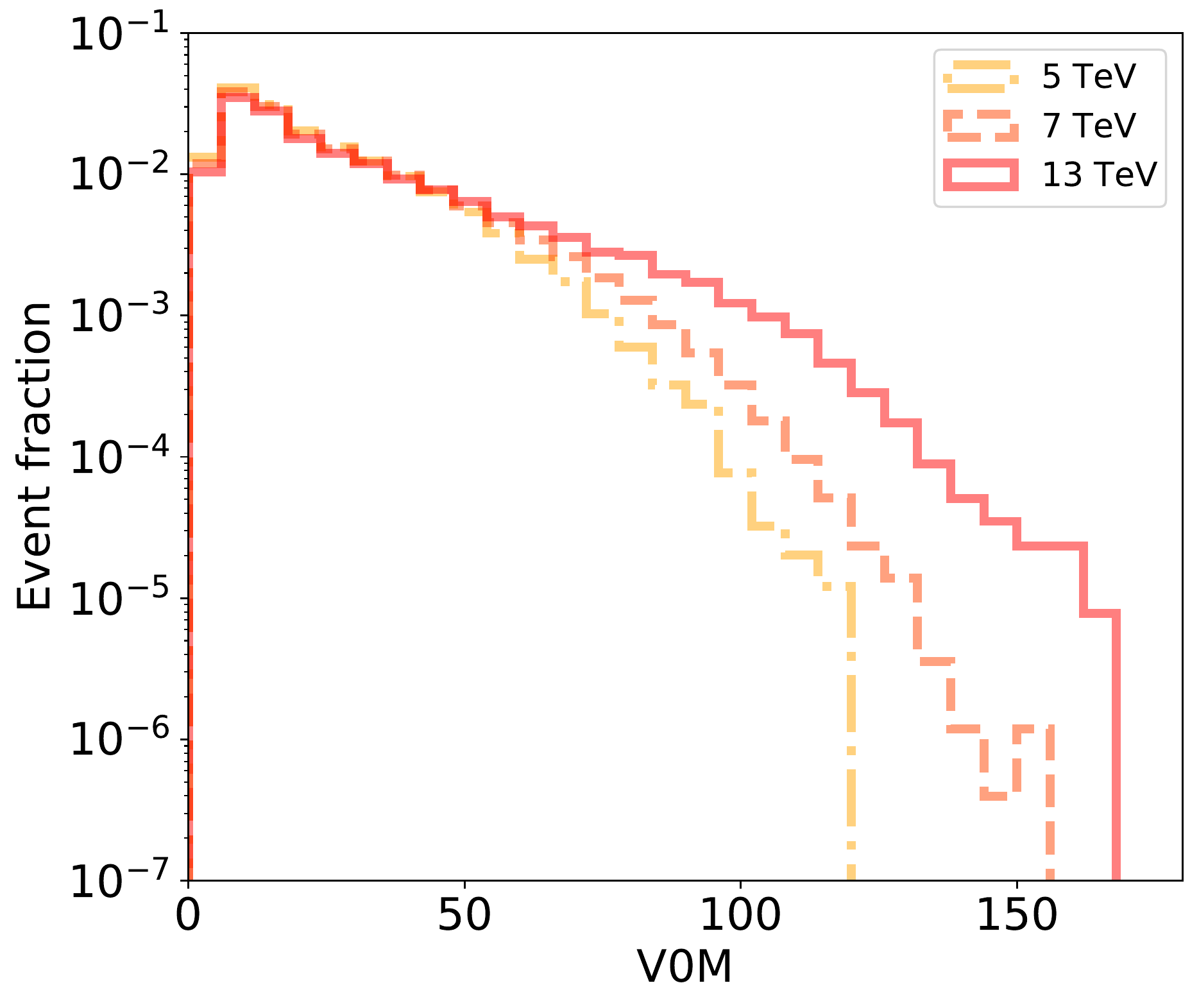}
    \includegraphics[bb=0 0 551 448, width=0.49\textwidth]{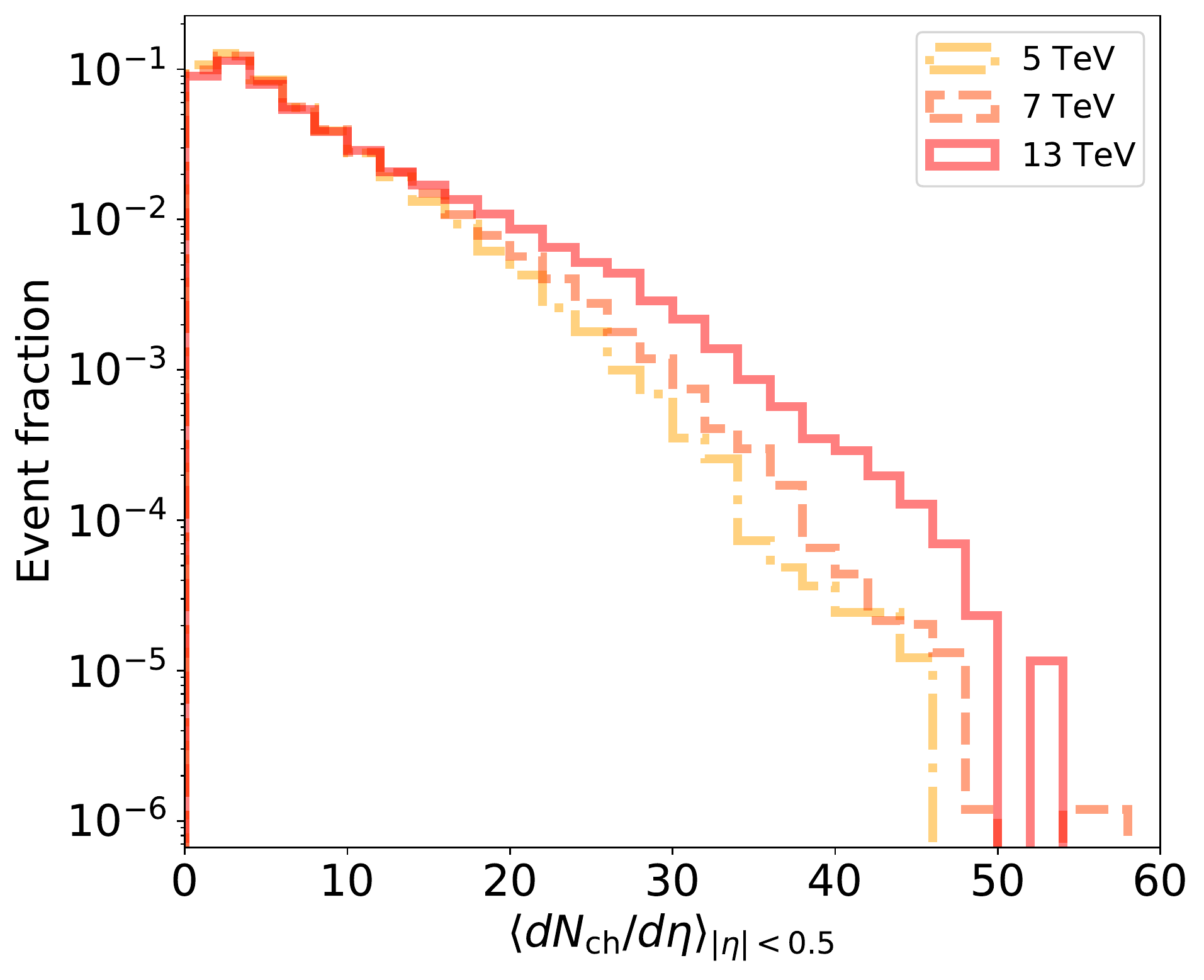}
    \caption{
    (Left) Comparisons of normalized event distributions as a function of V0M amplitude from $p$+$p$ collisions among \snn[proton]=5 (dash-doted lines), 7 (dashed lines), and 13 (solid lines) TeV in DCCI2. 
    (Right) Comparisons of normalized event distribution as a function of multiplicity from $p$+$p$ collisions among \snn[proton]=5 (dash-doted lines), 7 (dashed lines), and 13 (solid lines) TeV in DCCI2.
    }
    \label{fig:EVENT_DISTRIBUTION_PP_ENERGYDEPDENDENCE}
\end{figure}

\begin{figure}
    \centering
    \includegraphics[bb=0 0 559 456, width=0.49\textwidth]{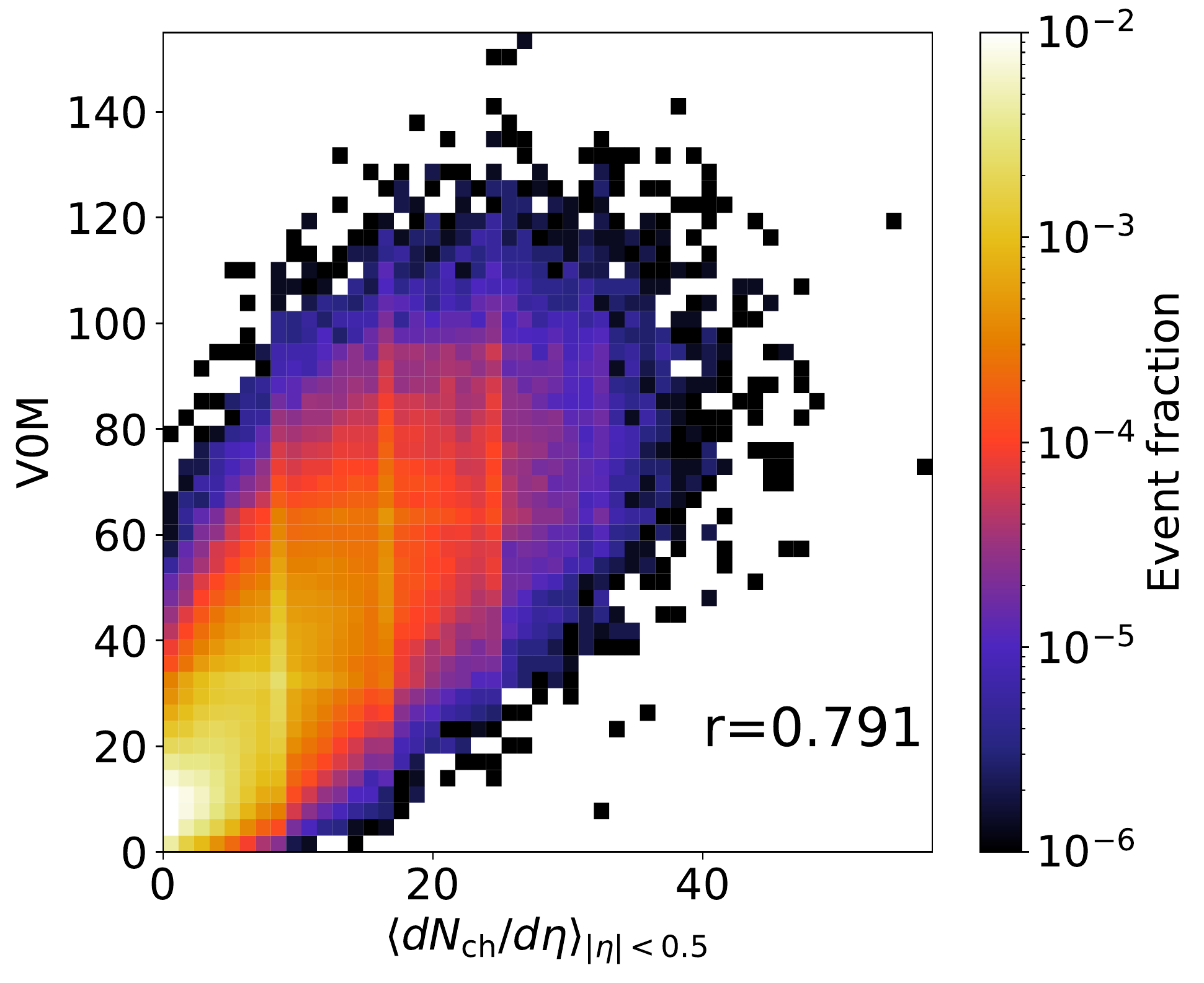}
    \includegraphics[bb=0 0 559 456, width=0.49\textwidth]{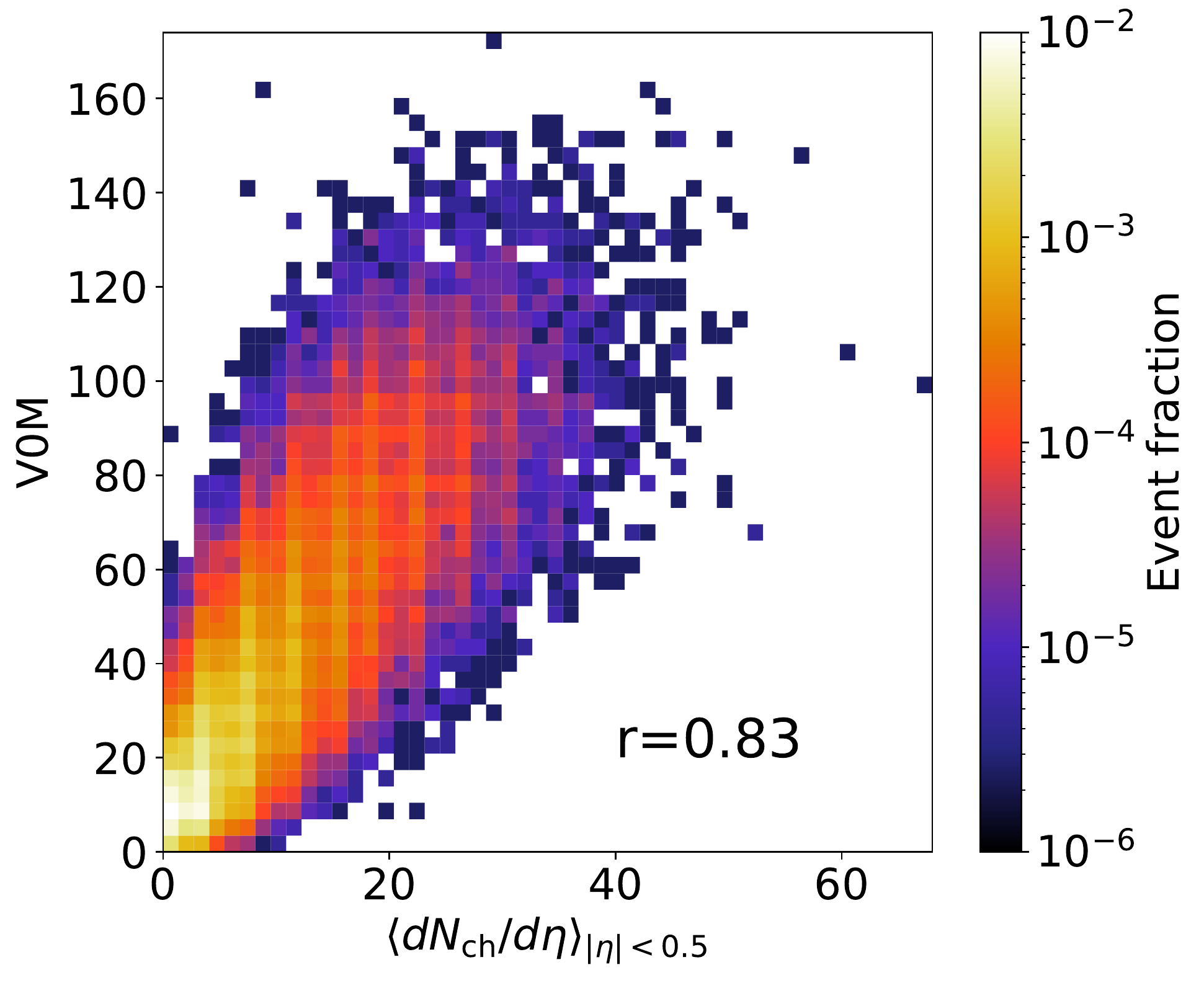}
    \caption{(Left) Correlation between multiplicity and V0M amplitude from $p$+$p$ collisions at \snn[proton]=7 TeV. Correlation coefficient, $r$, is shown in the plot. (Right) Same as the left figure but with $p$+$p$ collisions at \snn[proton]=13 TeV.}
    \label{fig:DNDETA_V0M_CORR_PP}
\end{figure}

\subsection{Event distribution}
As I discussed in Sec.~\ref{subsubsec:MultiplicityDistribution} and \ref{subsec:ParameterDetermination},
it is expected that event classification according to their multiplicity (at forward/backward rapidity)
allows us to explore effects of QGP formation.
In this sense, event distribution of V0M amplitude, which is the multiplicity at forward/backward rapidity used in centrality classification and multiplicity at mid-rapidity, \dndeta, are worth investigating.

%pp=> cross section, PBPB==> impact parameter distribution
%=========================================================
What can we know from event distribution of multiplicity at mid/forward-rapidity?
In $p$+$p$ collisions, it is nothing but a cross section as a function of multiplicity.
%FIGURE EXP
%===============
Figure \ref{fig:EVENT_DISTRIBUTION_PP7} shows normalized event distributions in $p$+$p$ collisions from DCCI2 at \snn[proton]=7 TeV.
Left and right figure show those as functions of V0M amplitude and multiplicity at mid-rapidity, respectively.
For the event distribution of V0M amplitude, 
vertical dashed lines and the percentages are shown to represent slices of events into each centrality class.
As one would notice, reproduction of centrality dependence of multiplicity, such as Fig.~\ref{fig:MULTIPLICITY_PP_PBPB}, requires the reproduction of the shape of event distribution.

Figure \ref{fig:EVENT_DISTRIBUTION_PP_ENERGYDEPDENDENCE} shows normalized distributions in $p$+$p$ collisions from DCCI2 at \snn[proton]= 5, 7, and 13 TeV.
Left and right figures correspond to event distributions of V0M amplitude and multiplicity as the same as Fig.~\ref{fig:EVENT_DISTRIBUTION_PP7}.
In both figures, it is obviously shown that event distribution shows longer tail toward large V0M amplitude and multiplicity regimes with higher collision energy,
which is the sake of multi-parton interactions implemented in \pythia.

Figure \ref{fig:DNDETA_V0M_CORR_PP} shows correlations between V0M amplitude and multiplicity, in other words multiplicity in forward/backward and mid-rapidity.
Corresponding correlation coefficient, $r$, is shown inside of the figure.
Left and right figures show results from $p$+$p$ collisions at \snn[proton]=7 and 13 TeV, respectively.
While there is a positive correlation in both collision energy, the correlation coefficients stays around $r\approx0.8$.
One also sees that the correlation is stronger at higher collision energy.
It should be also noted that event distribution at larger V0M amplitude/multiplicity shows wider dispersion compared to lower regimes.

\begin{figure}
    \centering
    \includegraphics[bb=0 0 611 518, width=0.49\textwidth]{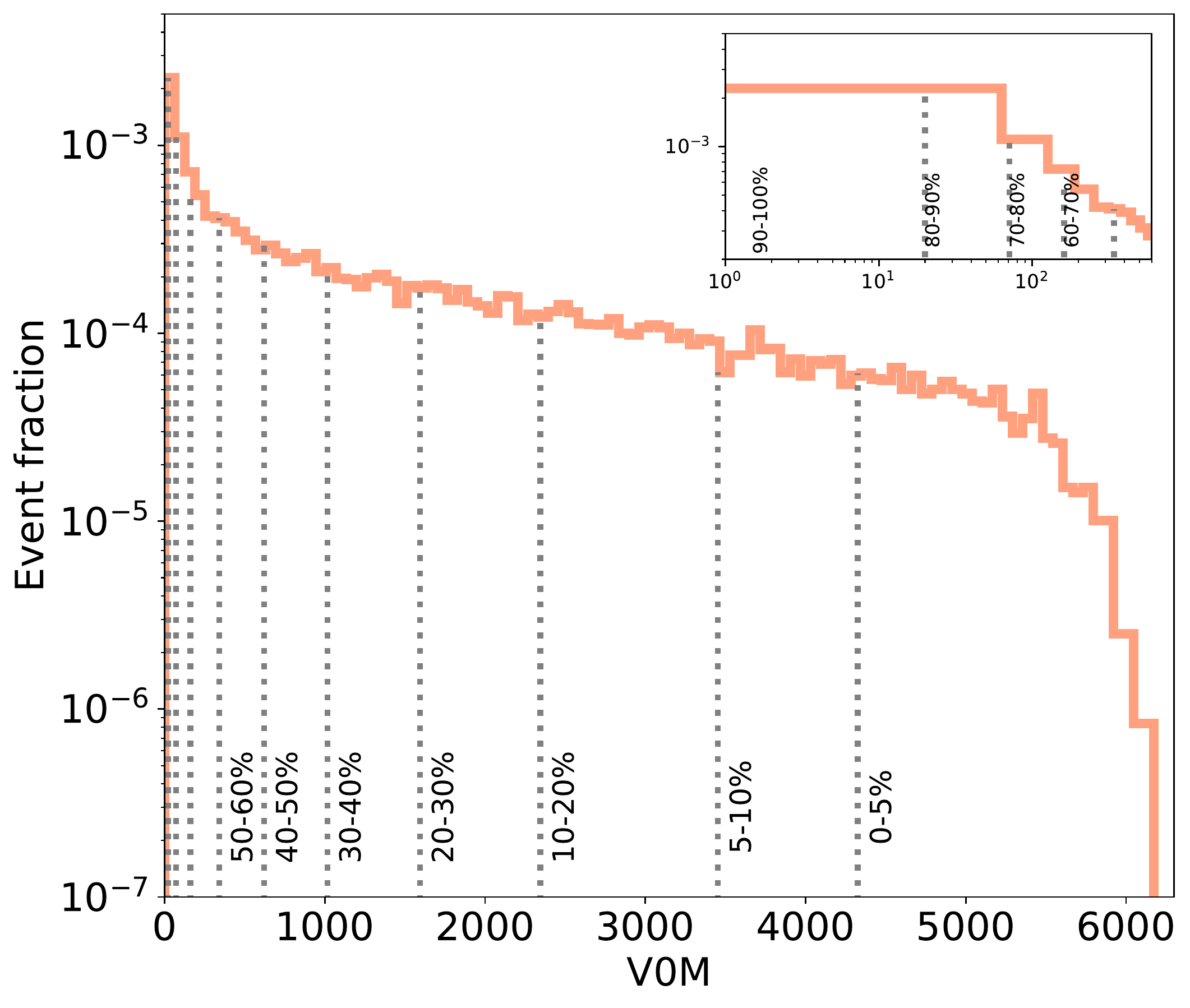}
    \includegraphics[bb=0 0 538 448, width=0.49\textwidth]{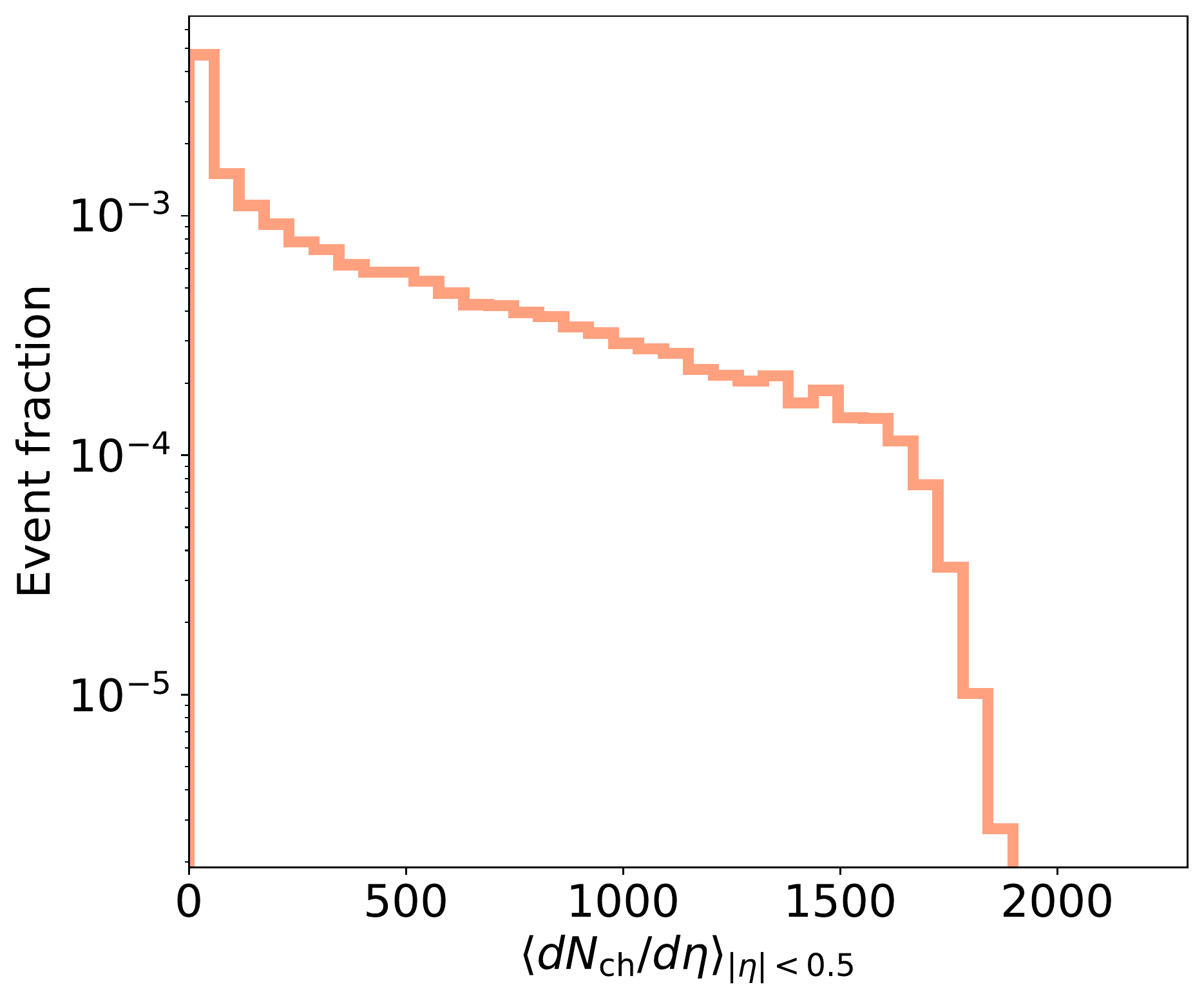}
    \caption{(Left) Normalized event distribution as a function of V0M amplitude in DCCI2. Vertical dashed lines and the percentages show slices of events into each centrality class. Inner plot at the top right shows blow-up at small V0M ranges with double log axes.
    (Right) Normalized event distribution as a function of multiplicity in DCCI2. 
    Both left and right figures are obtained from $Pb$+$Pb$ collisions at \snn=2.76 TeV}
    \label{fig:EVENT_DISTRIBUTION_PBPB}
\end{figure}

\begin{figure}
    \centering
    \includegraphics[bb=0 0 571 456, width=0.5\textwidth]{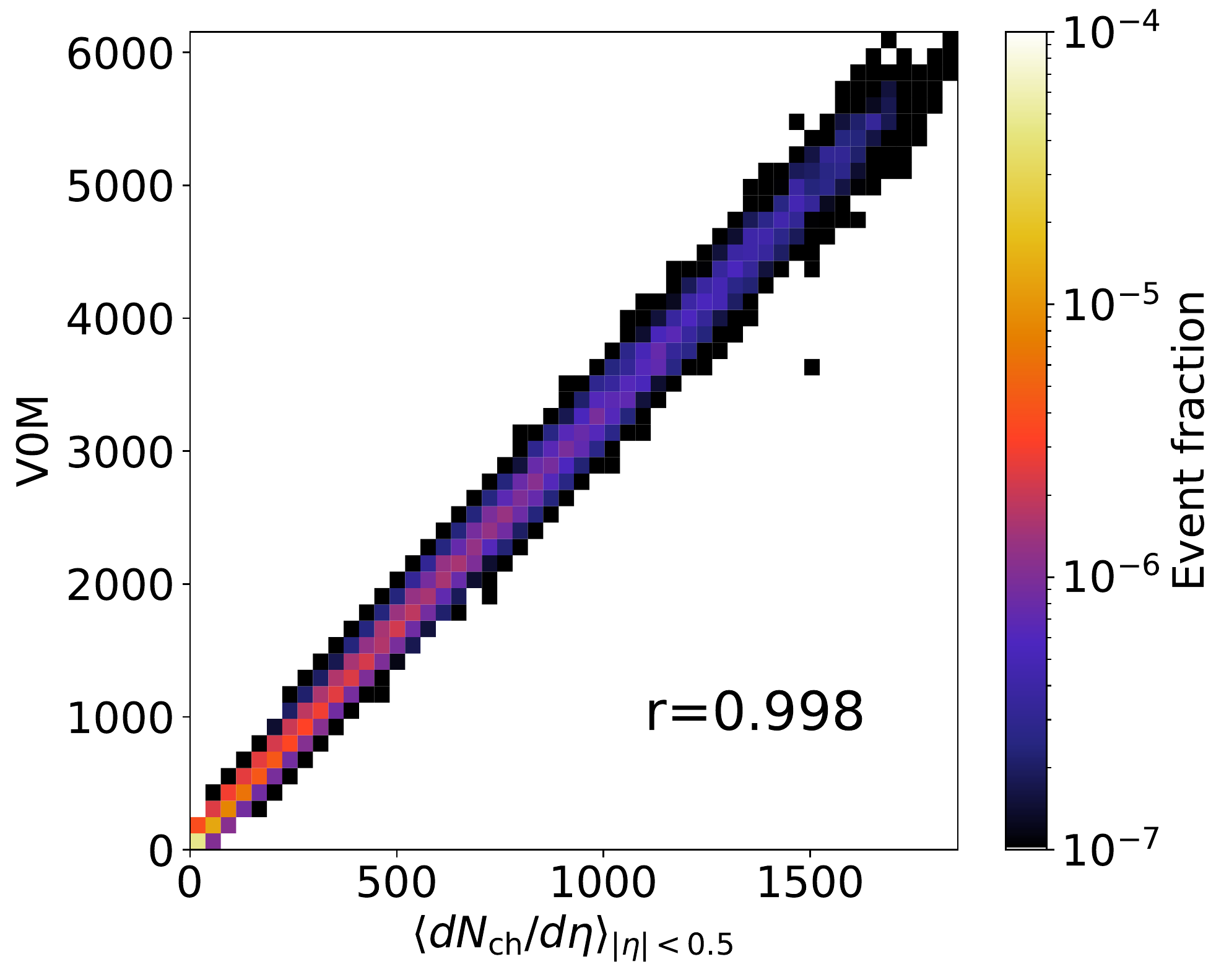}
    \caption{Correlation between multiplicity and V0M amplitude from $Pb$+$Pb$ collisions at \snn=2.76 TeV. Correlation coefficient, $r$, is shown in the plot.}
    \label{fig:DNDETA_V0M_CORR_PBPB}
\end{figure}

%Event distribution in $Pb$+$Pb$ collisions reflect impact parameter distribution
%because multiplicity is more driven by impact parameter of each event.
Figure \ref{fig:EVENT_DISTRIBUTION_PBPB} shows 
normalized event distributions in $Pb$+$Pb$ collisions from DCCI2 at \snn=2.76 TeV.
Left and right figures show those as functions of V0M amplitude and multiplicity at mid-rapidity, respectively.
For the left figure, vertical dashed lines and the percentages show slices of events into each centrality class. An inset plot at the top right shows blow-up at small V0M ranges with double-log axes.
One sees that the shape of event distributions is quite different from that in $p$+$p$ collisions: the distribution has a clear shoulder at large V0M amplitude and multiplicity.
%As a result, the dispersion of event distribution at large V0M amplitude and multiplicity is smaller in $Pb$+$Pb$ collisions.
Figure \ref{fig:DNDETA_V0M_CORR_PBPB} shows 
the resultant correlation coefficients between V0M amplitude and multiplicity, which turned to be strongly and positively correlated ($r\approx0.99$).

\section{Momentum distributions}
\label{sec:MomentumDistrubutions}

It is also interesting to compare the sizes of core and corona contributions in transverse momentum $p_T$ spectra of the final state particles. 
In this section, let me start from showing overall tendency in DCCI2 results.
According to the implementation of the core--corona picture by
Eq.~\refbra{eq:four-momentum-deposition}, initial low momentum partons are likely to generate QGP fluids and expected to contribute largely in the low-$p_T$ region. 
Meanwhile, high momentum particles are likely to traverse vacuum or fluids as mostly keeping their initial momentum and supposed to dominate the high-$p_T$ region. 

\subsection{Overall tendency}
\label{subsec:OverallTendency}
\begin{figure*}
\includegraphics[bb=0 0 630 597, width=0.5\textwidth]{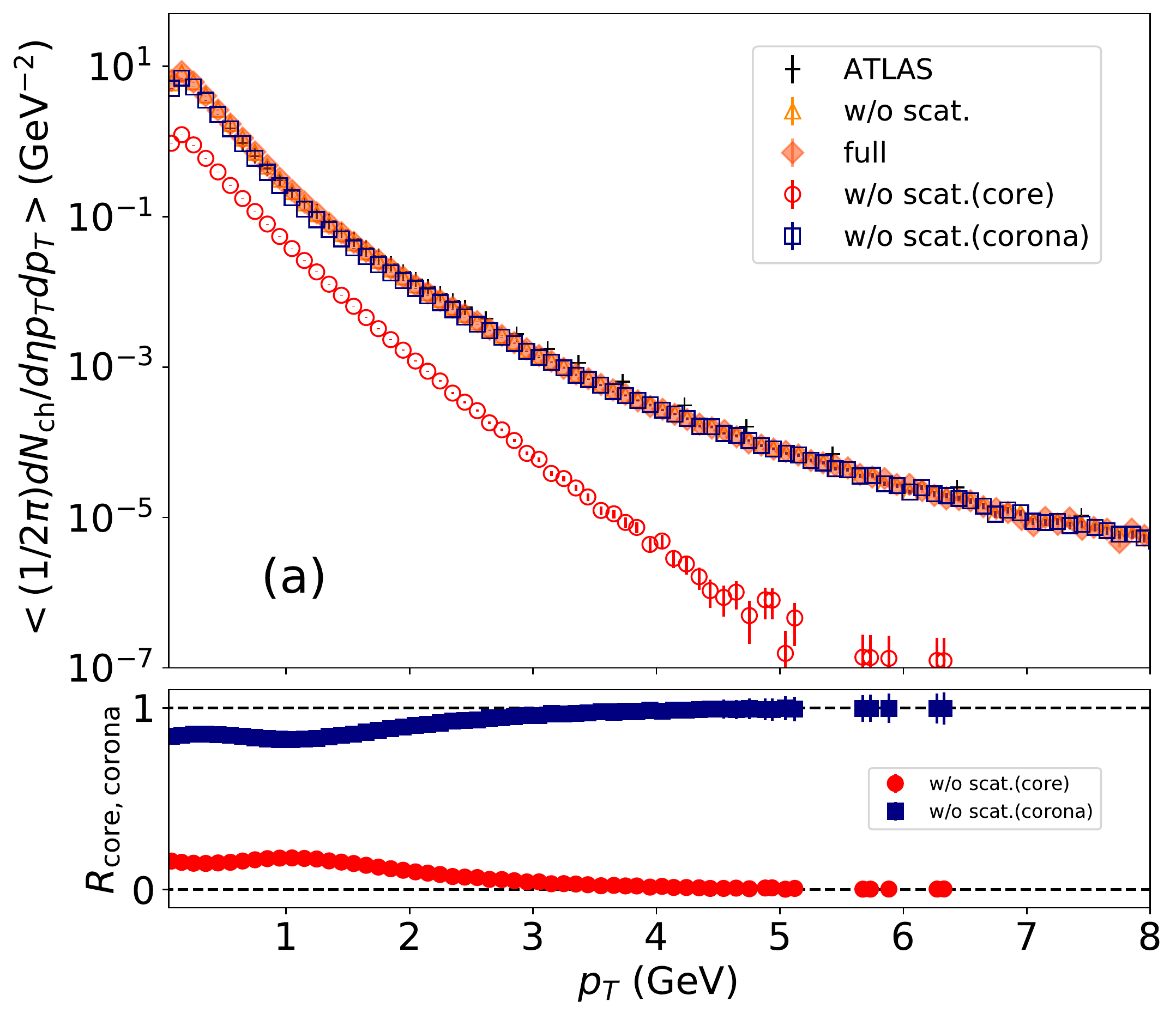}
\includegraphics[bb=0 0 630 597, width=0.5\textwidth]{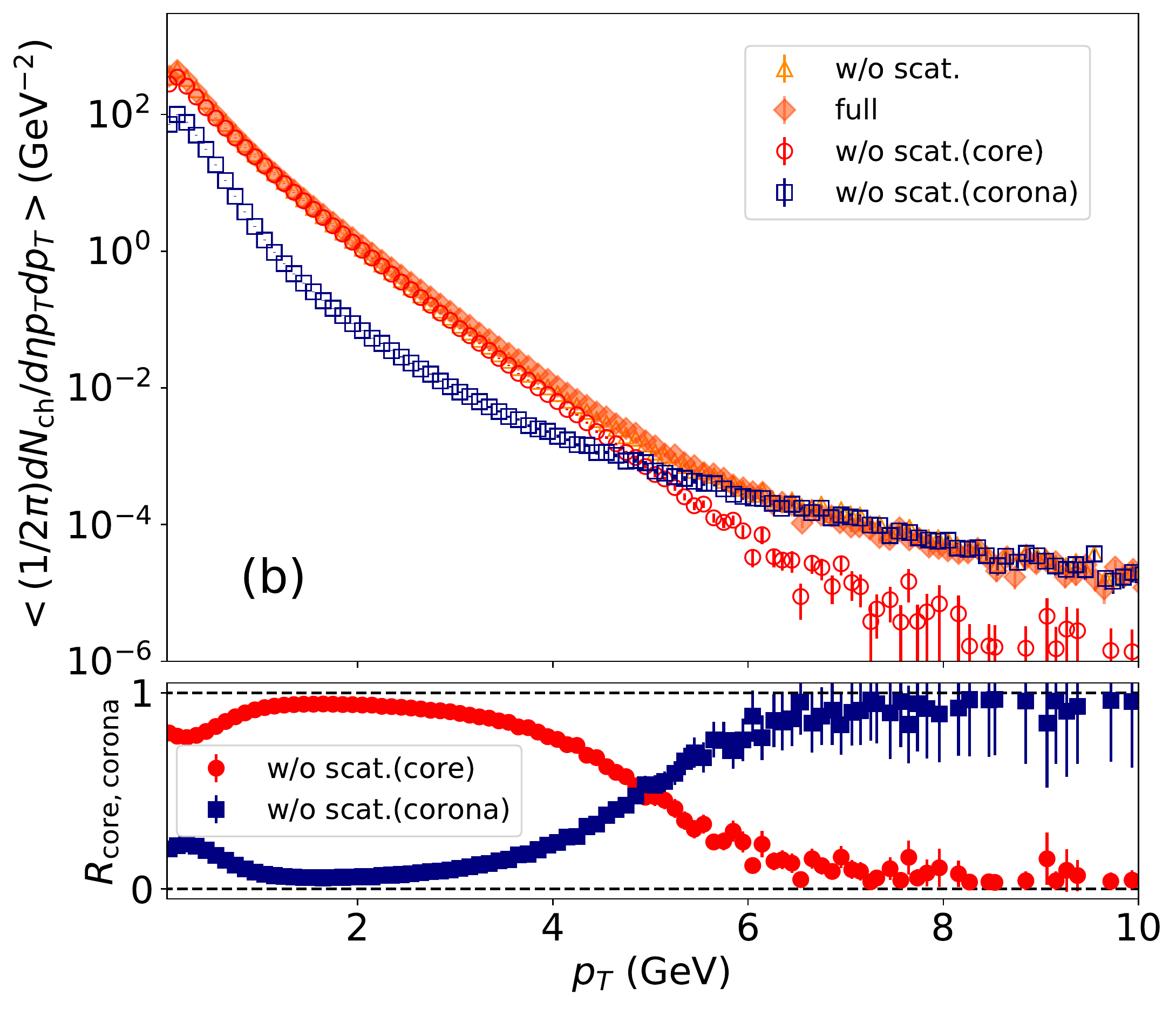}
\caption{(Color Online) (Upper) Transverse momentum spectra of charged particles at midrapidity from DCCI2 in (a) $p$+$p$ collisions at $\sqrt{s} = 7$ TeV  and (b) $Pb$+$Pb$ collisions at \snn = $2.76$ TeV.
An event average is taken with at least one charged particle having $p_T>0.5$ GeV and $|\eta|<2.5$ in both cases.
Results from full simulations (closed orange diamonds) and simulations without hadronic rescatterings (open orange triangles) are plotted and compared with the ATLAS data (black pluses) \cite{Aad:2010ac} only in $p$+$p$ collisions as a reference.
Results from core (open red diamonds) and from corona (open blue squares) are also plotted for simulations without hadronic rescatterings.
(Lower) Corresponding fractions of core (red circles) and corona (blue squares) components in the final hadron without hadronic rescatterings are shown as functions of transverse momentum. 
}
\label{fig:PTSPECTRA_PP_PBPB}
\end{figure*}
\begin{figure*}
\includegraphics[bb=0 0 556 454, width=0.5\textwidth]{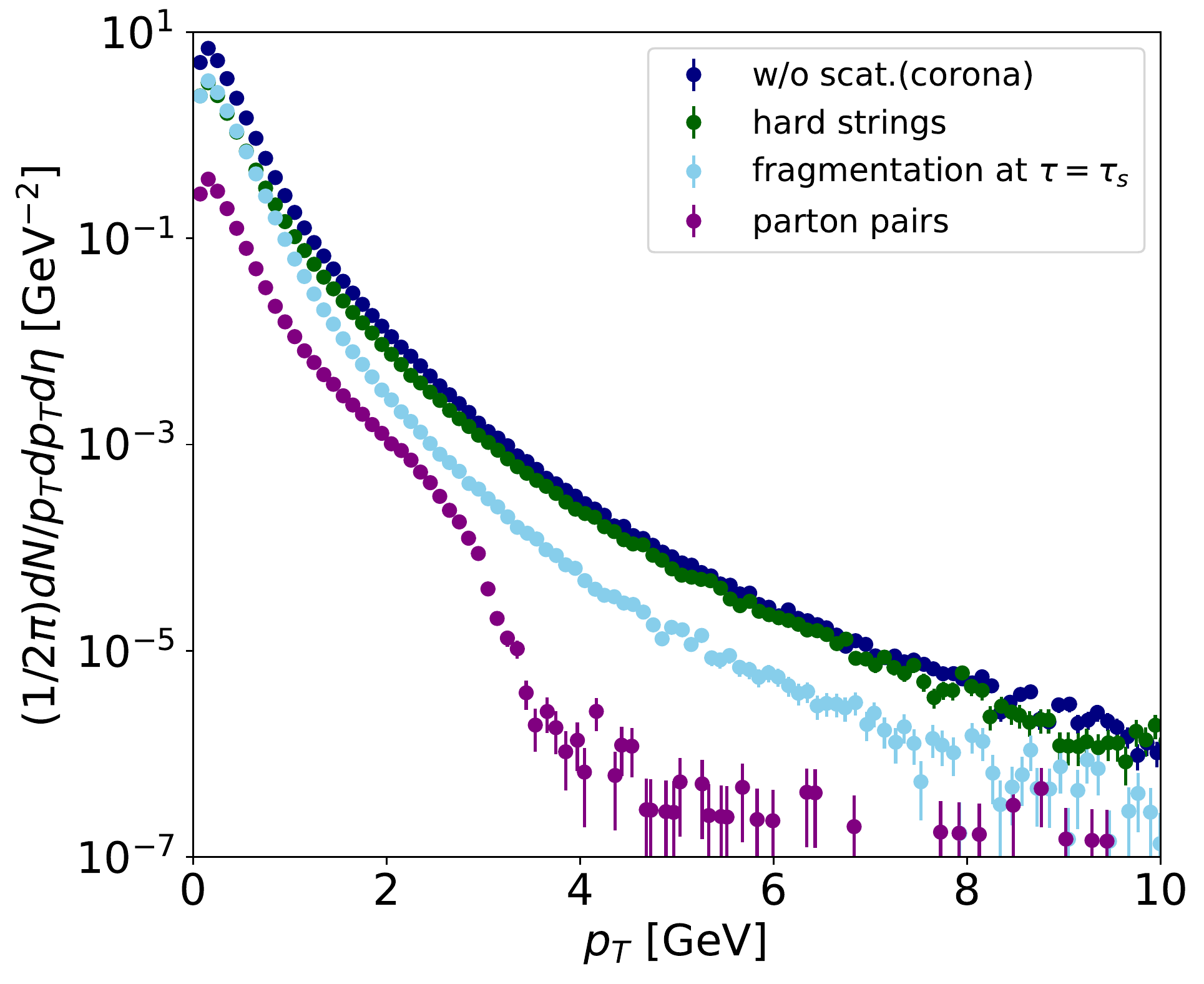}
\includegraphics[bb=0 0 556 454, width=0.5\textwidth]{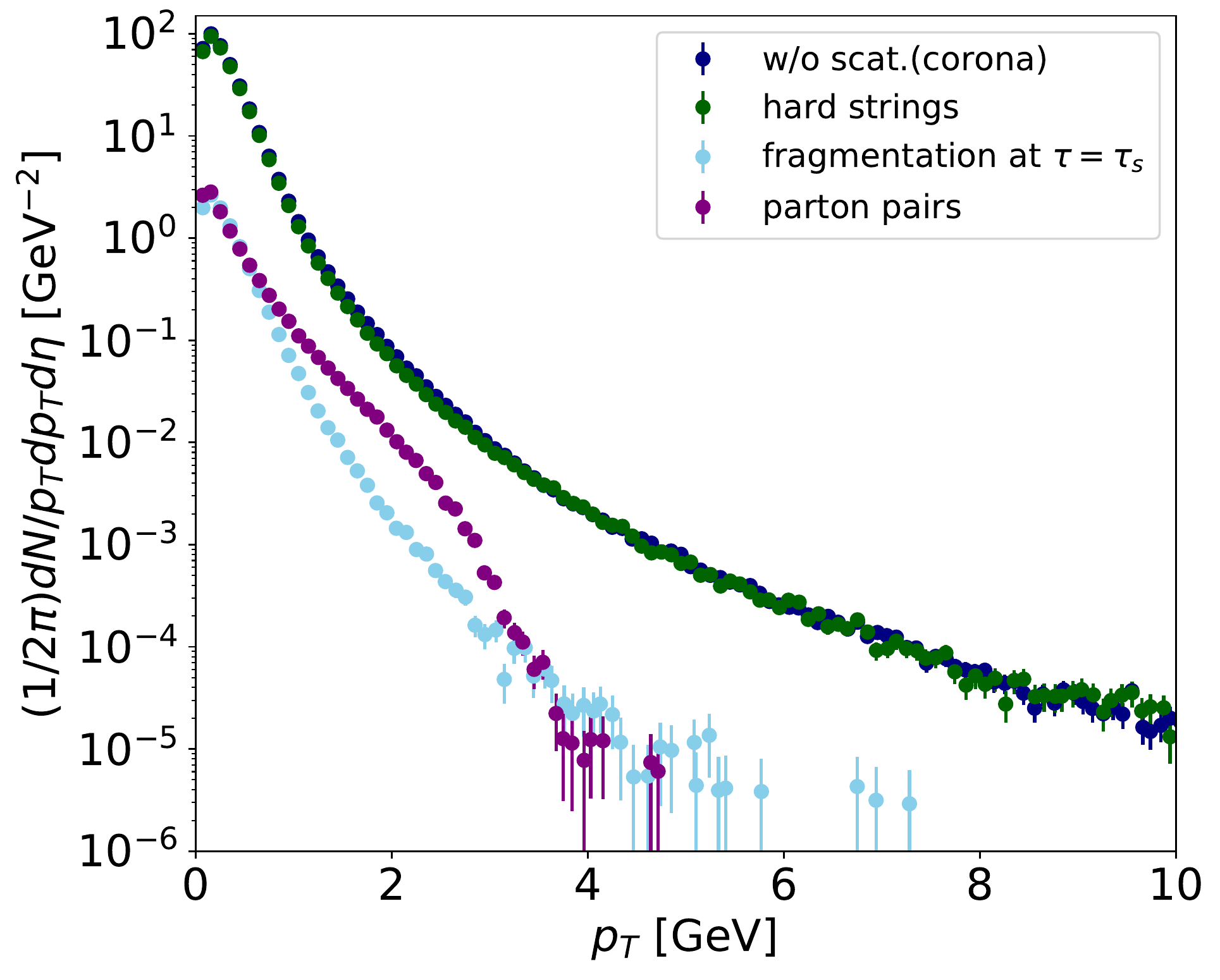}
\caption{(Color Online) Breakdown of transverse momentum spectra of charged particles from corona components at midrapidity in (a) $p$+$p$ collisions at $\sqrt{s} = 7$ TeV  and (b) $Pb$+$Pb$ collisions at \snn = $2.76$ TeV.
An event average is taken with at least one charged particle having $p_T>0.5$ GeV and $|\eta|<2.5$ in both cases.
The breakdown is shown by categorizing the corona contributions into 
(i) strings consisting of ``hard'' partons and coming out from fluids at $\tau > \tau_s$ (green), (ii) strings outside the fluids at $\tau=\tau_{s}$ including ones with ``hard'' partons outside the fluids as a whole and ones partially outside the fluids being cut at $\tau=\tau_s$ (sky blue),
and (iii) parton pairs coming out at $\tau > \tau_{s}$ (purple).
}
\label{fig:PTSPECTRA_corona_PP_PBPB}
\end{figure*}

Upper panels of Fig.~\ref{fig:PTSPECTRA_PP_PBPB} show the charged particle $p_T$ spectra at midrapidity $|\eta|<2.5$ in (a) $p$+$p$ and (b) $Pb$+$Pb$ collisions.
The kinematic cuts and event selections are the same as the ones used in the ATLAS experimental results \cite{Aad:2010ac}.
Event averages are taken with at least one charged particle having $p_T>0.5$ GeV and $|\eta|<2.5$ in both $p$+$p$ and $Pb$+$Pb$ collisions, which can be regarded as almost minimum-bias events.
Again, a comparison between results obtained from full simulations and ones from simulations without hadronic rescatterings is made here.
Each contribution from core and corona components to the final hadrons from simulations without hadronic rescatterings is shown as well.
In both $p$+$p$ and $Pb$+$Pb$ collisions, the $p_T$ spectra of final hadrons without hadronic rescatterings are represented as sums of contributions of
core and corona components over the whole $p_T$ regions.
One also sees that the effect of hadronic rescatterings on the $p_T$ spectra of charged particles is almost absent in both $p$+$p$ and $Pb$+$Pb$ collisions.
Since the charged particles are mainly composed of charged pions, their $p_T$ spectra are relatively insensitive to hadronic rescatterings.

The corresponding lower panels of Fig.~\ref{fig:PTSPECTRA_PP_PBPB} show the fractions of core and corona components for final hadrons without hadronic rescatterings as functions of $p_T$.
As an overall tendency, the dominance of the corona components at high $p_T$ regions is seen in both $p$+$p$ and $Pb$+$Pb$ collisions, which is exactly what I expect from the core--corona picture in the momentum space encoded in Eq.~\refbra{eq:four-momentum-deposition}.
In $p$+$p$ collisions, the contribution from the core components reaches $R_{\mathrm{core}} \approx 0.3$ around $p_T \approx 1.0$-$1.5$ GeV
and the contribution from the corona components is almost dominant 
over the whole $p_T$ range. 
On the other hand, the contribution from core components is dominant in low $p_T$ regions in $Pb$+$Pb$ collisions, 
while the dominant contribution is flipped to the corona components at $p_T \approx 5.5$ GeV towards high $p_T$ regions. 
Remarkably, 
only within $0.7 \lesssim p_{T} \lesssim 3.6$ GeV,
the core components highly dominate, $R_{\mathrm{core}} \gtrsim 0.9$.
The existence of corona components should be considered below $\approx 0.7$ GeV and above $\approx3.6$ GeV even in minimum-bias events.

In particular, there is a small peak in the fraction of corona components with $R_{\mathrm{corona}} \approx 0.2$ at most in $p_T\lesssim1$ GeV.
This contribution originates mainly from a feed-down from fragmentation of strings including surviving partons during the dynamical initialization stage.
This is a consequence of the dynamical core--corona initialization against initially generated partons. Thus there should be a kind of ``redshift" of the $p_T$ spectrum due to energy loss of traversing partons which contribute as corona components in the soft region.
%As we emphasized in Introduction, 
This result exactly illustrates ``soft-from-corona" that there exists a non-negligible contribution of non-equilibrated corona components in low $p_{T}$ region.
Therefore, 
in order to properly extract transport coefficients of the QGP fluids from, for example, an analysis of flow observables, hydrodynamic results should be corrected with corona components.

Figure~\ref{fig:PTSPECTRA_corona_PP_PBPB} shows breakdowns of the corona contributions obtained in Fig.~\ref{fig:PTSPECTRA_PP_PBPB} to reveal which types of partons contribute to the production at the low $p_T$. The kinematic cuts and event selection are the same as used in Fig.~\ref{fig:PTSPECTRA_PP_PBPB}.
Here, I categorize the corona contribution into three types of contributions based on the modification of color strings explained in Sec.~\ref{subsec:COLORSTRING_TREATMENT_TAU0}:
(i) strings consisting of ``hard'' partons and coming out from fluids at $\tau > \tau_s$, (ii) strings outside the fluids at $\tau=\tau_{s}$ including ones with ``hard'' partons outside the fluids as a whole and ones partially outside the fluids being cut at $\tau=\tau_s$,
and (iii) parton pairs coming out at $\tau > \tau_{s}$.
The total corona contributions, which are identical to the ones plotted in Fig.~\ref{fig:PTSPECTRA_PP_PBPB}, are also plotted for a comparison.
Figure~\ref{fig:PTSPECTRA_corona_PP_PBPB} (a) shows breakdowns in $p$+$p$ collisions.
One sees that the contribution from the ``hard'' partons at $\tau> \tau_s$ (Category (i)) is dominant for the all $p_T$ range while the contribution from strings fragmented at $\tau=\tau_s$ (Category (ii)) becomes more comparable at the low $p_T$ compared to the high $p_T$.
One notices that the contribution from parton pairs (Category (iii)) is much smaller than the total corona contributions
The ``cliff" structure of (iii) at $p_T \approx 3 $ GeV originates from the parameter $p_{T, \mathrm{cut}} (=3.0)$ GeV since surviving partons to pick up a thermal partons must have $p_T<p_{T, \mathrm{cut}}$.
Figure~\ref{fig:PTSPECTRA_corona_PP_PBPB} (b) shows breakdowns in $Pb$+$Pb$ collisions.
Similar to the results in $p$+$p$ collisions, the contribution from ``hard'' partons at $\tau> \tau_s$ (Category (i)) is dominant for the all $p_T$ range.
The rest two contributions (Categories (ii) and (iii)) contribute much less to the total corona contributions.
Therefore, it can be concluded that the corona 
contribution at the low $p_T$ originates from the hadronic contribution from 
strings consisting of ``hard'' partons and coming out from fluids at $\tau > \tau_s$.

\subsection{Centrality dependence of identified particle $p_T$ spectra}
\label{subsec:IdentifiedParticlePTSpectra}
Now that the overall trends have been seen, I discuss the centrality dependence and particle species dependence for further discussion.
%Why Centrality dependence ?
%==========================
I discussed in Sec.~\ref{section:p_TintegratedObservables} that centrality dependence can shed a light on effects of existence of core components.
The more central events become, the more core contributions appear in a system.
Thus, I already know that the absolute values of $p_T$ spectrum of each core and corona component would change,
and which means that the momentum range that core/corona component dominates would change too.
Also, it can be expected that there is centrality dependence of slopes of core spectra due to generated flows in radial directions,
which adds another non-triviality on interplay of core and corona components.

%Why PID?
%==========================
In addition to centrality cut, 
particle identification brings us further information of a system too.
In heavy-ion collisions, it has been understood that
if particles are thermally produced from QGP fluids with a certain
velocity, heavier particles acquire more momentum and are pushed towards high $p_T$.
Thus, within DCCI2, it can be expected that the momentum range that core/corona component dominates would be different among different particle species.

\subsubsection{Core/corona fraction in $p_T$ directions}

\begin{figure}
    \centering
    \includegraphics[bb = 0 0 628 542, width=0.45\textwidth]{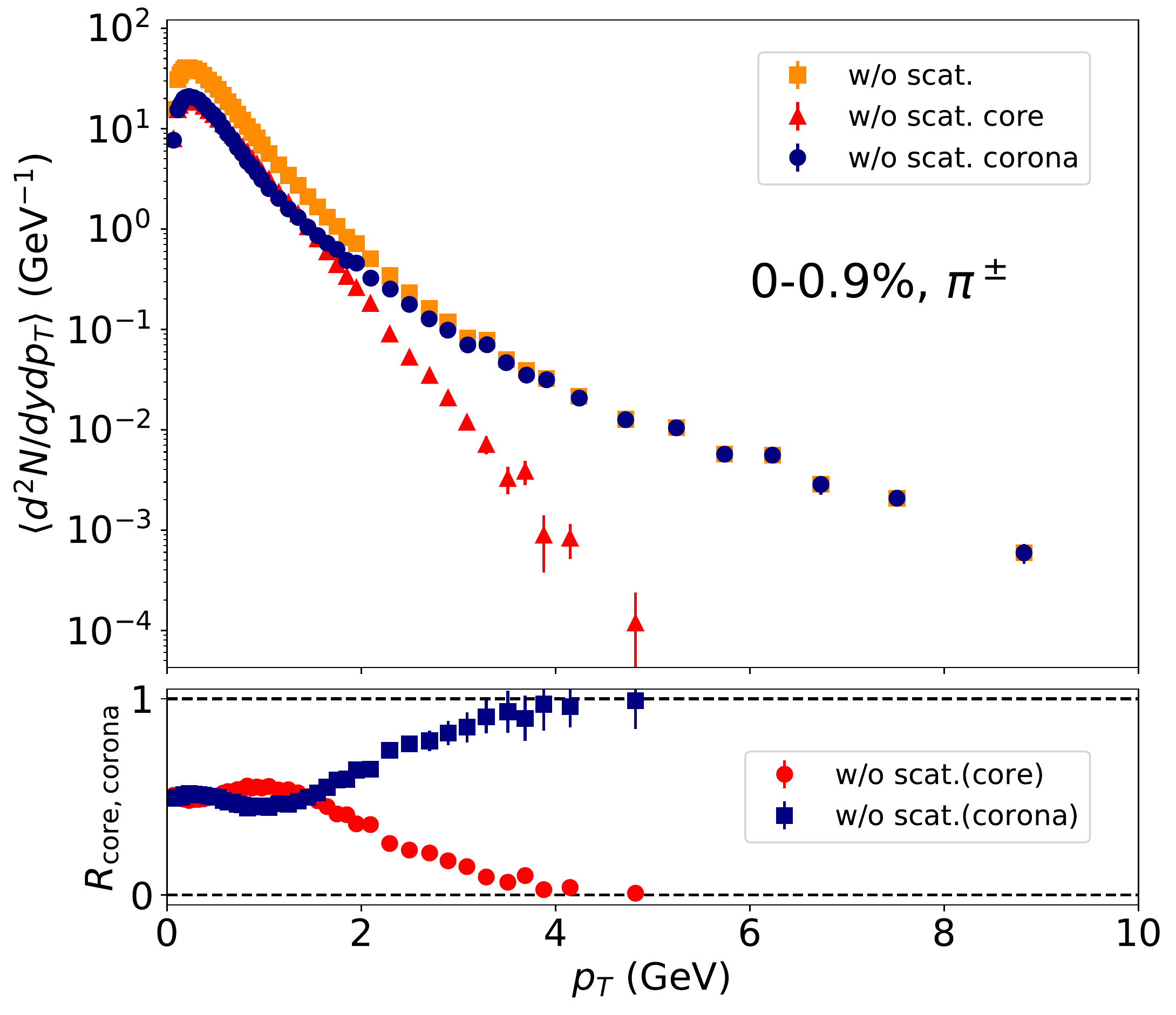}
    \includegraphics[bb = 0 0 628 542, width=0.45\textwidth]{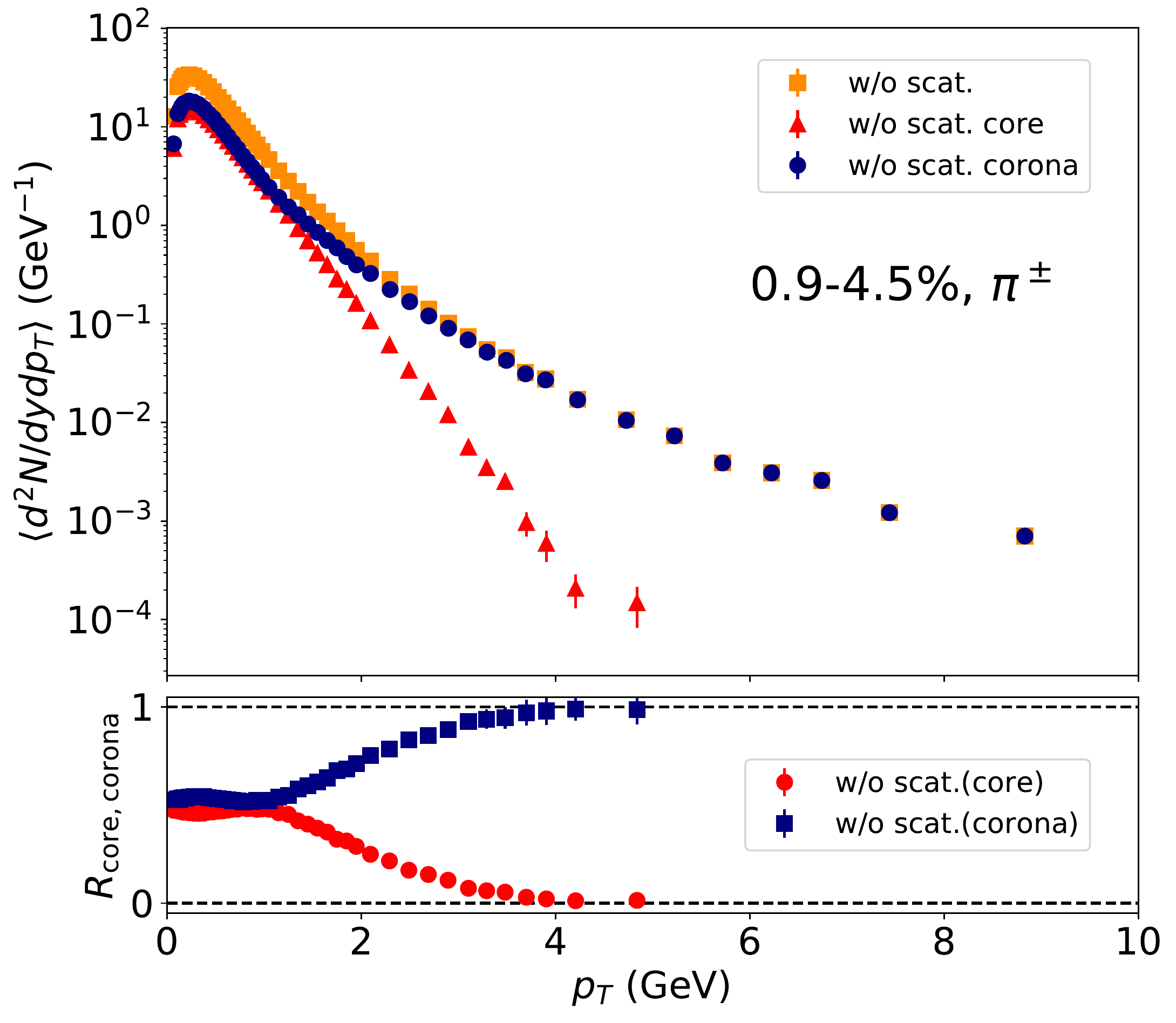}
    \includegraphics[bb = 0 0 628 542, width=0.45\textwidth]{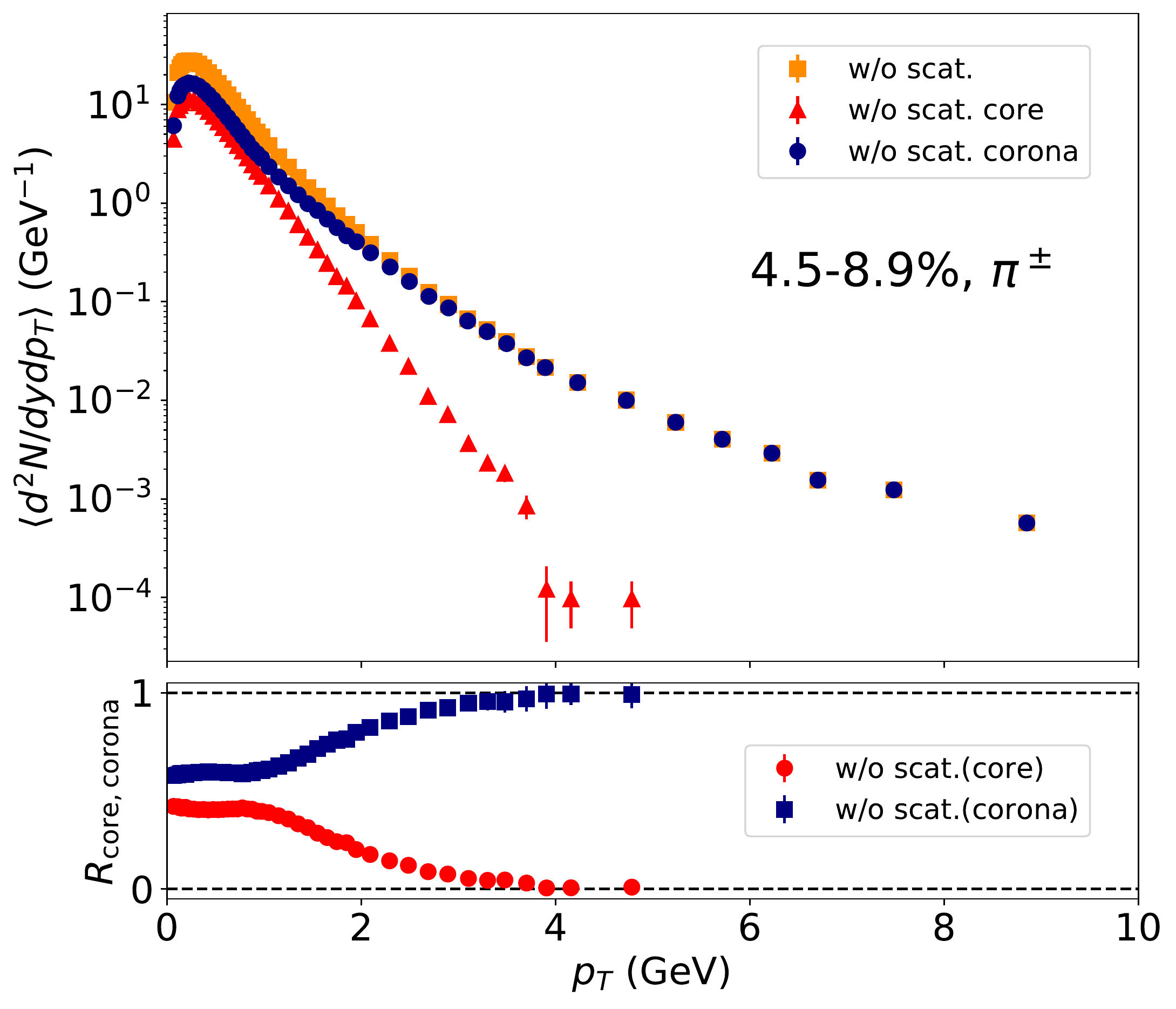}
    \includegraphics[bb = 0 0 628 542, width=0.45\textwidth]{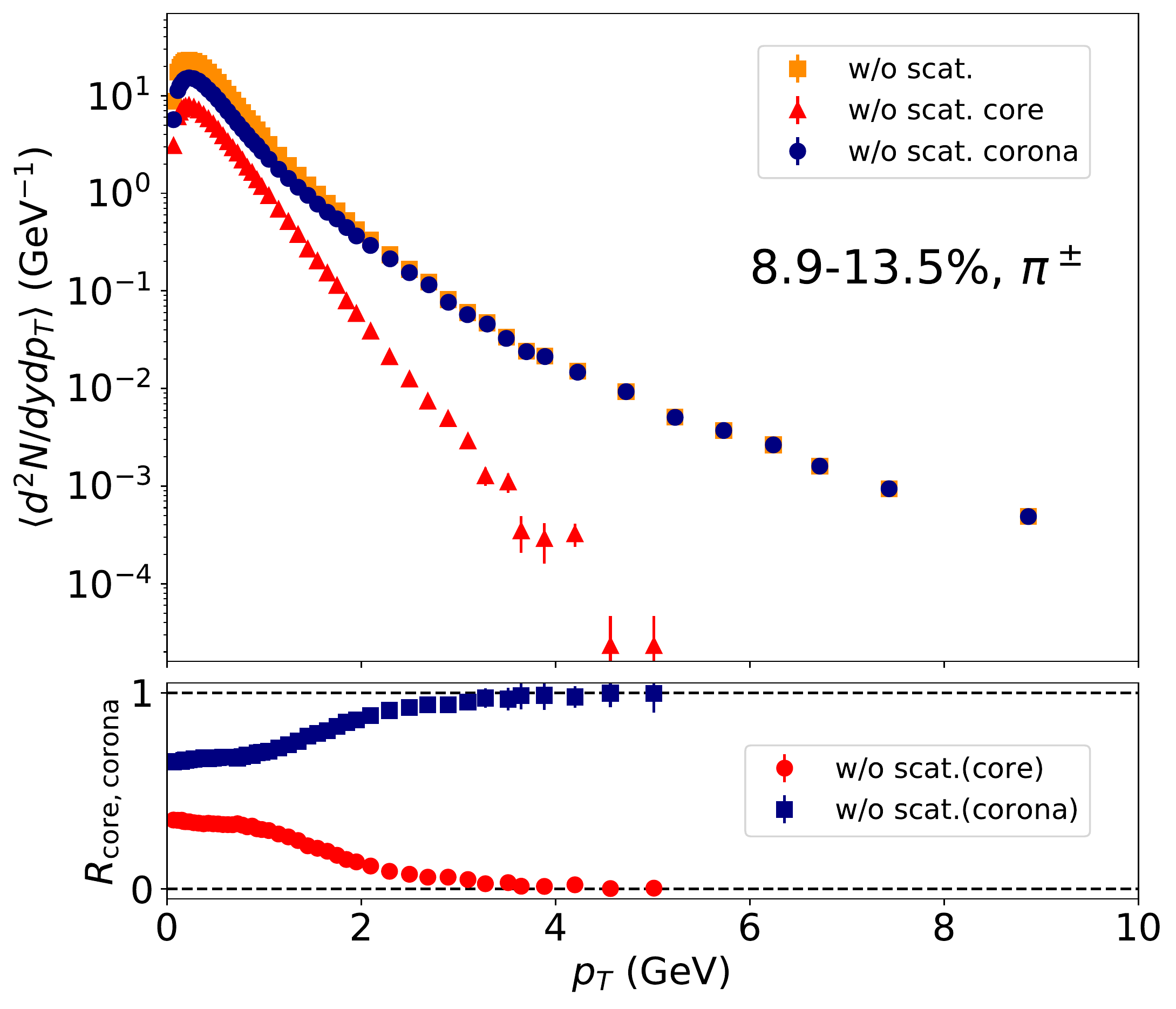}
    \includegraphics[bb = 0 0 628 542, width=0.45\textwidth]{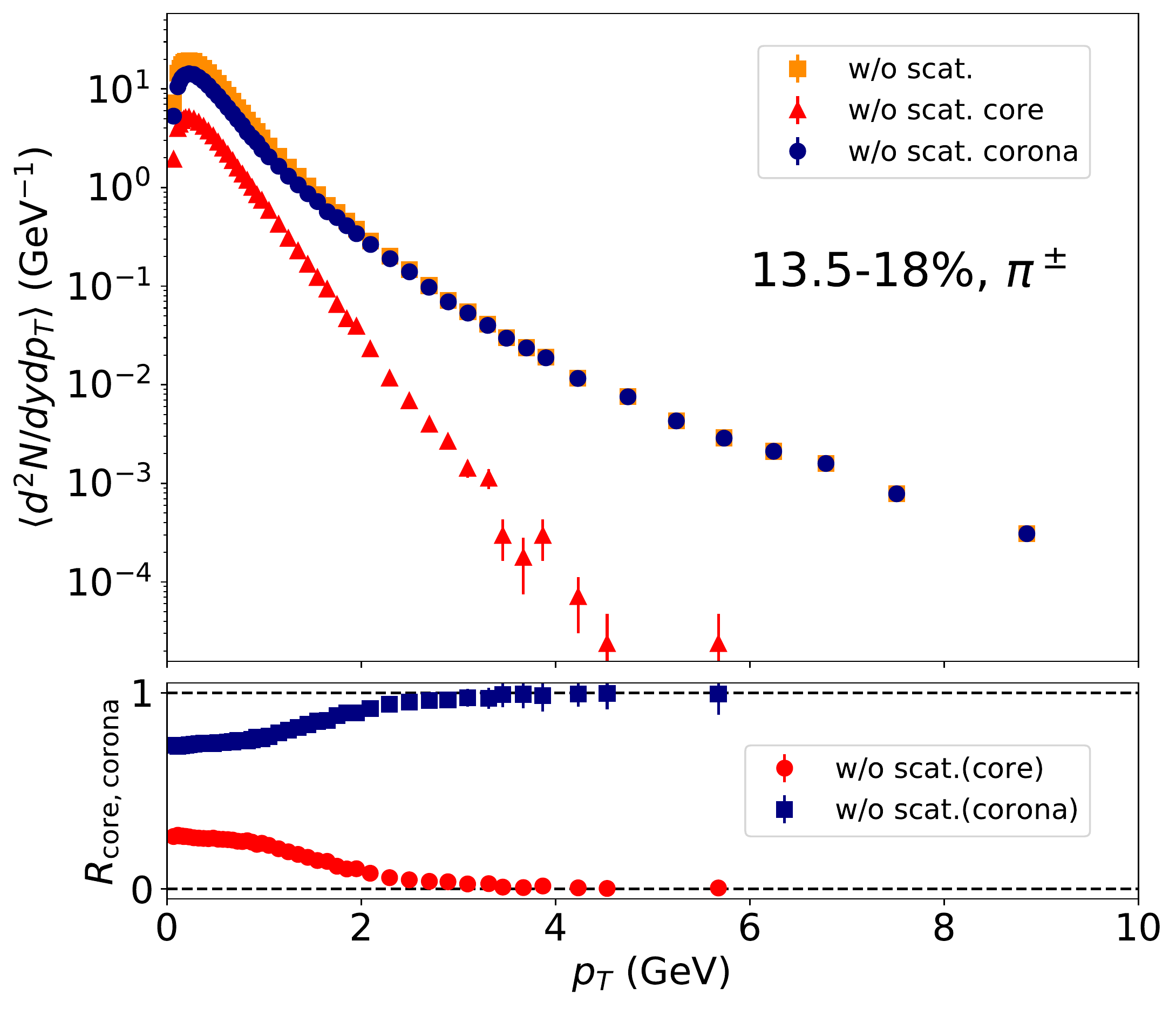}
    \includegraphics[bb = 0 0 628 542, width=0.45\textwidth]{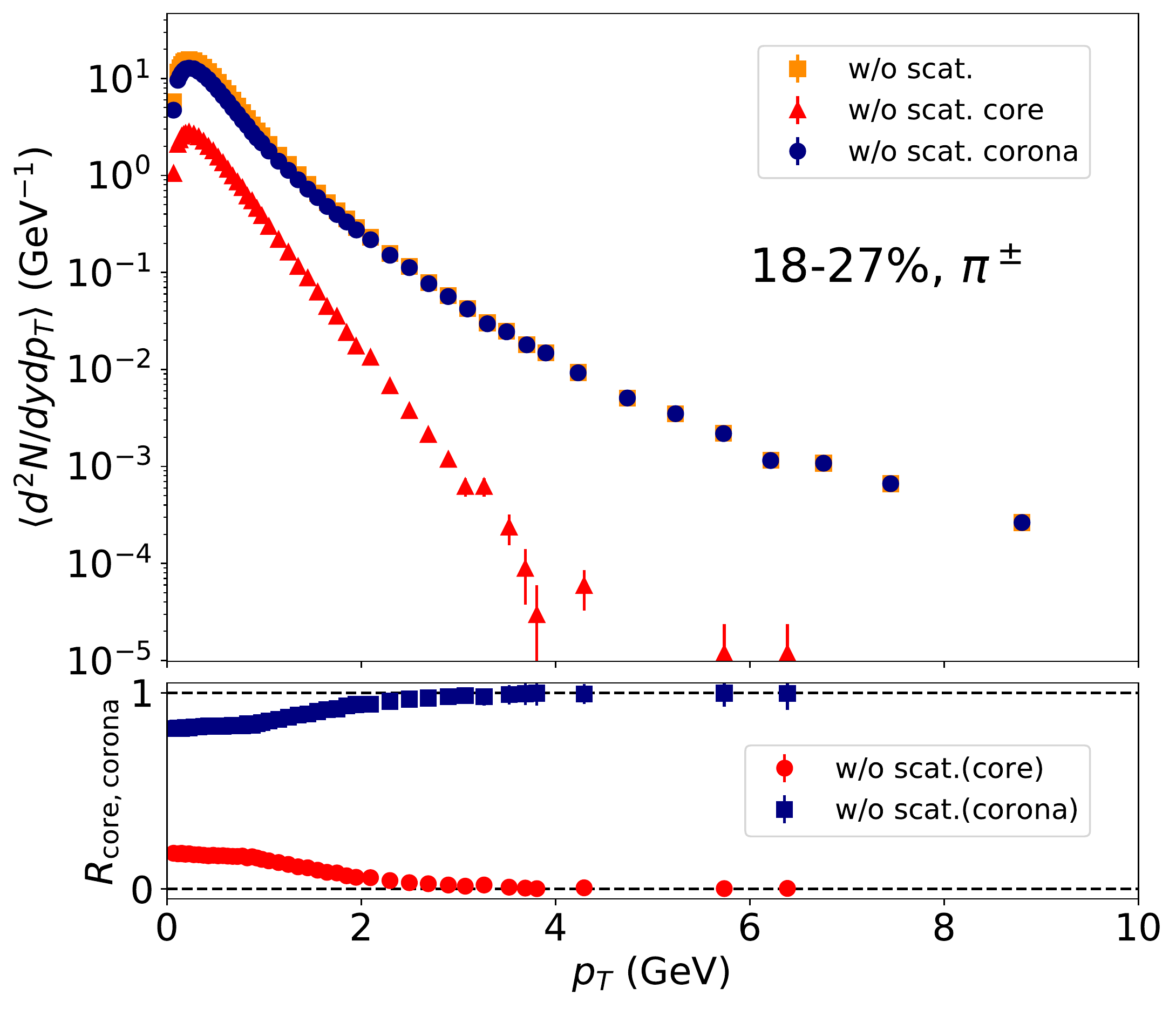}
    \caption{(Upper) Centrality dependence of $p_T$ spectra of charged pions ($\pi^+ + \pi^-$) in $p$+$p$ collisions at \snn[proton] = 13 TeV from DCCI2.
    Results with switching off hadronic rescatterings (orange squares) and their breakdown into core (red triangles) and corona contributions (blue circles) are shown.
    (Lower) Fraction of core (red circles) and corona (blue squares) components, $R_{\mathrm{core, corona}}$, in each $p_T$ bin. Centrality classes from 0-0.9\% to 18-27\% are shown.
    }
    \label{fig:PP13_PTSPECTRA_PI_CORECORONA_1}
\end{figure}

\begin{figure}
    \centering
    \includegraphics[bb = 0 0 628 542, width=0.45\textwidth]{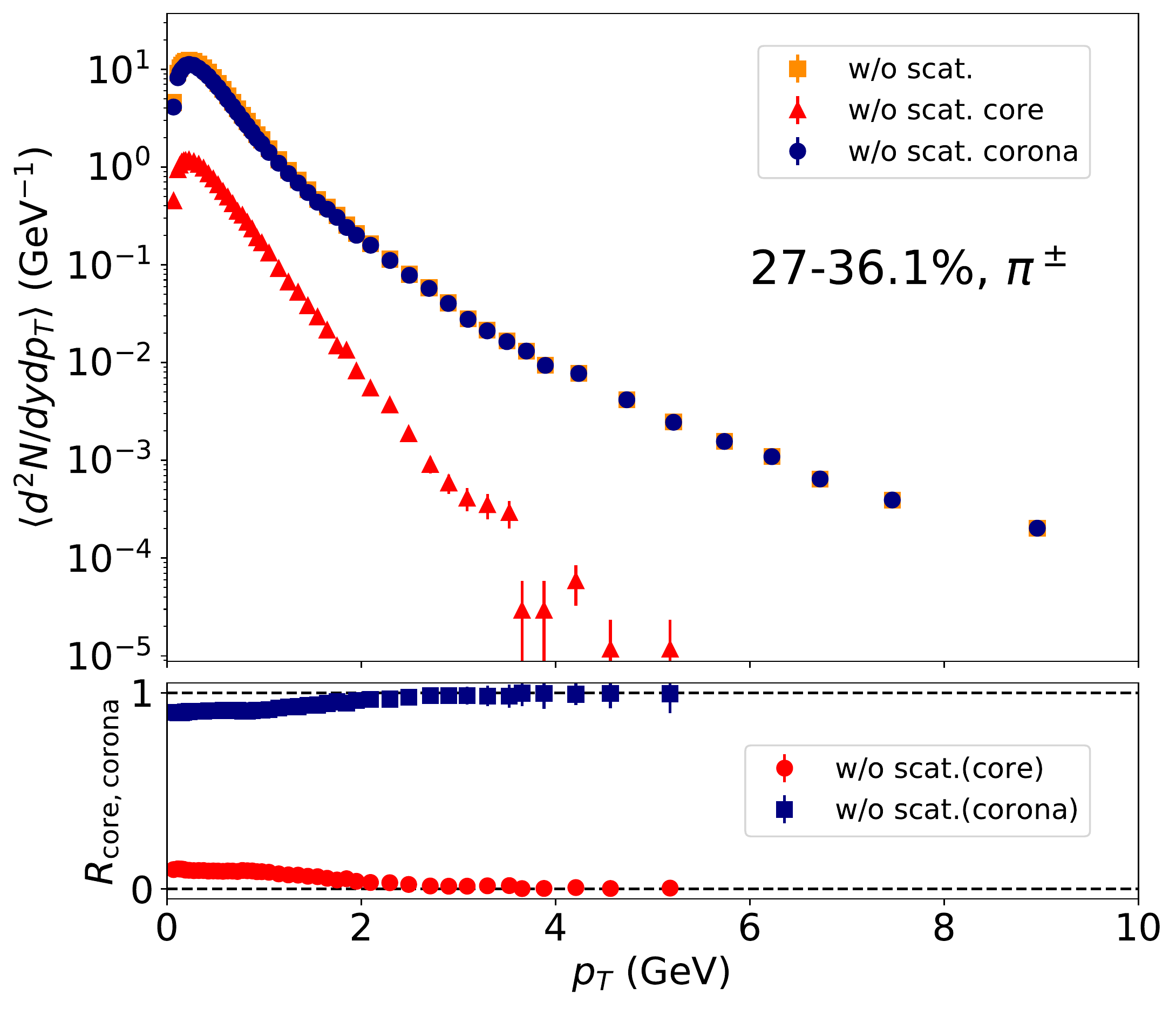}
    \includegraphics[bb = 0 0 628 542, width=0.45\textwidth]{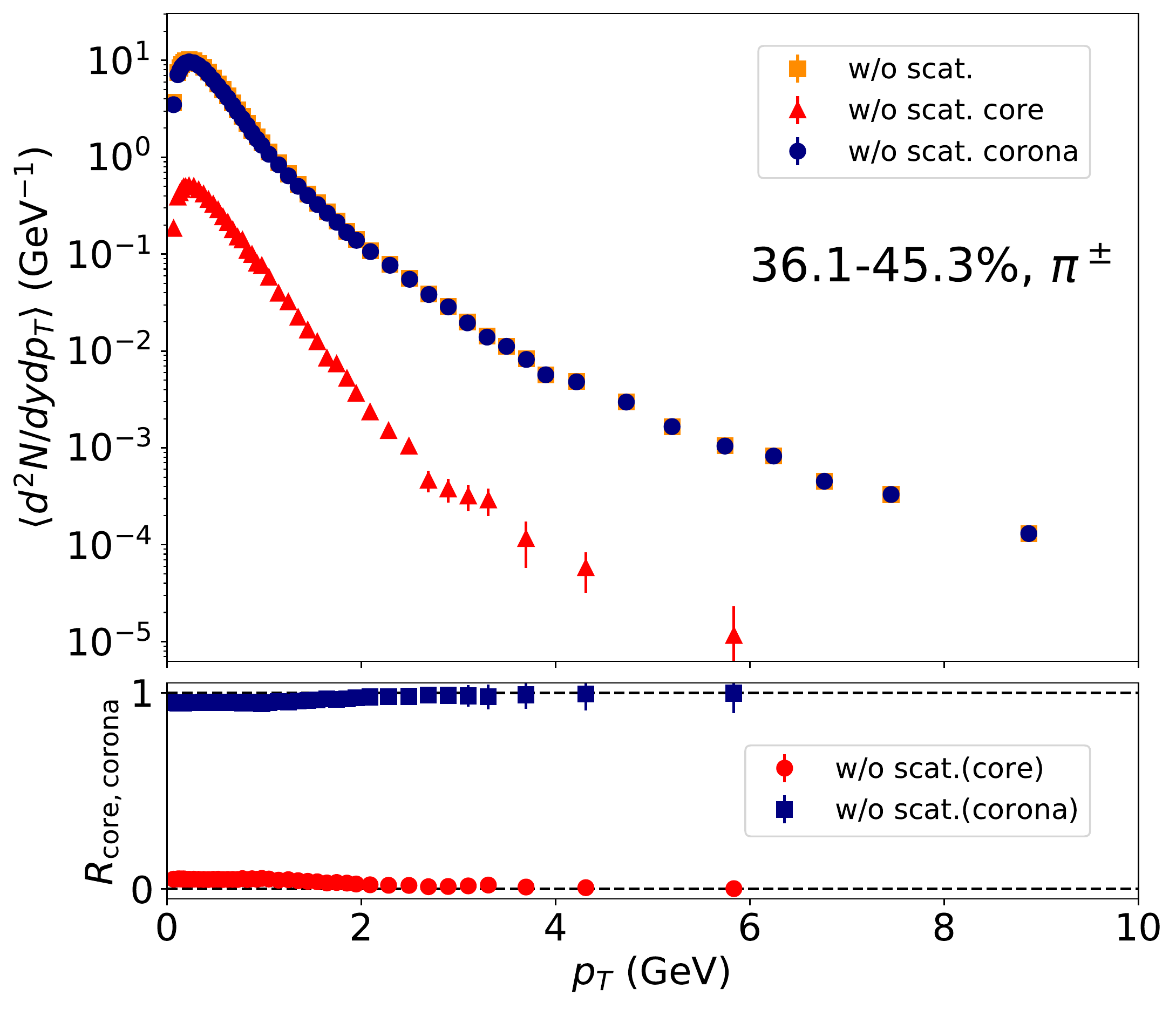}
    \includegraphics[bb = 0 0 628 542, width=0.45\textwidth]{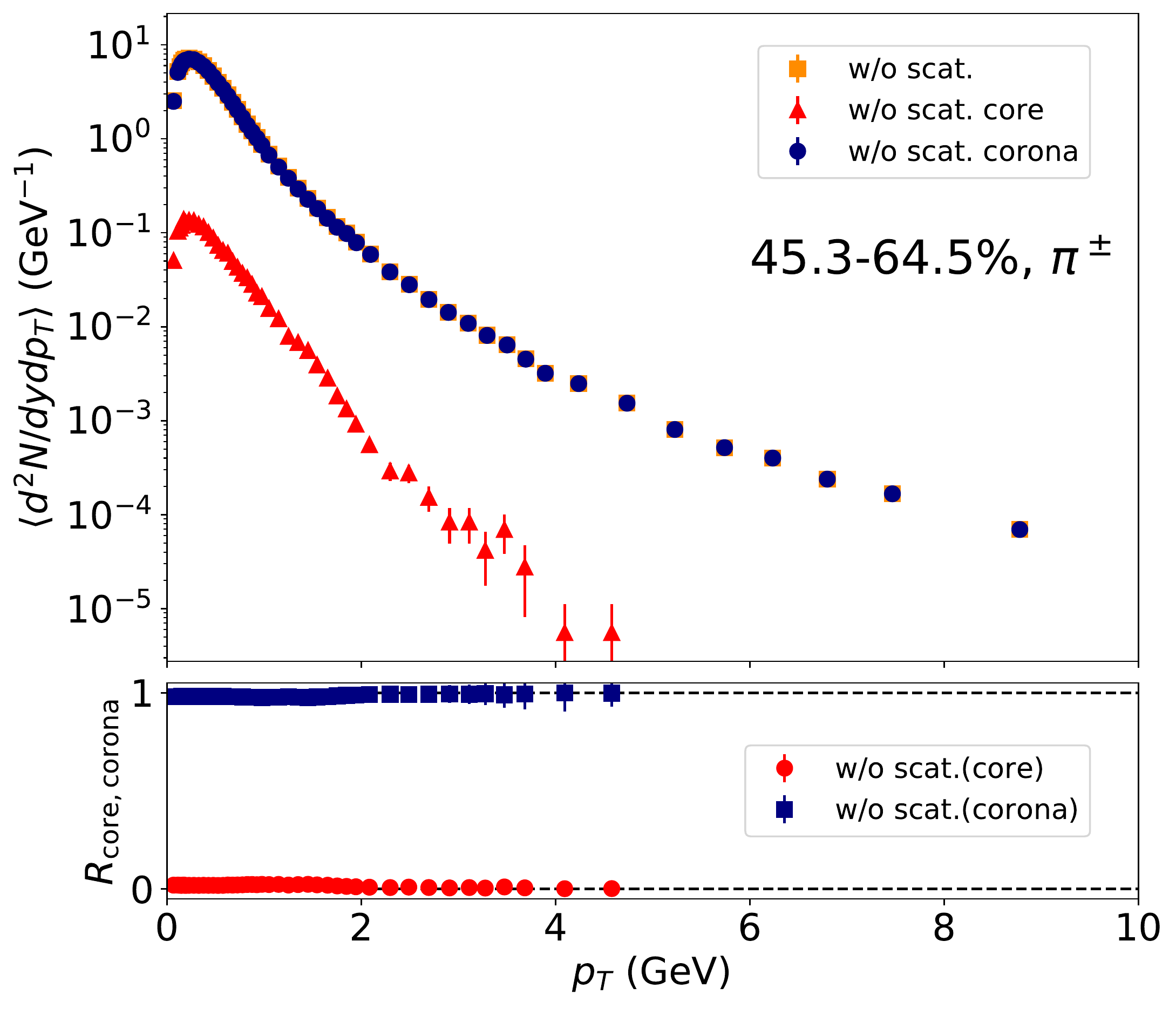}
    \includegraphics[bb = 0 0 628 542, width=0.45\textwidth]{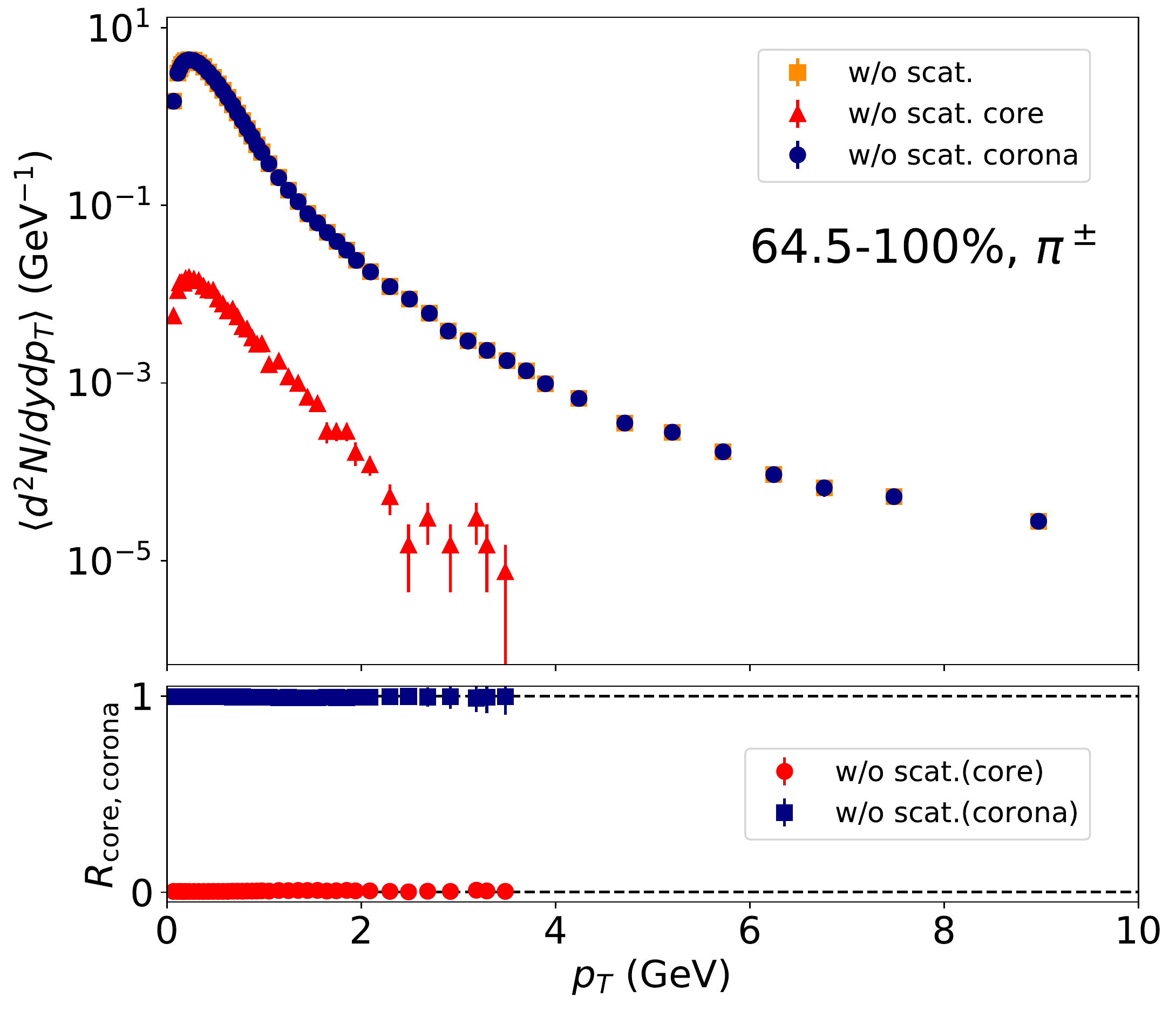}
    \caption{(Upper) Centrality dependence of $p_T$ spectra of charged pions ($\pi^+ + \pi^-$) in $p$+$p$ collisions at \snn[proton] = 13 TeV from DCCI2.
    Results with switching off hadronic rescatterings (orange squares) and their breakdown into core (red triangles) and corona contributions (blue circles) are shown.
    (Lower) Fraction of core (red circles) and corona (blue squares) components, $R_{\mathrm{core, corona}}$, in each $p_T$ bin. Centrality classes from 27-36.1\% to 64.5-100\% are shown.}
    \label{fig:PP13_PTSPECTRA_PI_CORECORONA_2}
\end{figure}

In this subsection, 
I aim to reveal in which momentum space we see core/corona dominance in different centrality class and particle species.
Figure \ref{fig:PP13_PTSPECTRA_PI_CORECORONA_1} and \ref{fig:PP13_PTSPECTRA_PI_CORECORONA_2} show 
centrality dependence of $p_T$ spectra of charged pions out of final hadrons from simulations with switching off hadronic rescatterings and their breakdown into core and corona contributions
in $p$+$p$ collisions at \snn[proton] = 13 TeV from DCCI2.
Hadronic rescatterings are switched off to distinguish if each hadron is a production from core or corona,
so it should be noted that these breakdowns are not identical with ones one would see in full simulations or experimental data.
However, with hadronic rescatterings, generally more flows are pronounced in heavy particles such as protons rather than pions.
Thus, compared to protons, it can be expected that these tendency seen in breakdowns into core and corona components of $p_T$ spectra would be likely to remain in charged pions even with hadronic rescatterings.
As an overall tendency about multiplicity-class dependence,
one sees that contribution of core components enhances at low $p_T$ towards high multiplicity classes.
The small enhancement of core components around $0.5<p_T<1.5$ GeV similar to the one we see in Fig.~\ref{fig:PTSPECTRA_PP_PBPB} is seen in results of high-multiplicity classes here too:
at the highest-multiplicity class ($0-0.9\%$), the core components overtakes corona around intermediate $p_T$ ($0.5<p_T<1.5$ GeV).
Therefore, the momentum range of $0.5<p_T<1.5$ GeV could be a sweet spot in which one can explore signals of particle productions from QGP fluids in high-multiplicity $p$+$p$ collisions.

\begin{figure}
    \centering
    \includegraphics[bb = 0 0 628 542, width=0.45\textwidth]{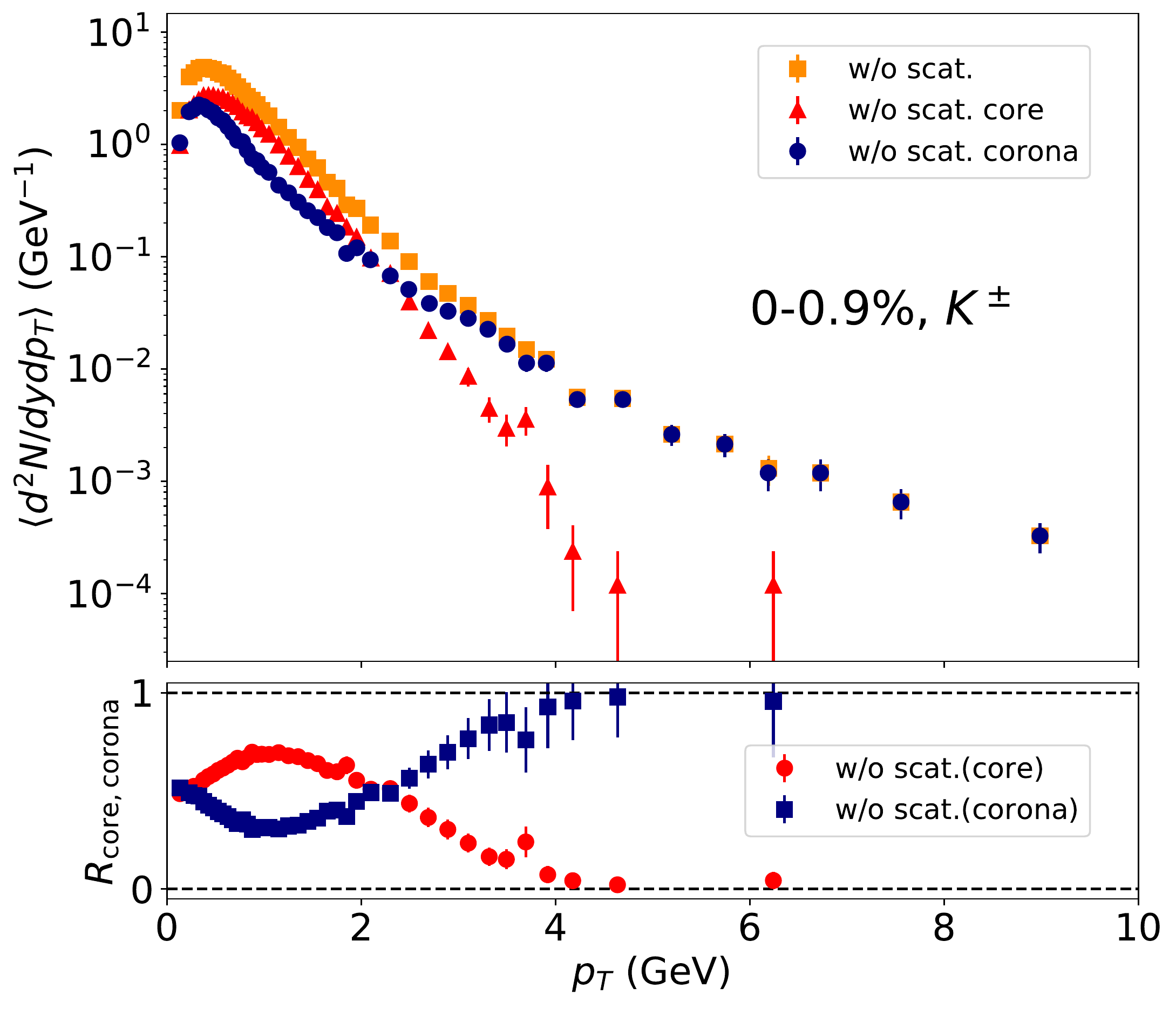}
    \includegraphics[bb = 0 0 628 542, width=0.45\textwidth]{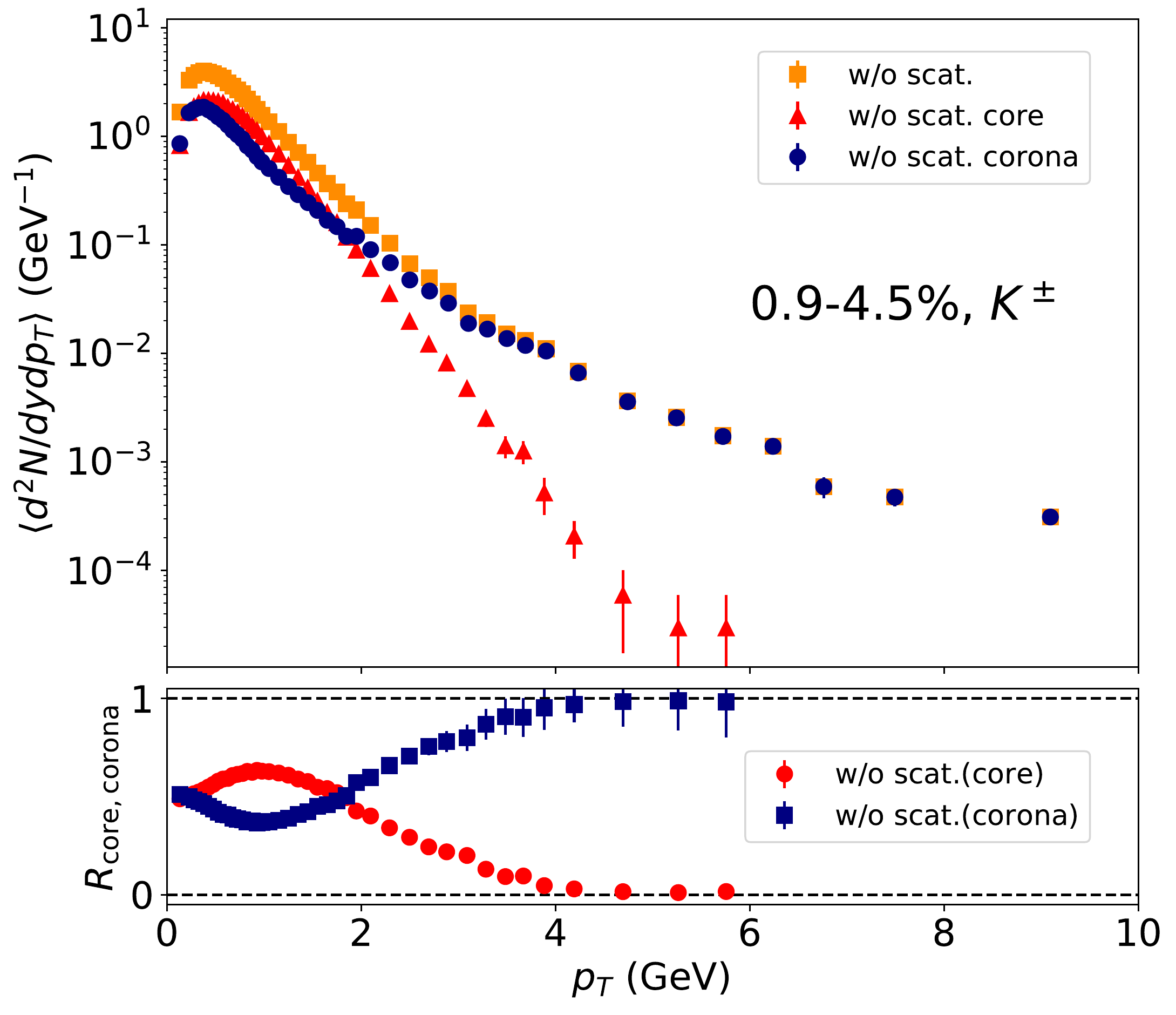}
    \includegraphics[bb = 0 0 628 542, width=0.45\textwidth]{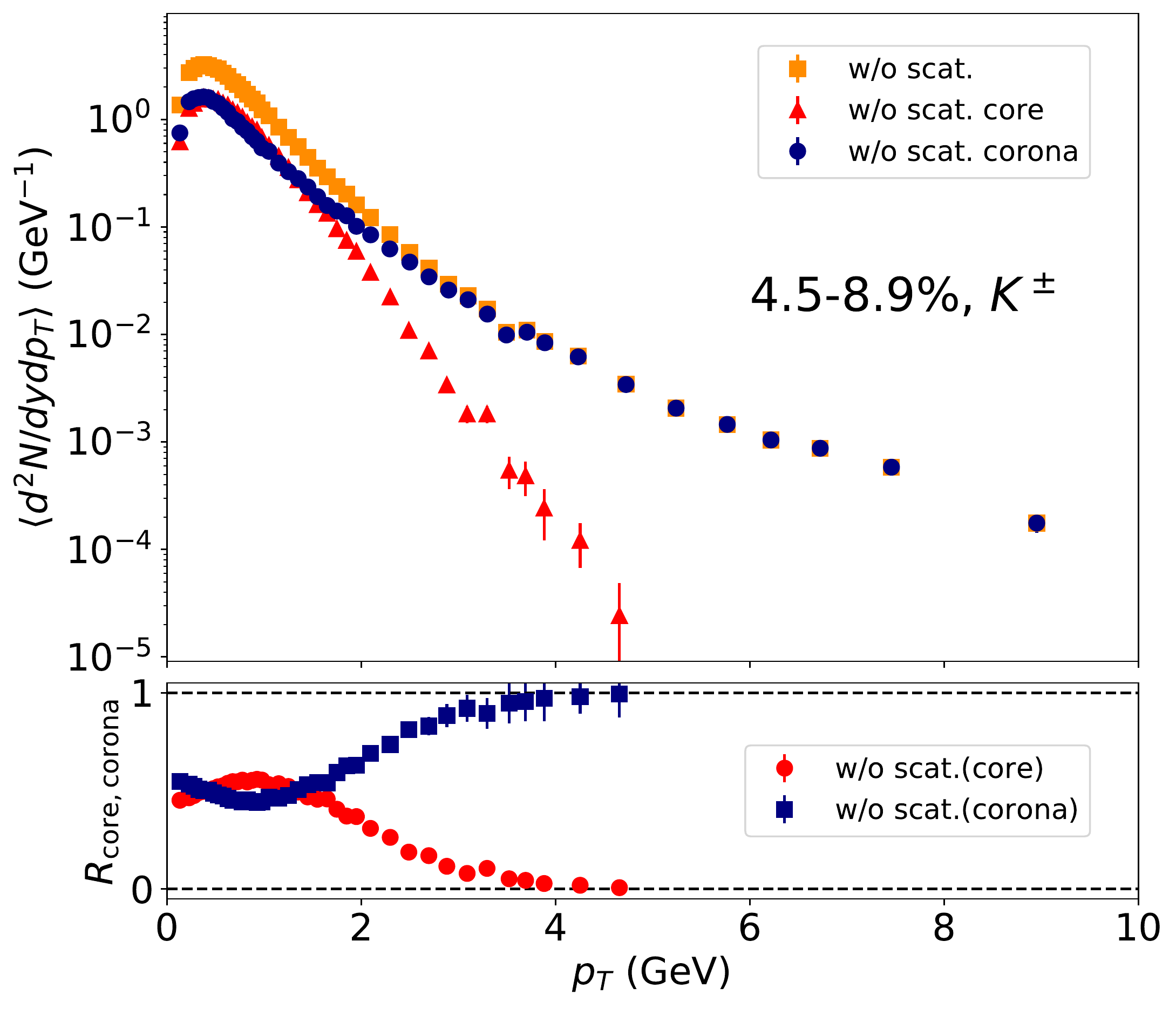}
    \includegraphics[bb = 0 0 628 542, width=0.45\textwidth]{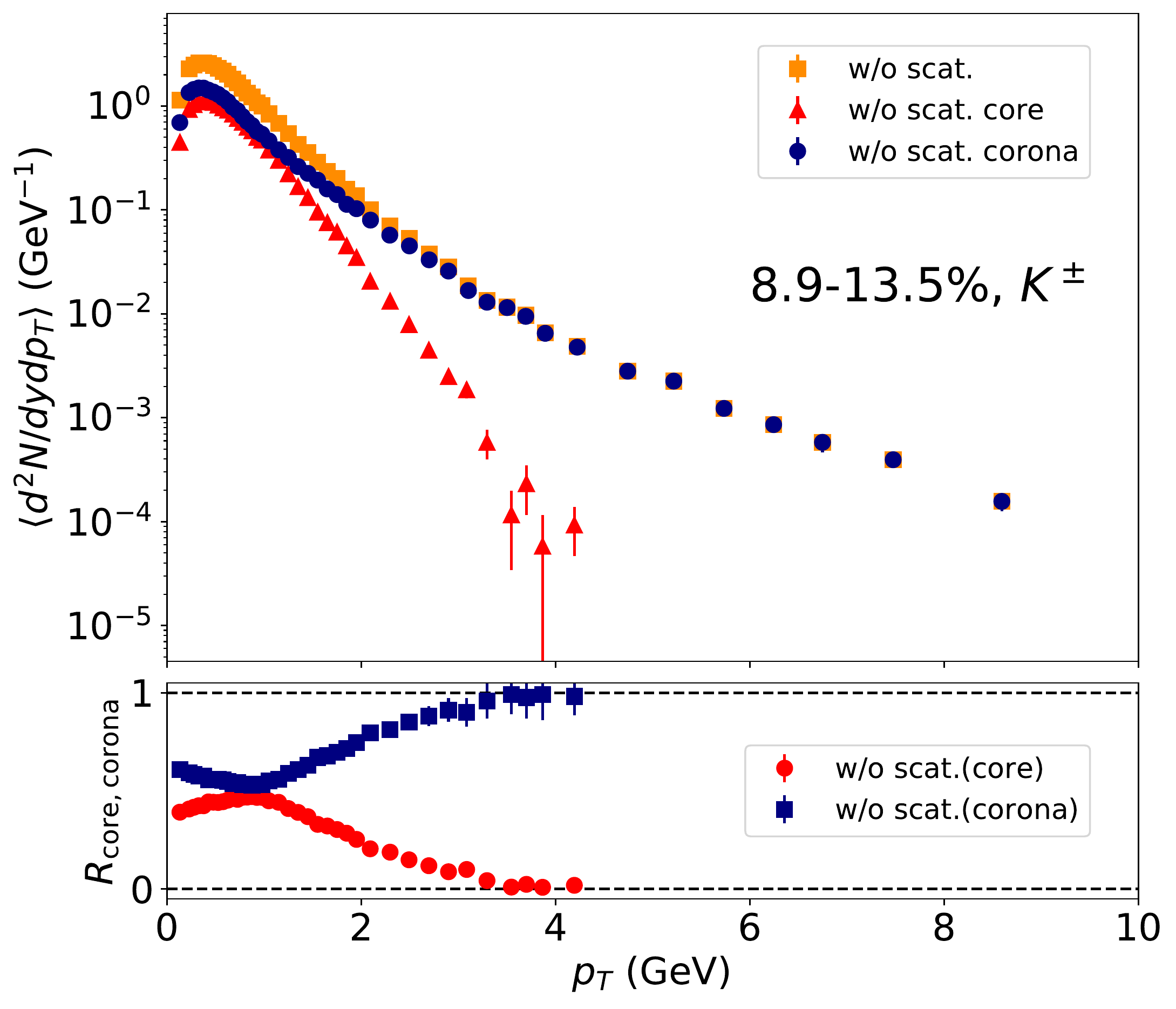}
    \includegraphics[bb = 0 0 628 542, width=0.45\textwidth]{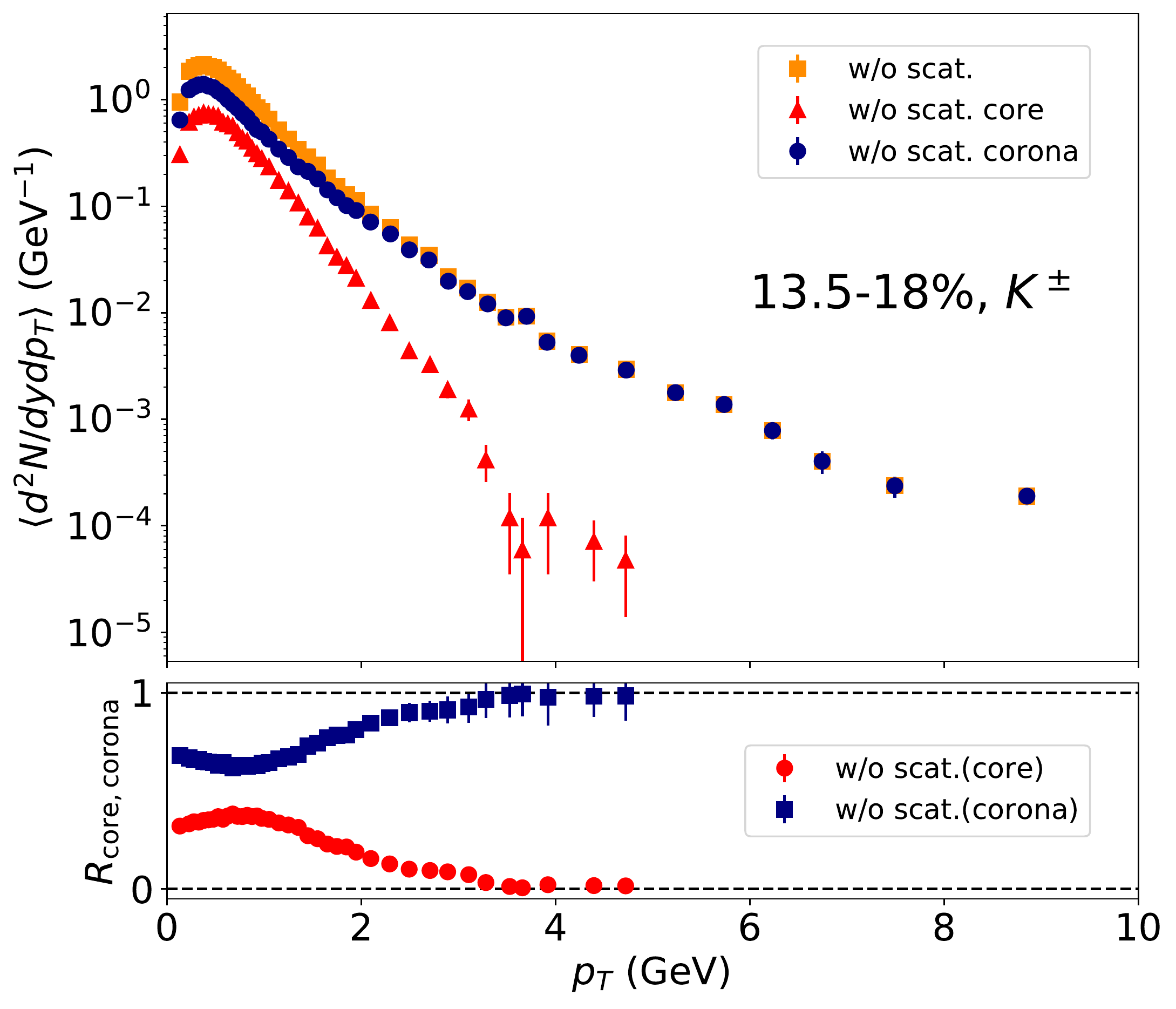}
    \includegraphics[bb = 0 0 628 542, width=0.45\textwidth]{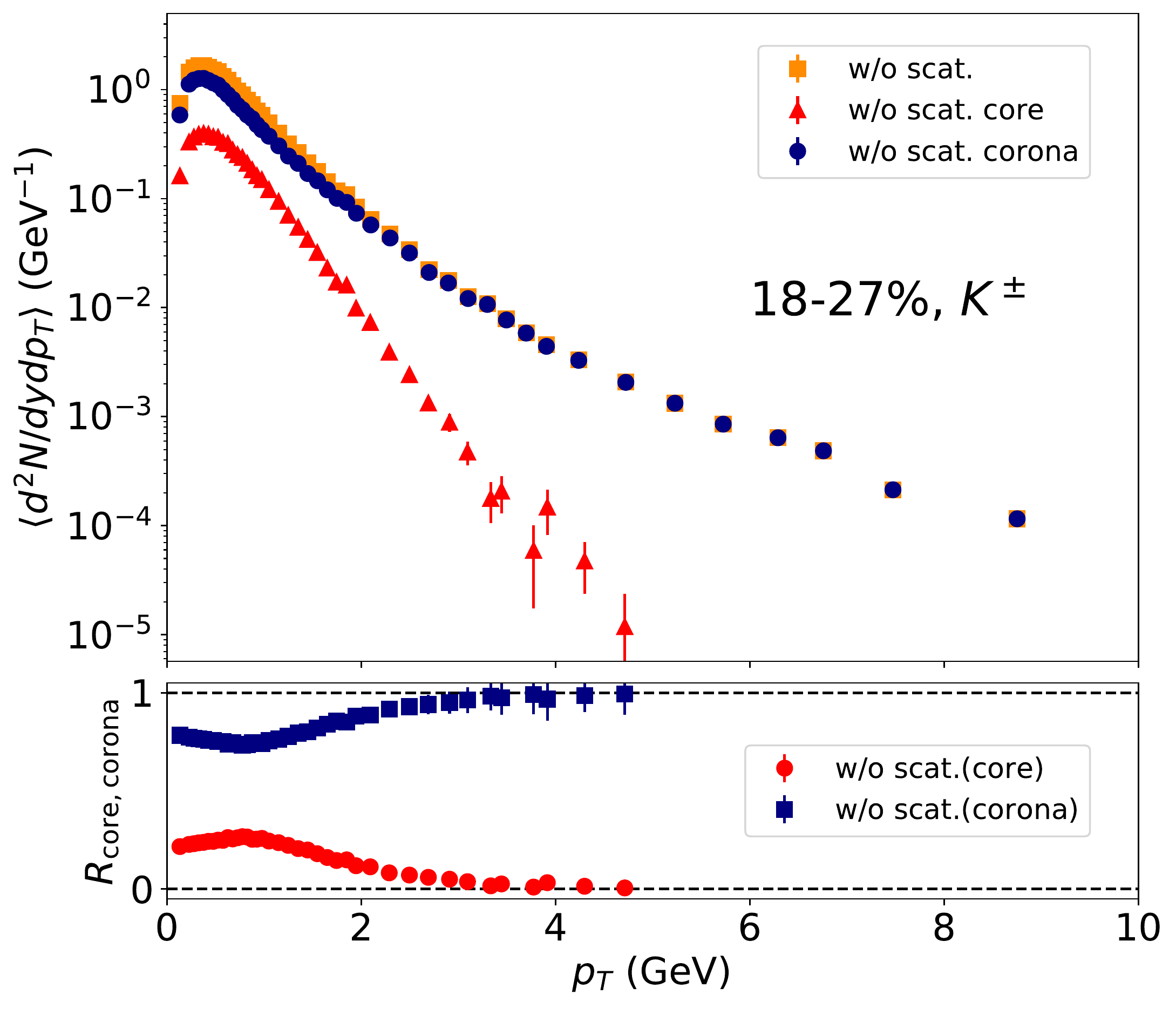}
    \caption{(Upper) Centrality dependence of $p_T$ spectra of charged kaons ($K^+ + K^-$) in $p$+$p$ collisions at \snn[proton] = 13 TeV from DCCI2.
    Results with switching off hadronic rescatterings (orange squares) and their breakdown into core (red triangles) and corona contributions (blue circles) are shown.
    (Lower) Fraction of core (red circles) and corona (blue squares) components, $R_{\mathrm{core, corona}}$, in each $p_T$ bin. Centrality classes from 0-0.9\% to 18-27\% are shown.
    }
    \label{fig:PP13_PTSPECTRA_K_CORECORONA_1}
\end{figure}

\begin{figure}
    \centering
    \includegraphics[bb = 0 0 628 542, width=0.45\textwidth]{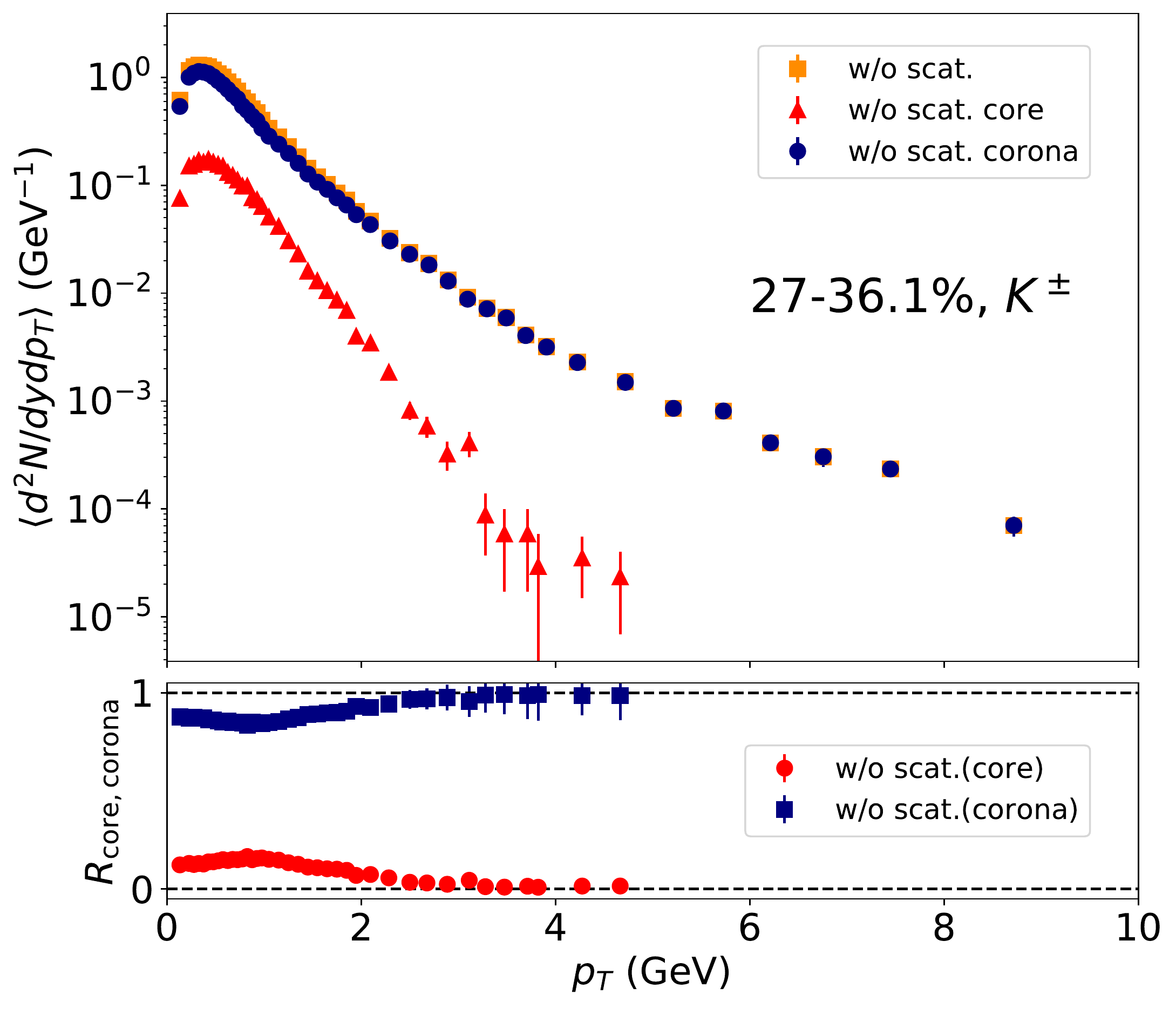}
    \includegraphics[bb = 0 0 628 542, width=0.45\textwidth]{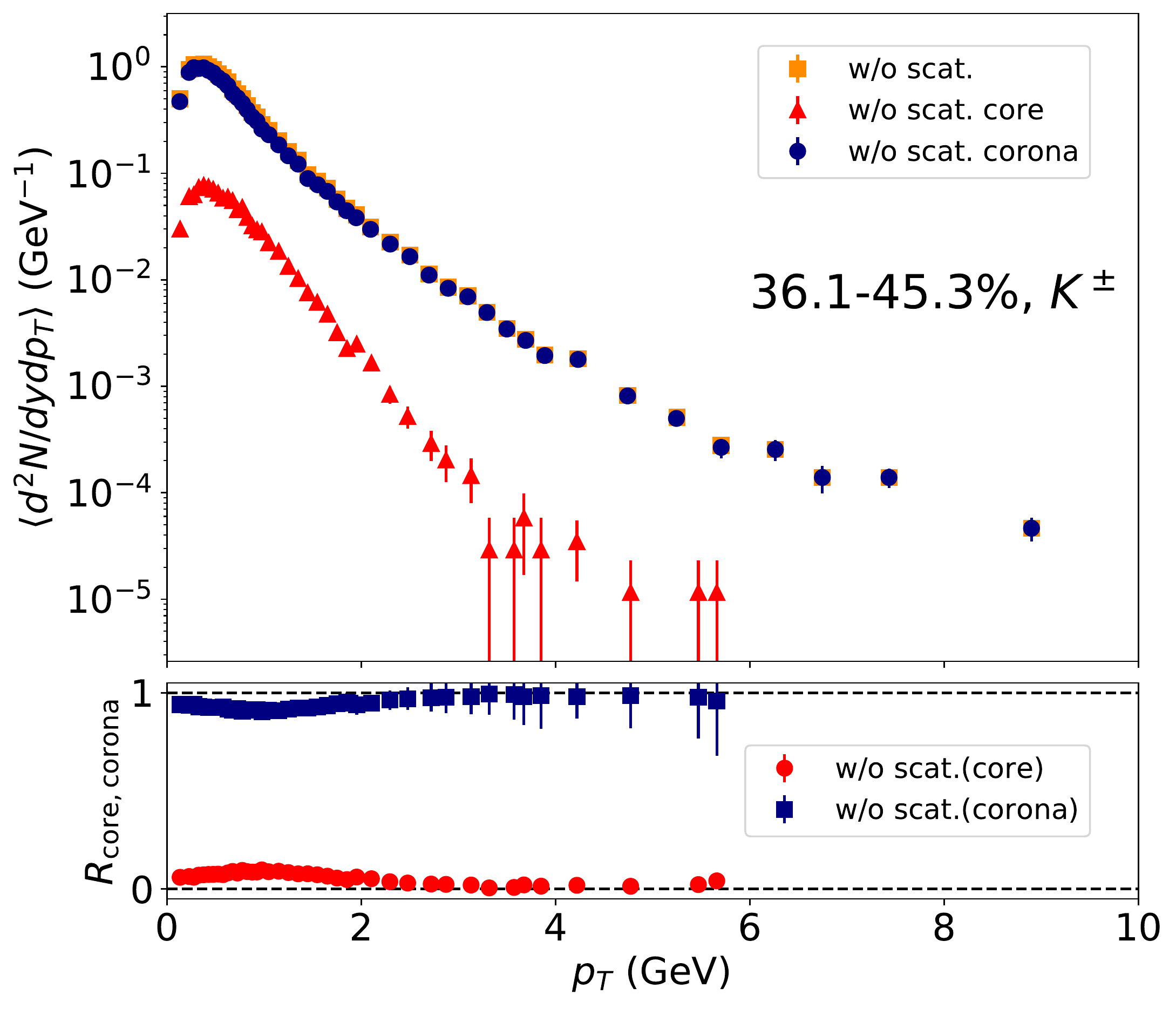}
    \includegraphics[bb = 0 0 628 542, width=0.45\textwidth]{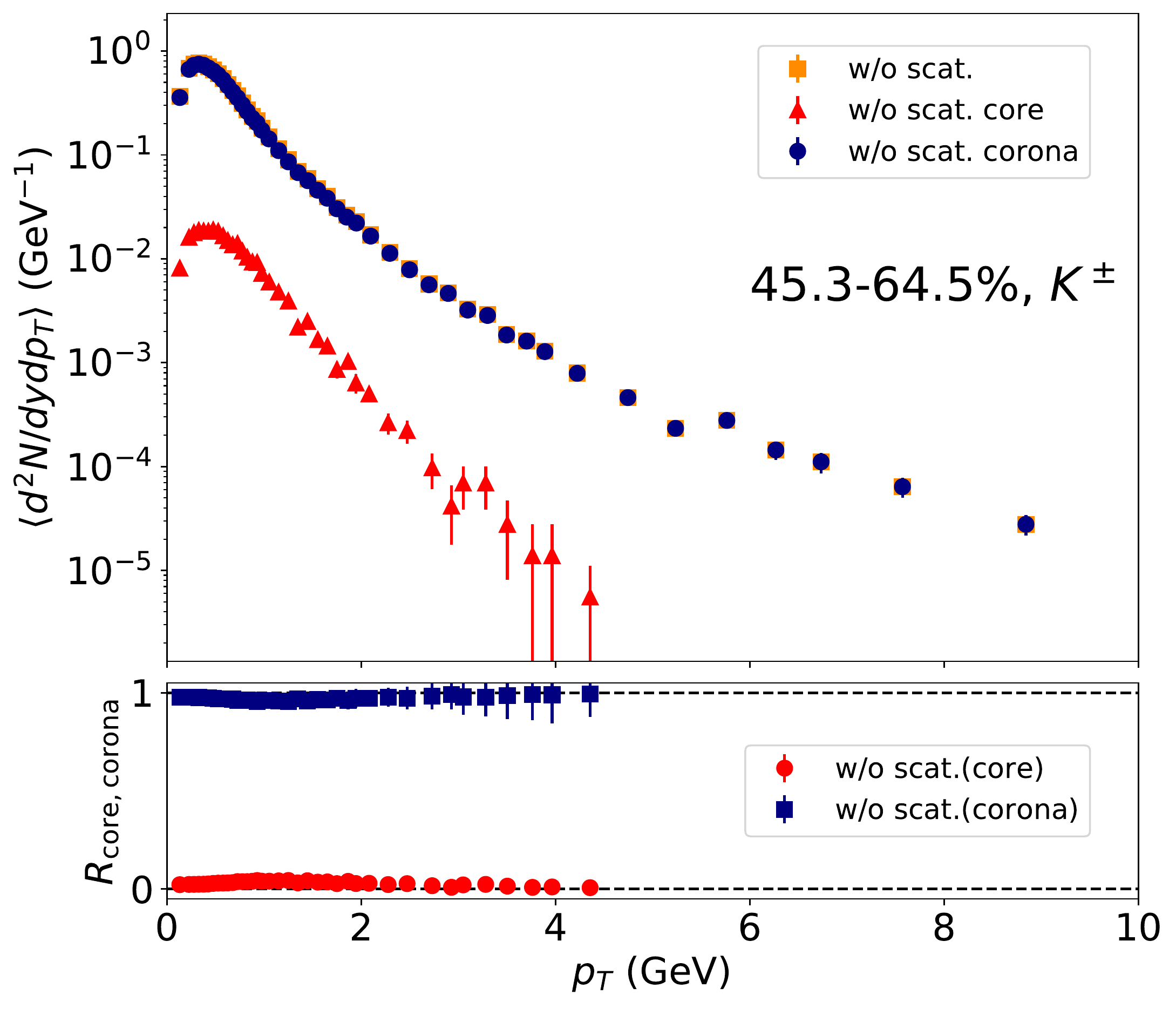}
    \includegraphics[bb = 0 0 628 542, width=0.45\textwidth]{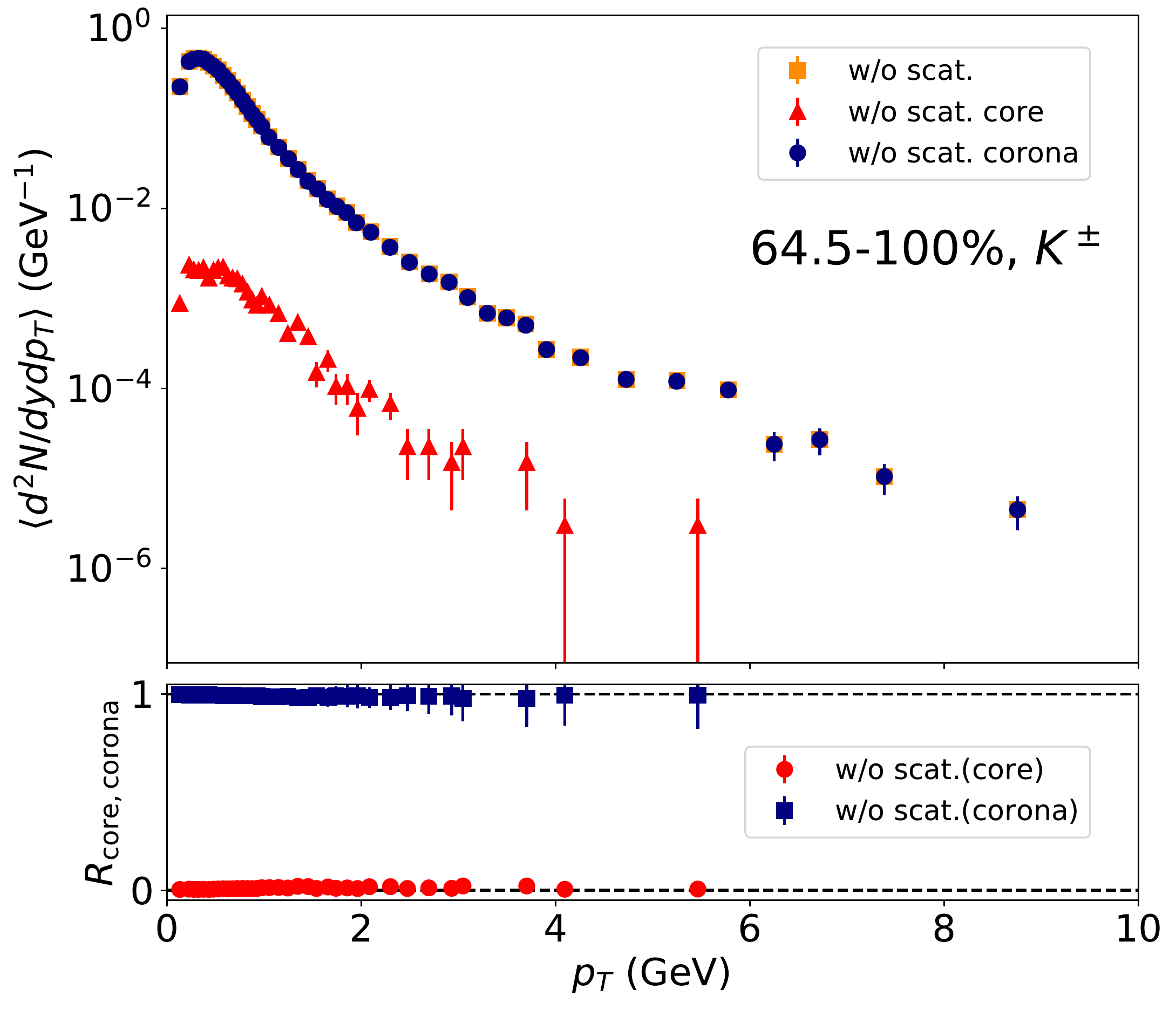}
    \caption{(Upper) Centrality dependence of $p_T$ spectra of charged kaons ($K^+ + K^-$) in $p$+$p$ collisions at \snn[proton] = 13 TeV from DCCI2.
    Results with switching off hadronic rescatterings (orange squares) and their breakdown into core (red triangles) and corona contributions (blue circles) are shown.
    (Lower) Fraction of core (red circles) and corona (blue squares) components, $R_{\mathrm{core, corona}}$, in each $p_T$ bin. Centrality classes from 27-36.1\% to 64.5-100\% are shown.}
    \label{fig:PP13_PTSPECTRA_K_CORECORONA_2}
\end{figure}

\begin{figure}
    \centering
    \includegraphics[bb = 0 0 628 542, width=0.45\textwidth]{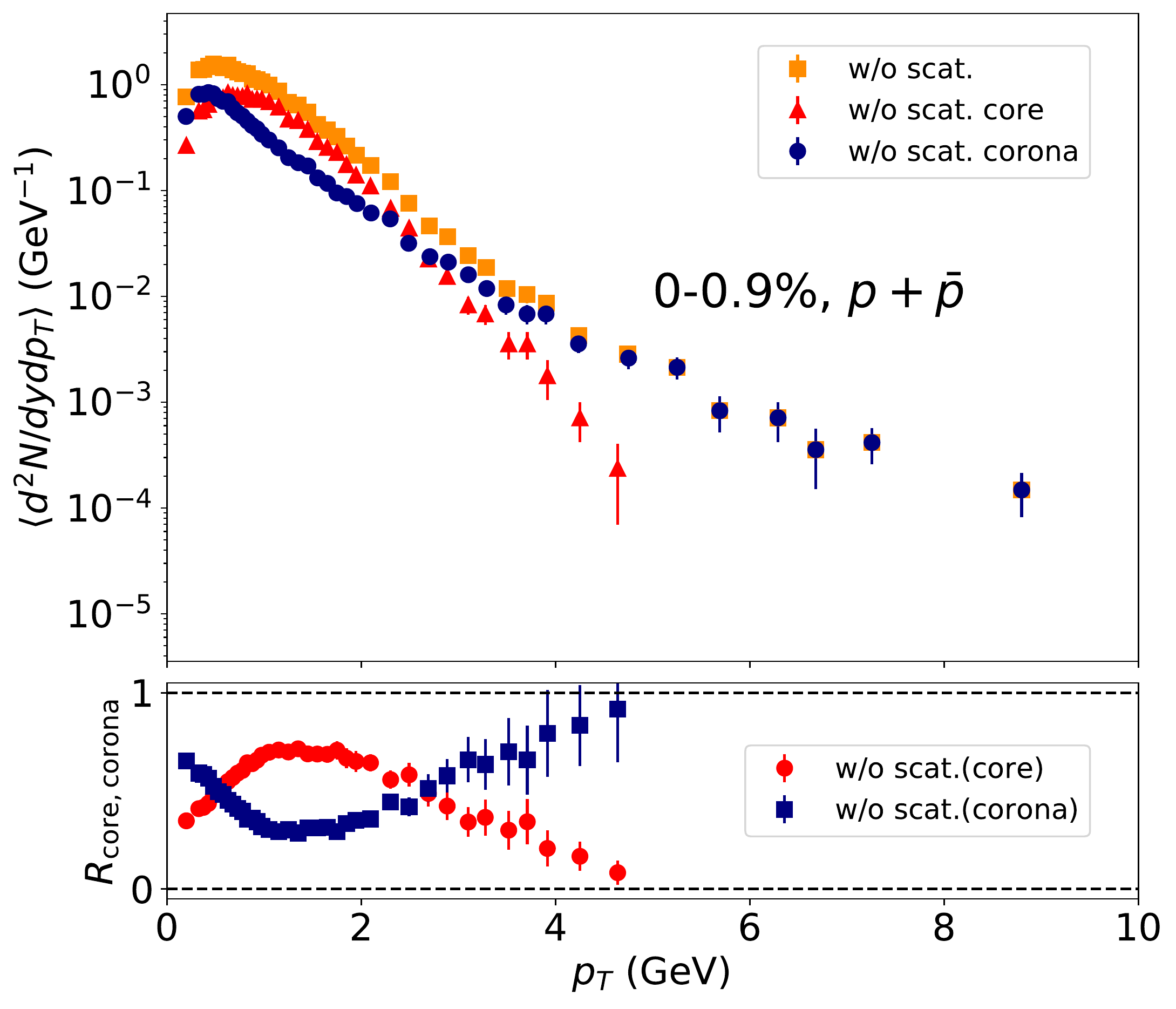}
    \includegraphics[bb = 0 0 628 542, width=0.45\textwidth]{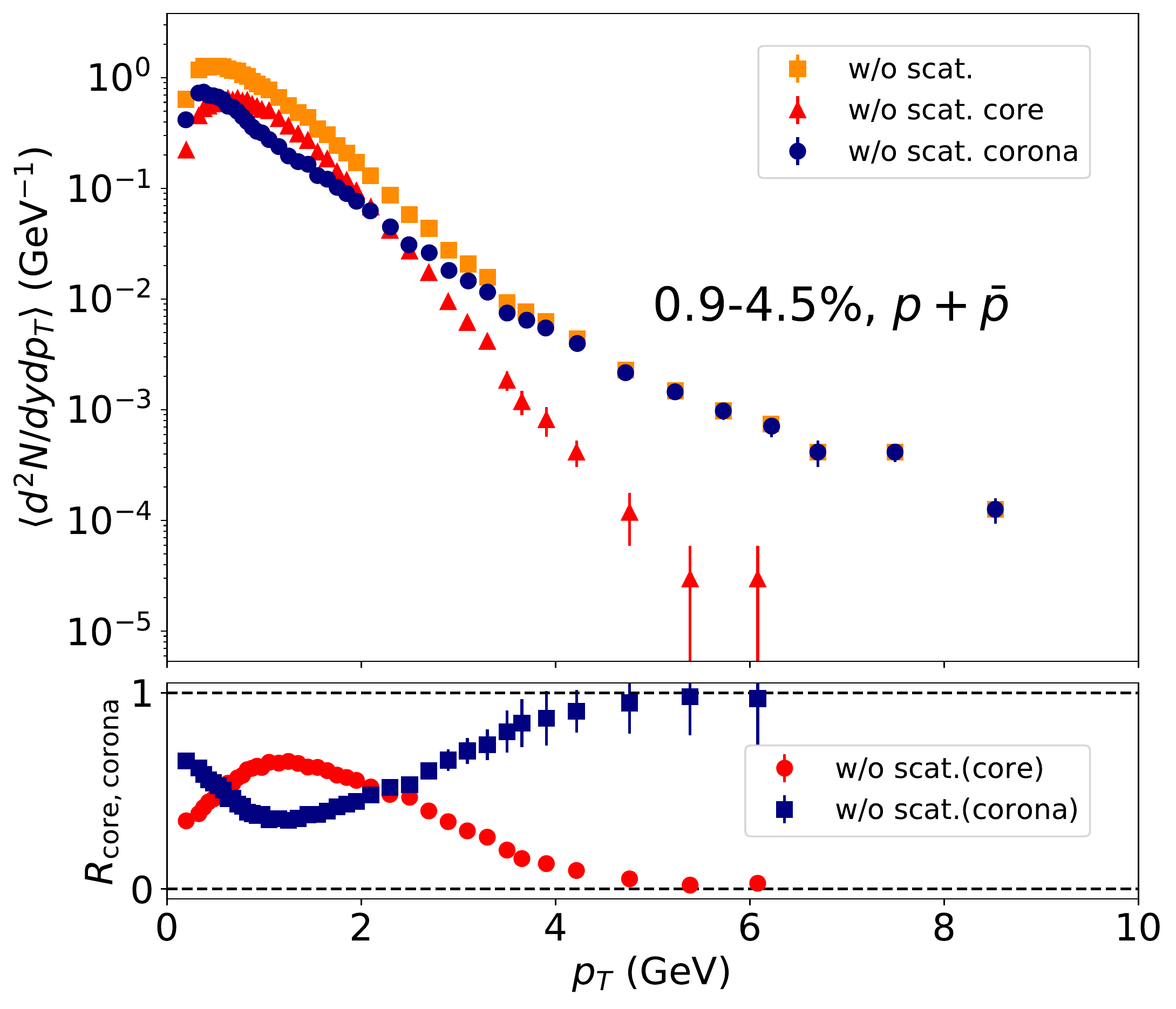}
    \includegraphics[bb = 0 0 628 542, width=0.45\textwidth]{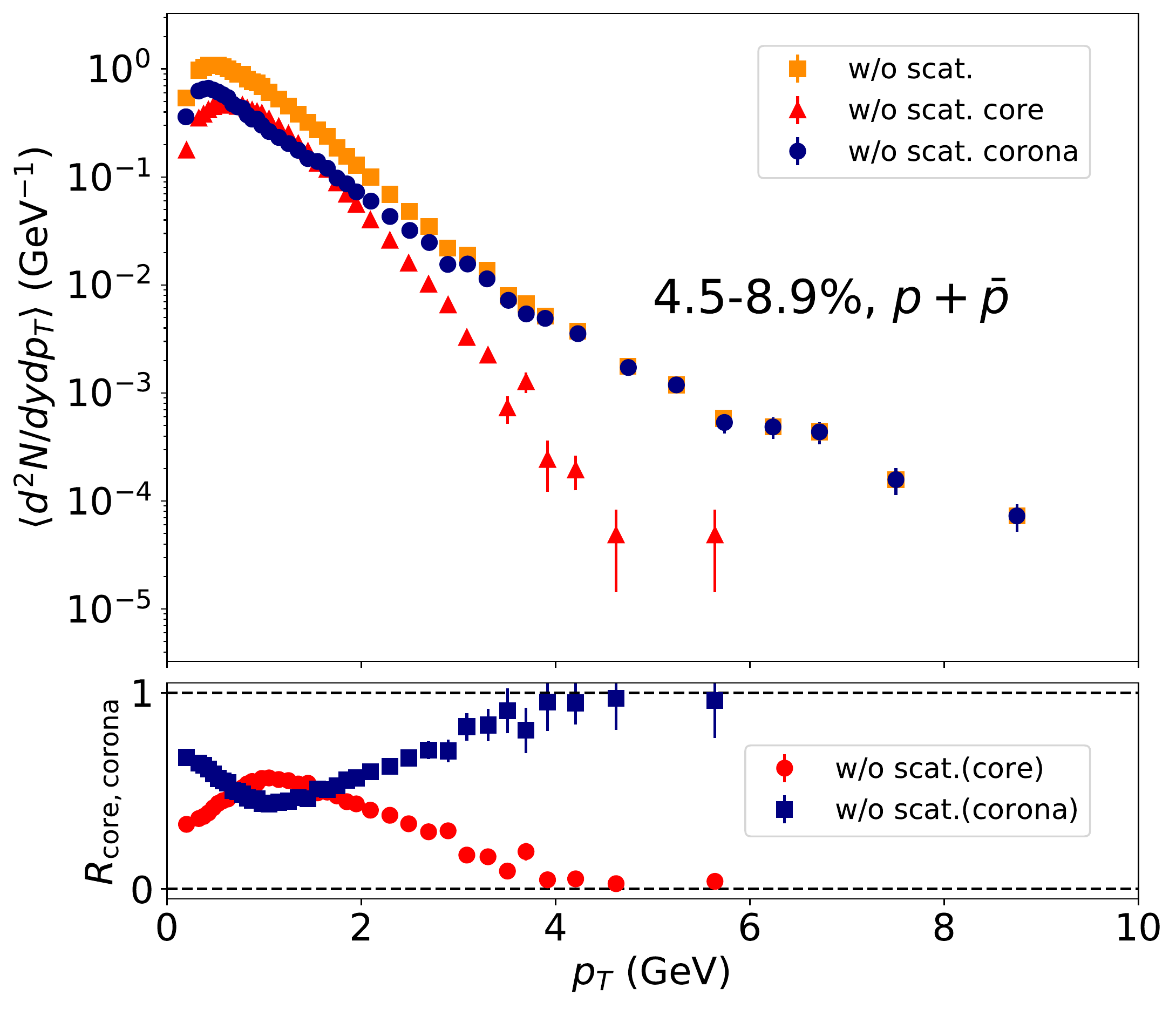}
    \includegraphics[bb = 0 0 628 542, width=0.45\textwidth]{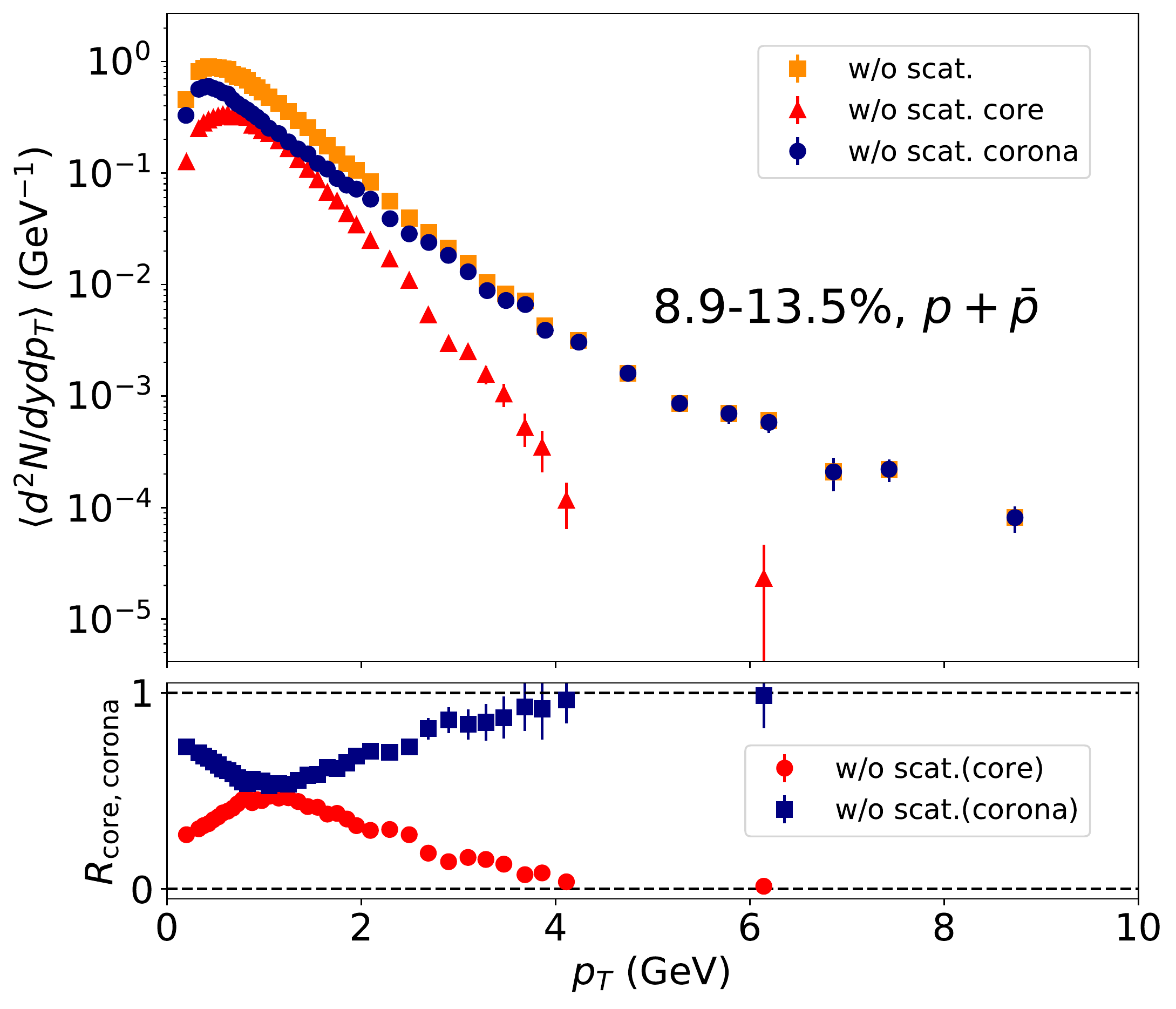}
    \includegraphics[bb = 0 0 628 542, width=0.45\textwidth]{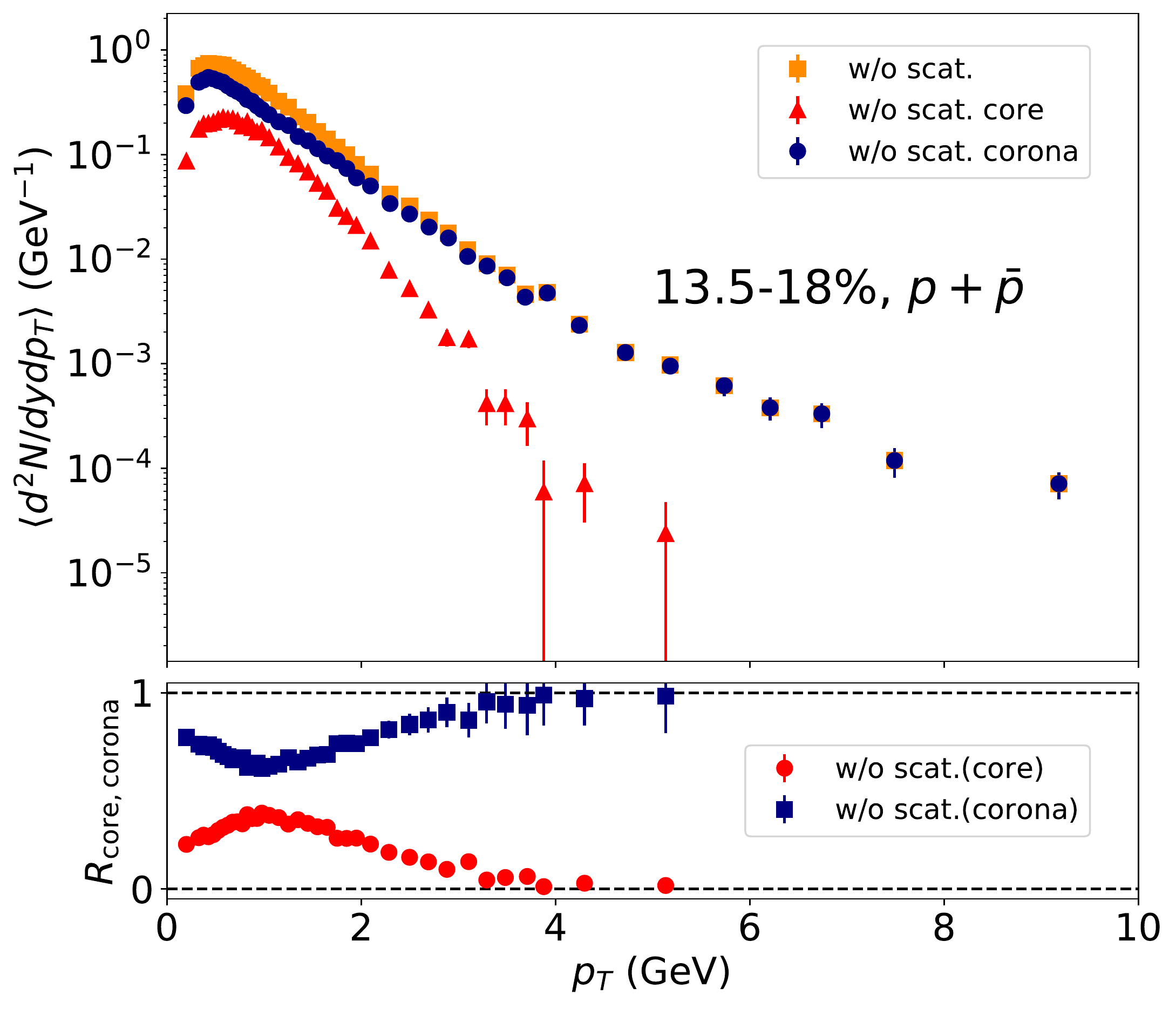}
    \includegraphics[bb = 0 0 628 542, width=0.45\textwidth]{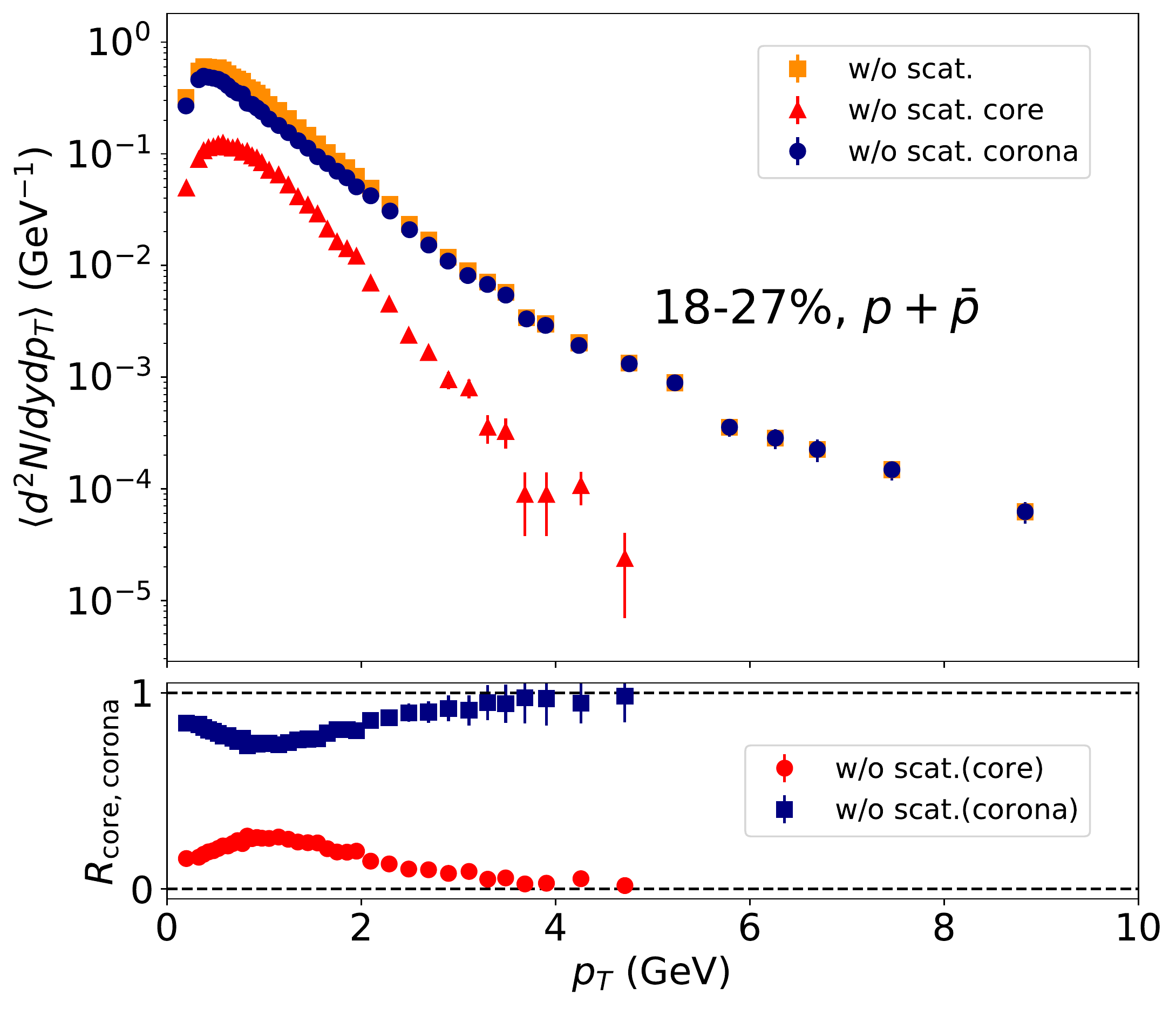}
    \caption{(Upper) Centrality dependence of $p_T$ spectra of protons and antiprotons ($p + \bar{p}$) in $p$+$p$ collisions at \snn[proton] = 13 TeV from DCCI2.
    Results with switching off hadronic rescatterings (orange squares) and their breakdown into core (red triangles) and corona contributions (blue circles) are shown.
    (Lower) Fraction of core (red circles) and corona (blue squares) components, $R_{\mathrm{core, corona}}$, in each $p_T$ bin. Centrality classes from 0-0.9\% to 18-27\% are shown.}
    \label{fig:PP13_PTSPECTRA_P_CORECORONA_1}
\end{figure}

\begin{figure}
    \centering
    \includegraphics[bb = 0 0 628 542, width=0.45\textwidth]{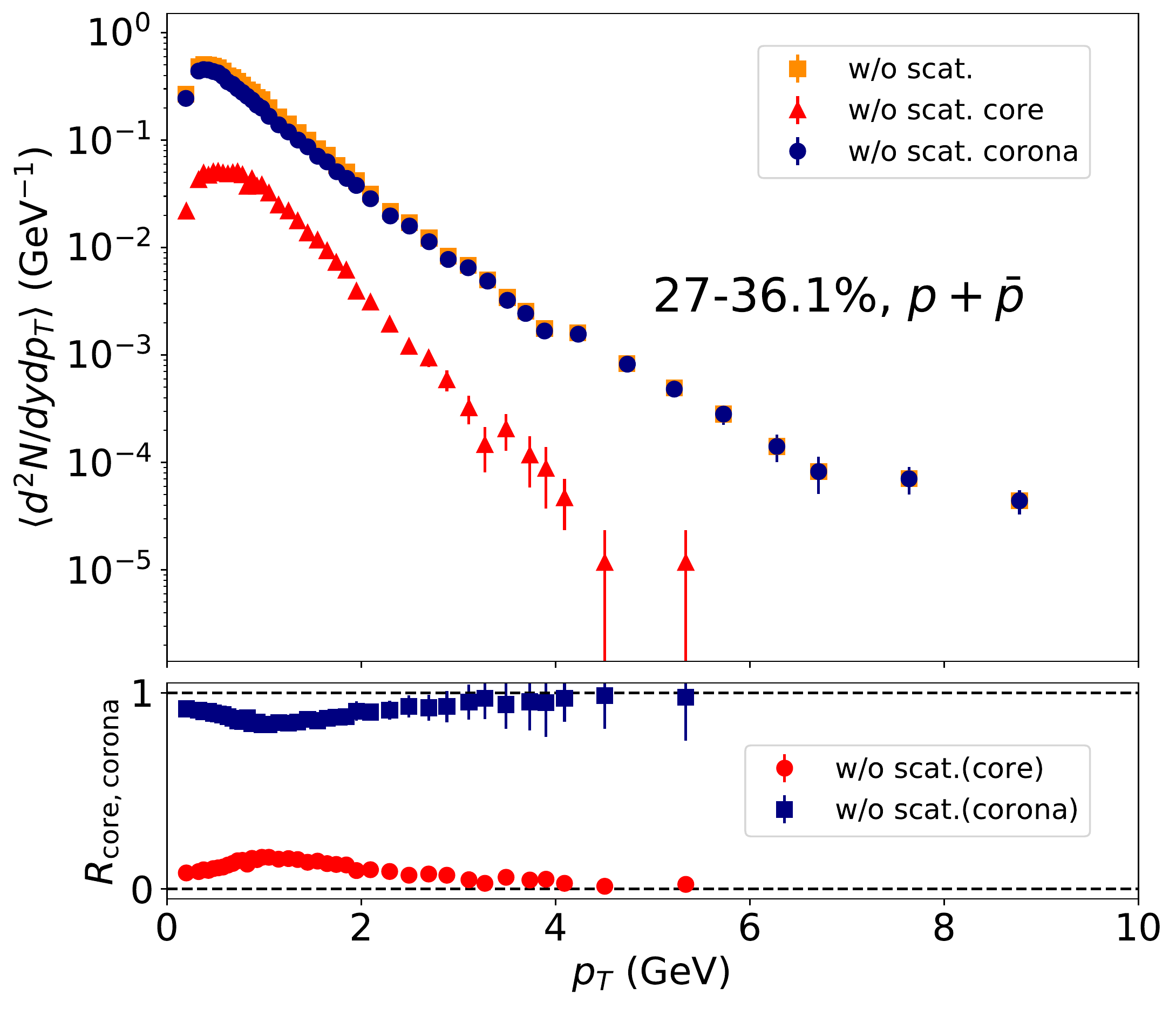}
    \includegraphics[bb = 0 0 628 542, width=0.45\textwidth]{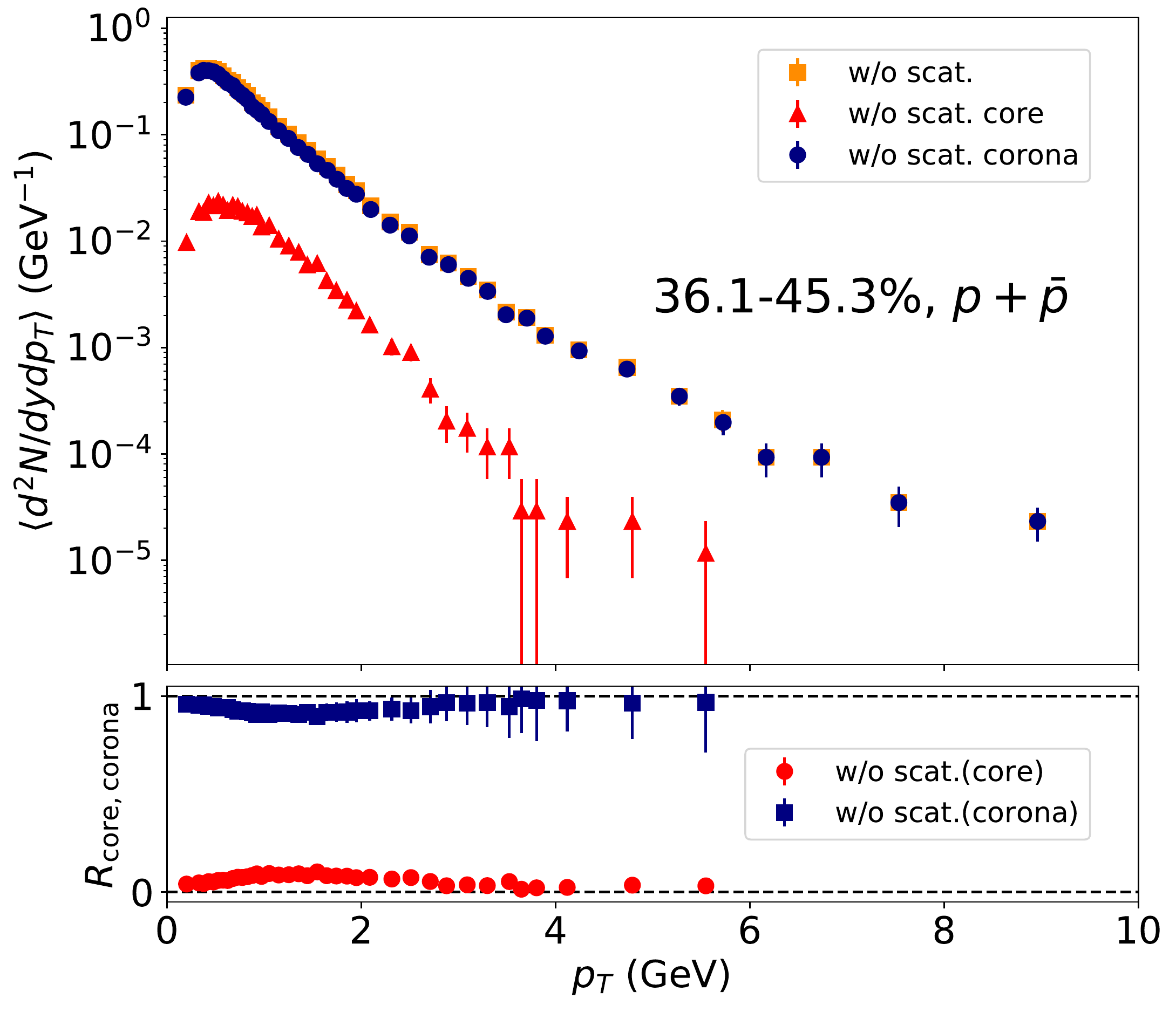}
    \includegraphics[bb = 0 0 628 542, width=0.45\textwidth]{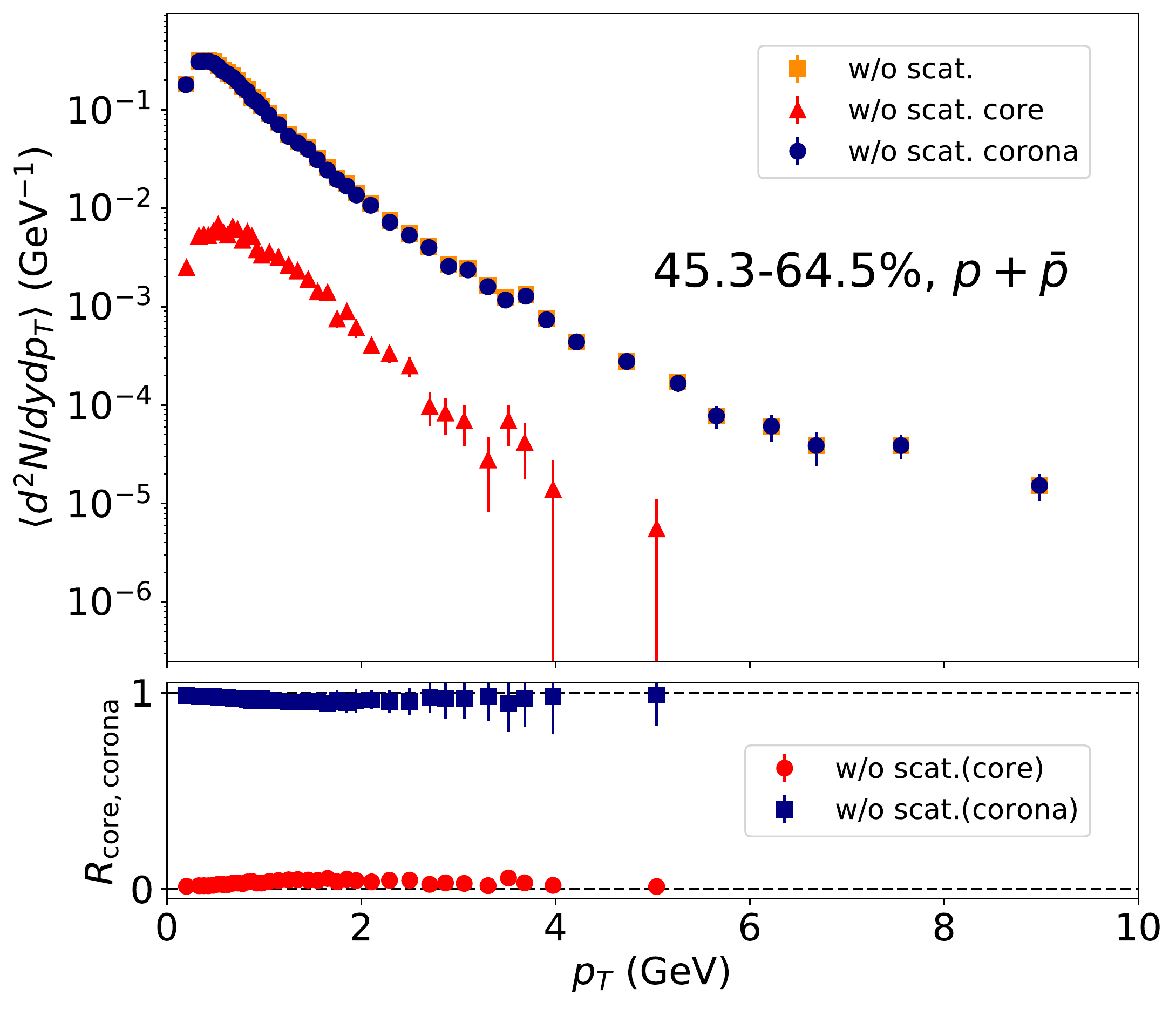}
    \includegraphics[bb = 0 0 628 542, width=0.45\textwidth]{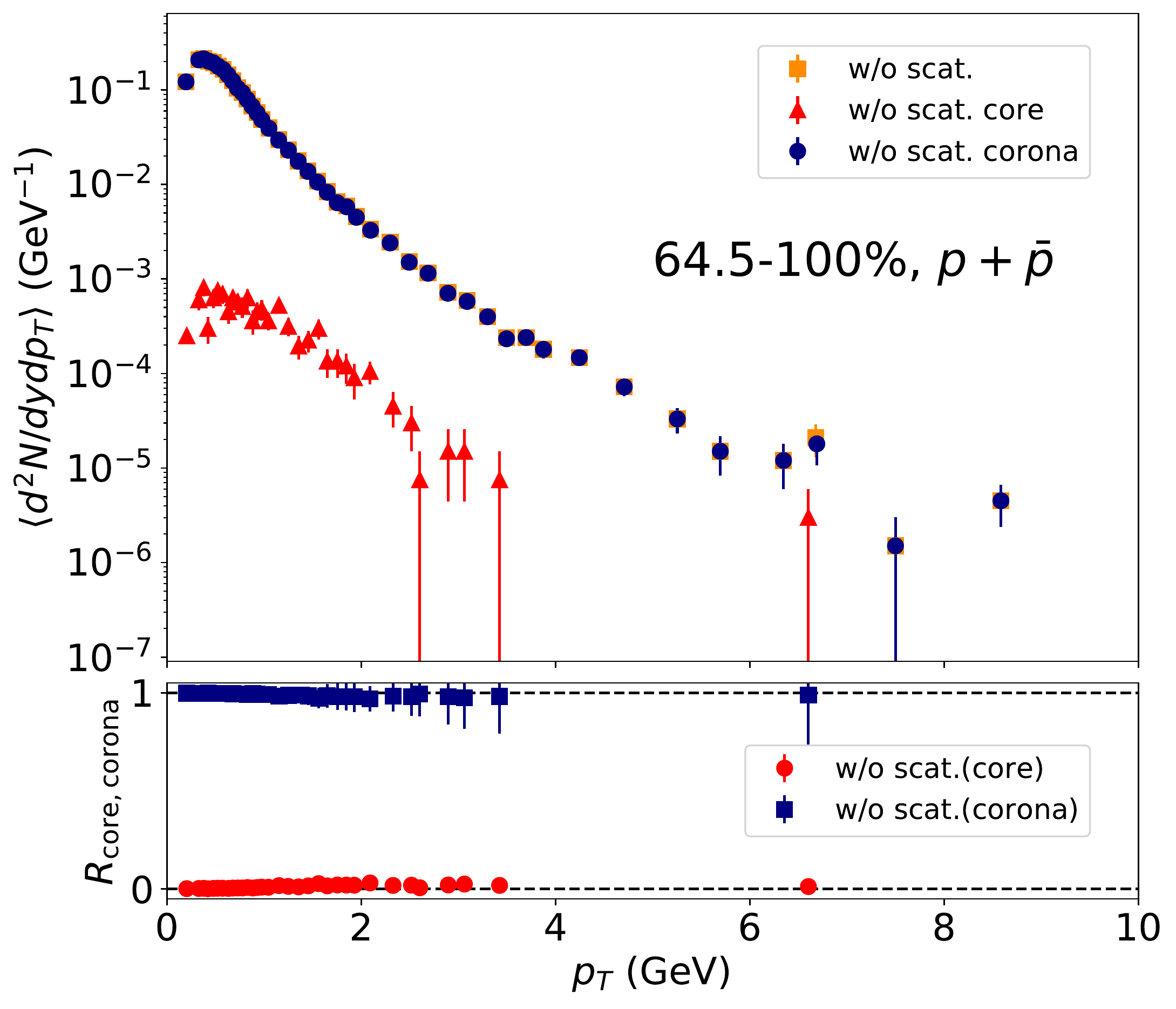}
    \caption{(Upper) Centrality dependence of $p_T$ spectra of protons and antiprotons ($p + \bar{p}$) in $p$+$p$ collisions at \snn[proton] = 13 TeV from DCCI2.
    Results with switching off hadronic rescatterings (orange squares) and their breakdown into core (red triangles) and corona contributions (blue circles) are shown.
    (Lower) Fraction of core (red circles) and corona (blue squares) components, $R_{\mathrm{core, corona}}$, in each $p_T$ bin. Centrality classes from 27-36.1\% to 64.5-100\% are shown.}
    \label{fig:PP13_PTSPECTRA_P_CORECORONA_2}
\end{figure}

Figures \ref{fig:PP13_PTSPECTRA_K_CORECORONA_1} and \ref{fig:PP13_PTSPECTRA_K_CORECORONA_2} show centrality dependence of $p_T$ spectra of charged kaons out of final hadrons from simulations with switching off hadronic rescatterings and their breakdown into core and corona contributions in $p$+$p$ collisions at \snn[proton] = 13 TeV from DCCI2.
Figures.~\ref{fig:PP13_PTSPECTRA_P_CORECORONA_1} and \ref{fig:PP13_PTSPECTRA_P_CORECORONA_2} show the corresponding results of protons.
The overall tendency on multiplicity-class dependence is the same as seen in charged pions:
core contributions enhances at a certain intermediate regime of $p_T$ at high-multiplicity classes.
As I discussed in Sec.~\ref{subsec:INTRO_TransverseMomentumDistribution},
heavier particles acquire larger momentum when those are produced from a QGP fluid with a certain velocity.
Especially, the effect is large in $p_T$ spectra of protons, and which is so called the effect of ``pion wind" in the late rescattering stage \cite{Hung:1997du,Bleicher:1999pu,Bratkovskaya:2000qy,Bass:2000ib,Oliinychenko:2021enj}.
This particle specie dependence is seen comparing results of charged pions, charged kaons, and protons at, for instance, $0$-$0.9\%$.
In charged pions, the core-component dominance is slightly seen and the crossing between $R_{\mathrm{core}}$ and $R_{\mathrm{corona}}$ in the lower panel happen at around $p_T\approx0.5$ and $\approx1.5$ GeV.
On the other hand in charged kaons and protons, the overtake becomes much clearer. Because of the flow pushing heavier particles towards higher $p_T$, the core-component dominance stays up to higher $p_T$ \cite{Hirano:2003pw}.
The resultant crossings of $R_{\mathrm{core, corona}}$ at higher $p_T$ take place around $p_T \approx 2.5$ GeV in $p_T$ spectra of protons.
Therefore, these result suggest that there is a possibility that heavier particle species produced in a certain intermediate regime of $p_T$ can be a good window to see signals of QGP formation in $p$+$p$ collisions.

\begin{figure}
    \centering
    \includegraphics[bb=0 0 628 538, width=0.49\textwidth]{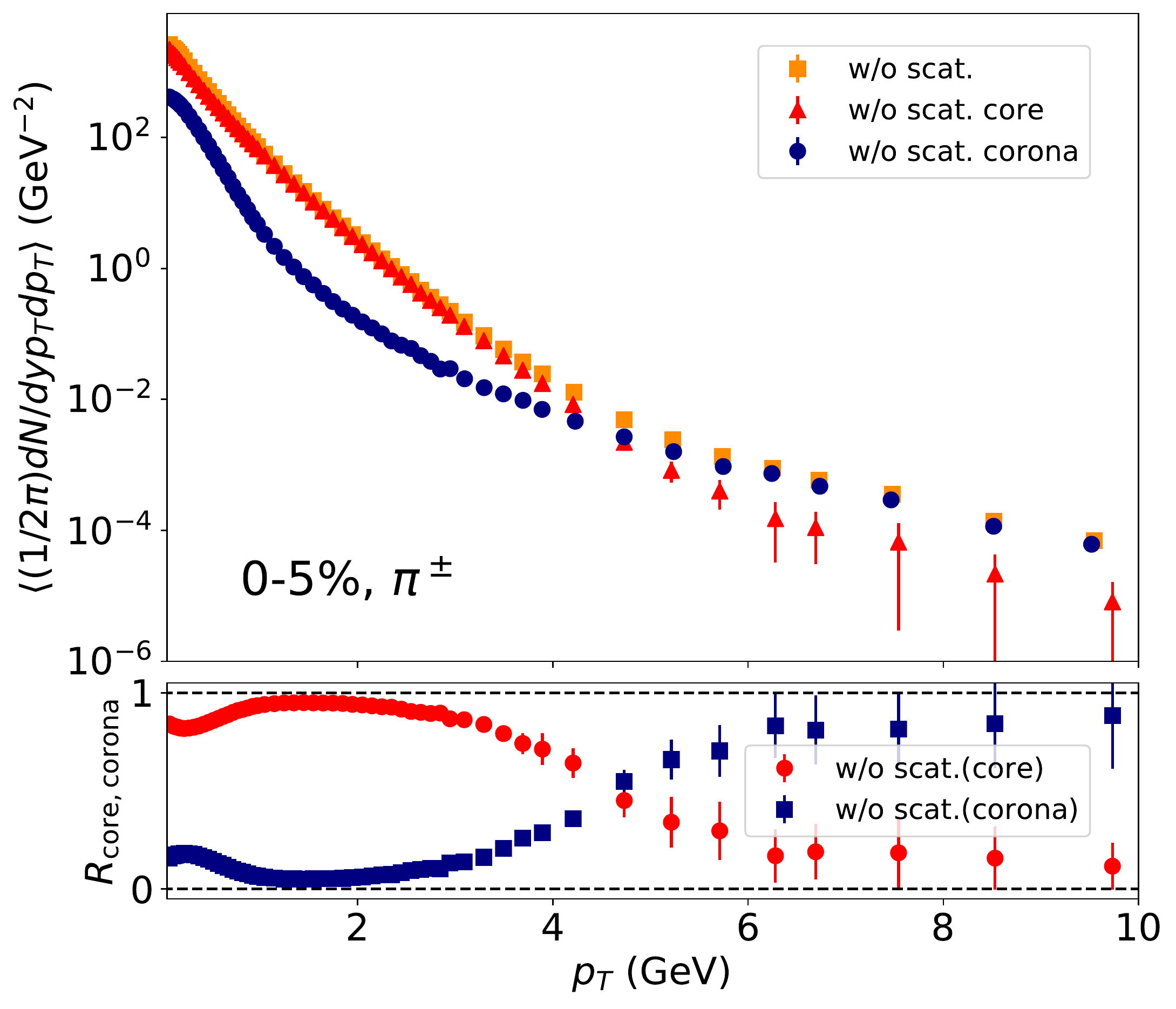}
    \includegraphics[bb=0 0 628 538, width=0.49\textwidth]{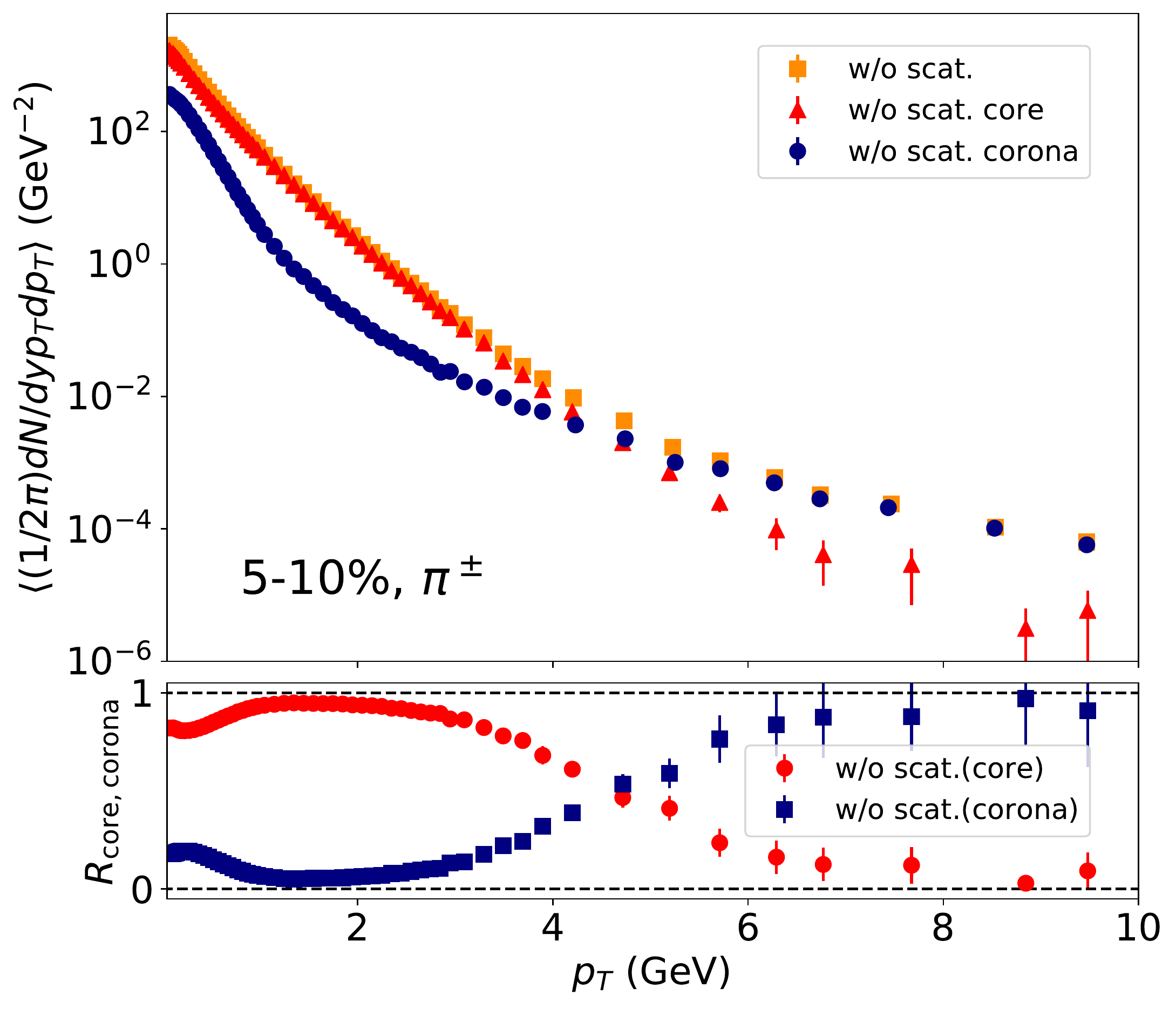}
    \includegraphics[bb=0 0 628 538, width=0.49\textwidth]{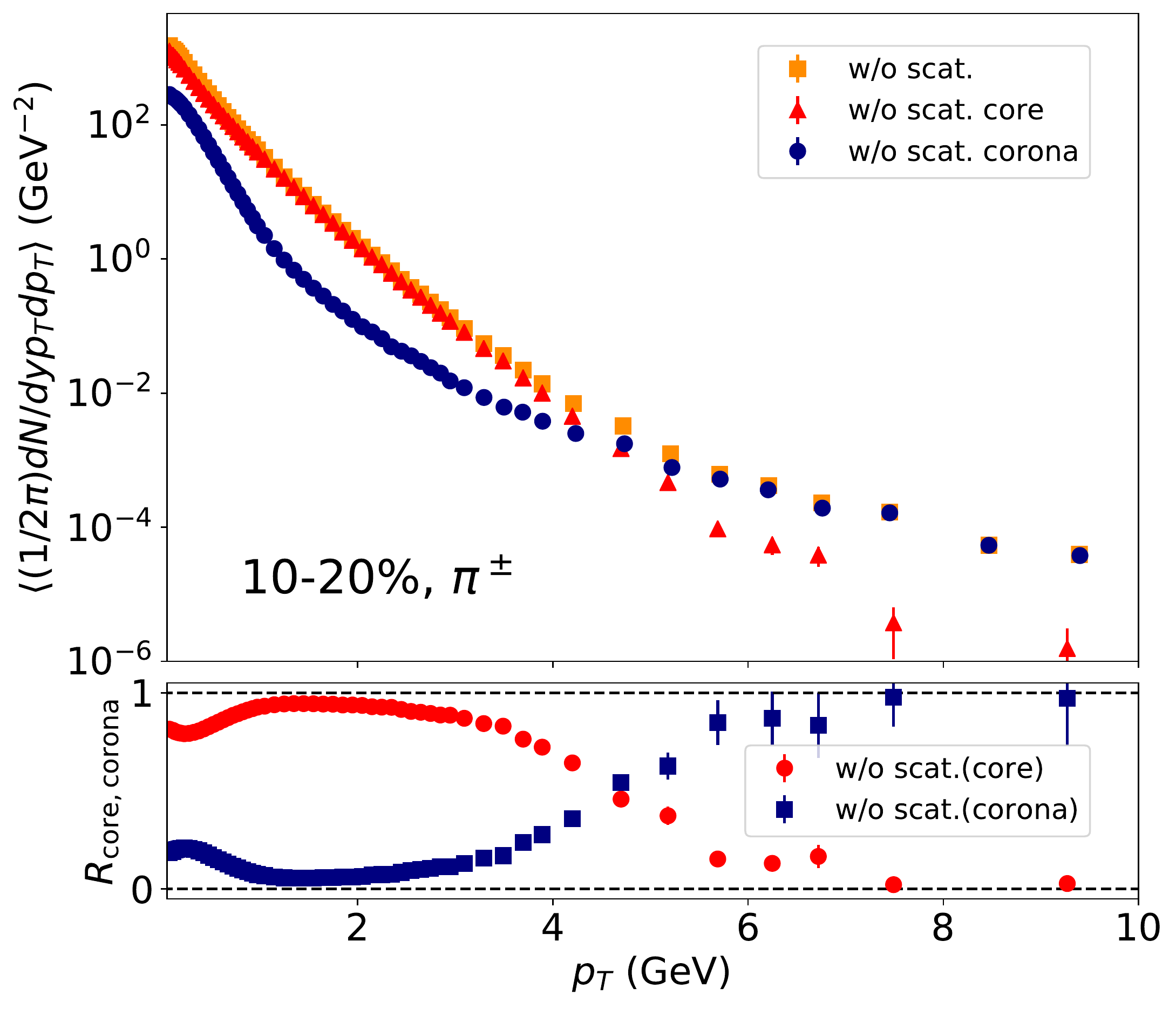}
    \includegraphics[bb=0 0 628 538, width=0.49\textwidth]{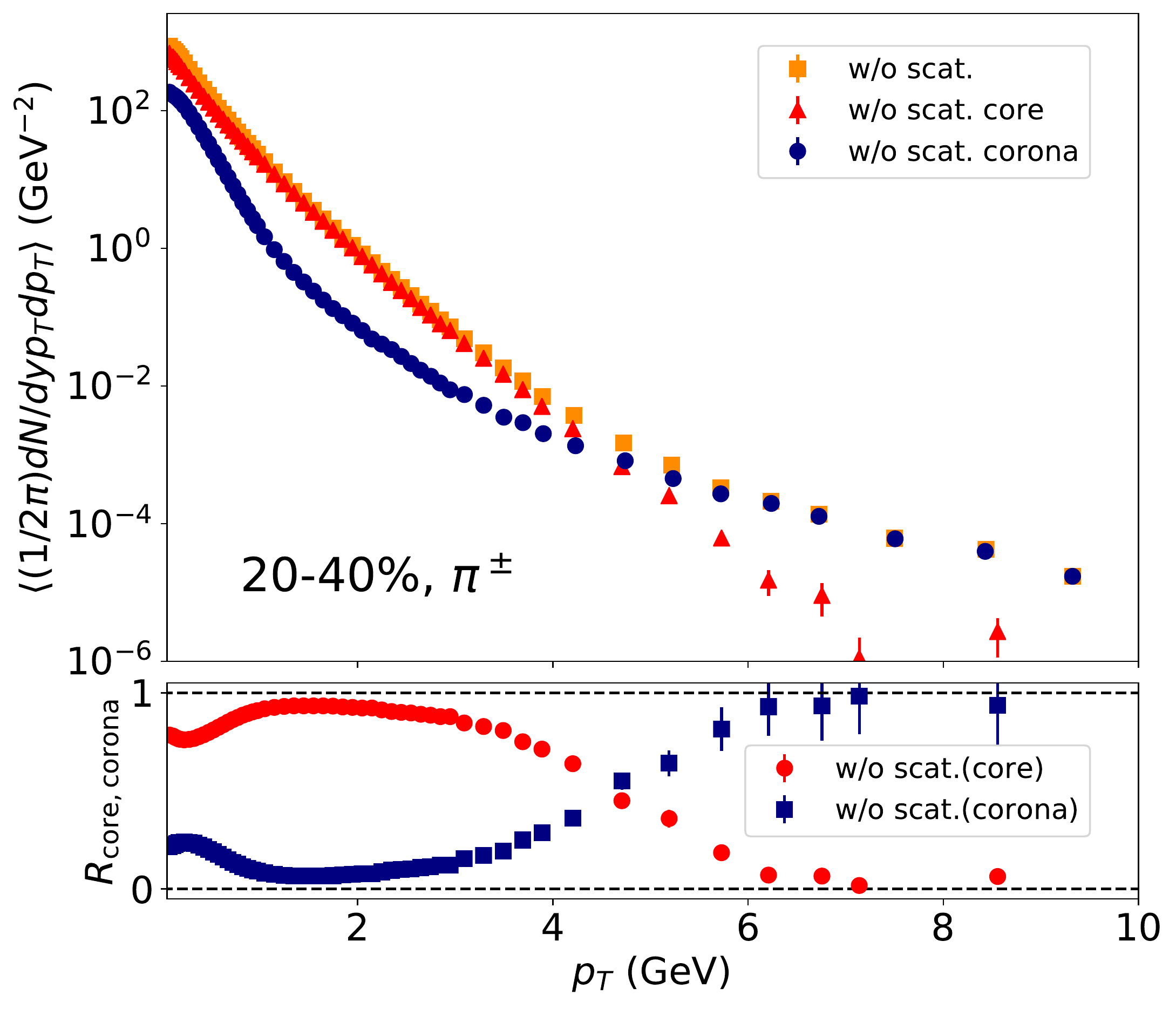}
    \includegraphics[bb=0 0 628 538, width=0.49\textwidth]{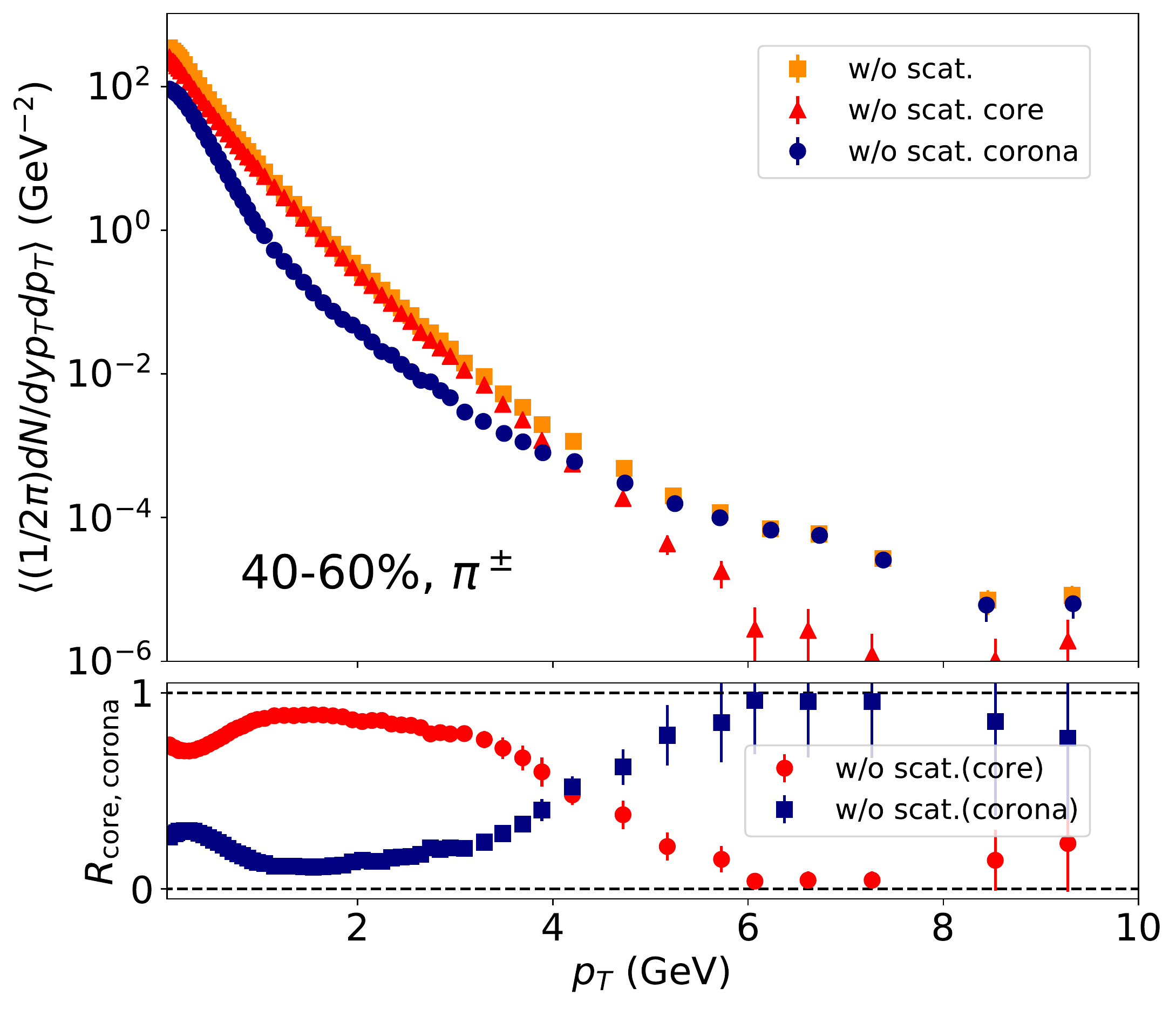}
    \includegraphics[bb=0 0 628 538, width=0.49\textwidth]{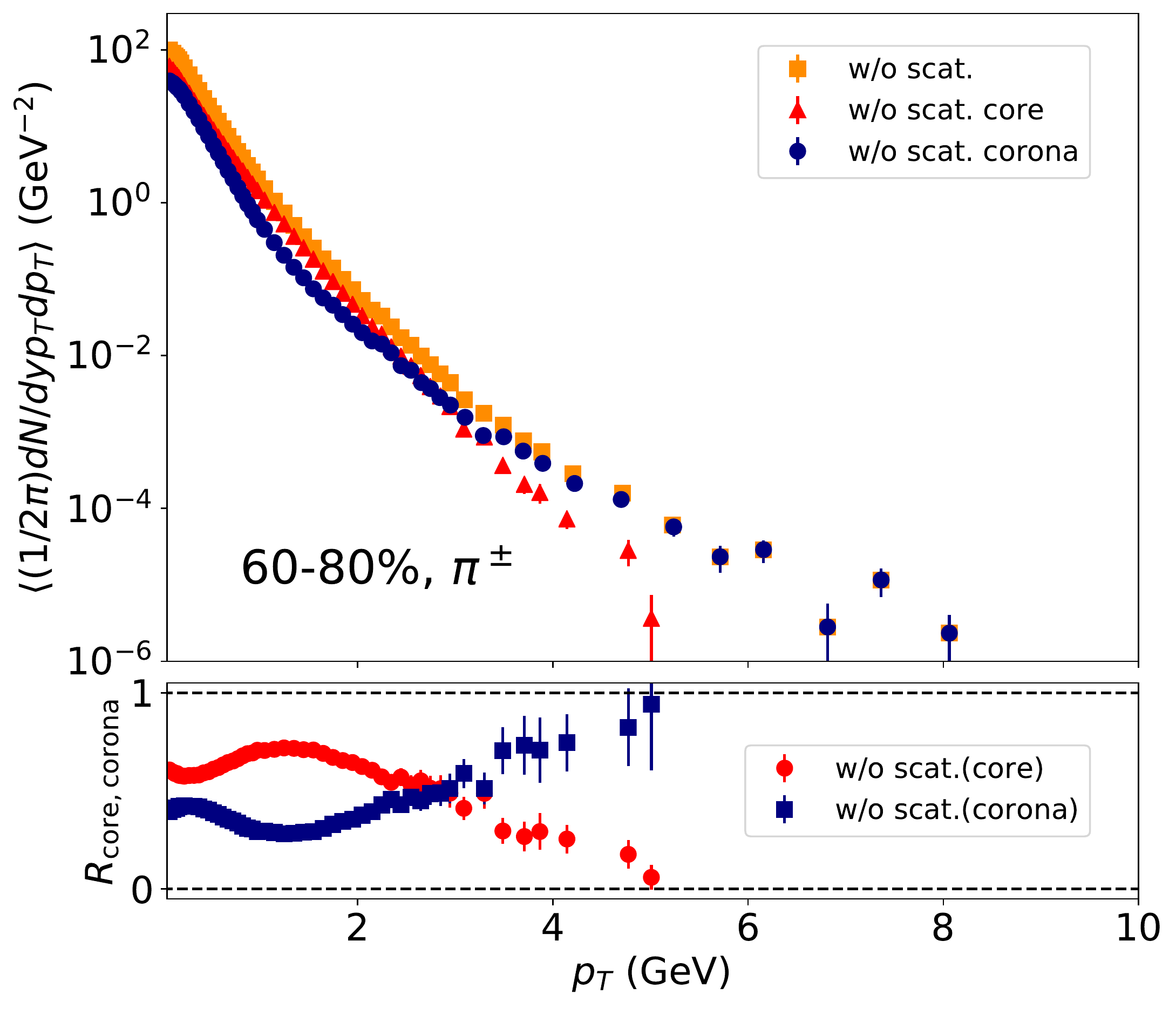}
    \caption{(Upper) Centrality dependence of $p_T$ spectra of charged pions ($\pi^+ + \pi^-$) in $Pb$+$Pb$ collisions at \snn = 2.76 TeV from DCCI2.
    Results with switching off hadronic rescatterings (orange squares) and their breakdown into core (red triangles) and corona contributions (blue circles) are shown.
    (Lower) Fraction of core (red circles) and corona (blue squares) components, $R_{\mathrm{core, corona}}$, in each $p_T$ bin. }
    \label{fig:PBPB2760_PTSPECTRA_PI_CORECORONA}
\end{figure}

\begin{figure}
    \centering
    \includegraphics[bb=0 0 628 538, width=0.49\textwidth]{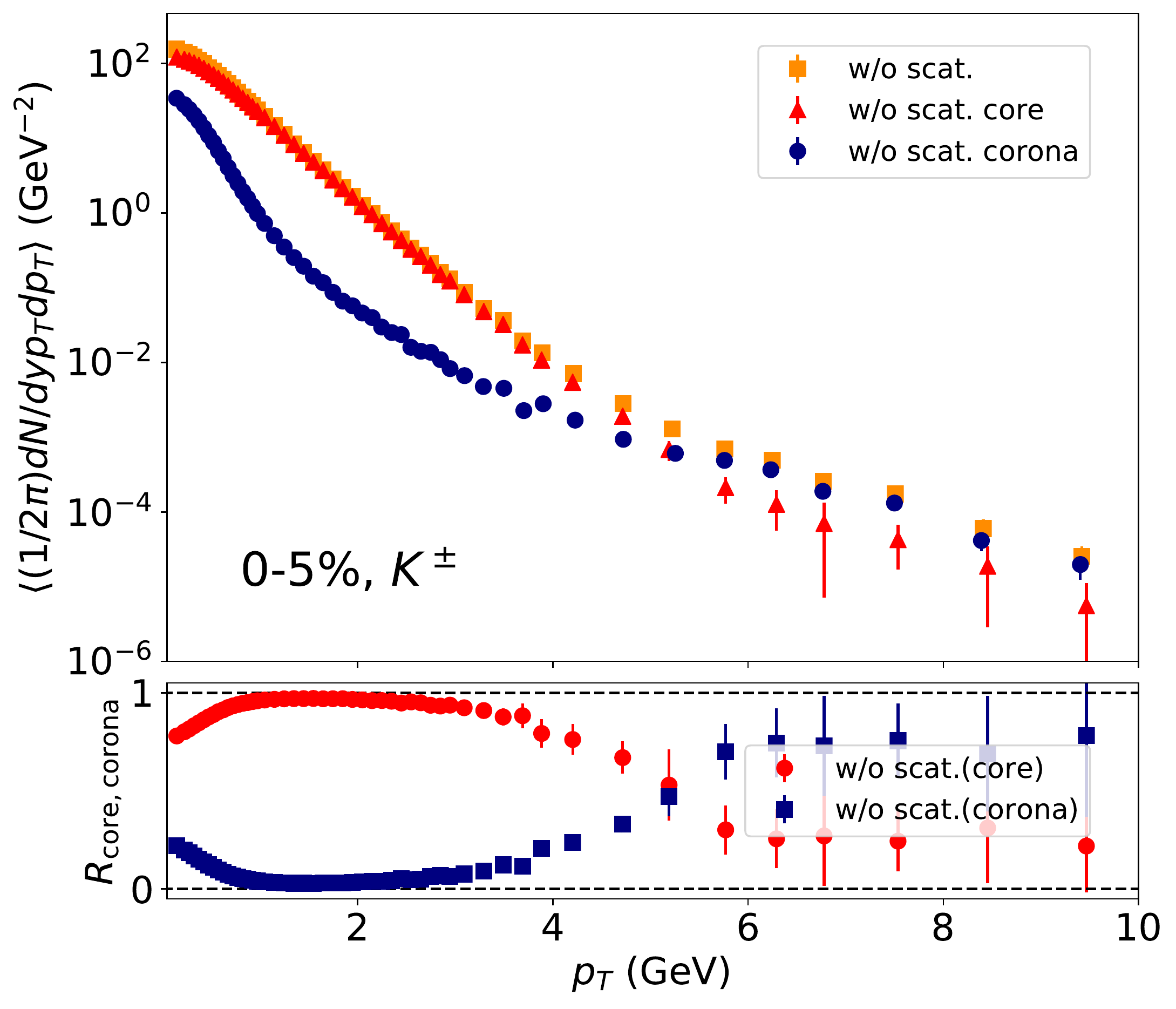}
    \includegraphics[bb=0 0 628 538, width=0.49\textwidth]{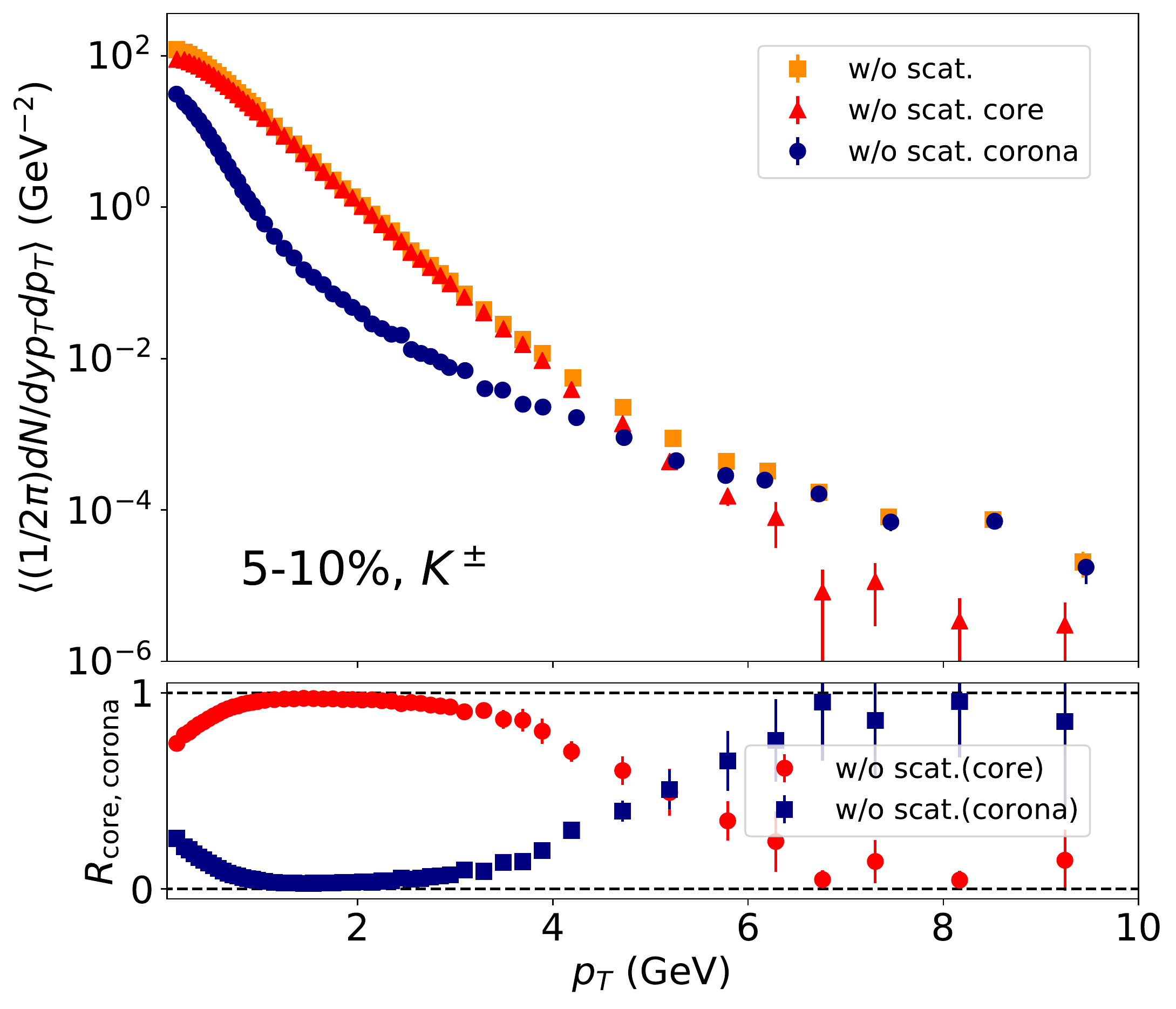}
    \includegraphics[bb=0 0 628 538, width=0.49\textwidth]{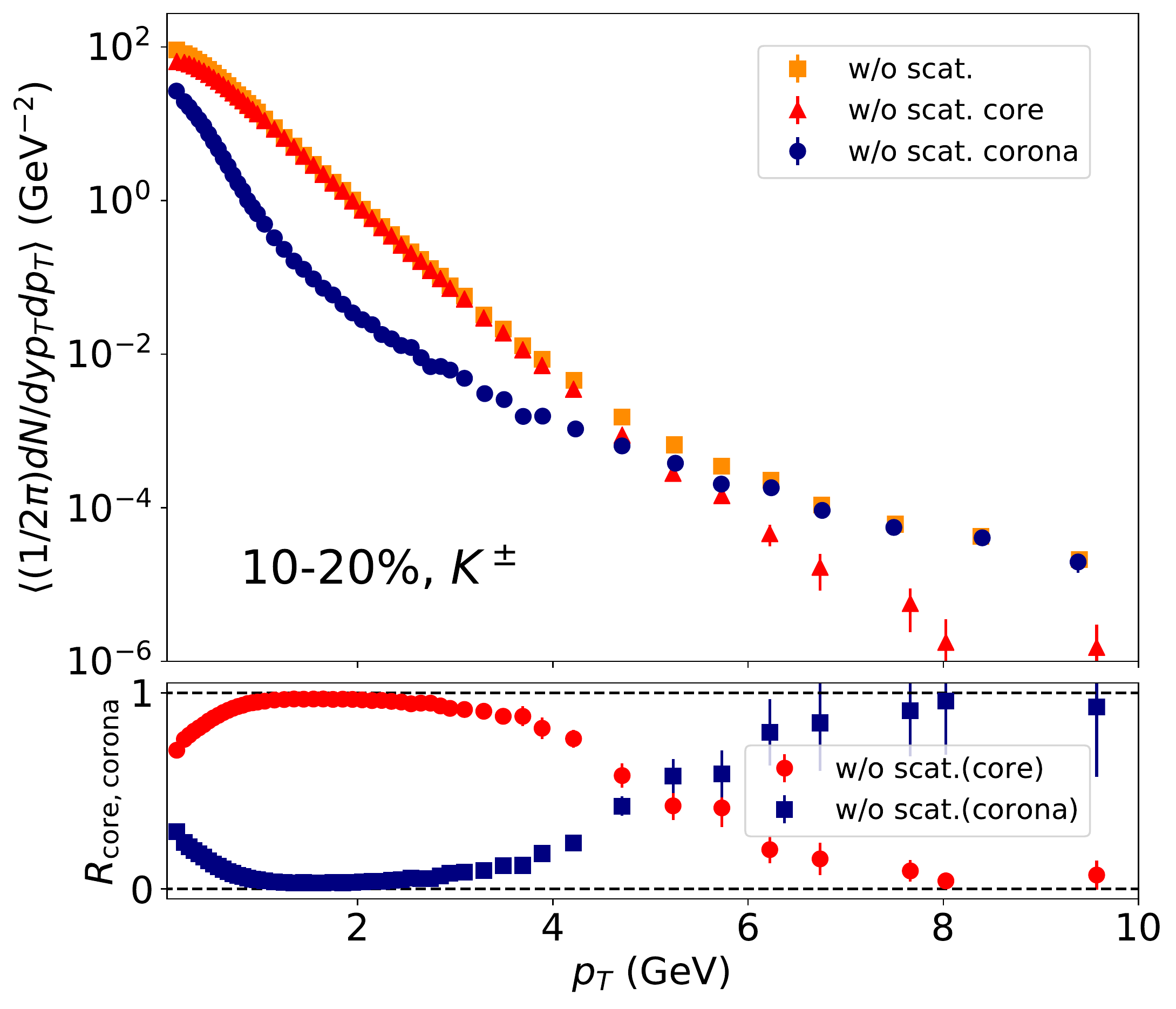}
    \includegraphics[bb=0 0 628 538, width=0.49\textwidth]{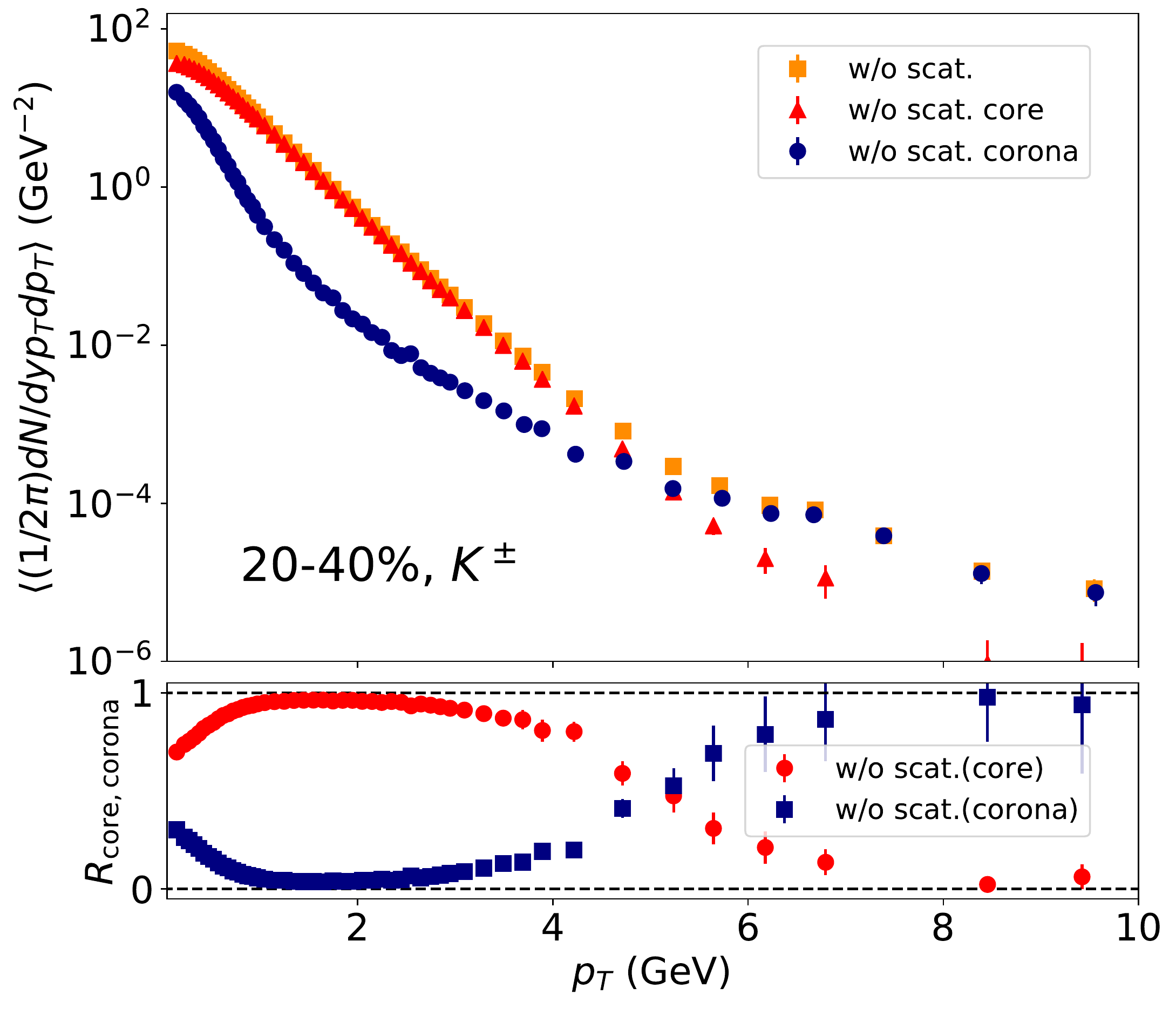}
    \includegraphics[bb=0 0 628 538, width=0.49\textwidth]{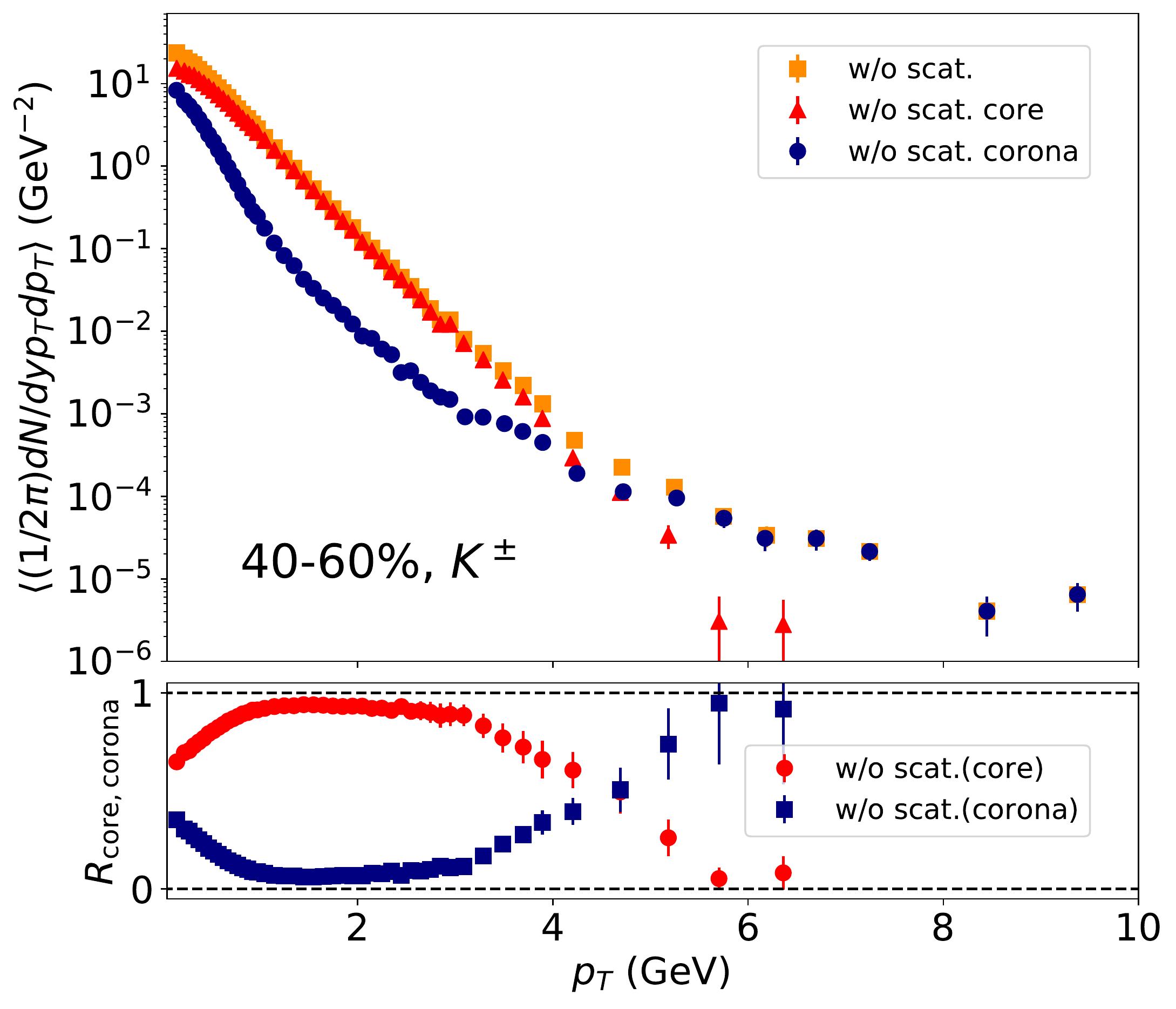}
    \includegraphics[bb=0 0 628 538, width=0.49\textwidth]{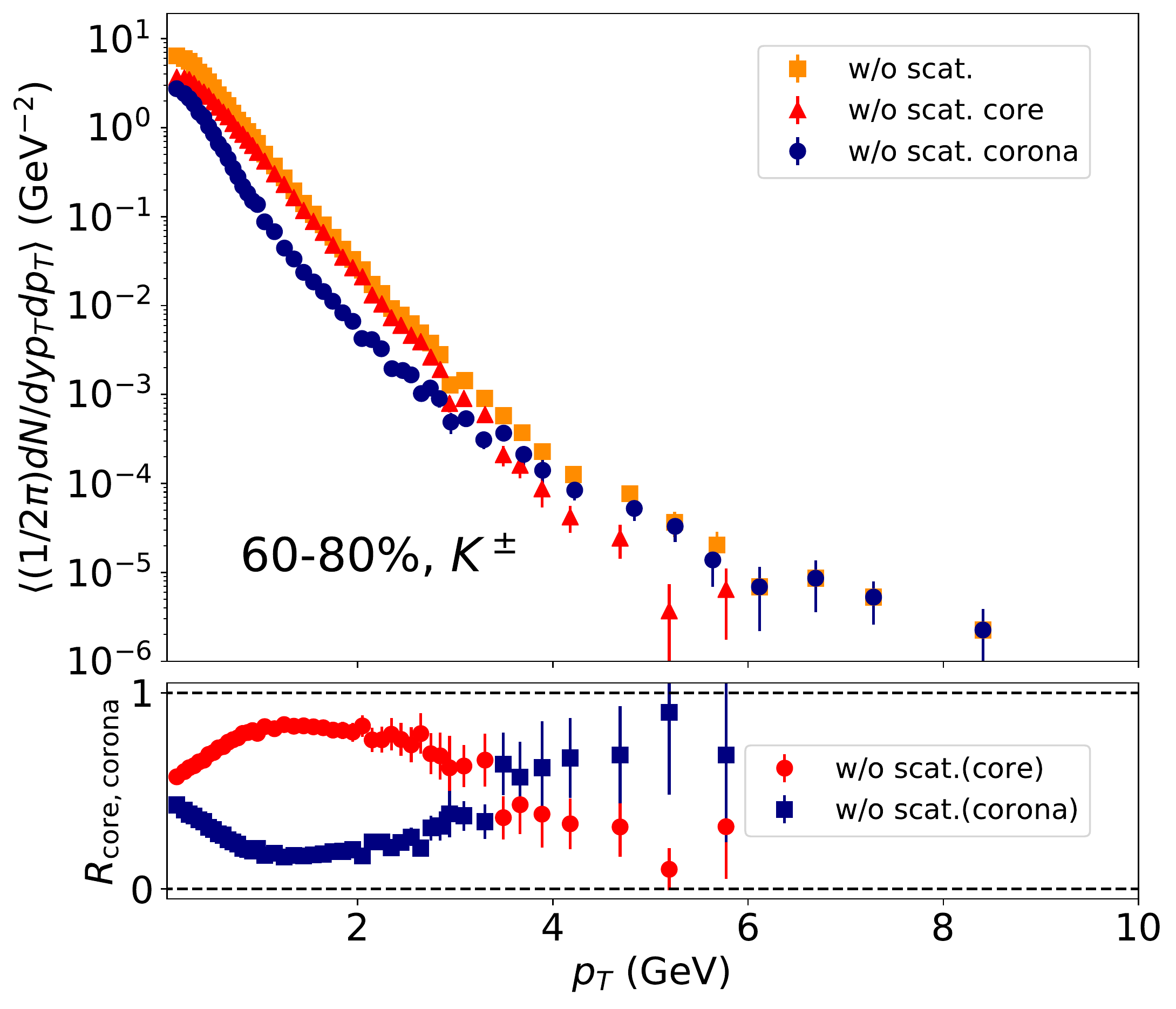}
    \caption{(Upper) Centrality dependence of $p_T$ spectra of charged kaons ($K^+ + K^-$) in $Pb$+$Pb$ collisions at \snn = 2.76 TeV from DCCI2.
    Results with switching off hadronic rescatterings (orange squares) and their breakdown into core (red triangles) and corona contributions (blue circles) are shown.
    (Lower) Fraction of core (red circles) and corona (blue squares) components, $R_{\mathrm{core, corona}}$, in each $p_T$ bin. }
    \label{fig:PBPB2760_PTSPECTRA_K_CORECORONA}
\end{figure}

\begin{figure}
    \centering
    \includegraphics[bb=0 0 628 538, width=0.49\textwidth]{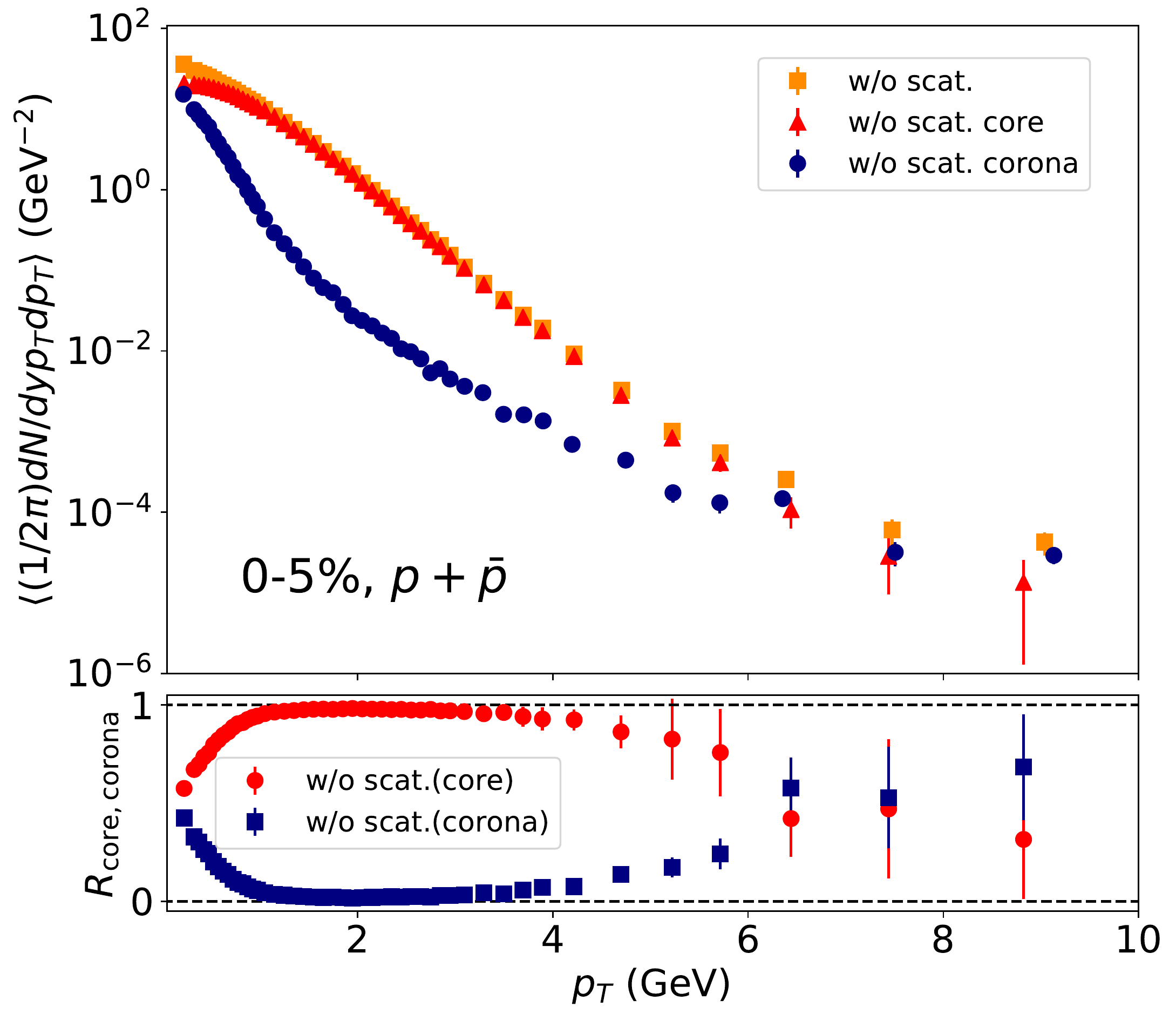}
    \includegraphics[bb=0 0 628 538, width=0.49\textwidth]{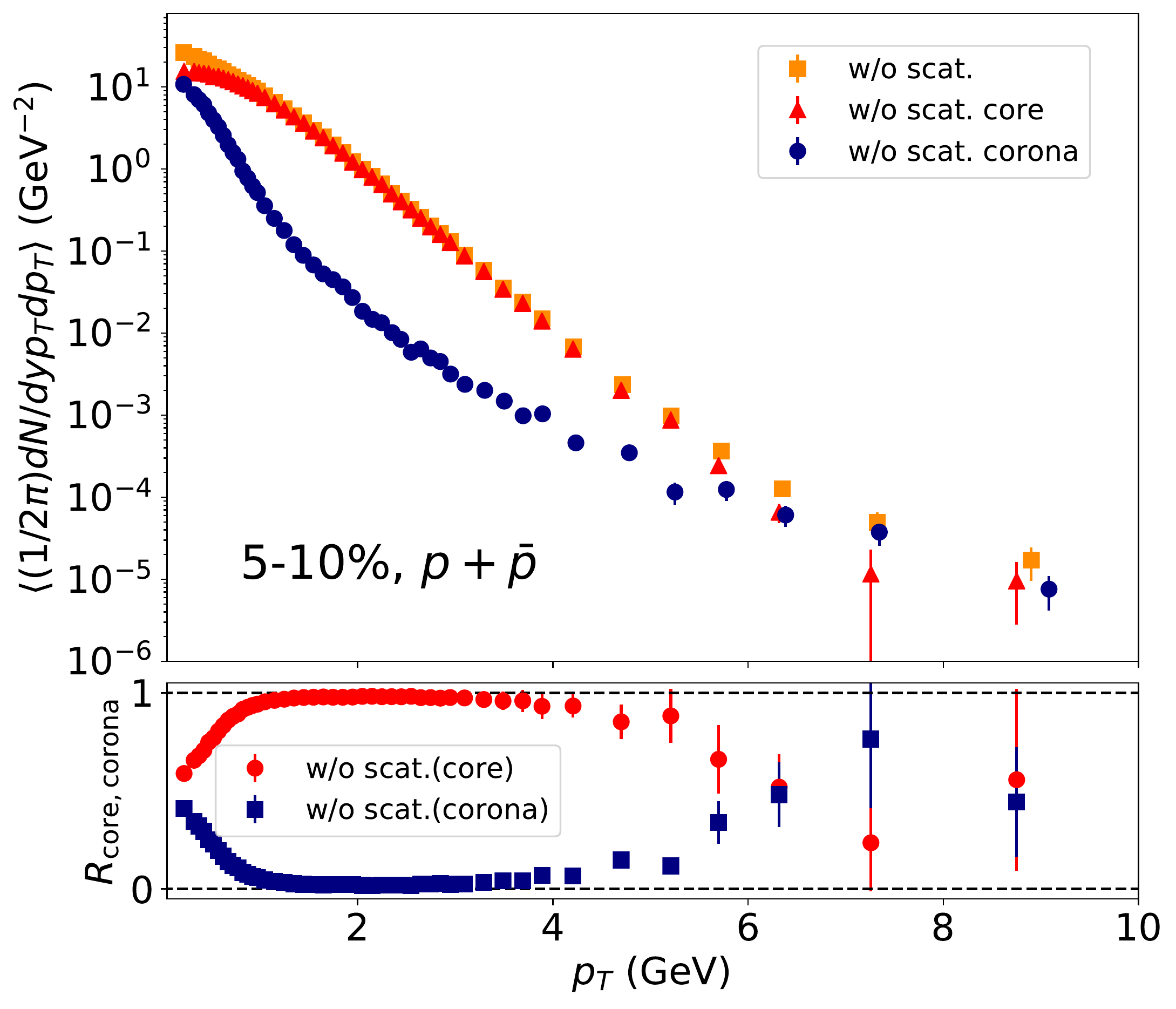}
    \includegraphics[bb=0 0 628 538, width=0.49\textwidth]{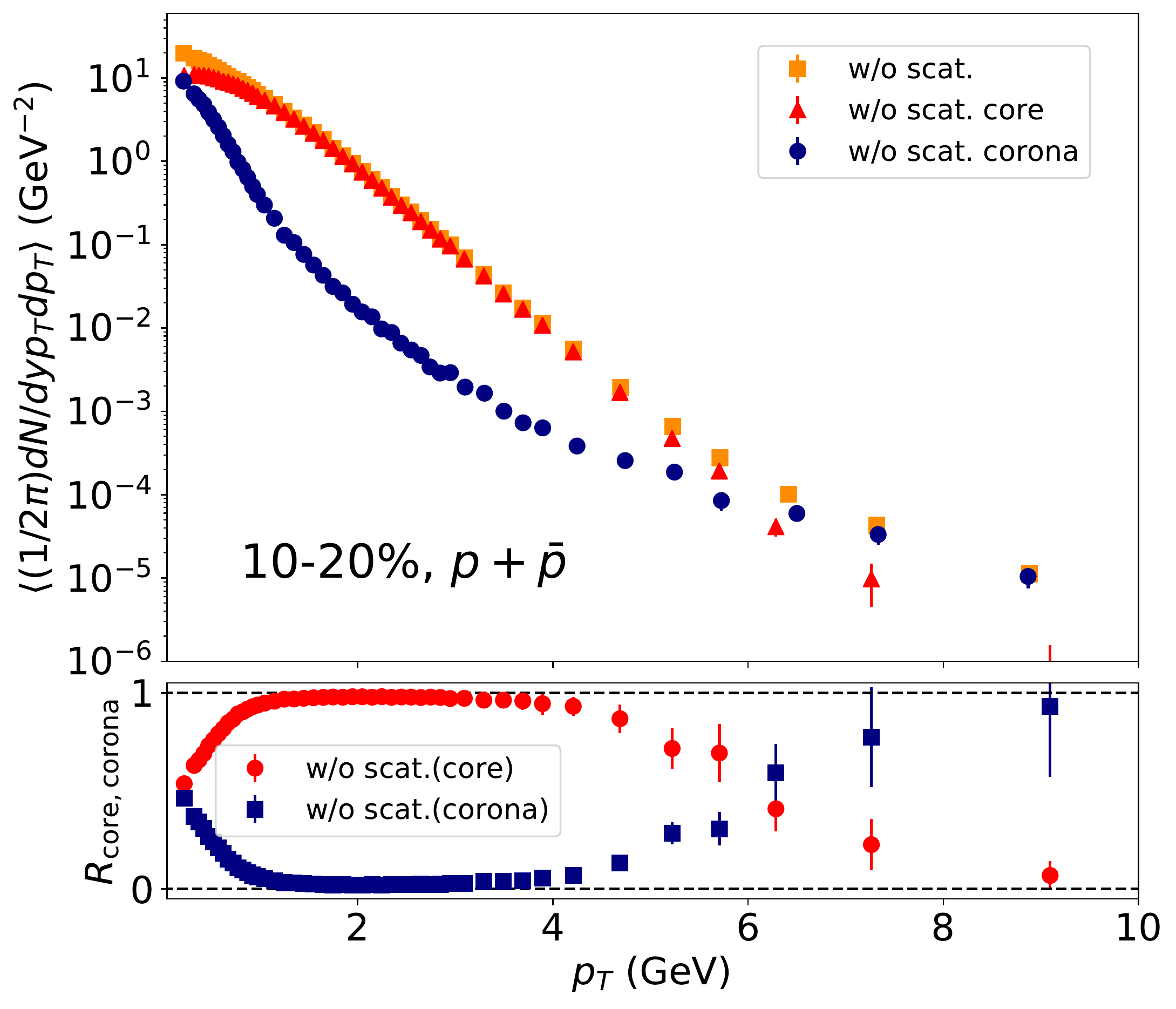}
    \includegraphics[bb=0 0 628 538, width=0.49\textwidth]{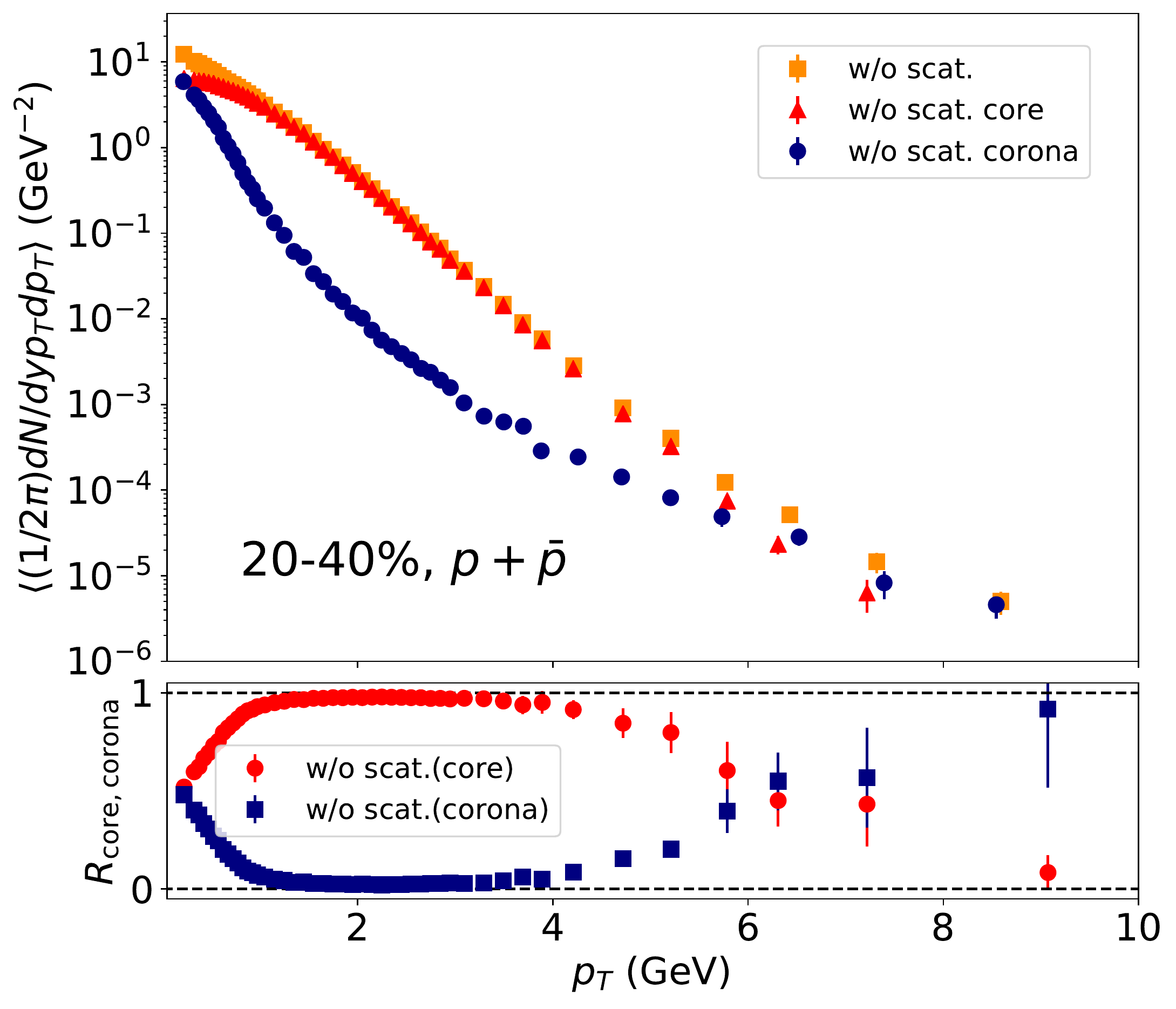}
    \includegraphics[bb=0 0 628 538, width=0.49\textwidth]{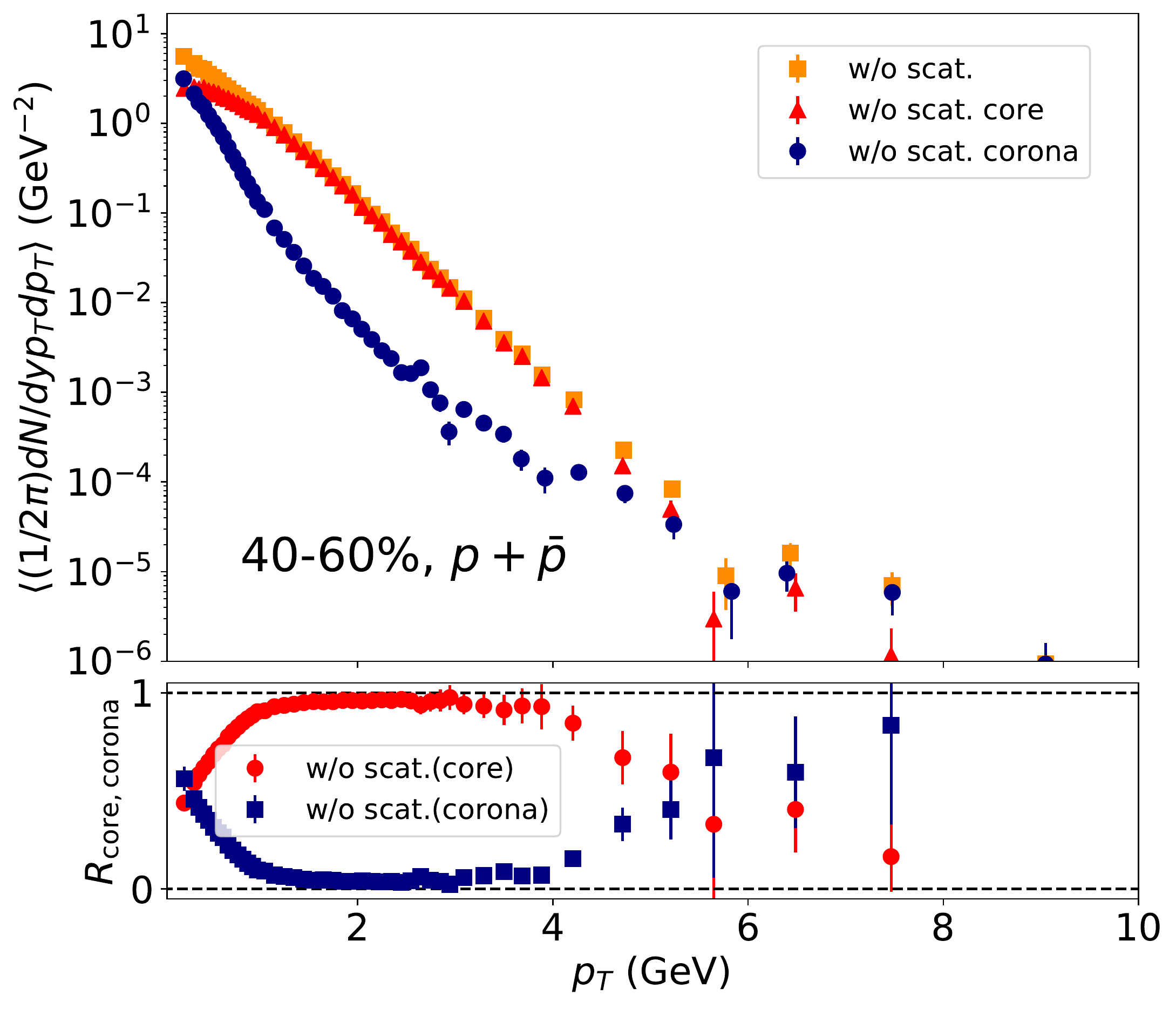}
    \includegraphics[bb=0 0 628 538, width=0.49\textwidth]{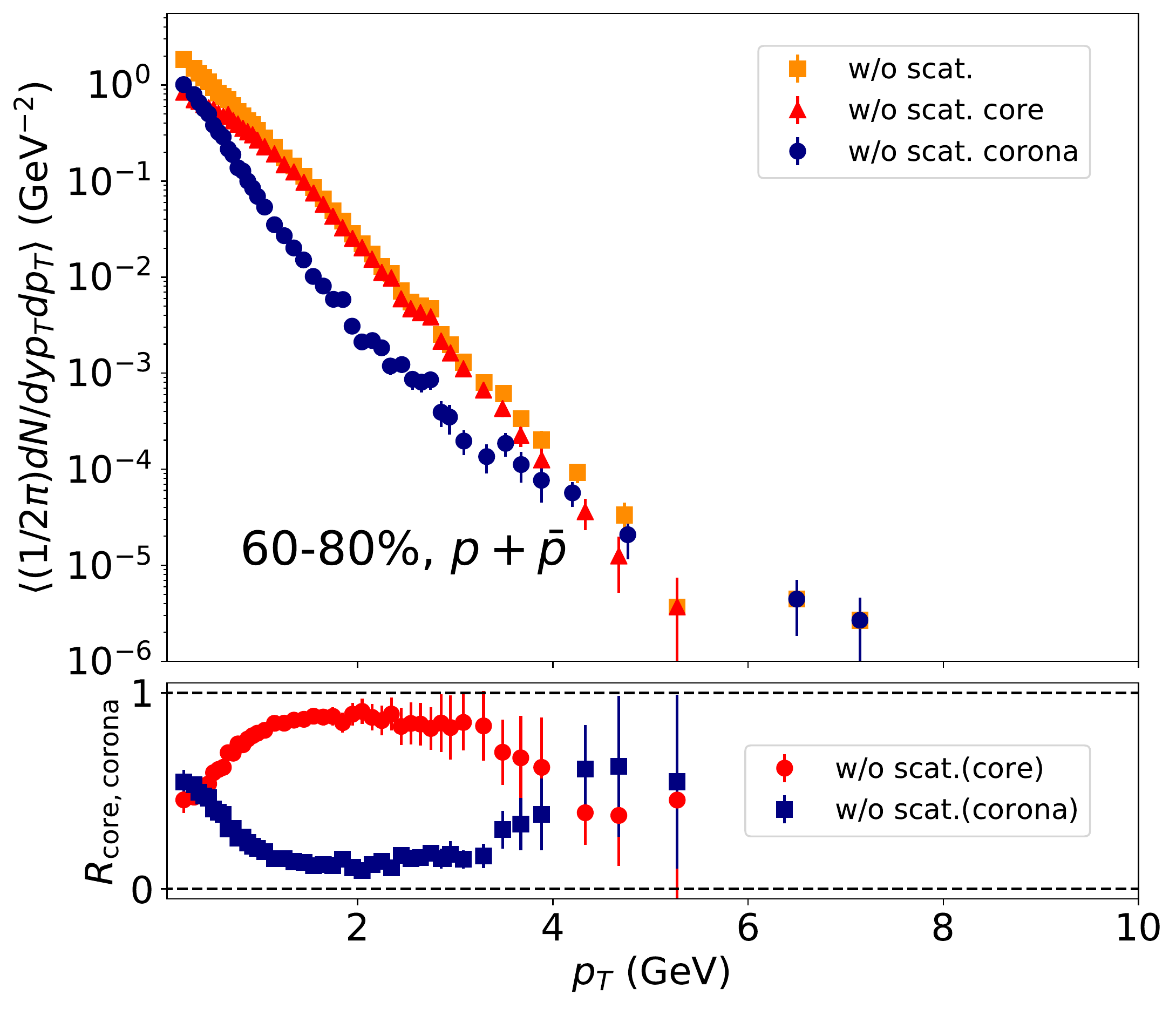}
    \caption{(Upper) Centrality dependence of $p_T$ spectra of protons and antiprotons ($p+ \bar{p}$) in $Pb$+$Pb$ collisions at \snn = 2.76 TeV from DCCI2.
    Results with switching off hadronic rescatterings (orange squares) and their breakdown into core (red triangles) and corona contributions (blue circles) are shown.
    (Lower) Fraction of core (red circles) and corona (blue squares) components, $R_{\mathrm{core, corona}}$, in each $p_T$ bin. }
    \label{fig:PBPB2760_PTSPECTRA_P_CORECORONA}
\end{figure}

Figures \ref{fig:PBPB2760_PTSPECTRA_PI_CORECORONA}, \ref{fig:PBPB2760_PTSPECTRA_K_CORECORONA}, and \ref{fig:PBPB2760_PTSPECTRA_P_CORECORONA} show centrality dependence of $p_T$ spectra of charged pions, charged kaons, and protons out of final hadrons from simulations with switching off hadronic rescatterings and their breakdown into core and corona contribution in $Pb$+$Pb$ collisions at \snn = 2.76 TeV from DCCI2.
Being different from $p$+$p$ collisions in which productions from core dominate in minimum-bias events,
productions from core are the major components in the total particle productions in minimum-bias $Pb$+$Pb$ collisions as we see in Fig.~\ref{fig:MULTIPLICITY_PP_PBPB}.
Note that the shape of $p_T$ spectra with full simulation or experimental data, especially for protons, is different from those results with switching off hadronic rescatterings.
This is because that the effect of pion wind is larger in high-multiplicity events if there are hadronic rescatterings in the late stage.
It can be also expected that there would be more particles whose origin cannot be identified due to mixing of core and corona components in inelastic scatterings compared to $p$+$p$ collisions.

Keeping those in mind, 
first the absolute values of core components become large at central events for overall $p_T$ regime in all particle species.
Second, one would see that the slopes of $p_T$ spectra of core components become flat at central events,
which originates from the existence of stronger radial flows at, roughly speaking, central events.
As a result of those, the overtakes of corona components at high $p_T$ happen higher $p_T$ compared at more central events.
It should be noted that the dominance of core contributions at intermediate $p_T$, the one seen in Fig.~\ref{fig:PTSPECTRA_PP_PBPB} (b), is seen in those identified $p_T$ spectra for wide range of centrality and every species.
As the same as the results in $p$+$p$ collisions, 
the core dominance in intermediate $p_T$ regime becomes stronger and the regime becomes wider for heavier particles and central events.
Within DCCI2, the protons from core components span around $0<p_T<6$ GeV at the most central events ($0$-$5\%$).
It should also be noted that the core components possibly are further pushed towards higher $p_T$ because of the dynamical initialization \cite{Okai:2017ofp}.
Because I dynamically fluidize four momentum of partons and generate initial conditions of hydrodynamic equations,
medium profile has initial velocities which produce larger radial flows compared to static initial profiles.

Here as well, the relative enhancement of corona contributions at very low $p_T$ regime is seen and becomes
larger for heavier particles.
This is because the relative enhancement of corona contribution at very low $p_T$ is more pronounced when core contributions get pushed towards higher $p_T$.

\begin{figure}
    \centering
    \includegraphics[bb = 0 0 564 567, width=0.45\textwidth]{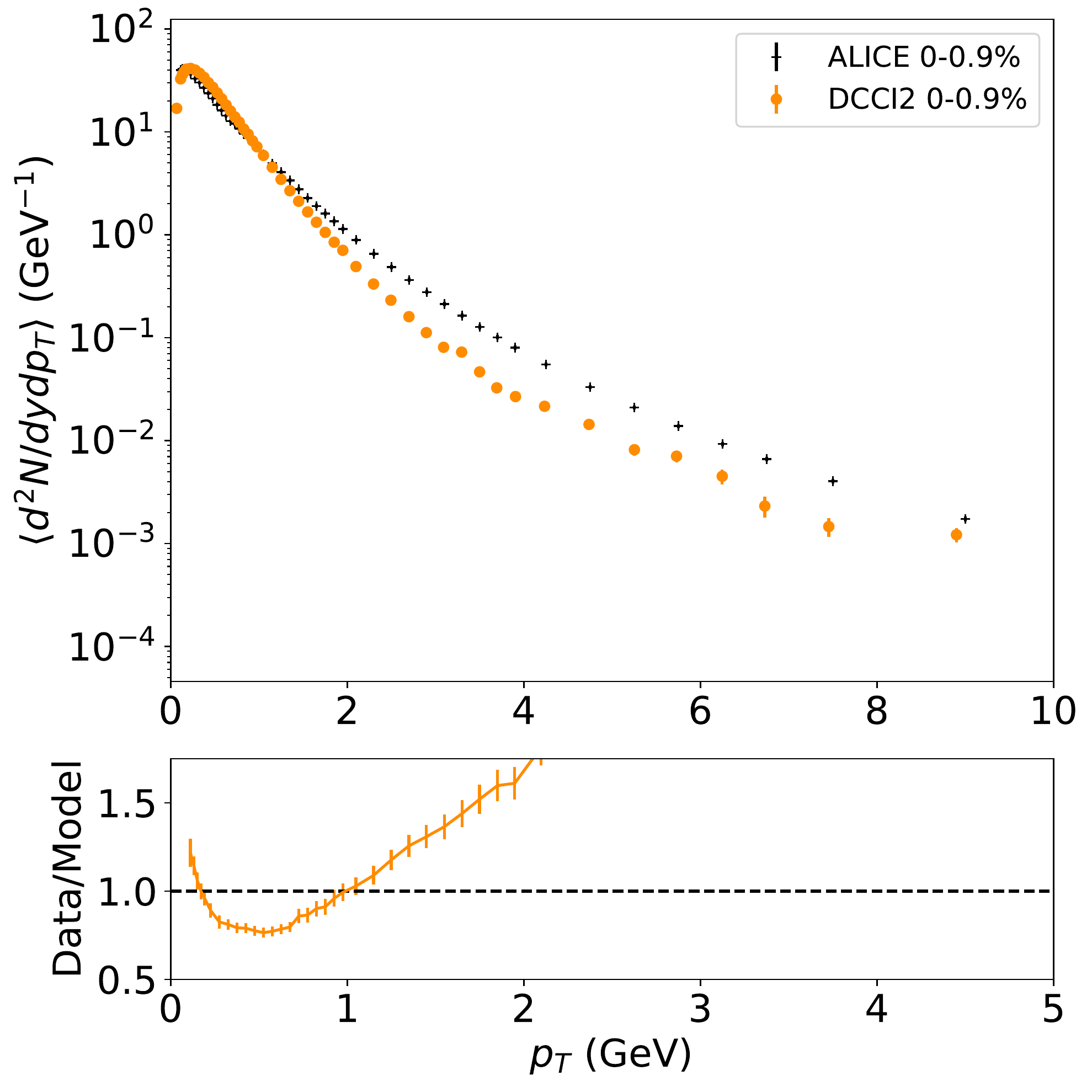}
    \includegraphics[bb = 0 0 564 567, width=0.45\textwidth]{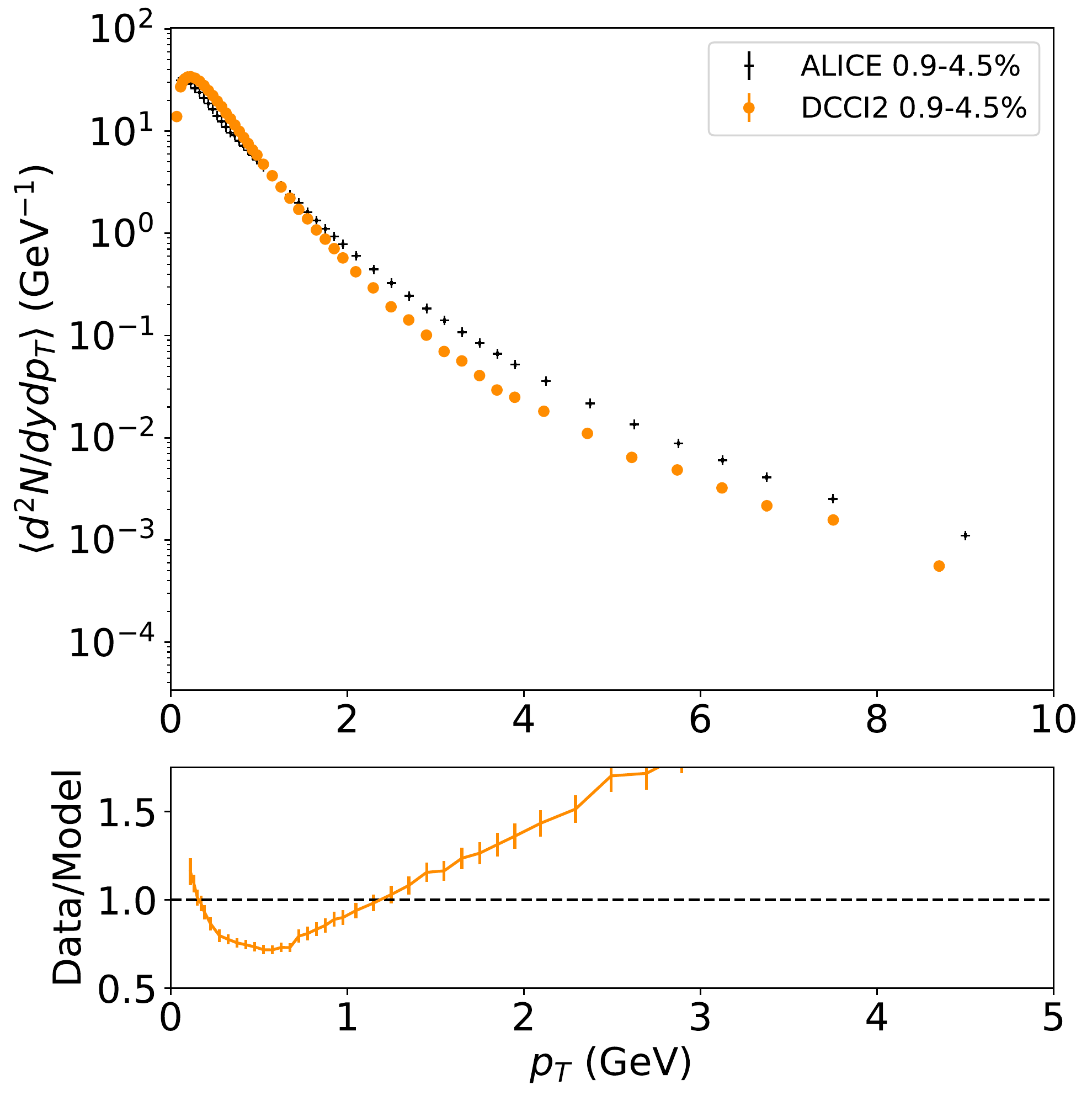}
    \includegraphics[bb = 0 0 564 567, width=0.45\textwidth]{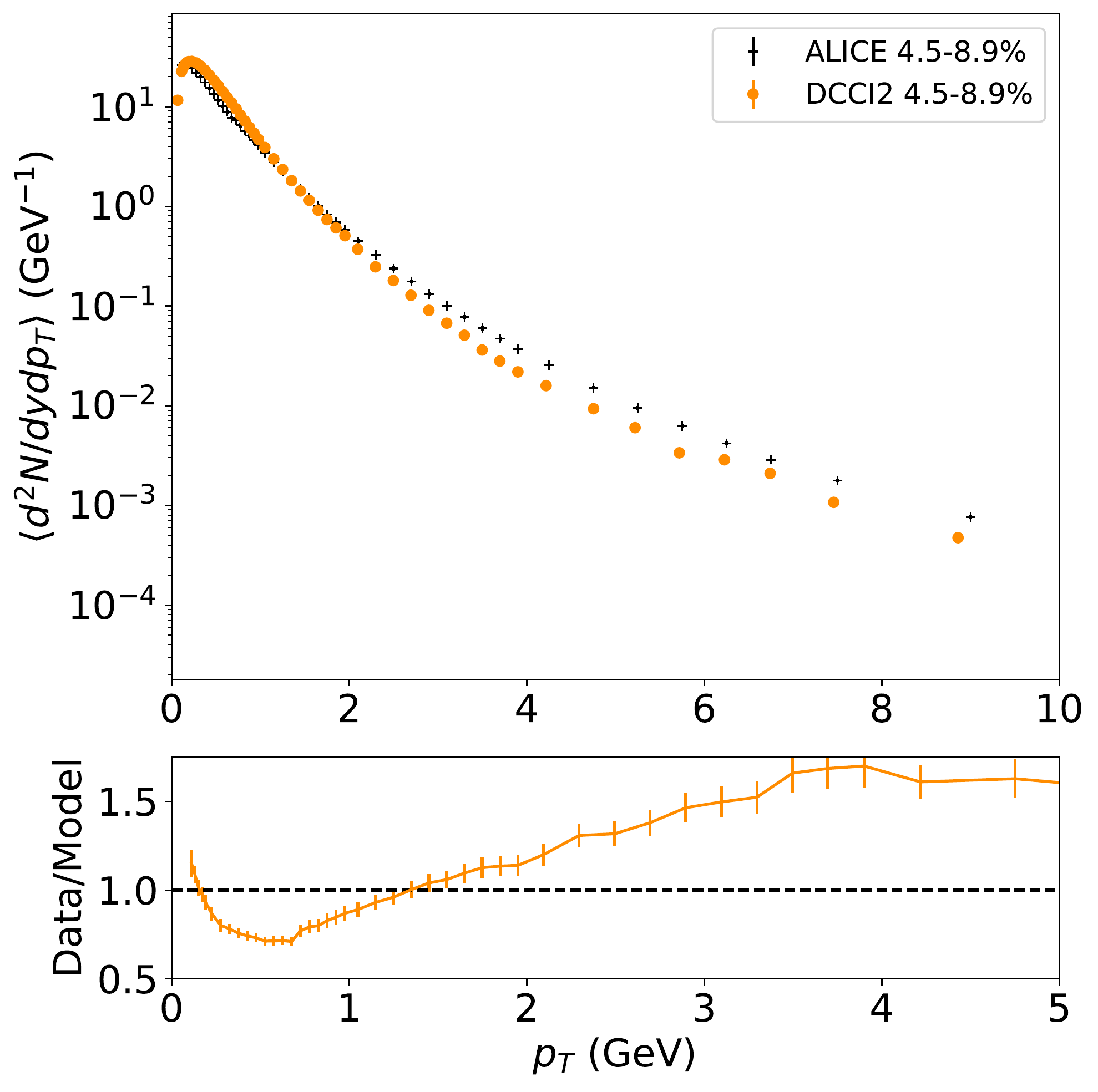}
    \includegraphics[bb = 0 0 564 567, width=0.45\textwidth]{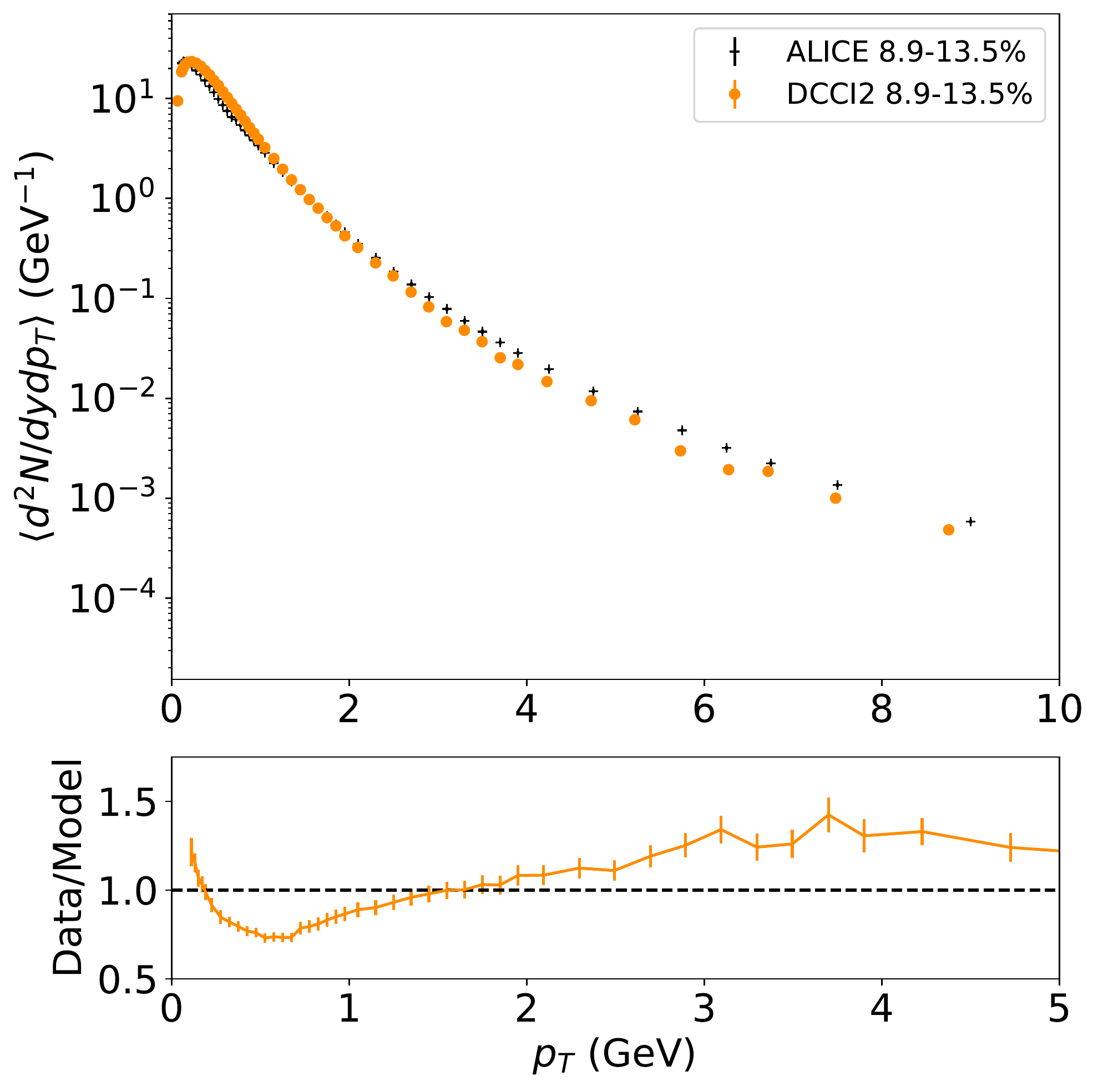}
    \includegraphics[bb = 0 0 564 567, width=0.45\textwidth]{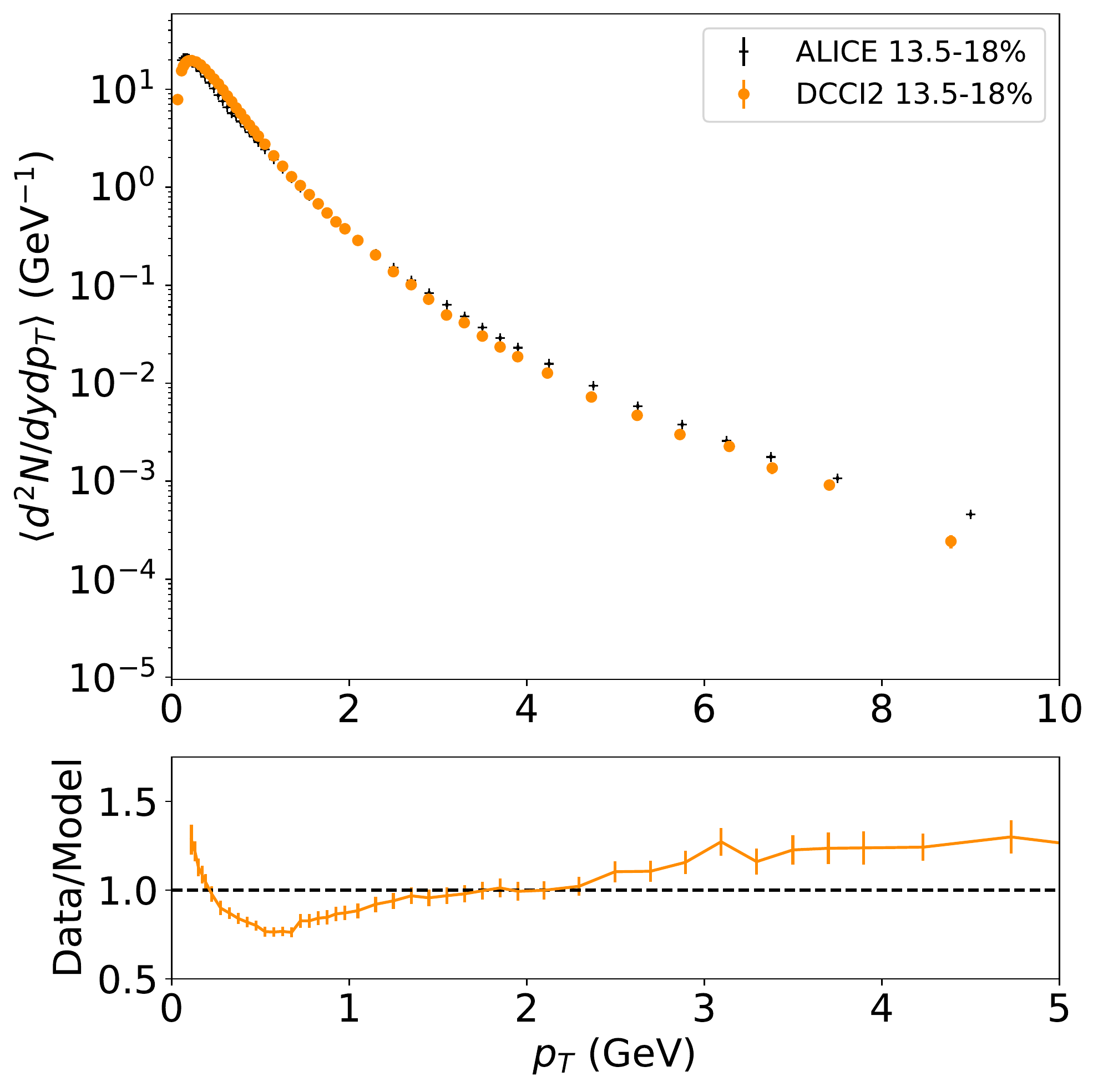}
    \includegraphics[bb = 0 0 564 567, width=0.45\textwidth]{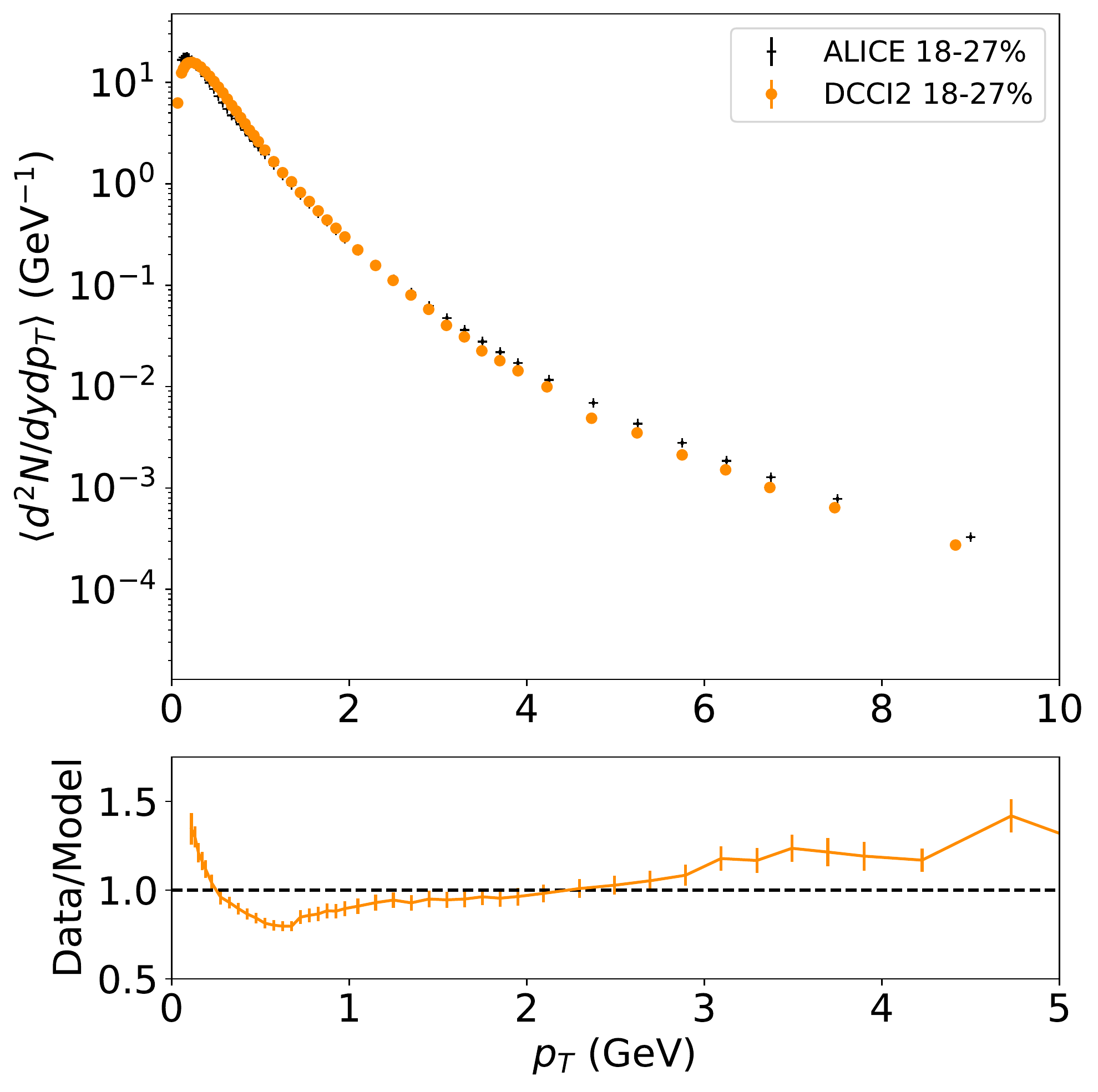}
    \caption{(Upper) Centrality dependence of transverse momentum spectra of charged pions ($\pi^+ + \pi^-$) from $p$+$p$ collisions at \snn[proton]=13 TeV in DCCI2. Comparison between results from DCCI2 (orange circles) and the ALICE experimental data (black crosses) are made.
    (Lower) Ratio of the ALICE experimental data to DCCI2 results at each $p_T$ bin. Centrality classes from 0-0.9\% to 18-27\% are shown.}
    \label{fig:PP13_PTSPECTRA_PI_1}
\end{figure}

\begin{figure}
    \centering
    \includegraphics[bb = 0 0 564 567, width=0.45\textwidth]{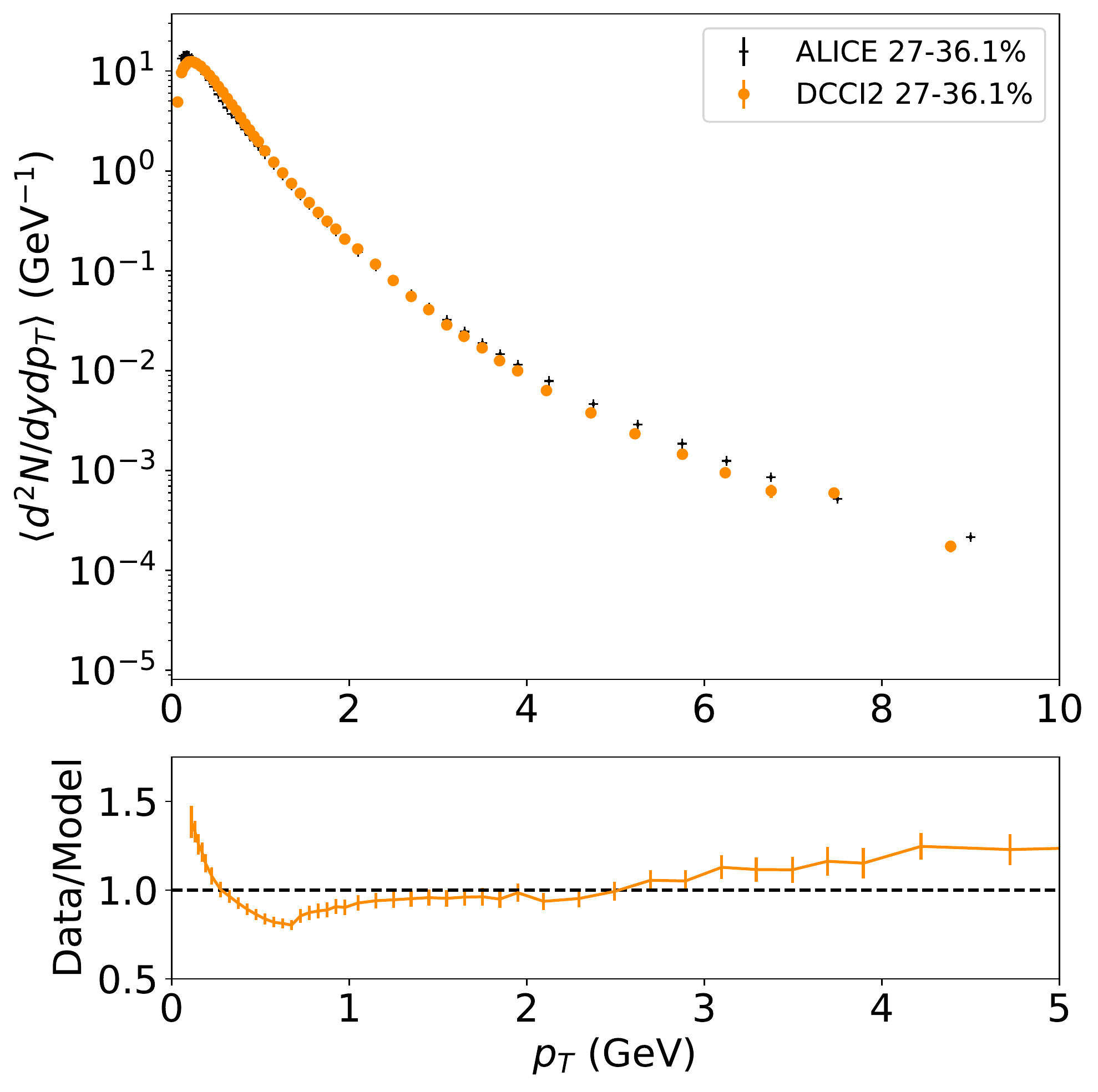}
    \includegraphics[bb = 0 0 564 567, width=0.45\textwidth]{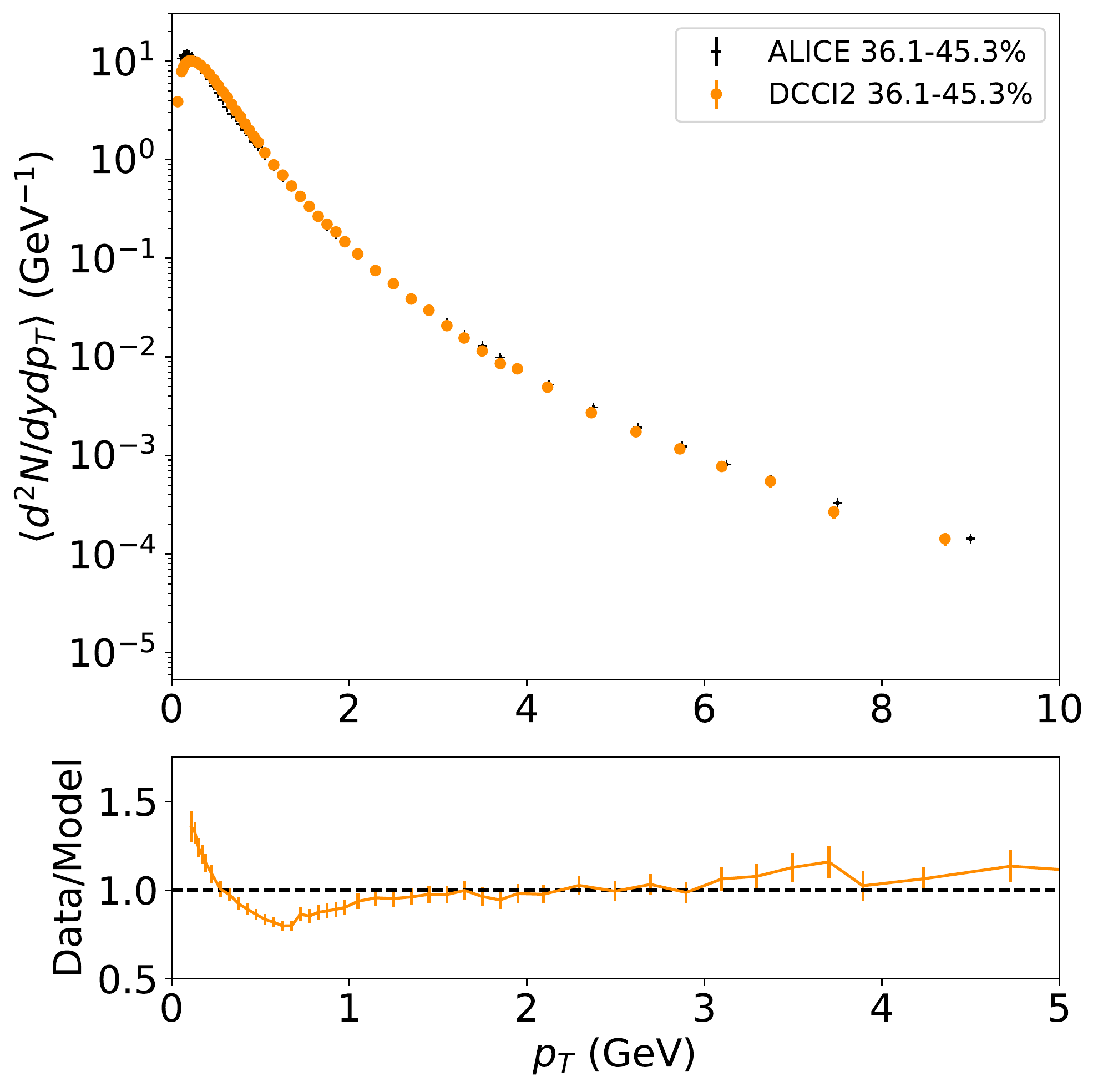}
    \includegraphics[bb = 0 0 564 567, width=0.45\textwidth]{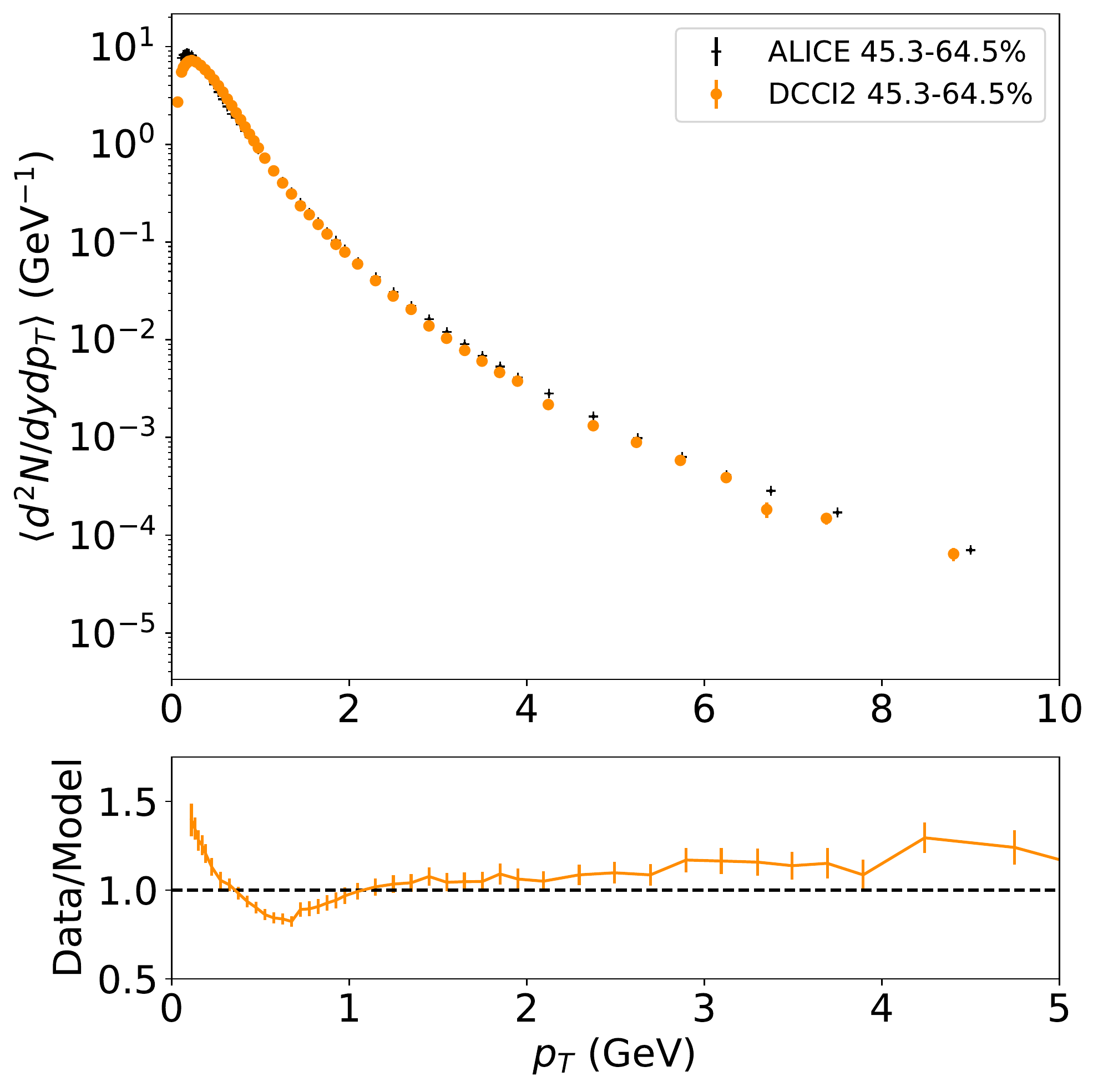}
    \includegraphics[bb = 0 0 564 567, width=0.45\textwidth]{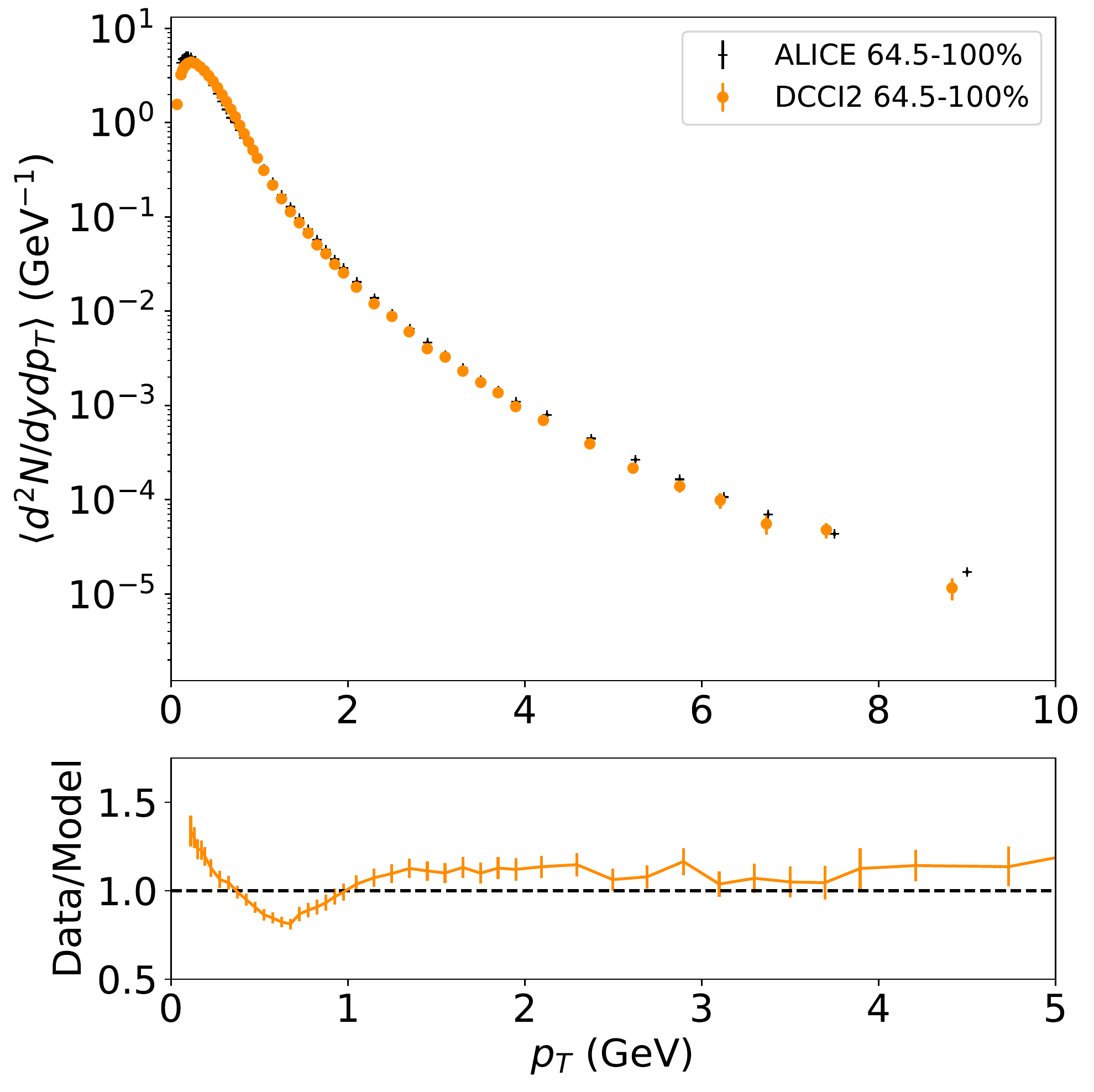}
    \caption{(Upper) Centrality dependence of transverse momentum spectra of charged pions ($\pi^+ + \pi^-$) from $p$+$p$ collisions at \snn[proton]=13 TeV in DCCI2. Comparison between results from DCCI2 (orange circles) and the ALICE experimental data (black crosses) are made.
    (Lower) Ratio of the ALICE experimental data to DCCI2 results at each $p_T$ bin. Centrality classes from 27-36.1\% to 64.5-100\% are shown.}
    \label{fig:PP13_PTSPECTRA_PI_2}
\end{figure}

\begin{figure}
    \centering
    \includegraphics[bb = 0 0 564 567, width=0.45\textwidth]{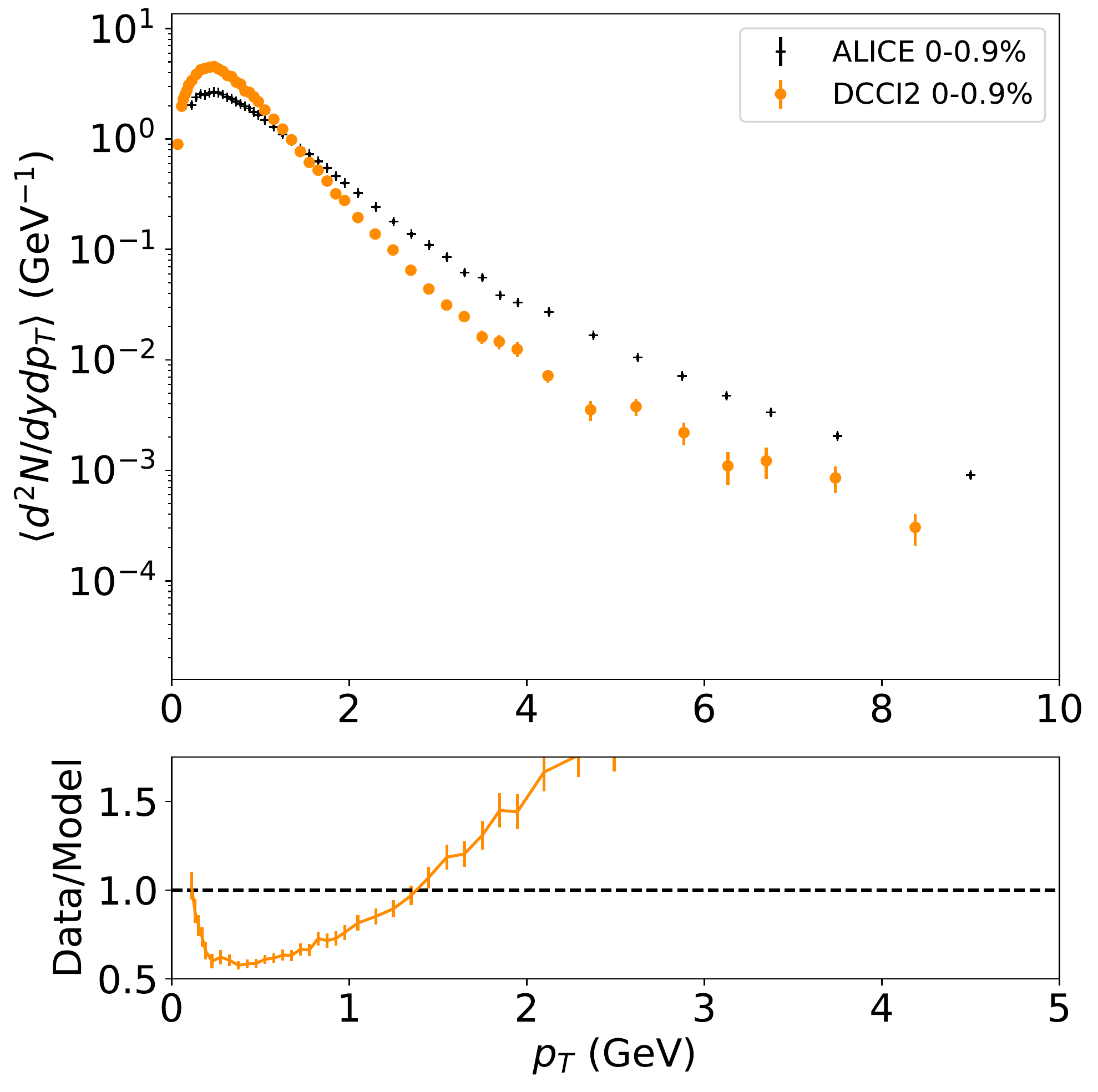}
    \includegraphics[bb = 0 0 564 567, width=0.45\textwidth]{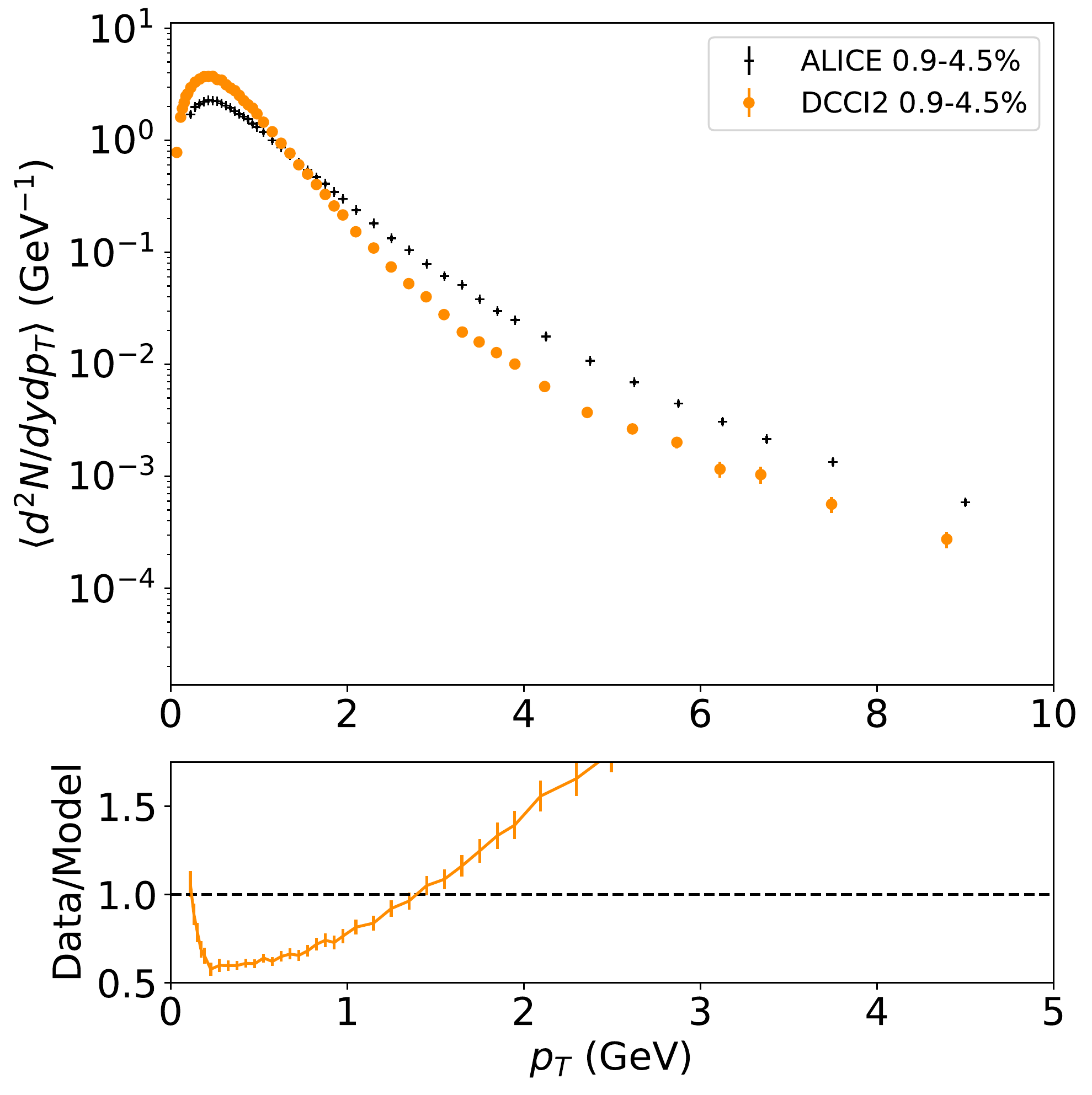}
    \includegraphics[bb = 0 0 564 567, width=0.45\textwidth]{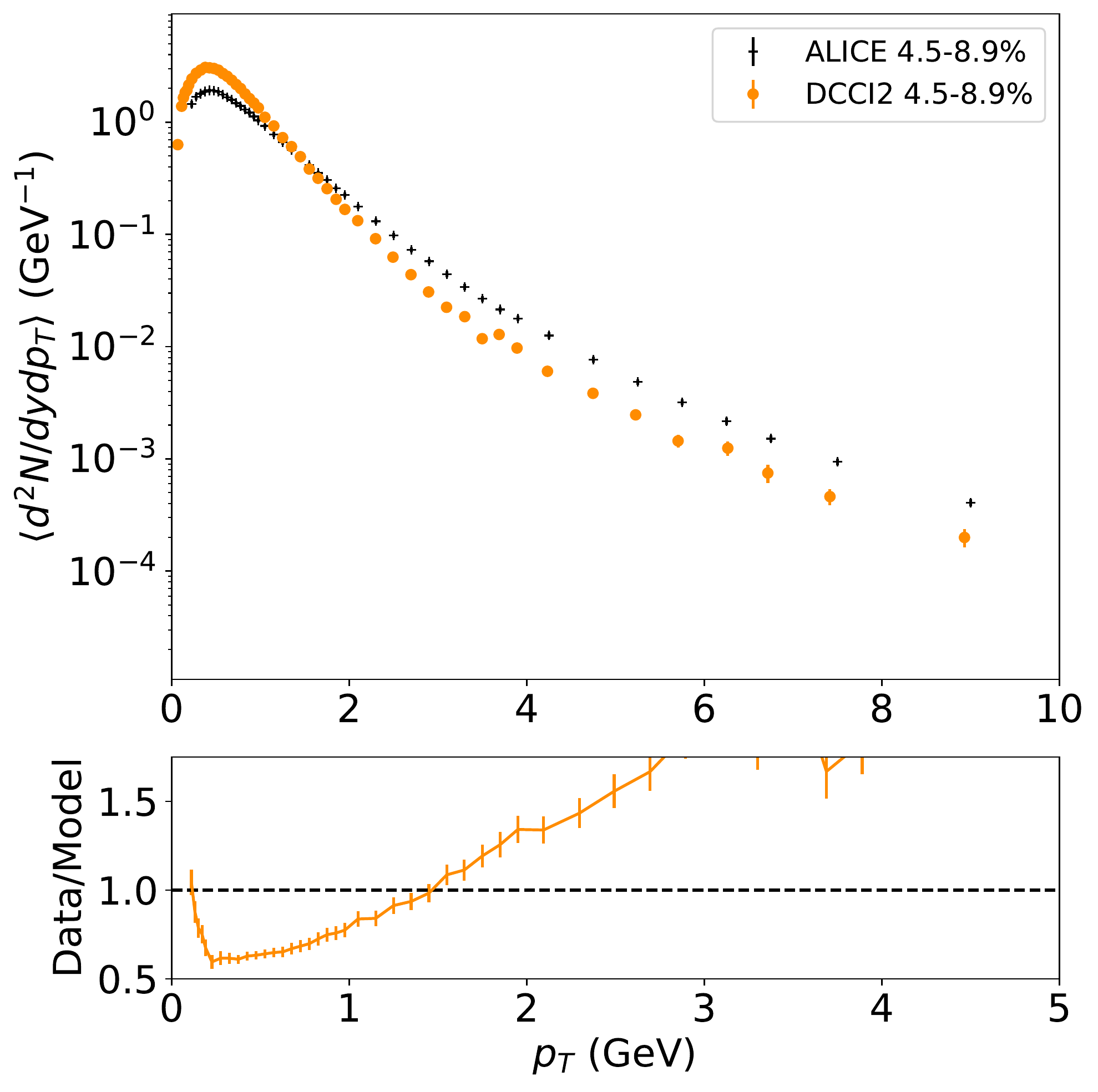}
    \includegraphics[bb = 0 0 564 567, width=0.45\textwidth]{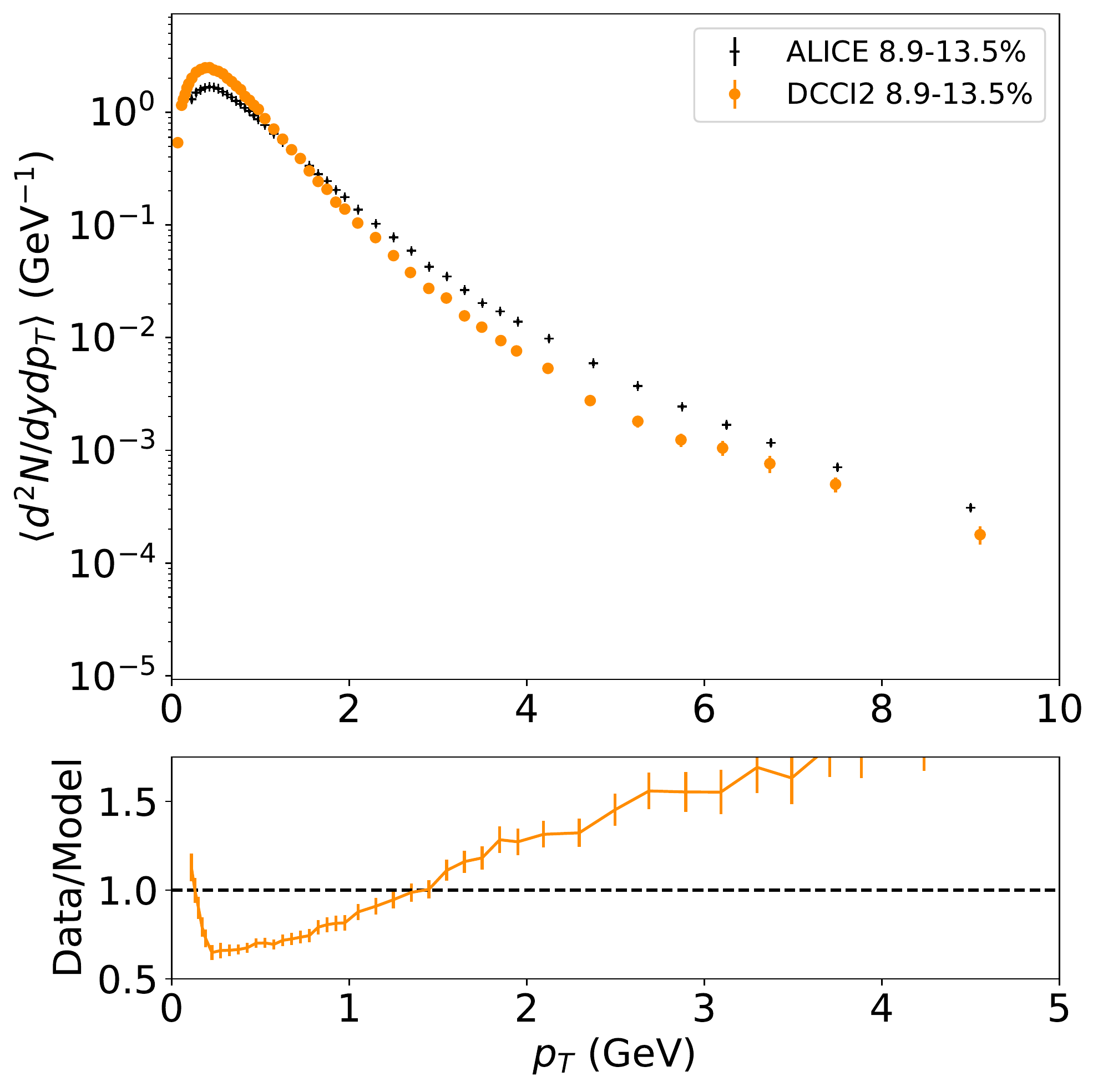}
    \includegraphics[bb = 0 0 564 567, width=0.45\textwidth]{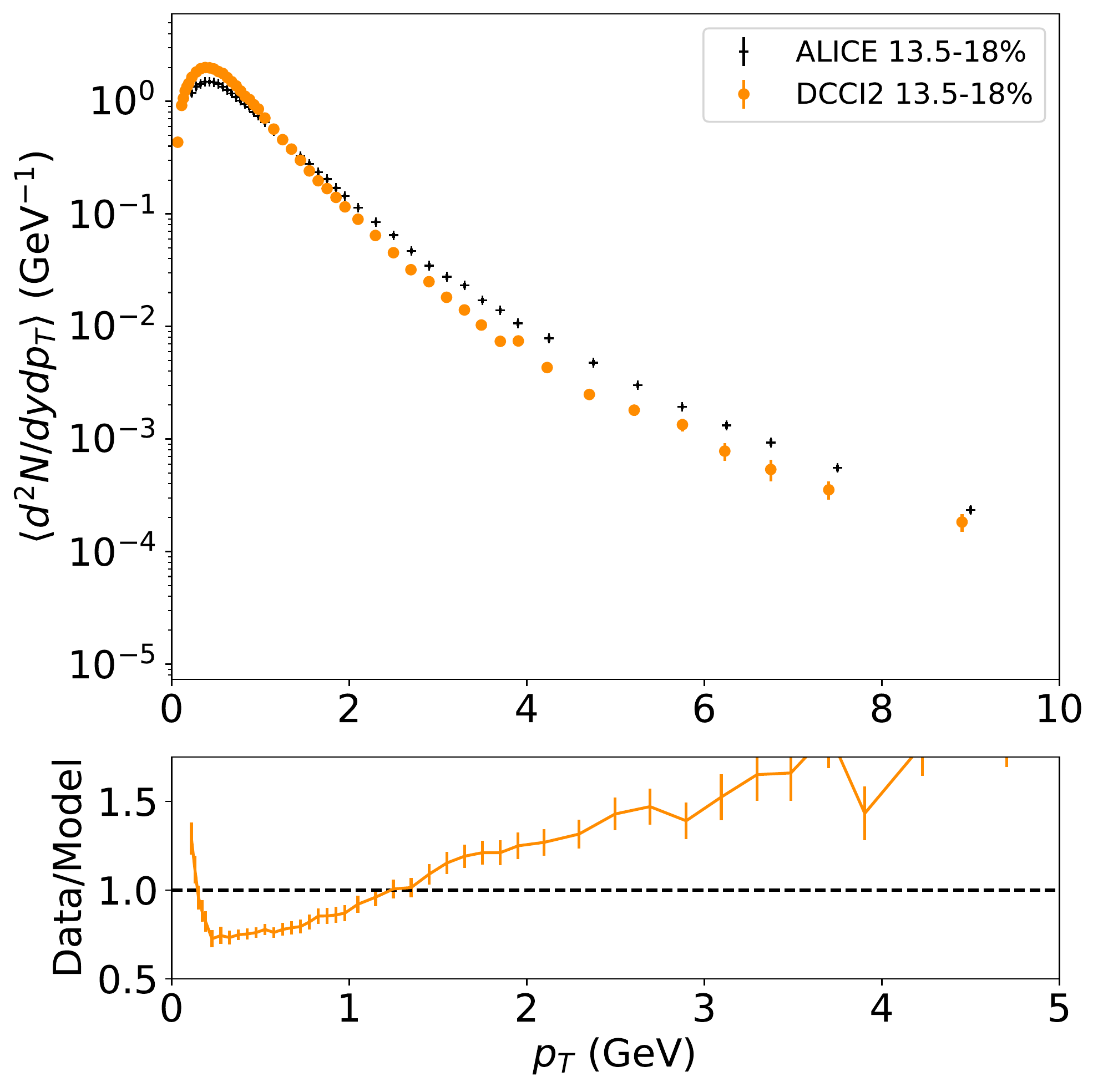}
    \includegraphics[bb = 0 0 564 567, width=0.45\textwidth]{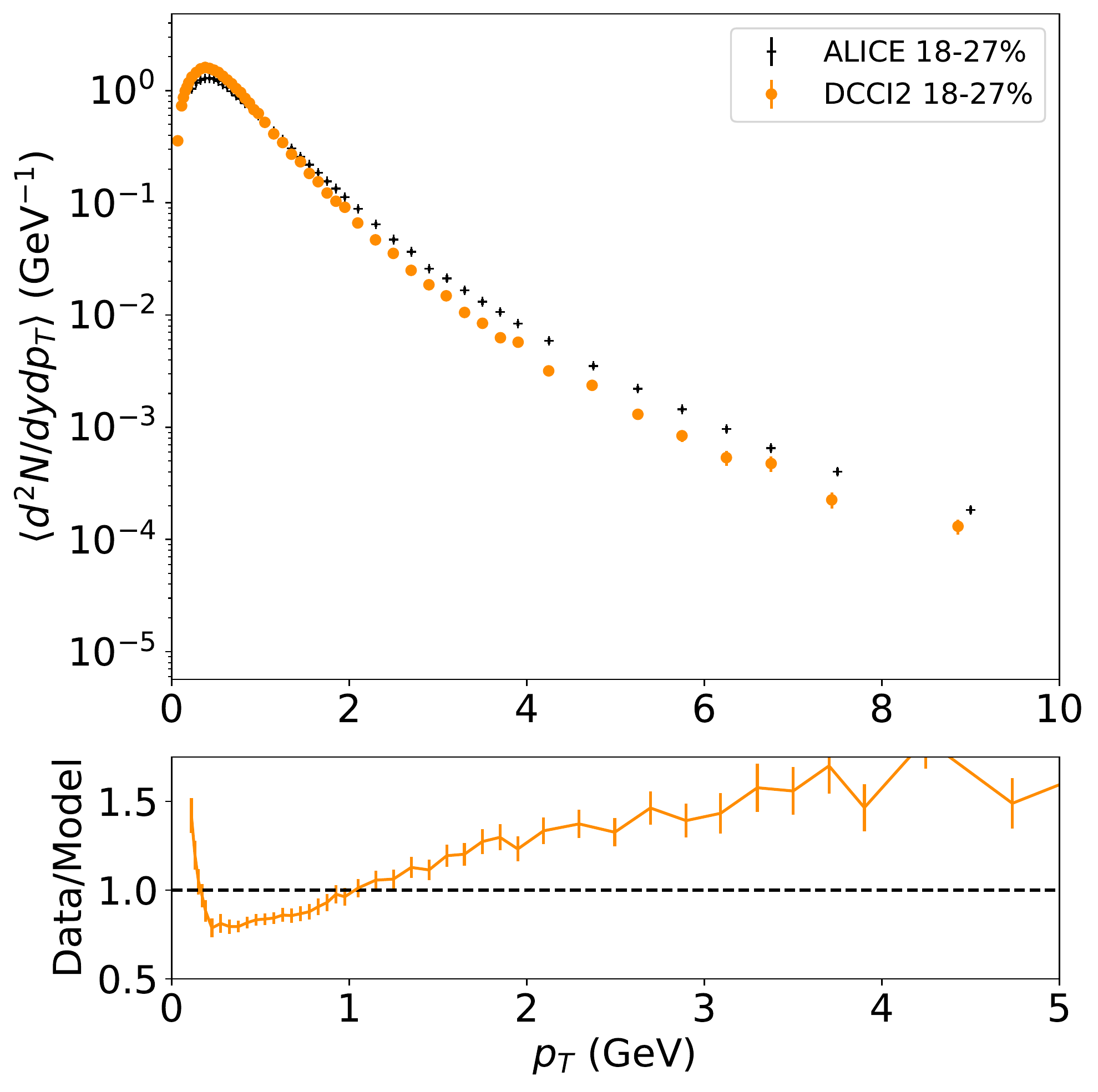}
    \caption{(Upper) Centrality dependence of transverse momentum spectra of charged kaons ($K^+ + K^-$) from $p$+$p$ collisions at \snn[proton]=13 TeV in DCCI2. Comparison between results from DCCI2 (orange circles) and the ALICE experimental data (black crosses) are made.
    (Lower) Ratio of the ALICE experimental data to DCCI2 results at each $p_T$ bin. Centrality classes from 0-0.9\% to 18-27\% are shown.}
    \label{fig:PP13_PTSPECTRA_K_1}
\end{figure}

\begin{figure}
    \centering
    \includegraphics[bb = 0 0 564 567, width=0.45\textwidth]{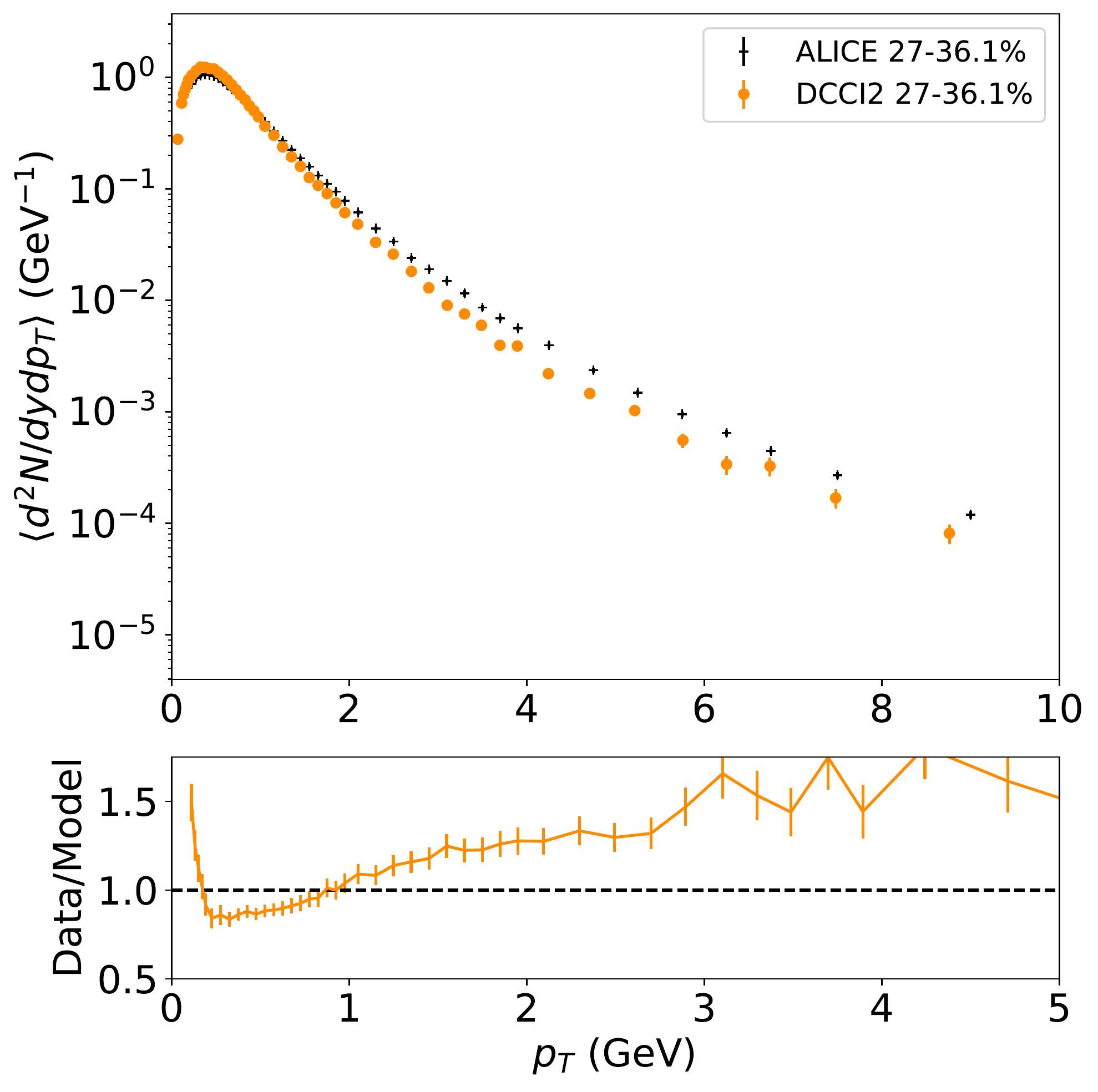}
    \includegraphics[bb = 0 0 564 567, width=0.45\textwidth]{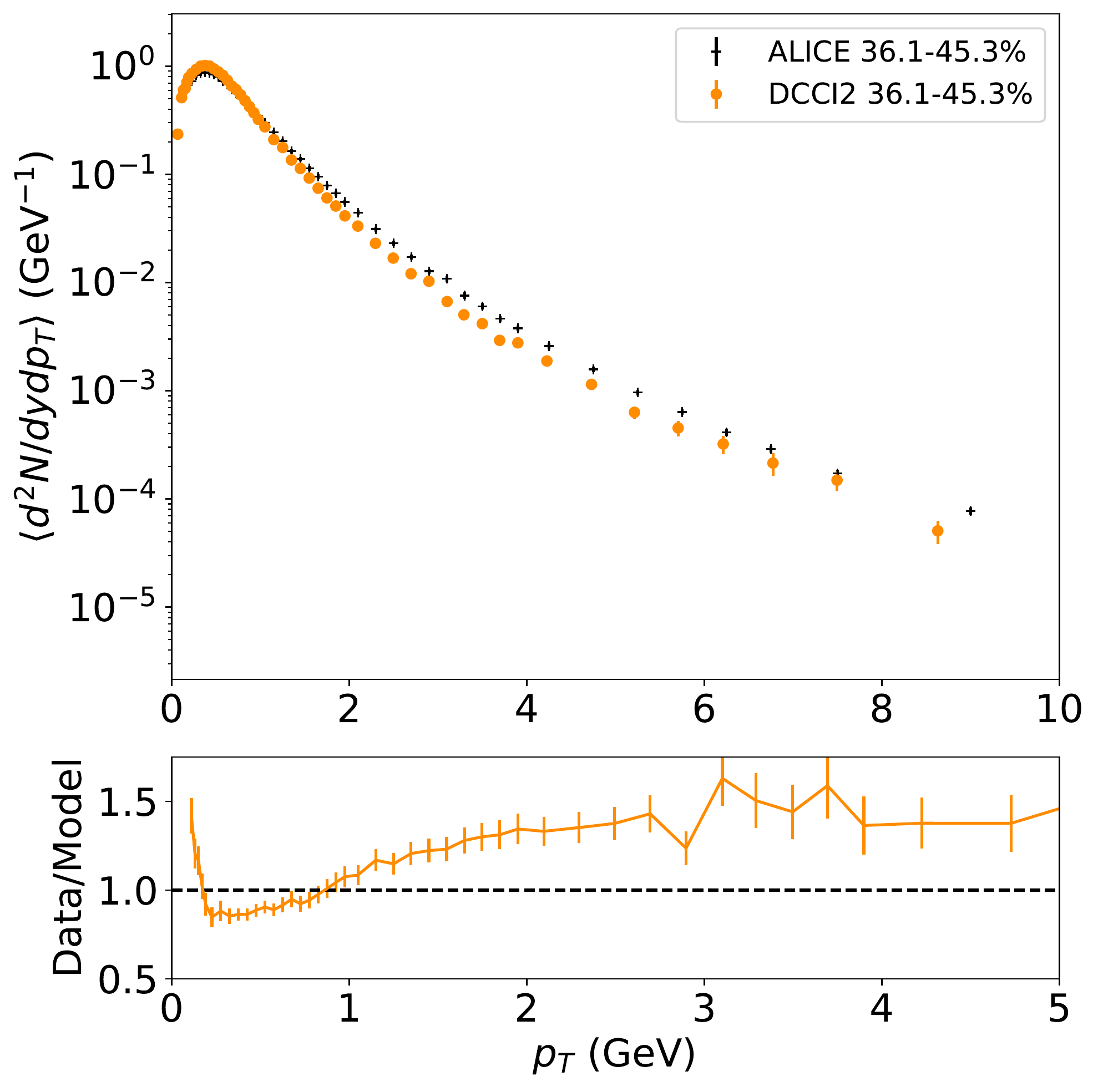}
    \includegraphics[bb = 0 0 564 567, width=0.45\textwidth]{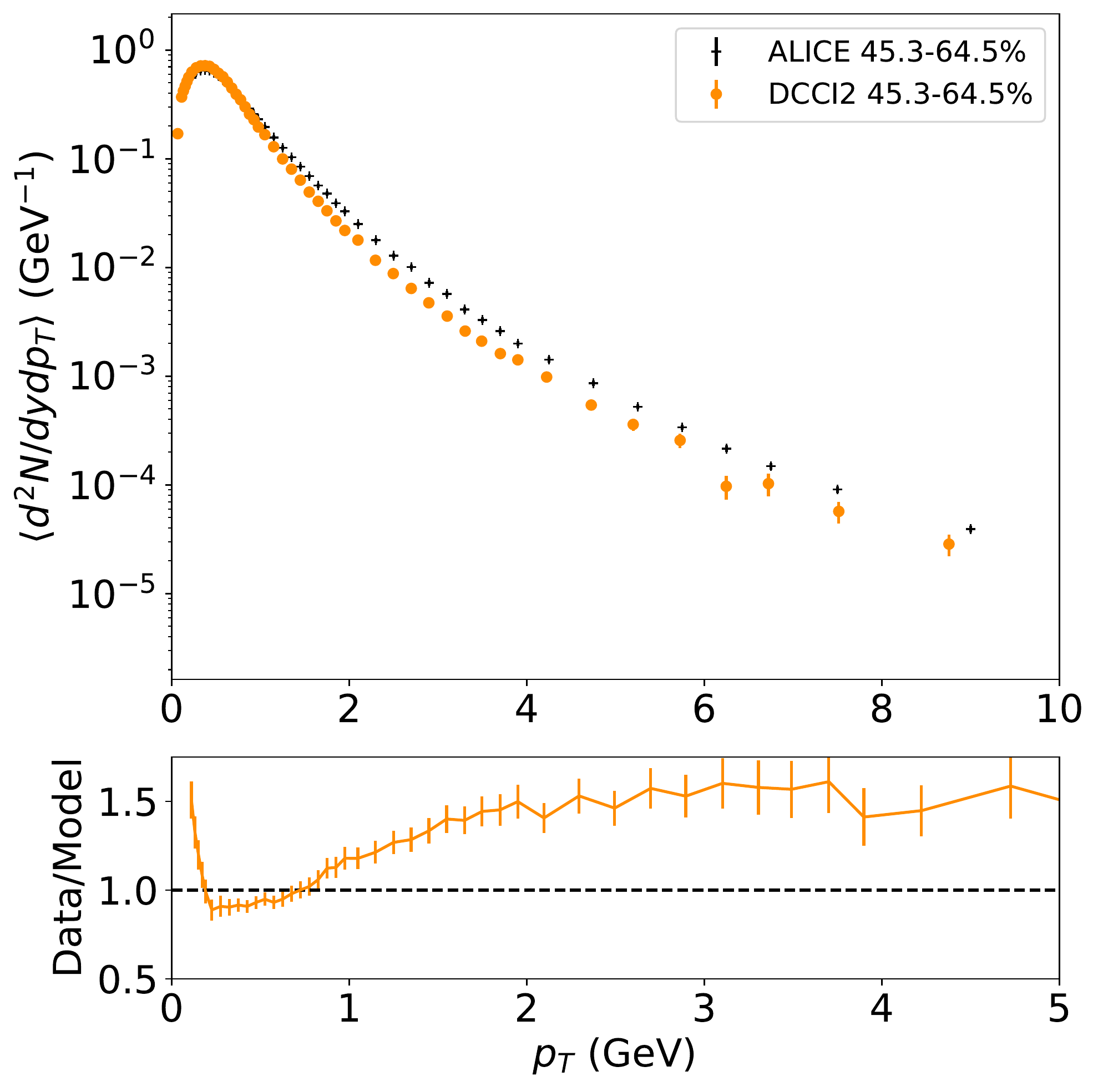}
    \includegraphics[bb = 0 0 564 567, width=0.45\textwidth]{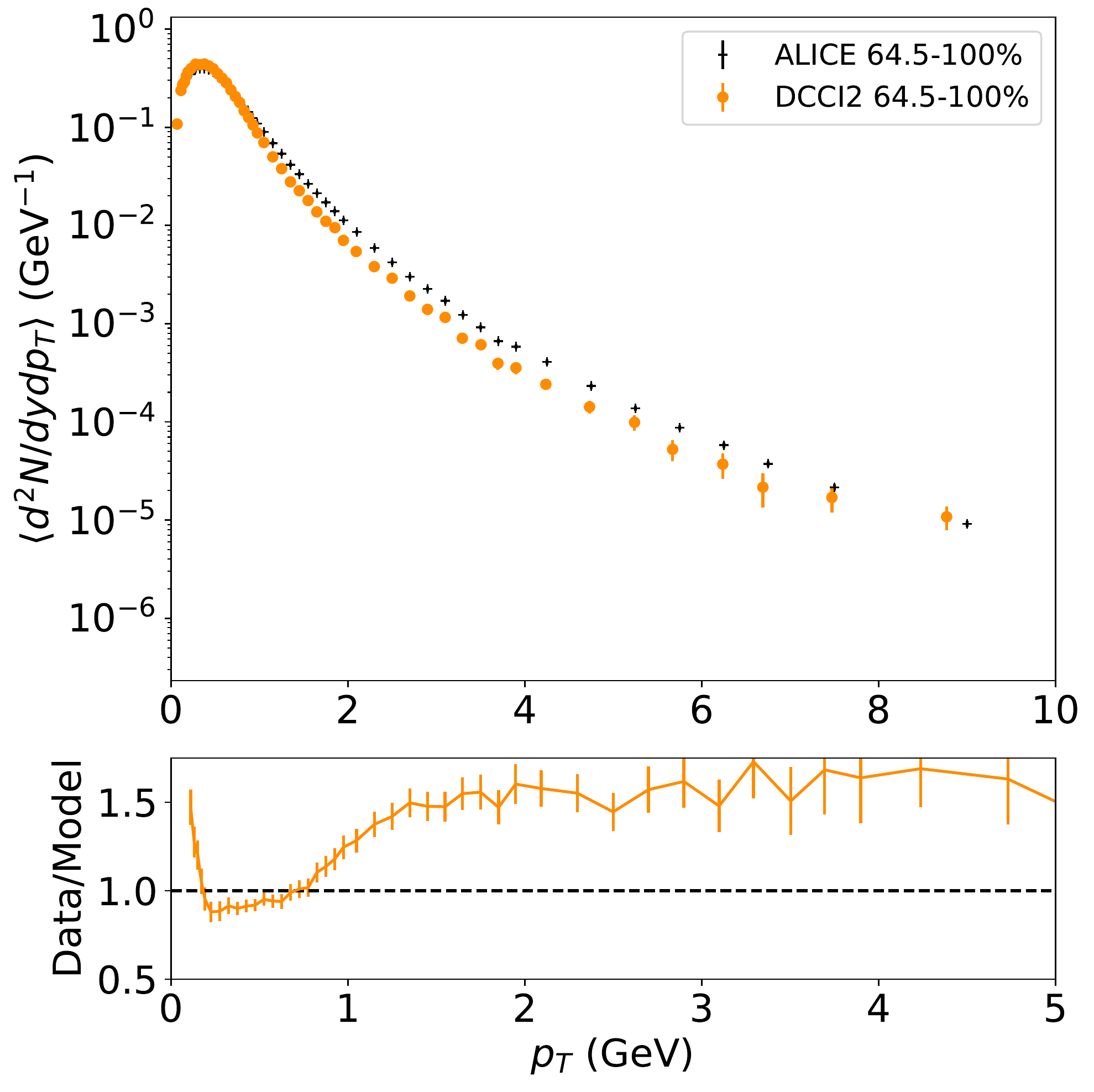}
    \caption{(Upper) Centrality dependence of transverse momentum spectra of charged kaons ($K^+ + K^-$) from $p$+$p$ collisions at \snn[proton]=13 TeV in DCCI2. Comparison between results from DCCI2 (orange circles) and the ALICE experimental data (black crosses) are made.
    (Lower) Ratio of the ALICE experimental data to DCCI2 results at each $p_T$ bin. Centrality classes from 27-36.1\% to 64.5-100\% are shown.}
    \label{fig:PP13_PTSPECTRA_K_2}
\end{figure}

\begin{figure}
    \centering
    \includegraphics[bb = 0 0 564 567, width=0.45\textwidth]{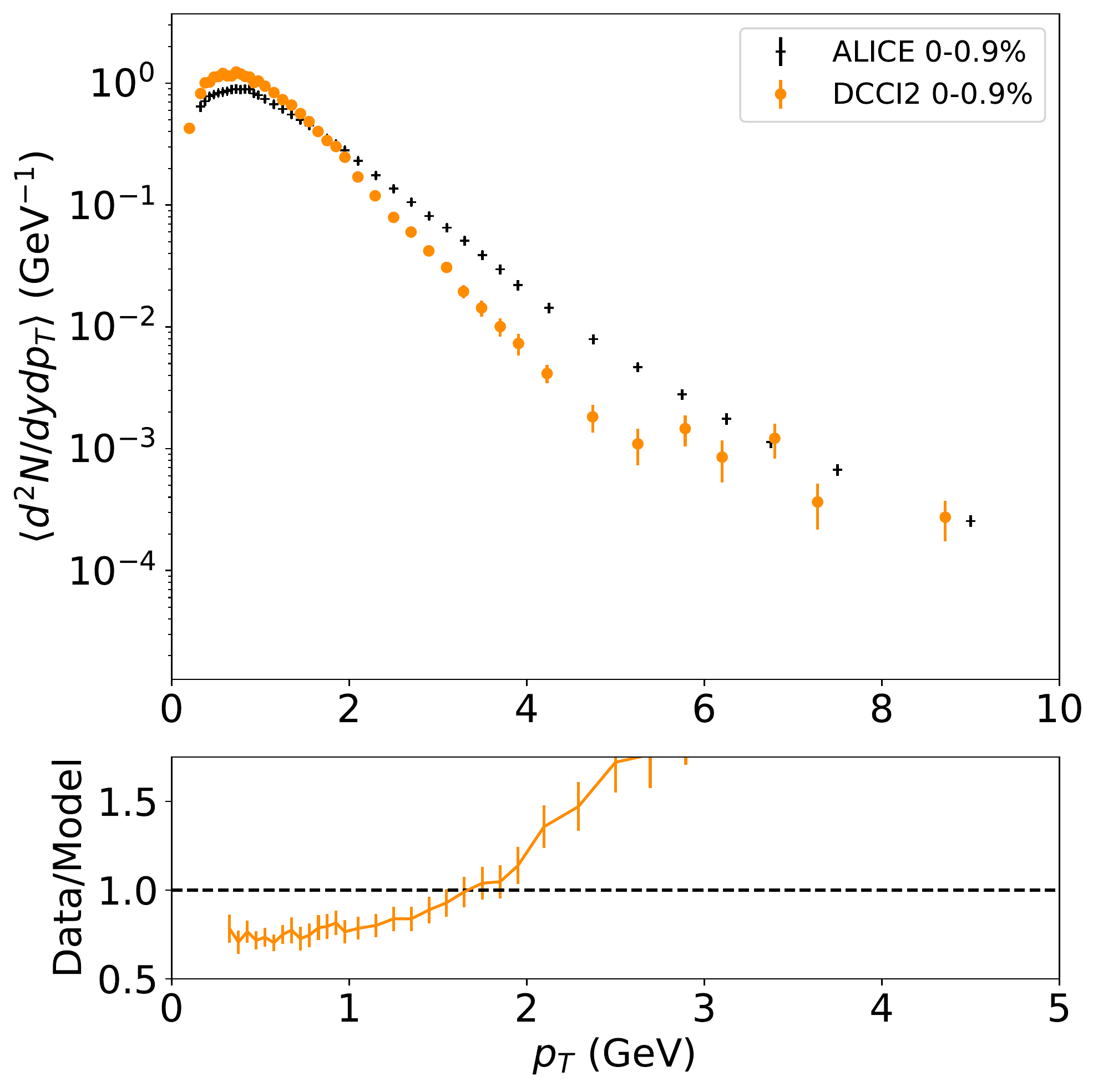}
    \includegraphics[bb = 0 0 564 567, width=0.45\textwidth]{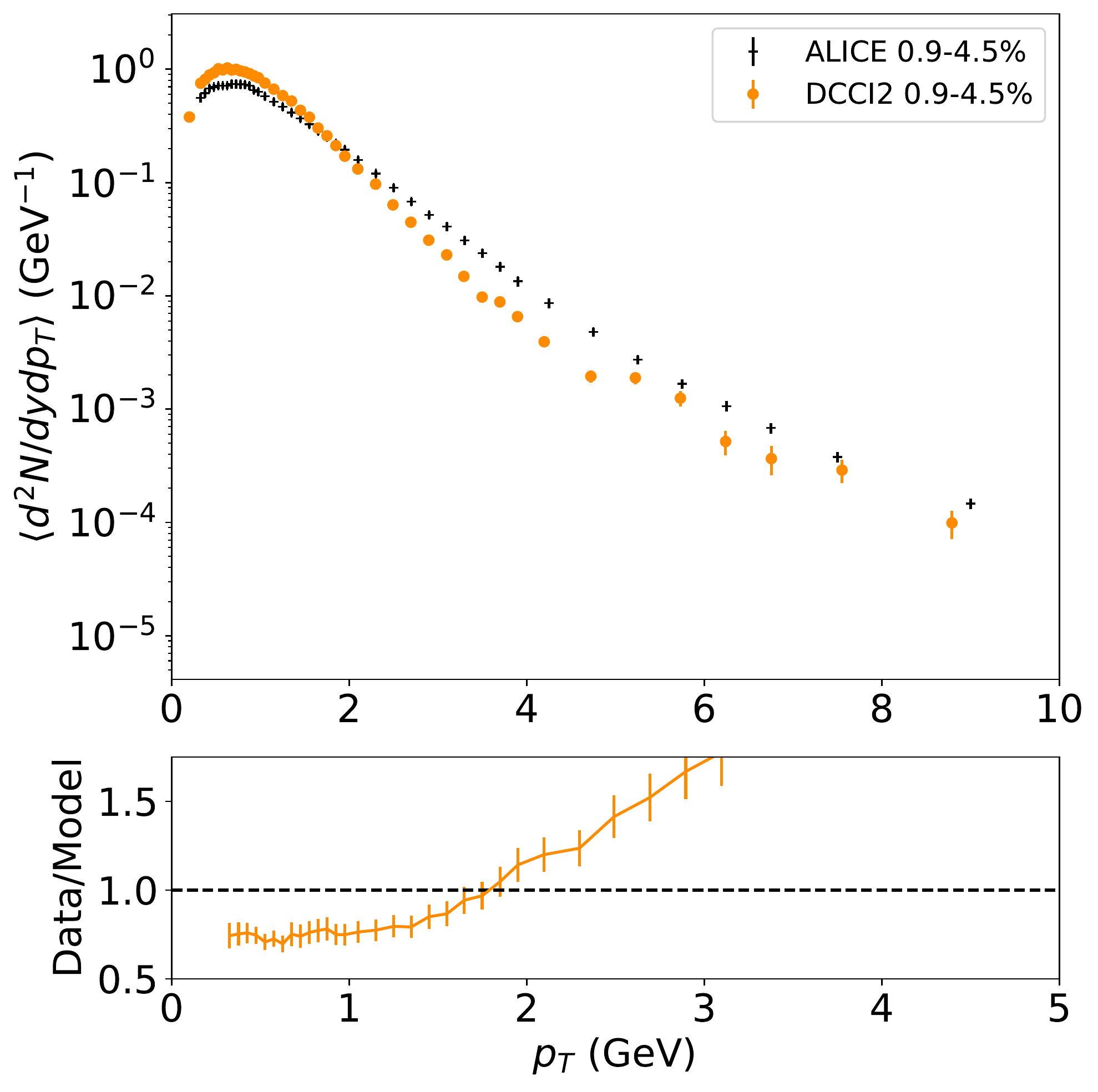}
    \includegraphics[bb = 0 0 564 567, width=0.45\textwidth]{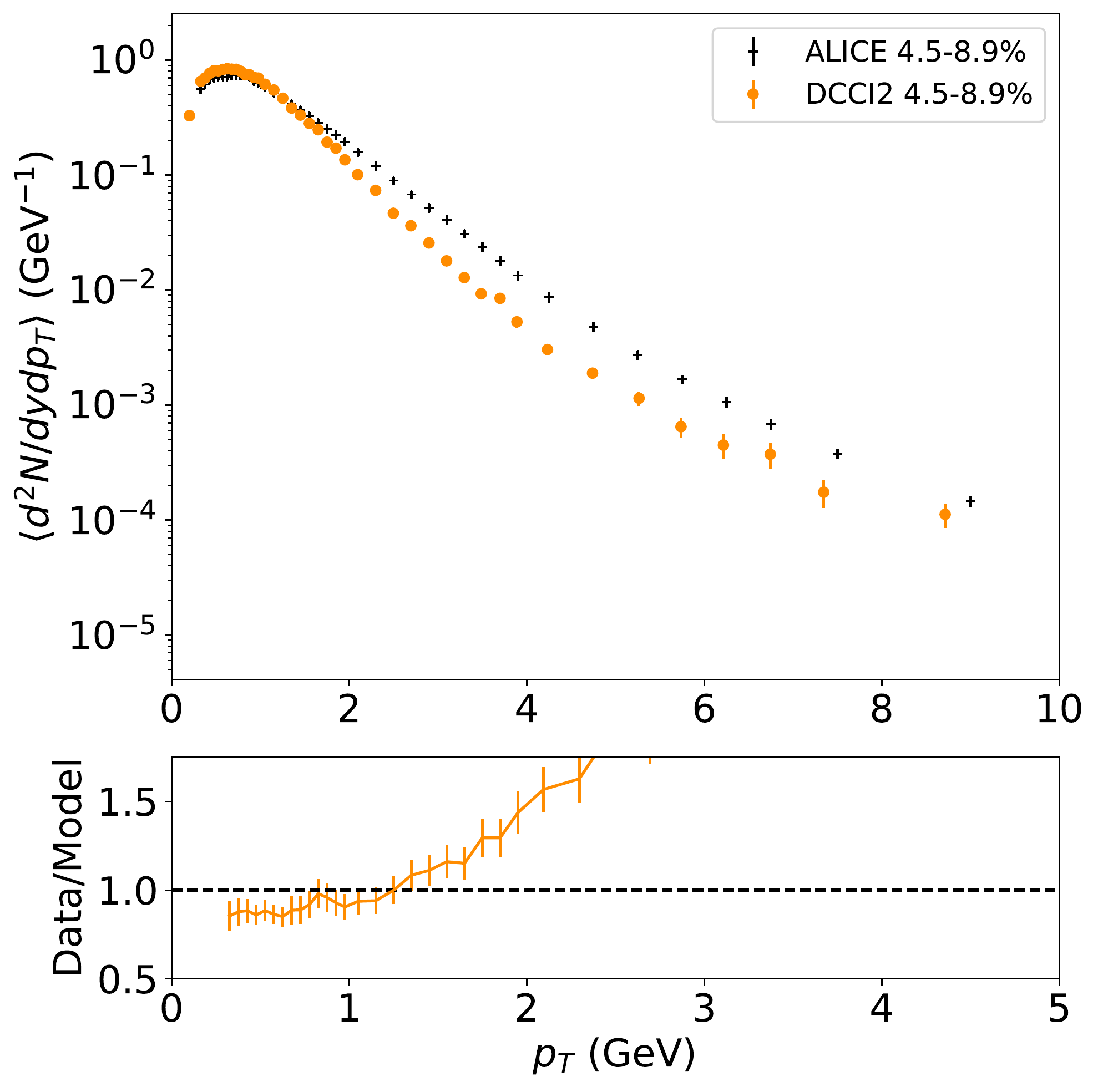}
    \includegraphics[bb = 0 0 564 567, width=0.45\textwidth]{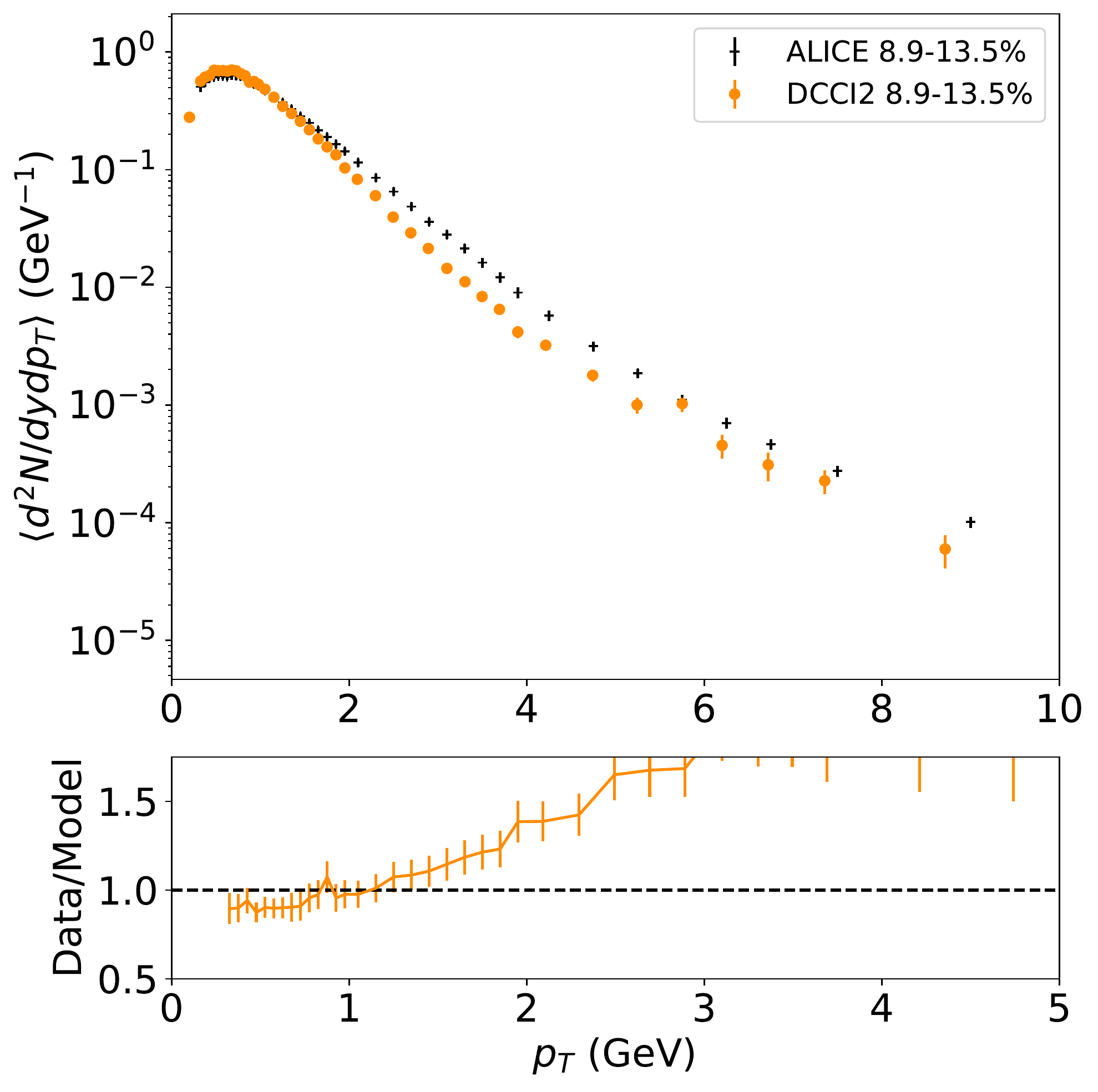}
    \includegraphics[bb = 0 0 564 567, width=0.45\textwidth]{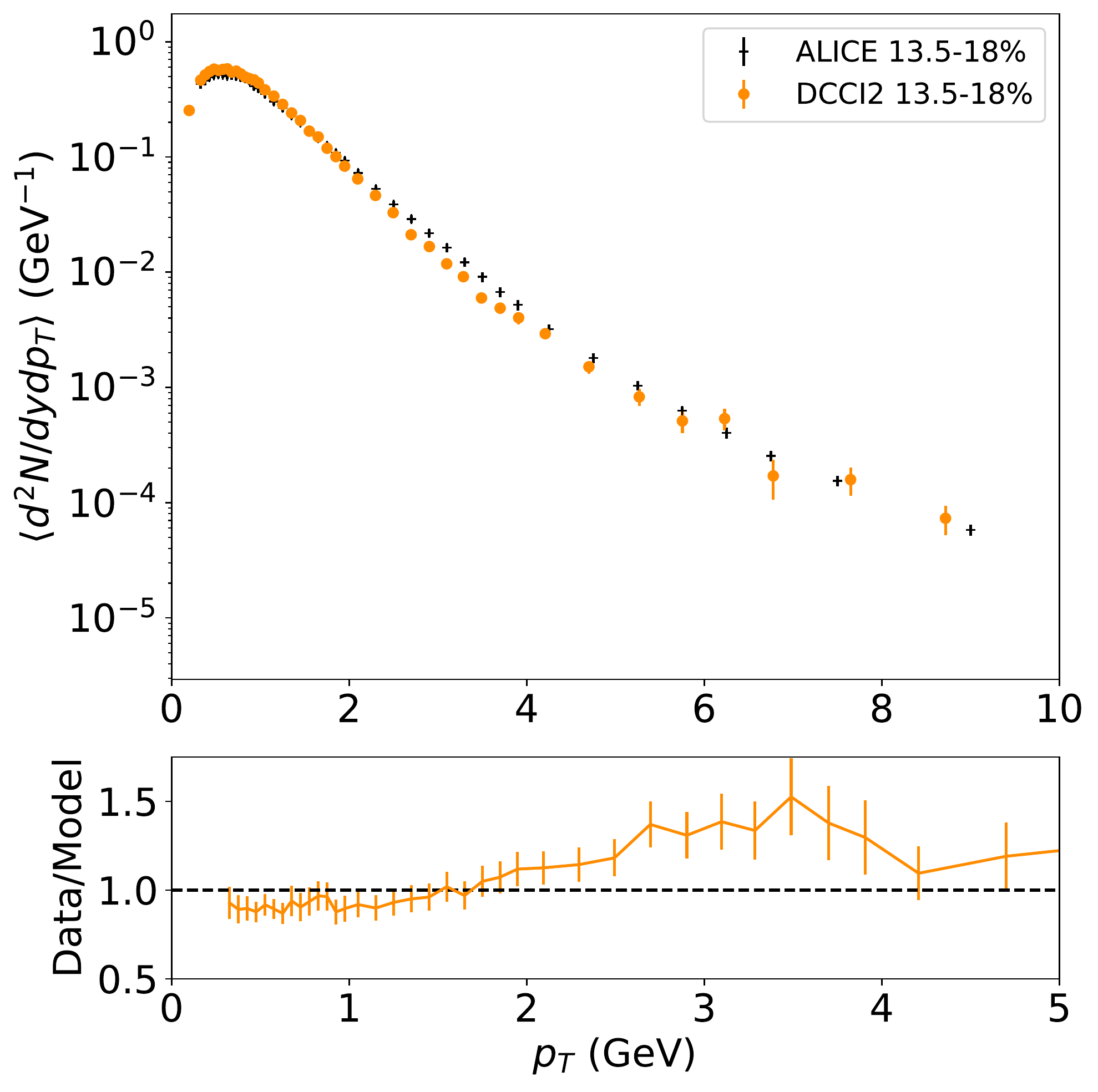}
    \includegraphics[bb = 0 0 564 567, width=0.45\textwidth]{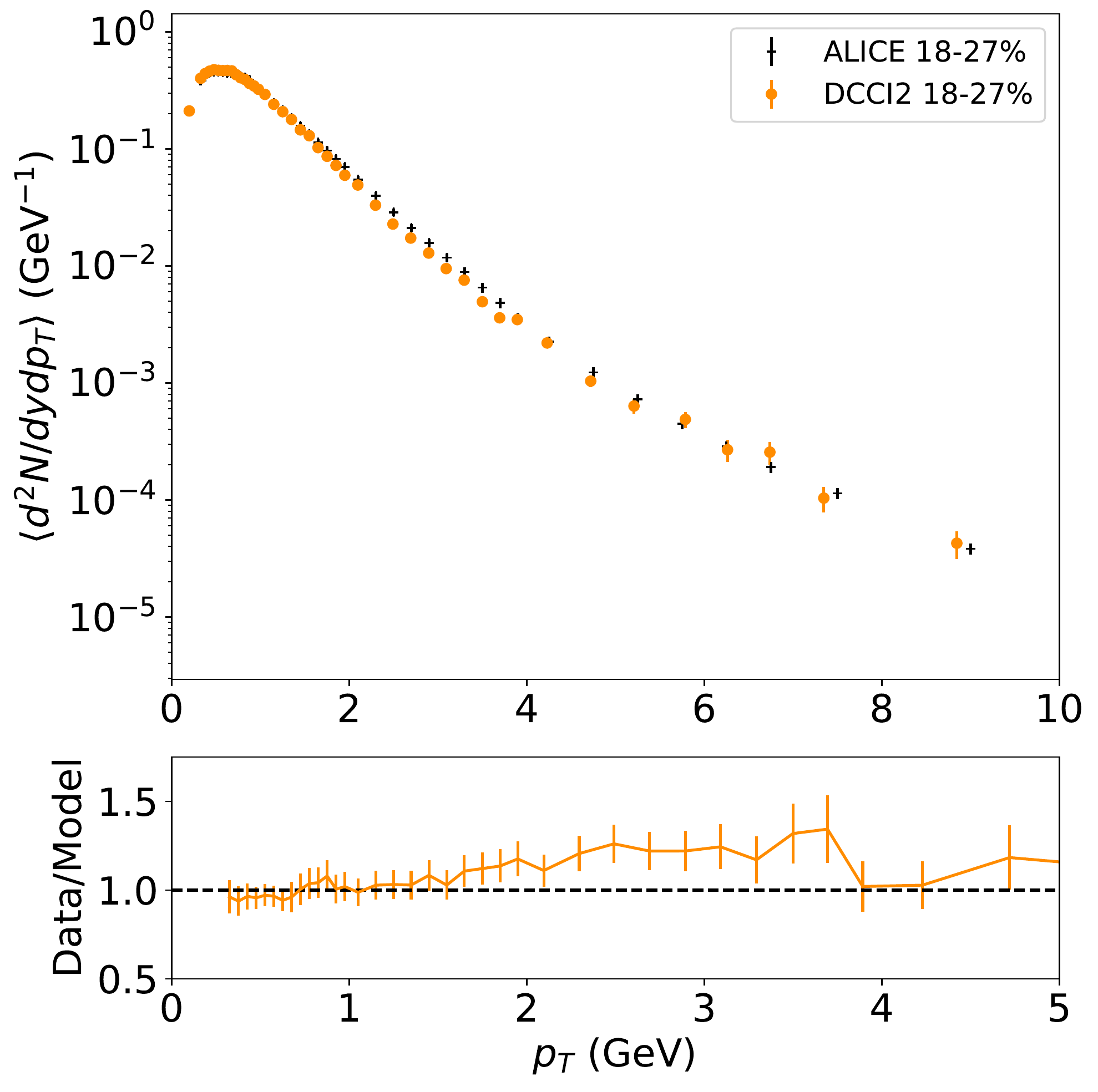}
    \caption{(Upper) Centrality dependence of transverse momentum spectra of protons and antiprotons ($p + \bar{p}$) from $p$+$p$ collisions at \snn[proton]=13 TeV in DCCI2. Comparison between results from DCCI2 (orange circles) and the ALICE experimental data (black crosses) are made.
    (Lower) Ratio of the ALICE experimental data to DCCI2 results at each $p_T$ bin. Centrality classes from 0-0.9\% to 18-27\% are shown.}
    \label{fig:PP13_PTSPECTRA_P_1}
\end{figure}

\begin{figure}
    \centering
    \includegraphics[bb = 0 0 564 567, width=0.45\textwidth]{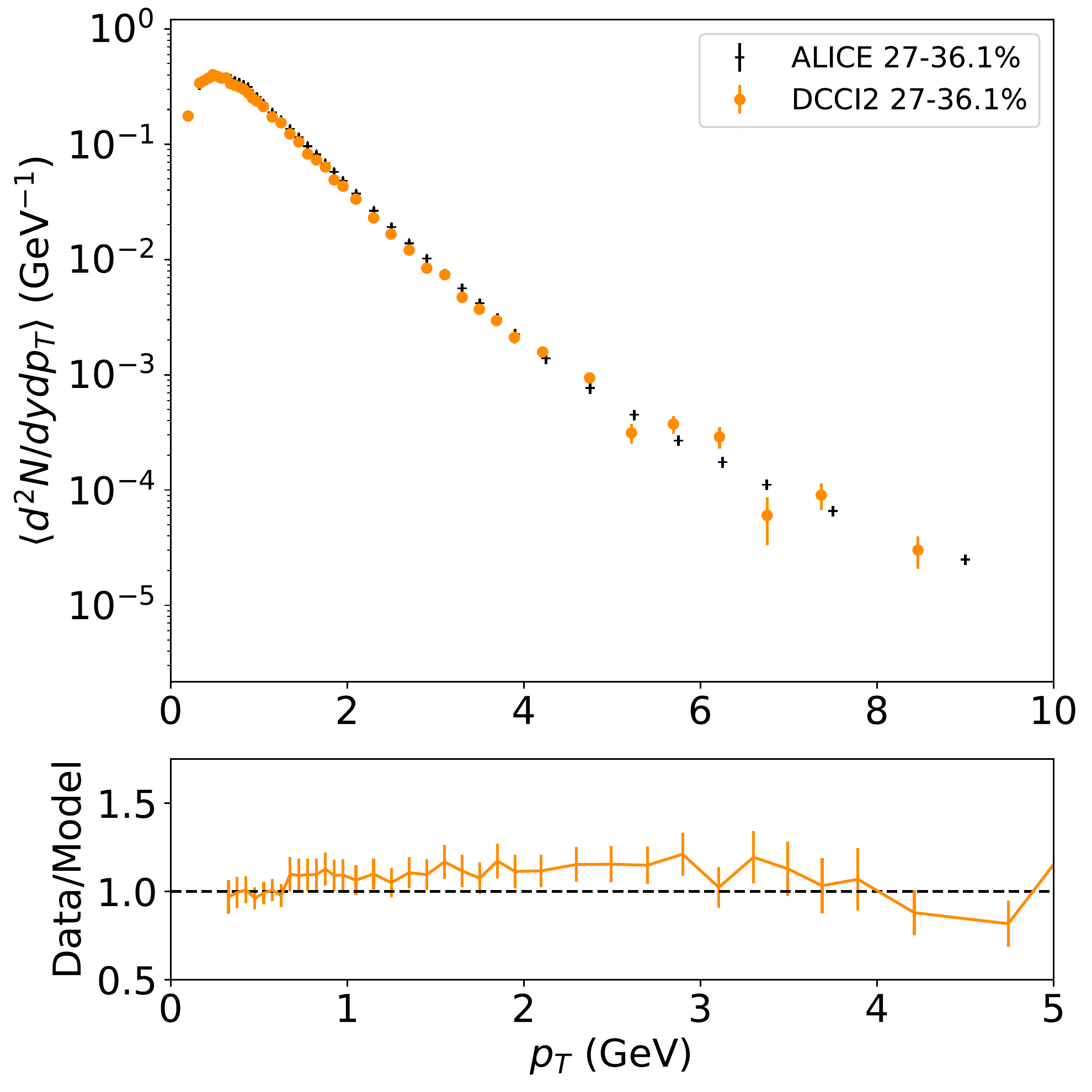}
    \includegraphics[bb = 0 0 564 567, width=0.45\textwidth]{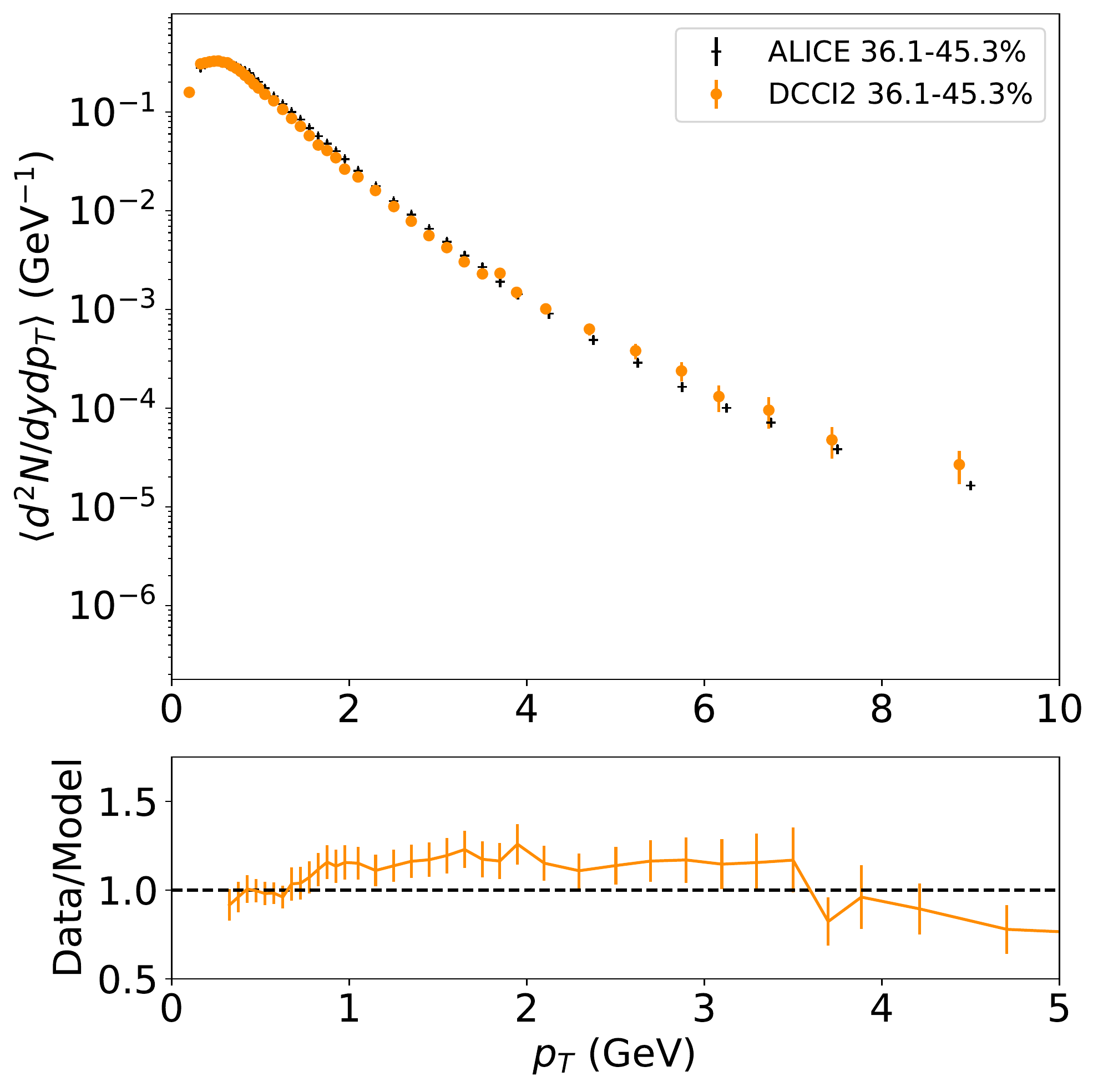}
    \includegraphics[bb = 0 0 564 567, width=0.45\textwidth]{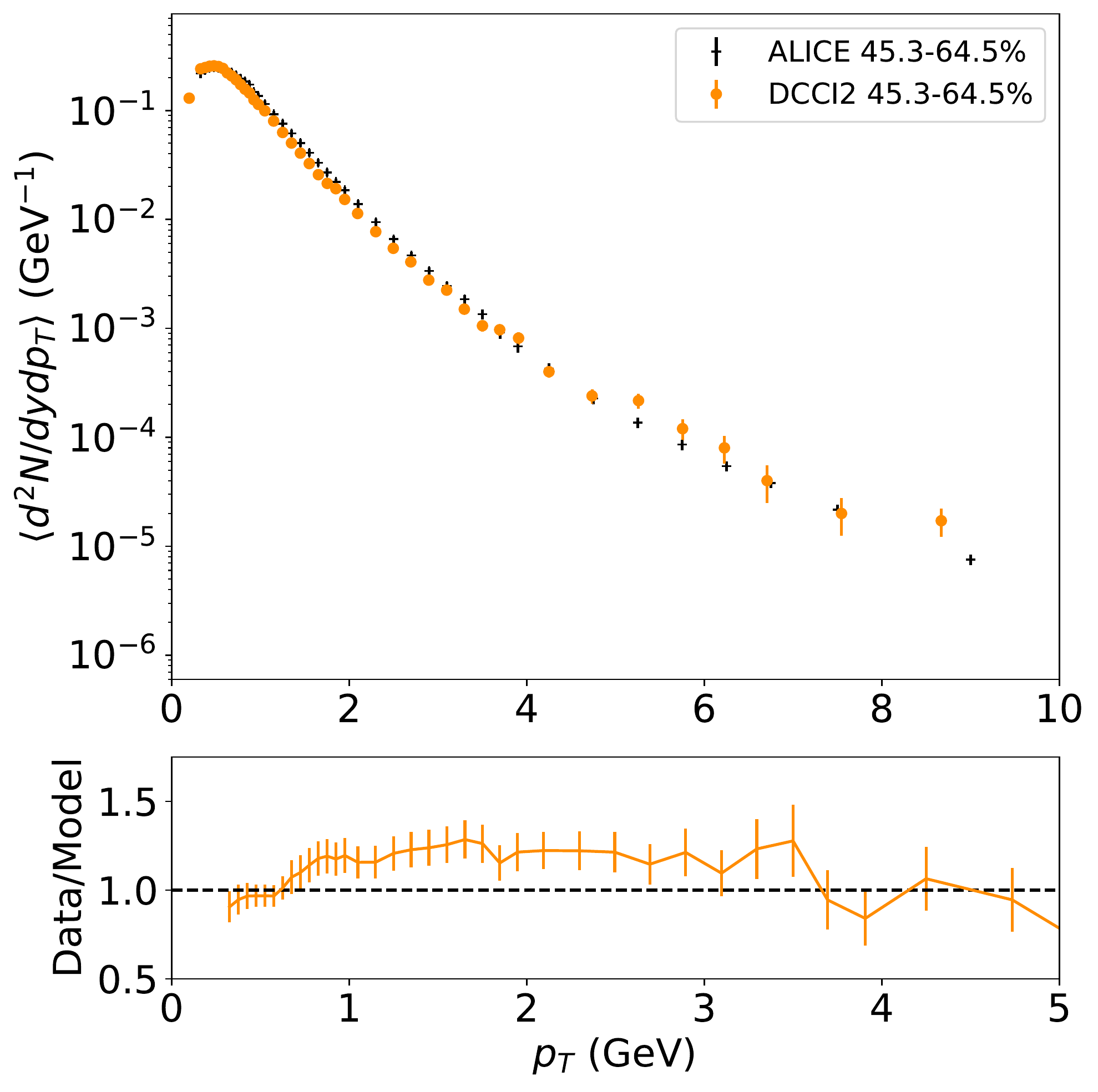}
    \includegraphics[bb = 0 0 564 567, width=0.45\textwidth]{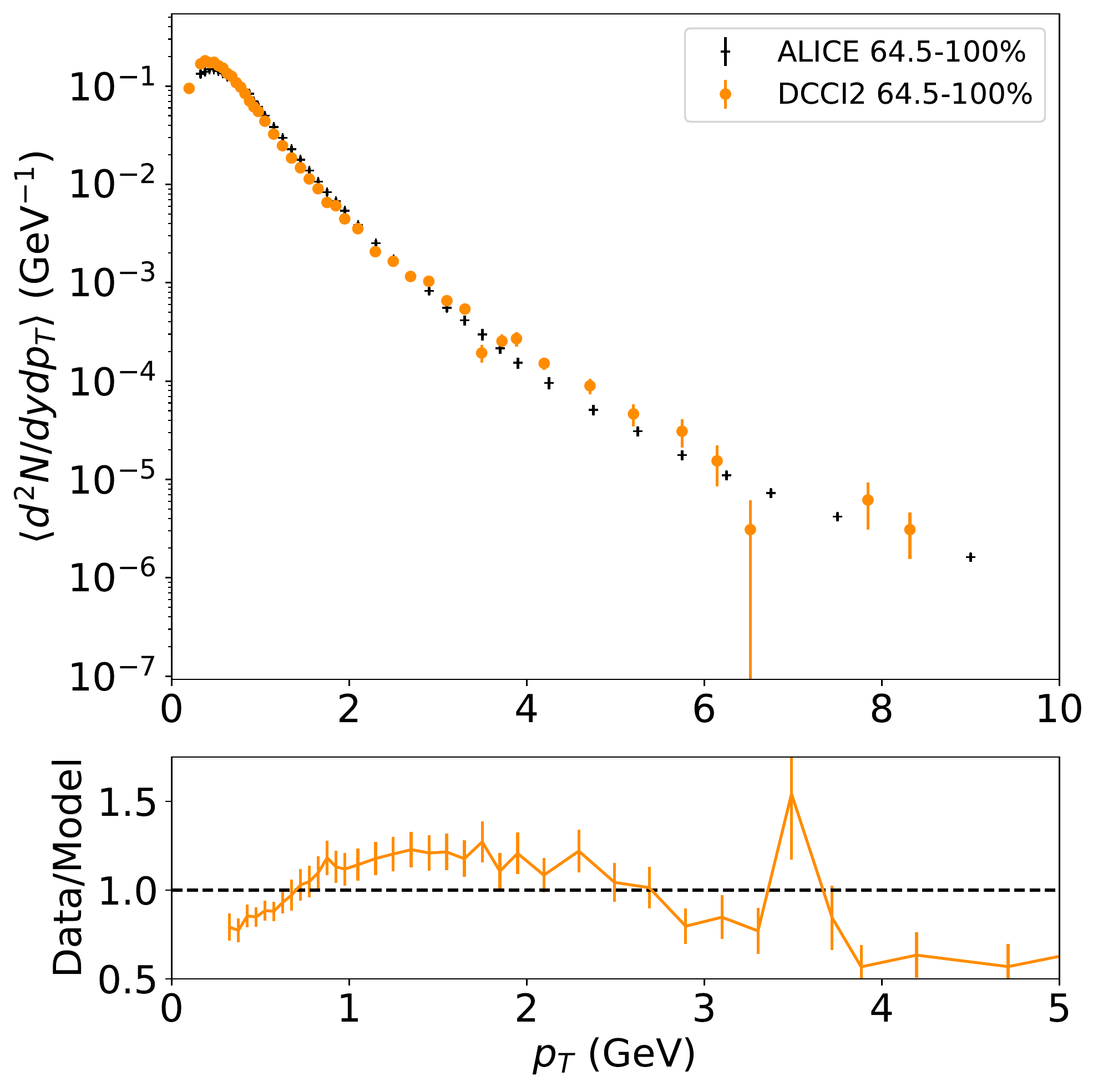}
    \caption{(Upper) Centrality dependence of transverse momentum spectra of protons and antiprotons ($p + \bar{p}$) from $p$+$p$ collisions at \snn[proton]=13 TeV in DCCI2. Comparison between results from DCCI2 (orange circles) and the ALICE experimental data (black crosses) are made.
    (Lower) Ratio of the ALICE experimental data to DCCI2 results at each $p_T$ bin. Centrality classes from 27-36.1\% to 64.5-100\% are shown.}
    \label{fig:PP13_PTSPECTRA_P_2}
\end{figure}

\begin{figure}
    \centering
    \includegraphics[bb=0 0 564 567, width=0.49\textwidth]{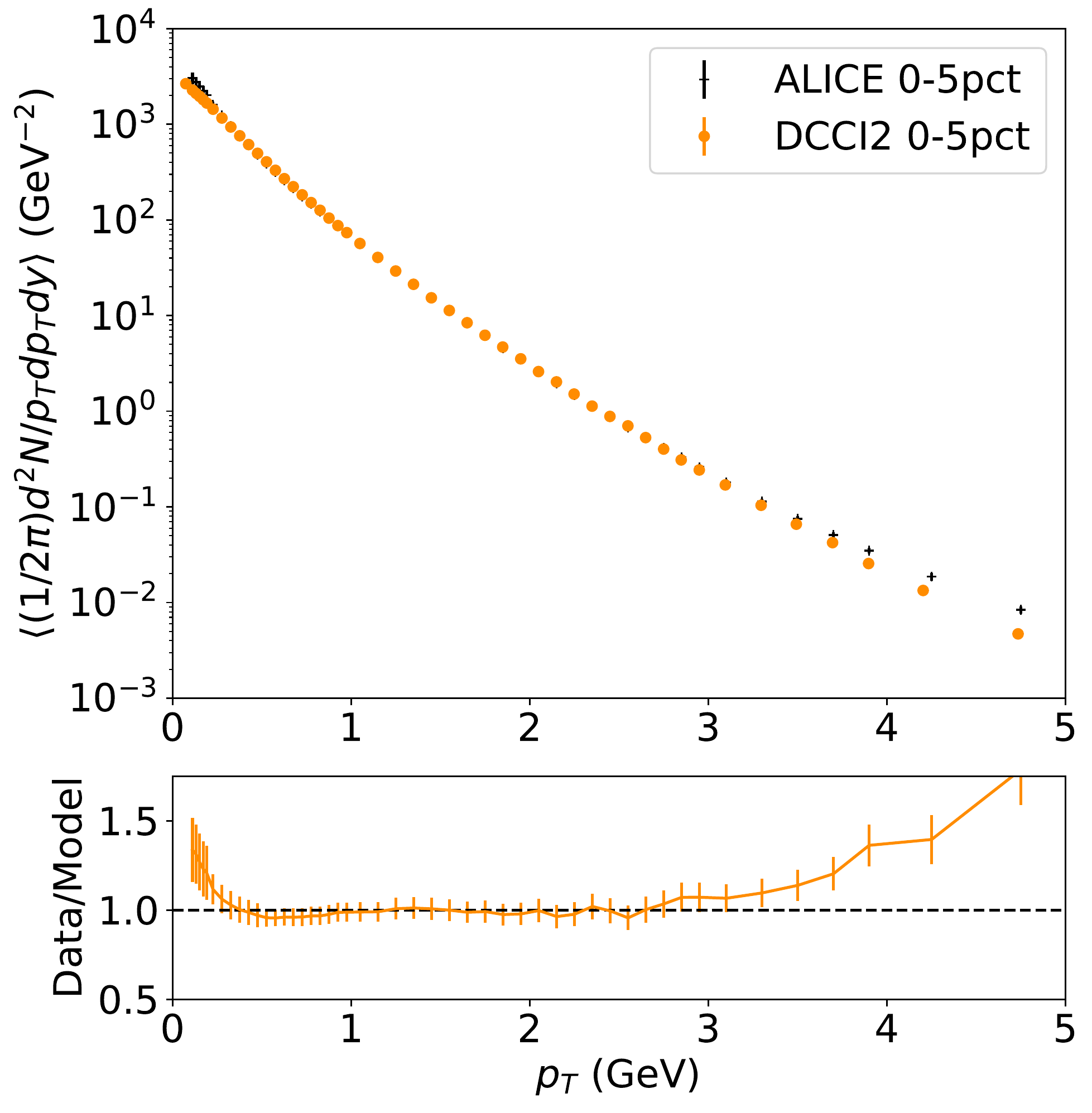}
    \includegraphics[bb=0 0 564 567, width=0.49\textwidth]{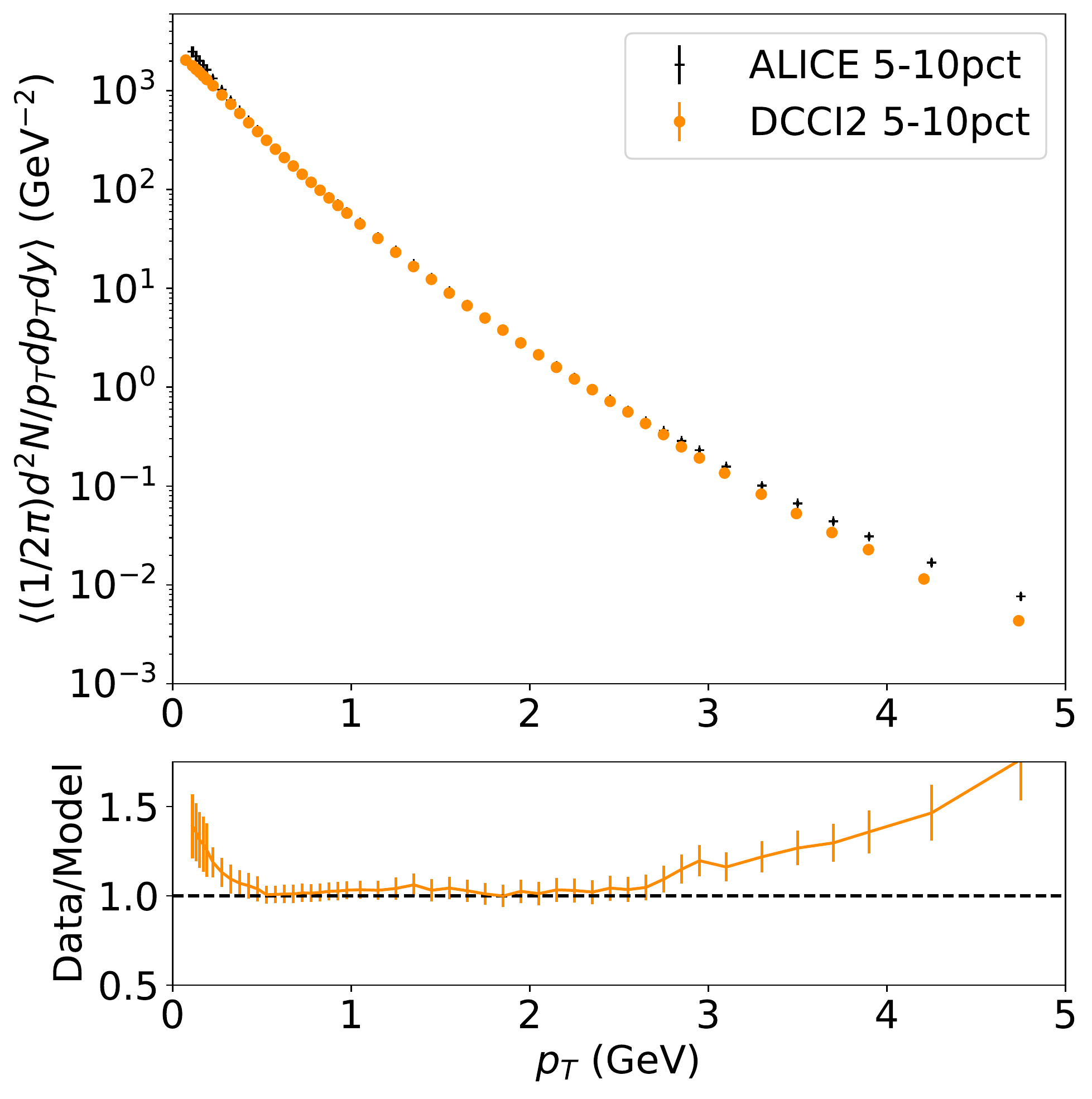}
    \includegraphics[bb=0 0 564 567, width=0.49\textwidth]{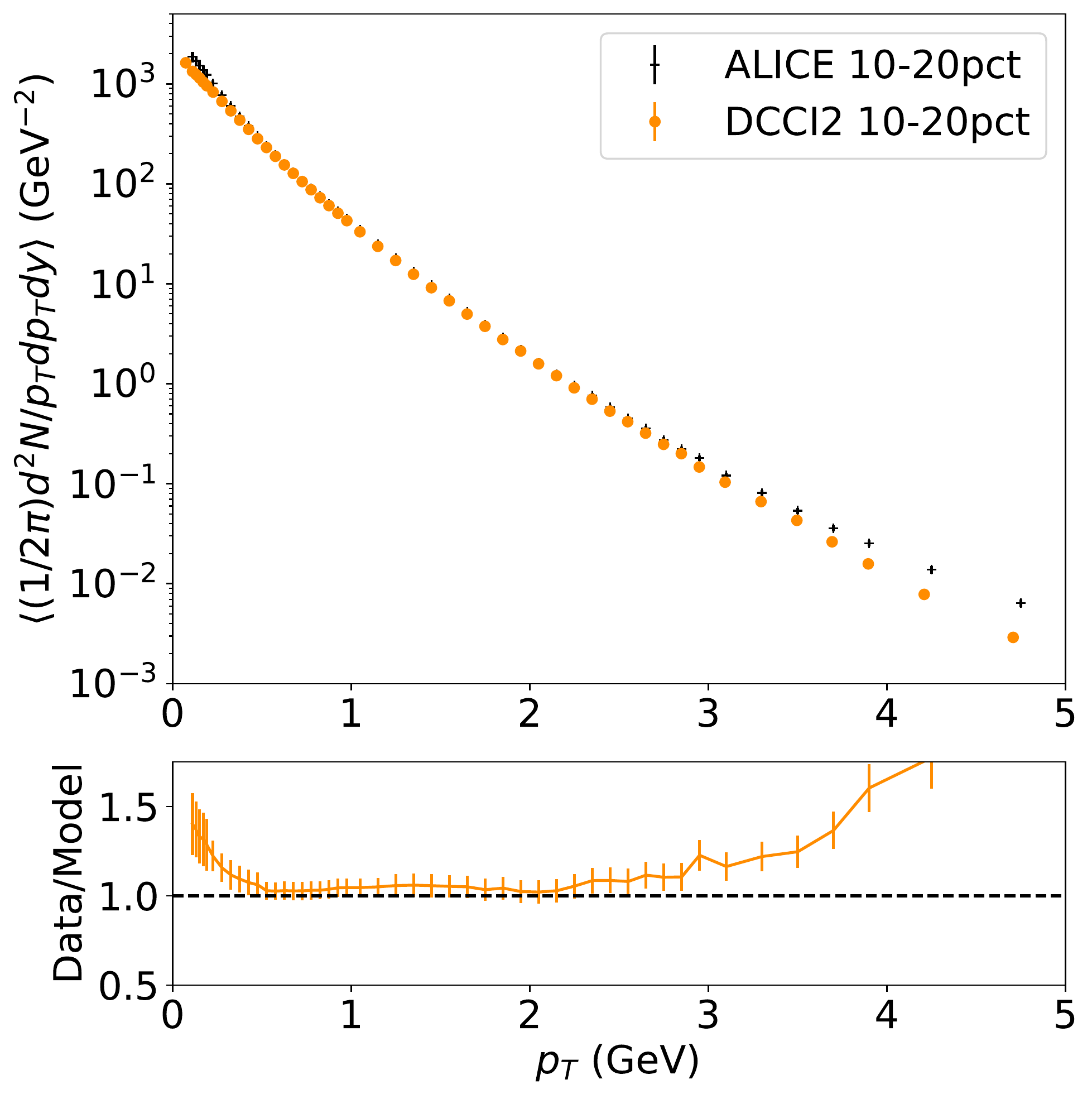}
    \includegraphics[bb=0 0 564 567, width=0.49\textwidth]{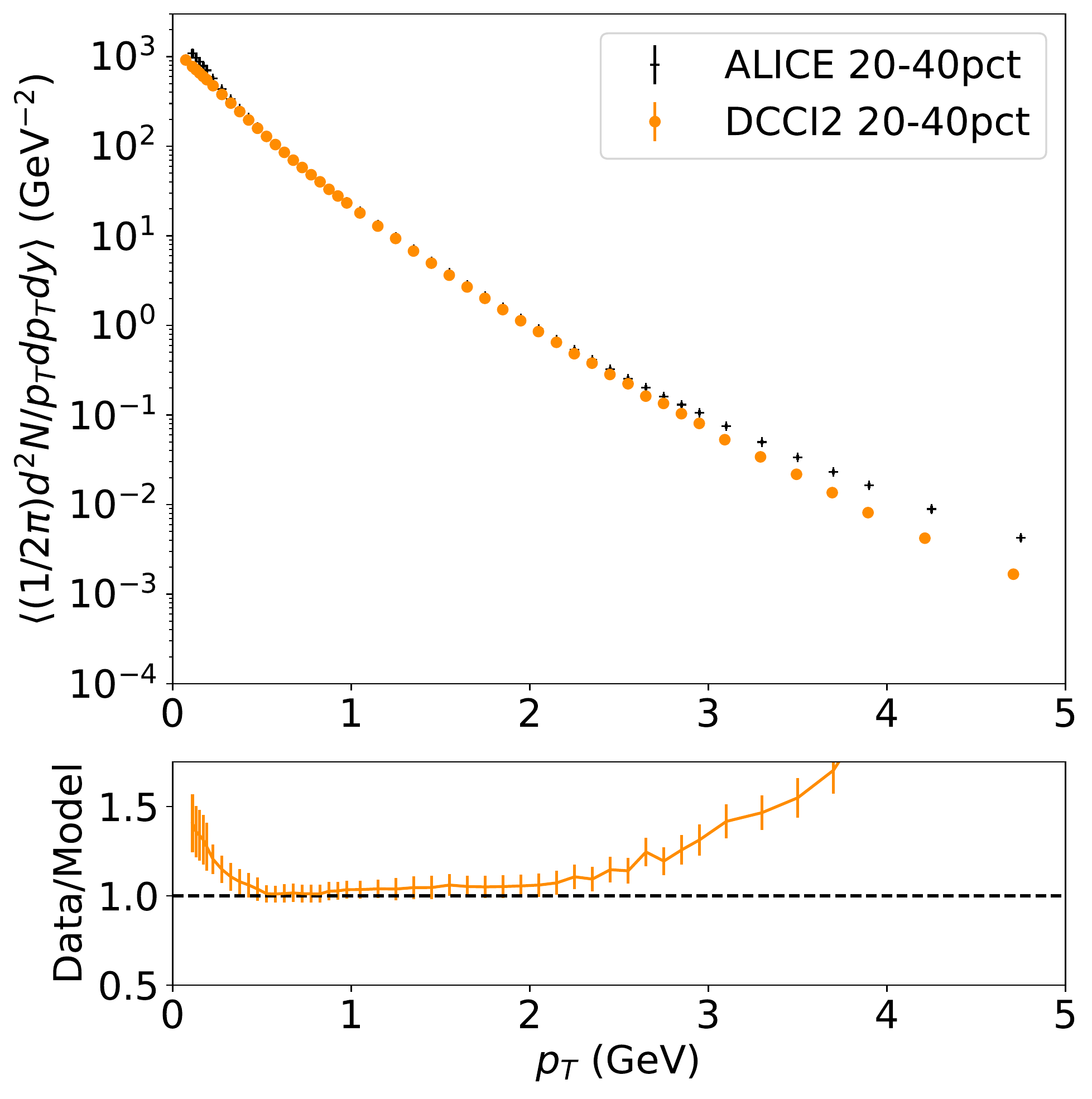}
    \includegraphics[bb=0 0 564 567, width=0.49\textwidth]{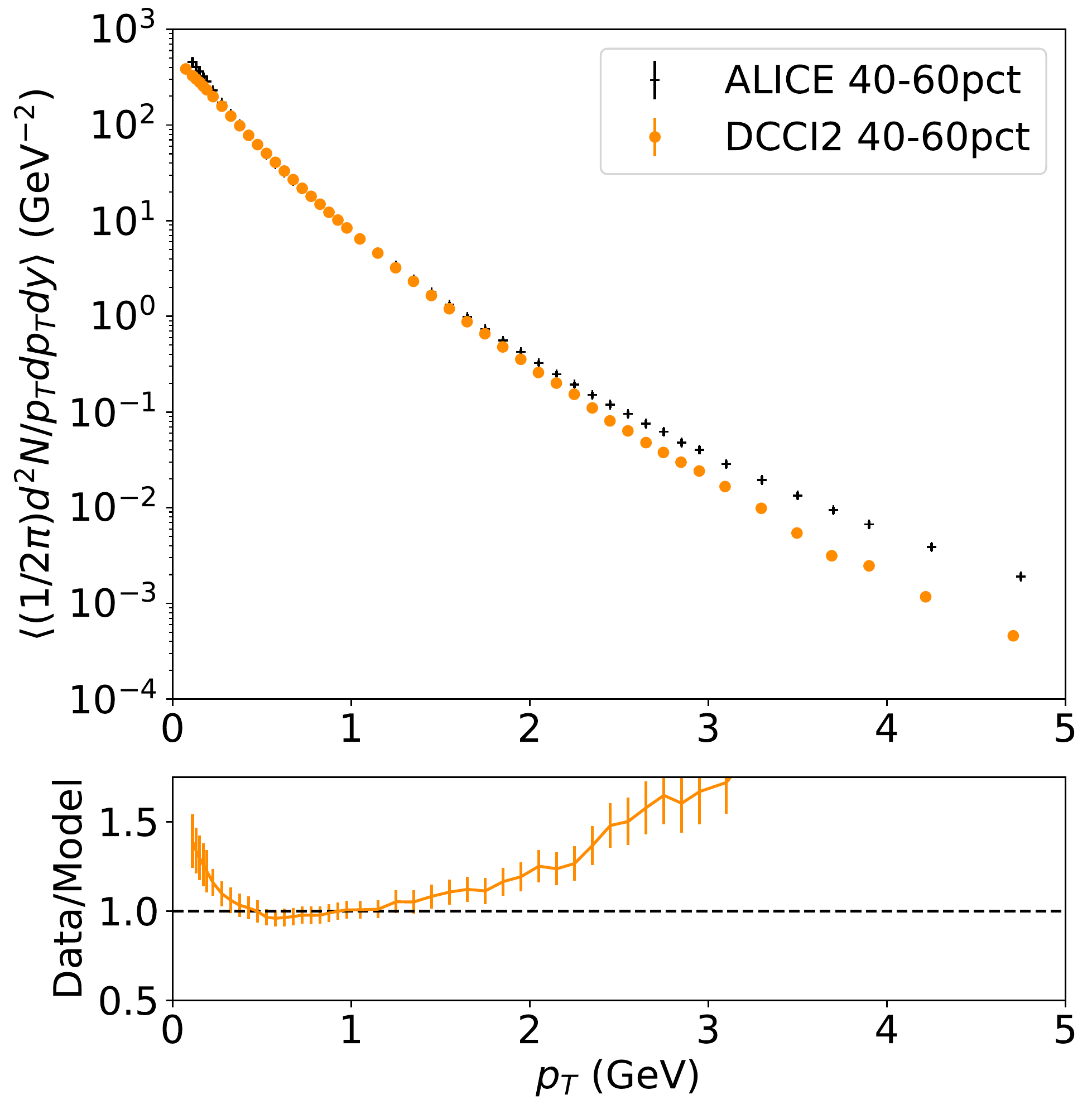}
    \includegraphics[bb=0 0 564 567, width=0.49\textwidth]{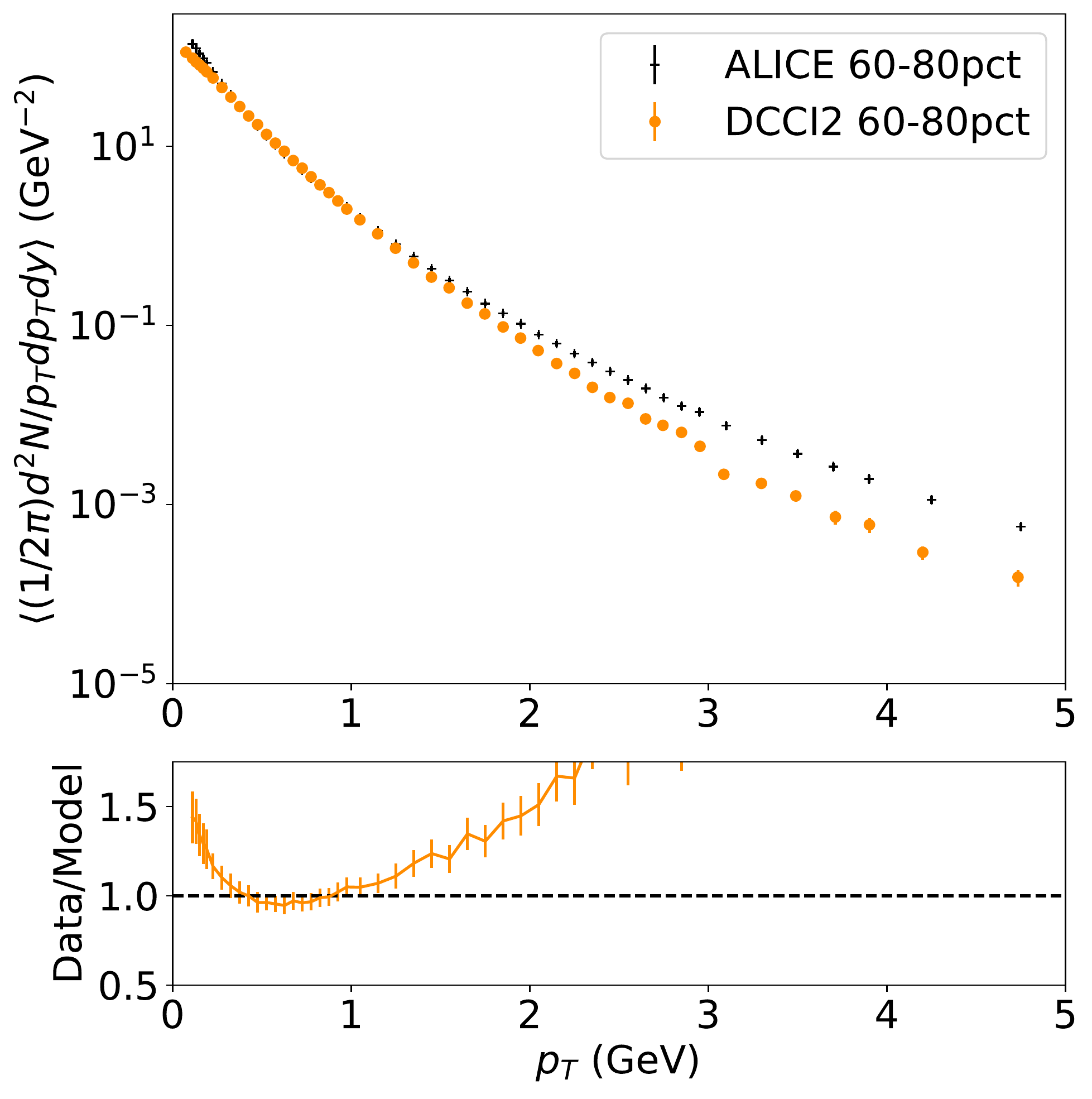}
    \caption{Centrality dependence of transverse momentum spectra of charged pions ($\pi^+ + \pi^-$) from $Pb$+$Pb$ collisions at \snn=2.76 TeV in DCCI2. Comparison between results from DCCI2 (orange circles) and the ALICE experimental data (black crosses) are made.
    (Lower) Ratio of the ALICE experimental data to DCCI2 results at each $p_T$ bin. }
    \label{fig:PBPB2760_PTSPECTRA_PI}
\end{figure}

\begin{figure}
    \centering
    \includegraphics[bb=0 0 564 567, width=0.49\textwidth]{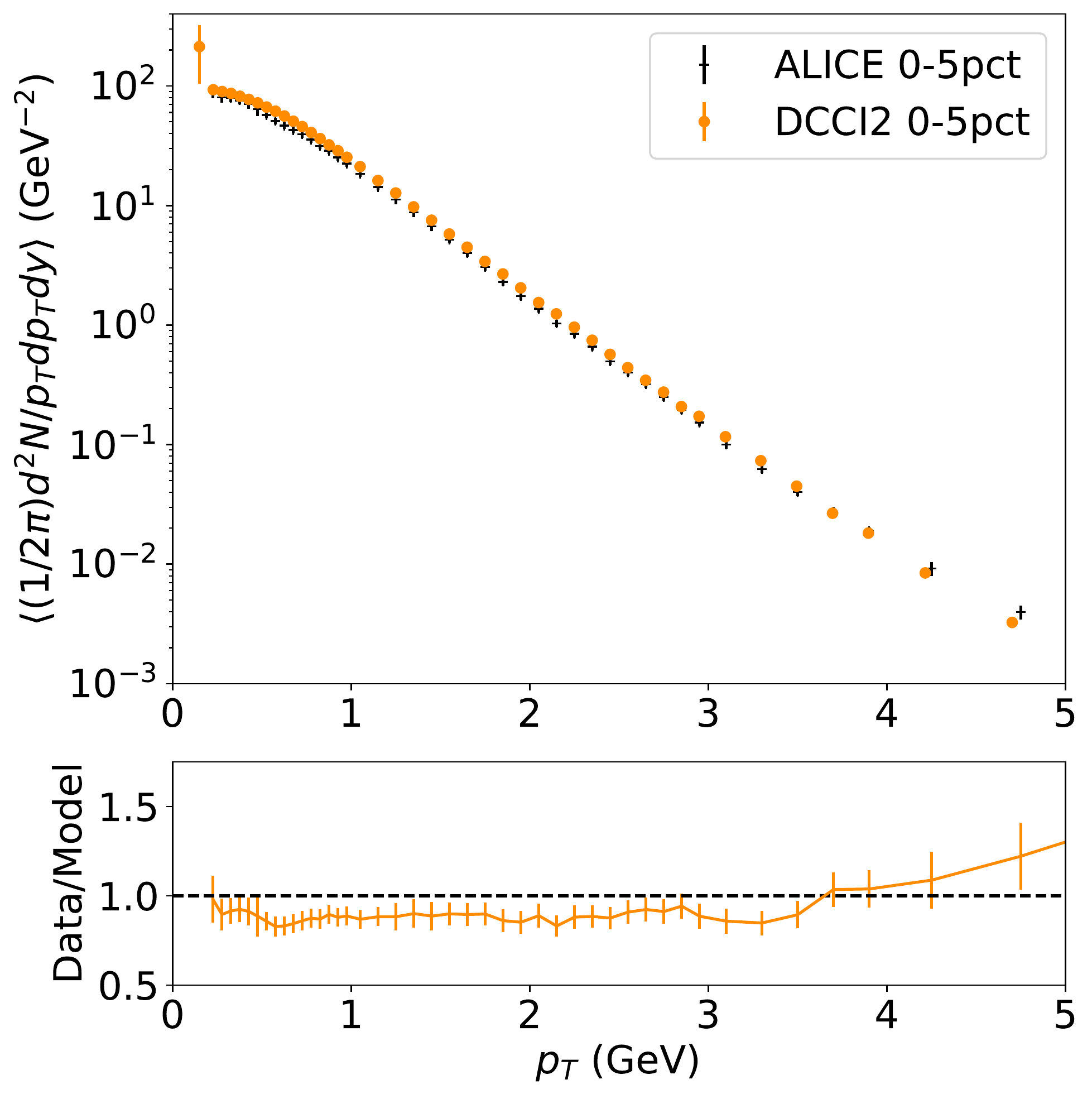}
    \includegraphics[bb=0 0 564 567, width=0.49\textwidth]{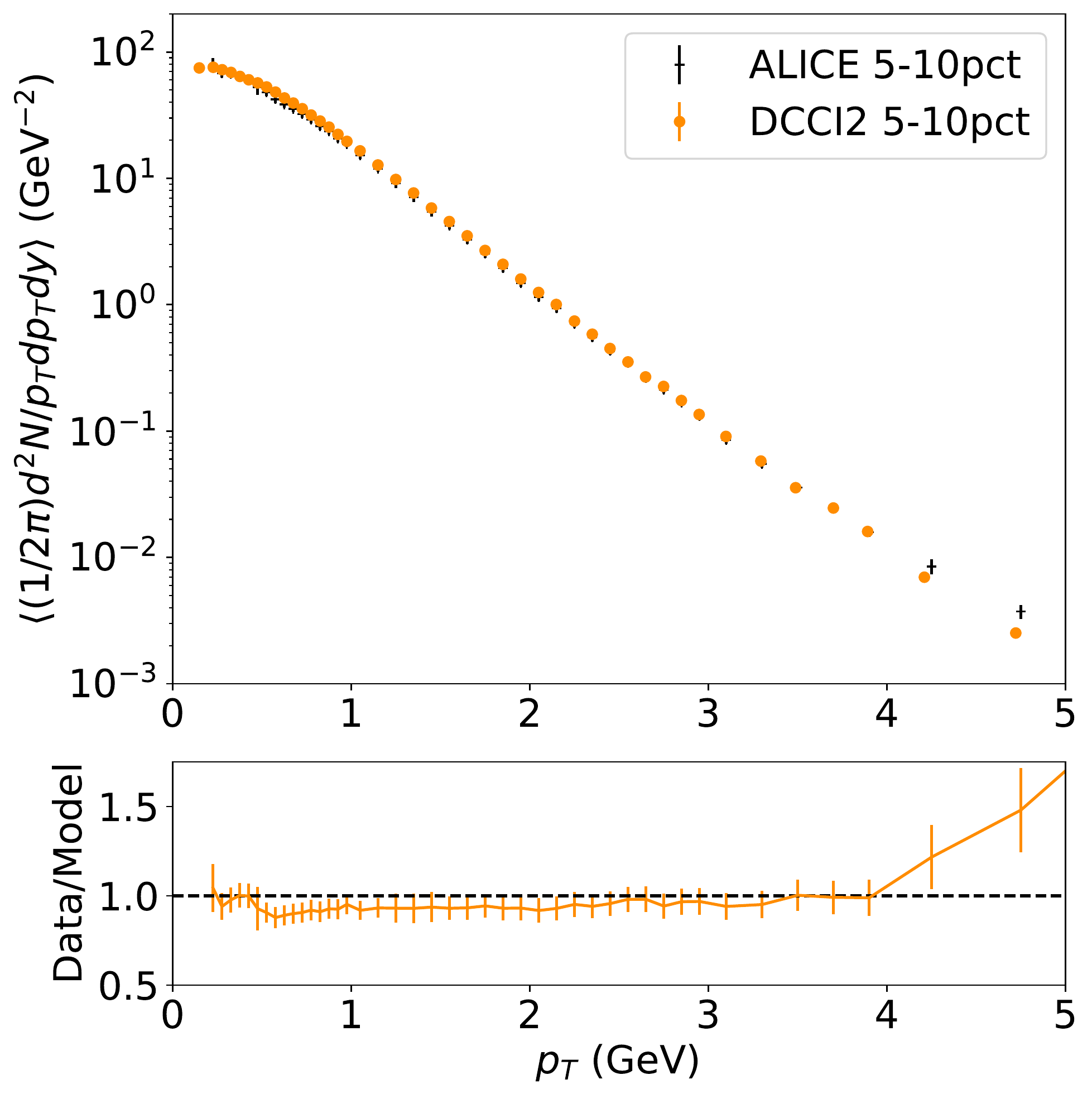}
    \includegraphics[bb=0 0 564 567, width=0.49\textwidth]{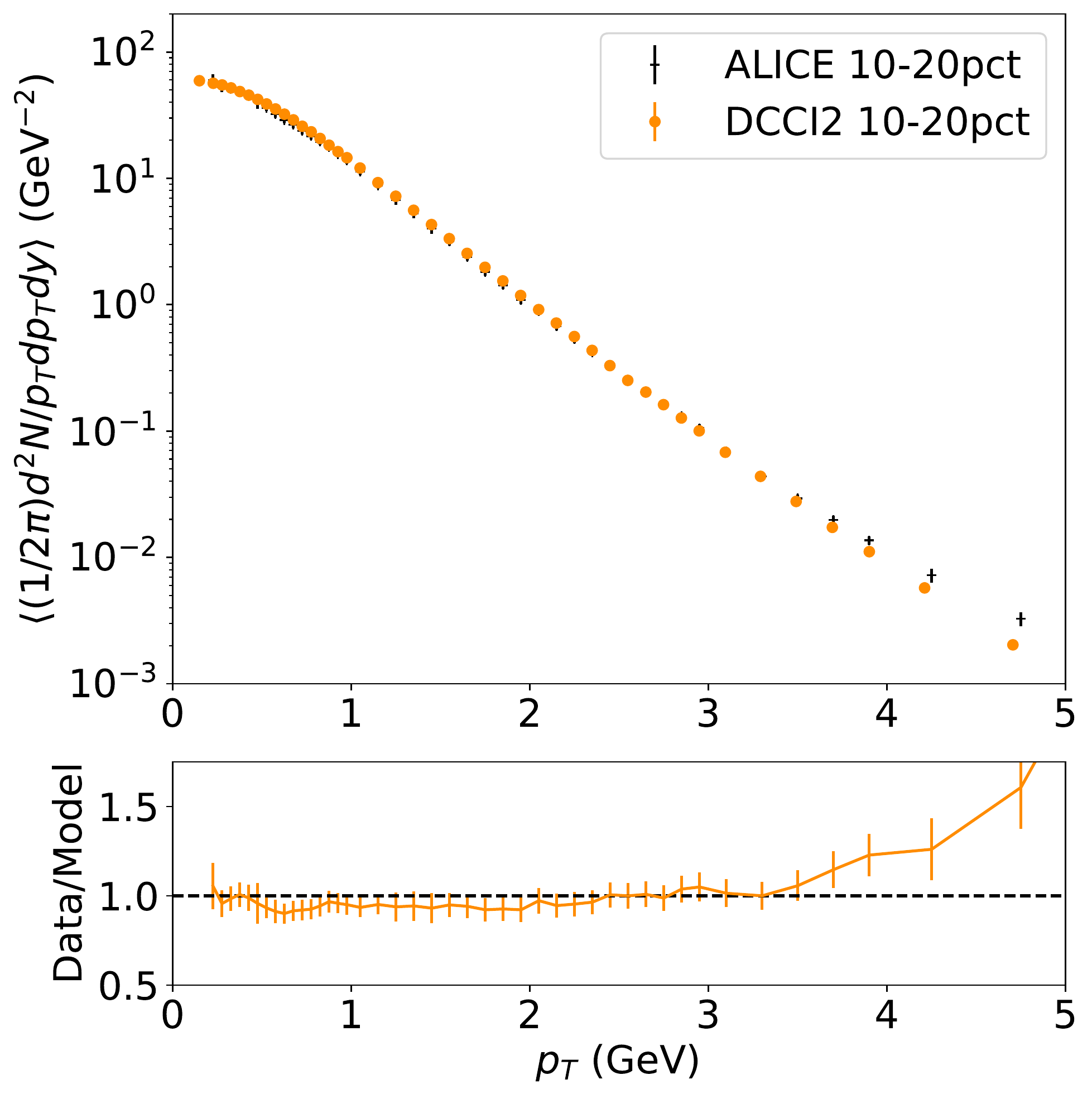}
    \includegraphics[bb=0 0 564 567, width=0.49\textwidth]{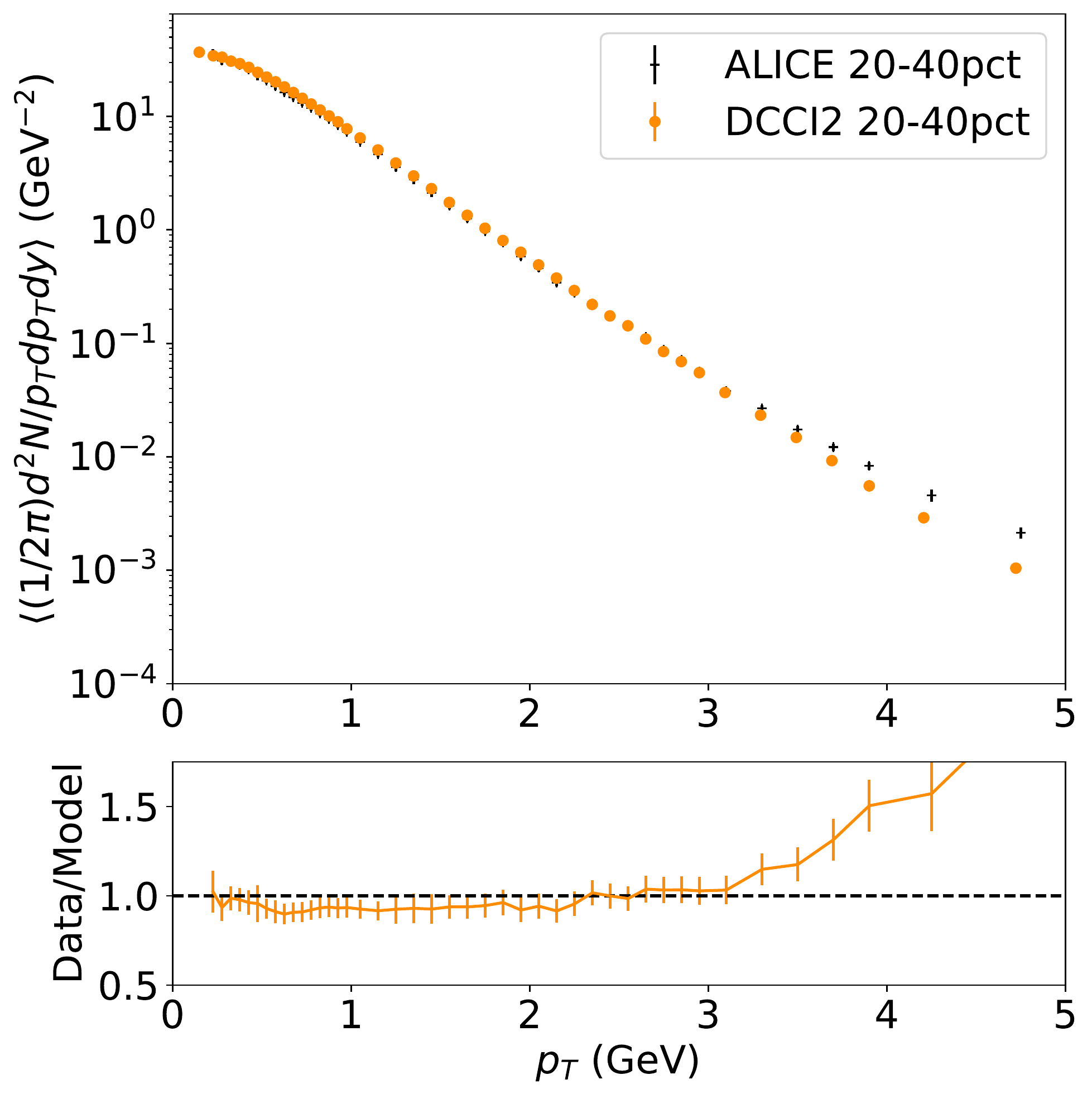}
    \includegraphics[bb=0 0 564 567, width=0.49\textwidth]{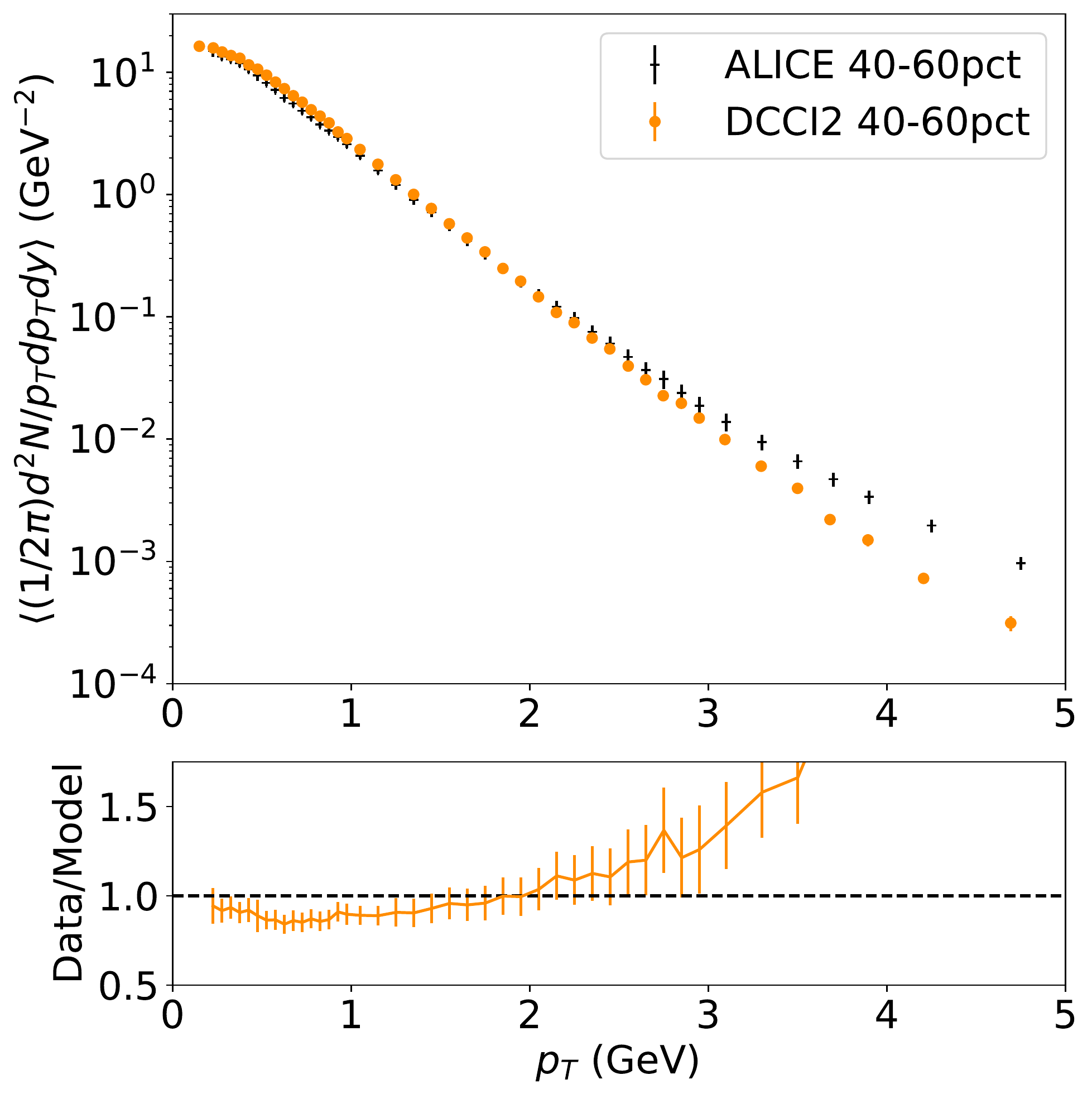}
    \includegraphics[bb=0 0 564 567, width=0.49\textwidth]{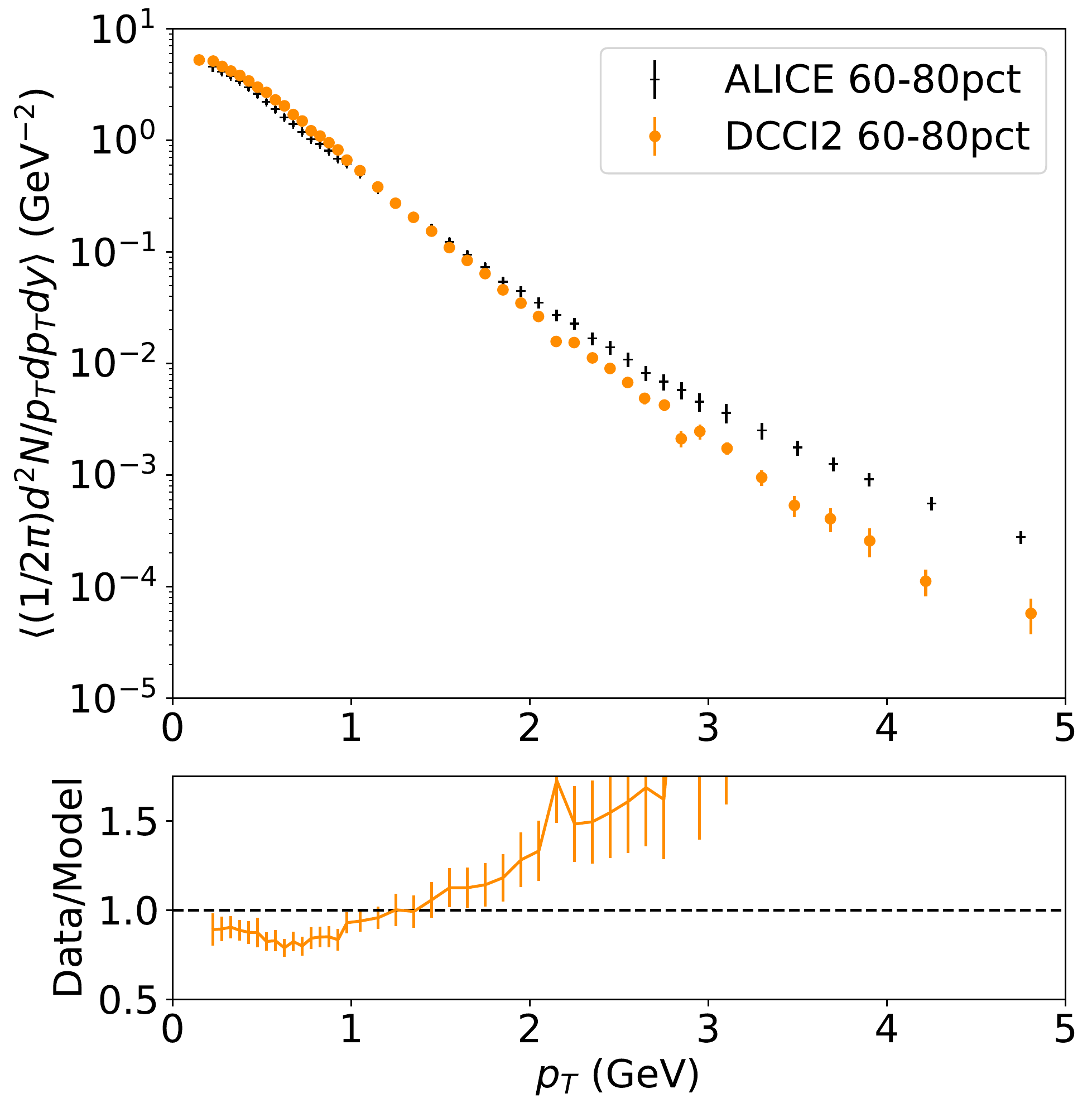}
    \caption{Centrality dependence of transverse momentum spectra of charged kaons ($K^+ + K^-$) from $Pb$+$Pb$ collisions at \snn=2.76 TeV in DCCI2. Comparison between results from DCCI2 (orange circles) and the ALICE experimental data (black crosses) are made.
    (Lower) Ratio of the ALICE experimental data to DCCI2 results at each $p_T$ bin.}
    \label{fig:PBPB2760_PTSPECTRA_K}
\end{figure}

\begin{figure}
    \centering
    \includegraphics[bb=0 0 564 567, width=0.49\textwidth]{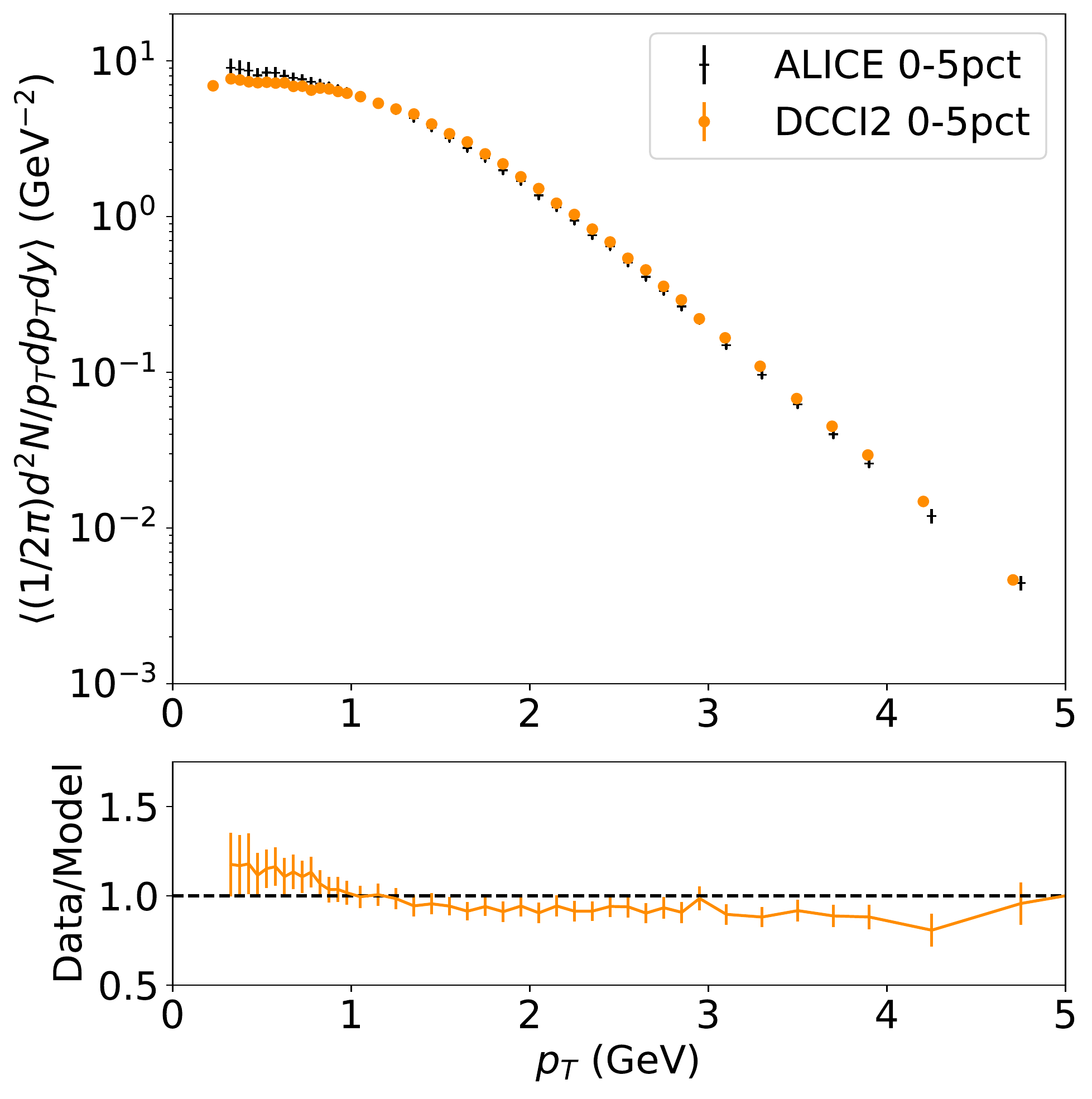}
    \includegraphics[bb=0 0 564 567, width=0.49\textwidth]{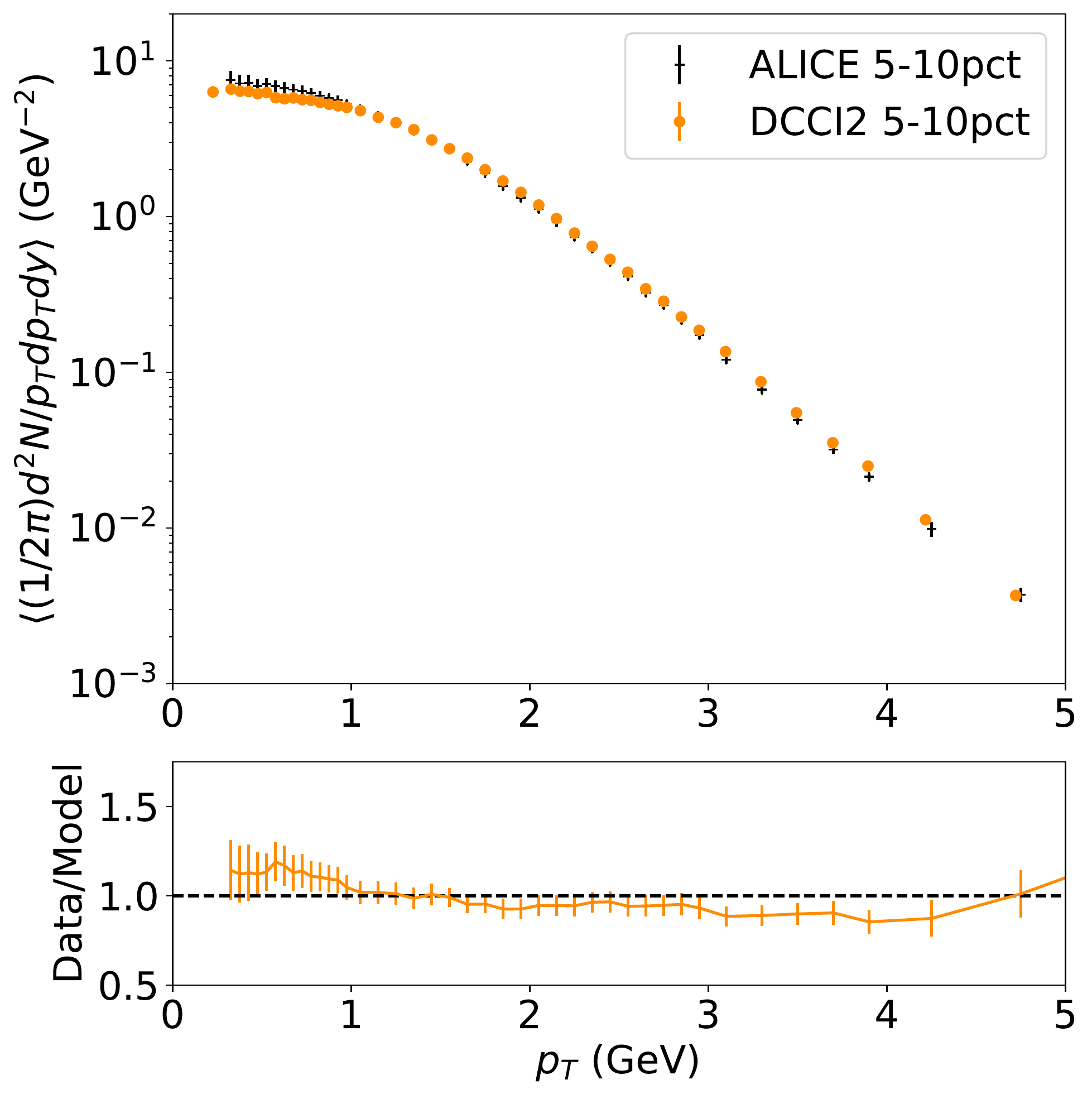}
    \includegraphics[bb=0 0 564 567, width=0.49\textwidth]{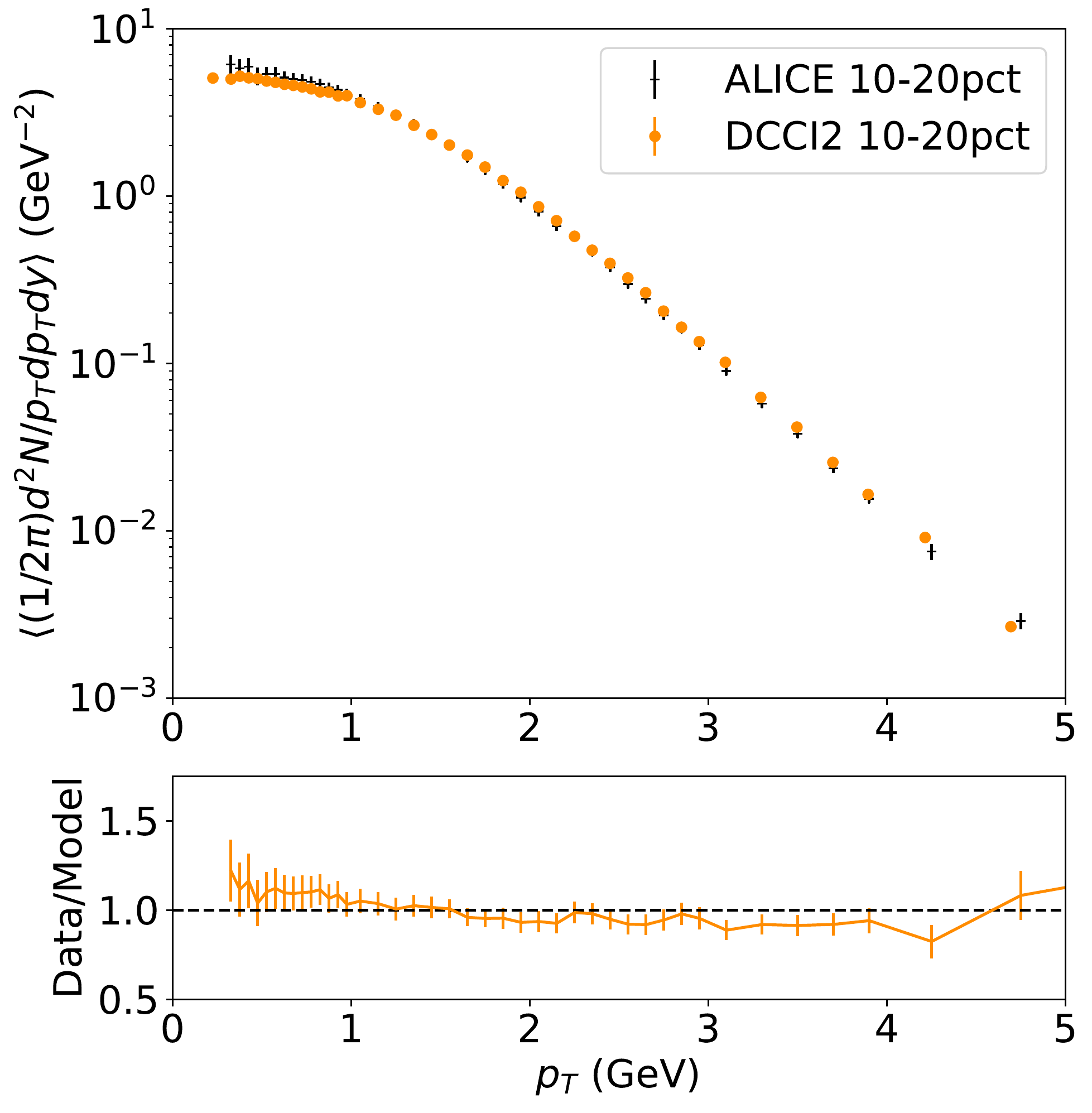}
    \includegraphics[bb=0 0 564 567, width=0.49\textwidth]{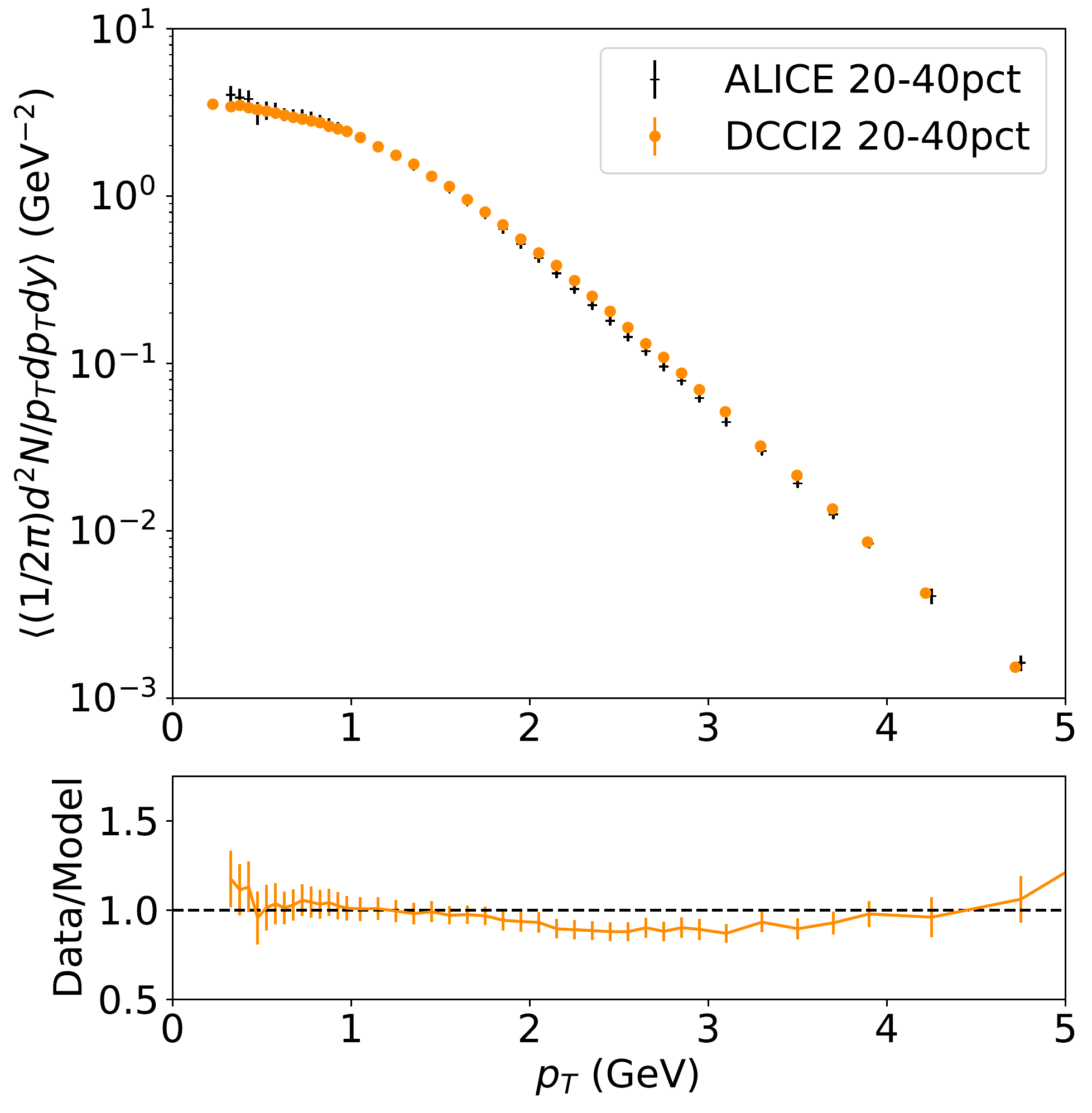}
    \includegraphics[bb=0 0 564 567, width=0.49\textwidth]{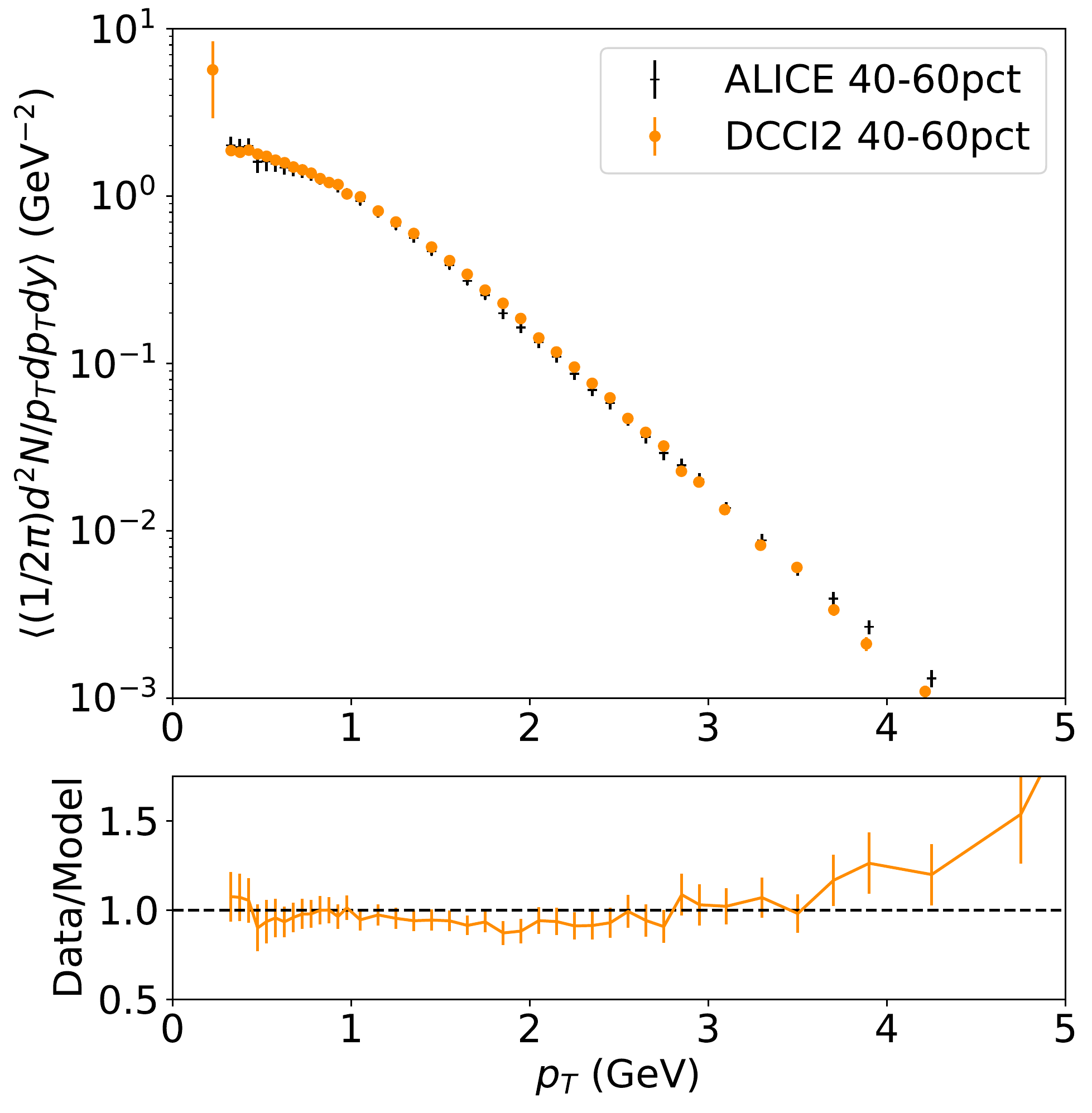}
    \includegraphics[bb=0 0 564 567, width=0.49\textwidth]{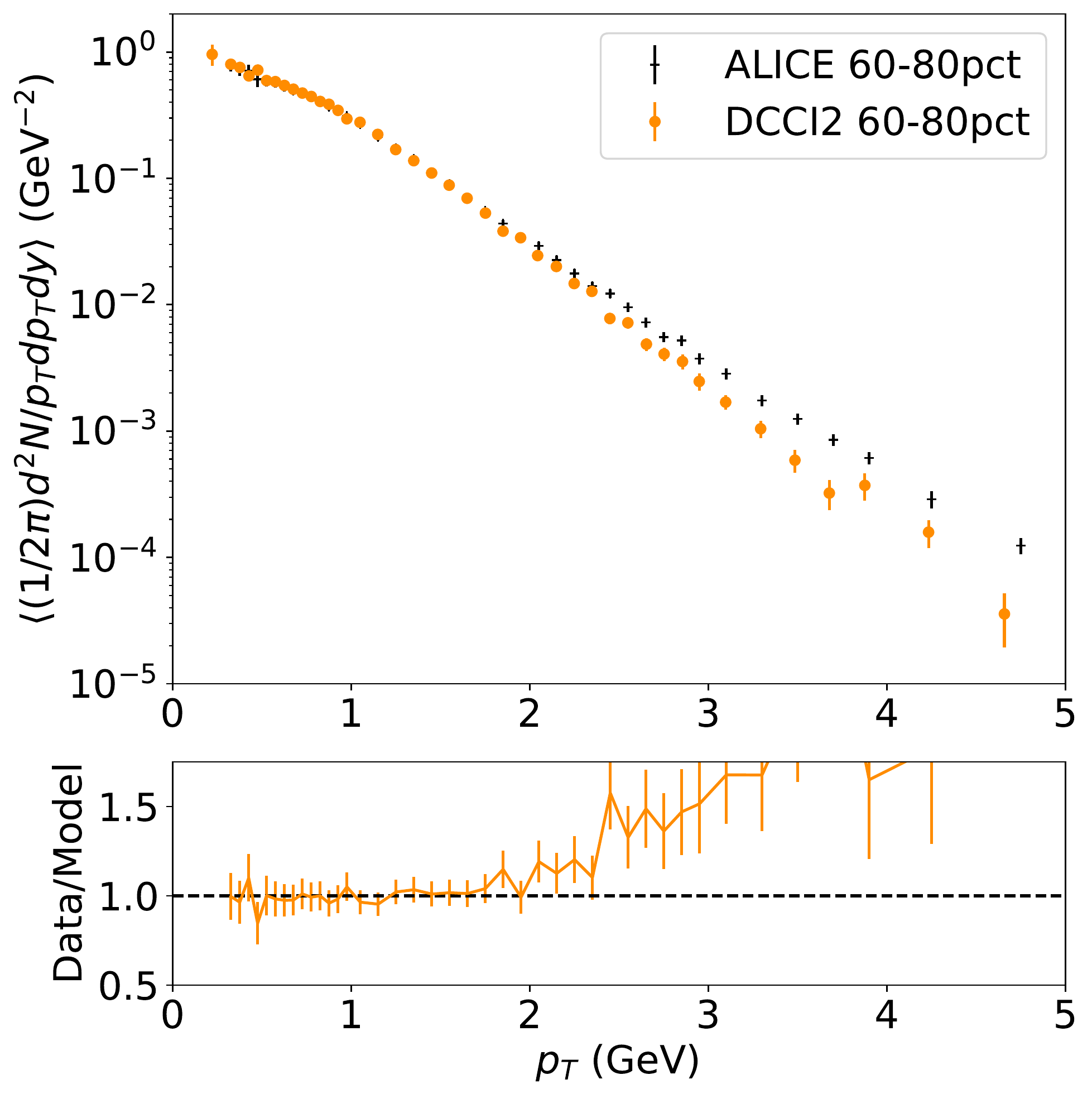}
    \caption{Centrality dependence of transverse momentum spectra of protons and antiprotons ($p + \bar{p}$) from $Pb$+$Pb$ collisions at \snn=2.76 TeV in DCCI2. Comparison between results from DCCI2 (orange circles) and the ALICE experimental data (black crosses) are made.
    (Lower) Ratio of the ALICE experimental data to DCCI2 results at each $p_T$ bin.}
    \label{fig:PBPB2760_PTSPECTRA_P}
\end{figure}

\subsubsection{Comparisons with experimental data}
In this subsection, I summarize comparisons of particle identified $p_T$ spectra in different multiplicity/centrality classes for $p$+$p$ collisions at \snn[proton] = 13 TeV and $Pb$+$Pb$ collisions at \snn = 2.76 TeV from DCCI2 with full simulation results, {\it{i.e.,}} with hadronic rescatterings.
Upper panels of Figs.~\ref{fig:PP13_PTSPECTRA_PI_1} and \ref{fig:PP13_PTSPECTRA_PI_2}
show centrality dependence of $p_T$ spectra of charged pions from DCCI2. 
Comparisons with ALICE experimental data \cite{ALICE:2020nkc} are made here.
The ratios of experimental data to DCCI2 results at each $p_T$ bin are shown in lower panels.
As an overall tendency, the results of DCCI2 clearly underestimate particle yields at high $p_T$ at high-multiplicity events.
Also, there is a difficulty in describing shapes of $p_T$ spectra:
there is a tendency that data to model ratios shown in lower panels show steep increase deviating from unity towards zero $p_T$ and, at the same time, they show a bump around or smaller than $p_T\approx0.5$ GeV except protons.

As for the underestimation at high $p_T$,
there are two possible factors that cause the underestimation.
First, I checked that there is a certain amount of energy loss of partons due to density of generated QGP fluids during dynamical core--corona initialization and it becomes larger at higher-multiplicity classes.
Second, \pythia \ is tuned to globally reproduce basic experimental data especially for charged particles, and, in DCCI2, generation of initial partons and string fragmentation of corona components are performed by \pythia \ with default parameters except $p_{T0, \mathrm{Ref}}$ which I discuss details in Sec.~\ref{subsection:Evolution_of_transverse_energy}.
Thus, if we have a certain modification on corona production more or less the results should deviate from experimental data.
Hence, the possible improvement of model description can be achieved by further parameter fitting on dynamical core-corona initialization in DCCI2 or on string fragmentation in \pythia.

On the other hand, regarding the difficulty in describing shapes at low $p_T$,
it should be emphasized that this regime is the very difficult part to reproduce experimental data because both core and corona components contribute to final hadron spectra.
There are some parameters that can change shapes of core spectra in DCCI2
\footnote{
For instance, transverse and longitudinal sizes of four momentum deposition in coordinate space.
Flow velocity is generated from pressure gradient in hydrodynamics, therefore
lumpiness of initially generated QGP in dynamical initialization sensitively affects the shape of $p_T$ spectra. 
}
, and I tuned them so that results roughly describe $p_T$ spectra of central $Pb$+$Pb$ collisions and adopt them universally from $p$+$p$ to $Pb$+$Pb$ and different multiplicity/centrality classes.
Meanwhile, I do not modify any parameters in \pythia \ as I mentioned.
In this sense,
I do not currently have possible solutions.
Further investigation on here should be done in future work.

Finally, I would like to mention some points on data to model comparisons on particle specie dependence.
There are different tendency on particle species in reproduction of experimental data.
Results of charged pions and protons show reasonable agreement with experimental data at small-multiplicity events while it is not the case for charged kaons:
there is a constant lack of particle yields above $p_T \approx 2$ GeV in DCCI2
where particle production is mostly from string fragmentation.

Figures \ref{fig:PBPB2760_PTSPECTRA_PI}, \ref{fig:PBPB2760_PTSPECTRA_K}, and \ref{fig:PBPB2760_PTSPECTRA_P} show comparisons between DCCI2 and experimental data of particle identified transverse momentum spectra of charged pions, charged kaons, and protons respectively from $Pb$+$Pb$ collisions at \snn = 2.76 TeV in different centrality classes \cite{ALICE:2015dtd}.
As an overall tendency, agreement with experimental data is better at central than peripheral events showing
lack of particle yields at high $p_T$ regime which should be compensated with corona productions at high $p_T$ at peripheral events.
The reasons for this problem is the same as in $p$+$p$ collisions:
there is a certain amount of energy loss of partons due to density of generated QGP fluids during dynamical core--corona initialization,
and \pythia \ provides a good agreement with experimental data as it is from low to high $p_T$.

\subsection{Core/corona correction on momentum distributions}
As I discussed in Secs.~\ref{subsec:ParameterDetermination} and \ref{subsec:OverallTendency}, 
both core and corona contributions appear over a wide range of multiplicity.
Moreover, each component contributes as a function of $p_T$ in a nontrivial way.
To further investigate how the effects of the interplay between core and corona components appear on observable, 
I first analyze
the mean transverse momentum $\langle p_T \rangle$ of charged particles at midrapidity as a function of
the number of charged hadrons generated at midrapidity $N_{\mathrm{ch}}$ in $p$+$p$ and $Pb$+$Pb$ collisions.

\begin{figure}
\begin{center}
\includegraphics[bb=0 0 527 619, width=0.49\textwidth]{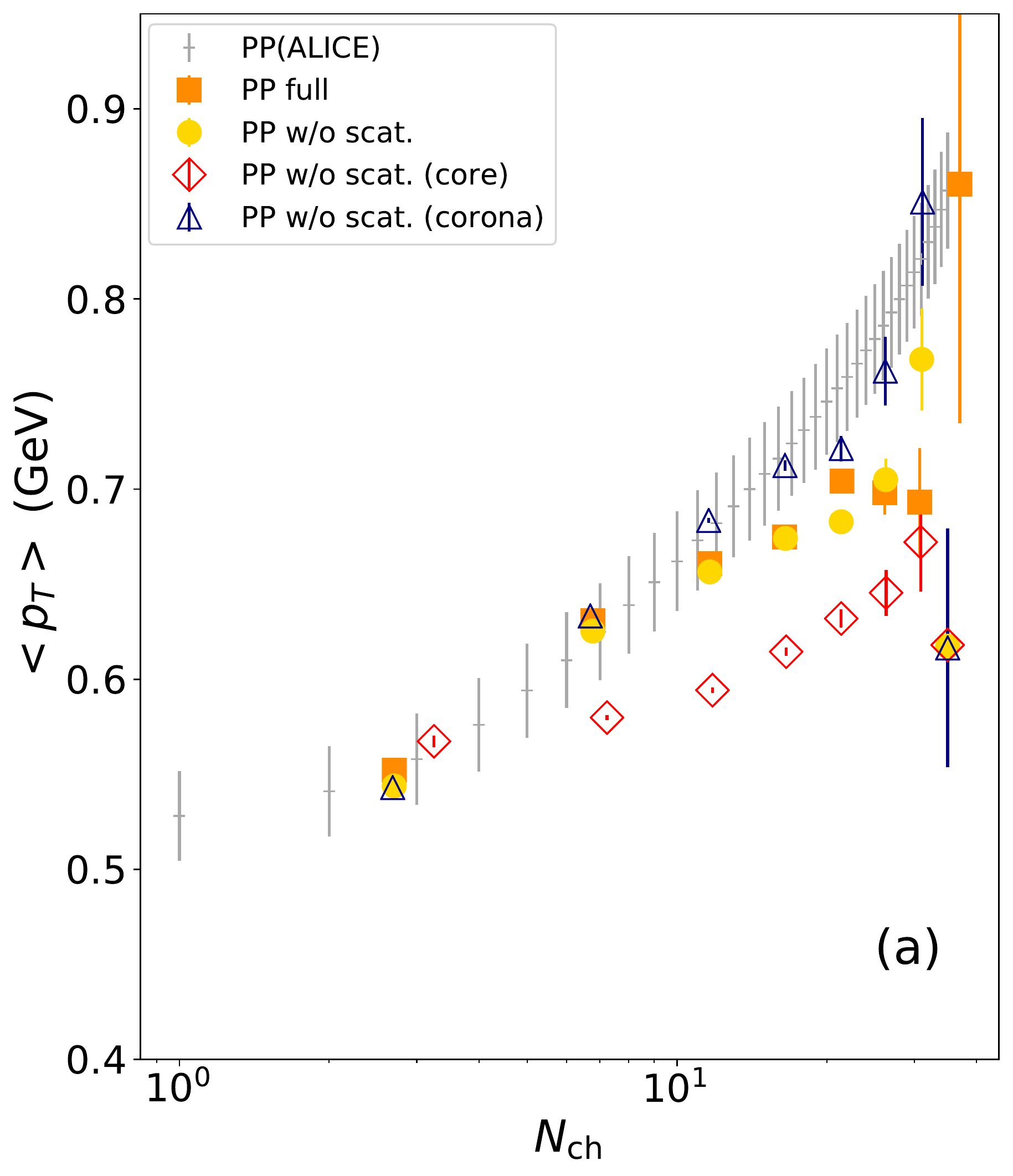}
\includegraphics[bb=0 0 527 619, width=0.49\textwidth]{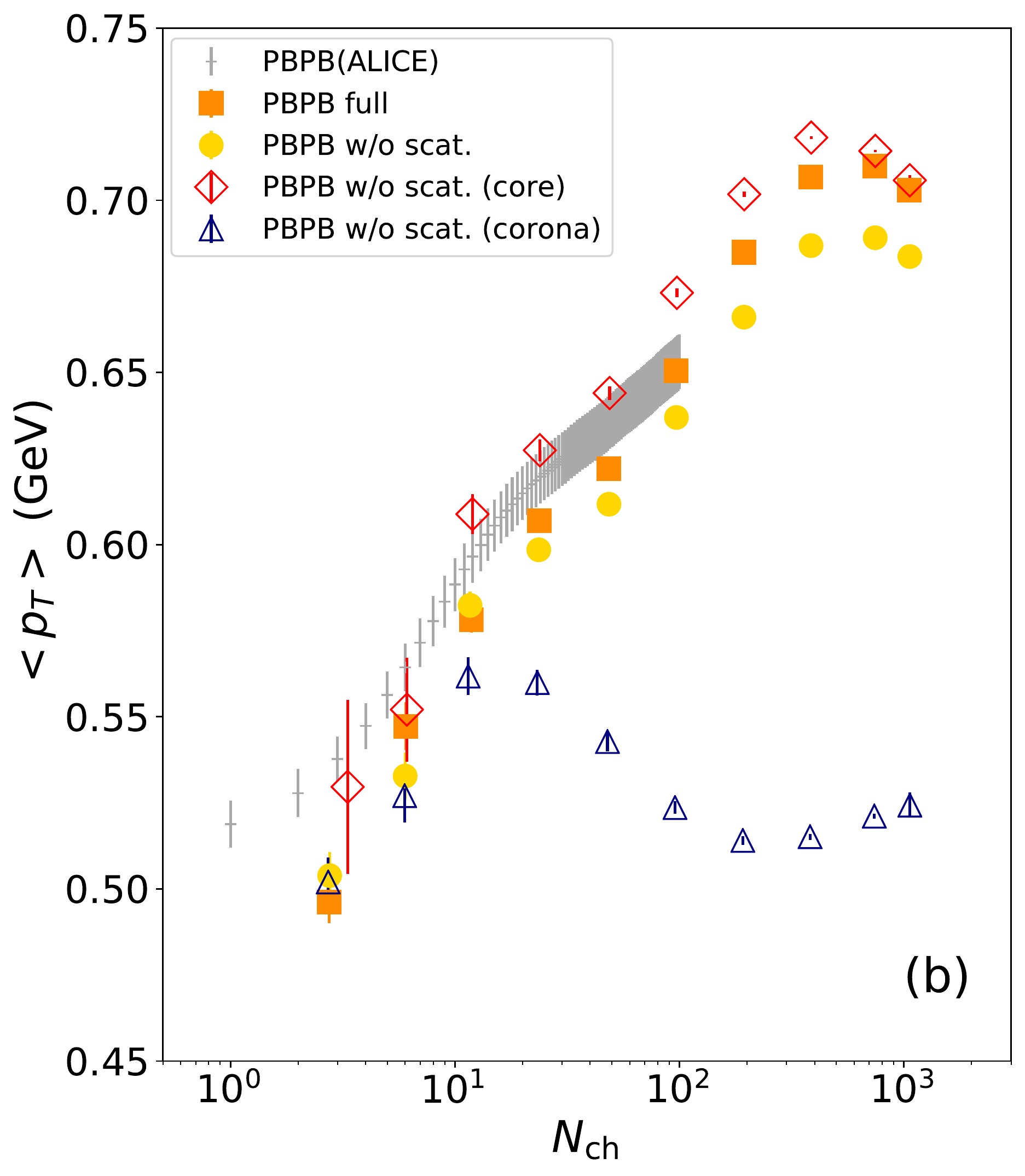}
\caption{(Color Online)
Mean transverse momentum of charged particles as a function of the number of charged particles in (a) $p$+$p$ collisions at \snn[proton] = 7 TeV and (b) $Pb$+$Pb$ collisions at \snn = $2.76$ TeV compared with the ALICE experimental data \cite{Abelev:2013bla}
(gray pluses). Results from full simulations (closed orange squares), from simulations without hadronic rescatterings (closed yellow circles), from core components (open red diamonds), and from corona components (open blue triangles) are shown for comparisons.
\label{fig:MEANPTMULTI_PP_PBPB_CORECORONA}
}
\end{center}
\end{figure}

Figure~\ref{fig:MEANPTMULTI_PP_PBPB_CORECORONA} shows the mean transverse momentum $\langle p_T\rangle$ of charged particles as a function of charged particle multiplicity $N_{\mathrm{ch}}$ in 
(a) $p$+$p$ collisions at $\sqrt{s} = 7 \ \mathrm{TeV}$ and (b) $Pb$+$Pb$ collisions at \snn $= 2.76 \ \mathrm{TeV}$.
Charged particles with $0.15<p_T<10.0$ GeV and $|\eta|<0.3$ are used for evaluation of $\langle p_T\rangle$,
while $N_{\mathrm{ch}}$ is obtained by counting charged particles with $|\eta|<0.3$ (without $p_T$ cut), which is the same kinematic range used in Ref.~\cite{Abelev:2013bla}.

For $p$+$p$ collisions in Fig.~\ref{fig:MEANPTMULTI_PP_PBPB_CORECORONA} (a), the result from DCCI2 qualitatively describes the steep enhancement of $\langle p_T\rangle$ along $N_{\mathrm{ch}}$ observed in the ALICE experimental data \cite{Abelev:2013bla}.
Almost no significant difference is seen between results from full simulations and the ones without hadronic rescatterings.
This means that the effect of hadronic rescatterings on $\langle p_T\rangle$ of charged particles is almost negligible due to a small number of final hadrons in $p$+$p$ collisions.
One also sees that 
the core and corona components show small difference of $\langle p_T \rangle$ below $N_{\mathrm{ch}} \approx 20$.
This is because, as seen in Fig.~\ref{fig:PTSPECTRA_PP_PBPB} (a),
there is no large difference for the slopes of $p_T$ spectrum of the core and corona components in low $p_T$ regions while the particle productions in the region would contribute to $\langle p_T \rangle$ significantly.

For $Pb$+$Pb$ collisions presented in Fig.~\ref{fig:MEANPTMULTI_PP_PBPB_CORECORONA} (b), the results from full simulations with DCCI2 reasonably describe the experimental data within the range of experimental data. 
A slight difference is seen between results from full simulations and the ones from simulations without hadronic rescatterings:
mean transverse momentum is slightly enhanced due to hadronic rescatterings and the effect becomes relatively clear as increasing $N_{\mathrm{ch}}$. 
On the other hand, the large difference is seen between the results from core and corona components.
The core components show larger $\langle p_T\rangle$ while the corona components show smaller values for almost the entire $N_{\mathrm{ch}}$. 
The larger $\langle p_T\rangle$ from core components originates from the flatter slope of $p_T$ spectrum,
while 
the smaller $\langle p_T\rangle$ from corona components originates from the steeper slope of $p_T$ spectrum in the low $p_T$ region seen in Fig.~\ref{fig:PTSPECTRA_PP_PBPB} (b).
The difference between the results without hadronic rescatterings and the ones from core components exactly exhibits there exists the sizable \textit{correction from non-thermalized matter} to the results obtained purely from hydrodynamics.
The correction is found to be visible for the entire  $N_{\mathrm{ch}}$ and to be $\approx 5$-$11\%$ in $N_{\mathrm{ch}}\lesssim 200$.

\begin{figure}
    \centering
    \includegraphics[bb=0 0 564 567, width=0.45\textwidth]{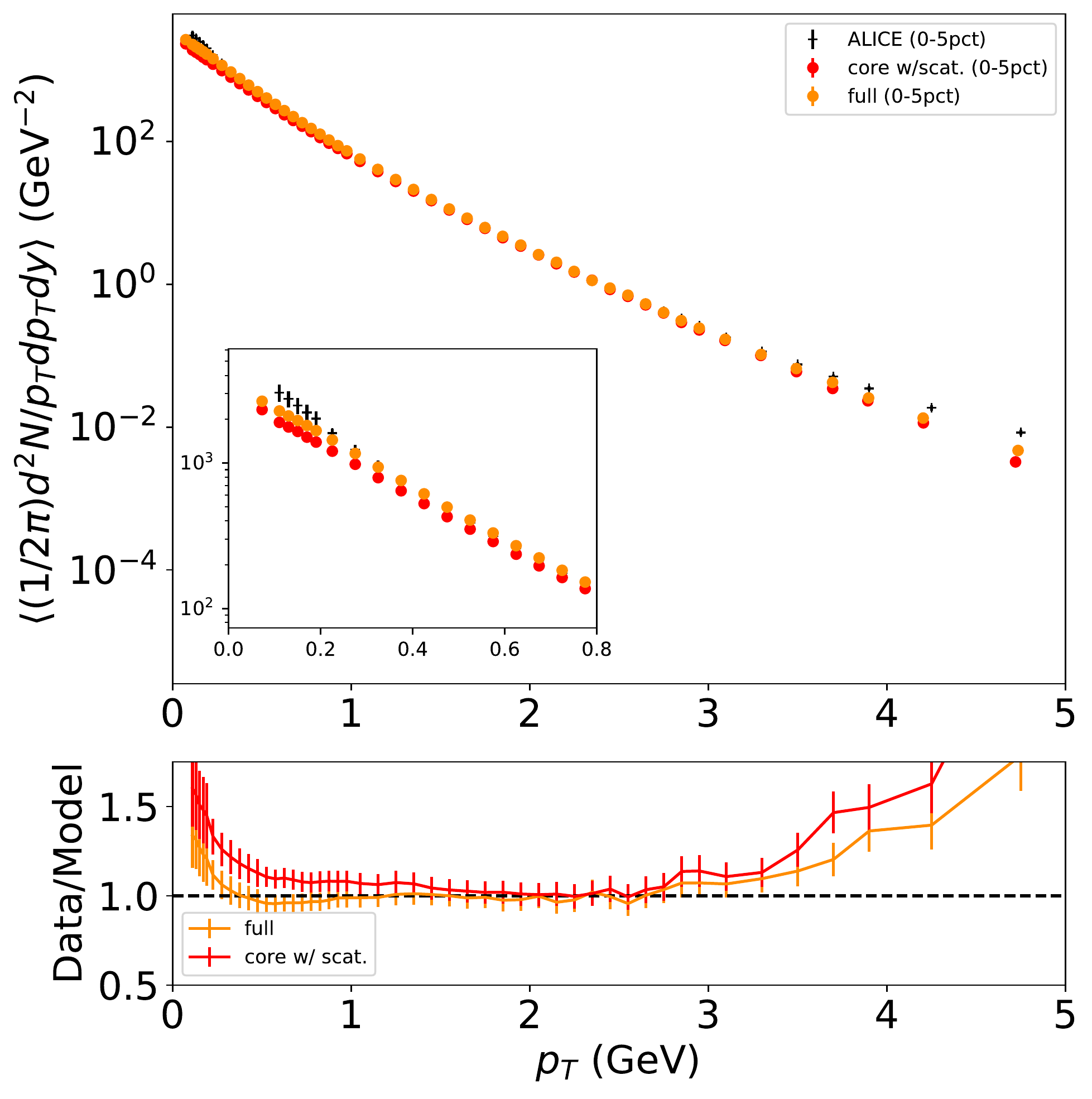}
    \includegraphics[bb=0 0 564 567, width=0.45\textwidth]{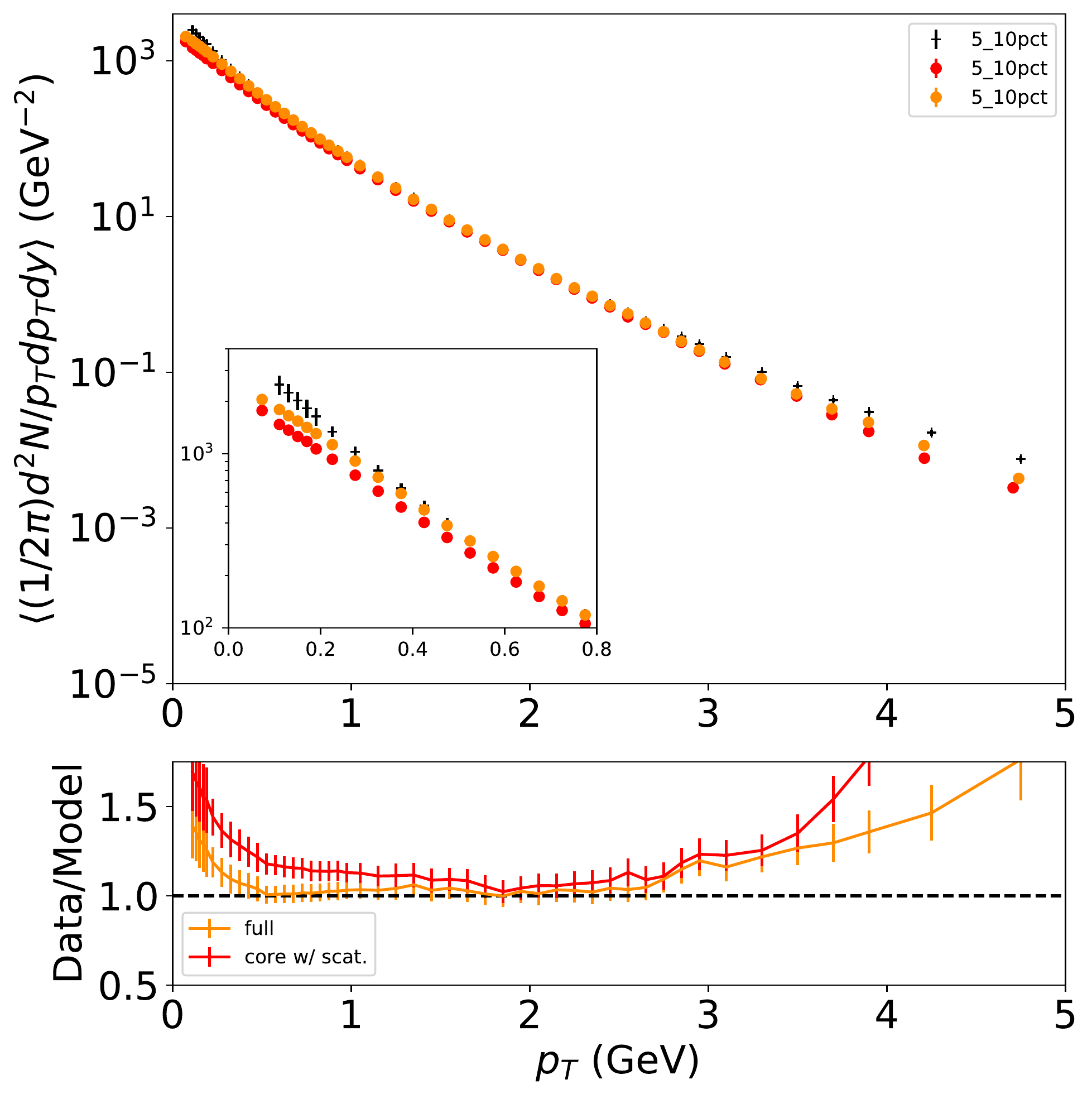}
    \includegraphics[bb=0 0 564 567, width=0.45\textwidth]{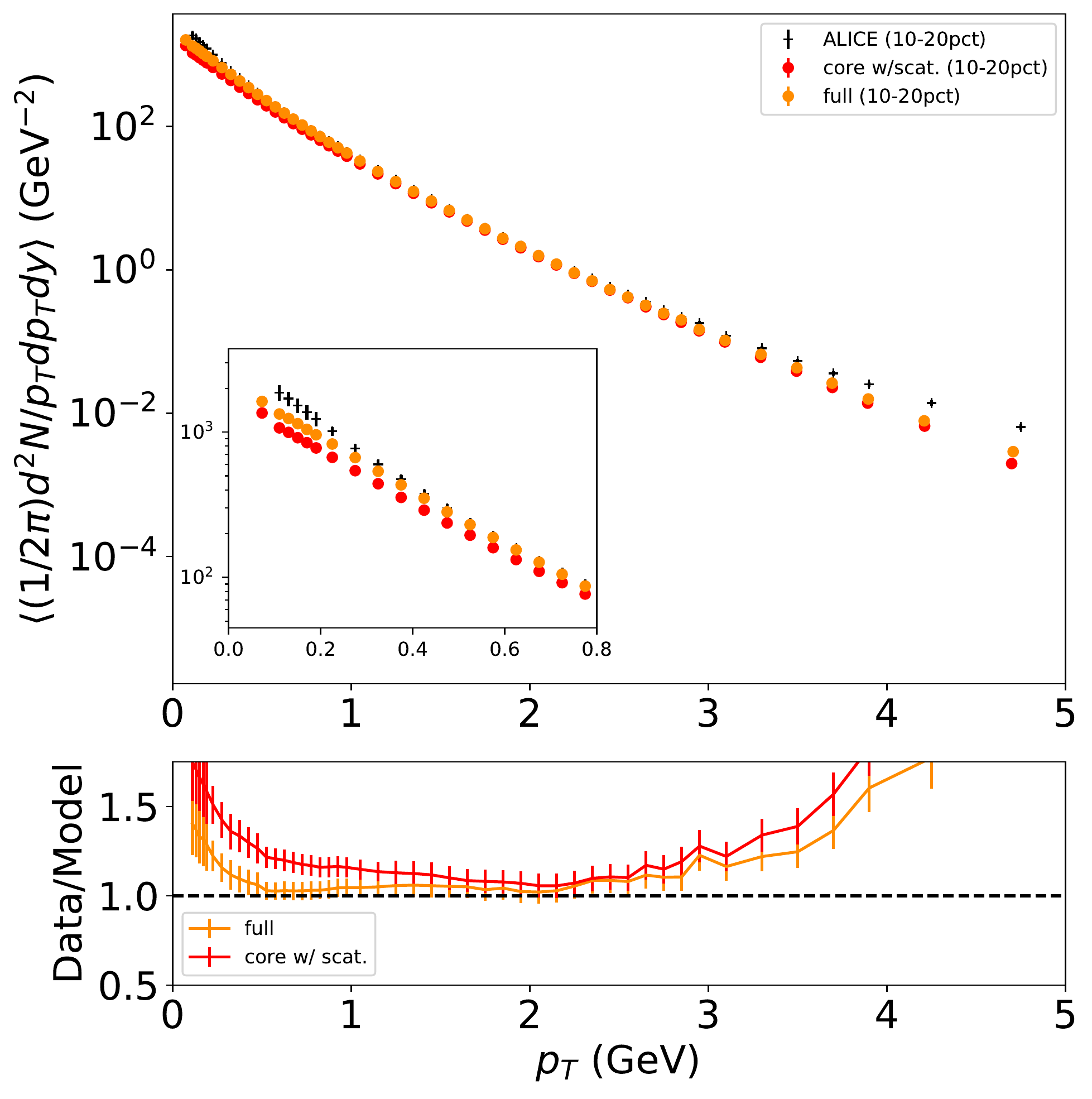}
    \includegraphics[bb=0 0 564 567, width=0.45\textwidth]{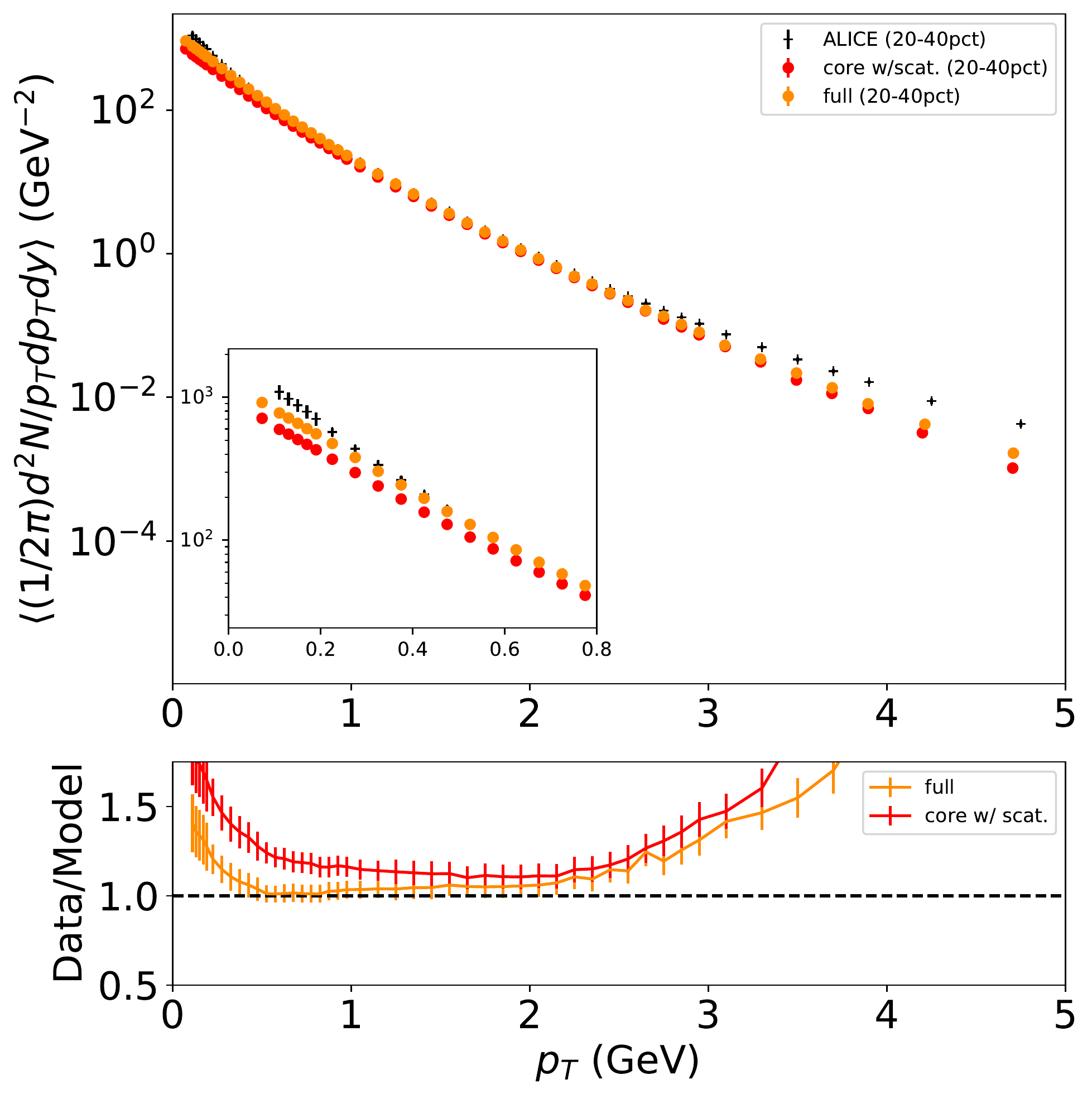}
    \includegraphics[bb=0 0 564 567, width=0.45\textwidth]{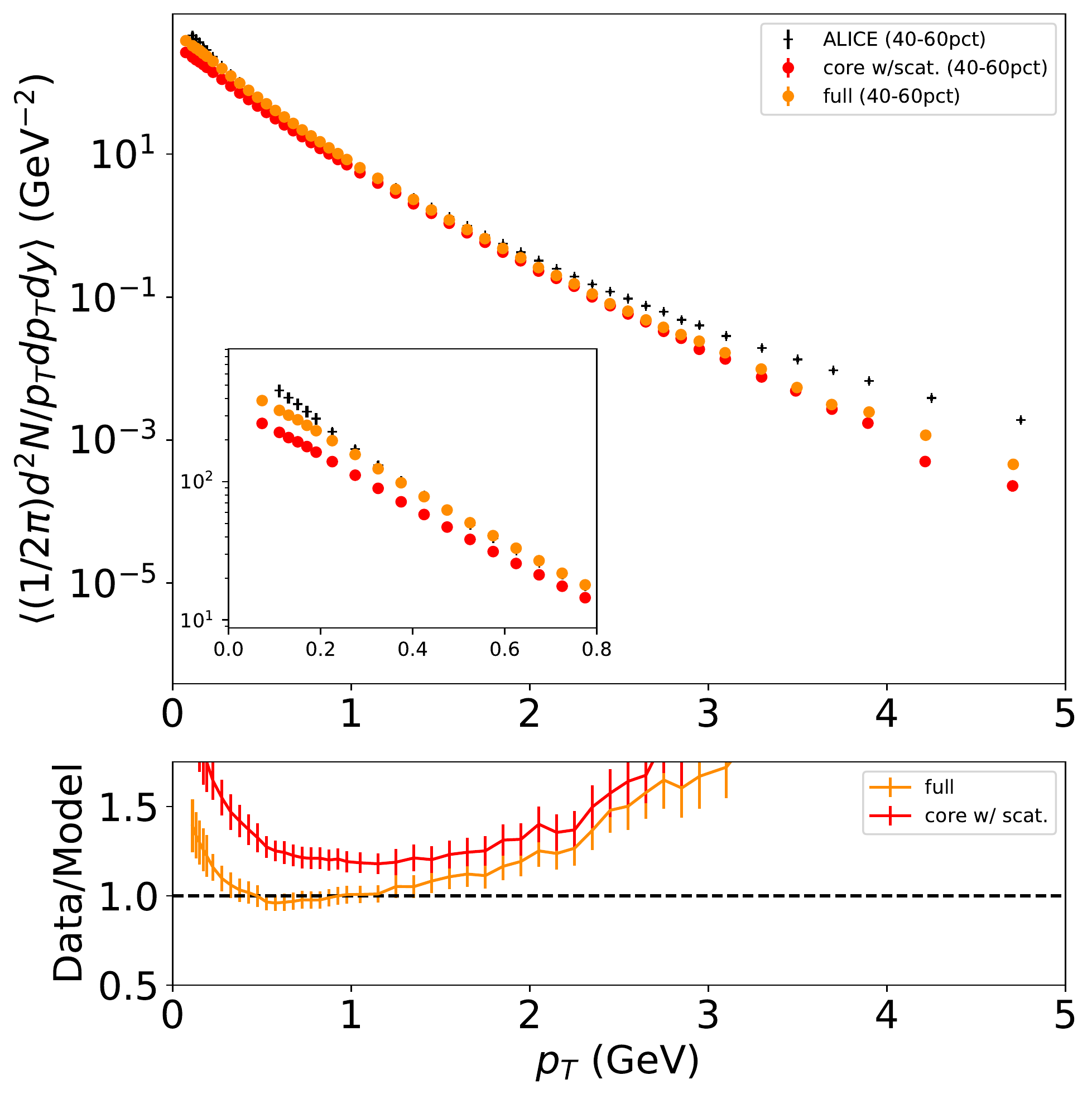}
    \includegraphics[bb=0 0 564 567, width=0.45\textwidth]{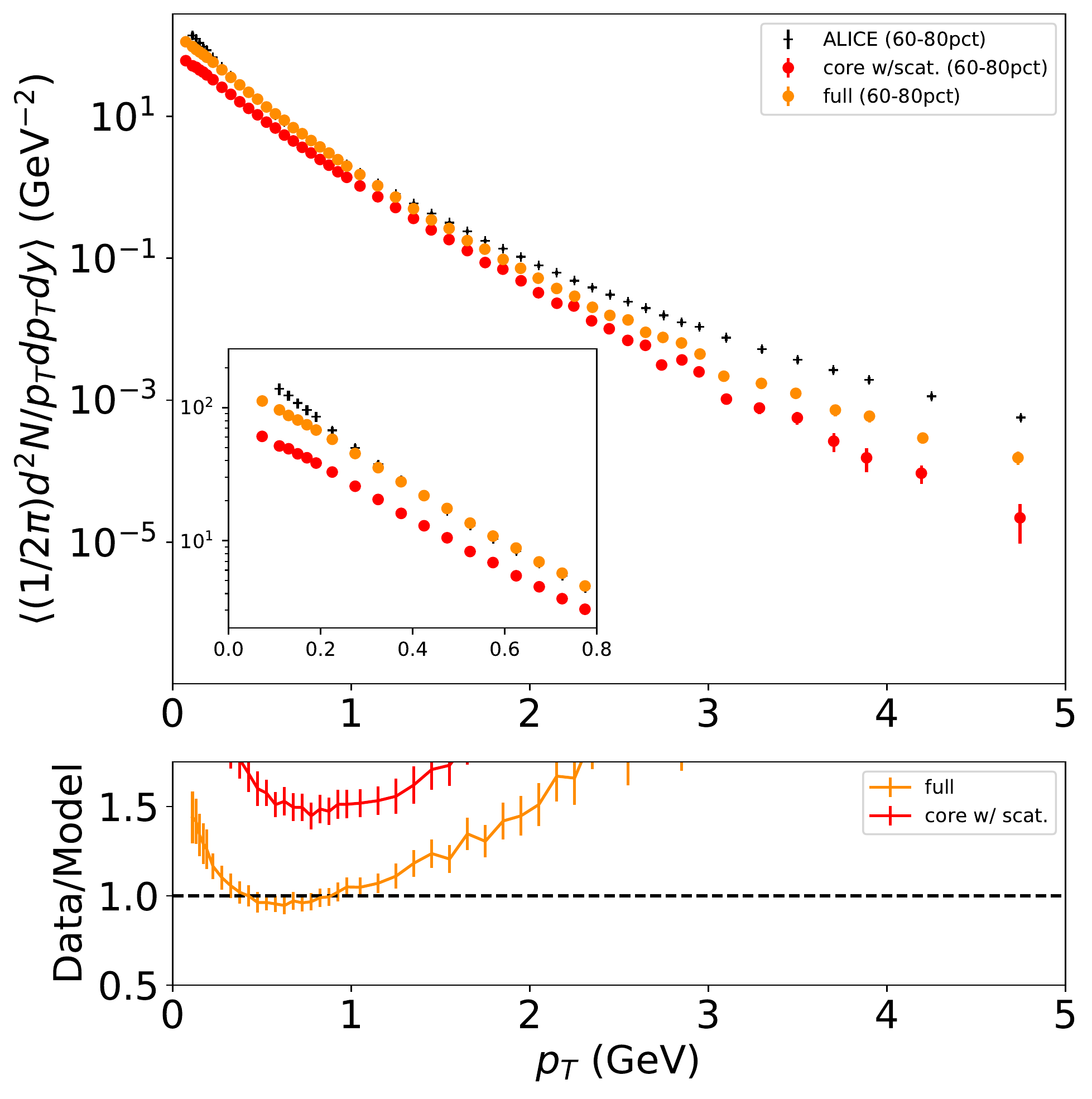}
    \caption{(Upper) Centrality dependence of transverse momentum spectra of charged pions ($\pi^+ + \pi^-$) from $Pb$+$Pb$ collisions at \snn = 2.76 TeV from DCCI2 with full simulations (orange) and core components with hadronic rescatterings (red). Comparisons to ALICE experimental data (black crosses) are also shown. Insets are blow-ups of very low $p_T$ (0.0-0.8 GeV) regime.
    (Lower) Ratio of the ALICE experimental data to DCCI2 results with full simulation (orange) and core components with hadronic rescatterings (red) at each $p_T$ bin.
    }
    \label{fig:PBPB2760_PTSPECTRA_PI_LOWPT}
\end{figure}

\begin{figure}
    \centering
    \includegraphics[bb=0 0 564 567, width=0.45\textwidth]{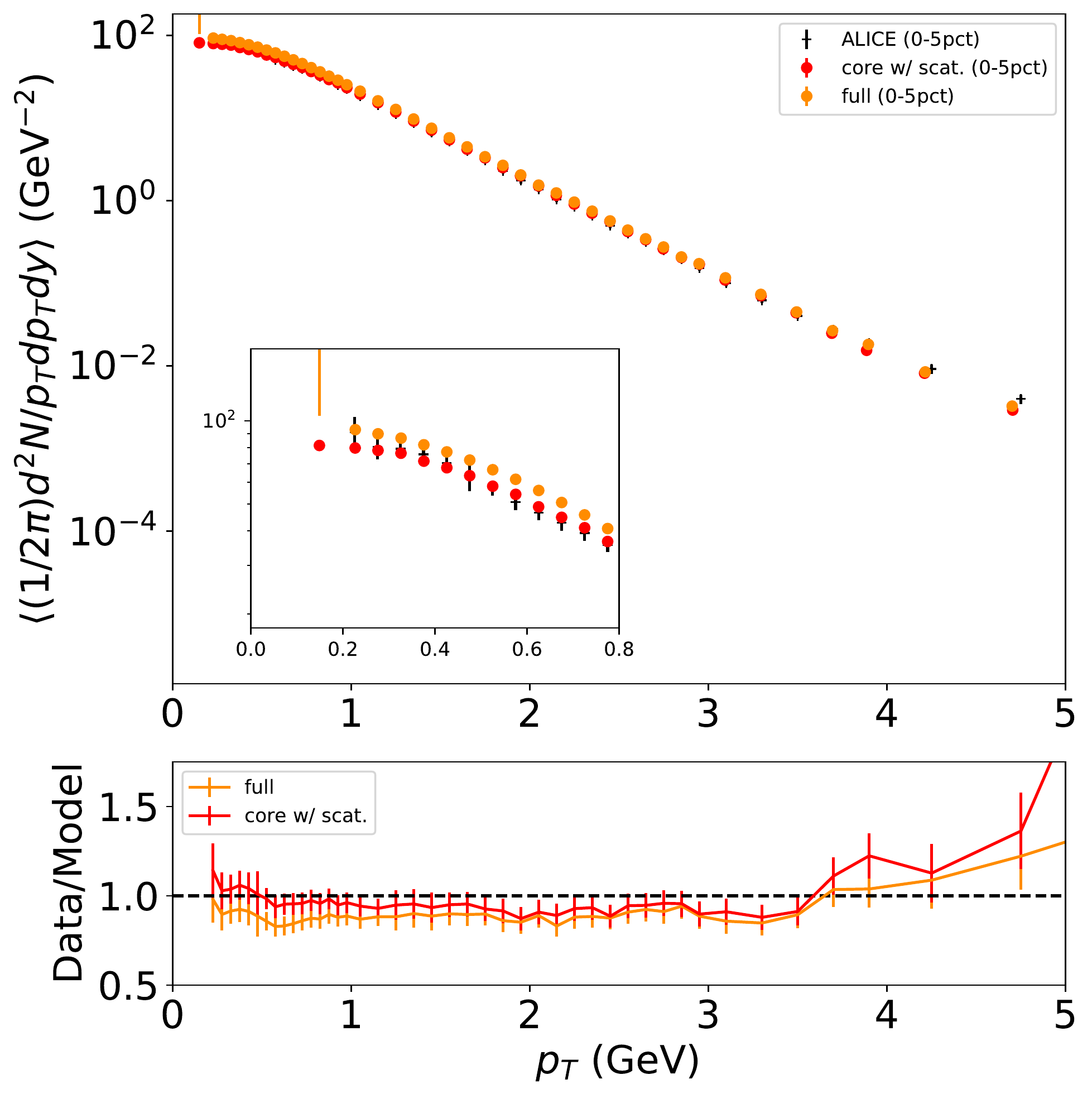}
    \includegraphics[bb=0 0 564 567, width=0.45\textwidth]{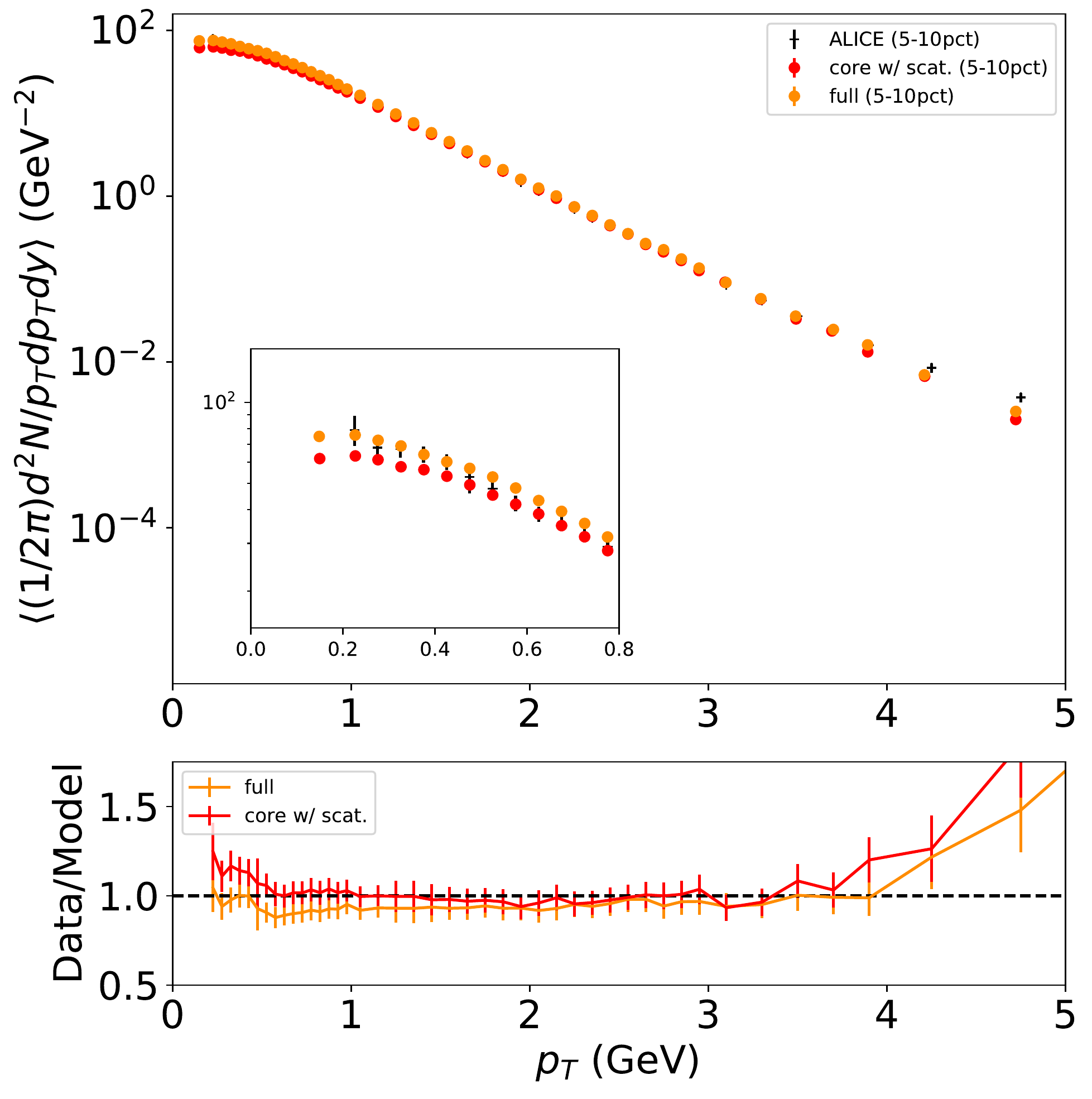}
    \includegraphics[bb=0 0 564 567, width=0.45\textwidth]{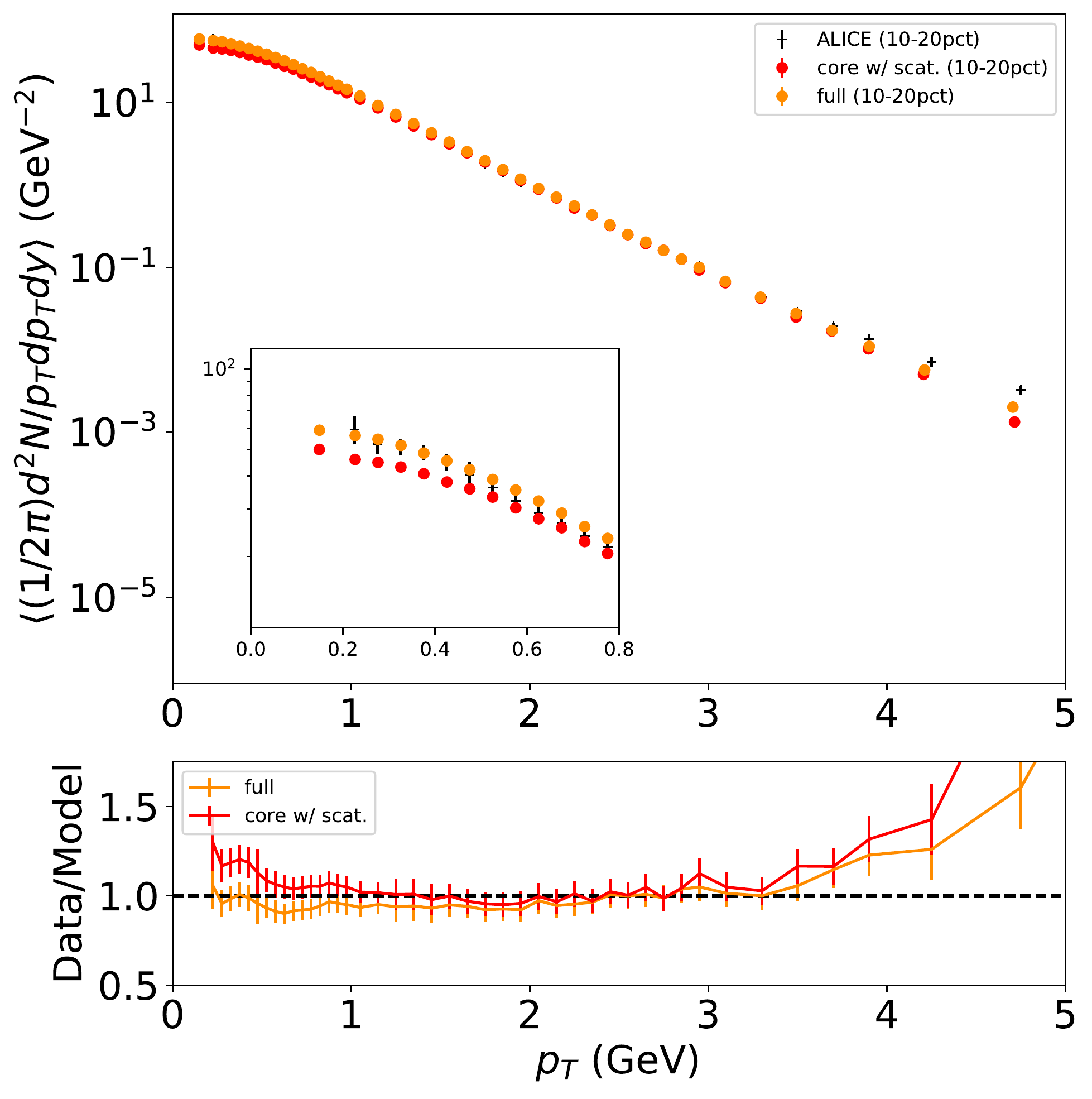}
    \includegraphics[bb=0 0 564 567, width=0.45\textwidth]{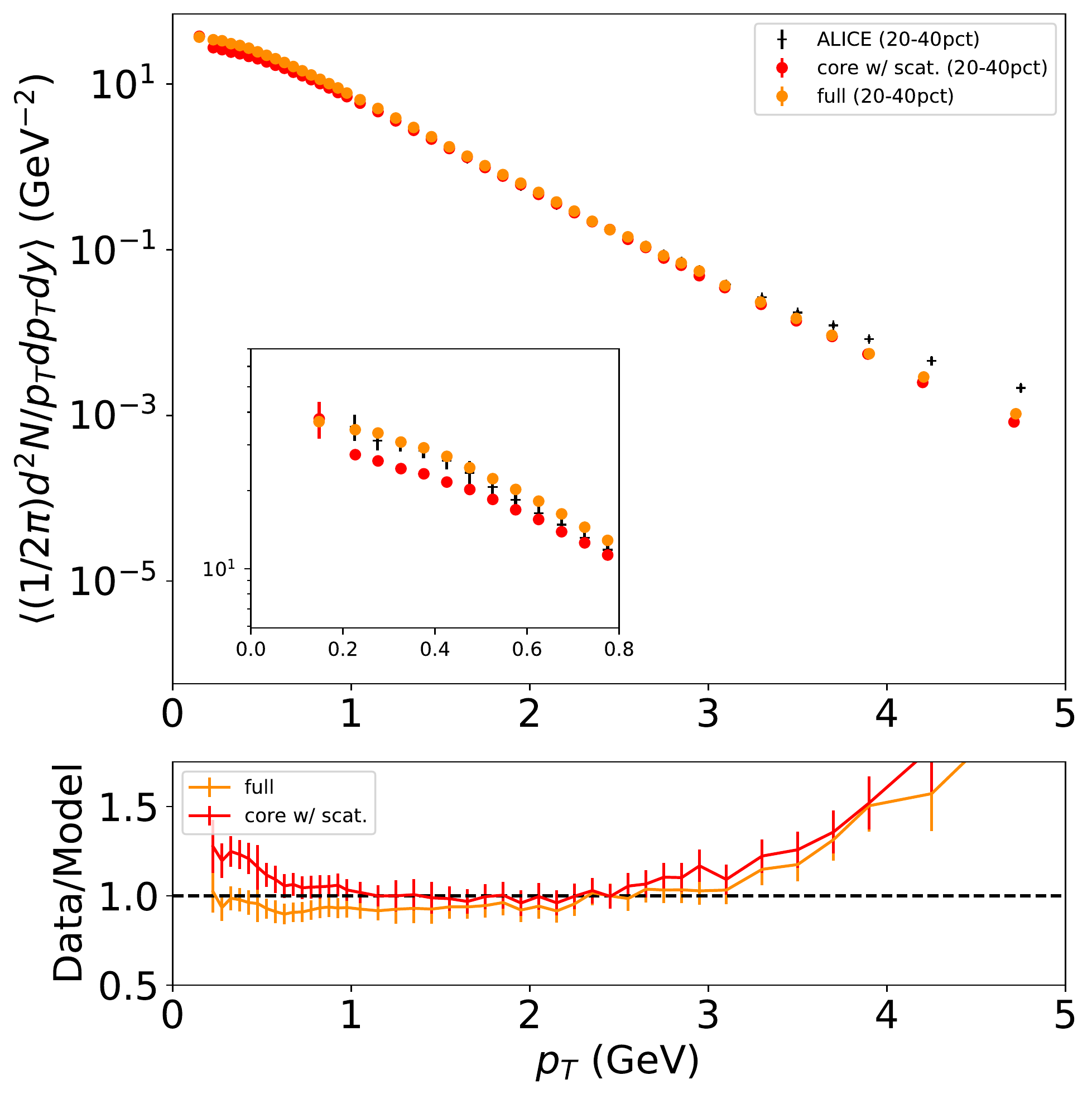}
    \includegraphics[bb=0 0 564 567, width=0.45\textwidth]{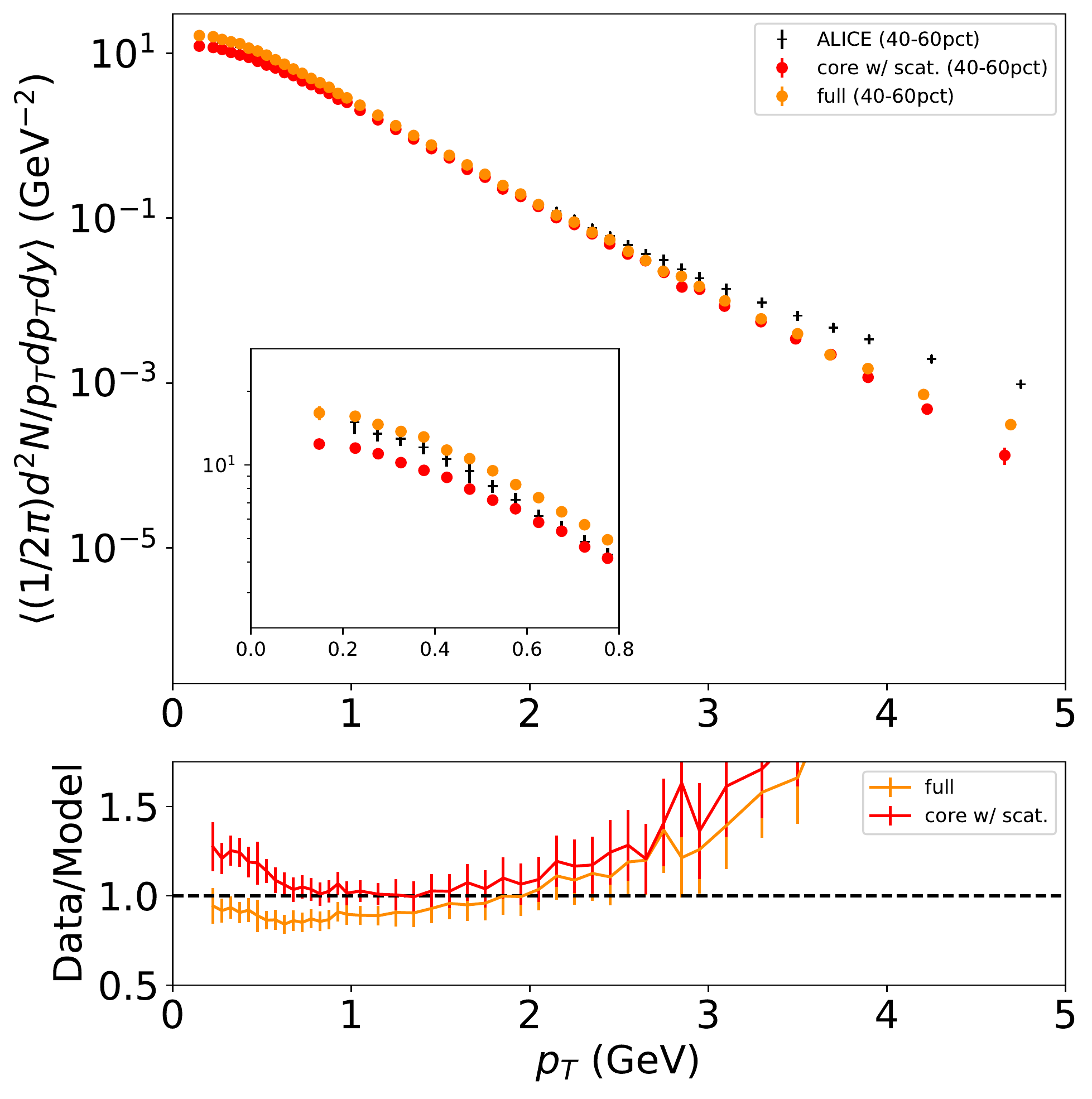}
    \includegraphics[bb=0 0 564 567, width=0.45\textwidth]{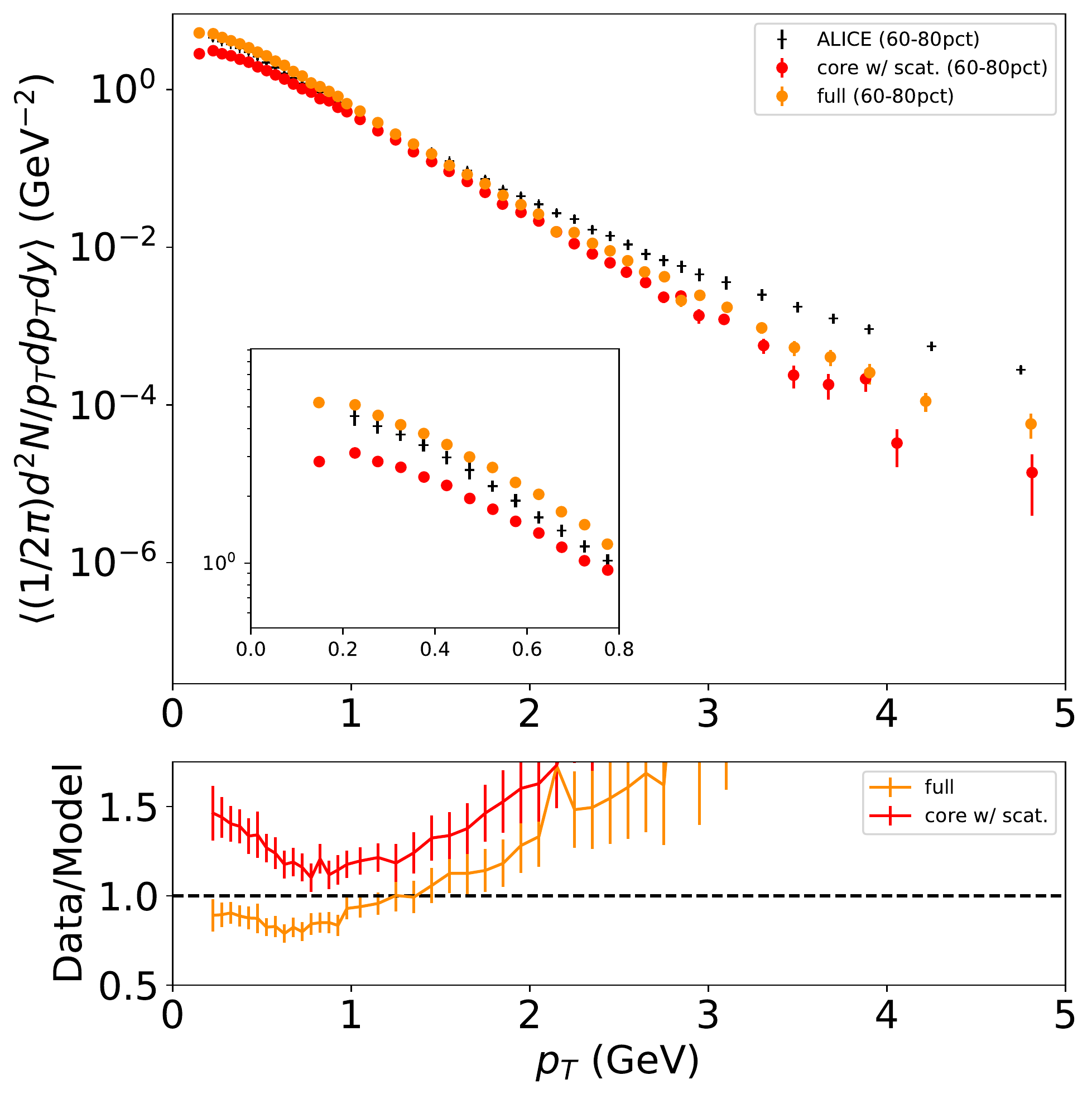}
    \caption{(Upper) Centrality dependence of transverse momentum spectra of charged kaons ($K^+ + K^-$) from $Pb$+$Pb$ collisions at \snn = 2.76 TeV from DCCI2 with full simulations (orange) and core components with hadronic rescatterings (red). Comparisons to ALICE experimental data (black crosses) are also shown. Insets are blow-ups of very low $p_T$ (0.0-0.8 GeV) regime.
    (Lower) Ratio of the ALICE experimental data to DCCI2 results with full simulation (orange) and core components with hadronic rescatterings (red) at each $p_T$ bin.}
    \label{fig:PBPB2760_PTSPECTRA_K_LOWPT}
\end{figure}

\begin{figure}
    \centering
    \includegraphics[bb=0 0 564 567, width=0.45\textwidth]{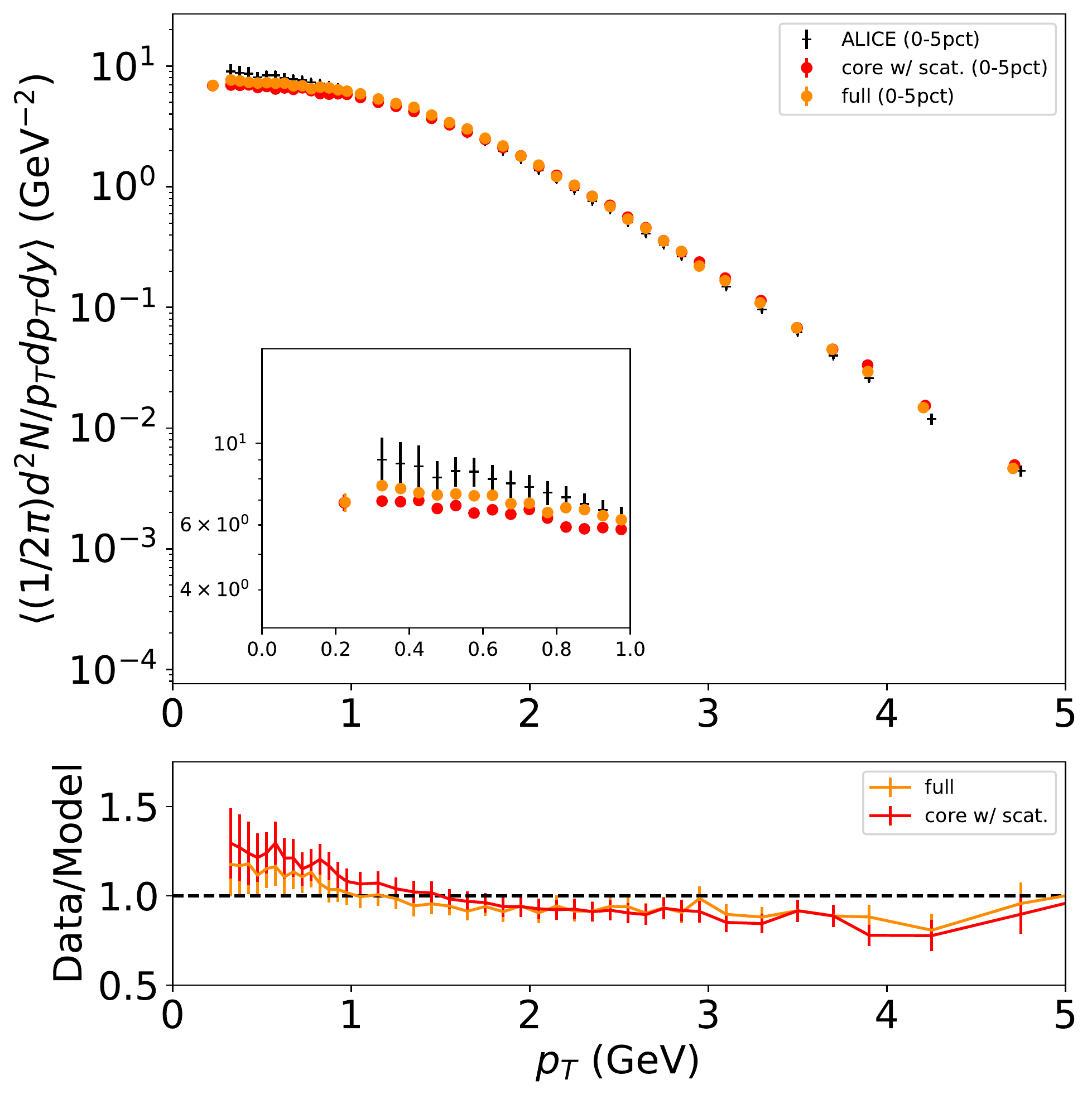}
    \includegraphics[bb=0 0 564 567, width=0.45\textwidth]{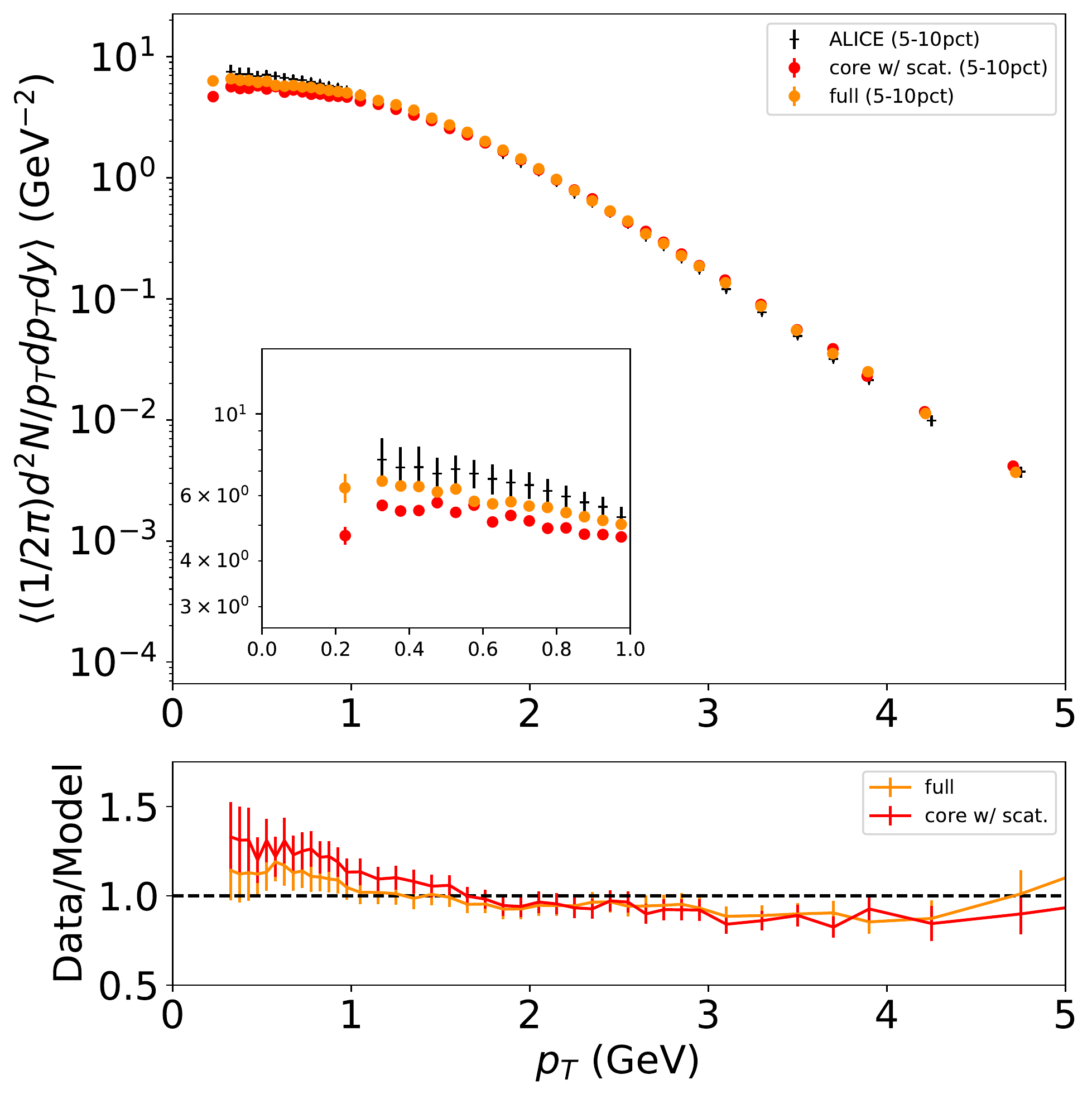}
    \includegraphics[bb=0 0 564 567, width=0.45\textwidth]{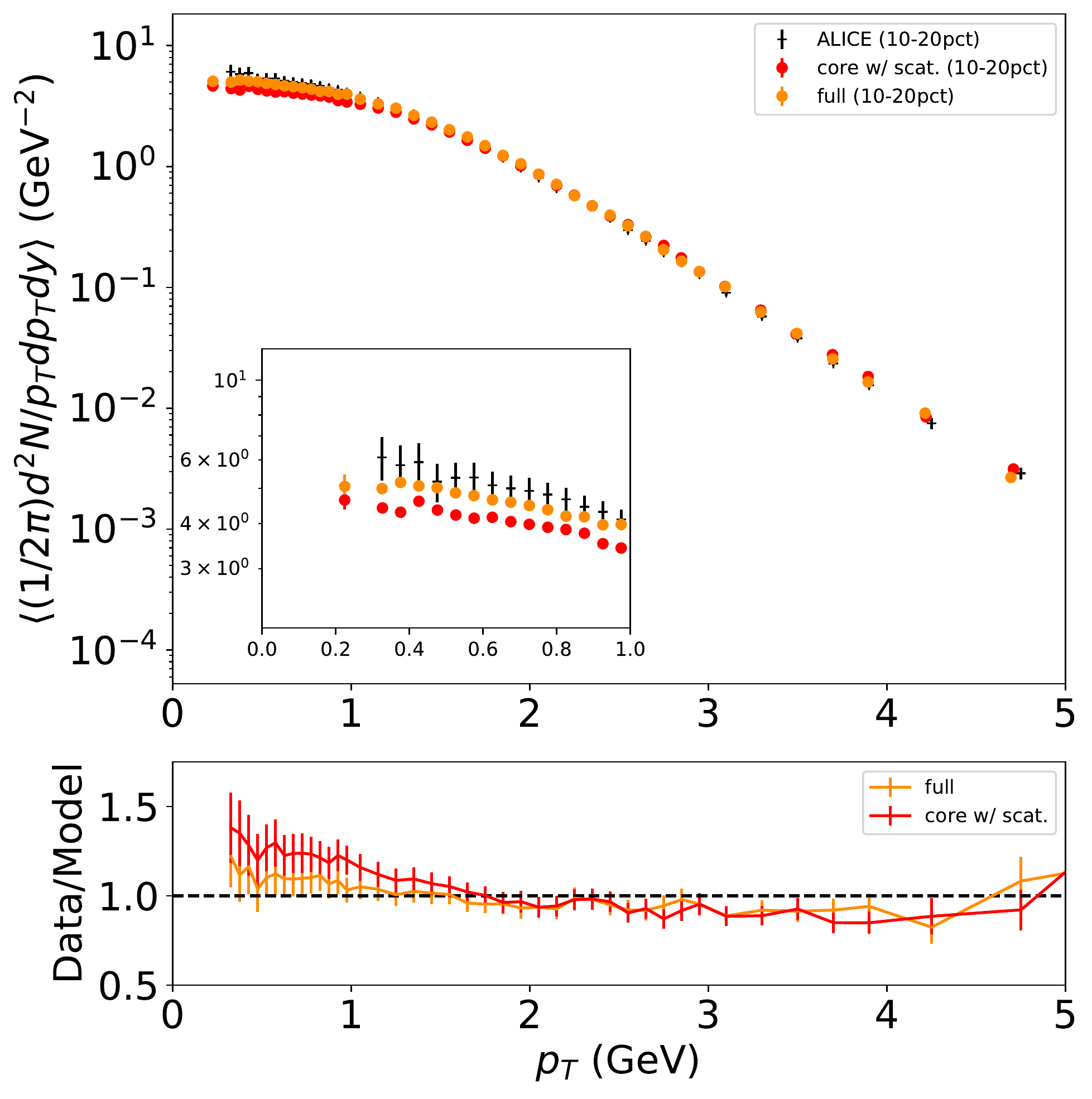}
    \includegraphics[bb=0 0 564 567, width=0.45\textwidth]{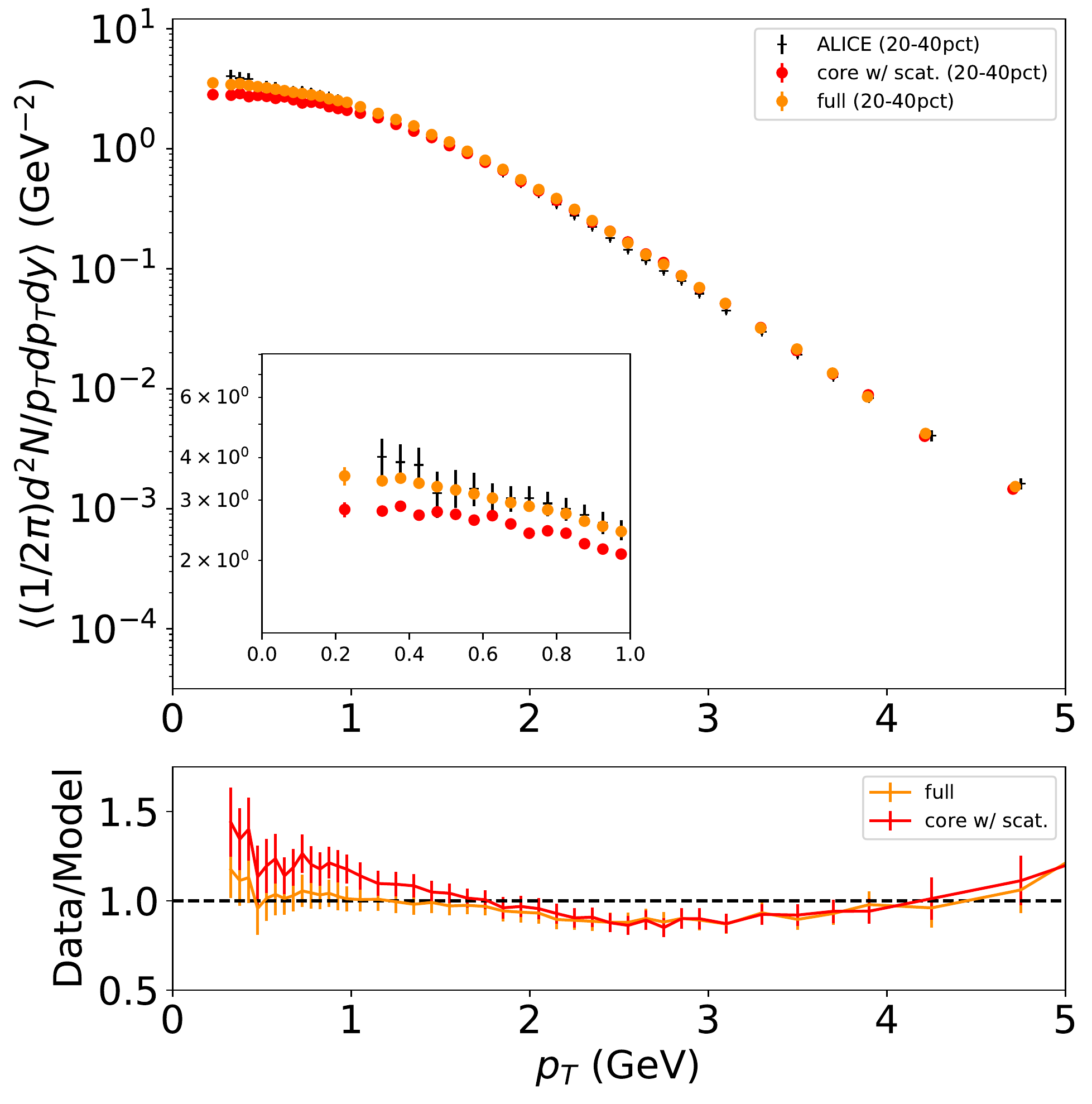}
    \includegraphics[bb=0 0 564 567, width=0.45\textwidth]{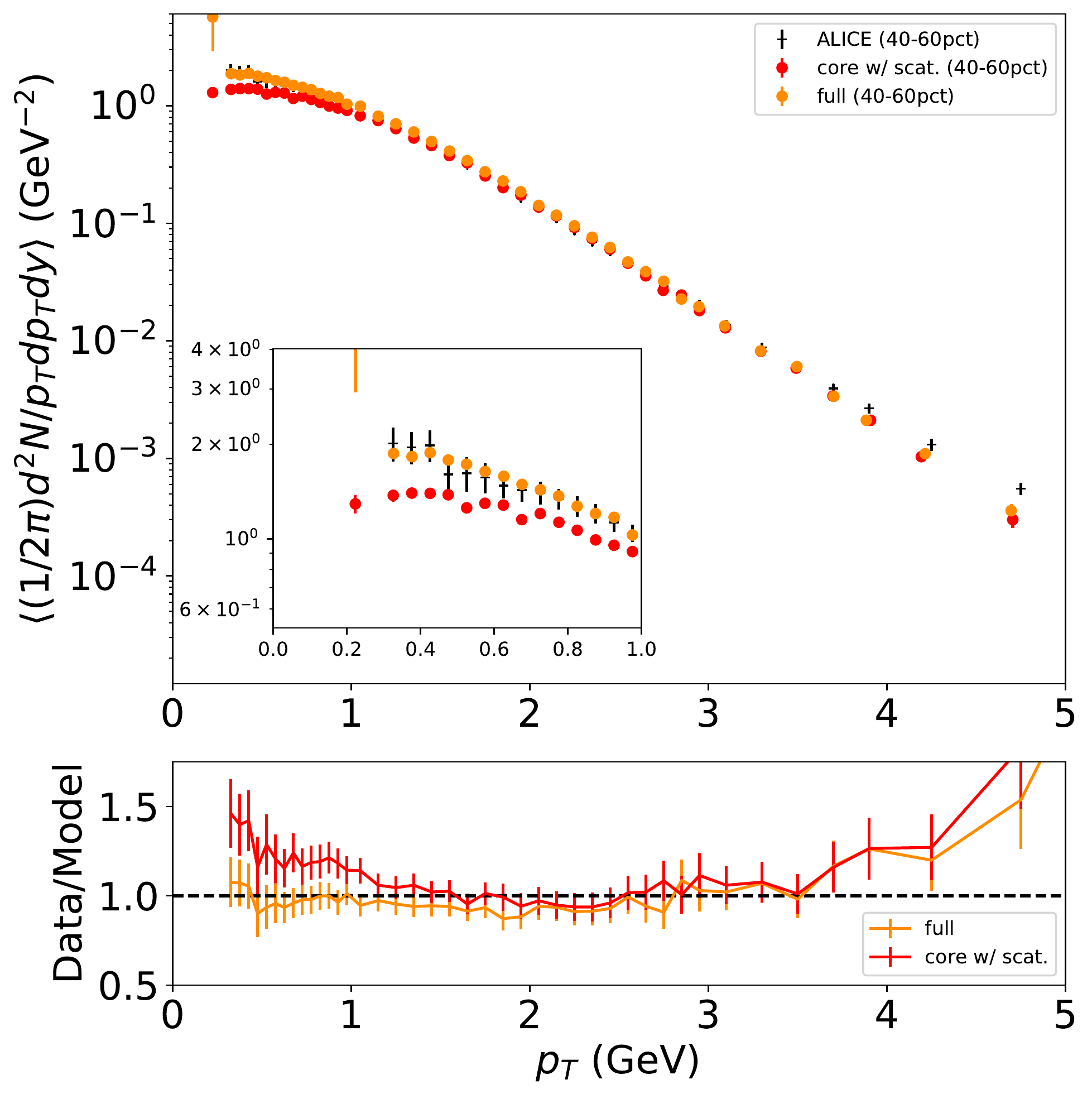}
    \includegraphics[bb=0 0 564 567, width=0.45\textwidth]{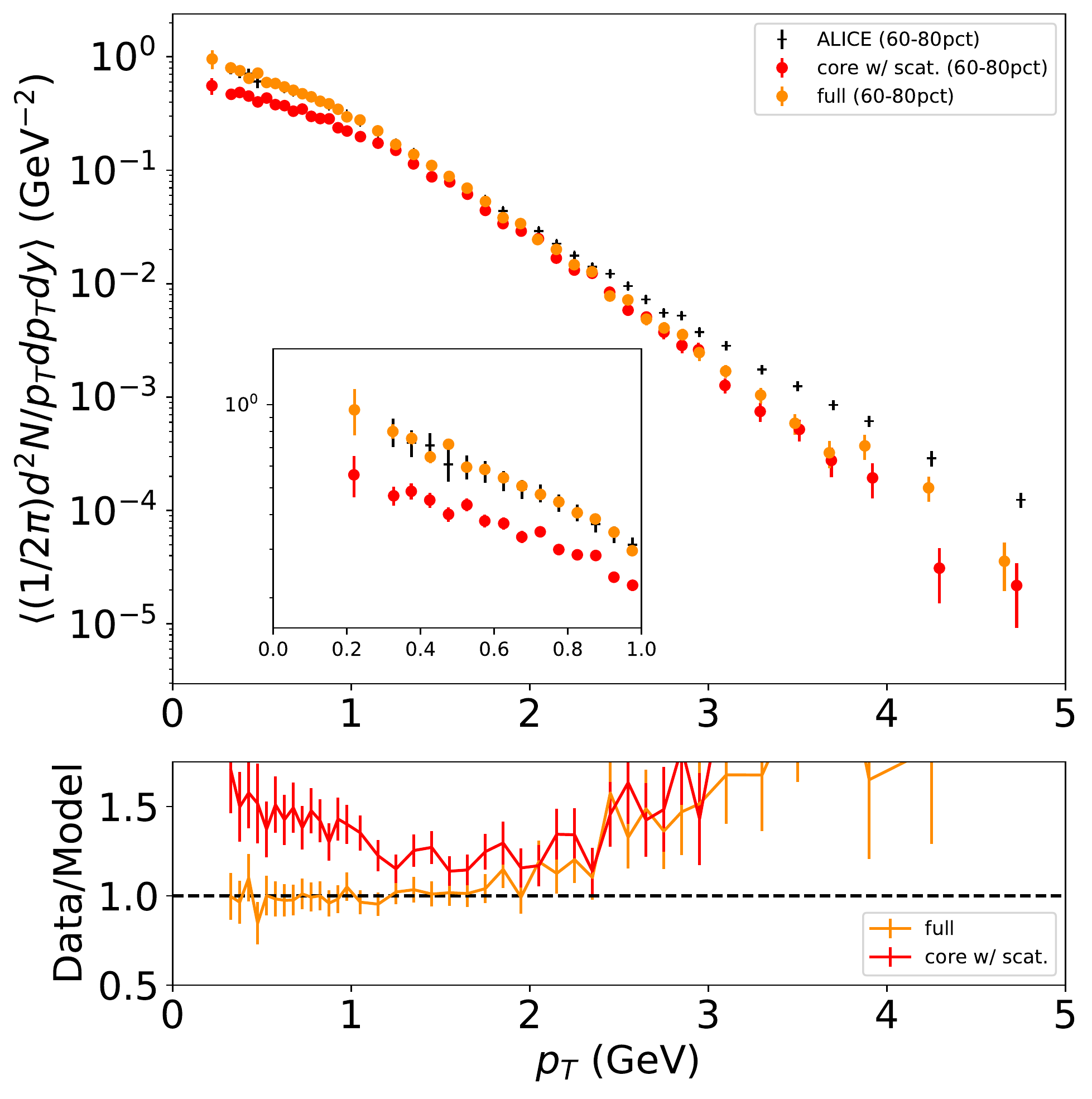}
    \caption{(Upper) Centrality dependence of transverse momentum spectra of protons and antiprotons ($p + \bar{p}$) from $Pb$+$Pb$ collisions at \snn = 2.76 TeV from DCCI2 with full simulations (orange) and core components with hadronic rescatterings (red). Comparisons to ALICE experimental data (black crosses) are also shown. Insets are blow-ups of very low $p_T$ (0.0-1.0 GeV) regime.
    (Lower) Ratio of the ALICE experimental data to DCCI2 results with full simulation (orange) and core components with hadronic rescatterings (red) at each $p_T$ bin.}
    \label{fig:PBPB2760_PTSPECTRA_P_LOWPT}
\end{figure}

To investigate effects of those correction from non-thermalized matter in $Pb$+$Pb$ collisions, 
I make comparisons of $p_T$ spectra from DCCI2 with full simulations and core components with hadronic rescatterings in upper panels of fig.~\ref{fig:PBPB2760_PTSPECTRA_PI_LOWPT}, \ref{fig:PBPB2760_PTSPECTRA_K_LOWPT}, and \ref{fig:PBPB2760_PTSPECTRA_P_LOWPT}.  
Also, the inset is shown to blow up very low $p_T$ regime.
In corresponding lower panels, ratio of the ALICE experimental data to DCCI2 results with full simulation and core components with hadronic rescatterings is shown at each $p_T$ bin.
The reason to perform hadronic rescatterings on core components to make comparisons is as follows.
In this comparison, corrections due to existence of corona components against core components where core components are highly dominant in $Pb$+$Pb$ collisions
and how they work in describing experimental data are exactly the aim of the present study.
As I mentioned in Sec.~\ref{subsec:IdentifiedParticlePTSpectra},
however, one cannot directly compare between results from full simulations/experimental data and results with switching off hadronic rescatterings because hadronic rescatterings do alter $p_T$ spectra.
In order to make reasonable comparisons as much as possible to see the corona corrections,
I perform hadronic rescatterings on particles only from core components.

The results show that at very low $p_T$ regime, roughly smaller than $p_T\approx0.5$ GeV,
the full simulation results show better agreement with experimental data compared to results of core components with hadronic rescatterings in every particle species and centrality classes.
In other words, the slopes at very low $p_T$ observed in experiment cannot be reproduced only with core components, {\it{i.e.,}} hydrodynamics with hadronic afterburner (a.k.a. hybrid model),
and the full simulation results show better description of the slopes by including corona component.
Therefore I conclude that the ``soft-from-corona" components are the key to precisely reproduce the multiplicity dependence of the mean transverse momentum.

It is discussed that the centrality dependence of {$\langle p_T\rangle$} is well described by hydrodynamic simulations introducing the finite bulk viscosity \cite{Ryu:2015vwa}.
While a recent Bayesian analysis supports the zero-consistent bulk viscosity by analyzing $p_T$-differential observables \cite{Nijs:2020ors}.
Both of them, however, still failed to reproduce pion $p_T$ spectra below $\approx 0.3$ GeV (see, \textit{e.g.}, Fig.~3 in Ref.~\cite{Ryu:2015vwa} and Fig.~21 in Ref.~\cite{Nijs:2020roc}. Another related discussion is in Ref.~\cite{Guillen:2020nul}), which would result in overestimation of {$\langle p_T\rangle$}.
The discrepancy between hydrodynamic results and experimental data in this low $p_T$ region becomes larger as going to peripheral collisions or small colliding systems \cite{Nijs:2020roc}. Therefore I suggest that the deviation between the model and the data in the low $p_T$ region could be filled with the corona components in those model calculations too.

\subsection{Multiplicity dependence of mean transverse mass}
\label{sec:MTSCALING}
In this section,
I analyze the mean transverse mass for various hadrons in high- and low-multiplicity $p$+$p$ and $Pb$+$Pb$ events and see its mass dependence.
It has been empirically known that $m_T$ spectra in small colliding systems exhibit the $m_T$ scaling, \textit{i.e.}, the slope of $m_T$ spectra being independent of the rest mass of hadrons \cite{Guettler:1976ce,Guettler:1976fc}. 
Here, $m_T=\sqrt{m^2 + p_T^2}$ is the transverse mass and $m$ is the rest mass of the hadron. 
In contrast, in heavy-ion collisions, the slope parameter increases with $m$
and, as a result, the $m_T$ scaling is violated, which is regarded as a sign of the existence of radial flow generated \cite{Bearden:1996dd,Xu:2001zj}.
Thus whether \textit{radial expansion exists in small colliding systems} due possibly to the QGP formation can be explored through the empirical scaling behavior and its violation in the mean transverse mass. 

I take two event classes, high-multiplicity ($0$-$10\%$) and low-multiplicity ($50$-$100\%$) events, in $p$+$p$ and in $Pb$+$Pb$ collisions \footnote{Due to the lack of statistics, I simply divide events into these classes regardless of collision system.}.
The multiplicity or centrality classification is performed in the same way as the one used in Fig.~\ref{fig:MULTIPLICITY_PP_PBPB}.
In the following, I analyze the mean transverse mass
of charged pions ($\pi^+$ and $\pi^-$), charged kaons ($K^+$ and $K^-$), protons ($p$ and $\bar{p}$), phi mesons ($\phi$), lambdas ($\Lambda$ and $\bar{\Lambda}$), cascade baryons ($\Xi^-$ and $\bar{\Xi}^+$), and omega baryons ($\Omega^-$ and $\bar{\Omega}^+$), in $|\eta|<0.5$ without $p_T$ cut.

\begin{figure}
\begin{center}
\includegraphics[bb=0 0 468 507, width=0.45\textwidth]{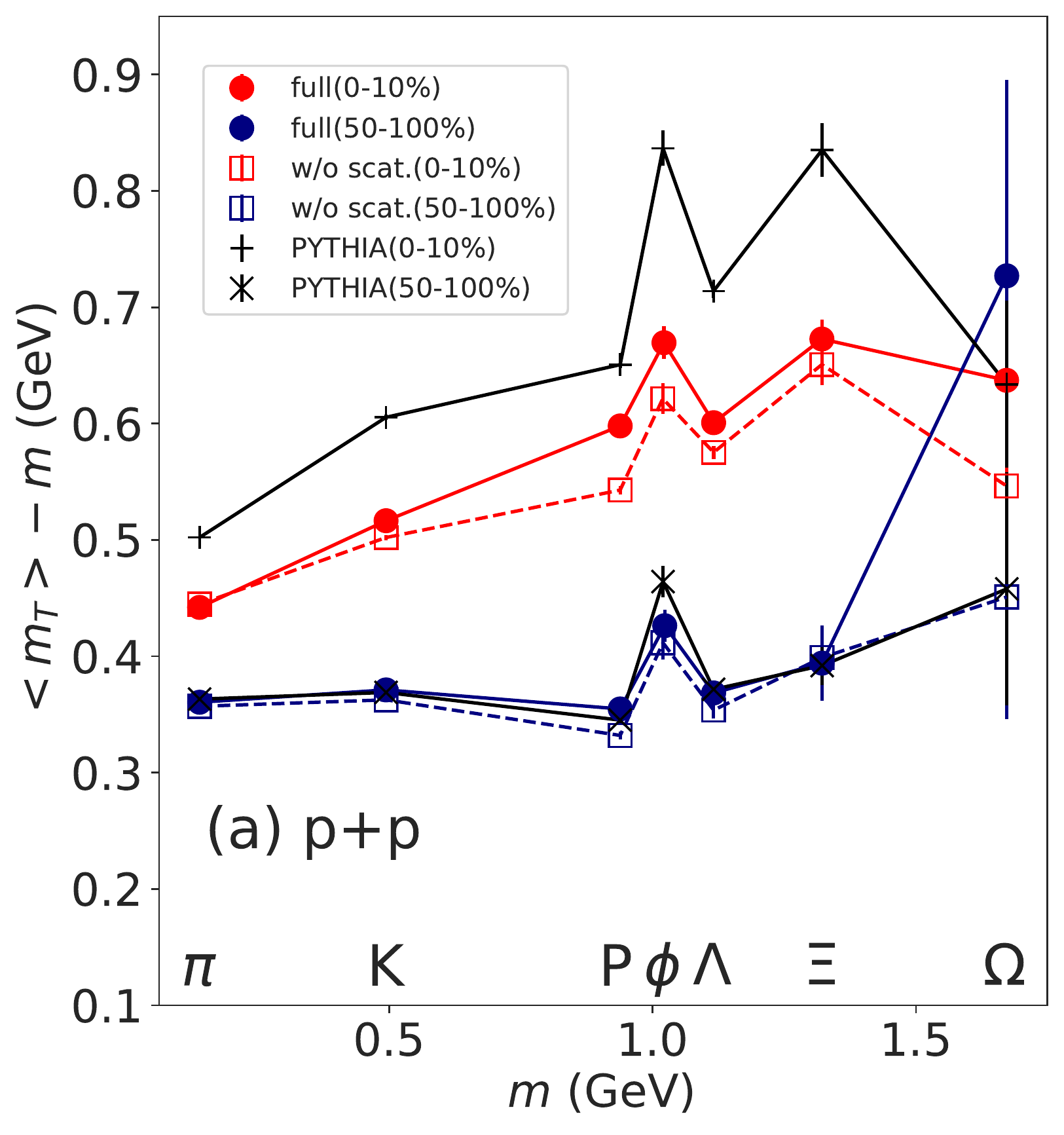}
\hspace{0.5cm}
\includegraphics[bb=0 0 468 507, width=0.45\textwidth]{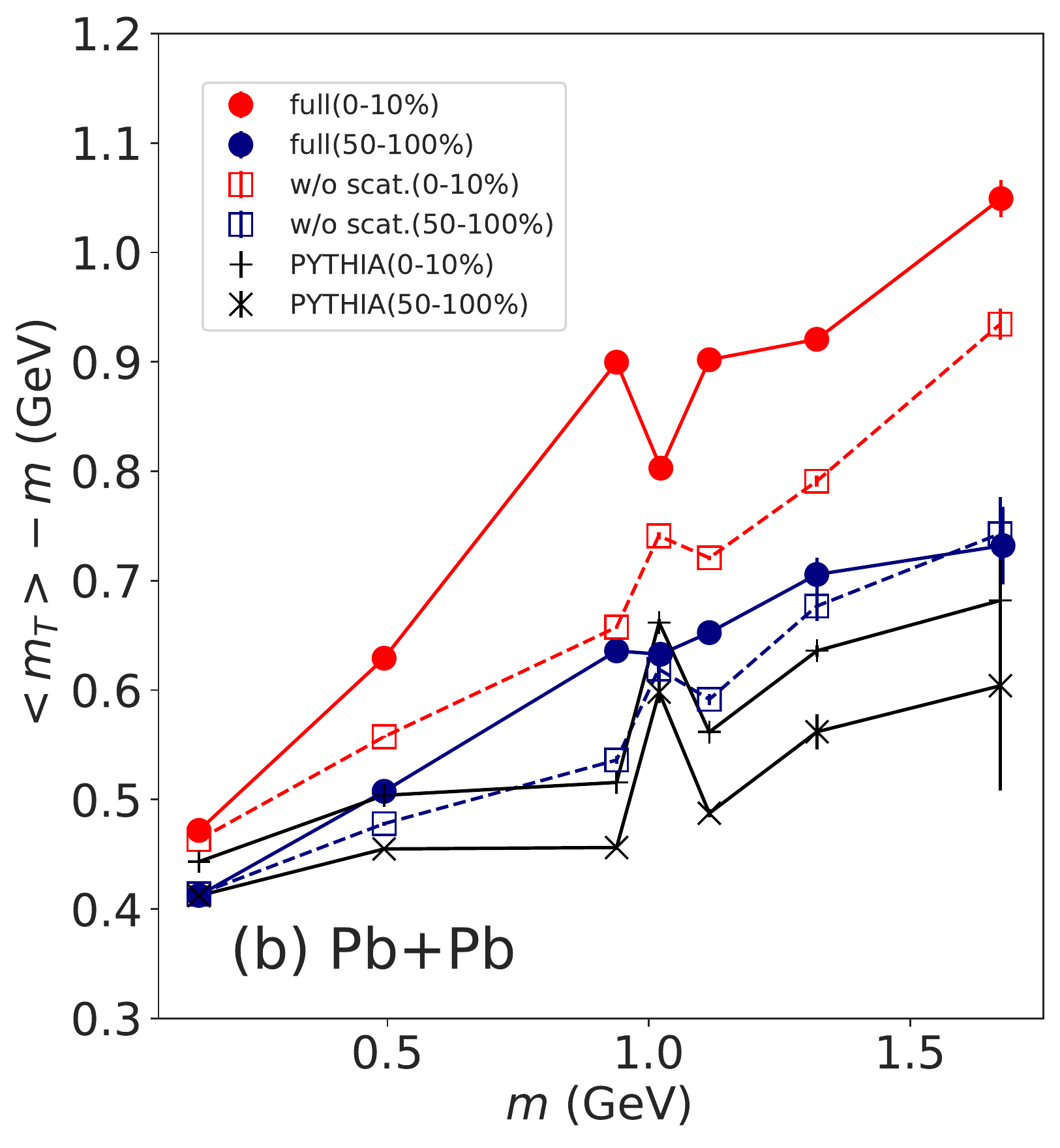}
\caption{(Color Online) Mean transverse mass, $\langle m_T \rangle -m$, as a function of rest mass of hadrons, $m$, from DCCI2 in (a) $p$+$p$ collisions at $\sqrt{s} = 7 \ \mathrm{TeV}$ and (b) $Pb$+$Pb$ collisions at \snn = $2.76 \ \mathrm{TeV}$. A comparison of the results from full simulations (closed symbols connected with solid lines) and the ones from simulations without hadronic rescatterings (open symbols connected with dashed lines) is made. Results of high-multiplicity events ($0$-$10\%$, red) and of low-multiplicity events  ($50$-$100\%$, blue)  are shown to see the effects of the fraction of core components. 
The result from \pythia8 and \pythia8 Angantyr (black symbols) with default parameters including color reconnection is plotted in $p$+$p$ and $Pb$+$Pb$ collisions, respectively, as references.
}
\label{fig:MTSCALING_PP_PBPB1}
\end{center}
\end{figure}

\begin{figure}
\begin{center}
\includegraphics[bb=0 0 468 507, width=0.45\textwidth]{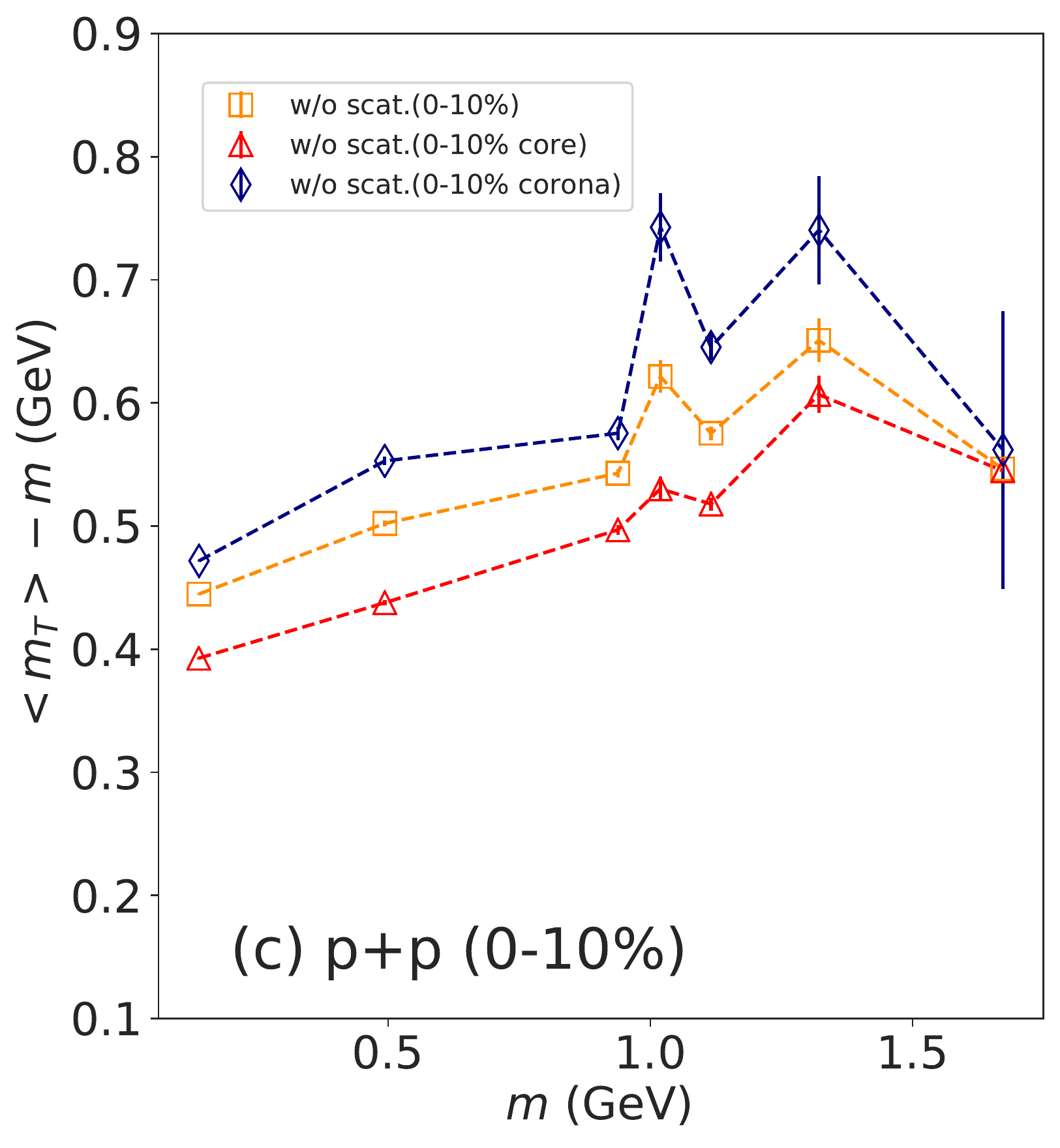}
\hspace{0.5cm}
\includegraphics[bb=0 0 468 507, width=0.45\textwidth]{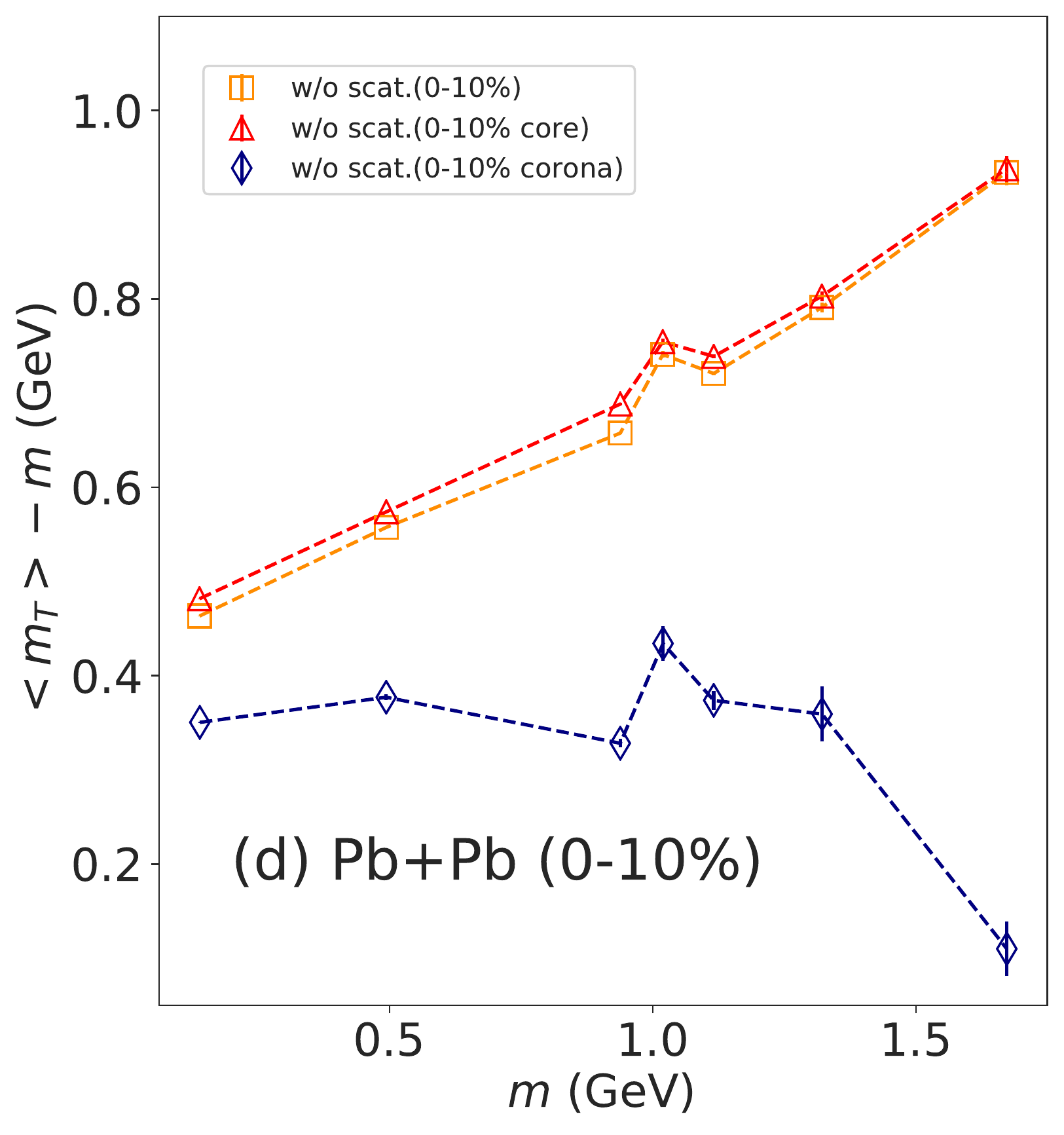}
\includegraphics[bb=0 0 468 507, width=0.45\textwidth]{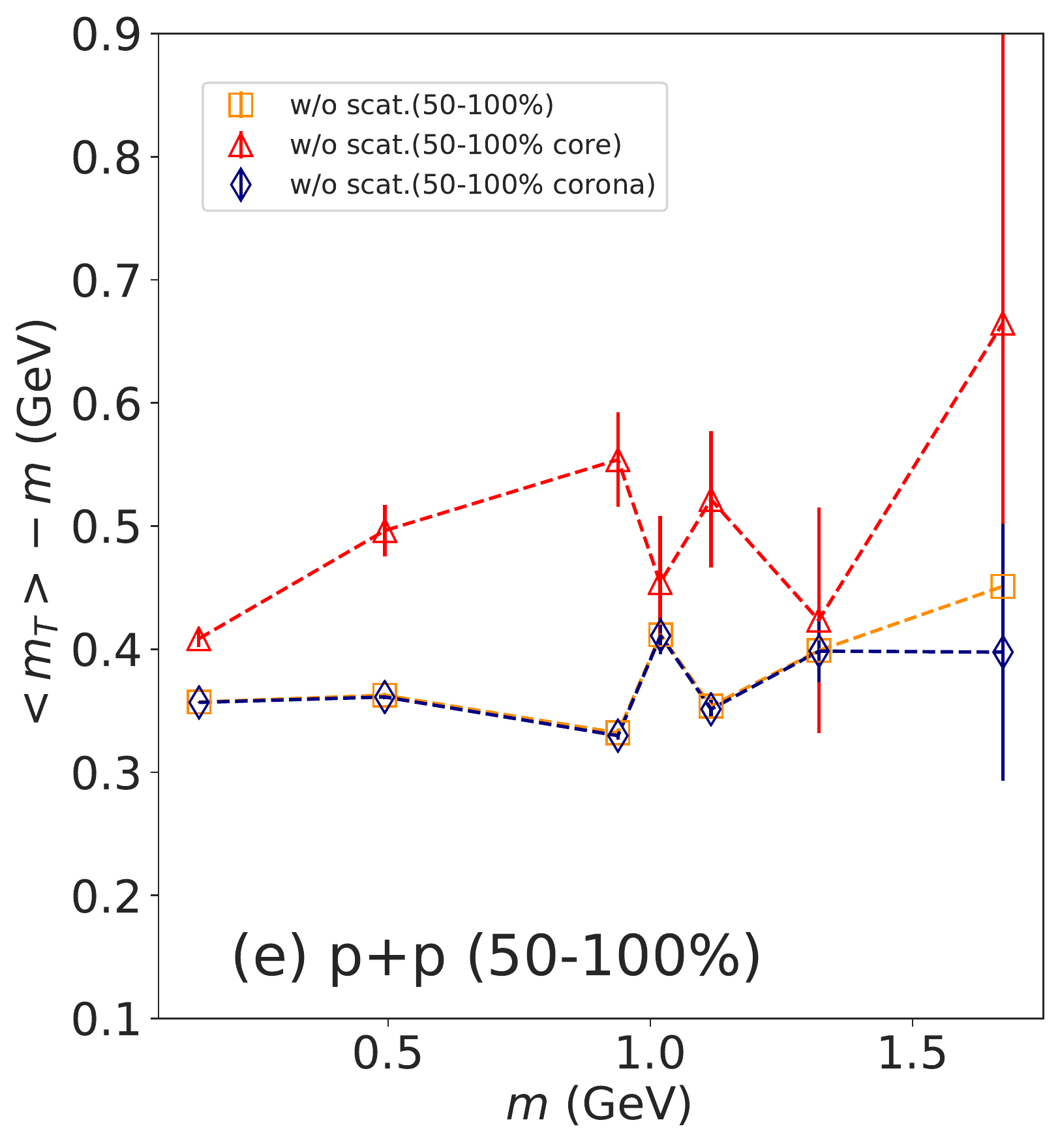}
\hspace{0.5cm}
\includegraphics[bb=0 0 468 507, width=0.45\textwidth]{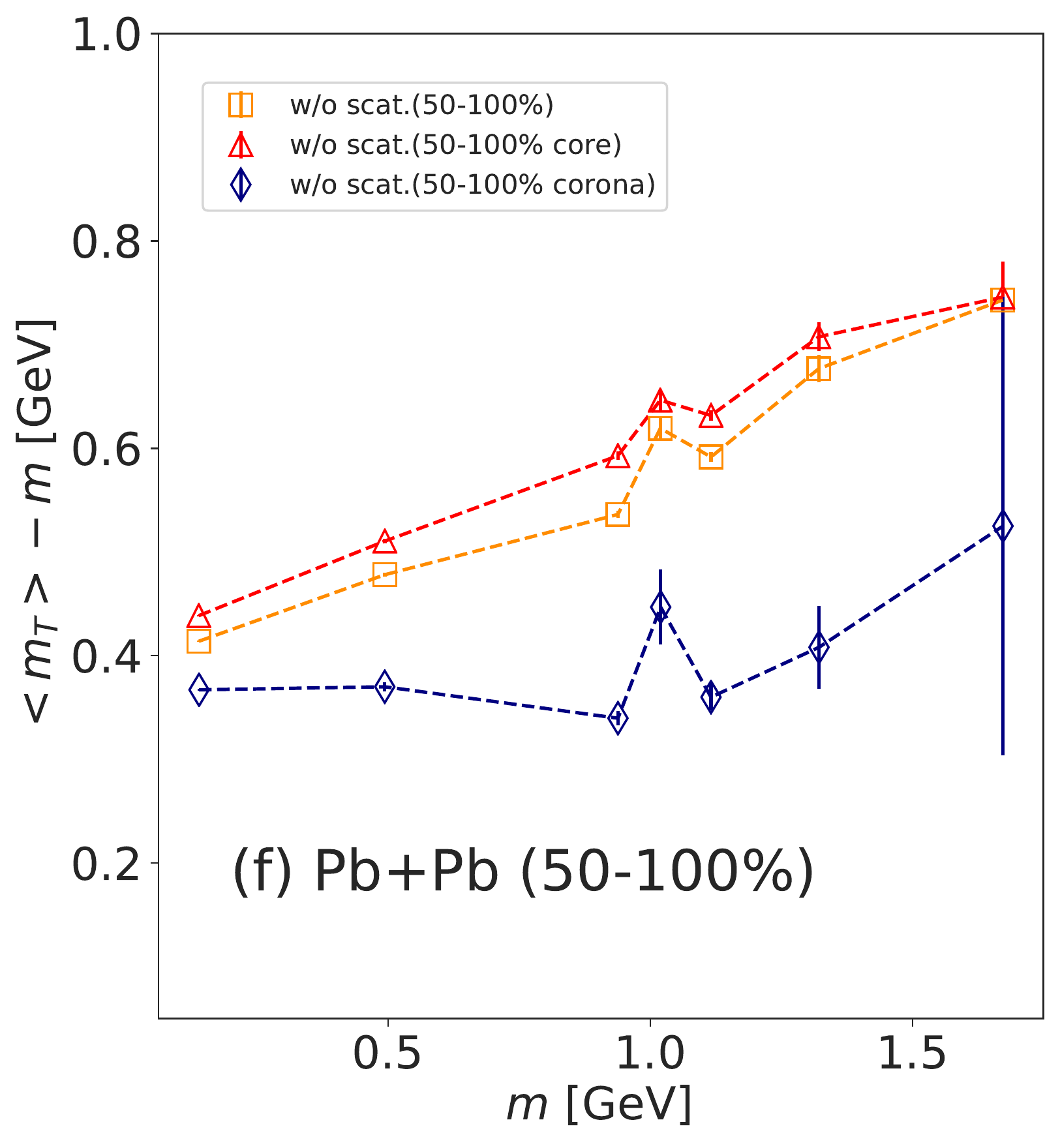}
\caption{ Mean transverse mass, $\langle m_T \rangle -m$, as a function of rest mass of hadrons, $m$, from DCCI2. 
Corresponding contributions of core and corona components in $0$-$10\%$ multiplicity and centrality classes in (c) $p$+$p$ and in (d) $Pb$+$Pb$ collisions, respectively.
Corresponding contribution of core and corona components in $50$-$100\%$ multiplicity and centrality classes in (e) $p$+$p$ and in (f) $Pb$+$Pb$ collisions, respectively. }
\label{fig:MTSCALING_PP_PBPB2}
\end{center}
\end{figure}

Figure~\ref{fig:MTSCALING_PP_PBPB1} (a) shows the mean transverse mass, $\langle m_T \rangle -m$, as a function of the rest mass of hadrons, $m$, in high-multiplicity ($0$-$10\%$) and low-multiplicity ($50$-$100\%$) $p$+$p$ collisions at $\sqrt{s}= 7$ TeV.
To pin down the effect of hadronic rescatterings, the results from full simulations and simulations without hadronic rescatterings are compared with each other.
The result from \pythia8 with the default settings including color reconnection is also plotted as a reference. 
Overall, $\langle m_T \rangle-m$ in high-multiplicity events ($0$-$10\%$) tends to exhibit an almost linear increase with increasing $m$ except for phi mesons.
On the other hand, such a clear mass dependence is not seen in low-multiplicity events ($50$-$100\%$), which is consistent with the $m_T$ scaling.
It should also be mentioned that the violation of linear increase of phi mesons in core components appears after resonance decays against direct hadrons (not shown). 
An apparent flow-like linear mass dependence is seen in results from \pythia8 in high-multiplicity events as well, which is due to the color reconnection \cite{Ortiz:2013yxa}.
The almost linear increasing behavior in DCCI2 is caused by both radial flow for core components from hydrodynamic expansion and color reconnection for corona components obtained with \pythia8.
As a result, the results from both DCCI2 and \pythia8 have similar tendencies.
Therefore, it is difficult to discriminate each effect by merely seeing the mean transverse mass.
The effect of hadronic rescatterings is almost absent for pions. This comes from an interplay between small $pdV$ work in the late hadronic rescattering stage and approximate conservation of pion number~\cite{Hirano:2005wx}. 
The small effects of hadronic rescatterings are seen for phi mesons and omega baryons because they do not form resonances in scattering with pions unlike other hadrons~\cite{Hirano:2007ei,Takeuchi:2015ana}.

Figure~\ref{fig:MTSCALING_PP_PBPB1} (b) shows the mean transverse mass as a function of hadron rest mass for $0$-$10\%$ and $50$-$100\%$ centrality classes in $Pb$+$Pb$ collisions at \snn = $2.76$ TeV. 
The almost linear increasing behavior of $\langle m_T \rangle-m$ appears even in $50$-$100\%$ centrality class as one can expect from the centrality dependence of the fraction of core components in Fig.~\ref{fig:MULTIPLICITY_PP_PBPB} (b).
The larger enhancement of the mean transverse mass due to hadronic rescatterings, in particular, for protons is seen in high-multiplicity events in comparison with the low-multiplicity events.
This is a manifestation of the famous ``pion wind" in the late rescattering stage \cite{Hung:1997du,Bleicher:1999pu,Bratkovskaya:2000qy,Bass:2000ib}.

Figure~\ref{fig:MTSCALING_PP_PBPB2} (c) shows each contribution of core and corona components to the final result without hadronic rescatterings in 0-10\% multiplicity class in $p$+$p$ collisions. The inclusive result here is identical to the one shown as the result without hadronic rescatterings in Fig.~\ref{fig:MTSCALING_PP_PBPB1} (a).
The difference between results of core and corona components is seen in protons, lambdas, and omega baryons.
The linear mass ordering of $\langle m_T \rangle-m$ from core components is slightly diluted 
for protons and lambdas in the inclusive result
due to the sizable contribution of corona components.
In contrast, the core result and the inclusive result are almost on top of each other since the contribution of corona components for omega baryons is smaller in 0-10\% multiplicity class compared to other particle species.

Figure~\ref{fig:MTSCALING_PP_PBPB2} (d) shows each contribution of core and corona components to the final result without hadronic rescatterings in 0-10\% centrality class in $Pb$+$Pb$ collisions. The linear increase except phi meson is seen very clearly for the core component, and the increase rate is more than the one from $p$+$p$ results shown in Fig.~\ref{fig:MTSCALING_PP_PBPB2} (c).

Figure \ref{fig:MTSCALING_PP_PBPB2} (e) shows the same variable with Fig.~\ref{fig:MTSCALING_PP_PBPB2} (c) but in 50-100\% multiplicity class in $p$+$p$ collisions.
Since the fraction of the core components is less than 10\% in this range of multiplicity class as shown in Fig.~\ref{fig:MULTIPLICITY_PP_PBPB} (a), the final result and the result from corona components are almost top of each other showing no significant dependence on hadron rest mass. 

Figure~\ref{fig:MTSCALING_PP_PBPB2} (f) shows the same variables with Fig.~\ref{fig:MTSCALING_PP_PBPB2} (d) but in 50-100\% centrality class in $Pb$+$Pb$ collisions. According to Fig.~\ref{fig:MULTIPLICITY_PP_PBPB} (b), the fraction of core components shows $R_{\mathrm{core}}\approx 0.8$ to $\approx 0.9$ in this centrality range. Eventually the result of core components is found to be slightly diluted by corona components.

\section{Anisotropic flow}
\label{sec:RESULTS_AnisotropicFlows}
Anisotropic flow is one of the observables that has been often investigated in high energy nuclear collisions to quantify {\it{collectivity}}.
As I mentioned in Sec.~\ref{subsec:Collectivity},
the word ``collectivity'' means that there is a correlation in azimuthal angle of momentum space for produced particles.
The physical origin of such a correlation has been, so far, understood as follows: in $Pb$+$Pb$ collisions, once there is a geometrical anisotropy in initial state of QGP fluids, the anisotropy in geometry is converted into that in momentum space due to hydrodynamic response. Final hadrons from the QGP fluids reflect the anisotropy even after hadronization, thus the momentum anisotropy in final hadrons is a signal of QGP formation.
On the other hand, in $p$+$p$ collisions, it has been understood that most of particles originate from fragmentation from hard/non-equilibrium particles.
Those hard particles called jets usually are produced in back-to-back due to momentum conservation
and have a strong correlation in momentum space.
Thus, it comes to ambiguous which physical origin momentum anisotropy implies in $p$+$p$ collisions.
In this sense, implication of anisotropy in $p$+$p$ collisions and $Pb$+$Pb$ collisions can be different. Therefore I explain results from $p$+$p$ and $Pb$+$Pb$ collisions separately in the following sequential subsections.

It would be worth mentioning the motivation to see anisotropic flows in DCCI2 here.
As I explained, there are mainly the two factors that pronounce
anisotropy in momentum space: hydrodynamic response of QGP fluids and jets from hard scatterings, and the anisotropy originates from these are often referred to as {\it{flow}} and {\it{non-flow}}, respectively
\footnote{
It should be noted that these are sorts of jargon in relativistic heavy-ion community, and there is not, at my best knowledge, solid definitions.
}
.
Generally speaking, it is assumed that flows are manifested in small to intermediate transverse momentum regime and have azimuthal correlations in long range towards (pseudo)rapidity directions.
On the other hand, non-flows are manifested in large transverse momentum regime and have azimuthal correlation in short range correlations.
If one wants to investigate the flow in experimental data, 
techniques called {\it{non-flow subtractions}} are adopted to, namely, subtract non-flow correlations,
and those techniques are built based on the aforementioned assumptions.
For instance, imposing a certain gap in long range direction between two-particle pair when one calculate two-particle correlation is expected to remove short range correlations of non-flows.
It is often simply expected that in core--corona picture,
the anisotropy from core and corona components corresponds to flow and non-flow.
In the following subsections,
I also investigate how each anisotropic flow changes when one imposes non-flow subtraction within DCCI2.

\subsection{Multiplicity dependence in $p$+$p$ collisions}
\label{subsec:AnisotropicFlows_MultiplicityDependenceInPPCollisions}

\begin{figure}
    \centering
    \includegraphics[bb=0 0 592 445, width=0.49\textwidth]{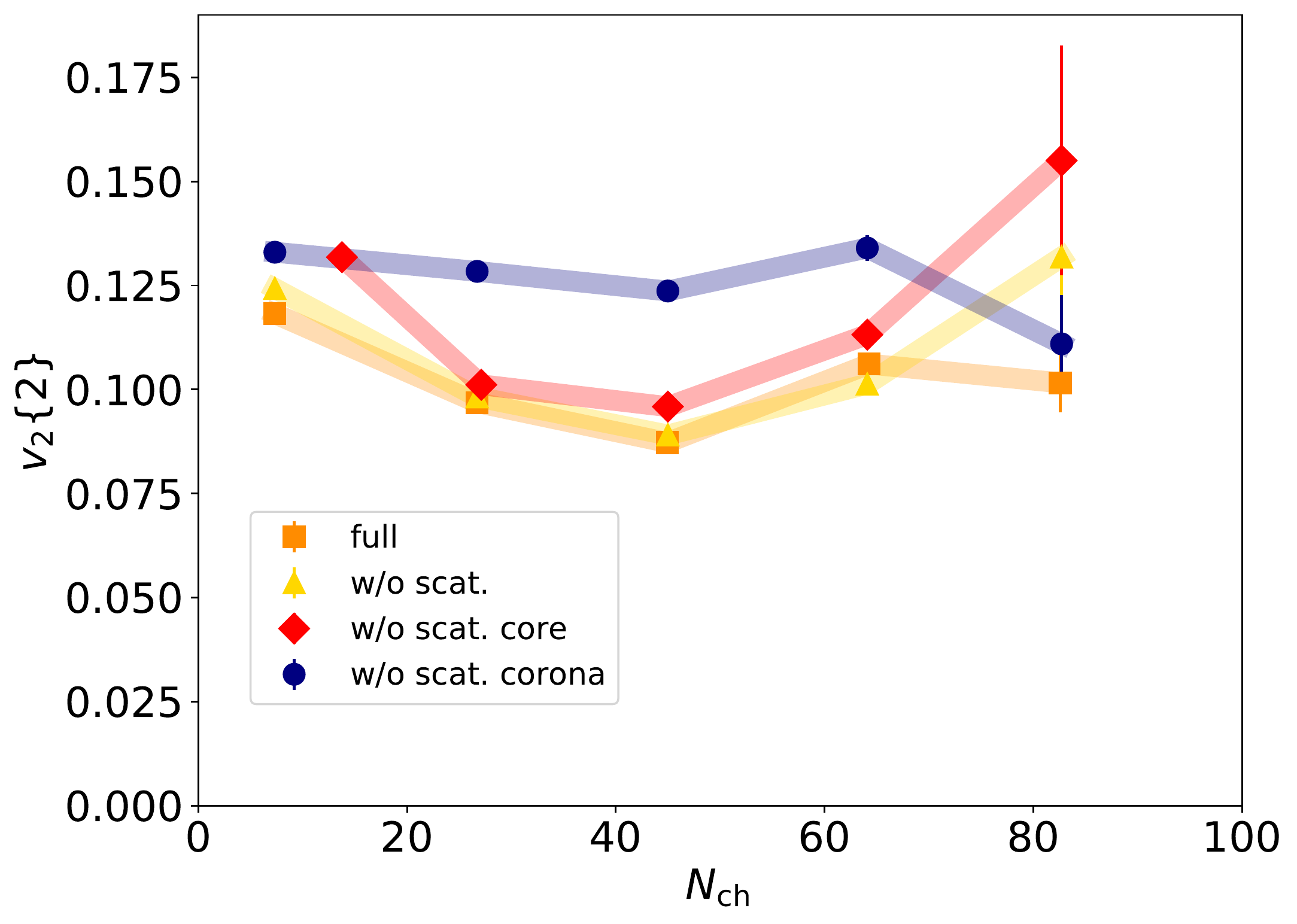}
    \includegraphics[bb=0 0 592 445, width=0.49\textwidth]{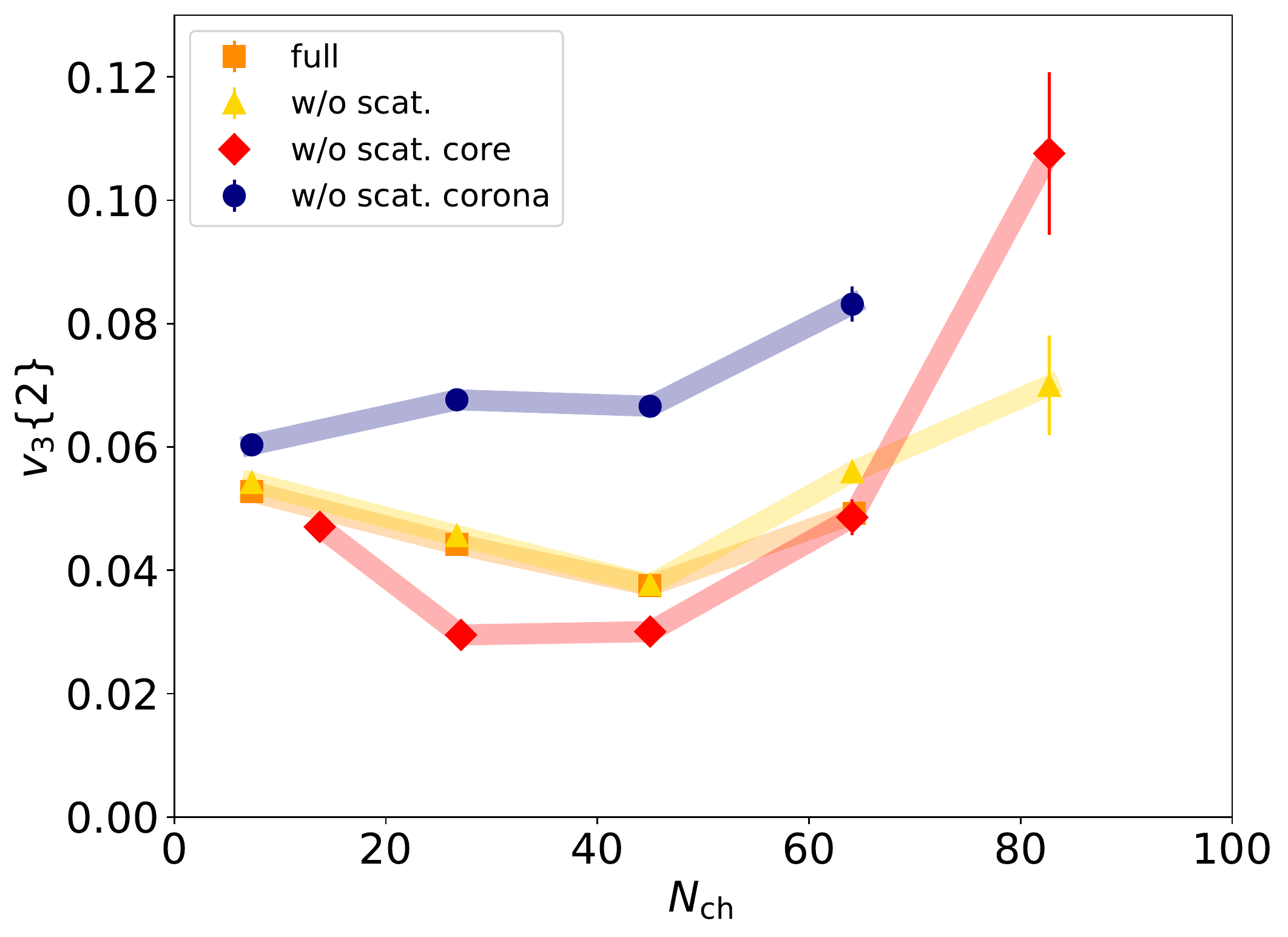}
    \caption{
    Second and third order of anisotropic flow coefficients obtained from two-particle correlation for charged hadrons without eta gap as a function of the number of produced charged particles in $p$+$p$ collisions at $\sqrt{s} = 13 \ \mathrm{TeV}$ are shown in left and right figures. Results from full simulations (orange squares) and simulations without hadronic rescatterings (yellow diamonds) are shown.  Results only from core (red diamonds) and corona (blue triangles) components as a breakdown of results of simulations without hadronic rescatterings are shown simultaneously.
    }
    \label{fig:PP13_V22Nch_V32Nch}
\end{figure}

Let me start from discussion on results from $p$+$p$ collisions.
Figure~\ref{fig:PP13_V22Nch_V32Nch} shows the second- and third- order anisotropic flow coefficient of charged particles obtained from two-particle cumulants, $\vtwtw$ and $\vthree$, as functions of $N_{\mathrm{ch}}$ in $p$+$p$ collisions at $\sqrt{s}=13$ TeV. 
Kinematic cuts for all of $\vtwtw$, $\vthree$, and $N_{\mathrm{ch}}$ are $0.2<p_T<3.0$ GeV and $|\eta|<0.8$ as used in Ref.~\cite{Acharya:2019vdf}.
One sees that $\vtwtw$ obtained from both core and corona components is larger than that from simulations without hadronic rescatterings for overall $N_{\mathrm{ch}}$.
This suggests that the event plane angle of core components might be different from that of corona components, which dilutes $\vtwtw$ of core and corona components with each other.
In contrast, $\vthree$ shows different tendency: the results from the summation of core and corona give the values of $\vthree$ lying in-between the results from core and corona components.
It might be possible to understand this tendency that the third-order of anisotropic flow is clearly manifested in corona components
\footnote{
One of the possible origins of the third-order anisotropic flow in corona component is a 
Bremsstrahlung from a back-to-back jets.
}
while those in core components vaguely appears and weaken those seen in the corona components.

\begin{figure}
    \centering
    \includegraphics[bb=0 0 592 445, width=0.49\textwidth]{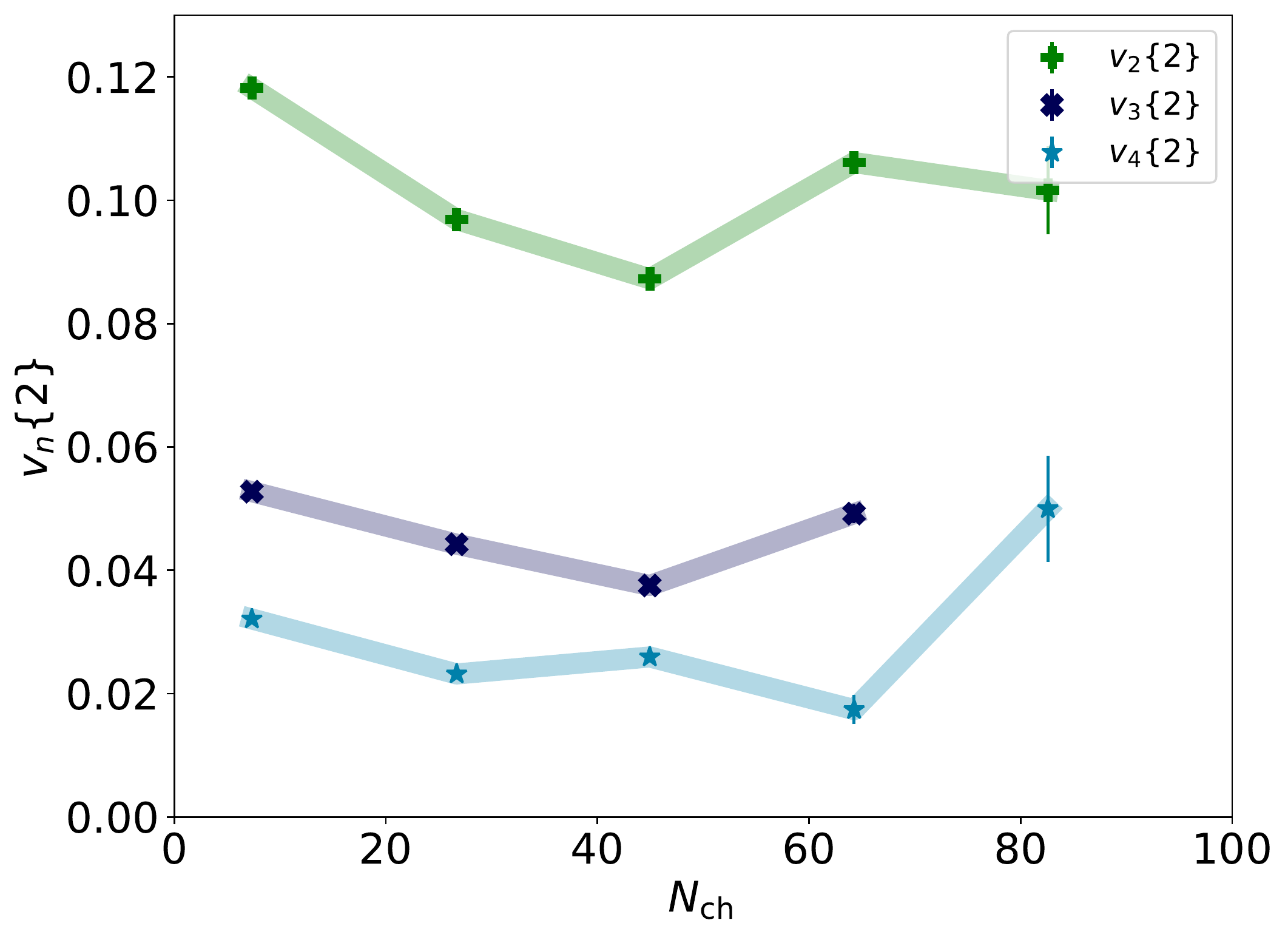}
    \caption{
    Comparisons among second- (green crosses), third- (navy x-s), and fourth- (blue stars) order of anisotropic flow coefficients obtained from two-particle correlation for charged hadrons as functions of the number of produced charged hadrons in $p$+$p$ collisions at \snn[proton] = 13 TeV from full simulations.
    }
    \label{fig:PP13_VN2Nch}
\end{figure}

Figure \ref{fig:PP13_VN2Nch} shows
comparisons among second-, third-, and fourth- order of anisotropic flow coefficients obtained from two-particle correlation as functions of the number of produced charged hadrons in $p$+$p$ collisions at \snn[proton] = 13 TeV from full simulations.
One sees that the absolute values of anisotropic flow coefficients decrease in ascending order of the order of coefficients.
It should be noted that, within DCCI2, initial momentum anisotropy is embedded in
initial matter profile due to dynamically deposited energy and momentum from initial partons. 
Hence, the origin of anisotropy pronounced even in core component is not limited to geometrical effects.
To investigate the origin of anisotropic flows,
it has been proposed that a sign change of correlation of anisotropic flow and mean $p_T$ in event-by-event is a proxy of the transition of origin of anisotropic flow in final hadrons between
initial momentum anisotropy or geometry of initial condition.
Further investigation can be performed through such observables in the future.

\begin{figure}
\centering
    \includegraphics[bb=0 0 592 445, width=0.49\textwidth]{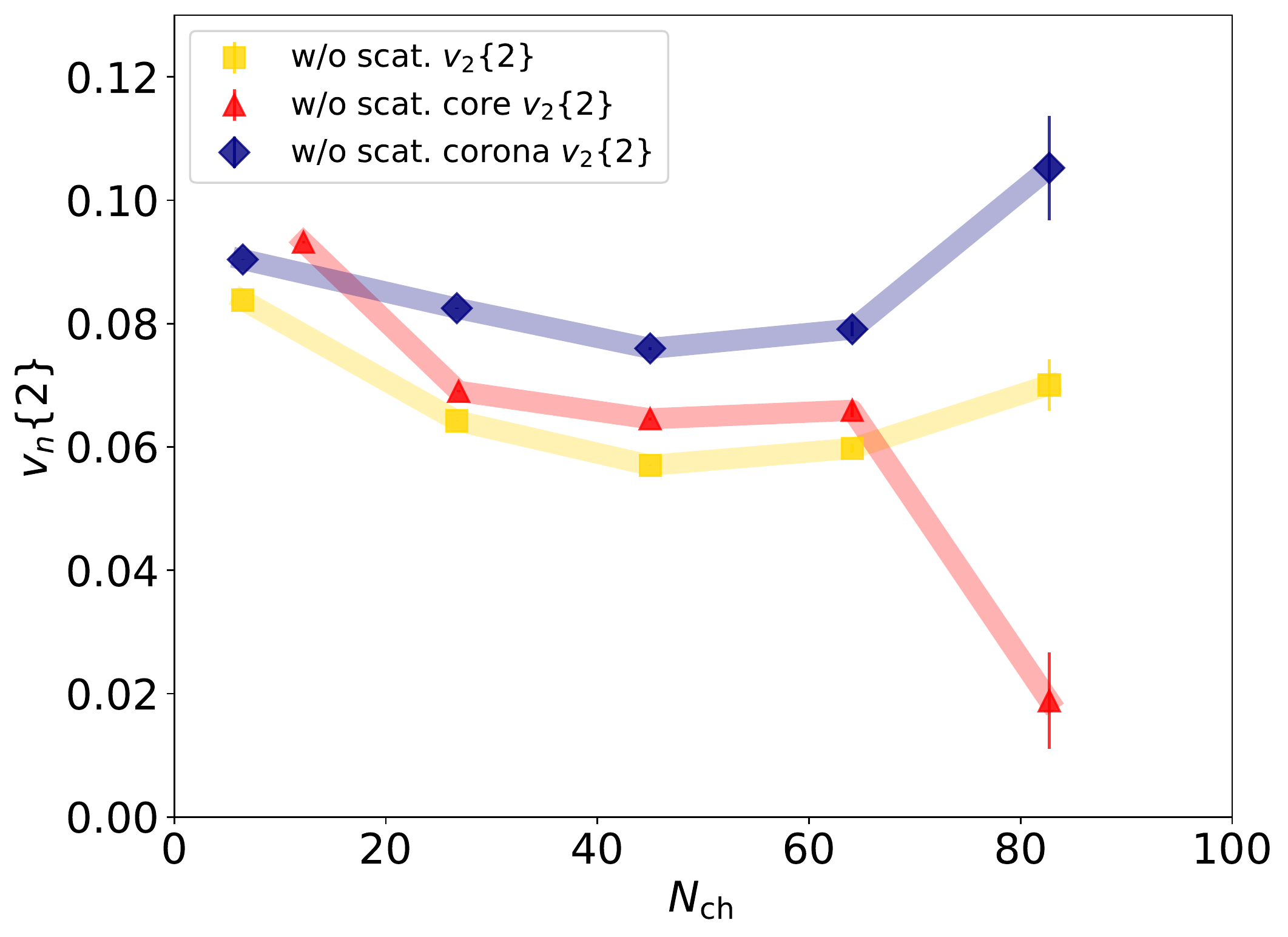}
    \includegraphics[bb=0 0 592 445, width=0.49\textwidth]{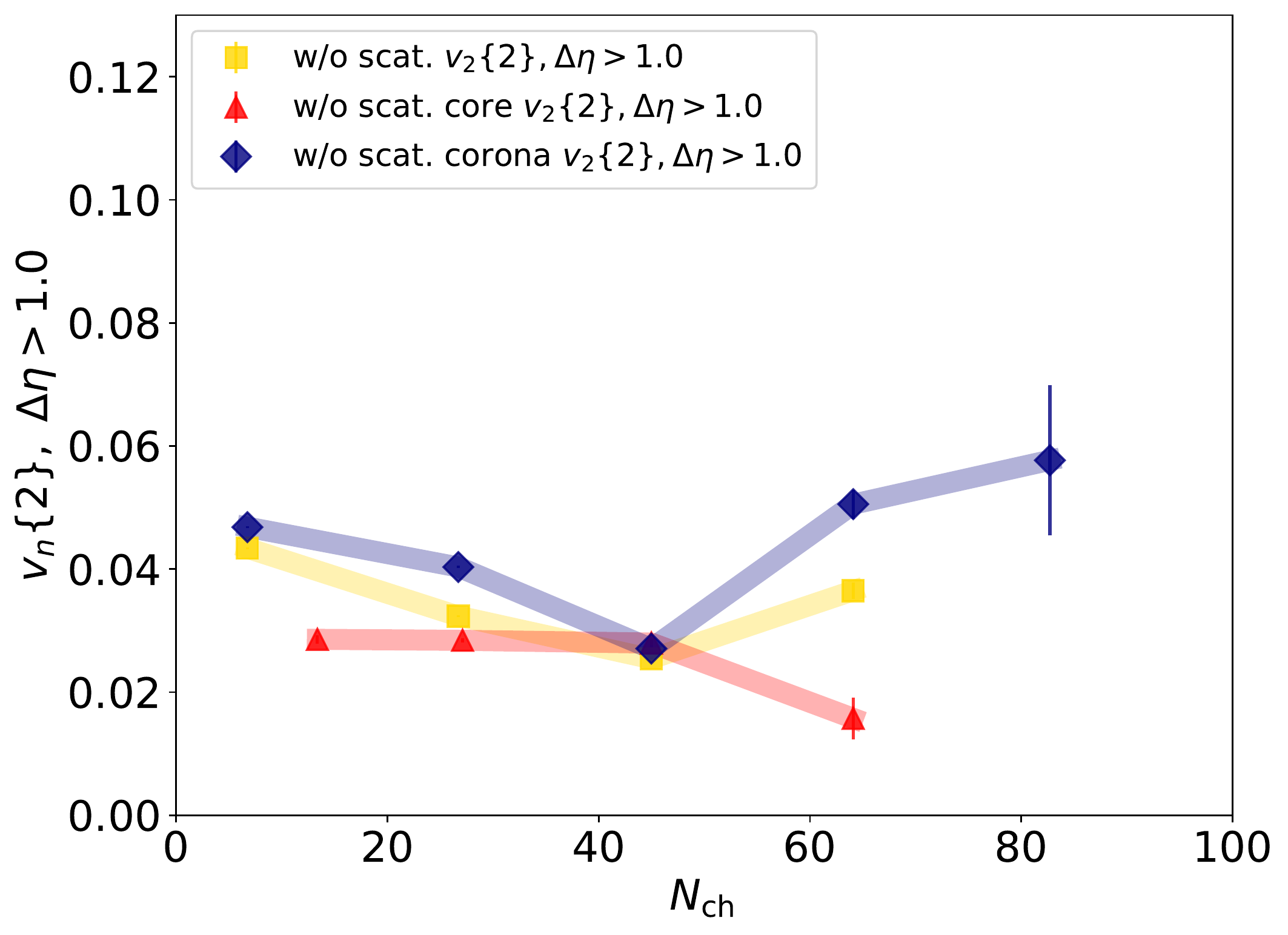}
    \caption{Comparison of with and without eta gap ($\Delta \eta > 1.0$) in second-order of anisotropic flow coefficient obtained from two-particle correlation for charged hadrons as a function of the number of produced charged particles in $p$+$p$ collisions at $\sqrt{s} = 13 \ \mathrm{TeV}$ are shown in left and right figures, respectively. Results from simulations without hadronic rescatterings (yellow diamonds) and its breakdown into core (red diamonds) and corona (blue triangles) components are shown.}
    \label{fig:PP13_V22Nch_etaGap}
\end{figure}

Figures.~\ref{fig:PP13_V22Nch_etaGap} (left) and (right) show comparisons of with and without eta gap in second-order of anisotropic flow coefficients obtained from two-particle correlations as a function of the number of produced charged particles in $p$+$p$ collisions at \snn[proton] = 13 TeV, respectively.
In order to see the effects of imposition of eta gap with abundant statistics, 
correlations are obtained with particles produced within $|\eta|<3.0$ instead of $|\eta|<0.8$.
When eta gap is imposed, each one particle used to calculate 2-particle correlation is taken from 2-sub region independently, and eta gap is imposed in-between the 2-sub regions. Detailed explanation are in Appendix.~\ref{sec:APPENDX_SubEvent},
As I aforementioned at the beginning of this section,
a certain eta gap is expected to remove short range correlation of non-flows
where non-flows are often expected to be the correlations in the corona.
In contrast to the expectation,
one sees that both $\vtwtw$ obtained from core and corona are suppressed
simultaneously due to imposition of eta gap of $\Delta \eta > 1.0$.
Thus, within DCCI2, such non-flow subtraction also subtract correlation from core components.
This is because QGP fluids are dynamically generated via four-momentum deposition of initial partons and there is a certain momentum correlation between core and corona components, 
which I discuss more details in Sec.~\ref{subsec:RESULT_RidgeStructure} by seeing two-particle correlations in 2-D map of $\Delta \phi$ and $\Delta \eta$.

\begin{figure}
    \centering
    \includegraphics[bb= 0 0 597 463, width=0.48\textwidth]{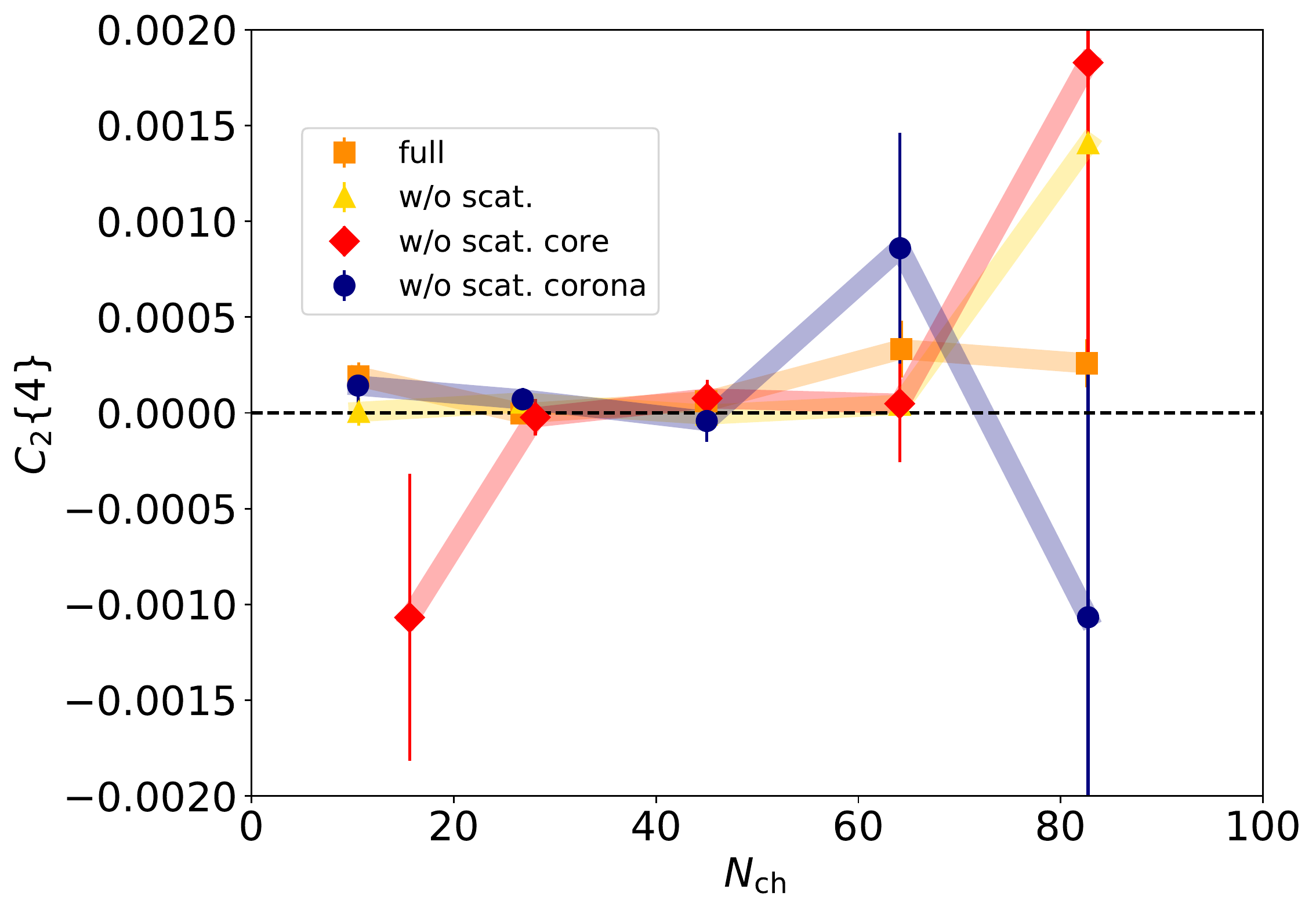}
    \hspace{7pt}
    \includegraphics[bb= 0 0 597 463, width=0.48\textwidth]{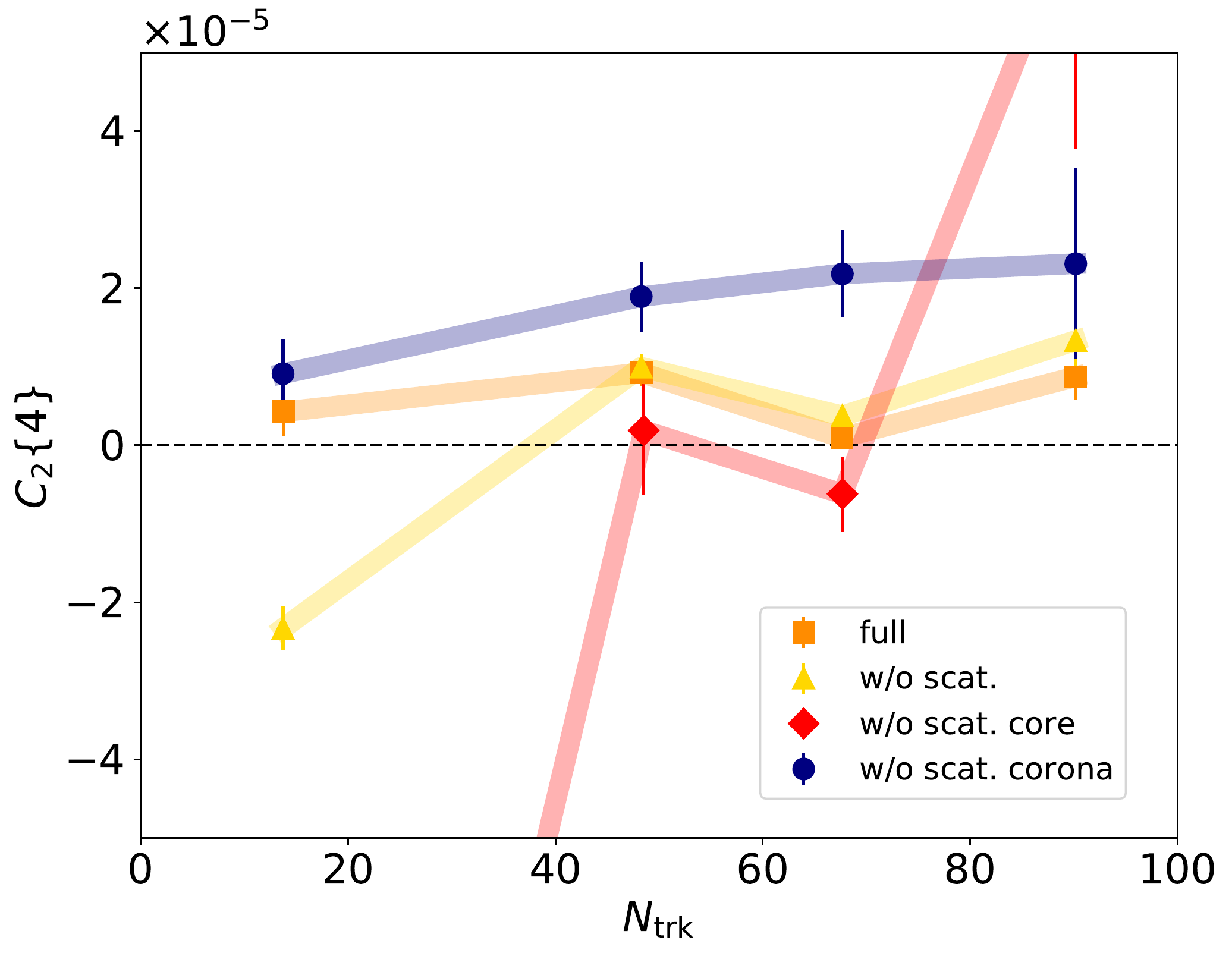}
    \caption{
    Second order of cumulant obtained from four-particle correlation for charged hadrons as a function of the number of produced charged hadrons in $p$+$p$ collisions at \snn[proton] = 13 TeV.
    Kinematic ranges are the same as used in (left) the ALICE experiment in which 3-sub event method is imposed and (right) the CMS experiment.
        }
    \label{fig:PP13_C24ALICE_CMS}
\end{figure}

I close discussion on $p$+$p$ collisions with a proxy of QGP formation in small systems nu showing second order of cumulant obtained from four-particle correlation, $\ctwofour$.
It is said that $\ctwofour$ is a better observable rather than $c_2\{2\} = (\vtwtw)^2$
because the former can truncate back-to-back 2-particle correlations, which are assumed to be non-flow and inherit in the latter.
Here, it should be noted that negative $\ctwofour$ is the proxy of
hydrodynamic behavior because of its definition in Eq.~\ref{eq:MultiParticle_vn4}.
Figure \ref{fig:PP13_C24ALICE_CMS} shows $\ctwofour$ as a function of the number of produced charged hadrons in $p$+$p$ collisions. 
Left and right figures are results obtained by adopting the same kinematic ranges and methods
used in the ALICE experiment and the CMS experiment in which
negative $\ctwofour$ is observed.
The results shown in the left panel using the ALICE kinematic range are ones that have been adopted
in calculations of anisotropic flows in this section so far.
On the other hand, $\ctwofour$ is calculated with, so called, 3-sub event method where
four particles are picked up from 3-sub rapidity ranges, $-0.8<\eta<-0.4$, $-0.4<\eta<0.4$, and $0.4<\eta<0.8$, to reduce non-flows by cutting short range correlations. For details see Appendix.~\ref{sec:APPENDX_SubEvent}. 
The CMS kinematic range adopted to calibrate the results shown in the right panel is as follows:
correlations are obtained by using particles produced with $0.3<p_T<3.0$ GeV
and within $|\eta|<2.4$ while $\Ntrk$ is the number of particles produced with $p_T>0.4$ GeV and within $|\eta|<2.4$.
The results from DCCI2 with full simulations show positive $\ctwofour$ for the entire $\Nch$ while much more statistics are required to make a solid conclusion.
To resolve this in detail, I discuss what is happening in DCCI2 with two-particle correlations in 2-D map in Sec.~\ref{subsec:RESULT_RidgeStructure}.

\subsection{Multiplicity dependence in $Pb$+$Pb$ collisions}
\label{subsec:AnisotropicFlows_MultiplicityDependenceInPBPBCollisions}
\begin{figure}
    \centering
    \includegraphics[bb=0 0 592 445, width=0.49\textwidth]{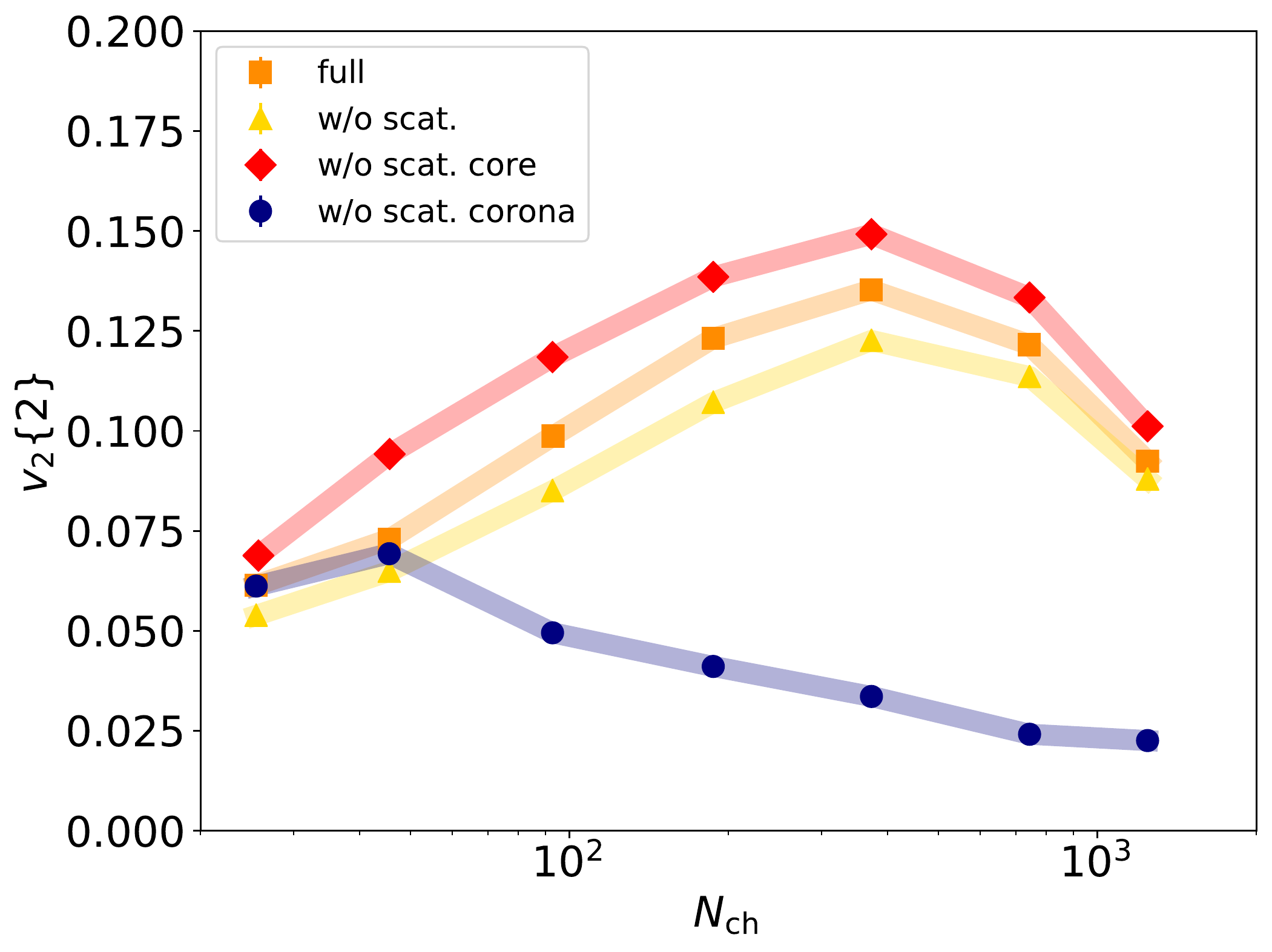}
    \includegraphics[bb=0 0 592 445, width=0.49\textwidth]{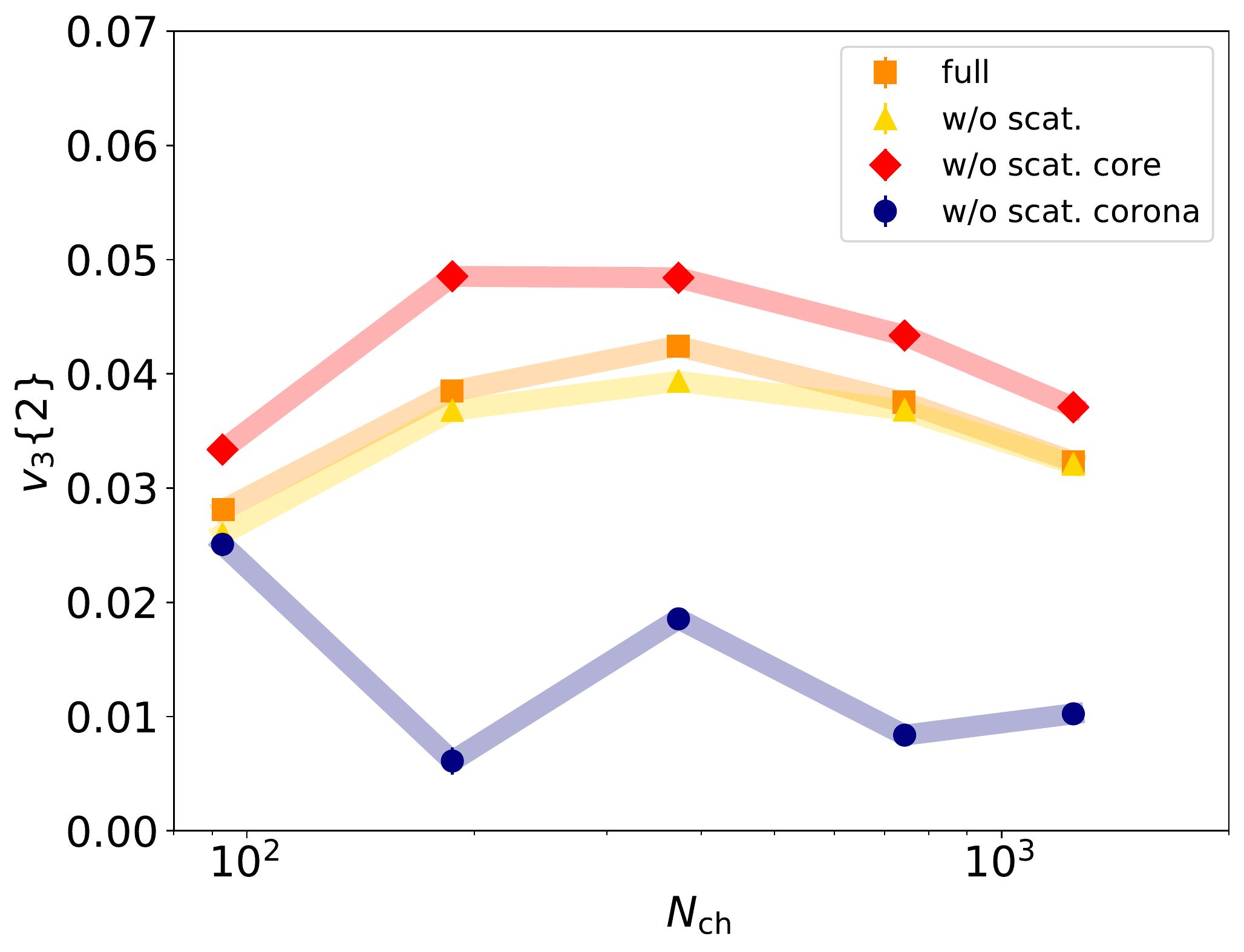}
    \includegraphics[bb=0 0 592 445, width=0.49\textwidth]{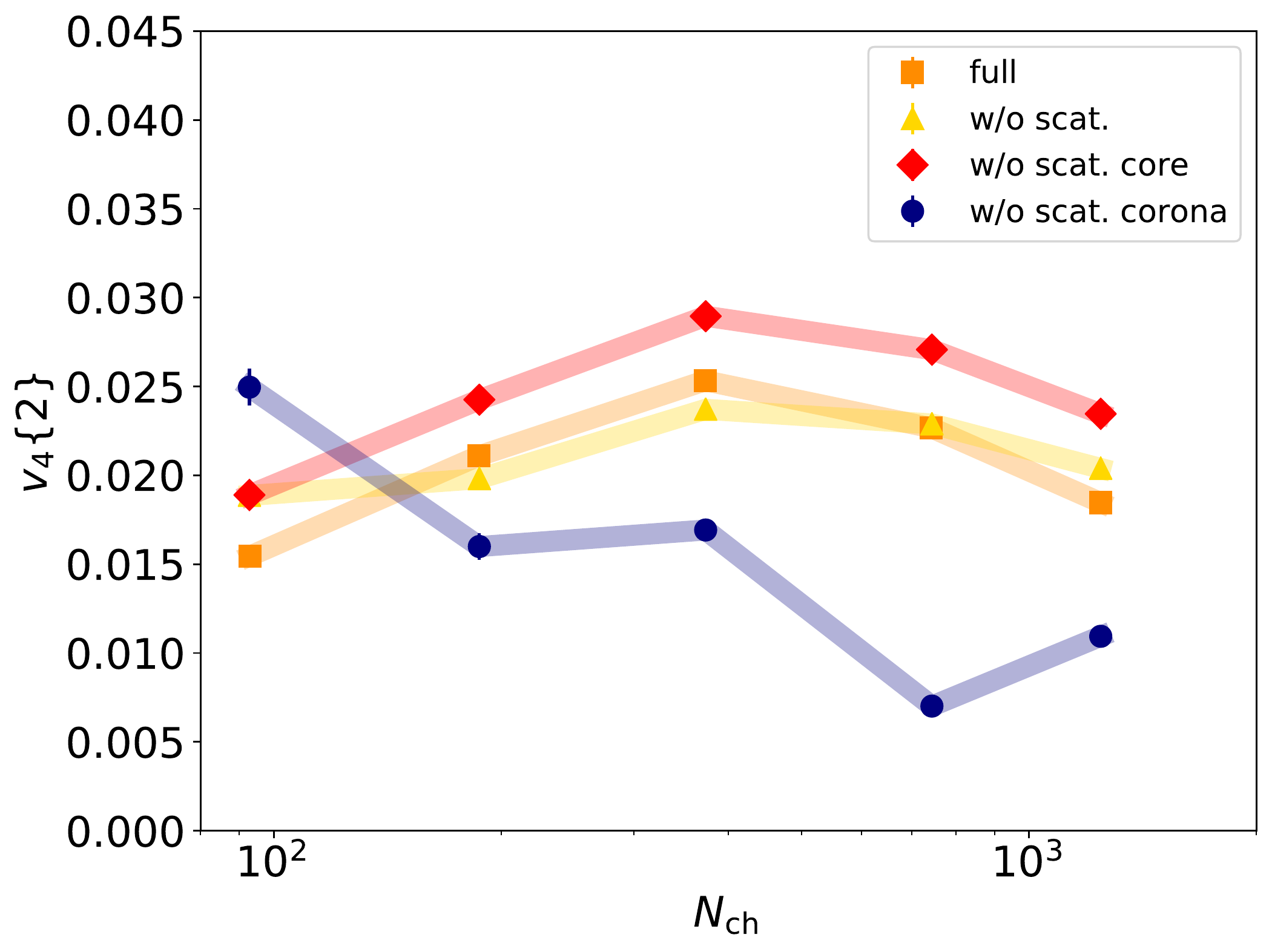}
    \caption{
Second, third, and fourth order of anisotropic flow coefficients obtained from two-particle correlation for charged hadrons as functions of the number of produced charged particles in $Pb$+$Pb$ collisions at \snn = 2.76 TeV are shown in upper left, upper right, and bottom figures. Results from full simulations (orange squares) and simulations without hadronic rescatterings (yellow triangles) are shown.  Results only from core (red diamonds) and corona (blue circles) components as a breakdown of results of simulations without hadronic rescatterings are shown simultaneously.
}
\label{fig:PBPB2760_V22Nch_V32Nch_V42Nch}
\end{figure}

Figure \ref{fig:PBPB2760_V22Nch_V32Nch_V42Nch} shows
second-, third-, and fourth- order of anisotropic flow coefficients obtained from two-particle correlation as functions of the number of produced charged particles in $Pb$+$Pb$ collisions at \snn = 2.76 TeV are shown for each.
There is a clear difference in behavior of $v_n\{2\}$ between core and corona components for $Pb$+$Pb$ collisions while there is not for $p$+$p$ collisions.
Let me first focus on results of $\vtwtw$.
From a comparison between the results with and without hadronic rescatterings, one can tell that a slight enhancement of $\vtwtw$ comes from generation of elliptic flow in the late hadronic rescattering stage \cite{Hirano:2005xf,Hirano:2007ei,Takeuchi:2015ana}.
As an over all tendency, $\vtwtw$ of core components shows maximum around
semi-central events, which is around $\Nch\approx 300-400$, and decreases from semi-central to central, which is similar to the tendency observed in experimental data \cite{Acharya:2019vdf}. 
This can be understood as follows: 
generated QGP fluids are expected to have an almond shape when one sees it in transverse plane around this centrality regime. The almond-shaped
QGP has azimuthally anisotropic geometry, which leads to azimuthal anisotropy
of pressure. Since fluid velocity is generated due to the pressure gradient, the azimuthal momentum anisotropy is inherited by the emitted particles from the pushed-out fluids.
On the other hand, $\vtwtw$ from corona components are quite small compared to those from core components.
The results show slow decrease from low to high multiplicity events.
Here again, one can see the correction from corona components in the comparison between the core result and the inclusive result in the case without hadronic rescatterings. 
The correction from corona components is found to be $\approx 15$-$30\%$ below $N_{\mathrm{ch}} \approx 400$, which originates from the small peak seen at very low $p_T$ region in the $p_T$ spectra in Fig.~\ref{fig:PTSPECTRA_PP_PBPB} (b) \footnote{The leftmost point of the contribution from core components is slightly shifted to large $N_{\mathrm{ch}}$ since there are some events  in which one cannot calculate two-particle cumulants due to less than two charged particles from the core parts are measured in a given kinematic window in this $N_{\mathrm{ch}}$ bin. Therefore the event average of $N_{\mathrm{ch}}$ for core components is biased to larger $N_{\mathrm{ch}}$.}.
This suggests that one would need to incorporate corona components in hydrodynamic frameworks to extract transport coefficients from comparisons with experimental data.

In both $p$+$p$ and $Pb$+$Pb$ results,
there are two factors that would give a finite anisotropy in corona components: color reconnection and feed-down from surviving partons.
The color reconnection effect implemented in default \pythia8 and \pythia8 Angantyr can arise collectivity \cite{Bierlich:2018lbp}.
With the color reconnection, dense color strings formed due to multiparton interactions interact with each other and eventually induce flow-like behavior of final hadrons.
Its effect can be enhanced due to more multiparton interactions in initial parton generation with DCCI2 compared to default \pythia8 and \pythia8 Angantyr. 
The detailed discussion on multiparton interactions in initial parton generation is made in Sec.~\ref{subsection:Evolution_of_transverse_energy}.
Under the dynamical core--corona initialization, partons originating from hard scatterings and emitted in the back-to-back directions tend to survive.
In contrast, soft partons, which originate from multiparton interactions and are randomly directed, tend to be converted into fluids.
Since the low $p_T$ charged hadrons come from such surviving partons through string  fragmentation, $\vtwtw$ of corona components could reflect that of their parents.
As a result, the corona components show larger anisotropy compared to results from the default \pythia8 \cite{Bierlich:2018lbp} and \pythia8 Angantyr.

I would like to mention some points from the comparison of anisotropic  flows among different orders.
In Fig.~\ref{fig:PBPB2760_V22Nch_V32Nch_V42Nch},
note that the scale of vertical axes are different in the three figures for each.
The overall tendency is quite similar for all of $\vtwtw$, $\vthree$, and $\vfour$.
From comparisons between full simulations and simulations without hadronic rescatterings, one sees that the effect of hadronic rescatterings are comparatively manifested in $\vtwtw$ rather than $\vthree$ and $\vfour$.
It is also noticeable that the fraction of anisotropy of corona components
against that of core components becomes closer in fourth- compared to second- or third- order.

\begin{figure}
    \centering
    \includegraphics[bb=0 0 592 445, width=0.49\textwidth]{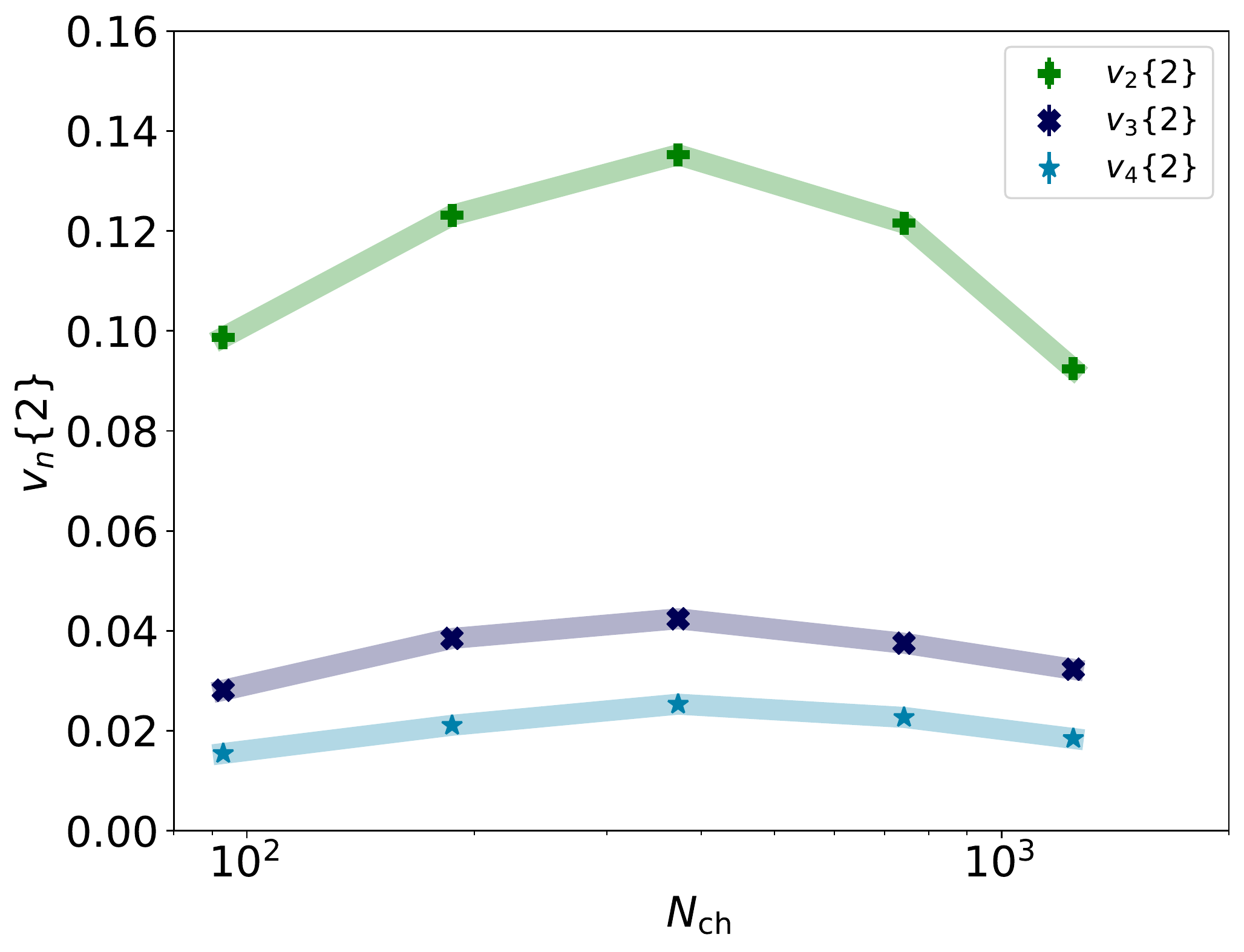}
    \caption{
    Comparisons among second (green diamonds), third (navy squares), and fourth- (blue stars) order of anisotropic flow coefficients obtained from two-particle correlation for charged hadrons as functions of the number of produced charged particles in $Pb$+$Pb$ collisions at \snn = 2.76 TeV from full simulations are shown.
    }
    \label{fig:PBPB2760_VN2Nch}
\end{figure}

To see the overall behavior of anisotropic flows in different orders, 
a comparison of full simulation results is made as a function of the number of produced charged particles.
As it is the case in other existing hydrodynamic models, 
the magnitude decreases as the order of anistoropic flows ascends
similarly to the $p$+$p$ collision results in Fig.~\ref{fig:PP13_VN2Nch}.
Again, it would be worth mentioning that, within DCCI2, initial momentum anisotropy is embedded in initial matter profile due to dynamically deposited energy and momentum from initial partons. 
Hence, it can be possible that the anisotropy pronounced even in $Pb$+$Pb$ collisions
might reflect initially generated anisotropy.

\begin{figure}
    \centering
    \includegraphics[bb=0 0 592 445, width=0.49\textwidth]{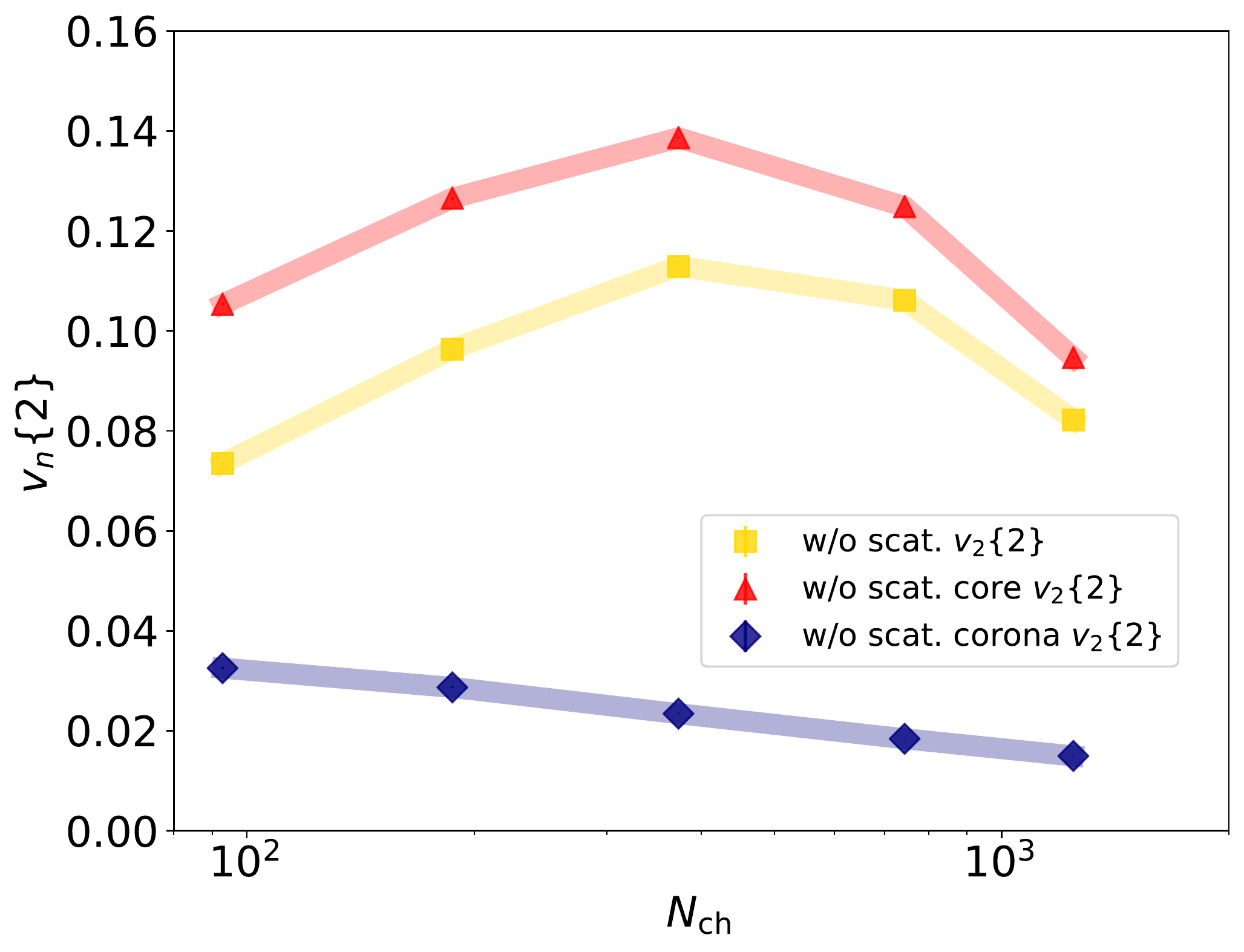}
    \includegraphics[bb=0 0 592 445, width=0.49\textwidth]{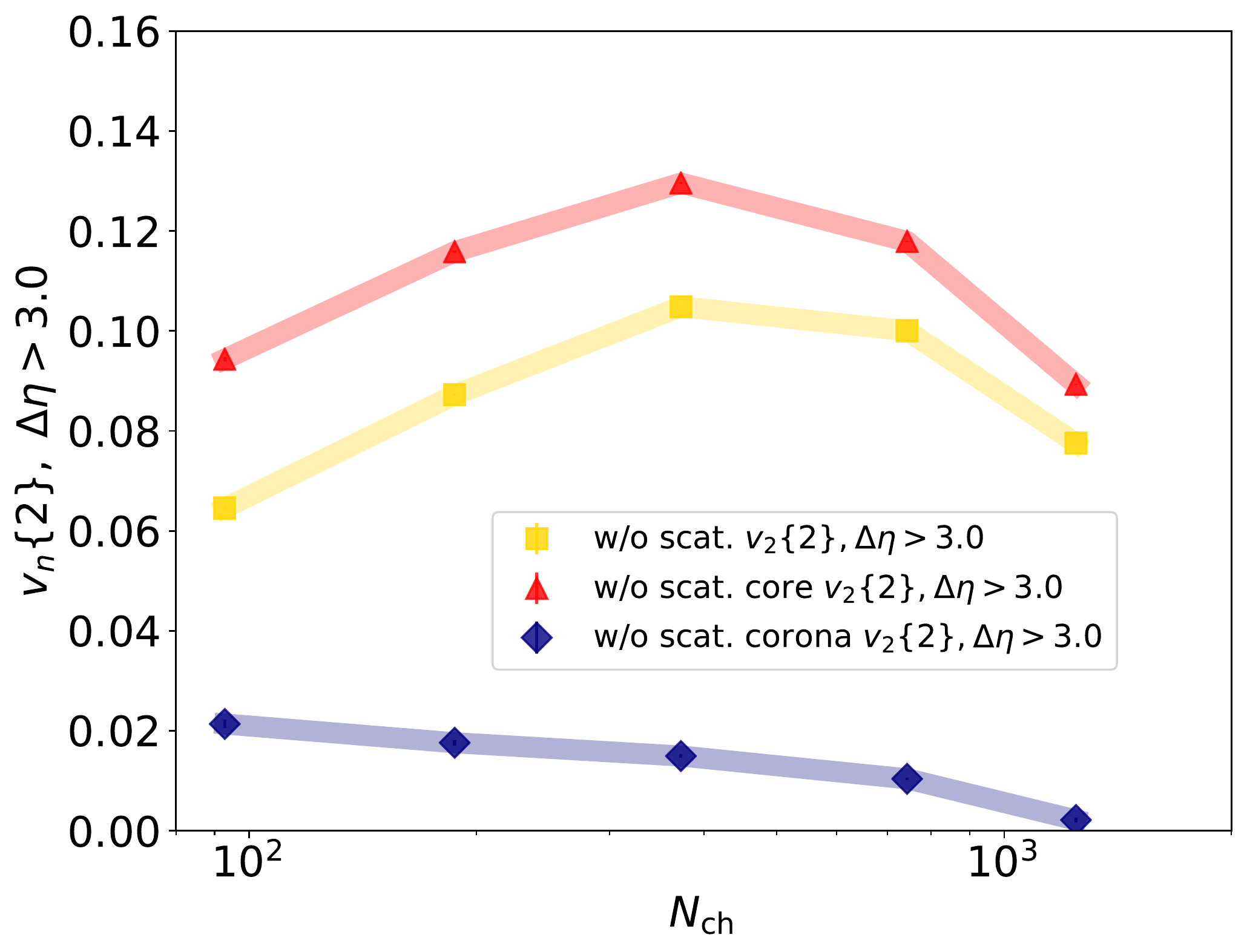}
    \caption{
    Second order of anisotropic flow coefficients obtained from two-particle correlation for charged hadrons with and without eta gap ($\Delta \eta > 3.0$) as functions of the number of produced charged particles in $Pb$+$Pb$ collisions at \snn = 2.76 TeV are shown in left and right figures. Results from simulations without hadronic rescatterings (yellow squares) and its breakdown into core (red triangles) and corona (blue diamonds) components are shown simultaneously.
    }
    \label{fig:PBPB2760_V22Nch_etaGap}
\end{figure}

\begin{figure}
    \centering
    \includegraphics[bb=0 0 634 453,  width=1.0\textwidth]{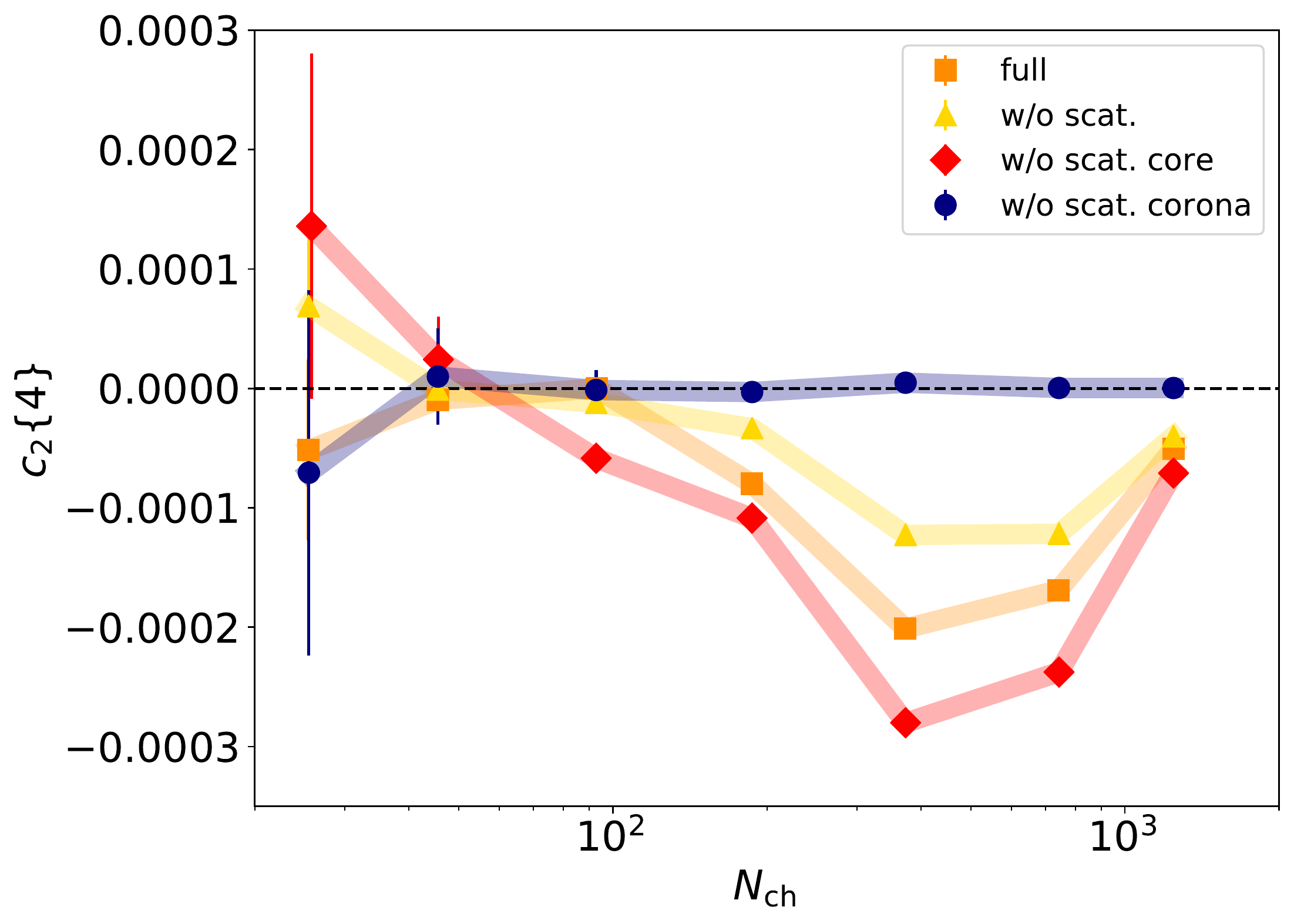}
    \caption{
    Second order of cumulant obtained from four-particle correlation for charged hadrons as a function of the number of produced charged hadrons in $Pb$+$Pb$ collisions at \snn = 2.76 TeV. Results from full simulations (orange squares) and simulations without hadronic rescatterings (yellow triangles) are shown.  Results only from core (red diamonds) and corona (blue circles) components as a breakdown of results of simulations without hadronic rescatterings are shown simultaneously.
    }
    \label{fig:PBPB2760_C24Nch}
\end{figure}

As I discuss the eta gap dependence of anisotropic flows in $p$+$p$ collision  in Fig.~\ref{fig:PP13_V22Nch_etaGap}, 
the corresponding results in $Pb$+$Pb$ collisions are shown in Fig.~\ref{fig:PBPB2760_V22Nch_etaGap}.
In order to take a large eta gap, $\vtwtw$ is obtained with particles produced within $|\eta| < 3.0$.
In the left figure, $\vtwtw$ without eta gap is shown, while that with eta gap of $\Delta \eta > 3.0$ is shown in the right.
The results from simulations without hadronic rescatterings, 
which include both core and corona components, show a slight decrease with eta gap compared to the one without eta gap especially in low multiplicity events, which is although negligible.
On the other hand, its breakdowns, the results both from core and corona components, show negligible decrease as well
despite of the expectation that anisotropic flow from corona components would be suppressed by imposing eta gap.

Figure \ref{fig:PBPB2760_C24Nch} shows $\ctwofour$ as a function of the number of produced charged particles in $Pb$+$Pb$ collisions at \snn = 2.76 TeV. 
There is no eta gap imposed because we know that the effect of eta gap is not so significant from results in Fig.~\ref{fig:PBPB2760_V22Nch_etaGap}.
Compared to the corresponding $p$+$p$ collision results,
one sees trends of each component clearly.
First, from the comparison of results between full simulations and simulations without hadronic rescatterings, 
one sees that the absolute value of $\ctwofour$ is enhanced in full simulations due to the effect of hadronic rescatterings, 
which indicates that the momentum anisotropy is grown due to the hadronic interactions.
Second, the core component shows negative $\ctwofour$, which manifests clear signal of hydrodynamic behavior.
On the other hand, one from corona components is zero consistent for the entire multiplicity range: non-flow as back-to-back 2-particle correlations is successfully subtracted in $\ctwofour$.
Here, the point that we should focus on is $\ctwofour$ from results without hadronic rescatterings. 
As we have been discussed, comparison of observables between core components and results without hadronic rescatterings extracts the corrections to pure hydrodynamic results
due to the existence of corona components in the system, especially in $Pb$+$Pb$ collisions.
In Fig.~\ref{fig:PBPB2760_C24Nch}, one sees the corona correction:
the absolute value of $\ctwofour$ of core components is decreased to the one from results without hadronic rescatterings while the result from corona components shows zero-consistent $\ctwofour$.
Simply speaking, multi-particle correlation is the correlation per the number of permutation of $m$-particle.
Even if there are particles that does not contribute to correlation, in this case they are the corona components, the permutation of $m$-particle in the denominator is calculated from all particles including corona components.
From this result, one lesson that we can learn is that
anisotropic flows obtained from pure hydrodynamic models cannot directly be compared with
experimental data if there is a certain portion of non-equilibrium components.

\subsection{Transverse momentum dependence}
\label{subsec:AnisotropicFlows_TransverseMomentumDependence}
We have been, so far, seeing $p_T$ integrated anisotropic flows.
Especially in $Pb$+$Pb$ collisions, we found that the corona components at low $p_T$ regime
make an non-negligible effects on anisotropic flows from core components.
To investigate characteristics of anisotropic flow from each core and corona components,
I analyze anisotropic flows differentiating by $p_T$.

\begin{figure}
    \centering
    \includegraphics[bb=0 0 543 454, width=0.49\textwidth]{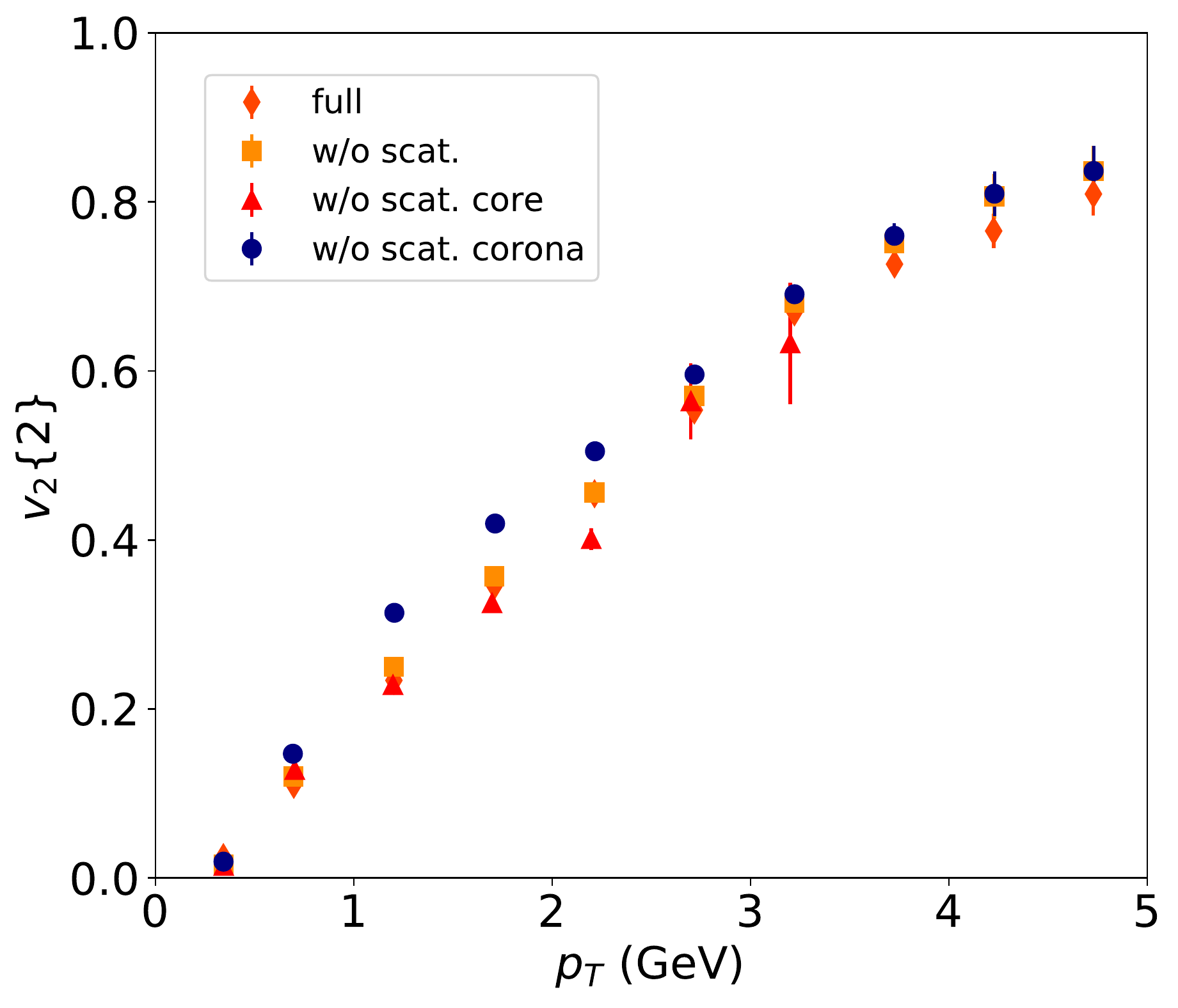}
    \includegraphics[bb=0 0 543 454, width=0.49\textwidth]{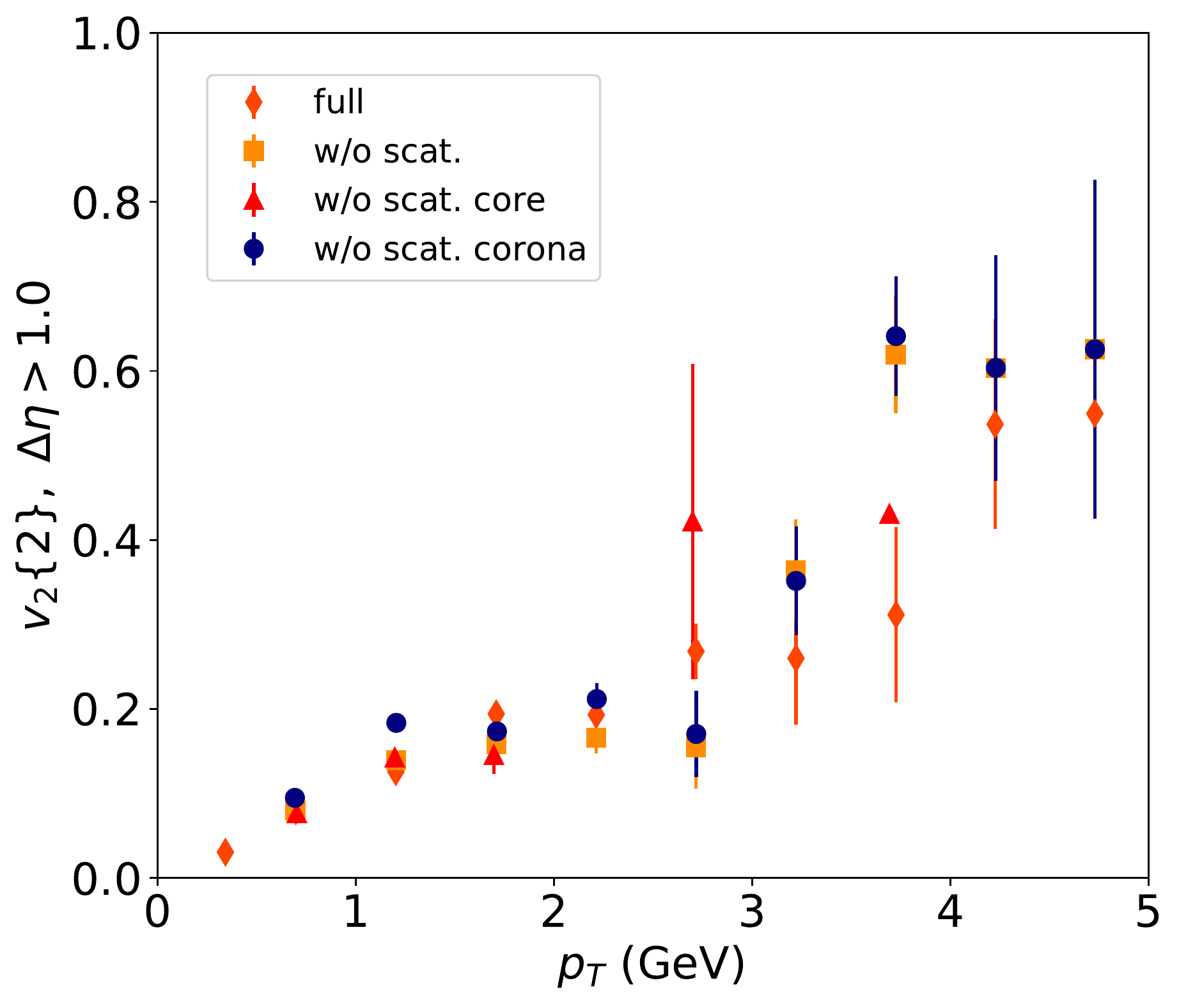}
    \caption{
    Second order of anisotropic flow coefficients obtained from two-particle correlation with and without eta gap as functions of transverse momentum from INEL$>0$ $p$+$p$ collisions at \snn[proton]=13 TeV are shown in left and right figures, respectively.
    Results from full simulations (orange diamonds) and simulations without hadronic rescatterings (yellow squares) are shown.  Results only from core (red triangles) and corona (blue circles) components as a breakdown of results of simulations without hadronic rescatterings are shown simultaneously.
    }
    \label{fig:PP13_V2PT_INEL_etaGap}
\end{figure}

Figure \ref{fig:PP13_V2PT_INEL_etaGap} shows second order of anisotropic flow coefficients obtained from two-particle correlation as a function of $p_T$ from INEL$>0$ $p$+$p$ collisions at \snn[proton] = 13 TeV.
The left and right figures show results without and with eta gap, respectively, while the eta gap imposed here is $\Delta \eta > 1.0$.
Note that anisotropic flow is obtained within each $p_T$ bin.
As an overall tendency, firstly, there is no significant difference of the results among full simulations, simulations without hadronic rescatterings, and core components while corona components show slightly larger $\vtwtw$.
On the other hand, once eta gap is imposed on $\vtwtw$ to eliminate short range correlation, $\vtwtw$ as a function of $p_T$ show significant suppress for intermediate to high $p_T$ regime,
and which is seen in both core and corona components.
Thus, it can be said that the distance of correlation becomes short in rapidity regardless of whether they are core or corona components, which indicates that there can be a strong anisotropic correlation between core and corona.

\begin{figure}
    \centering
    \includegraphics[bb=0 0 543 454, width=0.49\textwidth]{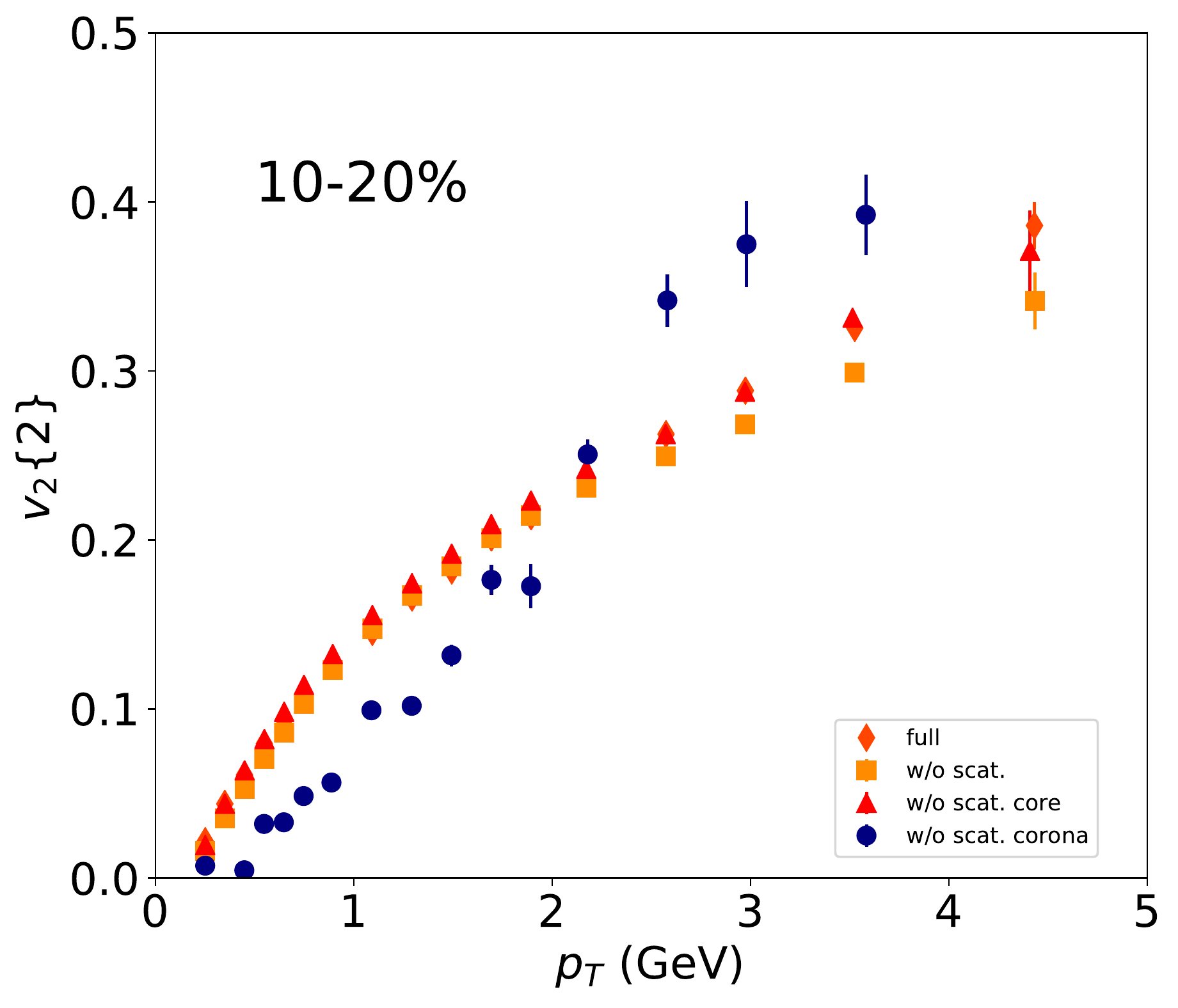}
    \includegraphics[bb=0 0 543 454, width=0.49\textwidth]{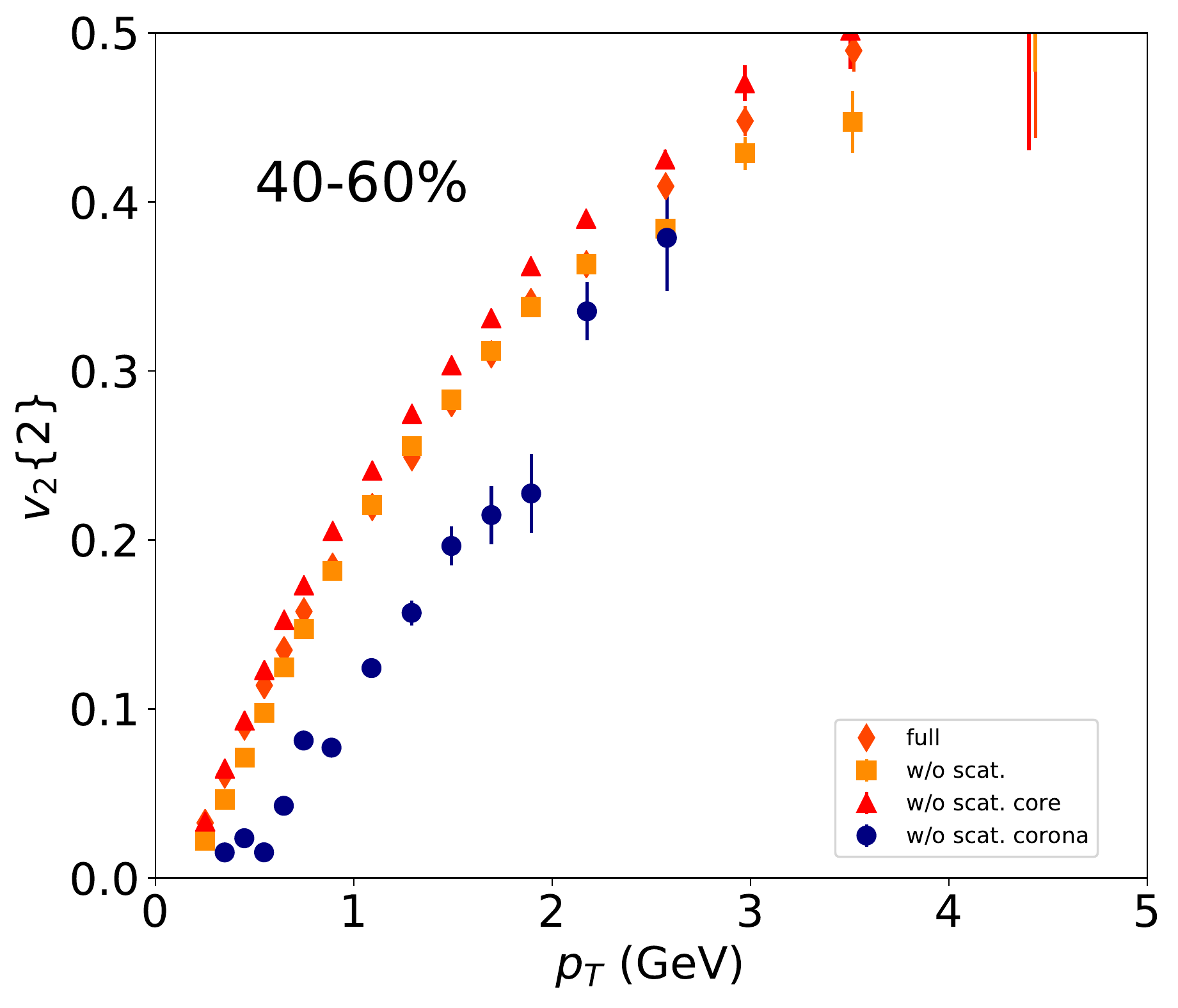}
    \includegraphics[bb=0 0 543 454, width=0.49\textwidth]{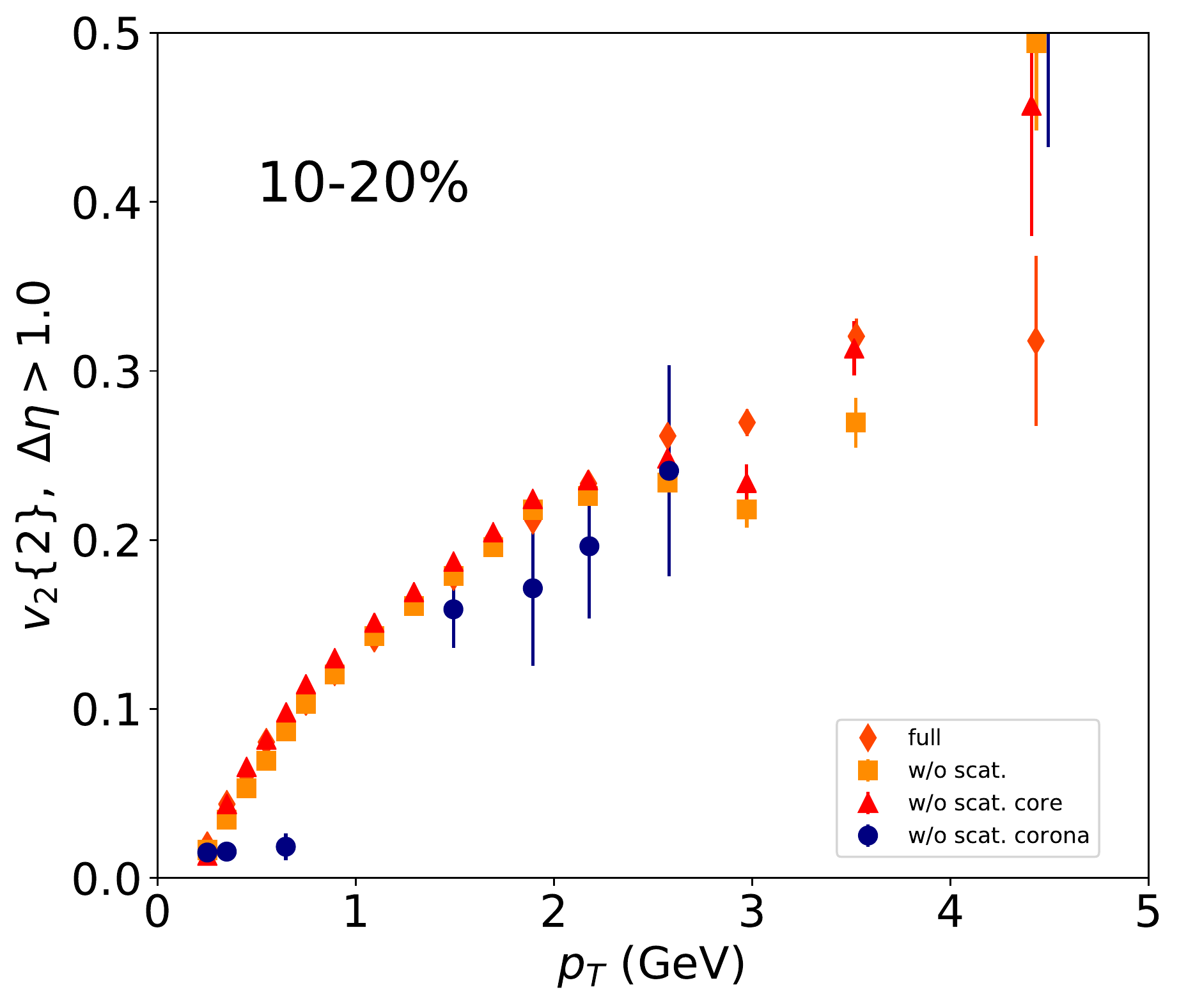}
    \includegraphics[bb=0 0 543 454, width=0.49\textwidth]{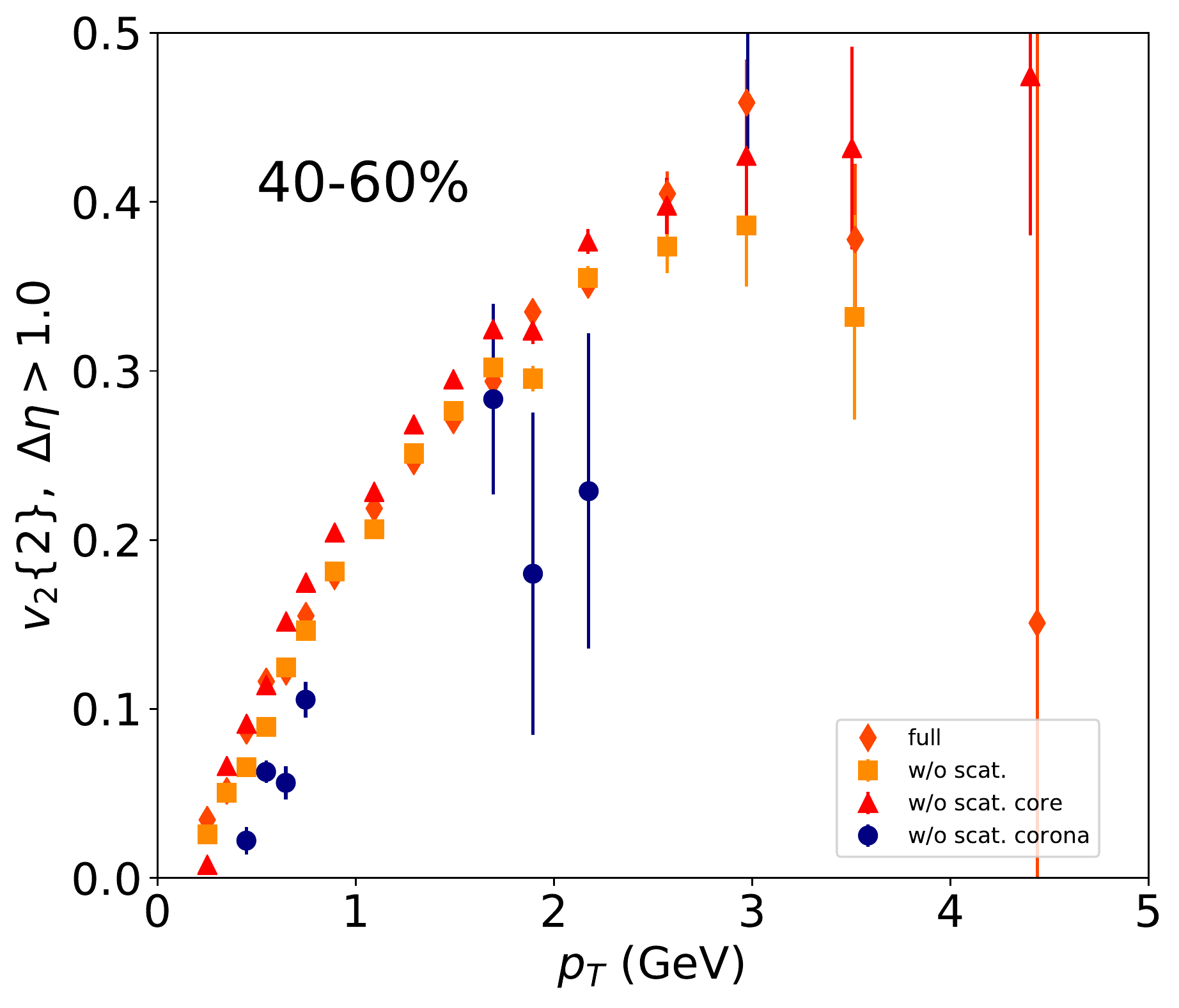}
    \caption{
    Second order of anisotropic flows obtained from two-particle correlation as a function of transverse momentum in $0-10\%$ and $40-60\%$ in centrality without eta gap are shown in left and right of upper panels, respectively.
    Corresponding results with eta gap ($\Delta \eta > 1.0$) are shown in lower panels.
     Results from full simulations (orange diamonds) and simulations without hadronic rescatterings (yellow squares) are shown.  Results only from core (red triangles) and corona (blue circles) components as a breakdown of results of simulations without hadronic rescatterings are shown simultaneously.
    }
    \label{fig:PBPB2760_V2PT_centrality_etaGap}
\end{figure}

\begin{figure}
    \centering
    \includegraphics[bb=0 0 531 454, width=0.9\textwidth]{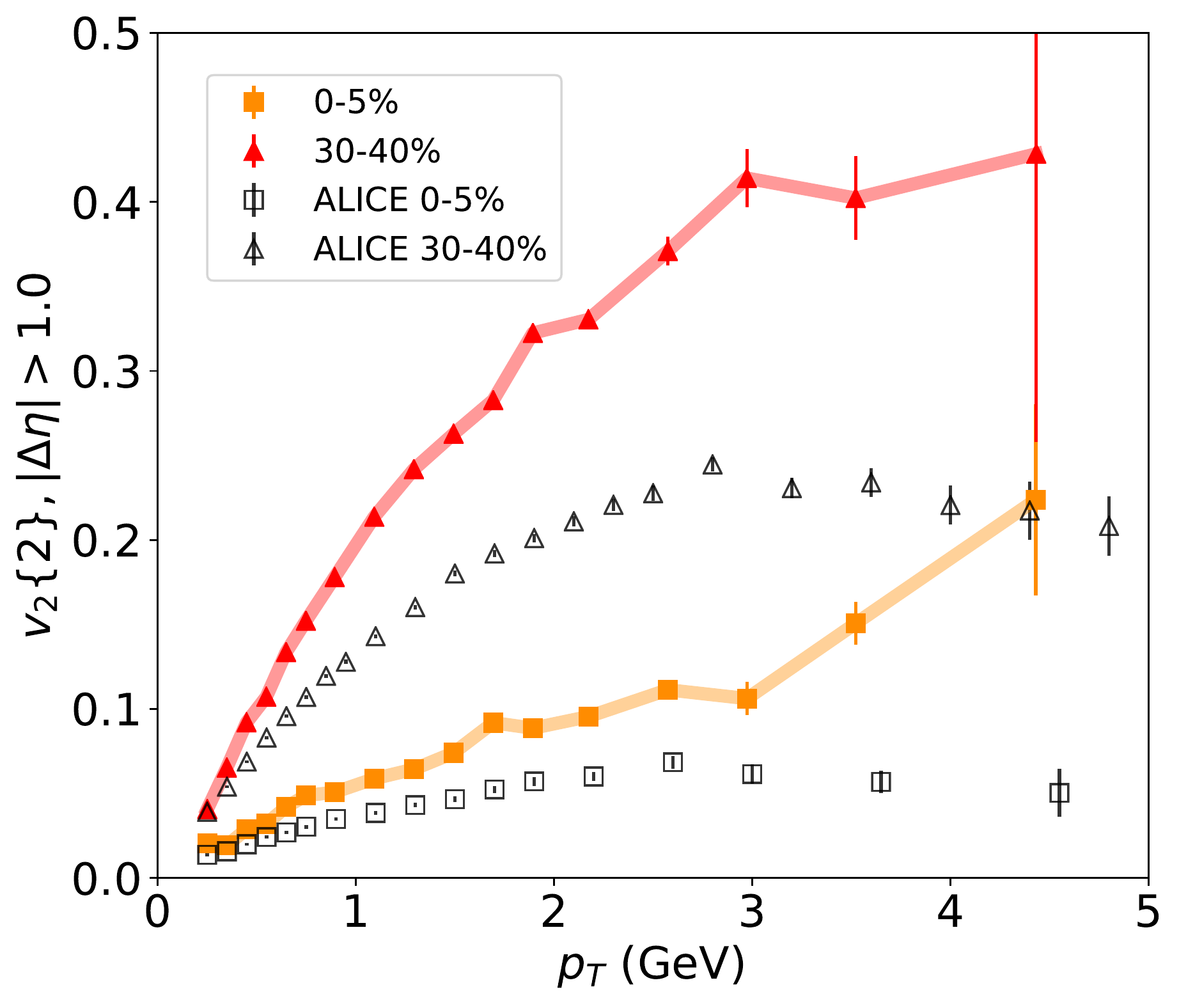}
    \caption{
    Second order of anisotropic flow coefficients obtained from two-particle correlation with eta gap ($\Delta \eta > 1.0$) as a function of transverse momentum from $Pb$+$Pb$ collisions in $0-5\%$ (orange squares) and $30-40\%$ (red triangles) in centrality. Comparisons with the ALICE experimental data \cite{ALICE:2011ab} (open black symbols) for each centrality is made. 
    }
    \label{fig:PBPB2760_V2PT_COMPARISONwithALICE}
\end{figure}

Figure \ref{fig:PBPB2760_V2PT_centrality_etaGap} shows the corresponding results with above in $Pb$+$Pb$ collisions.
Upper and lower figures are results without and with eta gap, respectively.
Results in 0-20\% and 40-60\% in centrality are shown in left and right panels, respectively.
One sees that there is a clear different behavior in the corona components while statistics is not sufficient for results with eta gap:
for results without eta gap, $\vtwtw$ from corona component is smaller in low $p_T$ (around $p_T<2$) while one shows rapid increase towards larger $p_T$ regime compared to core component.
We can infer some points from the above results:
first, the corona components in very low $p_T$, that behave as a correction to core components, has smaller anisotropy in momentum space.
Second, the back-to-back 2-particle anisotropy from non-flow is manifested especially above $p_T \approx 2$ GeV in results without eta gap.

Finally I show data to model comparison of $\vtwtw$ with eta gap ($\Delta \eta> 1.0$) as a function of $p_T$ in 0-5\% and 30-40\% in centrality of $Pb$+$Pb$ collisions in Fig.~\ref{fig:PBPB2760_V2PT_COMPARISONwithALICE}.
The results from DCCI2 clearly overestimate the experimental data in both centrality.
The overestimation is expected to be reconciled by introducing shear viscosity in hydrodynamic calculations because shear stress induced by gradient of flow in a perpendicular direction of the direction of flow suppress the growth of anisotropy.
I admit that the introduction of such dissipative currents in hydrodynamic simulation is necessary to reproduce the experimental data and it is indispensable to explore the QGP transport.
I leave this update for future work.

%\begin{figure*}[htbp]
%\begin{center}
%\includegraphics[bb=0 0 539 611, width=0.48\textwidth]{chapters/figure/results/Fig8a.pdf}
%\includegraphics[bb=0 0 539 611, width=0.48\textwidth]{chapters/figure/results/Fig8b.pdf}
%\caption{(Color Online)
%Second order of anisotropic flow coefficient obtained from two-particle correlation for charged hadrons as a function of the number of produced charged particles in (a) $p$+$p$ collisions at $\sqrt{s} = 7 \ \mathrm{TeV}$ and (b) $Pb$+$Pb$ collisions at $\sqrt{s_{NN}} = 2.76 \ \mathrm{TeV}$. Results from full simulations (orange squares) and simulations without hadronic rescatterings (yellow diamonds) are shown with closed symbols. While results from core (red diamonds) and corona (blue triangles) components from simulations without hadronic rescatterings are plotted with open symbols.
%}
%\label{fig:V22MULTI_PP_PBPB_CORECORONA}
%\end{center}
%\end{figure*}

\subsection{Ridge structure}
\label{subsec:RESULT_RidgeStructure}
In anisotropic flow coefficients, 
we can quantify amplitude of each order of anisotropy.
In this section, I discuss details of two-particle correlation by plotting as a functions of $\Delta eta$ and $\Delta \phi$.
Details of the technique used in this calculation is explained in Appendix.~\ref{sec:APPENDX_TwoParticle}.

As I discussed in Sec.~\ref{subsec:Collectivity}, the ridge structure, 
two-particle correlation at $\Delta \approx 0$ along wide $\Delta \eta$ range,
is a proxy of hydrodynamic response due to QGP formation.
The ridge structure has been observed not only in heavy-ion but also in $p$+$p$ collisions.
Thus, the general question is, ``what is the origin of the ridge structure observed in the experimental data of $p$+$p$ collisions at high multiplicity?''.
Giving an answer within DCCI2 is one of the big goals in this study.
However, as we see from Sec.~\ref{subsec:AnisotropicFlows_MultiplicityDependenceInPPCollisions} to 
\ref{subsec:AnisotropicFlows_TransverseMomentumDependence},
DCCI2 has not reached to the level in which I can compare the anisotropic flow results from the model with experimental data.
Admitting that fact,
I discuss whether we can see the ridge structure in high-multiplicity events in $p$+$p$ collisions.
In addition, I show if the simple expectation: 
core components as hydrodynamic response should give a long range correlation 
and that corona component should give a short range correlation, 
is realized within DCCI2
in the followings.

\subsubsection{$p$+$p$ collision results}

\begin{figure}
    \centering
    \includegraphics[bb= 0 0 826 692, width=1.0\textwidth]{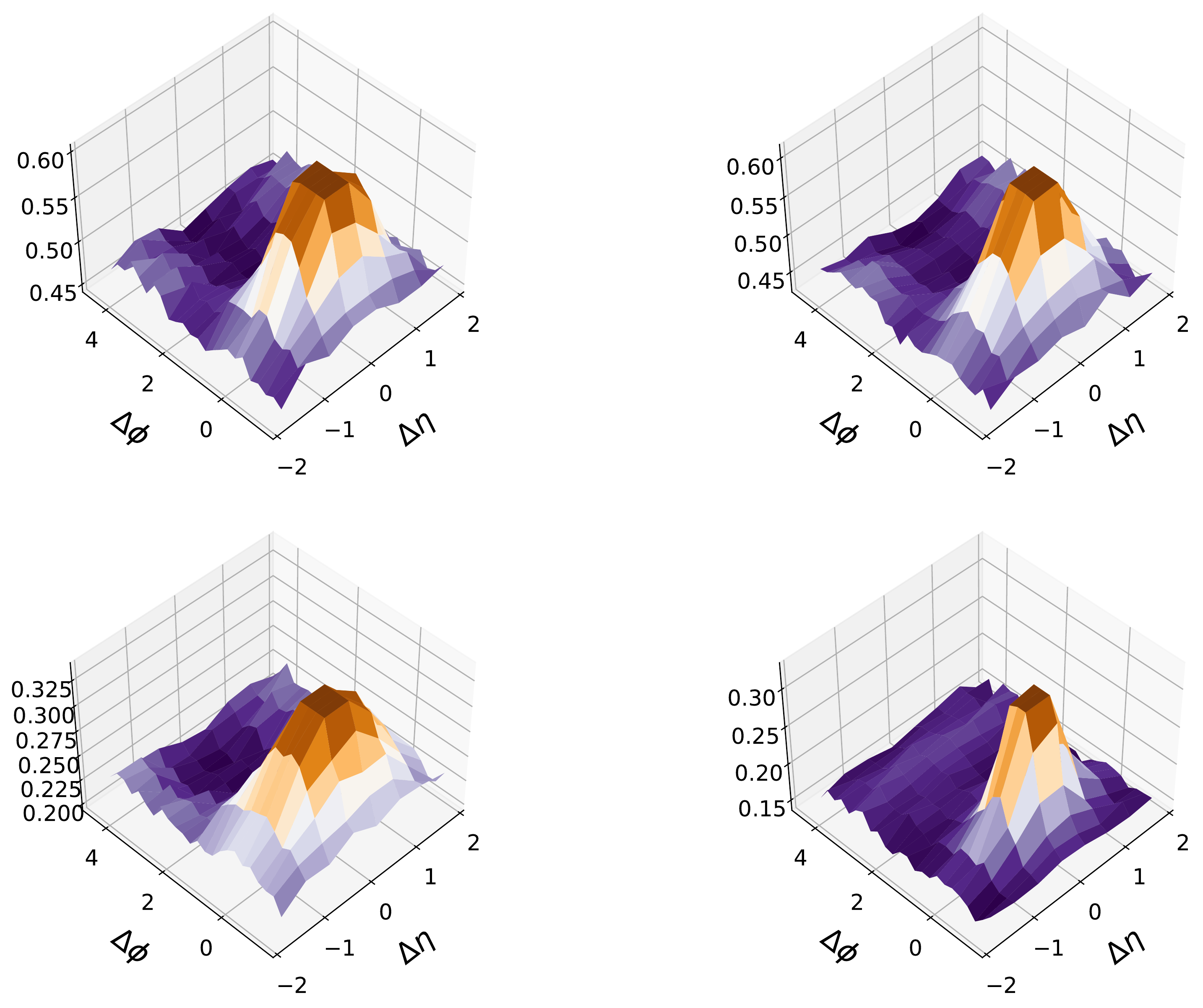}
    \caption{Two-particle correlations as functions of $\Delta \eta $ and $\Delta \phi$ obtained from charged particles produced within $|\eta|<1.5$ and with $1<p_T<2$ GeV for both triggers and associates in 0-0.95\% of V0M multiplicity class of $p$+$p$ collisions at \snn[proton]=13 TeV. Results from full simulation (left), simulations without hadronic rescatterings (right) are shown at the top, while breakdowns of produced particles from simulations without hadronic rescatterings into core (left) and corona (right) components are shown at the bottom. The centrality and the $p_T$ range are determined based on the method used in the ALICE paper \cite{ALICE:2021nir}.}
    \label{fig:PP13_RIDGE0-0.95pct}
\end{figure}
\begin{figure}
    \centering
    \includegraphics[bb= 0 0 826 692, width=1.0\textwidth]{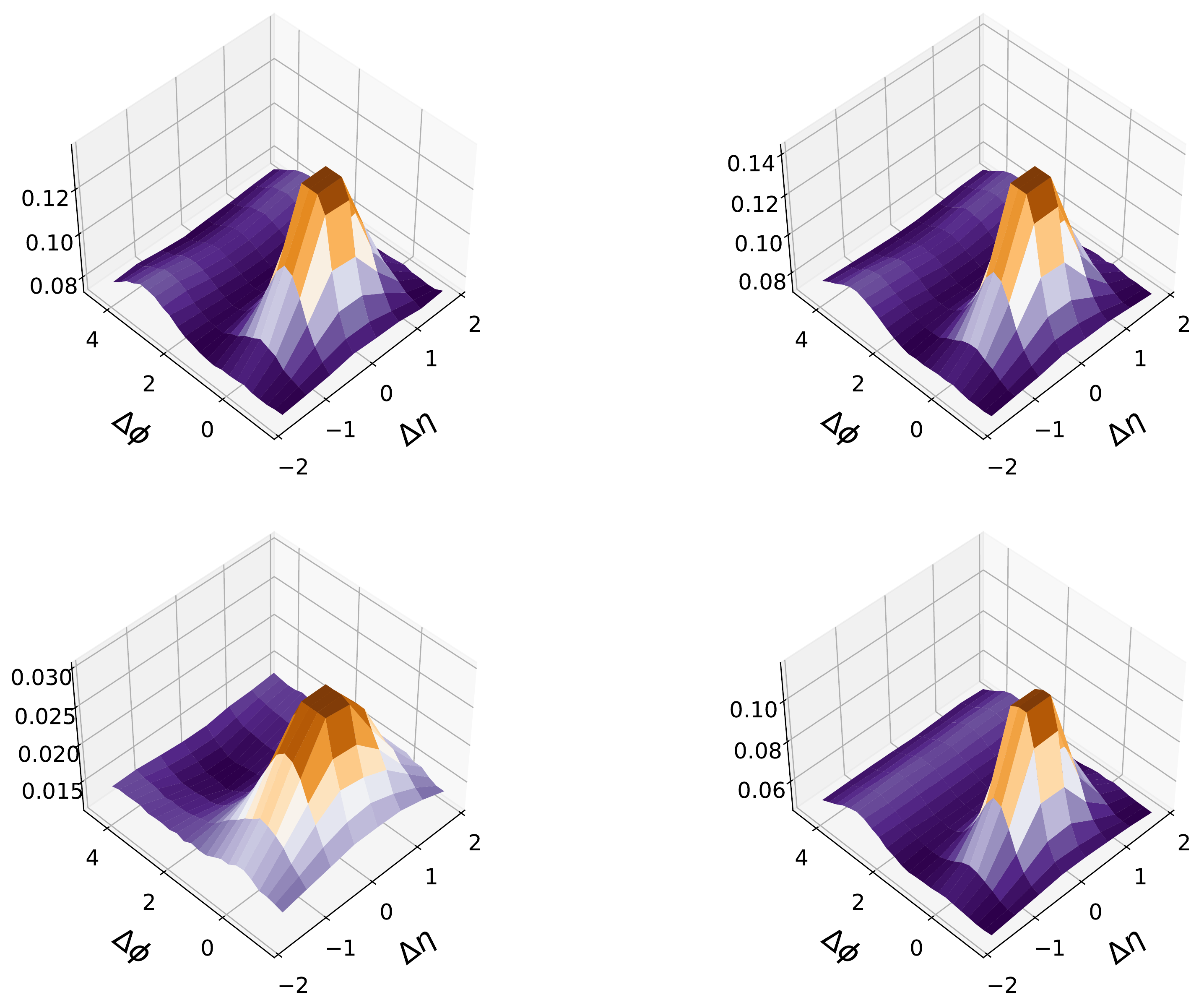}
    \caption{Two-particle correlations as functions of $\Delta \eta $ and $\Delta \phi$ obtained from charged particles produced within $|\eta|<1.5$ and with $1<p_T<3$ GeV for both triggers and associates in 0-100\% of V0M multiplicity class of $p$+$p$ collisions at \snn[proton]=13 TeV. Results from full simulation (left), simulations without hadronic rescatterings (right) are shown at the top, while breakdowns of produced particles from simulations without hadronic rescatterings into core (left) and corona (right) components are shown at the bottom. The centrality and the $p_T$ range are determined based on the method used in the ALICE paper \cite{ALICE:2021nir}.}
    \label{fig:PP13_RIDGE0-100pct}
\end{figure}

I discuss multiplicity dependence of two-particle correlation in $p$+$p$ collisions at \snn[proton]=13 TeV
with 2 different centrality/multiplicity cuts: the events are classified following the ALICE \cite{ALICE:2021nir} and CMS collaborations \cite{CMS:2016fnw}, respectively.
First I start with the ALICE kinematic settings.
Figure \ref{fig:PP13_RIDGE0-0.95pct} shows
two-particle correlation functions of $\Delta \eta$ and $\Delta \phi$
from high-multiplicity events (0-0.95\% of V0M multiplicity class) of $p$+$p$ collisions at \snn[proton]=13 TeV. Results from full simulation and simulations without hadronic rescatterings are shown in left and right at the top, while breakdowns of produced particles from simulations without hadronic rescatterings into core and corona components are shown in left and right at the bottom.
On the other hand, the corresponding results in low-multiplicity events (0-100\% of V0M centrality)
are shown in Fig.~\ref{fig:PP13_RIDGE0-100pct}.
It should be noted that, to gain statistics, two-particle correlations are obtained from charged particles produced within $|\eta|<1.5$ instead of $|\eta|<0.8$, the eta range covered by the detector at mid-rapidity introduced in the ALICE collaboration.
Hence the results shown here cannot be compared directly to the results from the ALICE experiment.
From the comparisons of full simulations between the two centrality classes, 
one sees that the ridge structure, correlation around $\Delta \phi \approx 0$ along $\Delta \eta $ axis, is slightly enhanced in high-multiplicity class:
for instance the small peak around $\Delta \phi \approx 0$ and $\Delta \eta \approx -1$ is seen more clearly in high-multiplicity compared to low-multiplicity events.
One also sees that there is not so significant difference of results between full simulations and simulation without hadronic rescatterings, which is consistent with the observation for the $p_T$ integrated $\vtwtw$.
The difference between the two-particle correlation of core and corona components is
relatively clear in both multiplicity classes.
The near side ridge-like structure appears in core components in both multiplicity classes
while such structure is not that manifested in corona components.
With these results, I conclude that the ridge-like structure in near-side can be seen up to $|\Delta \eta| \approx 1$, which originates from the core components, within DCCI2 with the current parameter set.
However, it should be noted that the away-side peak at $\Delta \approx \pi$ stretched along $\Delta \eta$ is so small that the structure is almost absent
in high-multiplicity events of full simulation results
while that is observed in experimental data.
This means that back-to-back correlation of mini-jets that originally should be produced due to momentum conservation dismiss due to the dynamical core--corona initialization.
As we see in the comparison of $p_T$ spectra between DCCI2 and experimental data in Sec.~\ref{subsec:IdentifiedParticlePTSpectra},
results from DCCI2 underestimate experimental data above $p_T\approx 1$ GeV at the highest multiplicity class.
Thus, these results in total suggest that there is too much energy loss of partons
at this $p_T$ regime,
which should be reconciled in future updates of DCCI.

%\begin{figure}
%    \centering
%    \includegraphics[bb= 0 0 826 692, width=0.7\textwidth]{chapters/figure/results/PP13_RIDGE85-95Ntrk.pdf}
%    \caption{Caption}
%    \label{fig:my_label}
%\end{figure}
\begin{figure}
    \centering
    \includegraphics[bb= 0 0 826 692, width=1.0\textwidth]{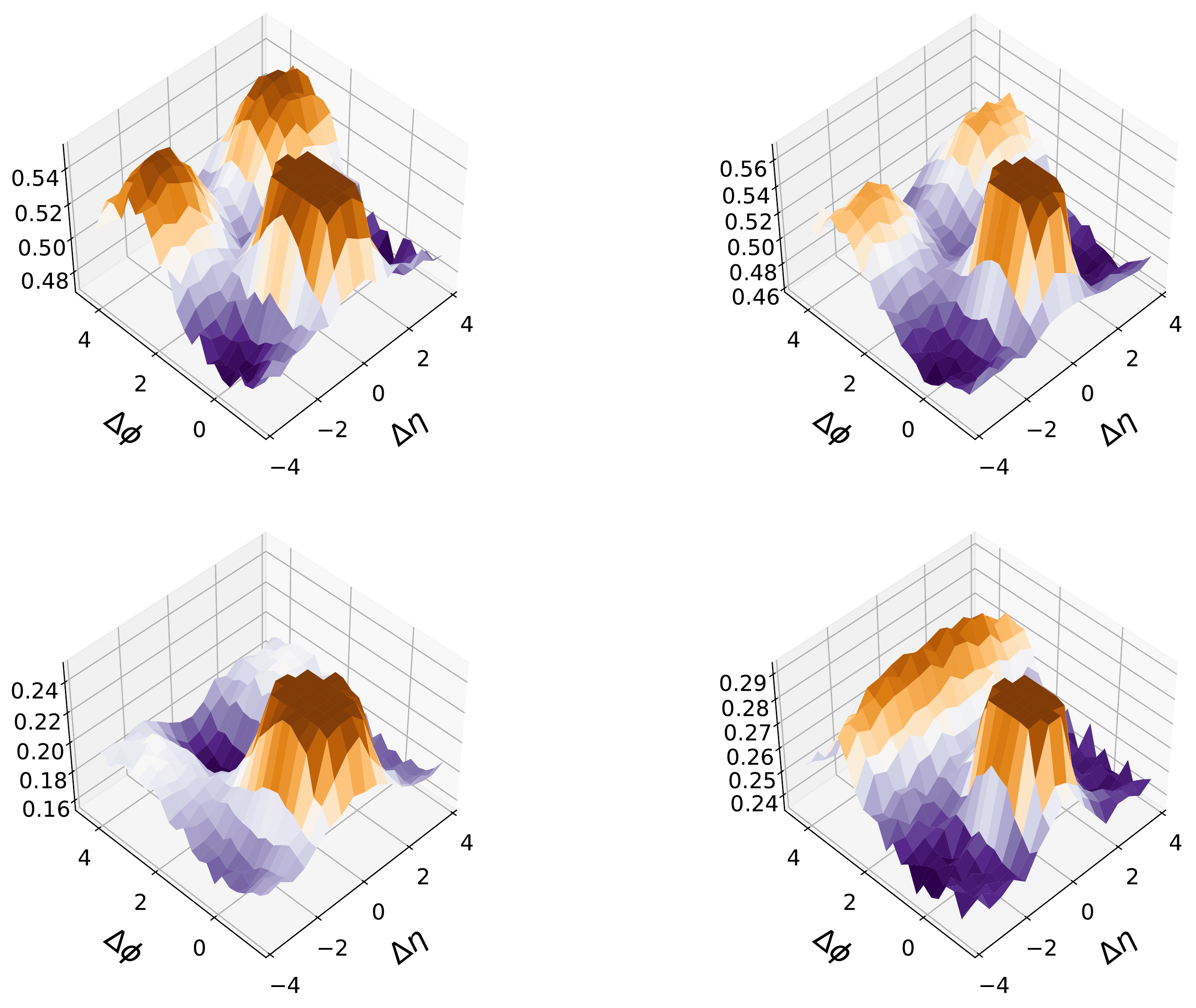}
    \caption{Two-particle correlations as functions of $\Delta \eta $ and $\Delta \phi$ obtained from charged particles produced within $|\eta|<2.4$ and with $1<p_T<3$ GeV for both triggers and associates in high-multiplicity $p$+$p$ collisions ($60<\Ntrk<85$) at \snn[proton]=13 TeV. Results from full simulation (left), simulations without hadronic rescatterings (right) are shown at the top, while breakdowns of produced particles from simulations without hadronic rescatterings into core (left) and corona (right) components are shown at the bottom. The same kinematic cuts and event classification used in the CMS paper \cite{CMS:2016fnw} are adopted.}
    \label{fig:PP13_RIDGE60-85Ntrk}
\end{figure}

\begin{figure}
    \centering
    \includegraphics[bb= 0 0 826 692, width=1.0\textwidth]{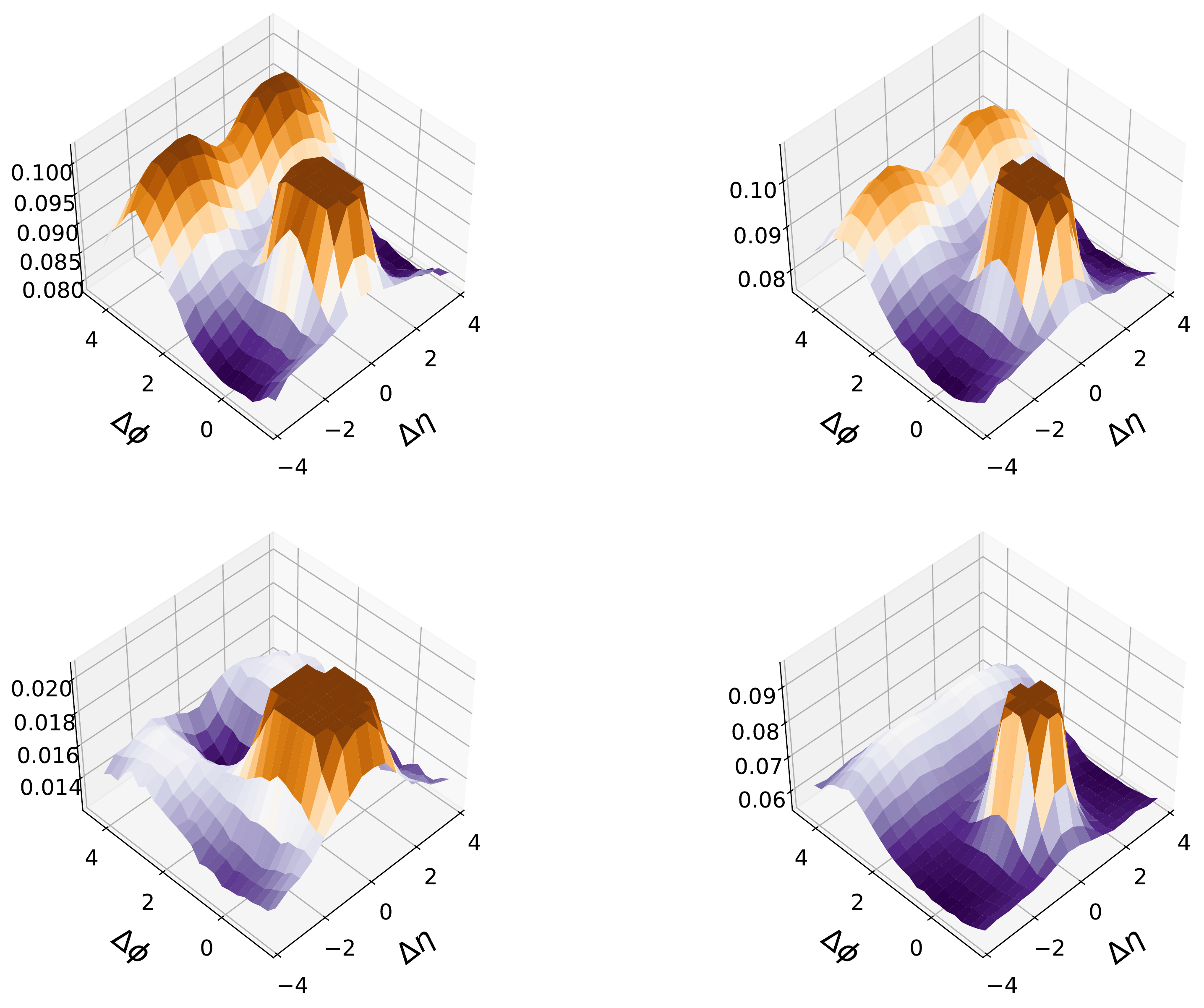}
    \caption{Two-particle correlations as functions of $\Delta \eta $ and $\Delta \phi$ obtained from charged particles produced within $|\eta|<2.4$ and with $1<p_T<3$ GeV for both triggers and associates in almost minimum-bias $p$+$p$ collisions ($0<\Ntrk<\infty$) at \snn[proton]=13 TeV. Results from full simulation (left), simulations without hadronic rescatterings (right) are shown at the top, while breakdowns of produced particles from simulations without hadronic rescatterings into core (left) and corona (right) components are shown at the bottom. The same kinematic cuts and event classification used in the CMS paper \cite{CMS:2016fnw} are adopted.}
    \label{fig:PP13_RIDGE0-infNtrk}
\end{figure}

Second, I investigate the corresponding results with the CMS kinematic cuts.
Figure \ref{fig:PP13_RIDGE60-85Ntrk} and \ref{fig:PP13_RIDGE0-infNtrk} show
two-particle correlation functions as functions of $\Delta \eta$ and $\Delta \phi$ 
obtained from charged particles produced within $|\eta|<2.4$ and with $1<p_T<3$ GeV
in high-multiplicity ($60<\Ntrk<85$) and minimum-bias ($0<\Ntrk<\infty$) events, respectively.
Note that the range of $\Delta \eta$ is different from the one in Figs.~\ref{fig:PP13_RIDGE0-0.95pct} and \ref{fig:PP13_RIDGE0-100pct}
because the two-particle correlation function is now obtained within wider eta range.
Only in high-multiplicity events in Fig.~\ref{fig:PP13_RIDGE60-85Ntrk}, one sees that there is the double peak structure appearing at $\Delta \phi \approx \pi$ in the results of full simulations and simulations without hadronic rescatterings,
which is not observed in experimental results.
Because the double peak structure is seen in results of core components but corona components,
it can be expected that the structure seen in full simulations mostly comes from the correlation in core components.
On the other hand, the results of minimum-bias events in Fig.~\ref{fig:PP13_RIDGE0-infNtrk}
show that the strong double peak structure that we see in Fig.~\ref{fig:PP13_RIDGE60-85Ntrk} is diluted.
The result of core components shows much smaller correlation only in near-side by 2-order compared to those from corona components.
It should also be mentioned that the hadronic rescatterings enhances correlations in both multiplicity events,
which indicates that hadronic interactions pronounce flows 
even in $p$+$p$ collisions.

\begin{figure}
    \centering
    \includegraphics[bb=0 0 936 303, width=1.0\textwidth]{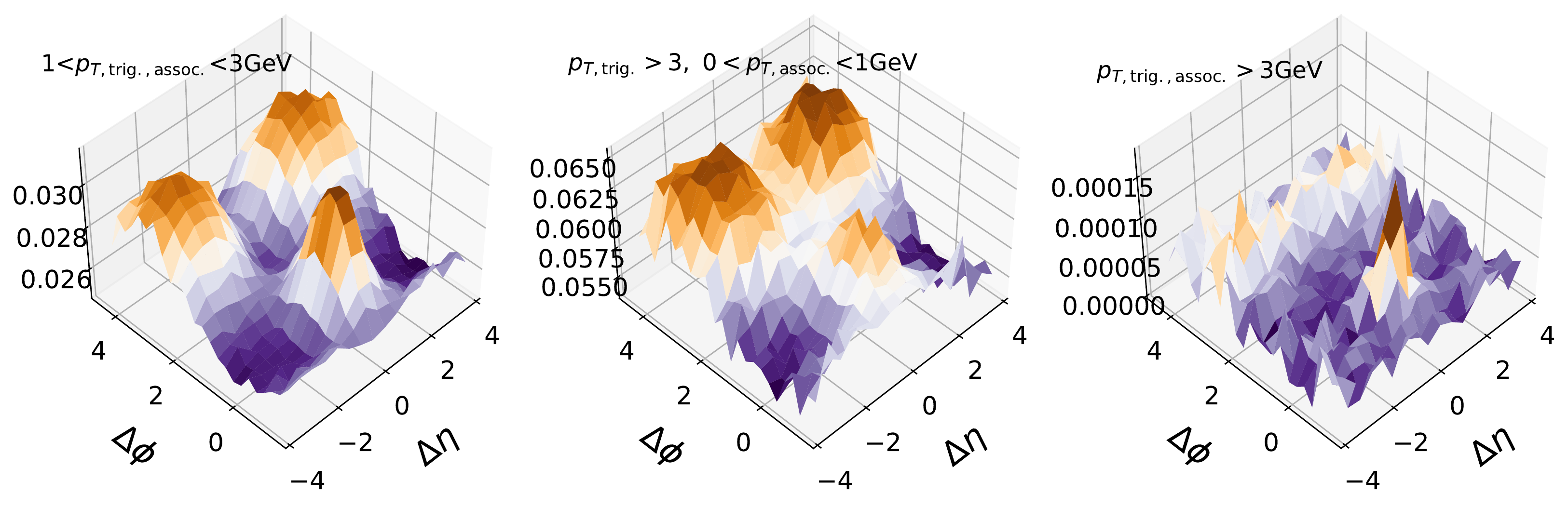}
    \caption{Kinematic cut dependence of two-particle correlations as functions of $\Delta \eta $ and $\Delta \phi$ obtained from charged particles produced within $|\eta|<2.4$ in minimum-bias (events with at least one charged particle produced within $|\eta|<1.0$) $p$+$p$ collisions at \snn[proton]=13 TeV.
    The two-particle correlation is obtained for associates limited to core components by triggering corona components. (Left) Transverse momentum ranges of both triggers and associates are $1.0<p_T<3.0$ GeV. (Middle, right) Trigger-particle kinematic cuts dependence of distribution of core components with regarding to corona components.}
    \label{fig:PP13_RIDGE_CORONATRIG_CORETRIG}
\end{figure}

Figure \ref{fig:PP13_RIDGE_CORONATRIG_CORETRIG} shows 
kinematic cut dependence of two-particle correlation function as functions of $\Delta \eta $ and $\Delta \phi$ obtained from charged particles produced within $|\eta|<2.4$ in minimum-bias (events with at least one charged particle produced within $|\eta|<1.0$) $p$+$p$ collisions at \snn[proton]=13 TeV.
The two-particle correlation is obtained for associates limited to core components by triggering corona components. Transverse momentum ranges of both triggers and associates are $1.0<p_T<3.0$ GeV in left figure. 
Comparing the results in left figure and Fig.~\ref{fig:PP13_RIDGE60-85Ntrk},
the double peak structure and the near-side peak are more obvious in the former result.
Thus, this means that within this $p_T$ range there is a tendency of that fluidization takes place along corona mini-jets when a mini-jet is produced
with a certain rapidity distance.
In the fluidization mechanism in Fig.~\ref{eq:four-momentum-deposition},
partons with high $p_T$ tend to keep their energy and momentum,
hence the probability that 
mini-jets produced with almost no rapidity gap are fluidized
is highly suppressed because the larger $p_T$ a particle obtains, the more mid-rapidity it is produced.
The middle and right figures show comparisons of trigger-particle kinematic cuts dependence of distribution of core components with regarding to corona components.
As one sees, the peaks seen in the left figure remain in the middle but disappear in the right.
Thus, the fluidization explained just above produces low $p_T$ core components.

\subsubsection{$Pb$+$Pb$ collision results}

\begin{figure}
    \centering
    \includegraphics[bb= 0 0 907 667, width=1.0\textwidth]{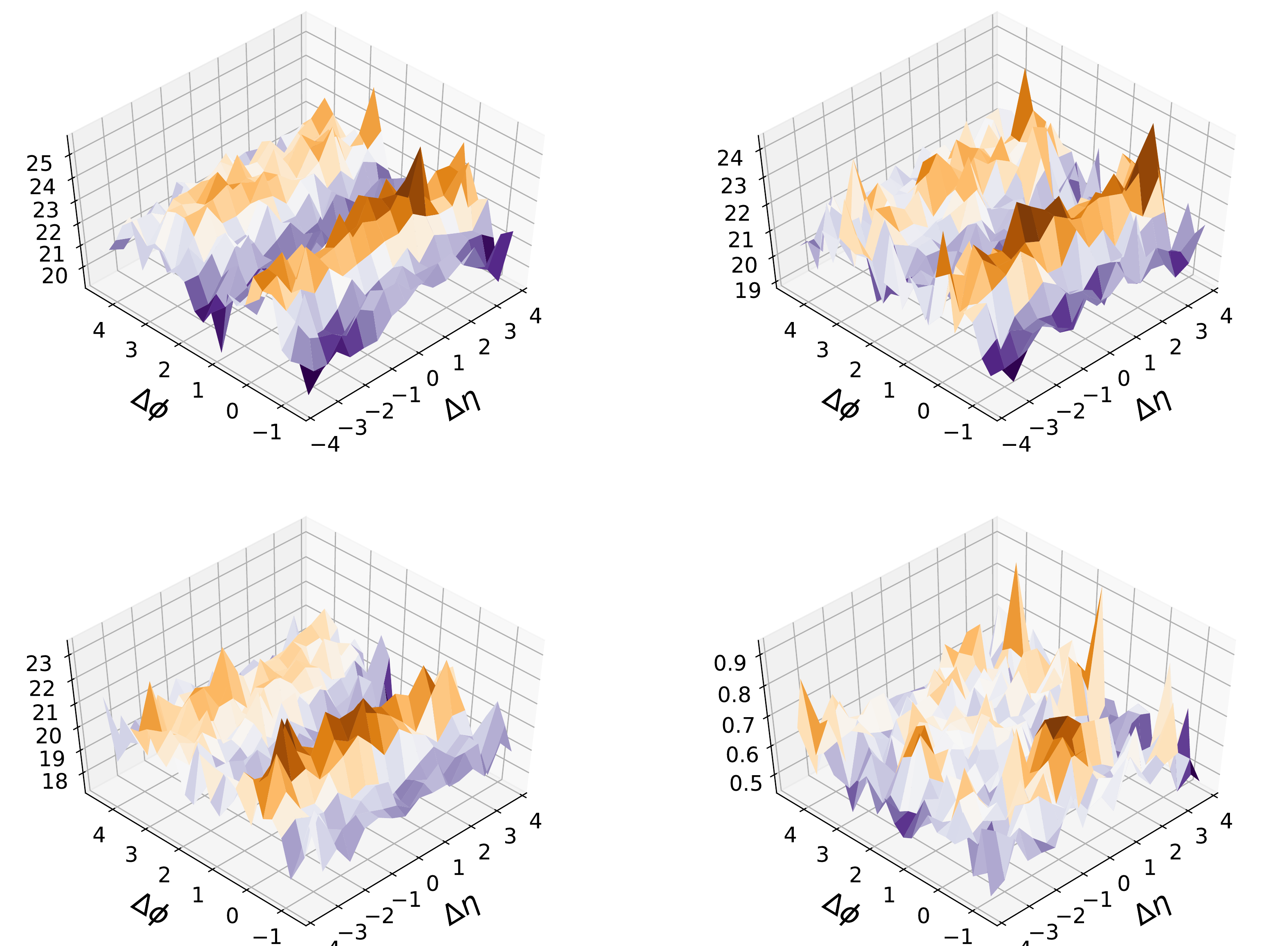}
    \caption{Two-particle correlations as functions of $\Delta \eta $ and $\Delta \phi$ obtained from charged particles produced within $|\eta|<2.4$ and with $3.0<p_T<3.5$ for triggers and $1.0<p_T<1.5$ for associates in 5-10\% centrality of $Pb$+$Pb$ collisions at \snn=2.76 TeV. Results from full simulation (left), simulations without hadronic rescatterings (right) are shown at the top, while breakdowns of produced particles from simulations without hadronic rescatterings into core (left) and corona (right) components are shown at the bottom. The same kinematic cuts used in the CMS paper \cite{CMS:2012xss} are adopted.}
    \label{fig:PBPB2760_RIDGE_5_10pct}
\end{figure}
\begin{figure}
    \centering
    \includegraphics[bb= 0 0 907 667, width=1.0\textwidth]{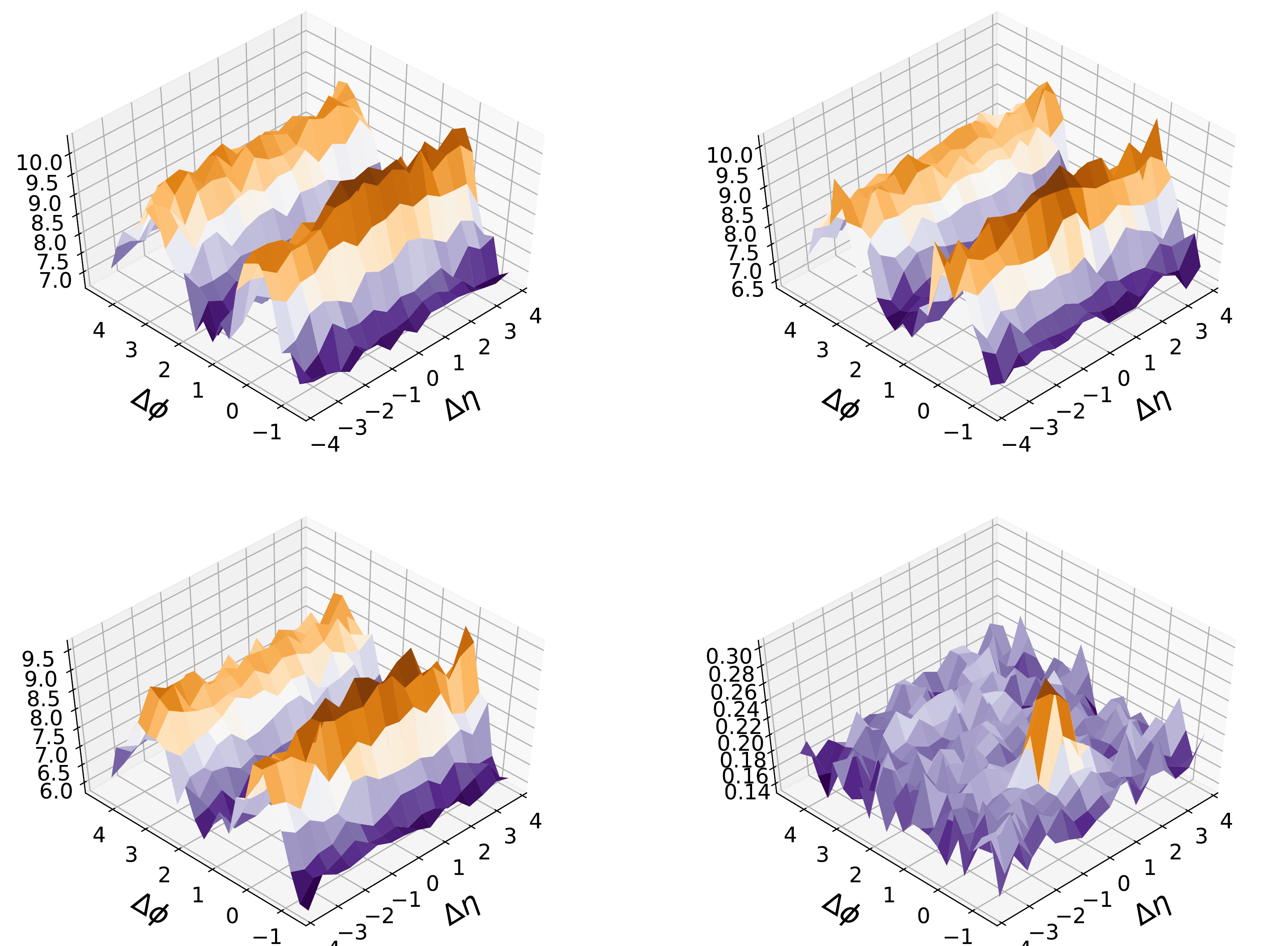}
    \caption{Two-particle correlations as functions of $\Delta \eta $ and $\Delta \phi$ obtained from charged particles produced within $|\eta|<2.4$ and with $3.0<p_T<3.5$ for triggers and $1.0<p_T<1.5$ for associates in 20-40\% centrality of $Pb$+$Pb$ collisions at \snn=2.76 TeV. Results from full simulation (left), simulations without hadronic rescatterings (right) are shown at the top, while breakdowns of produced particles from simulations without hadronic rescatterings into core (left) and corona (right) components are shown at the bottom. The same kinematic cuts used in the CMS paper \cite{CMS:2012xss} are adopted.}
    \label{fig:PBPB2760_RIDGE_20_40pct}
\end{figure}

Contrast to $p$+$p$ collisions, it has been understood and concluded that ridge structures observed in $Pb$+$Pb$ collisions originate from hydrodynamic responses against geometric anisotropy of initial profile of matter produced just after a collision.
Here, I discuss whether I can conclude so even in DCCI2 
where both core and corona components are incorporated and initial profiles are dynamically generated, which are different from other hydro-based dynamical models.
Figures \ref{fig:PBPB2760_RIDGE_5_10pct} and \ref{fig:PBPB2760_RIDGE_20_40pct} show two-particle correlations as functions of $\Delta \eta $ and $\Delta \phi$ obtained from charged particles produced within $|\eta|<2.4$ and with $3.0<p_T<3.5$ for triggers and $1.0<p_T<1.5$ for associates in 5-10\% and 20-40\% of centrality from $Pb$+$Pb$ collisions at \snn=2.76 TeV. The same kinematic cuts used in the CMS paper are adopted.
In contrast to $p$+$p$ collision results, ridge structure spanning long range in $\Delta \eta$ is seen in $Pb$+$Pb$ collisions, and that is more clearly seen in peripheral (20-40\%) compared to central (5-10\%) collisions.
This is consistent with the picture that almond-shaped matter generated early time in peripheral collisions produces more second order of anisotropic flow in momentum as a hydrodynamic response compared to central collisions.
It should be also noted that the ridge structure is seen only in core but in corona components.

%\begin{figure}
%    \centering
%    \includegraphics[bb= 0 0 907 667, width=1.0\textwidth]{chapters/figure/results/PBPB2760_RIDGE_40-60pct.pdf}
%    \caption{Caption}
%    \label{fig:my_label}
%\end{figure}
%\begin{figure}
%    \centering
%    \includegraphics[bb= 0 0 907 667, width=1.0\textwidth]{chapters/figure/results/PBPB2760_RIDGE_60-80pct.pdf}
%    \caption{Caption}
%    \label{fig:my_label}
%\end{figure}

\section{Evolution of transverse energy}
\label{subsection:Evolution_of_transverse_energy}
As shown in Fig.~\ref{fig:MULTIPLICITY_PP_PBPB}, I reproduced centrality dependence of charged particle multiplicity in $p$+$p$ and $Pb$+$Pb$ collisions within DCCI2.
Although the default \pythia8 (or Angantyr model in heavy-ion modes) works reasonably well, reproduction of multiplicity within DCCI2 can be attained only after the considerable change of a parameter $p_{\mathrm{T0Ref}}$ from its default value as mentioned in Sec.~\ref{subsection:Parameter_set_in_DCCI2}.
A nontriviality in DCCI2 stems from different competing mechanisms of how the transverse energy changes during the evolution of the system.
In this subsection, I discuss the effects of string formation/fragmentation and longitudinal $pdV$ work on the transverse energy and explain why I needed to change this parameter in DCCI2.

The transverse energy per unit rapidity $dE_{T}/d\eta$ is a basic observable in high-energy nuclear collisions and contains rich information on the dynamics of an entire stage of the reactions.
The transverse energy changes mainly in the initial and the expansion stages of the reactions.
In the initial stage, the two energetic hadrons and/or nuclei form color flux tubes between them as they pass through each other.
The chromo-electric and magnetic fields in the color flux tubes possess the energy originating from the kinetic energy of colliding hadrons or nuclei.
The decays of color flux tubes into partons and subsequent rescatterings among them are supposed to lead to the QGP formation \cite{Kajantie:1985jh,Gatoff:1987uf,Eskola:1992bd}.
Thus, how much energy is deposited in the reaction region is a fundamental problem of the QGP formation and depends on the initial dynamics of high-energy nuclear collisions.
On the other hand, in the expansion stage, the $pdV$ work associated with the longitudinal expansion after the QGP formation reduces the energy produced in the initial reaction region \cite{Gyulassy:1983ub,Ruuskanen:1984wv}.
The amount of reduced energy is sensitive to viscosity and other transport properties of the QGP~\cite{Gyulassy:1997ib}.
In addition, it can be possible to access the information on the initial state \cite{Eskola:1999fc}.
Therefore $dE_{T}/d\eta$ can be a good measure to scrutinize modeling in the initial and the expansion stages of the reactions.

In \pythia8, partons are first generated through hard scatterings and then, together with partons from initial and final state radiations, form hadron strings which eventually fragment into hadrons.
The transverse energy per unit rapidity of final hadrons is always larger than that of initially generated partons around midrapidity.
To understand this enhancement around midrapidity, suppose a hadronic string formed from a di-quark in the forward beam rapidity region and a quark in the backward beam rapidity region as an extreme case. 
Although the partons lie only around beam rapidity regions and the transverse energy of them vanishes around midrapidity, that string fragments into hadrons almost uniformly in rapidity space.
Thus, the emergence of the transverse energy at midrapidity is a consequence of the formation of a color string between such partons around beam rapidity.
Since parameters in \pythia8 are so tuned to reproduce the final hadron spectra, the initial parameters are highly correlated with parameters in the fragmentation as a whole.
Therefore the default parameter set should not be used if the subsequent hydrodynamic evolution, which reduces the transverse energy from its initial value of generated partons, is incorporated in DCCI2.

\begin{figure*}[htpb]
\begin{center}
\includegraphics[bb=0 0 1175 336, width=1.0\textwidth]{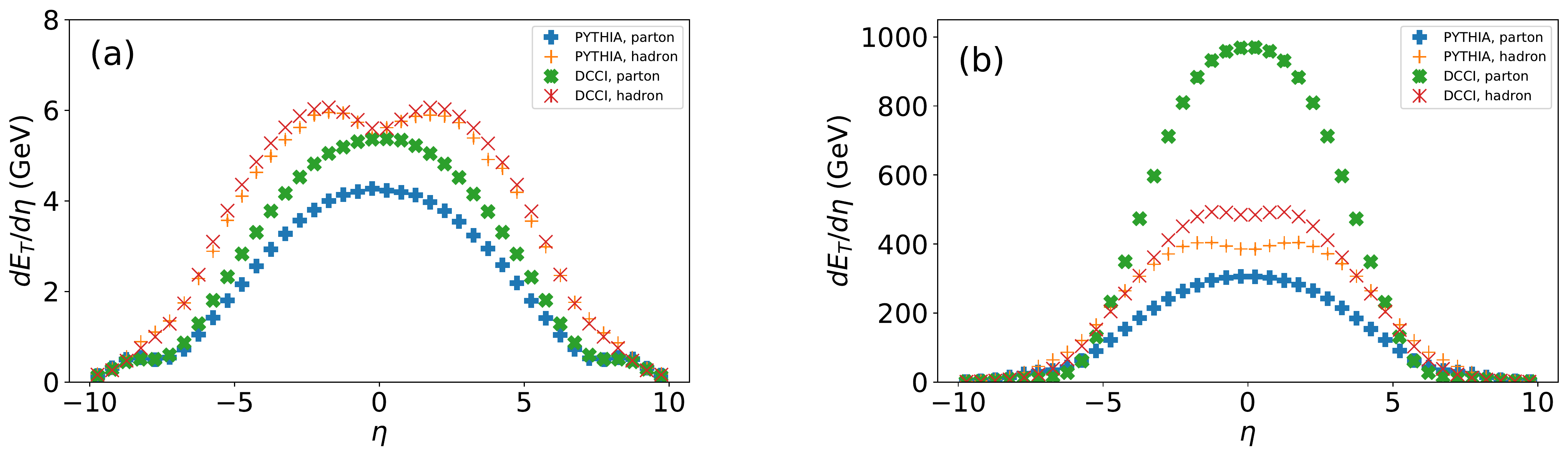}
\caption{(Color Online)
Pseudo-rapidity distribution of transverse energy in (a) minimum-bias $p$+$p$ collisions at $\sqrt{s} = 7$ TeV 
and (b) minimum-bias $Pb$+$Pb$ collisions at \snn = $2.76$ TeV from DCCI2 and \pythia8.
A comparison of transverse energy distribution between parton and hadron levels is made for results from DCCI2 and \pythia8.
Results from the parton level in DCCI2 (green crosses) and \pythia8 (blue pluses) and the ones from the hadron level (red crosses) in DCCI2 and \pythia8 (orange pluses) are shown for comparison. For $Pb$+$Pb$ collisions, \pythia8 Angantyr is used to obtain the results.
}
\label{fig:SHIFTOFDETDETA}
\end{center}
\end{figure*}

\begin{figure}
    \centering
    \includegraphics[bb=0 0 580 454, width = 0.5\textwidth]{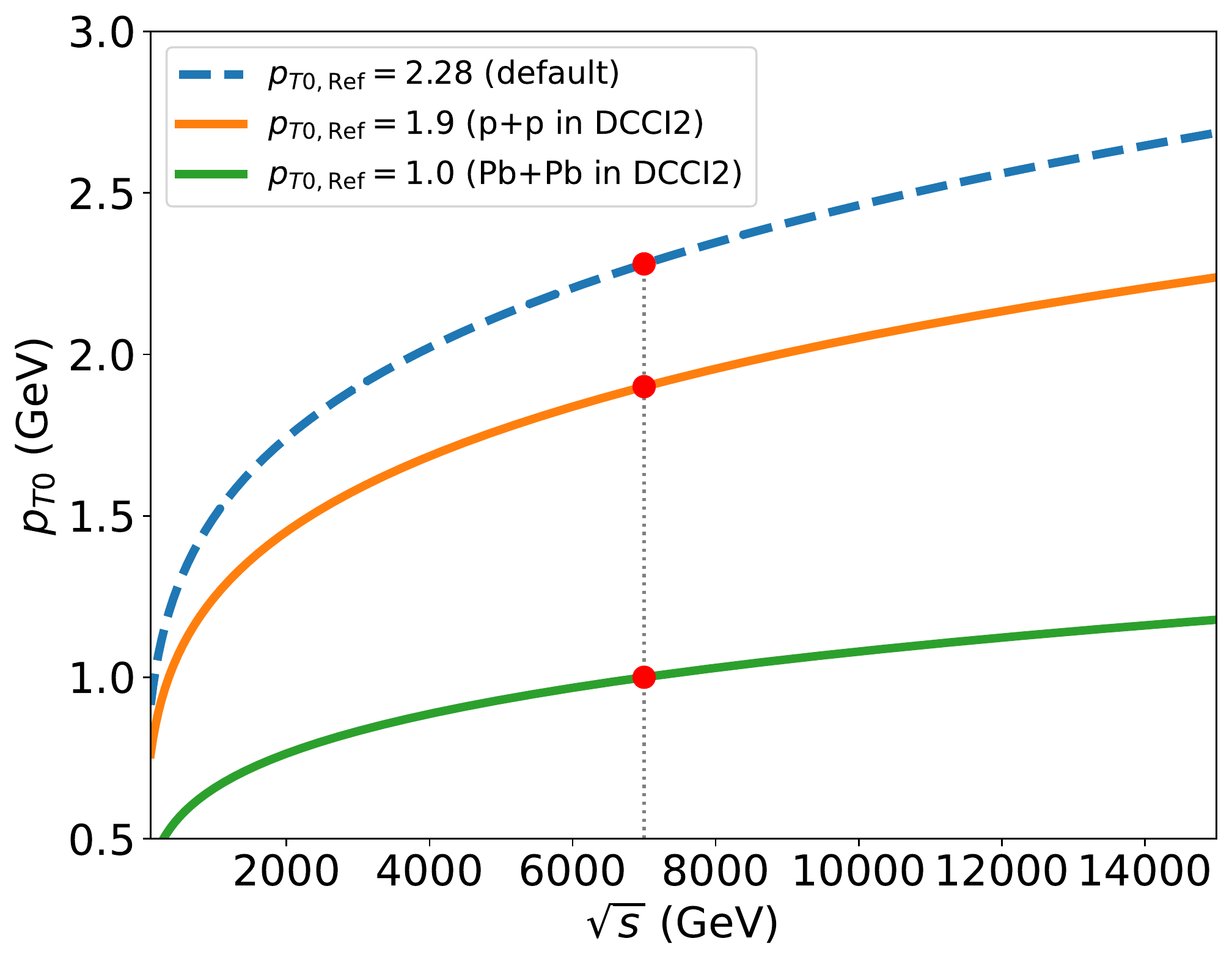}
    \caption{The infrared cut-off of the MPI, $p_{T0}$, as a function of center-of-mass energy of nucleon-nucleon collisions. Comparisons of $p_{T0}$ in default value used in \pythia8 (dashed blue line), $p$+$p$ (solid orange line), and $Pb$+$Pb$ collisions (solid green line) in DCCI2 are shown. The red circle is each $p_{T0}$ at \snn[proton] = 7 TeV, which corresponds to $p_{\mathrm{T0Ref}}$.}
    \label{fig:PT0REF_in_DCCI2}
\end{figure}

Figure \ref{fig:SHIFTOFDETDETA} shows $dE_{T}/d\eta$
of the initial partons before the string hadronization or the dynamical initialization, and that of the hadrons in the final state from both \pythia8 and DCCI2 in minimum-bias (a) $p$+$p$ collisions at $\sqrt{s} = 7 $ TeV and (b) $Pb$+$Pb$ collisions at \snn = $2.76$ TeV.
For the results from \pythia8 in Fig.~\ref{fig:SHIFTOFDETDETA} (a), 
the transverse energy per unit pseudorapidity, $dE_{T}/d\eta$, of the final hadrons is always larger than that of the initial partons in the whole rapidity region except around the beam rapidity.
Since the final hadron yield is dominated by the corona components in minimum-bias $p$+$p$ collisions in DCCI2, the above behavior is qualitatively the same as the ones from \pythia8.
The difference of the results between DCCI2 and \pythia8 appears in the absolute value of $E_T/d\eta$ in parton level,
which is a consequence of partial formation of QGP fluids. 
To obtain the same amount of the transverse energy in the final state in DCCI2, the transverse energy must be deposited initially $\approx 1.5$ times as large as that in \pythia8 at midrapidity to reconcile the reduction of transverse energy due to $pdV$ work.
This is exactly possible by considerably decreasing the parameter $p_{\mathrm{T0Ref}}$.

For the results from \pythia8 Angantyr in Fig.~\ref{fig:SHIFTOFDETDETA} (b),
the behavior of the transverse energy in $Pb$+$Pb$ collisions is again the same as in $p$+$p$ collisions.
In contrast, the transverse energy in initial state  from DCCI2 shows $\approx 3$ times as large as that in \pythia8 Angantyr.
The reason is the same as mentioned on the results of DCCI2 in $p$+$p$ collisions.
The decrease of the transverse energy due to $pdV$ work is clearly seen in $Pb$+$Pb$ collisions.

The parameter $p_{\mathrm{T0Ref}}$ is used to regulate infrared divergence of the QCD cross section.
The actual infrared cut-off is  $p_{T0}$ which is given as a function of center-of-mass energy of nucleon+nucleon collisions, \snn[proton].
The $p_{\mathrm{T0Ref}}$ gives the reference $p_{T0}$ value at a reference \snn[proton], which is 7 TeV by default.
Figure \ref{fig:PT0REF_in_DCCI2} shows the infrared cut-off $p_{T0}$ as a function of \snn[proton].
Comparisons of $p_{T0}$ in default value used in \pythia8 (dashed blue line), $p$+$p$ (solid orange line), and $Pb$+$Pb$ collisions (solid green line) in DCCI2 are shown. The red circle is each $p_{T0}$ at \snn[proton] = 7 TeV, which corresponds to $p_{\mathrm{T0Ref}}$.
The collision energy dependence of $p_{T0}$ is calm within the range of the LHC energy. Thus, one can see that $p_{\mathrm{T0Ref}}$ determines the overall factor of $p_{T0}$.

One can also interpret $p_{\mathrm{T0Ref}}$ as a parameter $p_{\perp\mathrm{min}}$ to separate soft from hard scales \footnote{In actual simulations in \pythia, a parameter $p_{\mathrm{T0Ref}}$ provides a scale to make a smooth turnoff of hard scattering rather than the sharp separation \cite{Sjostrand:2017cdm}.}, and controls the number of multiparton interactions in \pythia\ \cite{Sjostrand:1987su,Sjostrand:2017cdm}.
The smaller the separation scale $p_{\perp\mathrm{min}}$ is, the larger the number of multiparton interaction $\langle n_{\mathrm{MPI}}(p_{\perp\mathrm{min}}) \rangle = \sigma_{2\rightarrow 2}(p_{\perp\mathrm{min}})/\sigma_{\mathrm{nd}} $ is. 
Here $\sigma_{2 \rightarrow 2}$ and $\sigma_{\mathrm{nd}}$ are the perturbative QCD $2\rightarrow 2$ cross section and the inelastic non-diffractive cross section, respectively.
By increasing $\langle n_{\mathrm{MPI}}(p_{\perp\mathrm{min}}) \rangle$ as decreasing $p_{\mathrm{T0Ref}}$, initial partons are generated more and bring the sufficient amount of transverse energy in the final hadron state as shown in Fig.~\ref{fig:SHIFTOFDETDETA} (b). 
In this work, I use $p_{\mathrm{T0Ref}}=$1.8 and 0.9 for $p$+$p$ and $Pb$+$Pb$ collisions as mentioned in Sec.~\ref{subsec:ParameterDetermination}, which are smaller than the default values,
2.28 and 2.0 for
{\texttt{MultipartonInteractions:pT0Ref}} and {\texttt{SpaceShower:pT0Ref}} 
in \pythia.

So far, we have found that 
hydrodynamics and string fragmentation have different evolution of transverse energy. Thus, the multiplicity of final hadrons from such a two-component model is sensitive to a fraction of each component in a system.
To, at least, reproduce multiplicity in DCCI2, I need to change the parameter $p_{\mathrm{T0Ref}}$ from its default value.
However, as I mentioned in Sec.~\ref{subsec:ParameterDetermination} in chapter 3, the other parameter $\sigma_0$ in Eq.~\refbra{eq:four-momentum-deposition} 
has a non-trivial correlation with $p_{\mathrm{T0Ref}}$,
which means that I need to tune both the parameters at the same time. 
Suppose that one firstly tries to reproduce multiplicity by tuning $p_{\mathrm{T0Ref}}$. 
Since a small $p_{\mathrm{T0Ref}}$ gives a rise to the number of multiparton interactions and 
deposited transverse energy is enhanced, it affects to enhance final hadron multiplicity.
On the other hand,
since the number of initial partons produced in midrapidity increases,
this causes more fluidization in dynamical core--corona initialization.
Once a fraction of the core is enhanced, the multiplicity of final hadrons can also decrease since the initial transverse energy deposited in midrapidity region is used for $pdV$ work.
As a result of competition between two effects, multiplicity cannot linearly enhance or decrease by decreasing or increasing $p_{\mathrm{T0Ref}}$. 
Secondly, suppose that one tries to reproduce particle yield ratios as functions of multiplicity by tuning $\sigma_0$. 
Since changing $\sigma_0$ means changing a fraction of core and corona, final multiplicity is easily altered, too.
This is the reason why I need to fix both parameters by taking into account multiplicity and particle yield ratios at the same time.

Note that, if I made viscous corrections in the hydrodynamic evolution, the resultant change of  $p_{\mathrm{T0Ref}}$ from its default value could have been modest due to the less reduction of transverse energy \cite{Gyulassy:1997ib}, which is beyond the scope of the present paper, but which should be investigated in the future work.

A string melting version of A Multi Phase Transport (AMPT) model  \cite{Lin:2001zk,Lin:2004en} and the hydrodynamic models using it for generating initial conditions \cite{Pang:2012he, Xu:2016hmp} avoid this issue of the transverse energy in an ``ad hoc" way: The hadrons decaying from a  string are re-decomposed into their constitutive quarks and antiquarks, and then form high-energy density partonic matter. 
Although it is possible to count the energy stored along the string contrary to considering the generated partons directly, this prescription lacks gluons from melting strings.
Therefore, I do not pursue this idea in the present paper.

\chapter{Conclusion}
\thispagestyle{fancy}

\label{sec:CONCLUSION}
I studied the interplay between core and corona components establishing the DCCI2, which describes the dynamical aspects of core--corona picture under the dynamical initialization scheme from $p$+$p$ to $A$+$A$ collisions.
To develop the DCCI2, I put an emphasis on the reconciliation of open issues of dynamical models, mainly relativistic hydrodynamic models, toward a comprehensive description of a whole reaction of high-energy nuclear collisions.
One of the important achievements is to generate the initial profiles of hydrodynamics by preserving the initial total energy and momentum of the collision systems.
This is achieved by adopting hydrodynamic equations with source terms on initial partons obtained from \pythia8, one of the widely accepted general-purpose Monte-Carlo event generators.
Consequently, in addition to the equilibrated matter (core) described by relativistic hydrodynamics, I also consider the existence of non-equilibrated matter (corona) through dynamical initialization with the core--corona picture.

I have updated DCCI that I established in the Master's thesis.
The updates include sophistication of four-momentum deposition of initial partons in dynamical core--corona initialization, samplings of hadrons from hypersurface of fluids with \ISthreeD, hadronic afterburner for final hadrons from both fluids and non-equilibrated partons with a hadronic transport model \jam, and modification on color string structures of non-equilibrated partons due to the co-existence with fluids in coordinate space.

Discussion on the interplay between core and corona components is made once 
fixing major parameters so that our model reasonably describes multiplicity as a function of multiplicity or centrality class and omega baryon yield ratios to charged pions as functions of multiplicity.
First I extracted the fractions of core and corona components to the final hadron yields as functions of multiplicity and centrality classes. I found that, as increasing multiplicity, the core components become dominant at $\langle dN_{\mathrm{ch}}/d\eta \rangle_{|\eta|<0.5} \approx 20$, which roughly corresponds to the highest multiplicity classes in $p$+$p$ collisions at $\sqrt{s} = 7$ and $13$ TeV and $\approx80\%$ centrality class in $Pb$+$Pb$ collisions at $\sqrt{s_{NN}} = 2.76$ TeV.
%Corona component in low PT
Next, I showed the fractions of core and corona components in $p_T$ spectra with and without particle identification.  
In the charged particle spectra from minimum-bias $Pb$+$Pb$ collisions, the fraction of core components is dominant below $p_T\approx 5.5$ GeV, while that of corona components is dominant above that.
Interestingly, I found that there was an enhancement in the fraction of corona contribution with $R_{\mathrm{corona}} \approx 0.2$ at most in $p_T \lesssim 1$ GeV even in minimum-bias $Pb$+$Pb$ collisions. From this, the fraction of the corona contribution is anticipated to increase in peripheral collisions.
This brings up a problem in all conventional hydrodynamic calculations in which low $p_T$ soft hadrons are regarded purely as core components.
Since the fraction of each component would exist finite for a wide range of multiplicity and, as a result, there should be interplay between them, I suggest that 
both small colliding systems and heavy-ion collisions 
should be investigated in a unified theoretical framework by incorporating both core and corona components.
The PID and centrality classification revealed the dynamics in more detail.
In central collisions, the radial flow is also enhanced
while the $p_T$ integrated core components increase.
As a result, the core component dominance at intermediate $p_T$ (around $1<p_T<3$ GeV) is clearly pronounced at central collisions.
Because heavier particles are pushed towards higher $p_T$ due to the radial flow,
the core dominance at the intermediate $p_T$ and the corona enhancement at the very low $p_T$ become obvious in heavier particle spectra.
Significantly,
the fraction of core components is almost $100\%$ from $0$-$60\%$ in intermediate $p_T$ for proton spectra, and the corona component reaches around $50\%$ at the very low $p_T$ from central to peripheral collisions.

It is known that transverse momentum spectra solely from hydrodynamics or hybrid (hydrodynamics followed by hadronic cascade) models do not perfectly reproduce the experimental data below $p_T \approx 0.5$ GeV \cite{Abelev:2013vea},
although hydrodynamics is believed to provide a better description in the low $p_T$ region in general. 
To investigate if the deviation between pure hydrodynamic results and the data in the low $p_T$ region can be filled with the corona components,
I made comparisons of $p_T$ spectra at the very low $p_T$ (smaller than $p_T \approx 1$ GeV) between experimental data and DCCI2 results from full simulations and those only from core components with hadronic rescatterings.
The slopes of $p_T$ spectra at the very low $p_T$ observed in the experiment cannot be reproduced only with core components with hadronic rescatterings, and the full simulation results show a better description of the slopes by including corona component. 
Therefore I conclude that there is possibly  the ``soft-from-corona'' components in heavy-ion collisions,
and which would be indispensable to be considered for an adequate interpretation of the particle productions obtained from experimental data.

To investigate the effects of co-existence of core and corona components on observables, I showed $\langle p_T \rangle$ and $v_{2}\{2\}$ as functions of $N_{\mathrm{ch}}$. 
In particular, in $Pb$+$Pb$ collisions, 
I found that the finite contribution of corona components at midrapidity gives a certain correction to the results obtained purely from core components, which is described by hydrodynamics.
The correction is $\approx 5$-$11$\% for $\langle p_T \rangle$ below $N_{\mathrm{ch}}\approx 200$,
while it is 
$\approx 15$-$30$\% for $\vtwtw$ below $N_{\mathrm{ch}}\approx 400$.
The former correction leads to the reasonable agreement of $\langle p_T \rangle$ with the experimental data. 
This suggests that one needs to incorporate corona components in hydrodynamic frameworks to extract transport coefficients from comparisons with experimental data.
%Mention the C24 obtaind from PBPB
In the results of $\ctwofour$ as a function of $\Nch$ obtained from $Pb$+$Pb$ collisions,
I saw that the $\ctwofour$ obtained from core components is diluted due to the existence of the corona components, while the corona components show the zero-consistent $\ctwofour$.
This result indicates that anisotropic flows obtained from experimental data cannot be directly compared with pure hydrodynamic calculations if there is a certain portion of non-equilibrium components.

I also explored the effects of radial flow based on violation of $m_T$ scaling with hadron rest mass by classifying events into high- and low-multiplicity ones.
Noteworthy, I found that it is difficult to discriminate the radial flow originating from hydrodynamics from collectivity arising from color reconnection in \pythia8.

%Conclude ridge results.
The investigation of ridge structure within DCCI2 was performed to reveal its origin.
The analysis of $p$+$p$ collision results is performed with the event classification used in the ALICE and CMS experiment, respectively.
With the method used in the ALICE experiment,
I saw that the ridge-like structure appears in high-multiplicity events, and it originates from the core components.
On the other hand, such a ridge-like structure was not seen with the method used in the CMS experiments, which is inconsistent with the CMS experimental results.
I also checked that there is a strong momentum correlation between core and corona components,
which would be a consequence of the dynamical initialization.

I finally discussed the evolution of transverse energy in the DCCI2.
In string fragmentation, final transverse energy is larger than initial transverse energy as producing hadrons around midrapidity. While in hydrodynamics, transverse energy just decreases from its initial value during the evolution due to the longitudinal $pdV$ work.
To obtain the same amount of transverse energy in the final state in DCCI2 with default \pythia8 Angantyr in minimum-bias $Pb$+$Pb$ collisions, it is necessary to have three times larger initial transverse energy than the one of default \pythia8 Angantyr.

For a more quantitative discussion on transport properties of the QGP fluids, I admit an absence of viscous corrections to fully equilibrium distribution in our analysis.
I leave this as one of our future works.
Nevertheless, I emphasize that the corrections from corona components mean the ones from ``far from'' equilibrium components which should exist nonetheless and would more significantly affect the final hadron distributions than the viscous corrections.

%Detailed analyses of centrality dependent particle identified $p_T$ spectra from DCCI2, which require high statistics, and its comparison with the experimental data will be made in a future publication.

With this model, I anticipate that it would be interesting to explore 
planned $O$+$O$ collisions at LHC \cite{Citron:2018lsq} since the collision system can provide data around the ``sweet spot'' 
in which 
the core components are to be dominant, and yet corrections from corona components cannot be ignored at all 
\cite{Brewer:2021kiv}.
In addition, investigation of strangeness enhancement in forward or backward regions might provide some insights into ultra-high-energy cosmic ray measurements \cite{Anchordoqui:2016oxy,Baur:2019cpv}.
Incorporation of a dynamical description of kinematic and chemical pre-equilibrium stage \cite{Kurkela:2018wud,Kurkela:2018oqw} and investigation of medium modification of jets \cite{Tachibana:2019hrn,Luo:2021iay} are in our interests as well.
I leave the discussion on those topics as future work.

\appendix
%\chapter{Quantum Chromo Dynamics (QCD)}
%\input{chapters/appendixA}
%
%\chapter{Coordinates}
%\input{chapters/appendixB}
%
%
\chapter{Analysis methods}
\thispagestyle{fancy}

Details of analysis, which I skipped explanations in the body of this thesis, are explained in this Appendix.

\section{Two-particle correlation}
\label{sec:APPENDX_TwoParticle}

Two-particle correlation function is a distribution of distance of two-particle pairs in the momentum space consist of one {\it{trigger}} particle and another {\it{associate}} particle.
Definition of the two-particle correlation function introduced in Sec.~\ref{subsec:Collectivity} and shown in Sec.~\ref{subsec:RESULT_RidgeStructure}
is given by
\begin{align}
    C(\Delta \eta, \Delta \phi) = B(0,0) \frac{S(\Delta \eta, \Delta \phi)}{B(\Delta \eta, \Delta \phi)},
    %\frac{1}{N_{\mathrm{trig}}} \frac{d^2N_{\mathrm{pair}}}{d\Delta \eta d\Delta \phi} = B(0,0) \frac{S(\Delta \eta, \Delta \phi)}{B(\Delta \eta, \Delta \phi)},
    \label{eq:TwoParticleCorrelationRidge}
\end{align}
where $S(\Delta \eta, \Delta \phi)$ and $B(\Delta \eta, \Delta \phi)$ are raw two-particle correlations obtained within the same events and constructed from 
randomly chosen two events, respectively.
It should be noted that there are several definitions besides Eq.~\refbra{eq:TwoParticleCorrelation} used in the CMS collaboration.
However, the idea for those are the same:
extraction of two-particle correlation within each event contrast to backgrounds.

The raw two-particle correlations within the same events are obtained as
\begin{align}
    S(\Delta \eta, \Delta \phi) =  \frac{1}{N_{\mathrm{trig}}} \frac{d^2N_{\mathrm{pair}}^{\mathrm{same}}}{d\Delta \eta d\Delta \phi},
    \label{eq:Same_TwoParticle}
\end{align} 
where $N_{\mathrm{trig}}$ is the number of trigger particles, and
$d^2N_{\mathrm{pair}}^{\mathrm{same}}/d\Delta \eta d\Delta \phi$ is the distribution of two-particle pair obtained within a single event as a function of $\Delta \eta$ and $\Delta \phi$.
For example, if kinematic cut is different for trigger and associate particles, $N_{\mathrm{pair}}^{\mathrm{same}}$ is $N_{\mathrm{trig}}\times N_{\mathrm{assoc}}$ where $N_{\mathrm{assoc}}$ is the number of associate particles.
The two-particle correlations constructed randomly chosen from two events are obtained as 
\begin{align}
    B(\Delta \eta, \Delta \phi) =  \frac{1}{N_{\mathrm{trig}}} \frac{d^2N_{\mathrm{pair}}^{\mathrm{mix}}}{d\Delta \eta d\Delta \phi}.
    \label{eq:Mix_TwoParticle}
\end{align}
Here, $N_{\mathrm{pair}}^{\mathrm{mix}}$ is the number of two-particle pairs
where associate particles are randomly chosen pair-by-pair within the same centrality/multiplicity class.
Note that $N_{\mathrm{pair}}^{\mathrm{mix}}$ can differ depending on how many the dummy associate is chosen.
However, one can choose a certain proper number of dummies here
because $B(0,0)$ in Eq.~\refbra{eq:TwoParticleCorrelation} can re-scale obtained $B(\Delta \eta, \Delta \phi)$.
In the results shown in Sec.~\ref{subsec:RESULT_RidgeStructure},
20 dummy associates are picked up.
It should also be noted that $N_{\mathrm{trig}}$ is the same as in Eq.~\refbra{eq:Same_TwoParticle}.

\begin{figure}
    \centering
    \includegraphics[bb=0 0 740 721, width=0.49\textwidth]{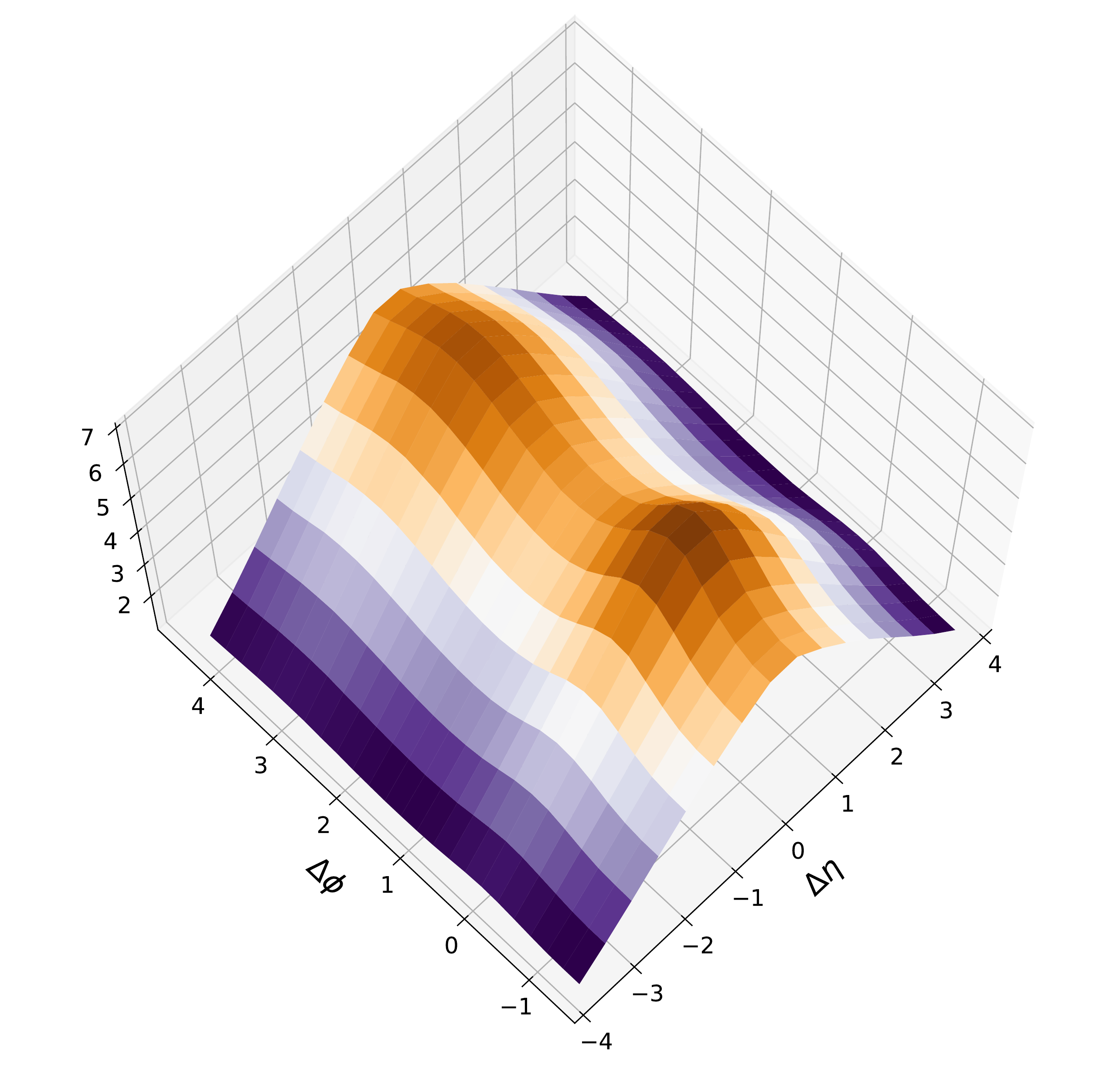}
    \includegraphics[bb=0 0 740 721, width=0.49\textwidth]{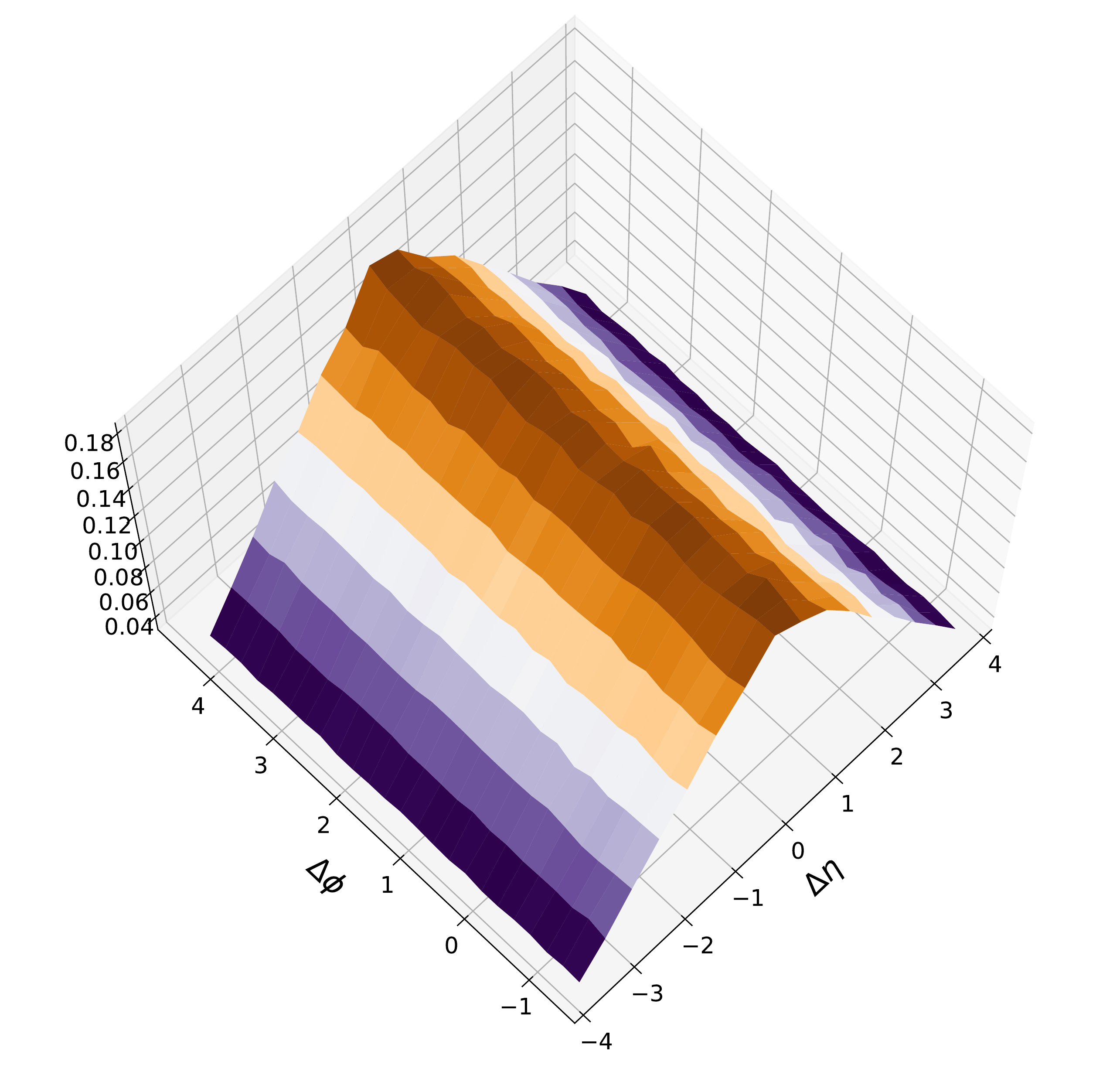}
    \includegraphics[bb=0 0 740 721, width=0.49\textwidth]{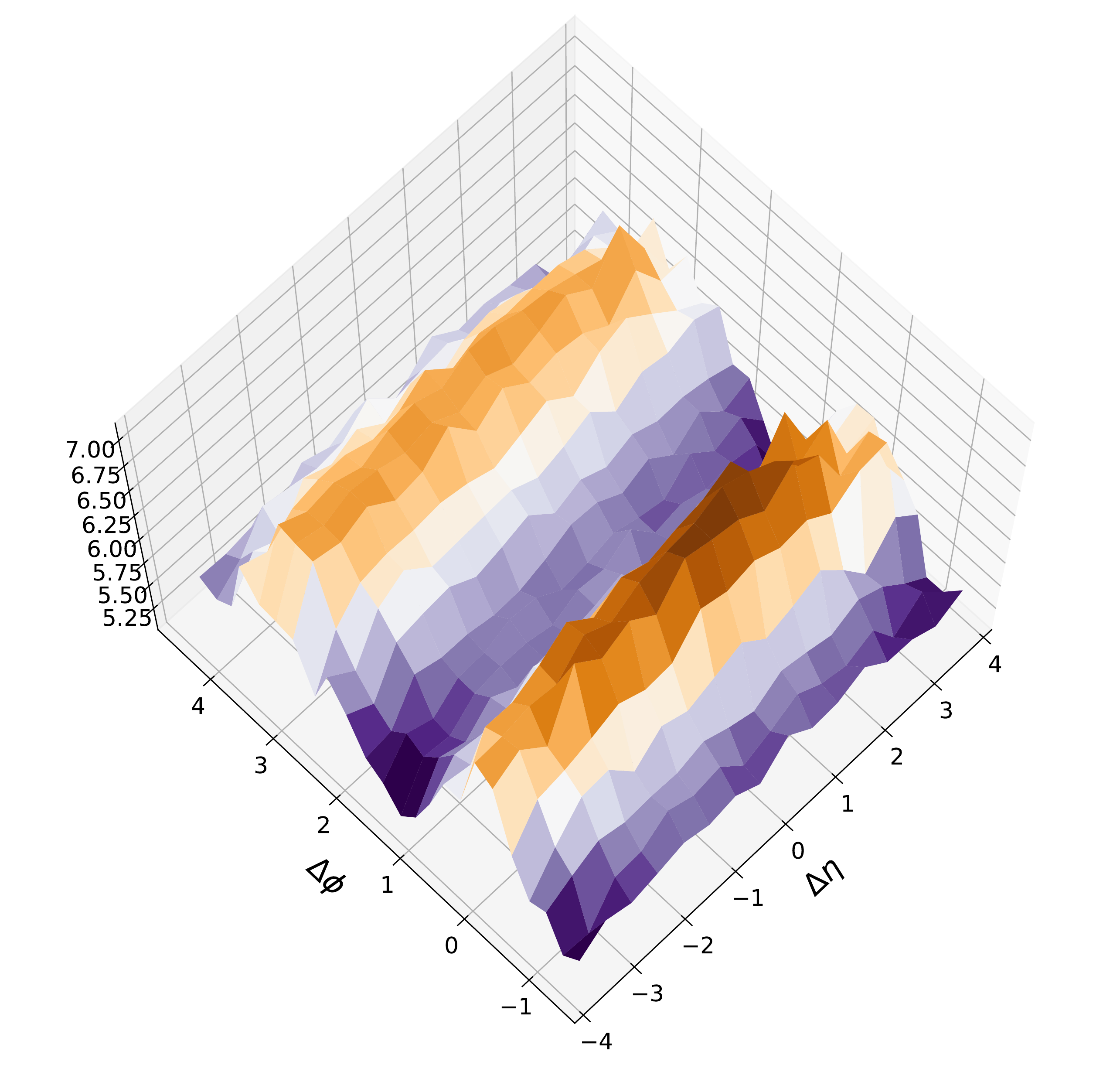}
    \caption{Top: examples of $S(\Delta \eta, \Delta \phi)$ (left) and $B(\Delta \eta, \Delta \phi)$ (right) obtained from Pb+Pb collisions. Bottom: Two-particle correlation function, $C(\Delta \eta, \Delta \phi)$, obtained from above $S(\Delta \eta, \Delta \phi)$ and $B(\Delta \eta, \Delta \phi)$.}
    \label{fig:ExampleSB}
\end{figure}

Figure \ref{fig:ExampleSB} shows examples of $S(\Delta \eta, \Delta \phi)$, $B(\Delta \eta, \Delta \phi)$, and two-particle correlation function, $C(\Delta \eta, \Delta \phi)$, obtained with the $S(\Delta \eta, \Delta \phi)$ and $B(\Delta \eta, \Delta \phi)$ in Pb+Pb collisions.
One would notice the meaning of subtracting background from this.
In high-energy nuclear collisions, because particles are produced around midrapidity, there should be an obvious strong correlation in midrapidity.
In this method, by dividing the raw two-particle correlation with background, 
such an obvious correlation in longitudinal direction is suppressed 
to clearly see the azimuthal anisotropy.

\section{Subevent method}
\label{sec:APPENDX_SubEvent}

The {\it{subevent method}} is proposed to subtract the back-to-back dijet correlation
\cite{Jia:2017hbm}.
It can be expected that the dijets tend to have short range correlation in longitudinal direction.
Under the subevent method, 
one first prepares some pseudorapidity ranges which have a certain longitudinal distance for each.
Then, in one single event, one can obtain subsets of particles produced in 
each of those pseudorapidity range, and the subsets of particles are called subevents.
Multi-particle correlations are calculated 
by picking up particles from those subevents.

In this section,
I explain some details of the calibration of the 2-subevent method for 2-particle correlation and the 3-subevent method for 4-particle correlation
which are calculated in this thesis.
For more details, see {\it{i.e.,}} Ref.~\cite{Jia:2017hbm,Gajdosova:2018zrt}.

\subsubsection{2-subevent method for 2-particle correlation}
In the 2-subevent method, one usually impose {\it{eta gap}}, $\Delta \eta$,
between the 2 subevents.
For instance, within a certain acceptance $|\eta|<\eta_{\mathrm{accept.}}$, the 2 subevents are obtained in $-\eta_{\mathrm{accept.}}<\eta<- \eta_{\mathrm{gap}}/2 $ and $ \eta_{\mathrm{gap}}/2 <\eta<\eta_{\mathrm{accept.}}$ with an eta gap $\Delta \eta>\eta_{\mathrm{gap}}$.
For the later explanation, I name these two subsets as $A$ and $B$.

Equation \refbra{eq:TwoParticleCorrelation} in Sec.~\ref{subsec:Collectivity} of chapter 1 shows the definition of 2-particle correlation.
On the other hand, 
the 2-particle correlation with 2-subevent method is obtained as,
\begin{align}
\label{eq:TwoParticleCorrelation_2sub}
    \langle 2 \rangle = \frac{1}{N_A N_B}\sum_{i_A j_B}^{N_A,N_B} e^{in(\phi_{i_A} -\phi_{j_B} )} = \frac{Q_{n,A}Q_{n,B}^*}{N_AN_B}, 
\end{align}
where $N_A$ and $N_B$ are the number of particles produced in each subevent of $A$ and $B$.
On the other hands, $Q_{n,A}$ and $Q_{n,B}$ are the Q-vector defined in Eq.~\refbra{eq:Qvector} and obtained only within a single subevent $A$ and $B$, respectively.
The change of the definition from Eq.~\refbra{eq:TwoParticleCorrelation} to Eq.~\refbra{eq:TwoParticleCorrelation_2sub} can be clearly understood once one thinks about its meaning:
in the 2-subevent, the number of particle pairs becomes $N_A N_B$ and the subtraction of self correlation is no longer needed because the 2-particle correlation is calculated by picking up 2-particle from each different subevent.

\subsubsection{3-subevent method for 4-particle correlation}
Under the 3-subevent for 4-particle correlation, the four particles are picked up from 3 rapidity ranges dividing the acceptance range into 3, for instance, $-\eta_{\mathrm{accept.}}<\eta<\eta_{B}$, $\eta_B<\eta<\eta_C$, and $\eta_C <\eta<\eta_{\mathrm{accept.}}$, 
where $\eta_B<0<\eta_C$.
From the backward (negative) to forward (positive) rapidity, I name those subevents as $B$, $A$, and $C$.

The 4-particles are picked up as follows. Two of four particles are picked up from the subevent $A$ located at the center and one particle is picked up from each $B$
and $C$ subevent, respectively. 
Here, although there are two ways to calculate 4-particle correlation from 3-sub event
considering the symmetry,
I adopt the one which can reduce the effects of dijet correlations \cite{Jia:2017hbm}.
The explicit form is expressed as
\begin{align}
    \langle 4 \rangle &= \frac{1}{N_A(N_A-1)N_BN_C} \sum_{i_A, j_A, k_B, l_C}^{N_A, N_A, N_B, N_C} e^{in(\phi_{i_A}+\phi_{j_A}-\phi_{k_B}-\phi_{l_C})} \nonumber \\
    &=\frac{(Q^2_{n,A}-Q_{2n,A})Q_{n,B}^*Q^*_{n,C}}{N_A(N_A-1)N_BN_C}.
\end{align}
For more discussions on multi-particle correlation, see Ref.~\cite{Bilandzic:2012wva}.

\chapter{String modification algorithm}
\thispagestyle{fancy}

In Sec.~\ref{subsec:STRING_CUTTING} in Chapter 2, I explained the color string modification due to the spacial overlap between strings and QGP fluids. 
Here, I would like to describe the algorithm used in the color string modification, which is the string tracing and string cutting.

\section{String tracing}
\label{sec:AppendixStringTracing}

\begin{figure}[htpb]
    \centering
    \includegraphics[bb=0 0 960 540, width=0.8\textwidth]{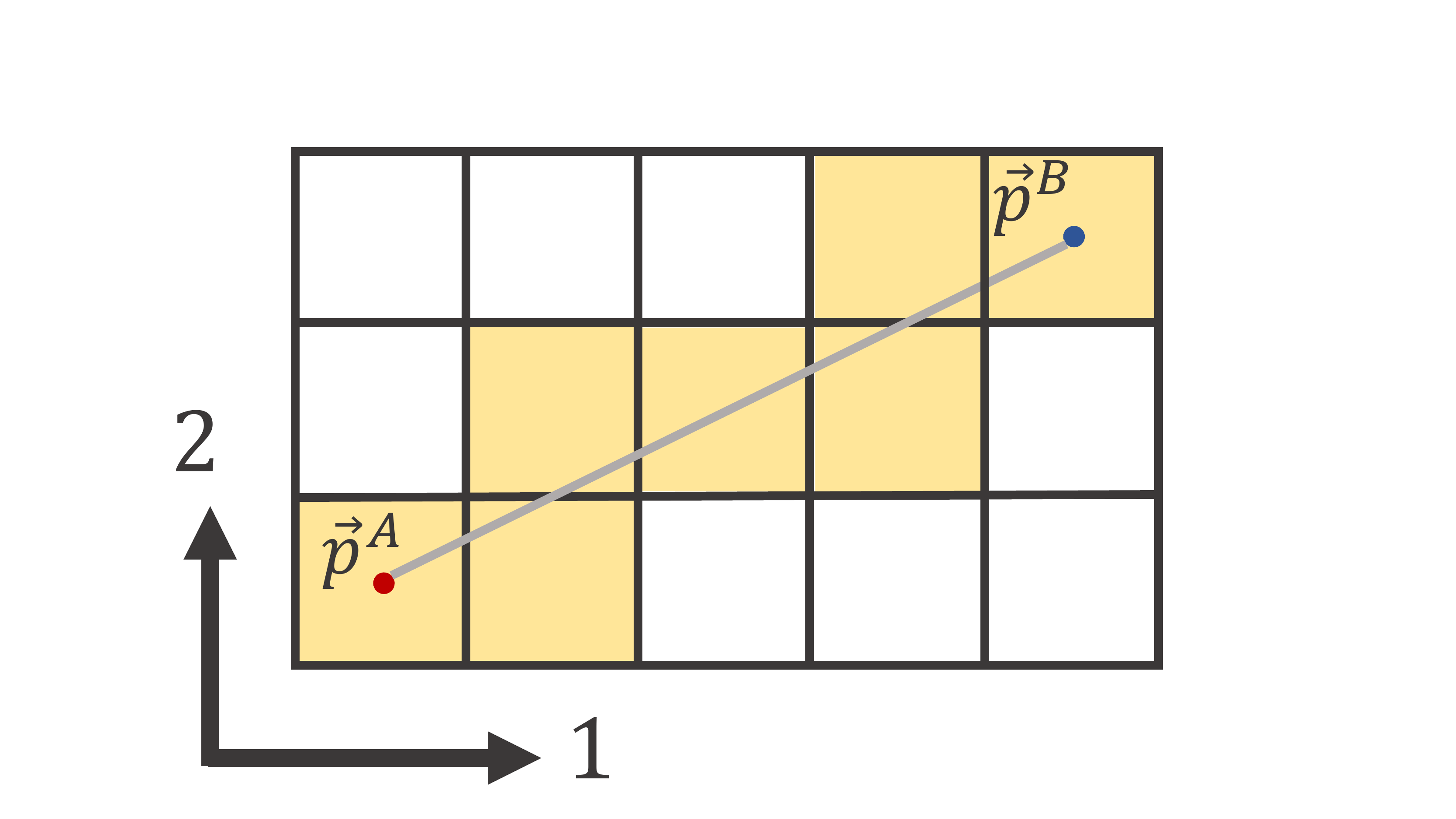}
    \caption{A color string consisting of a quark and an anti quark (red and blue) on a 2-dimensional discretized space.
    Cells that have spacial overlap with the string are highlighted with yellow.
    }
    \label{fig:Lattice}
\end{figure}

As the name tells, this algorithm is used to trace color strings. Under the string tracing, I need to check if there exist any crossing points between color strings and hypersurfaces.
In Fig.~\ref{fig:Lattice}, I show an example of the situation to deal with.
A color string consisting of a quark and an anti quark on a 2-dimensional plane of discretized space is shown.
The discretized space corresponds to the 
hydrodynamic-simulation space.
While the actual simulation is performed 
in the Milne coordinate, for space component $(x, y, \eta_s)$,
I will explain the algorithm with a general 3-D coordinate of cell number, {\it{i.e.,}} direction 1, 2, and 3,
because the algorithm is constructed in terms of cell number rather than coordinate with physical scale.
The two leading quark and anti-quark are located at 
$\vec{p}^A$ and $\vec{p}^B$, where these are vectors expressed with cell numbers.
The color string is linearly stretched in-between the leading quark and anti-quark.
It does not matter which of $\vec{p}^A$ and $\vec{p}^B$ corresponds to the quark and anti-quark,
so, hereafter, I call them just particle A and B.
Here, cells that have spacial overlap with the string are highlighted with yellow.
Thus, to look for crossing points between the color string and QGP fluids, one is required to check temperature of each yellow cell.

The problem we need to solve is 
how we trace these cells with a reasonable computational time.
The computational cost on this algorithm matters because the number of and the length of strings generated in one heavy-ion collision event are enormous.
The answer is as follows:
one needs to trace a color string
one by one in the smallest-slope direction 
of the vector connecting the particle A and B.
I will describe details in the followings.

Under this algorithm, there is one assumption which makes the calculation easy:
all partons are assumed to be located at the center of each cell and a color string stretches in between partons, which is not the case, for example in Fig.~\ref{fig:INIT_STRING},
because all partons can be at any point within the simulation space of hydrodynamics.
However, since we take small cell widths as much as possible, the difference can be small enough to be neglected.

\subsubsection{Determination of the direction to step}
In the hydrodynamic simulation,
it would be a usual case that
one numbers each cell as $0, 1, 2, 3 \cdots, N_i$ where $N_i-1$ is the number of discretized cells in $i$-direction.
Under this algorithm,
I treat these cell number as the coordinate.
Here I define that the center of a cell 
located at the $k_1$-th in direction 1, 
$l_2$-th in direction 2, 
and $m_3$-th in direction 3 corresponds to
a coordinate $(k_1, l_2, m_3)$.
In Fig.~\ref{fig:Lattice}, 
for instance, $\vec{p}^A$ and $\vec{p}^B$ are expressed as
\begin{align}
    p^A_1 = 0.0, \enskip p^A_2 = 0.0, \enskip p^A_3 = 0.0,
\end{align}
and 
\begin{align}
    p^B_1 = 4.0, \enskip p^B_2 = 2.0, \enskip p^B_3 = 0.0.
\end{align}

First one needs to find a cell that each parton is located.
The set of cell number is obtained in the actual simulation as
\begin{align}
    k_1 &= \mathrm{floor} (x/\Delta x) + N_x, \\
    l_2 &= \mathrm{floor} (y/\Delta y) + N_y, \\
    m_3 &= \mathrm{floor} (\eta_s/\Delta \eta_s) + N_{\eta_s}, \\
\end{align}
where $\mathrm{floor}(x)$ returns the largest integer that is not greater than $x$,
$\Delta x$, $\Delta y$, and $\Delta \eta_s$ are cell widths determined in Tab.~\ref{tab:PARAMETERSET}.
Note that the origin of the coordinate $(x,y,\eta_s)$ is set at the center of the simulation space.

Here, the unit vector, $\vec{r}$, starting from the particle A and pointing the particle B is obtained as
\begin{align}
    \vec{r} = \frac{\vec{p}^B-\vec{p}^A}{|\vec{p}^B-\vec{p}^A|}.
    \label{eq:rAB}
\end{align}
Here, I name the direction
in which a string has the smallest slope as
$d_{\mathrm{max}}$ for the later explanation.
One can determine $d_{\mathrm{max}}$ as 
\begin{align}
    |r_{\mathrm{d_{\mathrm{max}}}}| = \max (|r_1|, |r_2|, |r_3|).
    \label{eq:dmax}
\end{align}
In the case of Fig.~\ref{fig:Lattice} where 
$|r_3|=0$ is assumed, one can get
$d_{\mathrm{max}}=1$.

\begin{figure}[htpb]
    \centering
    \includegraphics[bb=0 0 960 540, width=0.48\textwidth]{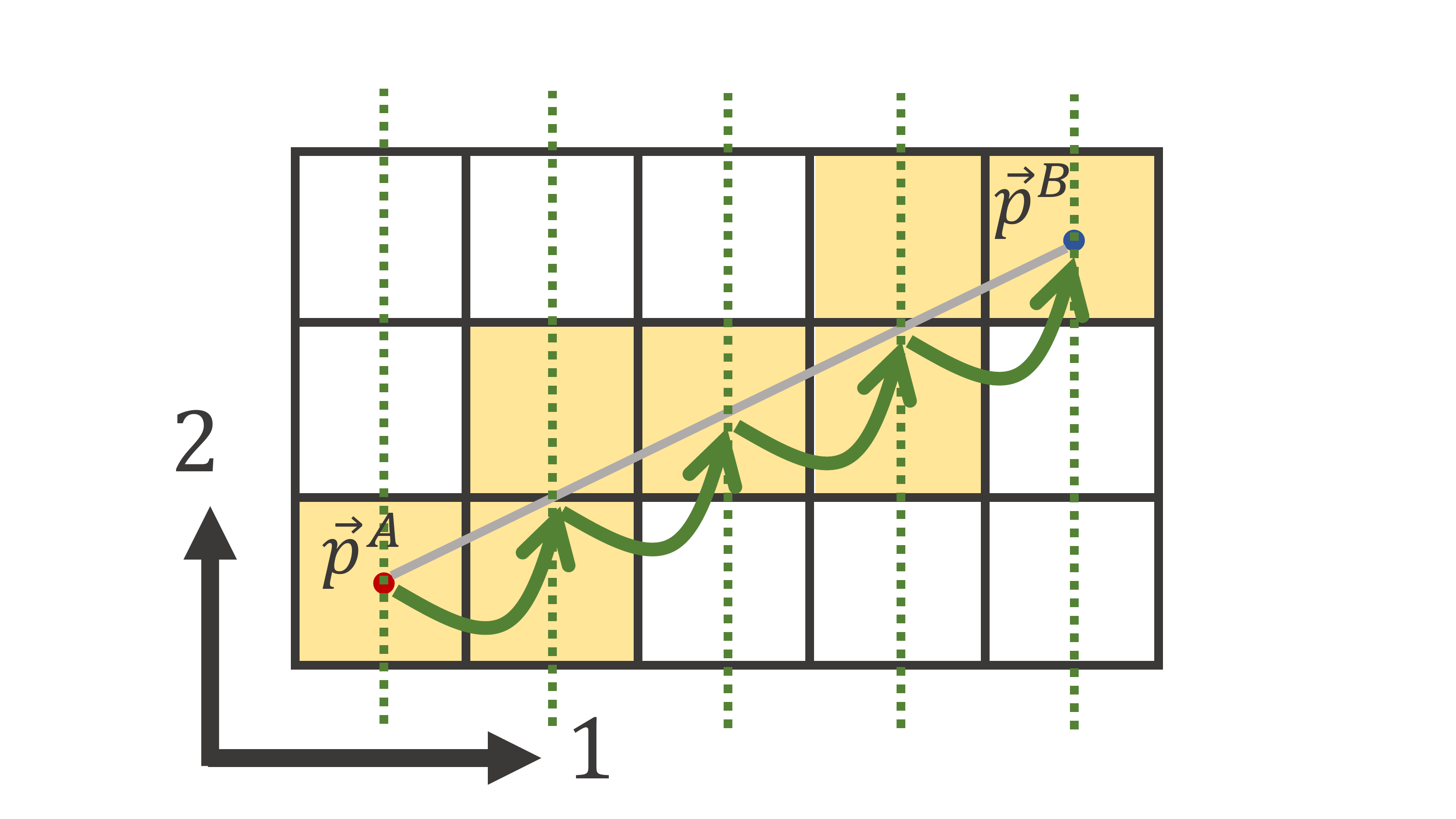}
    \includegraphics[bb=0 0 960 540, width=0.48\textwidth]{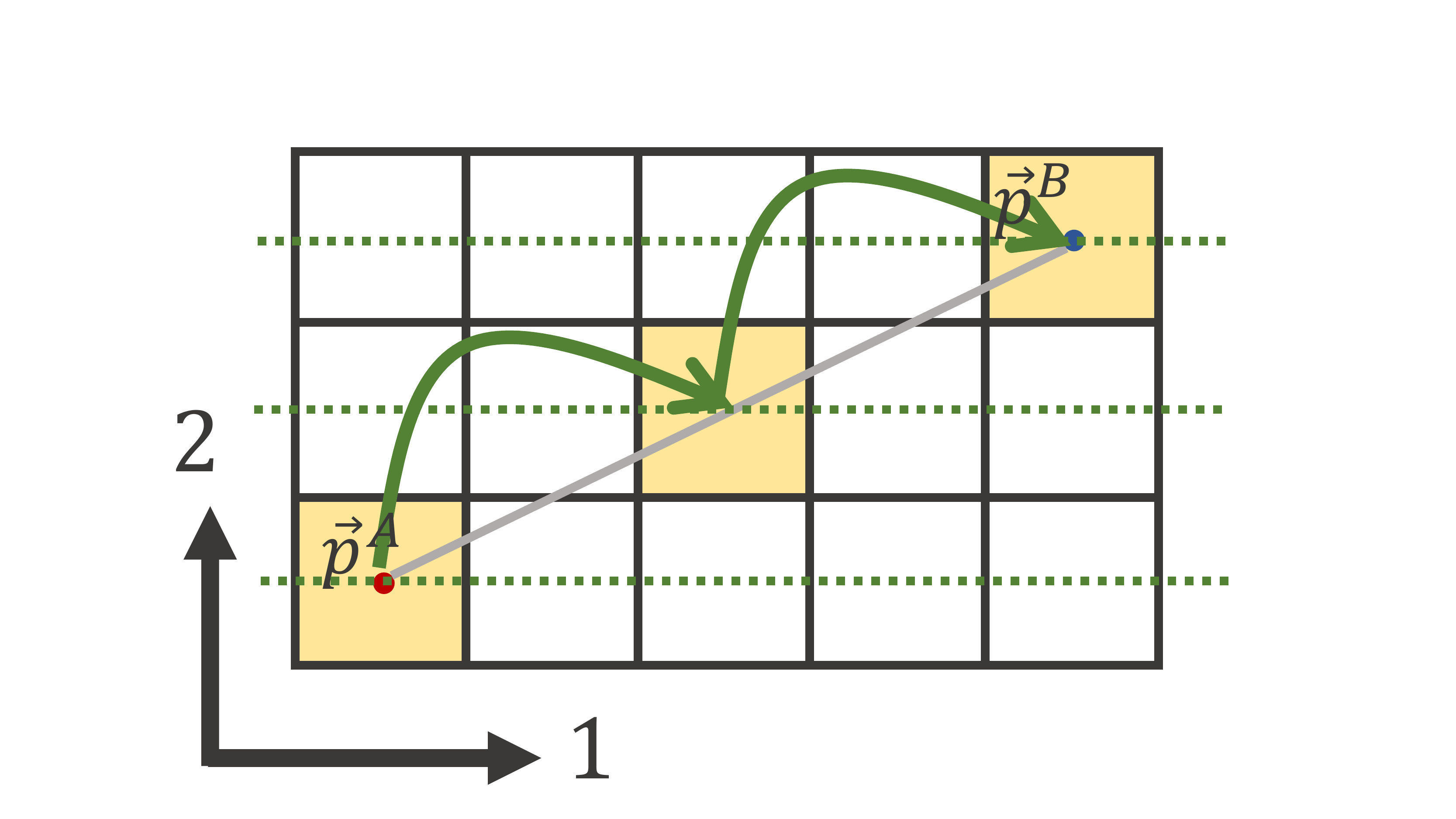}
    \caption{
    (Left) One-by-one step in the $d_{\mathrm{max}}$ direction (green arrows).
    Cells that can be checked with the algorithm are highlighted with yellow.
    (Right) One-by-one step in the $d\neq d_{\mathrm{max}}$ direction. 
    The dotted lines are grid lines in each direction.
    }
    \label{fig:Lattice23}
\end{figure}

So far, there is no conditions for 
$\vec{p}^A$ and $\vec{p}^B$,
but it is convenient to take the tracing direction always from small to large cell numbers.
Therefore, I define the starting point 
$\vec{S}$ as
\begin{align}
\vec{S} = \left\{
    \begin{aligned}
         \  & \vec{p}^A \ (p_{d_{\mathrm{max}}}^A <p_{d_{\mathrm{max}}}^B)  \\
            & \vec{p}^B \ (p_{d_{\mathrm{max}}}^B <p_{d_{\mathrm{max}}}^A).
    \end{aligned}
    \right. .
    \label{eq:Svec}
\end{align}
With these variable determined above, we consider one-by-one steps in $d_{\mathrm{max}}$ direction shown in the left of Fig.~\ref{fig:Lattice23}.
The green arrow is the step with a width of $1$ in $d=1$ which corresponds to $d_{\mathrm{max}}$ in this case.
The right figure of Fig.~\ref{fig:Lattice23} shows the one-by-one step in the $d=2$ direction which is not $d_{\mathrm{max}}$.
The comparison between the left and right figures would clearly explain the reason 
I take the $d_{\mathrm{max}}$ direction to see information of each cell.
For the later explanation,
I define the step with the width of $1$ in the $d_{\mathrm{max}}$ as $i_{\mathrm{cell}} = 0, 1, 2, 3 \cdots$,
{\it{i.e.,}}
then the position of the $d_{\mathrm{max}}$ direction becomes $p^A_1 + i_{\mathrm{cell}}$ after $i_{\mathrm{cell}}$ steps.

\subsubsection{Coordinates at each step}
\begin{figure}[htpb]
    \centering
    \includegraphics[bb=0 0 960 540, width=0.8\textwidth]{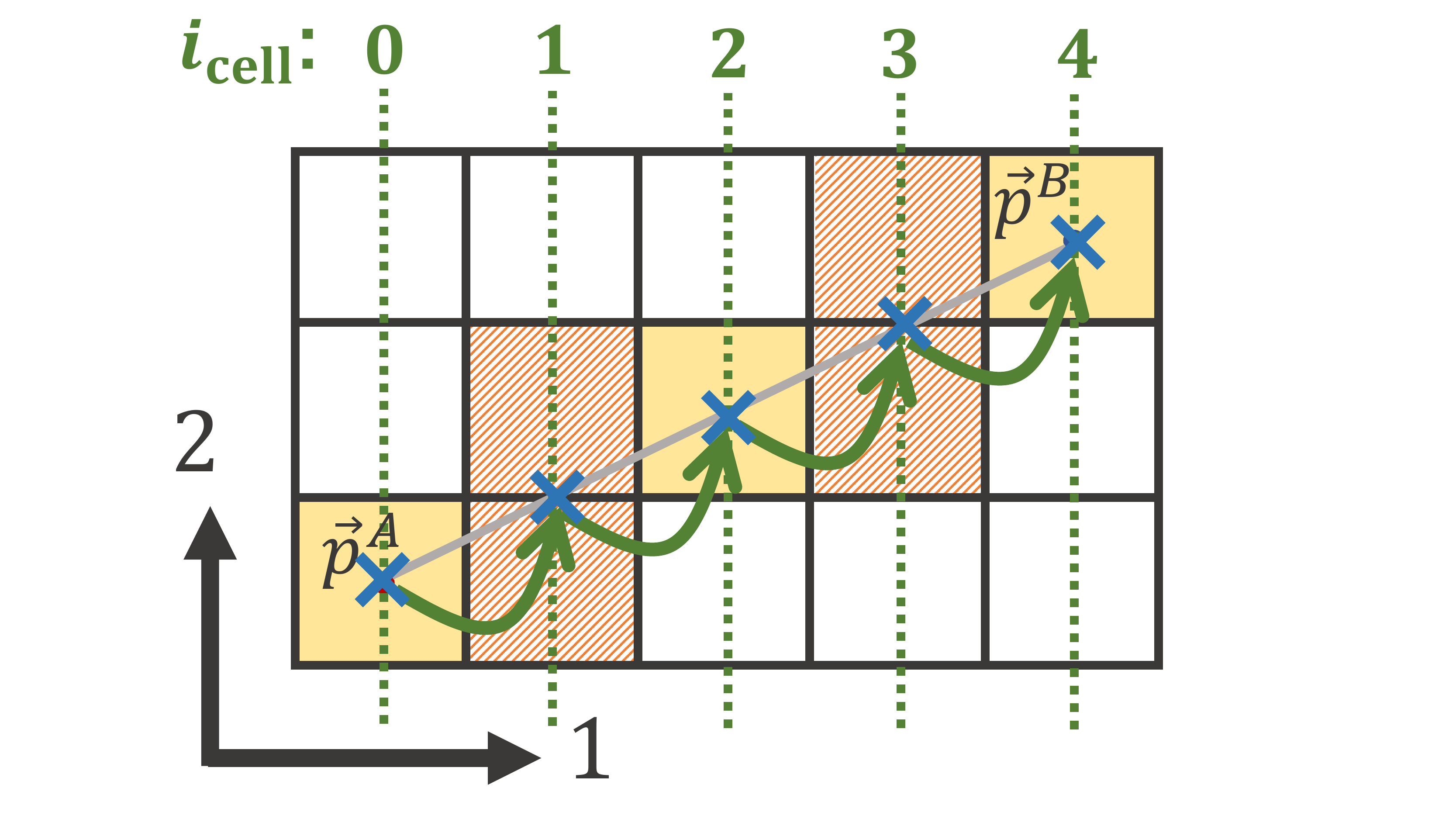}
    \caption{
    Each point of $\vec{p}^{\mathrm{now}}$ that is accessible with the one-by-one step (blue crosses).
    The crosses located at boundaries are assumed to span two cells sharing the boundary (filled with orange diagonal stripe pattern).
    }
    \label{fig:Lattice4}
\end{figure}

Now I need to obtain values of the other dimension on a color string.
In general expression, we can obtain the position after $i_{\mathrm{cell}}$ steps $\vec{p}^{\mathrm{now}}$  as
\begin{align}
\label{eq:pnow}
        \vec{p}^{\mathrm{now}}  = \vec{S} + \frac{\vec{r}}{r_{d_{\mathrm{max}}}} (i_{\mathrm{cell}}-S_{d_{\mathrm{max}}} ).
\end{align}
This gives one the full coordinates of $\vec{p}^{\mathrm{now}}$ that is accessible with the one-by-one step illustrated in Fig.~\ref{fig:Lattice4}.
Basically, I trace a color string and find crossing points between hypersurfaces and color strings
by checking information at each $\vec{p}^{\mathrm{now}}$.
However, as one sees in Fig,~\ref{fig:Lattice4},
there is another issue to consider:
some positions that is accessed with the one-by-one step are located at boundaries.
In order not to miss the crossing points, 
I check both cells that is sharing the boundary.

To realize that,
I need to judge where coordinates are located in a cell and which cell one should check in string tracing at every step with the following procedure.
There are three cases for steps:
1) a step is at the center of a cell, 
2) a step is on the boundary of a cell, and
3) otherwise.
Coordinates are defined to take $(k_1, l_2, m_3)$ at
the center of a cell 
located at the $k_1$-th in direction 1, 
$l_2$-th in direction 2, 
and $m_3$-th in direction 3.
The case 1) is satisfied when a value of $\vec{p}^{\mathrm{now}}$ in a direction have zero decimal part.
In this case, the coordinate exactly corresponds to the cell numbering.
The case 2) is satisfied when 
a twice value of $\vec{p}^{\mathrm{now}}$ in a direction have zero decimal part.
The cell numbering to check is obtained by rounding and flooring the value.
For the case 3), 
one can obtain the cell number to check by rounding the value.

More explicitly, I explain the case of Fig.~\ref{fig:Lattice4}.
In the direction of $d=1$, there is no ambiguity to determine cells to check since the one-by-one step is set to $d=1$.
Thus, all the coordinate of $p^{\mathrm{now}}_1$ obtained in Eq.~\refbra{eq:pnow}
has a one-to-one correspondence with cell numbering.
For the direction of $d=2$, the $\vec{p}^{\mathrm{now}}$ at $i_{\mathrm{cell}}=0, 2, 4$ correspond to the case 1),
while that of $i_{\mathrm{cell}}=1, 3$ correspond to the case 2).
This is because in the latter case, $p^{\mathrm{now}}_2(i_{\mathrm{cell}}=1) = 1.5$ and $p^{\mathrm{now}}_2(i_{\mathrm{cell}}=3) = 2.5$.
Here, one needs to check the cells located at $p^{\mathrm{now}}_2=1$ and $p^{\mathrm{now}}_2=2$ at $i_{\mathrm{cell}}=1$,
and $p^{\mathrm{now}}_2=2$ and $p^{\mathrm{now}}_2=3$ at $i_{\mathrm{cell}}=3$ by rounding and flooring the $p^{\mathrm{now}}_2$ on a boundary.

\subsubsection{Program flow}

I show the program flow of the string tracing that I have explained so far in List.~\ref{list:stringtracing}.
Note that this just illustrates the flow of program, not the program itself.
For a simplicity and generality, I omit accurate notation of each variables and formats of programming language
\footnote{
Since the actual program is written in C++, some notations are exactly expressed in C++.
}
.
I would like readers of this thesis to extract the main idea of the algorithm if one needs.

Line 8 and 9 correspond to Eqs.~\refbra{eq:rAB} and \refbra{eq:dmax}.
From line 11 to 14, Eq.~\refbra{eq:Svec} is obtained.
From line 18, the loop of one-by-one step begins.
At line 22, Eq.~\refbra{eq:pnow} is obtained.

\begin{lstlisting}[language=C++, caption=Program flow of the string tracing]

//...several definitions of each function are above.

StringTracting(){
  
  //1. Determination of the direction to step
  //=========================================
  r_vec = (p_Avec - p_Bvec)/absolute(p_Avec - p_Bvec)
  r_dmax = max(r_vec[0], r_vec[1], r_vec[2])

  if(p_Admax<p_Bdmax)
       S_vec =  p_Avec
  else 
       S_vec =  p_Bvec

  //One-by-one step
  //*+*+*+*+*+*+*+*+*+*+*+*+*+*+*+*+*+*+
  for(i_cell<i_cellMax){
  
    //2. Coordinates at each step
    //================================
    p_now = S_vec + (r_vec/r_dmax)*(i_cell - S_vec[dmax])
    
    
    //Check location of rho_now for each direction,
    //and find cell numbering to check
    for(i_dir<3){
    
        if (case 1)
           cell_numbering[i_dir].push_back(p_now[i_dir])
        else if (case 2)
           cell_numbering[i_dir].push_back(floor(p_now[i_dir]))
           cell_numbering[i_dir].push_back(round(p_now[i_dir]))
        else //case 3
           cell_numbering[i_dir].push_back(round(p_now[i_dir]))
    }//i_dir
    
    //Check if there is crossing points between hypersurface and color strings.
    //===============================
    for(k<cell_numbering[0].size()){
     for(l<cell_numbering[1].size()){
      for(m<cell_numbering[2].size()){
        check_temperature(k,l,m)
        if (crossing point)
             //Archive the information!
             CrossingPoint.append(info(k,l,m))
    }//m
     }//l
      }//k
      
      
  }//i_cell
  //*+*+*+*+*+*+*+*+*+*+*+*+*+*+*+*+*+*+

}



\end{lstlisting}
\label{list:stringtracing}

\section{Cutting and pairing}
With the information obtained at crossing points between hypersurface and color strings with the string tracing method explained above,
I modify configuration of color strings as explained in Sec.~\ref{subsec:STRING_CUTTING} in chapter 2.

In order to explain the algorithm to perform the color string modification, it should be instructive to first explain 
how \pythia \ handles colors (and anti-colors).
In \pythia, which partons are connected and confined as a color-singlet string is determined by assigning color and anti-color tags.
These tags are just random positive integers, not colors of ($R$, $G$, $B$) with $N_\mathrm{c}=3$.
Thus, a color and an anti-color tag having the same value can form a color singlet.
For instance, a quark having color and anti-color ($101$, $0$) and 
an anti-quark with ($0$, $101$) form a color singlet string.
This treatment is valid under the {\it{leading color limit}} ($N_\mathrm{c} \rightarrow \infty$) \cite{HOOFT1974461}.
In \pythia, all perturbative processes are handled under this limit \cite{Bierlich:2022pfr}, and this enables us to deal with color string modification with relative ease.

\subsubsection{String cutting algorithm}
Let me consider a simple case: one string that consists of a quark and anti-quark.
\begin{figure}[htpb]
    \centering
    \includegraphics[bb = 0 0 1280 960, width = 0.48\textwidth]{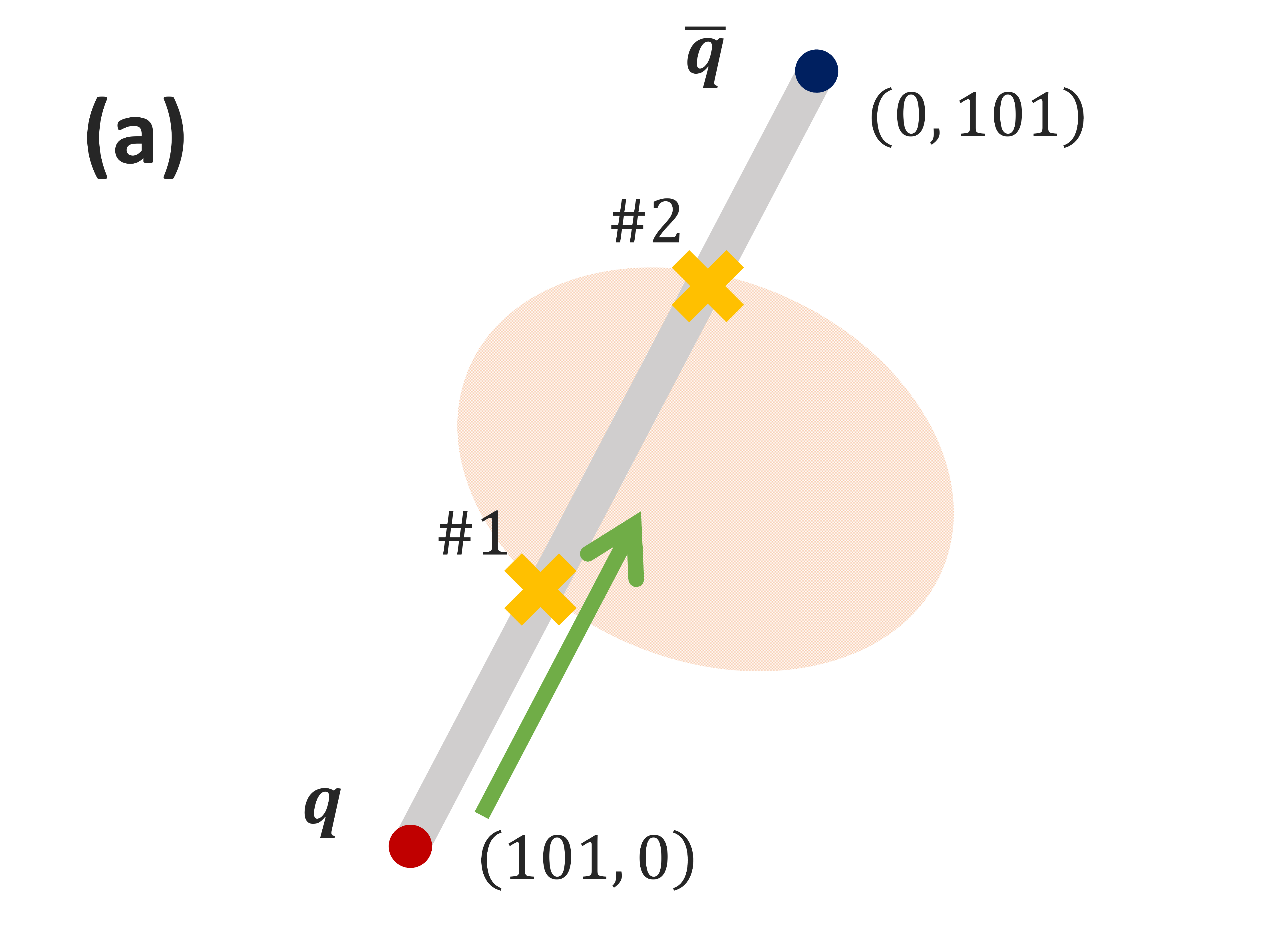}
    \includegraphics[bb = 0 0 1280 960, width = 0.48\textwidth]{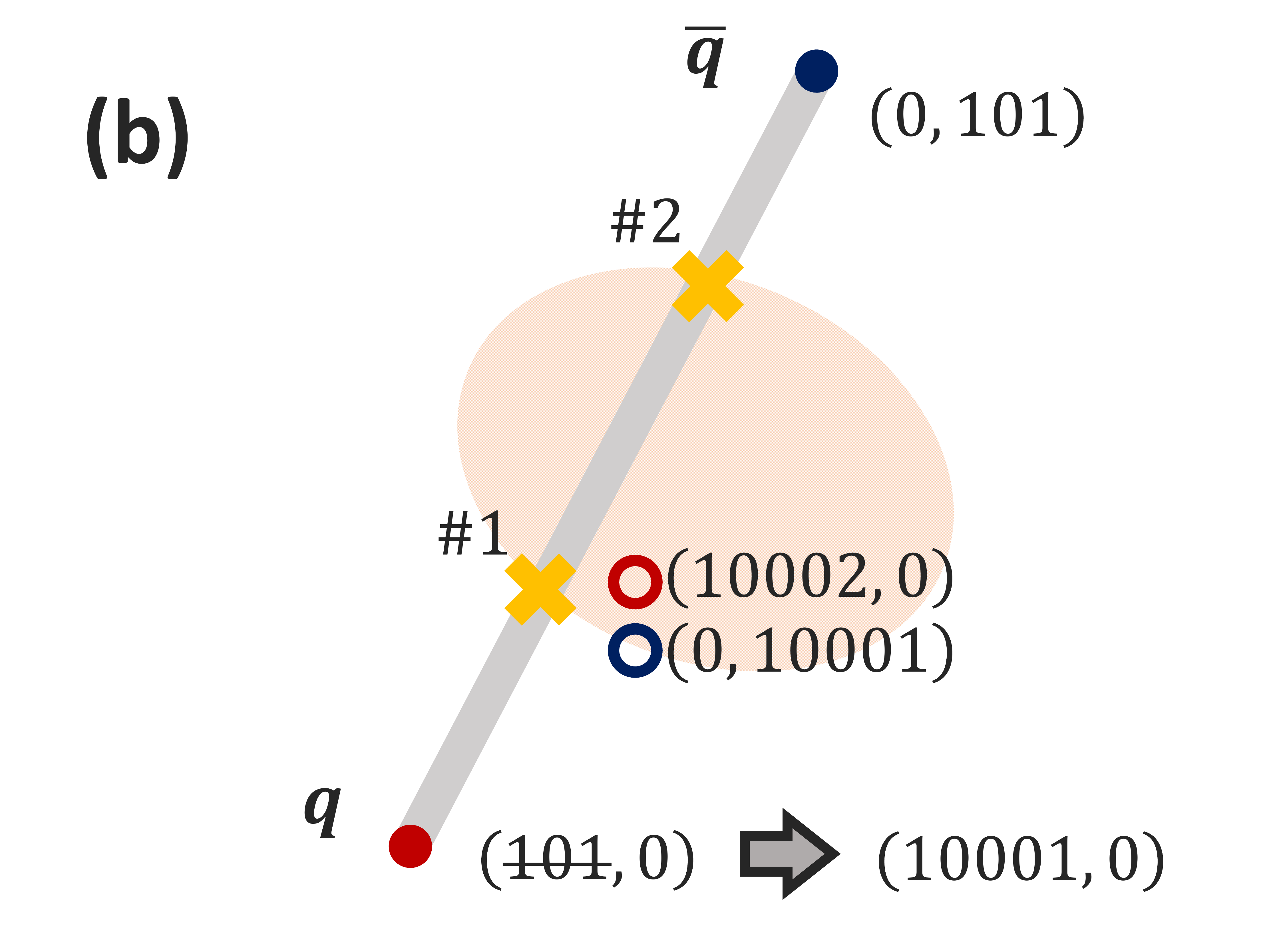}
    \includegraphics[bb = 0 0 1280 960, width = 0.48\textwidth]{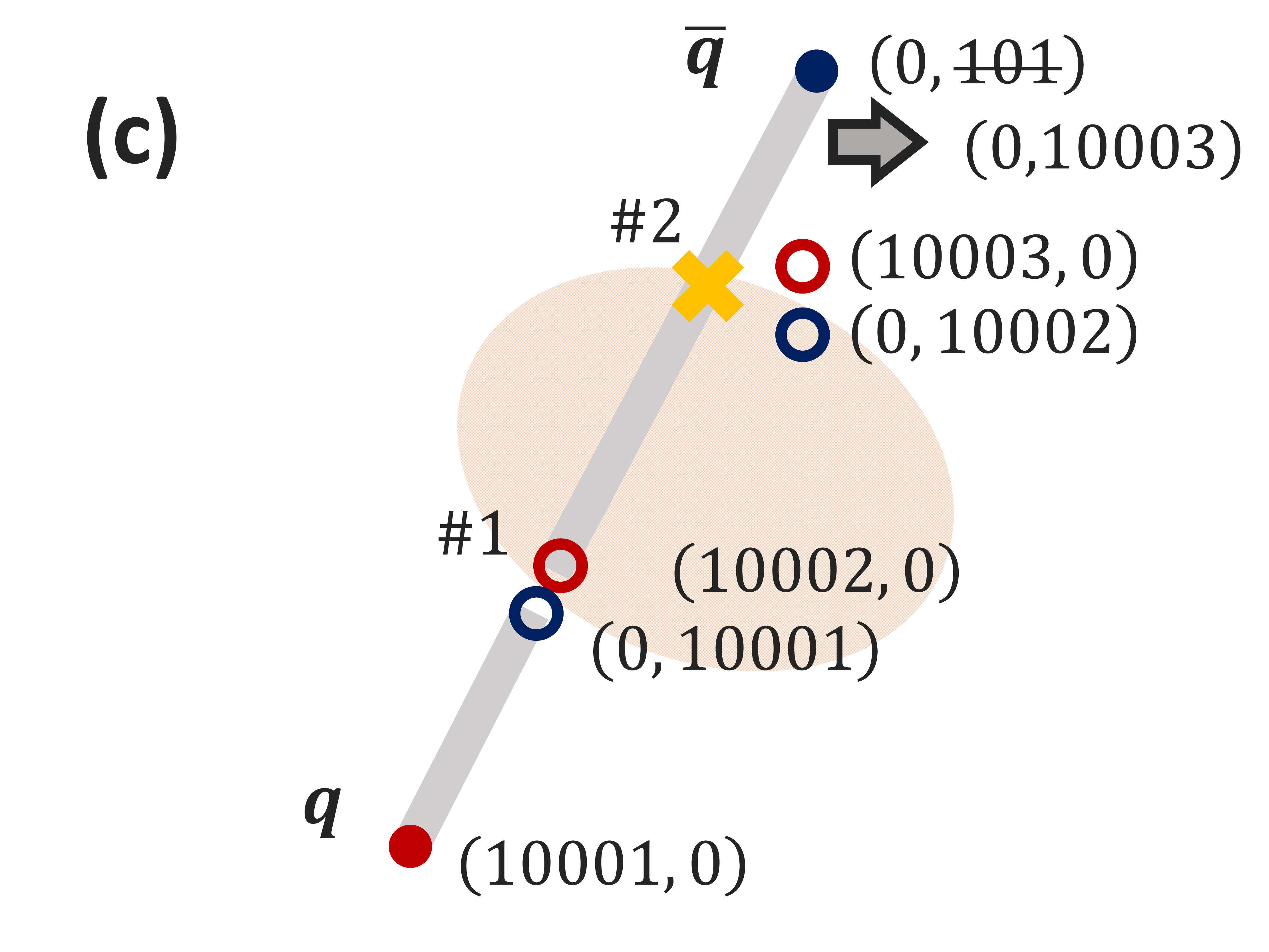}
    \includegraphics[bb = 0 0 1280 960, width = 0.48\textwidth]{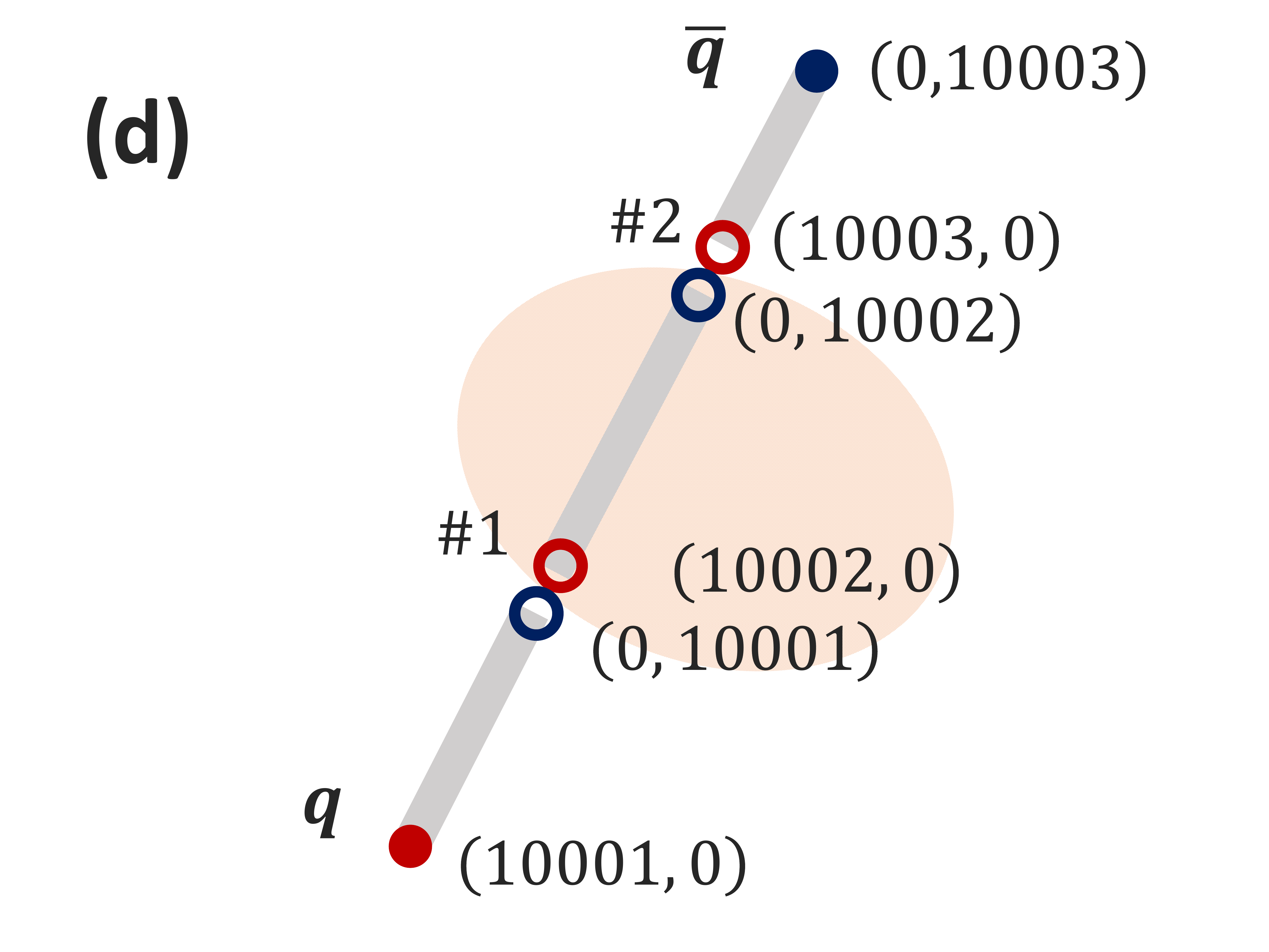}
    \caption{An example of the string cutting algorithm. A quark (red) and anti-quark (blue) is forming a string. A matter profile within hypersurface is illustrated with pale orange oval. Two crossing points (\#1 and \#2) between the string and the surface are expressed with yellow crosses.The process of the algorithm is shown from (a), (b), (c), to (d). The open red and blue circles are quarks and anti-quarks picked up from the crossing points.}
    \label{fig:StringCutting}
\end{figure}
Figure \ref{fig:StringCutting} shows the flow of the string cutting algorithm performed at $\tau = \tau_s$ with a simple example.
The process starts from (a) in Fig.~\ref{fig:StringCutting}.
With the string tracing algorithm explained in Sec.~\ref{sec:AppendixStringTracing} in this Appendix B,
one can find crossing points between the hypersurface and the string as shown with the yellow crosses
by tracing the string in the direction of the green arrow.
One can also obtain the information of the fluid cell of the hypersurface at the crossing point such as thermodynamic variables and flow velocity.

Figure \ref{fig:StringCutting} (b) shows how to handle the string cutting at the crossing point \#1.
Since which partons are confined and form which color strings are determined with color and anti-color tags,
one can ``cut'' a string by modifying the tags.
To cut the string at the crossing point \#1, 
I add a quark and an anti-quark pair sampled from the hypersurface at the crossing point.
Here, I assign tags ($10002$,$0$) and ($0$,$10001$) for the sampled quark and anti-quark, respectively
\footnote{
For the color and anti-color tags of sampled partons, in order to
avoid assigning the same value as the originally exist one, 
large integers ($\leq10^8$) are used in the actual simulations.
}.
To let the program know that the leading quark ($101$, $0$) and the sampled anti-quark ($0$,$10001$) form a string,
the original color tag assigned to the leading quark is changed to $10001$.

Figure \ref{fig:StringCutting} (c) shows the string cutting at \#2.
Here, color and anti-color tags ($10003$,$0$) and ($0$, $10002$) are assigned to the newly sampled parton pair.
Note that the same value for color and anti-color cannot be used more than once.
The sampled anti-quark has the same anti-color value as the color tag assigned to the quark at \#1.
Since this is the last crossing point appearing on this string,
the quark sampled at \#2 and the original leading anti-quark with ($0$, $101$) should form a string.
This is done by replacing the anti-color tag of the original leading anti-quark with the color tag of the quark sampled at \#2.

Finally Fig.~\ref{fig:StringCutting} (d) shows the color flow obtained just after the modification.
The middle piece of the string is discarded by identifying a string that consists only of a quark and an anti-quark assuming that such strings should be inside of the hypersurface.

\begin{figure}
    \centering
    \includegraphics[bb = 0 0 574 558, width=0.49\textwidth]{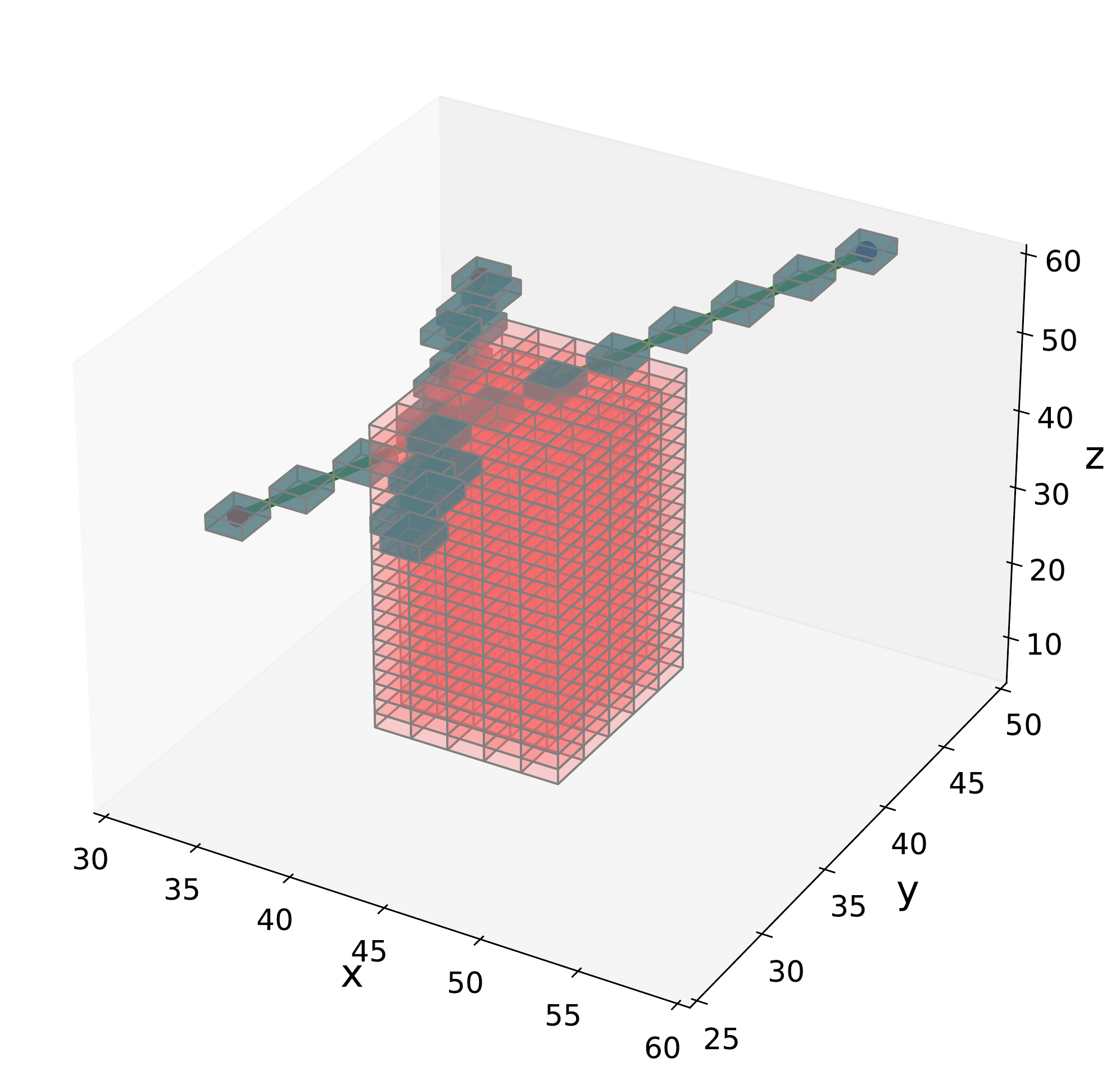}
    \includegraphics[bb = 0 0 574 558, width=0.49\textwidth]{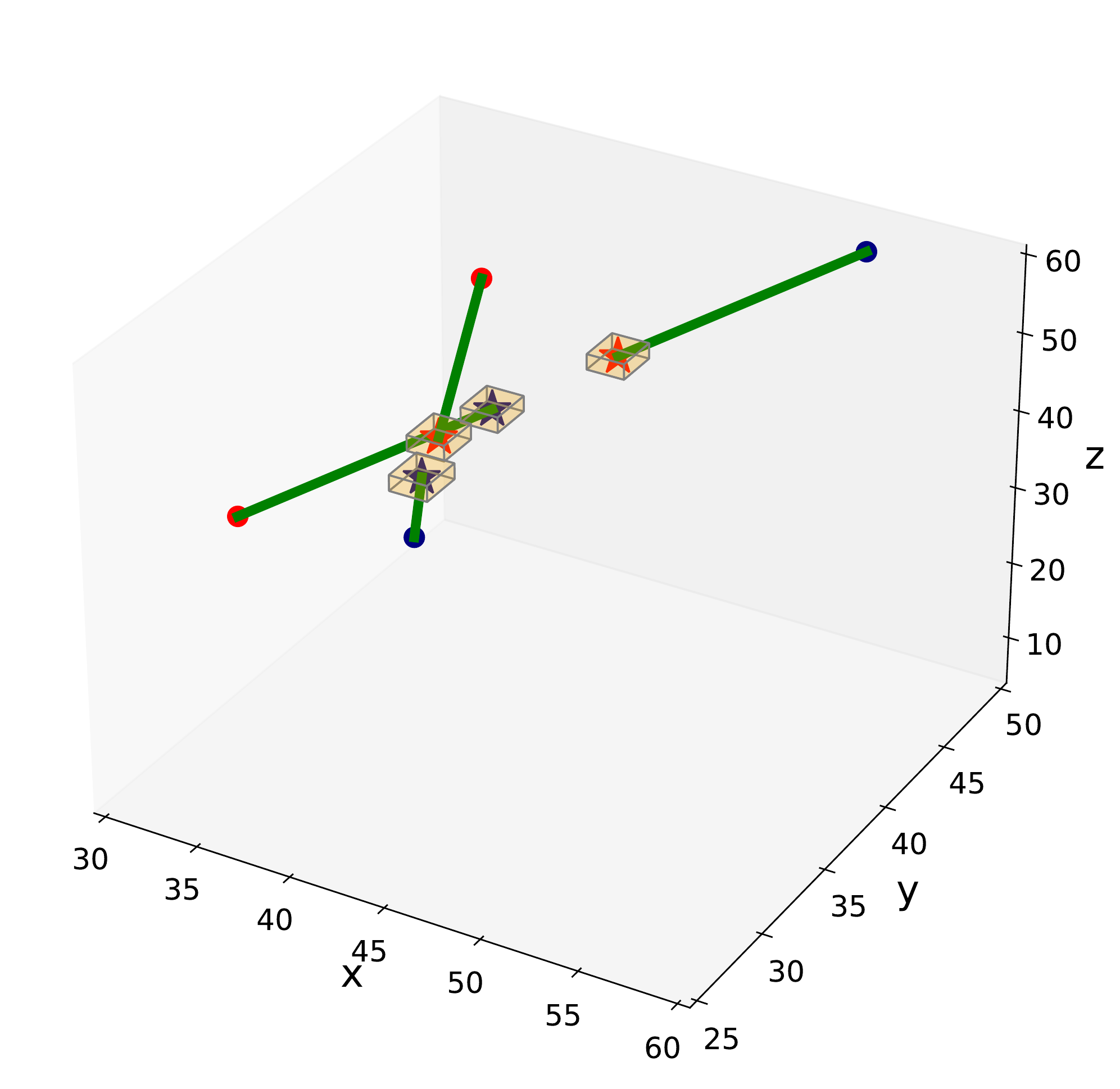}
    \caption{Example of numerical treatment of string cutting. Left: Two strings (blue blocks) are penetrating thermal medium having $T>T_{\mathrm{sw}}$ (pink blocks). Right: Modified strings (green lines) due to string cutting adopted in the left hand case. Thermal partons (stars) are picked up from crossing points between hypersurface and color strings (orange blocks). Note that these are not obtained from an actual simulation. The axes $x, y,$, and $z$ are numbering of each hydro cell in $x, y$, and $\eta_s$ direction.}
    \label{fig:STRINGCUT}
\end{figure}

Figure \ref{fig:STRINGCUT} shows 
an example of numerical treatment of the string cutting just explained above. Left figure shows the case that two strings are penetrating thermal medium having $T>T_{\mathrm{sw}}$. On the other hand, the right figure shows the modified strings due to string cutting adopted in the left hand case.
Four crossing points between hypersurface and color strings are found in this case, and 
four thermal partons are picked up from the crossing points,
which cuts the original two strings into four strings.

\subsubsection{Parton pairing algorithm}
\begin{figure}[htpb]
    \centering
    \includegraphics[bb= 0 0 1707 960, width = 0.8\textwidth]{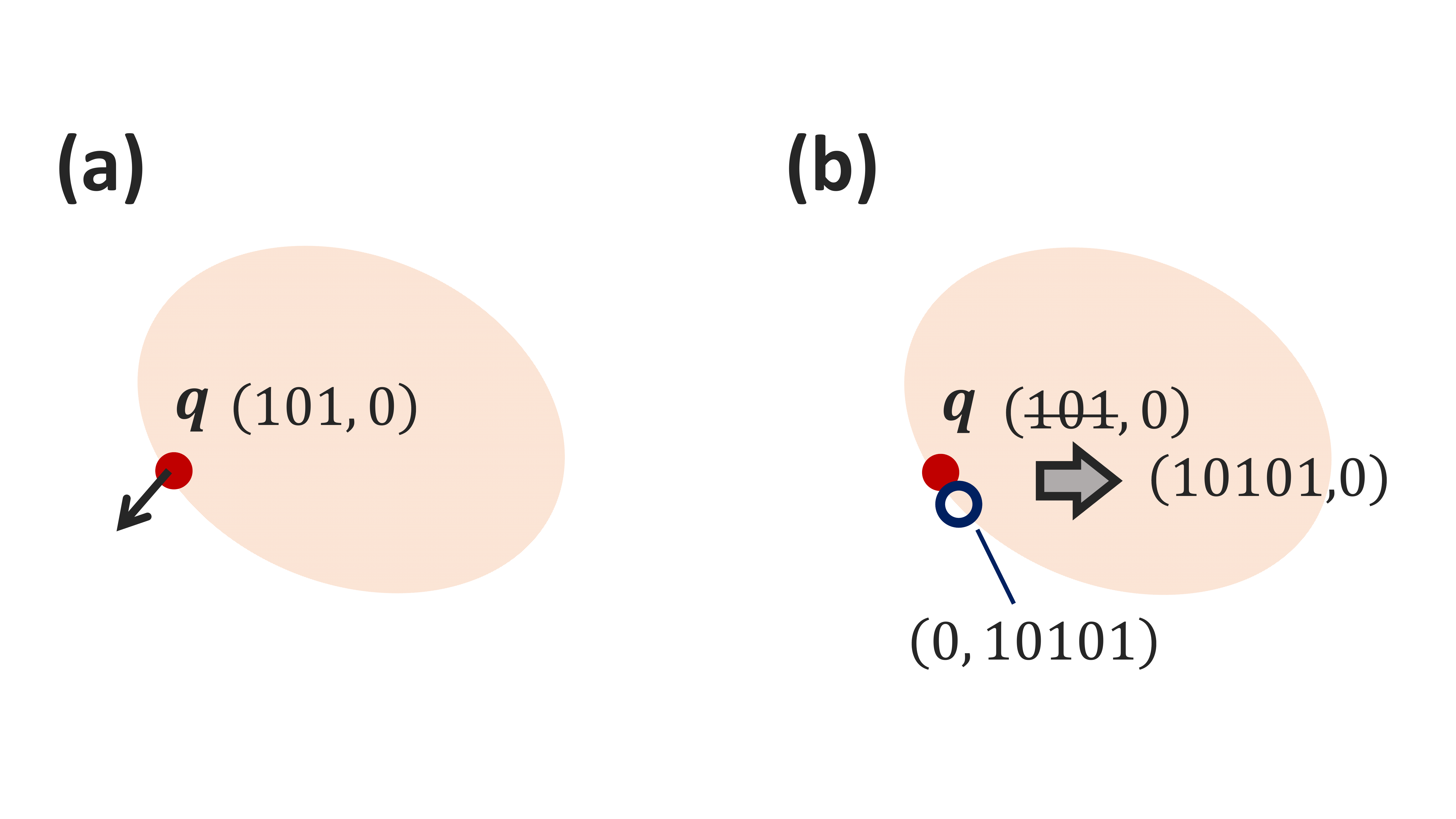}
    \caption{The parton pairing algorithm when a quark comes out from a hypersurface.
    (a) A quark with color and anti-color tag ($101$,$0$) is about to come out from the hypersurface. (b) An anti-quark with ($0$,$10101$) is sampled at the hypersurface, and the color tag of the quark is changed to $10101$.}
    \label{fig:PartonPair}
\end{figure}

The parton pairing algorithm is much simpler compared to the string cutting explained above
because one does not need to trace strings any more.
It is assumed that this kind of partons coming out from hypersurface with finite energy would form a string by picking up a single thermal parton in vacuum and are hadronized via string fragmentation.
This is realized as follows.
In Fig.~\ref{fig:PartonPair} (a),
a quark that comes out from a hypersurface and has a color and anti-color tag ($101$, $0$) is shown.
In order to make the string color singlet,
an anti-quark is picked up from the hypersurface in Fig.~\ref{fig:PartonPair} (b).
To avoid using the same value for color and anti-color tags,
new value is assigned.
The sampled anti-quark has ($0$, $10101$),
and the color tag of the quark that is about to come out from the hypersurface is changed from $101$ to $10101$.

\begin{figure}[htpb]
    \centering
    \includegraphics[bb= 0 0 1707 960, width = 0.8\textwidth]{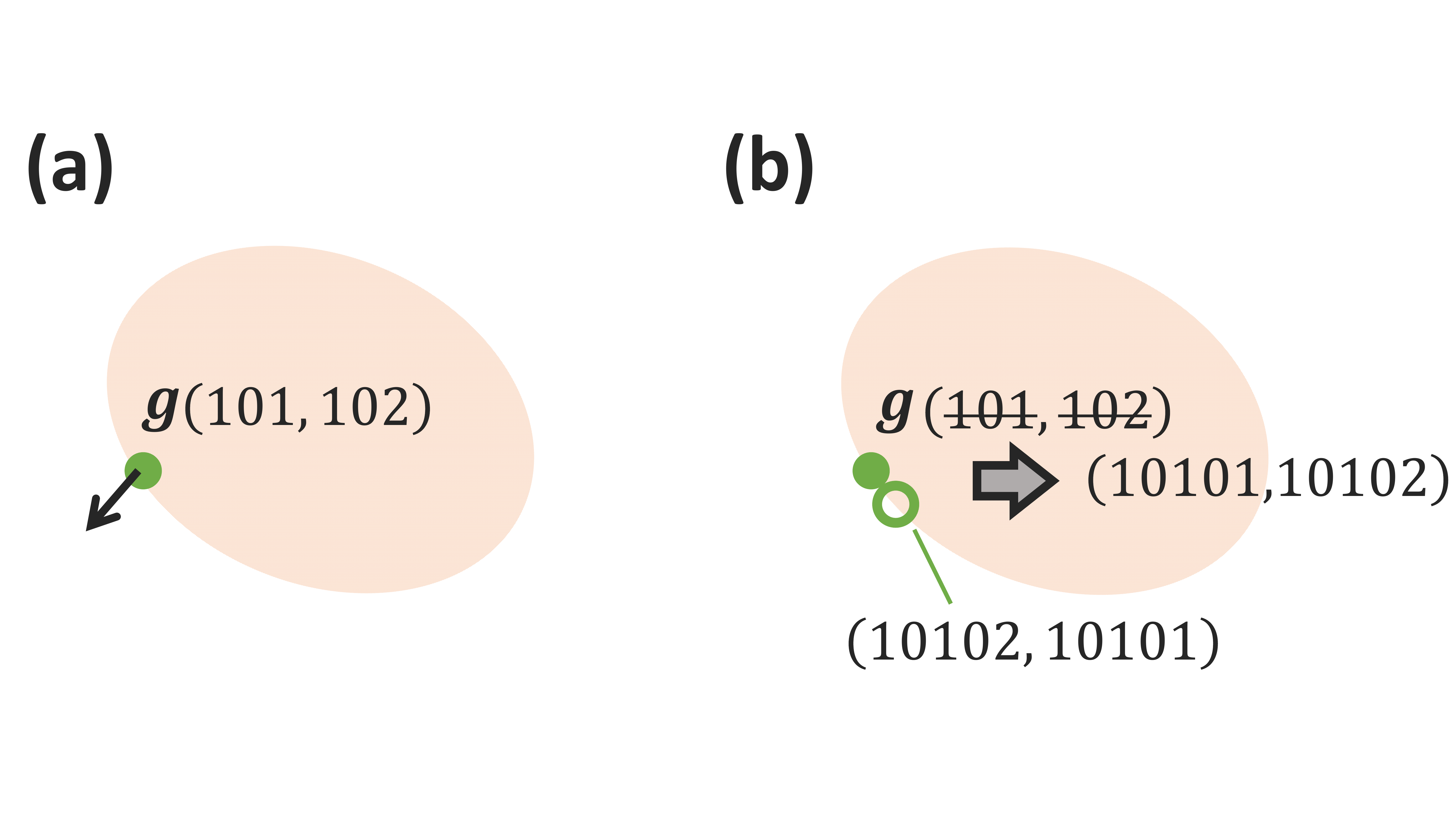}
    \caption{The parton pairing algorithm when a gluon comes out from a hypersurface.
    (a) A gluon with color and anti-color tag ($101$,$102$) is about to come out from the hypersurface. (b) A gluon with ($10101$,$10102$) is sampled at the hypersurface, and the color and anti-color tag of the gluon that was about to come out from the hypersurface is changed to ($10102$, $10101$).}
    \label{fig:PartonPair2}
\end{figure}

For the case that an anti-quark comes out from a hypersurface, 
the anti-quark picked up a quark from the hypersurface likewise the above case.
The case that a gluon comes out from medium is shown in Fig.~\ref{fig:PartonPair2} (a).
The color and anti-color tags of the gluon is ($101$, $102$).
For the case of gluons, it is assumed that
a gluon picks up another gluon from a hypersurface and form a color singlet string.
In Fig.~\ref{fig:PartonPair2} (b),
the gluon picked up from the hypersurface has color and anti-color tags, ($10102$, $10101$).
Here again, the color and anti-color tags of the gluon that comes out from the hypersurface are changed from ($101$, $102$) to ($10101$,$10102$), which form a color singlet.


\vspace{30pt}
Note that the examples shown so far are very simple cases.
In the actual simulations, for instance, 
junctions or gluon loops are produced.
The former is a Y-shaped color singlet object that end points of three strings are converged, which can look like a Y shape.
The latter is a color singlet object which only consists of gluons.
To make it color singlet, gluons form a loop.
In these cases, string modifications are performed in different ways from ones explained above because of the different configurations of the color flows.

%\chapter{Peacewise Parabolic Method}
%\input{chapters/appendixC}
%\input{chapters/appendix_numHydro}

%
%
%\chapter{Data analysis}
%\input{chapters/appendixD}

%\bibliographystyle{abbrv}
\bibliographystyle{ieeetr}
\bibliography{inspire.bib}

\end{document}